\documentclass[preprint2]{aastex}

\usepackage{epsf}
\usepackage{graphics}
\usepackage{color}

\def\eps@scaling{0.6}%

\newcommand\plotthree[3]{{%
 \typeout{Plottwo included the files #1 #2 #3}
 \centering
 \leavevmode
 \columnwidth=.30\columnwidth
 \includegraphics[width={\eps@scaling\columnwidth}]{#1}%
 \hfil
 \includegraphics[width={\eps@scaling\columnwidth}]{#2}%
 \hfil
 \includegraphics[width={\eps@scaling\columnwidth}]{#3}%
}}%

\newcommand{\be}{\begin{equation}}
\newcommand{\ee}{\end{equation}}

\newcommand{\ax}{$\alpha_{\rm X}$}
\newcommand{\lx}{$L_{\rm X}$}

\newcommand{\lbbb}{$L_{\rm BBB}$}
\newcommand{\auv}{$\alpha_{\rm UV}$}

\newcommand{\aox}{$\alpha_{\rm ox}$}

\newcommand{\rb}[1]{\raisebox{1.5ex}[-1.5ex]{#1}}
\newcommand{\msun}{$M_{\odot}$}

\newcommand{\plm}{$\pm$}

\newcommand{\swift}{{\it Swift}}
\newcommand{\xmm}{{\it XMM-Newton}}
\newcommand{\chandra}{{\it Chandra}}

\newcommand{\lledd}{$L/L_{\rm Edd}$}

\shorttitle{AGN Spectral Energy Distribution}
\shortauthors{Grupe et al.}

\begin{document}


\title{The Simultaneous Optical-to-X-ray Spectral Energy Distribution of Soft X-ray Selected AGN observed by \swift
}


\author{Dirk Grupe\altaffilmark{1},
\email{grupe@astro.psu.edu}
S. Komossa\altaffilmark{2},
Karen M. Leighly\altaffilmark{3},
\& Kim L. Page\altaffilmark{4}
}

\altaffiltext{1}{Department of Astronomy and Astrophysics, Pennsylvania State
University, 525 Davey Lab, University Park, PA 16802} 

\altaffiltext{2}{Max-Planck-Institut f\"ur extraterrestrische Physik, Giessenbachstr., D-85748 Garching,
Germany; email: skomossa@mpe.mpg.de}

\altaffiltext{3}{Homer L. Dodge Department of Physics and Astronomy, University of Oklahoma, 
440 West Brooks Street, Norman, OK 73019; email: leighly@nhn.ou.edu}

\altaffiltext{4}{Department of Astronomy, University of Leicester, Leicester, U.K.}



\begin{abstract}
We report \swift\ observations of a sample of 92 bright soft X-ray
selected active galactic nuclei (AGN). This sample represents the largest number of AGN
observed to study the spectral energy distribution (SED) of AGN with
simultaneous optical/UV and X-ray data. The principal motivation of
this study is to understand  the SEDs of AGN in the optical/UV to
X-ray regime  and to provide bolometric corrections  which are
important in determining the Eddington ratio \lledd. In particular, we
rigorously explore the dependence  of the UV-EUV contribution to the
bolometric correction on  the assumed EUV spectral shape. We find
strong correlations of the spectral slopes \ax\ and \auv\ with \lledd.  Although
Narrow-Line Seyfert 1 galaxies (NLS1s) have steeper \ax\ and higher
\lledd\ than Broad-Line Seyfert 1 galaxies (BLS1s), their optical/UV
to X-ray spectral slopes \aox\  and optical/UV slopes \auv\ are very
similar. The mean SED of NLS1s shows that in general this type of AGN
appears to be fainter in the UV and at hard X-ray energies than BLS1s. 
We find a strong correlation between \ax\ and \auv\ for AGN with X-ray
spectral slopes \ax$<$1.6.   For AGN with steeper X-ray spectra, both
this relation and the relation between \ax\ and \lledd\ break down.
At \ax$\approx$1.6,  \lledd\ reaches unity. We note an offset in the \auv -
\lledd\ relation between NLS1s and BLS1s.  We argue that \auv\ is a
good estimator of \lledd\ and suggest that \auv\ can be used to
estimate \lledd\ in high-redshift QSOs. Although NLS1s appear to be
highly variable in X-rays they only vary marginally in the UV.  
\end{abstract}


\section{Introduction}

Two of the key parameters of Active Galactic Nuclei (AGN) 
are their bolometric luminosity $L$ and their
Eddington ratio \lledd. Besides the mass of the central black hole,
\lledd\ is the parameter that is likely to control a variety of
observed AGN properties such as spectral slopes and some emission-line
properties  \citep[e.g.][]{bor92, sulentic00, boroson02, grupe04,
shemmer08}.  
The Eddington ratio \lledd\ may also be seen as
an indicator of the evolutionary stage of an AGN.  In this picture,
Narrow Line Seyfert 1 galaxies (NLS1s) with their high \lledd~ are
young members of the AGN family \citep{gru99, mat00, grupe04}.
The knowledge of the shape of the Spectral Energy
Distribution (SED) of an AGN is critical for measuring the bolometric
luminosity and \lledd. 

In this paper we study the SEDs of a sample of 92 AGN with {\em simultaneous} 
optical/UV and X-ray observations obtained by \swift. 
The main goal of our study is to estimate the UV-EUV contribution to the
SED based on these simultaneous multi-wavelength data. 
Another goal is to look for relations between properties of 
the SED such as the optical/UV and X-ray spectral slopes \auv\ and \ax\
and other observed properties, and determine how they are related to
the Eddington ratio \lledd.  
The aim is to find properties such as e.g.\ \auv\ that can be
used to estimate \lledd\ in a similar way as it has already been done
for \ax\ and \lledd\ \citep{grupe04, shemmer08}. If there is such a
relation, it could be applied to high-redshift quasars where the
soft X-ray band is shifted out of the observing window, and which are
typically X-ray faint making a robust X-ray spectral analysis
challenging or impossible.  Bolometric corrections, and their
dependence on AGN type, luminosity, Eddington ratio and redshift are
also a key ingredient in determining black hole mass functions from
(single-band) luminosity functions \citep[e.g.,][]{hopkins07}. Since
the bolometric  AGN luminosity is usually dominated by the UV to
soft-X-ray part of the SED, its careful  measurement, and an
evaluation of its uncertainties, in nearby, well-studied AGN is
important.

The presence of a soft X-ray excess over a flatter hard X-ray
component was first reported by \citet{arnaud85} in {\it EXOSAT}
spectra from the Seyfert 1 galaxy Mkn 841. It is commonly thought that
this soft X-ray excess is the high-energy tail of the Big-Blue-Bump
emission (BBB).  In Seyferts, this part of the SED is the most
energetic part of the entire electromagnetic spectrum.  A popular
interpretation of the BBB is thermal emission from the accretion disk
around the central black hole \citep[e.g.,][]{shields78}.  The thermal
UV photons which originate from the center of the accretion disk are
then thought to be modified by Comptonization by hot electrons in the
corona above the disk.  Some of the UV photons may end up in
the soft X-ray band \citep[e.g.,][]{czerny87,  ross92,
mannheim95, pounds95}.
 
However, as pointed out by \citet[e.g.,][]{gierlinski04} this simple
picture of the SED may be far more complicated. Given recent findings
of spectral complexity of AGN soft X-ray spectra, including the
presence of power law or black body-like soft excesses,
ionized reflection, ionized absorption, or partial
covering
\citep[e.g.,][]{fabian04,done07,gallo06,crummy06,grupe08,turner09}, it
is especially important to consider broad-band SEDs in efforts to
constrain the AGN continuum emission mechanisms. This is the approach
that we followed in our study.  Since the \swift\ X-ray telescope
spectra do not allow complex multi-component X-ray spectral fitting
due to the relatively small number of counts typically obtained during an observation,
we instead focus on power laws to assess the global spectral steepness
in the 0.3-10 keV band, and concentrate on the UV/X-ray
relationships.

A number of previous studies have focused on the measurements of SEDs, 
and the determination of bolometric corrections in different wave-bands. 
In order to determine the bolometric luminosity it is quite common 
to use a linear relation between an observed monochromatic luminosity
and  $L_{\rm bol}$. For example, \citet{elvis94} gave a relation with
$L_{\rm bol} = 5.6\times L_{2500\AA}$ and $13.2\times L_{\rm V}$. On
the other hand, \citet{marconi04} suggested that the conversion from a
monochromatic luminosity to $L_{\rm bol}$ is actually luminosity
dependent (see also Hopkins et al. 2007).  
 By using \xmm\ with simultaneous optical/UV and X-ray data,
\citet{vasudevan08} recently concluded that the bolometric correction
for the conversion between $L_{\rm 2-10 keV}$ is larger for high \lledd\
objects than for AGN which operate at lower \lledd.
\citet{walter93} studied the
UV and X-ray properties of a sample of 58 Seyfert 1 galaxies and found
that the soft X-ray excess found from ROSAT observations is
well-correlated with the strength of the BBB observed by IUE.  Based
on a sample of 76 bright soft X-ray selected {\it ROSAT} AGN,
\citet{gru98a} showed that the BBB extends as far as the optical band
and that sources with steeper X-ray spectra tend to have bluer optical
spectra, suggesting that Narrow Line Seyfert 1 galaxies (NLS1s) 
are the AGN with the strongest BBB component. However, from a
study of the IUE spectra of NLS1s, \citet{rodriguez97} came to the
conclusion that NLS1s have weaker UV emission than Broad Line Seyfert 
1s (BLS1s).  \citet{grupe04} found that there is a strong correlation
between the ROSAT X-ray spectral slope \ax\ and the Eddington ratio
$L/L_{\rm Edd}$.  This result was confirmed by \citet{williams04} 
who showed a correlation between the soft X-ray photon index
$\Gamma_{\rm X}$ and the ratio of the luminosity density  at 1 keV to
the Eddington luminosity. \citet{shemmer08} extended this relation
into the hard X-ray regime.  \citet{atlee09} recently presented a
sample of AGN based on our soft X-ray  selected AGN sample
\citep{gru98a,gru01} using {\it GALEX} Far-UV low-resolution spectra
and {\it ROSAT} All Sky Survey (RASS) data. They found that the
strength of the BBB  is correlated with the X-ray spectral slope, but
that there is no correlation with the shape of the UV continuum. 
They concluded that the properties of the accretion disk are
independent of the mass of the central black hole and the Eddington
ratio.  

 In the past, however, this type of SED studies were hampered by the lack of
simultaneous observations in the optical/UV and X-ray bands; the
observations available frequently had been performed years apart.
Since AGN are known to be variable in both the X-ray and the UV, the
lack of simultaneity adds considerable scatter to the data.  For a long time,
the only
sample study with simultaneous UV and X-ray observations was the one
performed by \citet{walter94}, who obtained simultaneous {\it IUE} and
{\it ROSAT} observations of a sample of 8 AGN and found no difference
to their results presented in \citet{walter93}.  This situation has
changed with the availability of the multi-wavelength observatories
XMM-Newton and \swift. Based on \xmm\ observations with the Optical
Monitor \citep[OM, ][]{mason01},  \citet{brocksopp06} presented a
multi-wavelength analysis of a sample of 23 Palomar-Green quasars and
correlated X-ray continuum with optical line properties. Recently,
\citet{vasudevan08} presented the bolometric corrections for 29 AGN
with simultaneous optical/UV and X-ray observations from the \xmm\ OM
and EPIC pn and \citet{vasudevan09} presented a sample of 26 AGN
selected from the 9 months \swift\ BAT survey \citep{markwardt05, winter09}
with \swift\ UVOT and XRT data.  In our paper, we present a sample of
92 bright soft X-ray selected AGN with simultaneous optical/UV and
X-ray data obtained with \swift. 

The \swift\ mission \citep{gehrels04} was launched on 2004 November
20. While its main purpose is to hunt and observe Gamma-Ray Bursts
(GRBs),  a significant part of \swift's observing time is used for fill-in
targets and targets-of-opportunity when no GRBs or guest investigator
targets are observed. Due to its multi-wavelength capacities and
its flexible scheduling, \swift\ is the ideal observatory  for 
multi-wavelength and/or monitoring observations of
AGN, as demonstrated by e.g \citet{grupe06} on the NLS1s RX
J0148.3--2758,  WPVS 007 \citep{grupe07, grupe08b}, PHL 1811 \citep{leighly07},
PG 1211+143 \citep{bachev09},
 and Mkn 335 \citep{grupe07b}. \swift\ is equipped
with three telescopes:  at the high energy end the Burst Alert
Telescope \citep[BAT, ][]{barthelmy05} operating in the 15-150 keV
energy range,   the X-Ray Telescope  \citep[XRT, ][]{burrows04}, which
covers the soft X-ray range between 0.3-10.0 keV, and at the long
wavelength end, the UV-optical Telescope \citep[UVOT,
][]{roming04}. The XRT uses a CCD detector identical to the EPIC MOS
on-board XMM \citep{tur01}.  The UVOT covers the range between
1700-6500\AA\ and  is a sister instrument  of \xmm's OM. The UVOT has
a similar set of filters to the OM \citep{mason01, roming04}. However,
the UVOT UV throughput is a factor of about 10 higher than that of the
OM. 

Although the BAT is performing a hard X-ray survey and has found
several hundreds of AGN so far \citep{markwardt05, tueller09,
  winter09, fabian09}, here we only use the \swift\ XRT and UVOT data
and the majority of the AGN in our sample are not detected (so far) in
the BAT survey.  We have started our project in
2005 and the data in this paper represent the status of our
study by the beginning of January 2010.  We focus on presenting the data 
and some simple statistical analyses. A more detailed statistical
analysis, a study of the relationships with the optical emission-line
properties, and a rigorous discussion of the implications for models
of NLS1 galaxies will be presented in a separate paper.  The outline
of the present paper is as follows: In \S\,\ref{observe} we describe
the sample selection, the \swift\ observations, and the data
reduction. In \S\,\ref{results} we present the results of the \swift\
XRT and UVOT data analysis, and these are then discussed in
\S\,\ref{discuss}.  Throughout this paper spectral indices are denoted
as energy spectral indices with $F_{\nu} \propto
\nu^{-\alpha}$. Luminosities are calculated assuming a $\Lambda$CDM
cosmology with $\Omega_{\rm M}$=0.27, $\Omega_{\Lambda}$=0.73 and a
Hubble constant of $H_0$=75 km s$^{-1}$ Mpc$^{-1}$. All errors are
1$\sigma$ unless stated otherwise.  

\section{\label{observe} Sample selection, observations and data reduction}

The AGN in the sample presented here were selected from the bright
soft X-ray selected AGN sample of \citet{gru01, gru04a}  which
contains a total of 110 AGN. 
This sample contains all Seyferts of the 
bright, soft X-ray sources by \citet{thomas98} which were selected from the 
{\it ROSAT} All-Sky Survey \citep[RASS;][]{voges99} which was using a
Position Sensitive Proportional Counter \citep[PSPC;][]{pfeffermann86}.
The X-ray sources in the \citet{thomas98} sample and consequently in the 
\citet{gru01} AGN sample had to be X-ray bright with a PSPC count rate $\ge$
0.5 counts s$^{-1}$, X-ray soft with a PSPC hardness ratio HR\footnote{The {\it ROSAT} PSPC hardness ratio was defined as
HR = (H-S)/(H+S) with S and H are the counts in the 0.1-0.4 keV and 0.5-2.4 keV energy ranges, respectively}$<$0.0, 
and a position at a high Galactic latitude with $|b|>20^{\circ}$.  
In addition to the AGN from the \citet{gru01} AGN sample, we added the NLS1s
RX~J0134.2$-$4258 and RX~J0136.9$-$3510 to our sample. These AGN had
been members of the original bright soft X-ray AGN ROSAT sample
\citep{gru98a,gru99} but were excluded from the final sample of 110
AGN because after reprocessing the RASS data their count rate dropped
below the cut off at 0.5 counts s$^{-1}$ in the {\it ROSAT} PSPC. 
The AGN of our sample were selected from the \citet{gru01} sample 
starting with the brightest sources first. There is some randomness in the sample selection for the sources which were
observed during the last year due to the nature of the fillin program. 
We have still about 20 AGN which need to be observed in the next years. 
As always throughout this paper, we have used
the FWHM(H$\beta$)=2000 km s$^{-1}$ criterion as defined by \citet{oster85} to 
separate between the 43 NLS1s and 49 BLS1s.

Table\,\ref{obj_list} lists all AGN presented here with their optical 
position, redshift, Galactic absorption column density derived from the 
\citet{dic90} HI maps, RASS
 0.2-2.0 keV spectral slope and  rest-frame
0.2-2.0 keV flux \citep{gru01},  $E_{\rm B-V}$ given by Schlegel et
al. (1998), and black hole mass estimated from FWHM(H$\beta$) and
$L_{5100\AA}$ using the \citet{kaspi00} relation\footnote{We are aware
of the uncertainties inherent in determining $M_{\rm BH}$ from 
extrapolation of reverberation data \citep[e.g.][]{bentz06,bentz09}, 
especially in NLS1s. However, this method is still the most reliable
way to estimate $M_{\rm BH}$.}.   

The table also contains the Balmer decrement  of the broad hydrogen
lines and the $E_{B-V}$ derived from H$\alpha$/H$\beta$. While the
majority of soft X-ray selected AGN do not show strong intrinsic
reddening, some do \citep[e.g.][]{goodrich89, gru98b}. In order to
correct for intrinsic reddening, we used the Balmer decrements
H$\alpha$/H$\beta$ of the broad-line components 
 measured by re-analyzing the
optical spectra published in \citet{gru99} and \citet{gru04a}.
 In
order to subtract the contributions by narrow lines we first used the
MIDAS\footnote{MIDAS is the European Southern Observatory's Munich Image Data Analysis System} 
command {\it deblend/line} to separate the broad and
narrow line components.  While this works quite well for most BLS1s,
it becomes unreliable for most NLS1s.  Therefore, for the NLS1s we
subtracted 10\%  and 30\% of the [OIII]$\lambda$5007\AA~line flux from
the total H$\beta$ and H$\alpha$ flux, respectively 
\citep[c.f.][]{cohen83}.
 As expected, most
NLS1s do not show significant reddening, while the Balmer decrements
of some BLS1s do suggest significant reddening in the optical/UV.  The
Balmer decrements were converted into $E_{\rm B-V}$ by assuming a zero
point H$\alpha$/H$\beta$=3.06 as derived by \citet{dong08} for a
sample of 446 blue AGN\footnote{We note that  we assume
 Galactic dust properties when converting $A_{\rm V}$ to $N_{\rm
H}$ and  vice versa.  We also assume that deviations from the assumed
zero-point broad Balmer decrement are a consequence of reddening
rather than evidence for a range of physical conditions (such as
optical depth) in the broad-line region.}.   All AGN in our sample for
which the Balmer decrement is found to be below this value were
considered to have no intrinsic reddening.  For 16 AGN, we do not have
a measurement of the H$\alpha$ line flux, and therefore we could not
determine the Balmer decrement and correct for intrinsic
reddening. Out of the remaining 76 AGN, 
38 have Balmer decrements H$\alpha$/H$\beta>$3.06. In 21 AGN 
significant reddening is expected from their H$\alpha$/H$\beta>$3.40.   
The applied reddening correction influenced the estimates
of several parameters such as \aox, \auv, and \lledd.  In almost all
figures, we show results with and without reddening correction.

The
majority of the \swift\ observations were performed as a \swift\  fillin
target program.  Additional AGN were observed as calibration targets
and ToOs.  In total, \swift\ has observed 92 bright soft X-ray
selected AGN so far. A summary of all XRT and UVOT observations is
given in Table\,\ref{obslog} listing the target ID, segment
number\footnote{Segments are used for \swift\ planning
purposes. Typically one segment covers a time span of one or two
days.}, the start and end times of each observing segment  and the
exposure times in the XRT and each of the UVOT filters. Note that most
AGN have been observed multiple times allowing us also to search for
UV and X-ray variability.   

For all observations\footnote{Except 3C 273 for which we used
observations made in Windowed Timing mode.} the XRT was operating in
photon counting mode \citep{hill04} and the data were reduced by the
task {\it xrtpipeline} version 0.11.6., which is included in the
HEASOFT package 6.4.  Source photons were selected in a circular  
region with a typical radius of 47$^{''}$ and background region of a close-by 
source-free region with r=188$^{''}$. In cases when the AGN was
brighter than 0.4 counts s$^{-1}$, in order to avoid the effects of
pileup, source photons from an inner radius of 10$^{''}$  were
excluded from the spectral analysis.  Photons with grades 0-12 were
selected.  The photons were extracted with {\it XSELECT} version
2.4. The spectral data were re-binned by using  {\it grppha} version
3.0.1 to 20 photons per bin. The spectra were analyzed using {\it
XSPEC} version 12.4.0 \citep{arnaud96}.  In 2007 August the substrate voltage of the XRT CCD was
raised from 0 to 6V in order to 
 lower the dark current and as a result
the detector can operate at slightly
higher temperatures \citep{godet09}.  Therefore, for XRT data taken
before 2007 September, we used the standard response matrix {\it
swxpc0to12s0\_20010101v010.rmf}, and used the {\it
  swxpc0to12s6\_20010101v010.rmf} response matrix for observations
taken after that date.   All spectral fits were performed in  the
observed 0.3--10.0 keV energy band. For each X-ray spectrum we created
an Ancillary Response Function (ARF) file using the \swift\ XRT task
{\it xrtmkarf} to correct for vignetting and bad CCD columns and
pixels. 

In total, 88 out of 92 AGN were observed at least once in all 6 UVOT filters or
in the three UV filters only. The rest had only sporadic observations
in one or two of the UVOT filters, because these AGN were observed as
calibration targets.  In the case of RX J2248.6--5109, UVOT was unable
to observe the AGN due to a bright star in the UVOT field-of-view.
Before analyzing the data, the data from each segment were co-added by
the UVOT task {\it uvotimsum}. Source counts were selected  within the
standard 5$^{''}$ radius for all UVOT filters according to the most
recent UVOT photometry calibration as described by \citet{poole08}. 
As we will discuss in \S\,\ref{uvot_photo}, with this source
extraction radius some contamination from host galaxy star light can
be expected for some of the nearby AGN.  Background photons were
selected in a source-free region close-by with a radius of 20$^{''}$. 
The data were analyzed with the UVOT software tool {\it
uvotsource}. This tool uses the count rate to flux/magnitude
conversion as described in \citet{poole08} assuming a GRB-like
power-law spectrum, which is also appropriate for AGN.  The magnitudes
as listed in Table\,\ref{uvot_res} and fluxes used for the spectral
energy distributions were all corrected for Galactic reddening using
the  $E_{\rm B-V}$ values by \citet{schlegel98} as listed in
Table\,\ref{obj_list}. The correction factor in each filter was
calculated with equation (2) in \citet{roming09}  
who used the standard reddening correction curves by \citet{cardelli89}.

\section{\label{results} Results}

\subsection{\label{note} A note on the Computation of the Eddington ratio}
The Eddington ratio is an important parameter in AGN studies. It determines
the time scale of black hole growth across cosmic times, and it is 
suspected to drive a number of the observed AGN spectral properties.
Its reliable determination is therefore of great interest. However,
both of the parameters that determine this ratio (the black hole
masses and bolometric luminosities) are not very easily accessible
observationally.   Direct determinations of black hole masses through
reverberation mapping still only exist for a small fraction of all AGN
\citep[e.g.,][]{peterson04,bentz09}, bolometric luminosities are
rarely based on SEDs which are measured in all wave-bands
simultaneously, and, in any case, the EUV part of the SED is not directly
observable in all low-redshift sources.  

Our study significantly improves on the determination of bolometric
luminosities and their uncertainties.  However, when reporting Eddington
ratios, we still make the common assumption that the BH mass scaling
relation derived for the reverberation-mapped AGN sample
\citep[e.g.,][]{kaspi00, peterson04, kaspi05} is applicable to all of
our sources.  This does not need to be the case; especially, since few
NLS1 galaxies have been reverberation-mapped, since radiation-pressure
corrections may be relevant in some cases \citep{marconi08}, and since
other processes might affect the Broad Line Region
 kinematics in individual galaxies
(e.g., jet-cloud interactions).  While the focus of this work is on
SEDs, we do show results involving the Eddington ratio, and would like
to emphasize, that the above-mentioned limitations have to be kept in
mind when interpreting those results.  

\subsection{X-ray Spectral Analysis}
Figure\,\ref{xrt_spec_example} shows three examples of \swift\ XRT
spectra with low, medium and high signal-to-noise ratios (RX J0117.5--3826
(segment 002, 188 counts), Fairall 1119 (segment 001, 600 counts), 
and RX J0128.1--1848 (segment 004, 4350 counts), respectively). 
Table\,\ref{xrt_res} summarizes the
results of the spectral analysis of the X-ray data.  All spectra were
fitted with an absorbed single power-law model in the 0.3-10 keV
energy range with the absorption column density fixed to the Galactic
value from \citet{dic90} as listed in Table\,\ref{obj_list}. 
This model fits the majority of the spectra quite well.  A few AGN
require additional components, and we fitted those spectra with an
absorbed broken power-law model.  All model fit parameters, the X-ray
fluxes, the count rates in the 0.3-10 keV band, and hardness
ratios\footnote{We define the hardness ratio as HR=(H-S)/(H+S) where S
and H are the   background-corrected number of counts in the 0.3-1.0
and 1.0-10.0 keV energy bands, respectively} are listed in
Table\,\ref{xrt_res}.   The X-ray fluxes are the rest-frame
absorption-corrected 0.2-2.0 keV fluxes.  We have selected this energy
band to allow for a direct comparison with the  RASS data given in
\citet{gru01}. As a consequence of the soft X-ray selection criterion
using the {\it ROSAT} PSPC hardness ratio \citep{gru98a, thomas98,
  gru01}, the AGN in our sample do not show any evidence for excess
absorption above the Galactic value, with the exception of a few
cases.  The new spectral fits confirm our previous findings
from the RASS data that these soft X-ray selected AGN are
intrinsically unabsorbed in X-rays \citep{gru01}, as the majority of
AGN can be fit sufficiently well by an absorbed single power-law model 
with the absorption column density fixed to the Galactic value.

\subsection{UVOT Photometry \label{uvot_photo}}

Table\,\ref{uvot_res} lists the magnitudes in each of the 6 UVOT filters (if available).
All magnitudes are corrected for Galactic reddening using the
$E_{B-V}$ values given by \citep{schlegel98} as listed in
Table\,\ref{obj_list}.  A correction for intrinsic reddening was not
applied at this point.  
Concerning source variability, previous studies have shown
\citep[e.g.][]{grupe07, grupe08} that the UVOT is relatively stable
between observations suggesting that any variability
$\Delta$mag$>$0.05 mag seen between epochs is real.

Does the UVOT photometry measure the intrinsic continuum from the AGN?
There are three possible reasons why it may not:  a contribution from
the host galaxy star light, intrinsic reddening, and the presence of
emission lines in the filter band-passes.  For our relatively low
redshift objects, the most likely contaminant is  \ion{Mg}{2}$\lambda2800$
which will contribute to the flux measured in the U filter if the
redshift of the AGN is around $z=0.2$. To a smaller extent, we can
also expect some contributions by  \ion{C}{4} and \ion{C}{3}] in the
W2 and M2 filters for objects with higher redshifts.   However, the
photometry of the majority of the AGN in our sample will not be
affected by emission lines.  Therefore, none of the 
UVOT values listed in our paper takes contributions by emission lines into 
account.

Host galaxy star light potentially presents a more serious problem.
Most of the AGN in our sample are rather low-luminosity AGN and the 
contribution in the V, B and even U bands can be significant. 
As pointed out by \citet{bentz06}, even for images with a spatial resolution of about 1$^{''}$
it is almost impossible to disentangle the nuclear from the host galaxy bulge emission.
The Point Spread Function of the UVOT images, however, is about 2$^{''}$ \citep{poole08}.
 As another
example, \citet{leighly09} showed that in the  \xmm\ OM, the \swift\
UVOT sister instrument, the host galaxy contribution in V can be 70\%
in the case of the low-luminosity NLS1 Mkn 493. The problem may not be
as severe in the UVOT, however.  The difference between the UVOT and
the OM is that the UVOT has less stray light than the OM  and the
source extraction radius in the UVOT is 5$^{''}$ instead of  12$^{''}$
which is used for the OM. We checked several nearby AGN including Mkn
493 and found that in the optical filters (V, B, and U) the
contribution by the host galaxy can be significant. We estimated the
host  galaxy contribution by decreasing the source extraction radius
in {\it uvotsource} to 3$^{''}$ and applying an aperture correction by
setting the {\it uvotsource} parameter {\it apercorr=curveofgrowth}.
In cases like Mkn 493 we measured a difference in the V filter
magnitudes of 0.25 mag between the 5$^{''}$ and 3$^{''}$ extraction
radii. In the UV filters, however,  we found only a difference of 0.05
mag. As for changes in the optical/UV slopes, we noticed that in our
most extreme case, Mkn 493, the difference is 0.15 in \auv. This is
similar to the measurement uncertainty in \auv\ for most of our
objects.  For distant sources, we cannot spatially resolve the host
from the AGN, but expect that the errors are no larger than the ones
we determined for the nearby sources. Therefore, we can conclude that
the star light from the host galaxy can affect our measurements of the
nuclear emission using the UVOT data.  However, for the majority of
our objects this contribution should not affect our results. Therefore,
all measurements have been performed using the standard 5$^{''}$
source extraction radius.  Note that in order to determine the optical/UV slope
\auv, we fitted a single power law model to the data of all 6 UVOT filters.
However, in cases where we noticed a flattening or an increase in the fluxes at
longer wavelengths which indicates galaxy contamination, we only fitted the
data of the u, w1, m2, and w2 filters.

The final effect that can influence the optical/UV slope is intrinsic
reddening.  The measured Balmer decrements indicate that a fair number 
of objects have intrinsic reddening of their broad lines.
One might expect that the reddening indicated 
by the Balmer decrements would be mirrored in a flattening of 
the optical/UV slope.
 \footnote{We note in passing that the majority of the
  objects of our sample show no strong excess X-ray absorption. 
  Objects with a dusty warm absorber may be reddened in the optical and UV but
  have little evidence for X-ray absorption \citep[e.g.,][]{brandt96, koba98}.
  However, additional evidence that this is not generally true for our
  objects is the lack of optical polarization \citep{gru98b} characteristic
  of dust \citep{leighly97}.}  
We have therefore searched for
evidence of continuum reddening associated with the Balmer decrement
by plotting the observed (but corrected for Galactic reddening)
optical/UV spectral slope \auv\ as a function of the Balmer  
decrement as listed in Table\,\ref{obj_list}
(Figure\,\ref{hahb_auv}). NLS1s  are displayed as blue triangles and
BLS1s as red circles.   This \auv - H$\alpha$/H$\beta$ relation is clearly dominated
by three outliers with H$\alpha$/H$\beta>$5.0: Mkn 766, IRAS 1334+24,
and NGC 4593. Excluding those three objects from the correlation
analysis we do not find any statistically significant correlation
between \auv\ and H$\alpha$/H$\beta$ ($r_{\rm l}$=0.133, P=0.272;
$r_s$=0.123, $T_{\rm s}$=1.00, P=0.309).  This result suggests that
the intrinsic reddening inferred from the Balmer decrement does not
significantly flatten \auv. Nevertheless, we applied a reddening correction to the
optical/UV spectra and list both the non-corrected values and those
corrected for intrinsic reddening.  
Although the mean, standard deviation and median of H$\alpha$/H$\beta$ of NLS1s are
3.10, 0.61, and 3.00, respectively compared with 3.43, 0.82, and 3.18 which seem to suggest
that BLS1 are generally more reddened than NLS1s, a KS test of the H$\alpha$/H$\beta$ distributions 
results in D=0.280 with a probability P=0.087 of a random result. This result shows that the distributions 
are not significantly different.

\subsection{Spectral Energy Distributions}

Table\,\ref{sed_res} summarizes the results from the analysis of the
SEDs.
For each object we typically selected the segment/observation with the
longest observing time to get the best signal-to-noise ratio. The
X-ray spectral slope \ax\ was taken from the power law fits to the XRT
data as listed in  Table\,\ref{xrt_res}. The UV/optical spectral
slopes \auv\ were determined from a single power-law model fit to the   
UVOT fluxes. We determined the optical-to-X-ray spectral slope
\aox\footnote{The X-ray loudness is defined by \citep{tananbaum79} as
  \aox=--0.384 $\log(f_{\rm 2keV}/f_{2500\AA}$).} by measuring the
flux densities at rest-frame 2500\AA\ and 2 keV. The X-ray
luminosities $L_{\rm X}$ were derived from the absorption-corrected
rest-frame 0.2-2.0 keV fluxes as given in Table\,\ref{xrt_res}. 

In the UV-to-X-ray part of the SED we fitted two different models to
the UVOT and XRT data.  The first (henceforth Model A) consists of  a
power-law  with an exponential cut-off describing the
optical-EUV part of the SED, added to an ``absorbed'' power law model
to describe the X-ray part of the SED. Note that this ``absorption'' 
merely serves as a mathematical description to ensure the appropriate
decrease of the X-ray power law at low energies in order to prevent it
from over-predicting the UV part of the SED  \citep[see also][]{gru04a}. 
The second (henceforth Model B) is a double broken power-law model;
that is, a  power law was fitted to the UV part of the SED and 
a second one separately to the X-ray part,
and a third power law was then used to connect these two models
across the EUV.  The break points in Model B are at 2000\AA\
and 0.3 keV, where the UVOT observing window ends and the XRT window
starts.  At 0.3 keV, the break point flux density was set to
that  of the unabsorbed power law model in X-rays. Model A  was
applied to both observed and the intrinsic-reddening-corrected UVOT
data.  In some cases where a very large Balmer decrement was found,
e.g., IRAS 1334+2438, this procedure may overestimate the luminosity
in the BBB, $L_{\rm BBB}$\footnote{Note that throughout this paper we use the
Big-Blue-Bump luminosity \lbbb\ rather than the bolometric luminosity
$L_{\rm bol}$. Although most of the SED continuum energy in a Seyfert galaxy is deposited
in the optical-to-X-ray band, we miss emission in the radio
band. However, we do not consider the radio contribution
significant in most of our AGN. Therefore $L_{\rm BBB}$ basically
represents $L_{\rm bol}$. This luminosity is used to determine \lledd.
}.  Model B  was  applied  only to the data not
corrected for intrinsic reddening.   

Since we do not know the shape of the BBB in the EUV band, we apply
both models (A and B) to estimate the SED.   This  approach allows us to get an
impression on the uncertainties in the bolometric luminosities, given
the uncertainties in the EUV shape of the SED.  The
reddening-corrected model A serves as a kind of upper limit on the EUV
luminosity, since the extrapolation of the reddening-corrected UV
spectrum produces a strong EUV bump. On the other hand, the piecewise
power law approach of model B introduces a break from the last observed
UV point to the first observed X-ray data point, and serves as a
reasonable lower limit to the EUV luminosity.  

The BBB luminosity $L_{\rm BBB}$ was estimated by
integrating over these continuum spectra  in the rest frame energy
range between 1 $\mu$m and 2 keV.  As mentioned before, the fits with
Model A give an upper limit on $L_{\rm BBB}$ while the fits with Model
B may be regarded as lower limits. Both values of $L_{\rm BBB}$ are
listed in Table\,\ref{sed_res}. We found that typically $L_{\rm BBB}$
calculated from Model A is a factor of 2 higher than that of Model
B. In other words, although we do not know the real shape of the BBB
in the EUV,  we may overestimate $L_{\rm BBB}$ by a factor of only a
few at the most and not by orders of magnitude.  The Eddington ratios
were calculated from $L_{\rm BBB}$ and the Eddington  luminosity. To
derive the latter, we used the black hole masses of \citet{grupe04}
as listed in Table\,\ref{obj_list}.
Note that \auv, \aox, $L_{\rm BBB}$, $L/L_{\rm Edd}$, and $L_{5100}$
are listed with and without the correction for intrinsic reddening.  

The mean SEDs, corrected and uncorrected for intrinsic reddening 
for NLS1s (blue lines) and BLS1s (red lines) are displayed in 
Figure\,\ref{aver_sed}. In general, NLS1s appear to be fainter 
in the UV and at hard X-ray energies than BLS1s. Nevertheless, their
mean and median \auv\ and \aox\ are very similar 
as also shown in Table\,\ref{mean}.
This table summarizes the mean, standard deviation and median of the whole sample, NLS1s and BLS1s
for the spectral slopes, luminosities, flux ratios and redshifts. 
Due to the sample
selection criteria, the distributions  are generally skewed and non-Gaussian and
therefore the medians are better estimators than the means.  BLS1s and
NLS1s show clear differences in their \ax,  $L/L_{\rm Edd}$, and X-ray
variability.  NLS1s show, as expected from previous studies,
significantly steeper X-ray spectra than BLS1s.  From previous studies
\citep{walter93, gru98a} we had expected that NLS1s would have bluer
optical-UV spectra.  As we will show later in Section\,\ref{corr}, we
do confirm that AGN with  steeper X-ray spectra have bluer optical-UV
spectra, but this relation holds only among those AGN that have
relatively flat X-ray spectral slopes in the first place.

Figure\,\ref{distr_ax} displays the distributions of the 0.3-10 keV
X-ray spectral slope \ax\ of NLS1s and BLS1s (solid blue line and
dotted red line, respectively).  NLS1s and BLS1s clearly show
different distributions in the X-ray spectral slopes \ax\,(a KS test
results in D=0.557 with a corresponding probability P$<0.0001$ of
being the same distribution). 

Figure\,\ref{distr_auv} shows the distributions of the optical/UV
slope \auv\ uncorrected and corrected for intrinsic reddening. We
found no significant difference in the observed \auv\ distributions
between NLS1s and BLS1s (D=0.140, P=0.757) but there is a slight
difference in the distribution in the reddening-corrected UV slope
$\alpha_{\rm UV-corr}$ (D=0.385, P=0.007), with the BLS1s having
bluer continua, resulting in steeper \auv.

Figure\,\ref{distr_aox} shows the distributions of the
optical-to-X-ray spectral slope \aox\ uncorrected and corrected for
intrinsic reddening. For the observed \aox, the distributions between
NLS1s and BLS1s are slightly different with D=0.292 and P=0.039. NLS1s
appear to be X-ray weaker at 2 keV than BLS1s which results in larger values 
of \aox\ in NLS1s.
However, this
difference disappears when the distributions for the \aox\ with the  
correction for intrinsic reddening are used. Here a KS test
gives D=0.104 and P=0.986. 

Figure\,\ref{distr_l_ledd} shows the distributions of the Eddington
ratio \lledd\ uncorrected and corrected for intrinsic reddening. The
distributions for NLS1s and BLS1s are different. A KS test
for the observed \lledd\ (uncorrected for intrinsic reddening) gives
D=0.443 with a corresponding probability P$<$0.001. However, for \lledd\
corrected for intrinsic reddening the samples are almost identical with
D=0.273 and a probability P=0.118.

Figure\,\ref{distr_lx} displays the distributions of the 0.2-2.0 keV
luminosity log $L_{\rm X}$  of  NLS1s  and BLS1s. These distributions
are essentially identical (D=0.125, P=0.851).
Figure\,\ref{distr_lbbb} displays the distributions of the
luminosities in the BBB emission uncorrected and corrected for
intrinsic reddening (left and right panels, respectively). 
As for the reddening-uncorrected luminosities the distributions
of NLS1s and BLS1s are almost identical (D=0.151 and
P=0.669). However, for the data corrected for intrinsic reddening,
BLS1s seem to have slightly more luminous Big Blue Bumps than NLS1s (D=0.299
and P=0.070). 

\subsection{Correlation analysis \label{corr}}

We have searched for correlations among the measured parameters. 
For all correlations we determined the linear correlation coefficient
$r_{\rm l}$ and the Spearman rank order correlation coefficient
$r_{\rm S}$ plus the Student's T-test value $T_{\rm S}$. The
probability P of a null correlation was determined also for both
correlation coefficients.  The results of the correlation analysis are
listed in Table\,\ref{correlations}. The part above the diagonal in
Table\,\ref{correlations} gives the linear correlation coefficient
$r_{\rm l}$, the corresponding probability P and number of sources
involved. The lower part of the table lists $r_{\rm S}$, $T_{\rm S}$,
number of sources, and the probability. Note that we did not do a
correlation analysis between reddening-corrected and uncorrected
properties, such as \aox\ and $\alpha_{\rm ox-corr}$ or \auv\ and 
$\alpha_{\rm UV-corr}$.  

One of the principal motivations of this study is to see whether there
is a relation between the X-ray spectral slope \ax\ and the optical/UV
slope \auv. This is motivated by earlier studies by e.g. 
\citet[][]{walter93} and \citet{gru98a} who found that AGN 
with bluer optical/UV spectra have steeper X-ray
spectra. Figure\,\ref{ax_auv} displays the 0.3-10  keV X-ray energy
spectral slope \ax\ versus the optical-UV slope \auv. 
The left side of Figure\,\ref{ax_auv} displays the relation with \auv\
corrected for only Galactic reddening. The right panel shows \auv\ 
corrected for both intrinsic and Galactic reddening.  For the UVOT
data uncorrected for intrinsic reddening we found linear and Spearman
rank order correlation coefficients $r_{\rm l}$=--0.15, $r_s=-0.19$
with a Student's T-test $T_s=-1.8$, and probabilities P=0.1606 and
0.077 of null correlations, respectively.  The trend does not improve
when we consider the reddening-corrected optical/UV slope 
$\alpha_{\rm UV-corr}$  ($r_{\rm l}=-0.24$, $r_{\rm s}=-0.14$, $T_{\rm
 s}$=--1.2). However, there seems to be a saturation in the
optical/UV slope when the X-ray spectral slope becomes steeper than
\ax$\approx$1.6.  As we will see below, this saturation is due to the
fact that at this X-ray spectral slope \lledd\ reaches unity. 
When we limited the sample to the 63 AGN with \ax$<$1.6 (63 AGN) we
found that there is a relatively strong correlation between \ax\ and
\auv\ with $r_{\rm l}=-0.36$ and $r_{\rm s}=-0.30$, $T_{\rm s}=2.5$
and probabilities P=0.0038 and 0.0151, respectively. Because NLS1s
typically have steeper X-ray spectra than BLS1s, this effect is
basically dominated by BLS1s. Excluding IRAS 1334+2438 which has a
highly reddened spectrum  (\auv=2.8, H$\alpha$/H$\beta$=6.2), we found
a linear correlation coefficient for the BLS1s of $r_{\rm l} =-0.46$
with a probability $P=0.0011$. For the Spearman rank order correlation
we found $r_{\rm s}=-0.48$ and $T_{\rm s}=-3.56$ with P=0.009. For the
NLS1s, however, there is only a marginal trend ($r_{\rm l}=-0.21$).
In summary, we can confirm the earlier results from ROSAT
\citep{walter93, gru98a} that \ax\ and \auv\ are correlated, but the
correlation appears to be dominated by AGN with \ax$<$1.6, i.e.,
mainly BLS1s.  

There are clear correlations between the optical-to-X-ray spectral slope \aox\
with the X-ray spectral slope \ax\ and the optical/UV slope \auv, such
that AGN with steeper X-ray spectra appear to be X-ray weaker ({\em at
2 keV}; note that their X-ray luminosities are similar) than those
with flatter \ax. Figure\,\ref{aox_ax} displays this relationship.
The linear and Spearman rank order correlation coefficients are
$r_{\rm l}$=+0.47 and $r_{\rm S}$=+0.47 ($T_{\rm s}$=5.0),
respectively; in both cases the probability of a null correlation is
P$<0.0001$. These correlations also hold for the UVOT data corrected
for intrinsic reddening.  The trend that AGN with steep X-ray spectra tend to
be X-ray weak {\em at hard X-rays}{\footnote{Note that this result
 may be an artifact of assuming a single power law to fit the whole
 X-ray spectrum. More complicated spectral models involving soft
 excesses on flatter power laws could reduce this trend.}} 
has also been found by \citet{atlee09}.  For comparison,
\citet{young09} found among their sample of SDSS quasars with
simultaneous X-ray observations that X-ray faint quasars have flatter
X-ray spectra.  Correlations are also found between 
\aox\ and the optical/UV spectral slope \auv\ (Figure\,\ref{aox_auv});
there is a  clear anti-correlation with $r_{\rm l}$=--0.63 and $r_{\rm
  s}$=--0.57 and $T_{\rm s}$=--6.4. For both correlation coefficients
the probability of a null-result is $P<0.0001$. 

Figure\,\ref{ax_auv_ledd} displays the relationships between the
Eddington ratio $L/L_{\rm Edd}$ and  \ax, \auv, and \aox. The X-ray
spectral slope \ax\ and the Eddington ratio $L/L_{\rm Edd}$ are
correlated with a Spearman rank order correlation coefficient $r_{\rm
  s}$=0.55 and a Student's T-test $T_{\rm s}$=+6.1 and the  linear
correlation coefficient is $r_{\rm l}$=0.51.  This correlation is
similar to those found previously by  \citet{grupe04} and \citet{shemmer08} 
who reported strong correlations between the Eddington ratio $L/L_{\rm
  Edd}$ and the soft and hard X-ray spectral slopes, respectively.   
The upper panel of Figure \ref{ax_auv_ledd} suggests that, however, at
an Eddington ratio $L/L_{\rm Edd}\approx 1$ the relationship
saturates.  This implies that \ax\ cannot be used to estimate
$L/L_{\rm Edd}$ universally; the relationship only works for X-ray
spectral indices with \ax\ up to $\approx$1.6, similar to the
\ax-\auv\ relation.  Again considering only the AGN with \ax$<$1.6, we
found correlation coefficients $r_{\rm l}=+0.64$, $r_{\rm s}=+0.60$ and
$T_{\rm s}=+5.7$ with probabilities of a random result P$<0.001$ in
both cases. For this subsample of objects with \ax$<$1.6,  we found
the following relation between \ax\ and log $L/L_{\rm Edd}$: 

\begin{equation}
\log L/L_{\rm Edd}~=~(1.65\pm0.26)\times \alpha_{\rm X}~-~(2.41\pm0.30)
\end{equation}
 
This relationship is shown as a dotted line in the upper left panel of
Figure\,\ref{ax_auv_ledd}.  It is dominated by BLS1s, because the
majority of the NLS1s in our soft X-ray selected sample accrete at the
Eddington limit or even at super-Eddington rates.  For the
super-Eddington accretors, the relationship breaks down; this is
similar to the behavior of the \ax-\auv\ relationship as shown below. 
It is interesting to mention that \citet{winter09} did not find a correlation between the 
hard X-ray 2-10 keV spectral slope and \lledd, although their sources are all low \lledd\
AGN. From our results and those of \citet{shemmer08}, we would have expected to see a 
correlation between \ax\ and \lledd. 

The middle panels in Figure\,\ref{ax_auv_ledd} display the anti-correlation
between the optical/UV spectral slope \auv\ and the Eddington ratio $L/L_{\rm
Edd}$.  AGN with bluer UV continua (\auv$<$1) tend to have higher
Eddington ratios. The correlation coefficients for a Spearman and
linear correlation analysis are $r_{\rm S}$=--0.57 with a T-test
$T_{\rm S}$=--6.4 and $r_{\rm l}$=--0.63 with a probabilities
P$<$0.001.  Note that there is an offset by 0.6 dex between NLS1s and
BLS1s in these plots: For a given \auv, NLS1s have a $\sim$4 times
higher Eddington ratio than BLS1s. This division becomes even
stronger when considering the data corrected for intrinsic reddening
(right panel).  Considering only the NLS1s, we found linear and
Spearman rank order correlation coefficients of $r_{\rm l}$=--0.73 and
$r_{\rm S}$=--0.70, $T_{\rm S}$, respectively.  For the BLS1s, the
linear correlation coefficient is $r_{\rm l}$=0.75 and the Spearman
rank order correlation coefficient $r_{\rm S}$=--0.67, $T_{\rm
  S}$=--5.9. In all these cases,  the probability of a random
correlation is P$<0.0001$. Clearly, \auv\ is strongly correlated with
\lledd.  Note, however, that in the model for the UV-EUV SED
employing a power law with exponential cut off (Model A),  \auv\ and
\lledd\ are not necessarily independent parameters.  Thus, a steeper
\auv\, when extrapolated,  will lead to a stronger bump and therefore
higher $L_{\rm BBB}$ and \lledd. If the \auv\ - \lledd\ relation is
real it must hold also when the double broken power law model is used
(Model B).  This is indeed the case. For the whole sample we  found a
linear correlation coefficient $r_{\rm l}=-0.46$ and a Spearman rank
order correlation coefficient $r_{\rm s}=-0.34$ with $T_{\rm s}=-3.4$
with probabilities P$<0.0001$ and 0.0010, respectively. Looking at the
NLS1 and BLS1 samples separately, the correlations are very strong
with $r_{\rm l}=-0.65$ and --0.62, respectively.  In both cases the
probabilities of a null correlation are P$<0.0001$. Therefore we
consider the anti-correlation between \auv\ and \lledd\ to be real. 

For the whole sample we derived the following relationship between
\auv\ and \lledd: 

\begin{equation}
\log L/L_{\rm Edd}~=~(-0.65\pm0.09)\times \alpha_{\rm UV}~+~(0.08\pm0.09).
\end{equation}

For the NLS1s alone, we find: 

\begin{equation}
\log L/L_{\rm Edd}~=~(-0.62\pm0.09)\times \alpha_{\rm UV}~+~(0.41\pm0.10).
\end{equation}

However, BLS1s show an offset by 0.58 dex:

\begin{equation}
\log L/L_{\rm Edd}~=~(-0.74\pm0.10)\times \alpha_{\rm
  UV}~-~(0.17\pm0.10). 
\end{equation}

The dotted lines in the left middle panel in Figure\,\ref{ax_auv_ledd}
show equations (3) and (4).  

For the reddening-corrected data  $\alpha_{\rm UV,corr}$ and $L/L_{\rm Edd-corr}$ 
we found the following relationships for the whole sample, the NLS1s,
and BLS1s, respectively: 

\begin{equation}
\log L/L_{\rm Edd-corr}~=~(-0.77\pm0.09)\times \alpha_{\rm UV-corr}~+~(0.17\pm0.07),
\end{equation}

\begin{equation}
\log L/L_{\rm Edd-corr}~=~(-0.99\pm0.14)\times \alpha_{\rm UV-corr}~+~(0.57\pm0.10),
\end{equation}

\begin{equation}
\log L/L_{\rm Edd-corr}~=~(-0.96\pm0.10)\times \alpha_{\rm UV-corr}~-~(0.08\pm0.08).
\end{equation}

The relationships for the NLS1s and the BLS1s are shown as dotted
lines in the right middle panel of Figure\,\ref{ax_auv_ledd}. 

Thus, \auv\ and $\alpha_{\rm UV-corr}$ can be used in order to
determine the Eddington ratio \lledd\ and $L/L_{\rm Edd-corr}$,
respectively.  However, one has to keep in mind that there is an
offset between BLS1s and NLS1s. 

Similar relationships also hold for the Eddington ratios derived from
the double broken power law fits to the SEDs as shown in Figure\,\ref{ax_auv_ledd_bknpo}.
The relationships for
the whole sample, the NLS1s, and the BLS1s are as follows:

\begin{equation}
\log L/L_{\rm Edd-bknpo}~=~(-0.47\pm0.10)\times \alpha_{\rm UV}~-~(0.30\pm0.10),
\end{equation}

\begin{equation}
\log L/L_{\rm Edd-bknpo}~=~(-0.41\pm0.09)\times \alpha_{\rm UV}~+~(0.12\pm0.09),
\end{equation}

\begin{equation}
\log L/L_{\rm Edd-bknpo}~=~(-0.58\pm0.10)\times \alpha_{\rm UV}~-~(0.69\pm0.10).
\end{equation}

Again, the slopes in the relations are similar, however, there is an
offset between NLS1s and BLS1s by  0.8 dex.

The lower panels in Figure\,\ref{ax_auv_ledd} show the relationship
between \aox\ and $L/L_{\rm Edd}$.  We found a rather strong
correlation between \aox\ and \lledd. For the observed \aox\ we found
a linear correlation coefficient $r_{\rm l}$=+0.51 and a Spearman rank
order correlation coefficient $r_{\rm s}$=+0.55.  This result,
however, is different from that reported by \citet{shemmer08} who did
not find such a correlation. Our result becomes even stronger when we
consider the reddening corrected values $\alpha_{\rm ox-corr}$ and
$L/L_{\rm Edd-corr}$.  
The relation between \aox\ and \lledd\ is given by 
\begin{equation}
\alpha_{\rm ox} = (0.11\pm0.02) log L/L_{\rm Edd} + (1.39\pm0.02)
\end{equation}

The slope in this relation is flatter than what has been reported recently  by
\citet{lusso10}. Note that the scatter in our relation is significantly smaller
than in the \citet{lusso10} sample.

Figure\,\ref{aox_l2500} displays the relationship between the
optical-to-X-ray spectral slope \aox\ as a function of the luminosity
density at 2500\AA, $l_{2500}$.  Also our sample shows the well-known
relation found by e.g. \citet{yuan98, strateva05, just07, gibson08}
that AGN with higher luminosity density at 2500\AA\ appear to be X-ray
weaker at 2 keV.  We found a relation between \aox\ and the luminosity
density at 2500\AA: 

\begin{equation}
\alpha_{\rm ox}~=~(0.114\pm0.014)\times \log\ l_{\rm 2500}~-~(1.177\pm0.305)
\end{equation}

with the luminosity density at 2500\AA\ given in units of W Hz$^{-1}$. 
This relationship is in excellent agreement with those given by
\citet{strateva05} and \citet{just07}. Dashed and solid lines in the
upper left panel of Figure\,\ref{aox_l2500} show our \aox-log $l_{\rm 2500\AA}$
relation and that of \citet{strateva05}, respectively. Our
relationship, however, deviates significantly from those given by
\citet{gibson08} and \citet{vasudevan09}
who found  much steeper slopes between \aox\ and log
$l_{\rm 2500\AA}$.  The most likely reason for this deviation  
is that the \citet{gibson08} sample contains more luminous AGN and
stretches only over one order of magnitude in log $l_{\rm 2500\AA}$
while ours and that of \citet{strateva05} stretch 4 and 5 orders,
respectively.  As shown in Figure\,\ref{aox_l2500}, there are also
relationships  between \aox~and  $L_{\rm BBB}$, $L_{\rm X}$, and
$L_{\rm 5100\AA}$. Also here we find correlations with \aox\ as listed 
in Table\,\ref{correlations}. 

Figure\,\ref{ax_l} displays the relations between the X-ray spectral
slope \ax\ with $L_{\rm X}$ and $L_{\rm BBB}$  in the left and right
panels,  respectively. In both cases we found marginal evidence 
that more luminous AGN have steeper X-ray spectra. Note that there is
an offset between NLS1s and BLS1s.  Thus, for a given \lx\ or \lbbb\,
NLS1s have steeper X-ray spectra; this may be a consequence of their
typically smaller black hole masses. 

Figure\,\ref{l5100_lbbb} displays the relationships between the
luminosity at 5100\AA\ and the 0.2-2.0 keV X-ray luminosity $L_{\rm
  x}$ with the  luminosity in the Big-Blue-Bump $L_{\rm BBB}$.  These
relationships give the bolometric corrections for optical and X-ray
luminosities.  The bolometric correction for the 5100\AA\ luminosity
$L{\rm 5100\AA}$ (shown in the left panel of figure\ \ref{l5100_lbbb})  
is given by;

\begin{equation}
\log L_{\rm BBB}~=~(1.32\pm0.06)\times \log L_{\rm 5100}~-~(10.84\pm2.21).
\end{equation}

The right panel of Figure\,\ref{l5100_lbbb} shows the correlation
between the rest frame 0.2-2.0 keV luminosity and $\log L_{\rm
BBB}$. Here we found the following relationship:

\begin{equation}
\log L_{\rm BBB}~=~(1.23\pm0.06)\times \log L_{\rm X}~-~(7.36\pm2.01).
\end{equation}

The left panel of Figure\,\ref{fwhb_ax} displays the well-known
relationship between FWHM(H$\beta$) and the X-ray spectral slope
\ax. The right panel shows  FWHM(H$\beta$) vs. \lledd. The
FWHM(H$\beta$) were taken from \citep{gru04a}.  As expected from
previous work \citep[e.g.][]{bol96, brandt97, gru99, lei99b, grupe04, zhou06}  
a clear anti-correlation between the width of H$\beta$ and \ax\ is
found.    The Spearman rank order correlation coefficient for the
FWHM(H$\beta$) - \ax\ relation is $r_{\rm s}$=--0.65 with a Student's
T-test    $T_{\rm s}$=--7.9 and a probability P$<0.0001$. The linear
correlation coefficient is $r_{\rm l}$=--0.58 with a probability P$<0.0001$. 
We also find a strong anti-correlation between FWHM(H$\beta$) and
$L/L_{\rm Edd}$, as seen in the right panel of Figure\,\ref{fwhb_ax}.
The correlation coefficients are $r_{\rm l}=-0.56$ and $r_{\rm
  s}=-0.50$ with $T_{\rm s}$=--5.3, and in both cases the probability
of a null result is P$<0.0001$.  Note that FWHM(H$\beta$) and \lledd\ 
are not independent properties. \lledd\ depends on FWHM(H$\beta$) because
we used the \citet{kaspi00} relation to determine $M_{\rm BH}$ which 
is then used to determine $L_{\rm Edd}$.  In any case, the correlation
between FWHM(H$\beta$) and \ax\ is a robust result, because both are
independent.  

Figure\,\ref{fwhb_auv_aox} shows the relationship of FWHM(H$\beta$)
with the optical UV slope \auv\ and \aox.  
We do not find a correlation between FWHM(H$\beta$) and \auv, even
among those AGN with \ax$<1.6$.

\subsection{Variability}

All of our objects are members of the {\it ROSAT} bright soft X-ray
selected sample of AGN \citep{gru01}, and thus have at least one {\it
  ROSAT} PSPC and one \swift\ observation. This enables us to search
for long-term variability on a time scale of more than 15 years. Also,
the majority of the AGN (86 out of 92) have been observed at least
twice by \swift, with the majority having more than two observations.    
Note that due to the nature of the fill-in target program this
sampling is not homogeneous. Some AGN have been observed twice within
a few days while for others the interval was more than a year.
Nevertheless, our study shows again that most AGN vary in X-rays by a
factor of 3 at the most on all timescales.  We note, however, 
X-ray variability by factors of more than 10 have been observed in
some AGN including Mkn 335, PG 1211+143, and RX J2217.9--5941
\citep[][respectively]{grupe07b, bachev09, grupe04b}.  

The AGN in our sample are also variable in the UV.  Except for the study
of variability using {\it IUE} and {\it HST} data by \citet{dunn06},
ours is the largest sample of data suitable to study UV
variability.  We found that the majority of AGN vary by up to 0.4
mag over the time scale of a few months, though as discussed below,
some AGN can vary by up to 1.5 mags within just a few months. 

Figure\,\ref{xray_cr_var} displays the short and long term X-ray flux 
variability in the AGN sample. The left panel shows a comparison of
the count rate from the brightest and faintest AGN with multiple
\swift\ observations.  The right panel shows the long-term flux
variability between the RASS and the \swift\ observations where we
again plot the data from the \swift\ observation with the largest
difference from the RASS observation.  The dashed lines in
Figure\,\ref{xray_cr_var} mark a variability by a factor of 3. While
the \swift\ observations in the left panel show that most AGN vary
over a time scale of about a year by a factor of 3 or less, the
comparison of the \swift\ data with the RASS data displays a slightly 
different picture: in general we see a trend that the AGN were 
slightly more luminous during the RASS observations. 
While during the RASS the mean and median 0.2-2.0 keV
luminosities were log $L_{\rm X}$=37.02 and 37.13 [W], respectively, 
during our \swift\ observations we found log $L_{\rm X}$=36.94 and
37.00 [W].  We note that this small differences is likely to be a
consequence of the sample selection by soft X-ray count rate. 

Figure\,\ref{xray_hr_var} shows the spectral variability among the AGN
in the sample. The left panel shows the changes in the hardness ratio
between the two \swift\ observations with the largest differences. The
right panel displays the long-term change of the 0.2-2.0 keV X-ray
spectral slope \ax\ between the RASS and the \swift\ observations. 
The left panel of Figure\,\ref{xray_hr_var} suggests that the majority
of AGN do not show strong  spectral variability on time scales of a
year or two. The long-term \ax\ between the RASS and the XRT
observations, at first glance, appears to indicate
a systematic flattening of the  X-ray
spectra over the last 15-18 years.  

A very obvious explanation of this apparent flattening is the fact,
that ROSAT spectra were only fit in the 0.1-2 keV regime,
where spectral complexity is known to be common, while {\em Swift}
spectra were fit over a much larger energy band, but still
assuming one single power law. Any spectral hardening beyond
a few keV would naturally produce most or all of
the observed trend. We argue that it is unlikely that
cross-calibration uncertainties contribute significantly to the
observed trend, even though that has apparently been identified as a
potential problem in comparison of {\it ROSAT} and {\it ASCA} data
\citep{iwasawa99}.  As shown by \citet{beuermann06} and
\citet{beuermann08} for the {\it ROSAT} PSPC and 
 \citet{godet09} for the \swift\ XRT, both instruments are
 well-calibrated and reliable. However, an additional selection effect 
 could contribute to the observed apparent flattening.  The objects
 were required to have {\it ROSAT} PSPC count rates higher than $0.5
 \rm \, counts \, s^{-1}$ as well as {\it ROSAT} PSPC hardness ratios
 HR$<$0.0 to be a member of our sample \citep{thomas98, gru98a}.
 Objects that were only temporarily soft may have been included in our
 sample.  In fact, we have observed dramatic spectral hardening in
 follow-up  observations of several objects in our sample including RX
 J0134.2--4258 and RX  J0148.3--2758
 \citep[][respectively]{grupe00, komossa00, grupe06}.  The apparent hardening of
 the X-ray spectra may thus be partly a consequence of the selection due
 to hardness ratio. Just as selection by the flux limit shows that a
 majority of our AGN appear slightly X-ray fainter than during the
 RASS observations.  Specifically, 52 AGN have lower \swift\ flux than
 RASS flux, while the opposite is true for only 40 AGN.

The left panel of Figure\,\ref{xray_var_cr_hr}  shows the \swift\ XRT
difference in hardness ratio as a function of corresponding count
ratio, while the right panel displays the flux ratio between the  RASS
and \swift\ observations as a function of the difference in the
\swift\ and RASS X-ray spectral slopes \ax. The purpose of these plots
is to test whether there is a general trend of a source spectrum to
become harder (or softer) when the source becomes fainter or
brighter. In neither case do we see any 
correlation between a change in the flux and any type of spectral 
variations. Note that this is a
general trend based on only two epochs. For an individual source this
picture might indeed be different when it is monitored as we have
shown in cases like RX J0148.3--2758  or Mkn 335 \citep{grupe06,
  grupe08}. 

Figure\,\ref{uvw2_var} displays the variability in the UVOT W2
filter. We selected  this filter because it is the bluest filter which
therefore shows a much stronger response to changes in the BBB emission
than any other filter. The left panel of Figure\,\ref{uvw2_var}
displays the difference between the two observations with the
brightest and faintest magnitudes in UVW2. The right panel shows the
difference between these magnitudes as a function of the ratio of the
count  rates in the 0.3-10.0 keV XRT X-ray band.  Clearly, the
majority of the AGN in our sample show some UV variability. Typically,
however, this variability is less than 0.4 mag, corresponding to a
flux ratio of 1.5.  Four AGN do show a variability in the UVW2 filter
by more than 0.5 mag: RX J0148.3--2758, ESO 242-G8, RX J0319.8--2627,
and RX J2349.4--3126.  The last object  has also been seen to show
strong spectral variability in X-rays from {\it ROSAT} observations
\citep{gru01}.   It is interesting to note that except for RX
J0148.3--2758 all these very variable AGN are BLS1s. RX J0319.8--2627
and RX J2349.4--3126 varied in W2 by more than 1.2 mag. 

The right panel of Figure\,\ref{uvw2_var} shows the XRT count rate
ratio between two epochs and the change in the UVOT W2 filter. Note
that here we took the UVW2 magnitudes from the same epochs as the
maximum and minimum count rates in the XRT. We found a clear
correlation between the amplitude of the UVW2 variability and the
ratio of the XRT count  rate ($r_{\rm l}$=--0.41, $r_{\rm s}$=--0.47,
$T_{\rm s}$=--4.55, and the probabilities of random results $P<0.0001$
for both correlation  coefficients. Thus, AGN that have
become brighter in X-rays in the second epoch also appear to be
brighter in the UV, generally speaking. There is one interesting trend
to note: The AGN with the strongest UV variability tend to be  BLS1.   

It has long been known that the variability time scales are strongly
correlated with the luminosity and the black hole mass
\citep[e.g. ][]{barr86, lawrence93, green93, nandra97, lei99a,
  oneill05, kelly08}.  We correlated the ratio in the XRT count rates
between two epochs with the 0.2-2.0 keV X-ray luminosity log $L_{\rm X}$
and the black hole mass as shown in Figure\,\ref{var_lx_mbh} (left and
right panels, respectively). While we found a clear trend that AGN
with smaller black hole masses show higher X-ray variability 
(85 AGN,
$r_{\rm l}=-0.22$, P=0.0406; $r_{\rm S}=-0.29$, $T_{\rm S}=-2.7$,
P=0.0084),  
we do not see this trend with the X-ray luminosity. Here
there seems to be a relatively large group of AGN, predominantly NLS1s
with luminosities around $L_{\rm X}$=37 [W] that show very strong
variability. Again, because of our inhomogeneous sampling due to the
nature of the fill-in program, as well as the soft X-ray selection,
one has to take the latter statement with caution. We cannot exclude
at all that there is not an anti-correlation in our sample as well
between the X-ray variability and the X-ray luminosity.    

\subsection{Individual target notes}

\subsubsection{Mkn 335}
Mkn 355 was discovered in May 2007 by \swift\ in an historical X-ray
low state \citep{grupe07b}. Consequently we started a monitoring
campaign with \swift\ and initiated a 20 ks ToO observation with
\xmm. The \xmm\ observation confirmed the X-ray low-state and revealed
the presence of strong soft X-ray emission lines in the RGS spectra
\citep{grupe08, longinotti08}. We are currently monitoring Mkn 335
with \swift. Since 2007 September it has returned to a high state
(Grupe et al. in prep). As discussed in \citet{grupe08} the cause of
the low-state was either a strong variable partial covering absorber
or reflection of X-rays on the accretion disk.  For the study
presented here we used the data from one of the high state
observations in September 2008 which seem to represent the 'normal'
state of Mkn 335.  Nevertheless, recent monitoring of Mkn 335 with
\swift\ starting in May 2009 showed it again in a very low flux state
(Grupe et al. in prep). 

\subsubsection{RX J0134.2--4258 and RX J0136.9--3510}
Although the NLS1s RX J0134.2--4258 and RX J0136.9--3510 are not
members of the sample of 110 bright soft X-ray selected AGN by
\citet{gru01} and \citet{grupe04}, they were added to our observing program.
 Both AGN are very interesting NLS1s.  RX J0134.2--4258 is one of a
few radio-loud NLS1s \citep{grupe00, komossa06} and had shown one of
the softest X-ray spectra of any AGN during the RASS. However, when it
was observed by the ROSAT PSPC about two years after the RASS, its
X-ray spectrum had dramatically hardened. An ASCA observation in
December  1997 confirmed this hard X-ray spectrum
\citet[]{grupe00}. While \citet{komossa97} and \citet{komossa00}
suggested that this variability is due to a variable `warm' ionized
absorber, \citet{grupe00} discussed that the spectral variability
maybe due to the absence and recovery of the accretion disk corona.
RX J0136.9--3510 also had one of the steepest X-ray spectra seen of any AGN
during the RASS. \citet{ghosh04} also discussed the presence of a
highly blueshifted Fe K$\alpha$ line seen by ASCA in this NLS1 and
\citet{jin09} reported on super-Eddington accretion flows. Our results presented
here also suggest super-Eddington accretion.

\subsubsection{RX J0148.3--2753}
The NLS1 RX J0148.3--2753 has been the target of several \swift\ fill
in and ToO observations. The results of the \swift\ observations
performed in 2005 May and December have been published by
\citet{grupe06}. This NLS1 is highly variable in X-rays by factors of
more than 5. We found that the flux variability is associated with
strong X-ray spectral variability. In this paper we only present the
data previously not published in \citet{grupe06}. 

\subsubsection{Mkn 110}

Mkn 110 has been a highly variable AGN in X-rays as well as at optical/UV 
wavelengths \citep[e.g.][]{gru01, kollat01, kollat06}.
Some caution has to be taken into account regarding the Balmer decrement. Because the fluxes in the optical emission lines are 
highly variable as show by \citet{kollat01, kollat06}, we do not know what the exact Balmer decrement was during the time of the \swift\ 
observation. For the correction for intrinsic reddening we used the optical spectrum published in \citet{gru04a}. 

\citet{veron07} discussed the nature of Mkn 110 and concluded that 
most-likely it is a BLS1s although its FWHM(H$\beta$) is less than 2000
km s$^{-1}$. Our continuum measurements presented here seem to support this assumption. In most plots Mkn 110 appears 
to be among BLS1s and not NLS1s.

\subsubsection{PG 1211+143}

The NLS1 PG 1211+143 was monitored by \swift\ simultaneously with
ground-based optical photometry between March and May 2007
\citep{bachev09}. It was caught  by \swift\ in a low state, about 10
times fainter than expected from previous ROSAT and \xmm\
observations.  At the end of our two-month monitoring campaign
PG1211+143 returned into its normal high state. For this paper we used
the observation from 2007 April 02 which was one of the high state
observations. 

\subsubsection{Mkn 766}

The NLS1 Mkn 766 was monitored by \swift\ for half a year between 2006
December and 2007 May (Grupe et al. in prep.) quasi-simultaneously with RXTE
and ground-based optical photometry.  For this paper we used the first of these
observations from 2006 December 21.

\subsubsection{3C 273}
The first quasar ever discovered, 3C 273 \citep{schmidt63}, has been
the target of many \swift\ observations since the start of the
mission.  3C 273 is a standard calibration source and also has been
the target of several monitoring campaigns.  However, not all
observations are suitable for our study. 3C 273 is the brightest AGN
in our sample and if observed in our standard observing modes  (XRT in
PC mode, UVOT in full frame) the observations are strongly compromised
by pileup/coincident losses. In order to obtain usable data, 3C 273
has  to be observed in Windowed Timing mode in XRT and in  the
5$^{'}\times5^{'}$ hardware window mode in the UVOT.    Because it is
beyond the scope of his paper to list all observations performed so
far, we picked some more recent observation from  the beginning of
2009 where 3C 273 was observed in WT in XRT and the UVOT
5$^{'}\times5^{'}$ hardware window. Other \swift\ observations of 3C
273 are listed in e.g. \citet[][]{pacciani09}.

\subsubsection{NGC 5548}

The Seyfert 1.5 galaxy NGC 5548 has been the target of two \swift\
monitoring campaigns, one in 2005 April/May \citep{goad06} and in 2007
June to  August (Grupe et al. 2010, in prep). The latter one was
simultaneous with Suzaku \citep{krongold09, liu09} in order to study the
variable 'warm' absorber in this AGN.  In the study presented here we
used the XRT and UVOT observation from segment 059 from the beginning
of the 2007 monitoring campaign on 2007 June 19. Note that NGC 5548 is
the AGN with the lowest Eddington ratio \lledd\ in our sample.  

\subsubsection{RX J2217.9--5941}
The NLS1 RX J2217.9--5941 was discovered during the RASS as a bright
X-ray AGN with a very soft X-ray spectrum \citep{gru98a}. However,
when it was observed in May 1997 and April 1998 by the {\it ROSAT}
High-Resolution Imager and in May 1998 by {\it ASCA} it appeared to have
become significantly fainter \citep{grupe01b}. Two follow-up
observations with \chandra\ confirmed this low state
\citep{grupe04b}. The spectra derived from the  \chandra\ data suggest
the presence of a partial covering absorber. Our \swift\ observations
also found this AGN in a low state.  Between the RASS observation and
the last \swift\ observation RX J2217.9--5941 has shown a decay in its
0.2-2.0 keV flux by a factor of about 30.  

\section{\label{discuss} Discussion}

In this paper we presented a sample of 92 bright soft X-ray selected
AGN with simultaneous optical/UV and X-ray observations using \swift\
in order to study the optical to X-ray spectral energy distribution in
AGN.  Many of these AGN exhibit strong variability in X-rays as well
as at UV wavelengths. Therefore, studying the SEDs of AGN, performing
the observations in the UV and in X-ray simultaneously is
critical. The NASA GRB Explorer mission \swift\ has the unique ability
to perform this type of multi-wavelength observations with relatively
short snap-shot observations. As a result, more than 95\% of the AGN
in our sample have been observed by \swift\ multiple times.  

The mean and median values of \aox\ and \auv\ of NLS1s and BLS1s
suggest that the SEDs are  relatively similar. However, NLS1s appear
to be  intrinsically fainter in the UV compared with
BLS1s. This result is most likely due to the accretion disk
temperatures in NLS1s and BLS1s. NLS1s have, generally speaking,
smaller black hole masses. Because of the $T \propto M^{-1/4}$
relation between the temperature of the accretion disk  and the mass
of the central black hole, NLS1s have hotter accretion disks than
BLS1s.  Therefore their BBB thermal spectra are shifted towards higher
energies making the  UV spectra appear fainter.  Our result reflects
very well the SED models used by \citet{kelly08} shown in their Figure
13.  This interpretation also explains the findings by \citet{atlee09}
and \citet{rodriguez97} that NLS1s appear to be fainter in the UV when
compared with other samples. 

One of our main interests in our study was to investigate whether AGN
with steep X-ray spectra also show blue optical/UV spectra as 
previously reported by e.g. \citet[][]{walter93} and \citet{gru98a}
using non-simultaneous data.   On a first glance there seems to be
only a slight correlation between \ax\ and \auv.  Certainly reddening
in the UV may produce scatter in this relationship, but that is not a
plausible explanation for the lack for correlation in the whole
sample.  The answer regarding the relationship between \ax\ and \auv\
seems to be a bit more complicated. Examining the BLS1s (filled
red circles in Figure\,\ref{ax_auv}) first, and excluding IRAS
1334+24, the AGN in the upper right 
corner which is known to have very strong reddening in the optical/UV
\citep{wills92, gru98b}, we find a fairly strong correlation between
\ax\ and \auv. Now the correlation coefficient is $r_{\rm l}=-0.49$
with a probability P=0.0011 of a random result. For a Spearman rank
order correlation we found $r_{\rm s}=-0.53$ and  $T_{\rm s}=-3.9$
with P=0.0004.  For the NLS1s, however, there appears to be nearly no
relationship between \ax\ and \auv\  ($r_{\rm l}=-0.21$).  The reason
for this appears to be related to the fact that NLS1s  operate close to
or above the Eddington limit.  This is not the case, however, for
BLS1s which operate at lower $L/L_{\rm Edd}$.  Looking back at the
entire sample,  we see that the relationship saturates at about
\ax=1.6. For objects with an X-ray spectrum  steeper than \ax=1.6 the
optical/UV slope remains nearly constant, and these AGN are dominated
by NLS1s.  When excluding all AGN with \ax$>$1.6, we find a clear
correlation that AGN with steeper X-ray spectra also show bluer
optical/UV continua ($r_{\rm l}=-0.43$,  P=0.0009; $r_{\rm S}=-0.36$,
$T_{\rm S}=-2.8$, P=0.0071).  
Our results, therefore, confirm the findings by \citet{walter93} and \citet{gru98a}.  

Another aspect of our study was to determine if there are observable 
properties that can be used to describe the SEDs and estimate
\lledd. As shown previously by \citet{grupe04} and \citet{shemmer08}, 
\lledd\ can be estimated from the X-ray spectral slope \ax. However,
we find that this relationship is only valid for AGN with X-ray
spectral slope \ax$<$1.6.  At about this point, $L/L_{\rm Edd}$
reaches unity.  For AGN with steep X-ray spectra, thus predominantly
NLS1s, this \ax - \auv\ relation breaks down.  As shown in the middle
panels of Figure\,\ref{ax_auv_ledd},  \auv\ can be used as an
estimator of  $L/L_{\rm Edd}$. This slope can actually predict
$L/L_{\rm Edd}$ even better than the X-ray spectral slope \ax\ because
it does not show the saturation that we see in the \ax-\lledd\
relation.  One interesting result of the \auv - $L/L_{\rm Edd}$
relation is that there is an offset between NLS1s  and BLS1s by about 0.6 dex. 
 So, for
a given optical/UV slope \auv, BLS1s show a lower Eddington ratio by a
factor of $\approx$4 compared with NLS1s (see Eqs. 2 to 10). The main
effect here is from the larger black hole masses found in BLS1s; on
average, BLS1s have a 10 times larger black hole mass than NLS1s. As
an alternative this result may also suggest that there is a physical
difference in the accretion disk properties of NLS1s and BLS1s.  As
shown by \citet[e.g. ][]{wandel88}, accretion discs with a 
blackbody spectrum modified by Comptonization 
have redder optical/UV spectra
for a given central black hole mass than those without
Comptonization. It could be that because NLS1s do accrete at high
\lledd\ a wind from the accretion disk causes the Comptonization layer
above the disk to be smaller than in BLS1s. Our conclusion differs
from that by \citet{atlee09} who suggested that the accretion disk
properties are independent  of the mass of the central black hole and
\lledd. 

The immediate application of these relations is for high-redshift
quasars. In that case it is difficult to estimate  $L/L_{\rm Edd}$
based on their X-ray data because their rest-frame soft X-ray 0.2-2.0
keV spectra are mostly shifted out of the observed energy window and
the majority of them are too X-ray faint to derive useful X-ray
spectra \citep[e.g. ][]{grupe06}.  Because rest-frame optical/UV
spectra exist for most of these AGN, one can use the spectral slope
\auv\ in order to  determine $L/L_{\rm Edd}$. When applying this
method to an AGN one has to make sure that one has the correct
classification of the AGN because of the offset between NLS1s and
BLS1s. 
 As a test, we applied our relation to the sample of
10 intermediate redshift QSOs presented by \citet{dietrich09} for
which rest-frame spectra of the H$\beta$ line region were taken. From
the rest-frame UV spectra we derived the Eddington ratios using the
relation for BLS1s (equation \#4)  and found that our estimates are in
good agreement with those \lledd\ values reported in
\citet{dietrich09}.  This suggests that our method is quite
reliable. However, it still requires a much larger sample of
intermediate and high redshift QSOs to verify this result.

Regarding the bolometric corrections, 
in Figure\,\ref{l5100_lbbb} we show the relationships between the
luminosity at 5100\AA\ and in the rest-frame 0.2--2.0 keV band vs. the luminosity
in the BBB \lbbb. In both panels the solid lines are the relations
found in our sample as given in equations (12) and (13). The dashed
line in the left panel shows the relation found by \citet{elvis94}. 
Our sample shows only a slight luminosity dependence of the bolometric
correction (i.e., $L_{\rm bbb}$) on 5100\AA\ or 0.2-10 keV luminosity.


One interesting result we found is  the trend that BLS1s appear to be
more  variable in the UV compared with NLS1s.  At first glance,
because they vary strongly in X-rays, one might expect that NLS1s 
would also be highly variable in the UV.  However, they are not.  This may be because
the UV in lower accretion rate objects comes from relatively close to
the black hole, where time scales are shorter, while the UV in NLS1s
comes from a more extended part of the accretion disk, where less
variability is expected. Another reason may be the effect of partial
covering absorption on the X-ray emission.  As we have seen in, e.g., 
Mkn 335 and WPVS 007 \citep{grupe07, grupe07b, grupe08, grupe08b, leighly09}, 
some NLS1s can vary strongly in X-rays, but only show minor
variability in the UV. In both cases, a possible explanation is the
presence of a partial  covering absorber that affects the spectrum at
X-ray energies,  but does not affect the UV range that much, assuming
it is   dust-free and compact.  

One of our main concerns regarding our study  was that the optical/UV
and X-ray data were taken simultaneously.  Nevertheless we found
similar results for the whole sample as have been reported earlier
from samples with non-simultaneous   observations, such as
\citet{walter93} or \citet{gru98a}. Is there still an advantage in
simultaneous data? The answer is 'yes'. There are several arguments
that make simultaneous observations crucial for this type of study:
1) Non-simultaneity smears out trends.  While this may not be a
problem for very large samples \citep[e.g., those considered
by][]{gru98a,atlee09}, in which the trends dominate the effect of
smearing, this would not apply for individual sources.  One of our
aims was to find relations between  observed parameters to get an
estimate of \lledd. As we have shown in some examples, this will not
work if the AGN is highly variable in the UV as well as in the  
X-rays (such as ESO242-G008, or RX J2349.4--3126) as shown in the SEDs
displayed in the appendix.  2) Observing the data simultaneously in
the optical/UV and in X-rays reduces the scatter in the relationships.
As shown above this is especially important if the sample contains
BLS1s, which tend to show stronger UV variability than NLS1s. One
example here is the \aox-log $l_{\rm 2500\AA}$ relation. While the
scatter in our relation is relatively small it appears to be
significantly larger in the sample of \citet[e.g.,][]{gibson08} and 
\citet{yuan98}. In other words, in a sample with simultaneous data, 
relations can be constrained better than with non-simultaneous data sets. 

Future work on our SED project includes obtaining \swift\ observations
for the entire sample of 110 bright soft X-ray selected AGN
\citet[]{gru01, grupe04} and extending the sample to hard X-ray selected
AGN by making use of existing data either from \swift\ or \xmm\
observations. The goal here is to increase the statistics on the
results and to avoid the selection effects that are currently in our
soft X-ray flux-selected AGN sample.  In a second paper we will also
perform a more detailed statistical analysis including a Principal
Component and cluster analysis  to examine the connections between
continuum and optical emission line properties, and will then carry
out a rigorous interpretation of these results with respect to
implications for NLS1 models.

\section{Conclusions}

We have derived SEDs for a sample of 92 AGN with {\em simultaneous}
optical-UV and X-ray observations with \swift.  88 of these AGN are
observed in at least 3 of the 6 UVOT filters.  This is the largest
sample of AGN with simultaneous multi-wavelength coverage, and one of
the largest with repeated UV observations. Our results can be summarized
as follows:  
\begin{itemize}
\item We provide bolometric corrections (Equations (12) and (13))
starting from the optical or X-ray luminosity, 
covering several decades in luminosity.
\item NLS1s appear to be fainter in the UV and at hard X-ray energies and show steeper X-ray spectra than BLS1s.
\item NLS1s and BLS1s have very similar \auv\ and \aox.
\item We found a strong correlation between \ax\ and \auv. 
   However, this relation saturates for AGN with \ax$>$1.6 (i.e., predominantly for NLS1 galaxies).
\item We confirm earlier findings by e.g. \citet{grupe04} and \citet{shemmer08} 
 that AGN with steeper X-ray spectral indices have higher \lledd.
Again, the relation saturates for AGN with \ax$>$1.6.
\item There is a clear anti-correlation between the optical/UV spectral slope \auv\ and \lledd. 
AGN with bluer \auv\ have higher \lledd. 
However, there is an
offset between NLS1s and BLS1s and for a given \auv\ NLS1s have
about 4 times  higher \lledd\ than BLS1s.
\item The relationships we found between \auv\ and \lledd\ for NLS1s and BLS1s 
can be used to estimate \lledd\ in, e.g., high-redshift QSOs.
\item Although NLS1s show strong X-ray variability, they vary only marginally in the UV.
\end{itemize}

\acknowledgments

We would like to thank the \swift\ team for performing the observations of our AGN sample, 
in particular the \swift\ PI Neil Gehrels for approving our various ToO requests for observing the AGN 
when they were found in an interesting flux state. Special thanks to   
the \swift\ science planners Judy Racusin, Sally
Hunsberger, Claudio Pagani, David Morris, Mike Stroh and Peter Brown, and our referee Lisa Winter 
for excellent comments and suggestions that significantly improved the paper.
This research has made use of the NASA/IPAC Extragalactic
Database (NED) which is operated by the Jet Propulsion Laboratory,
Caltech, under contract with the National Aeronautics and Space
Administration.
\swift\ at PSU is supported by NASA contract NAS5-00136.
This research was supported by NASA contract NNX07AH67G (D.G.).

\clearpage

\begin{figure*}

\def\eps@scaling{1.7}%

\plotthree{f1a.ps}{f1b.ps}{f1c.ps}
\caption{\label{xrt_spec_example} Examples of XRT spectra, from the left to the right, RX J0117.5--3826, Fairall 1119, and RX J0128.1--1848. All these spectra were fitted by an
absorbed single power law model.
}
\end{figure*}

\begin{figure}
\epsscale{0.75}
\plotone{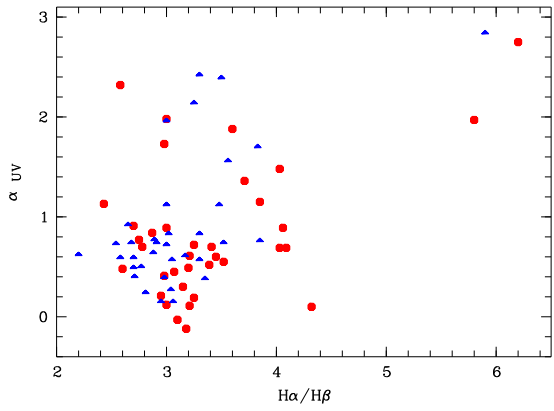}

\caption{\label{hahb_auv} Relation between the optical/UV slope \auv\ corrected for Galactic reddening and Balmer decrement H$\alpha$/H$\beta$. NLS1s are displayed as blue
triangles and BLS1s and filled red circles.
}
\end{figure}

\begin{figure*}
\epsscale{1.6}
\plottwo{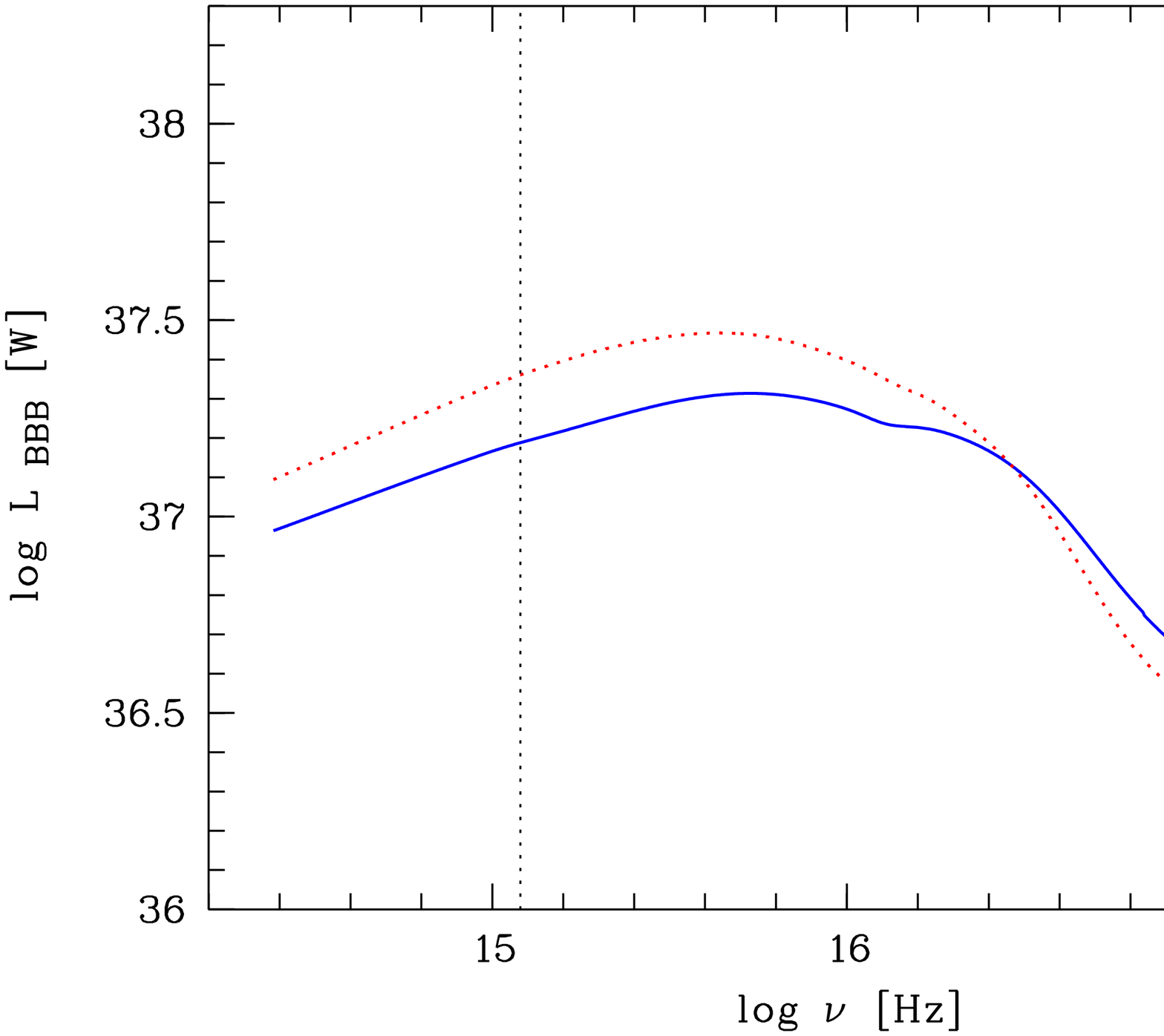}{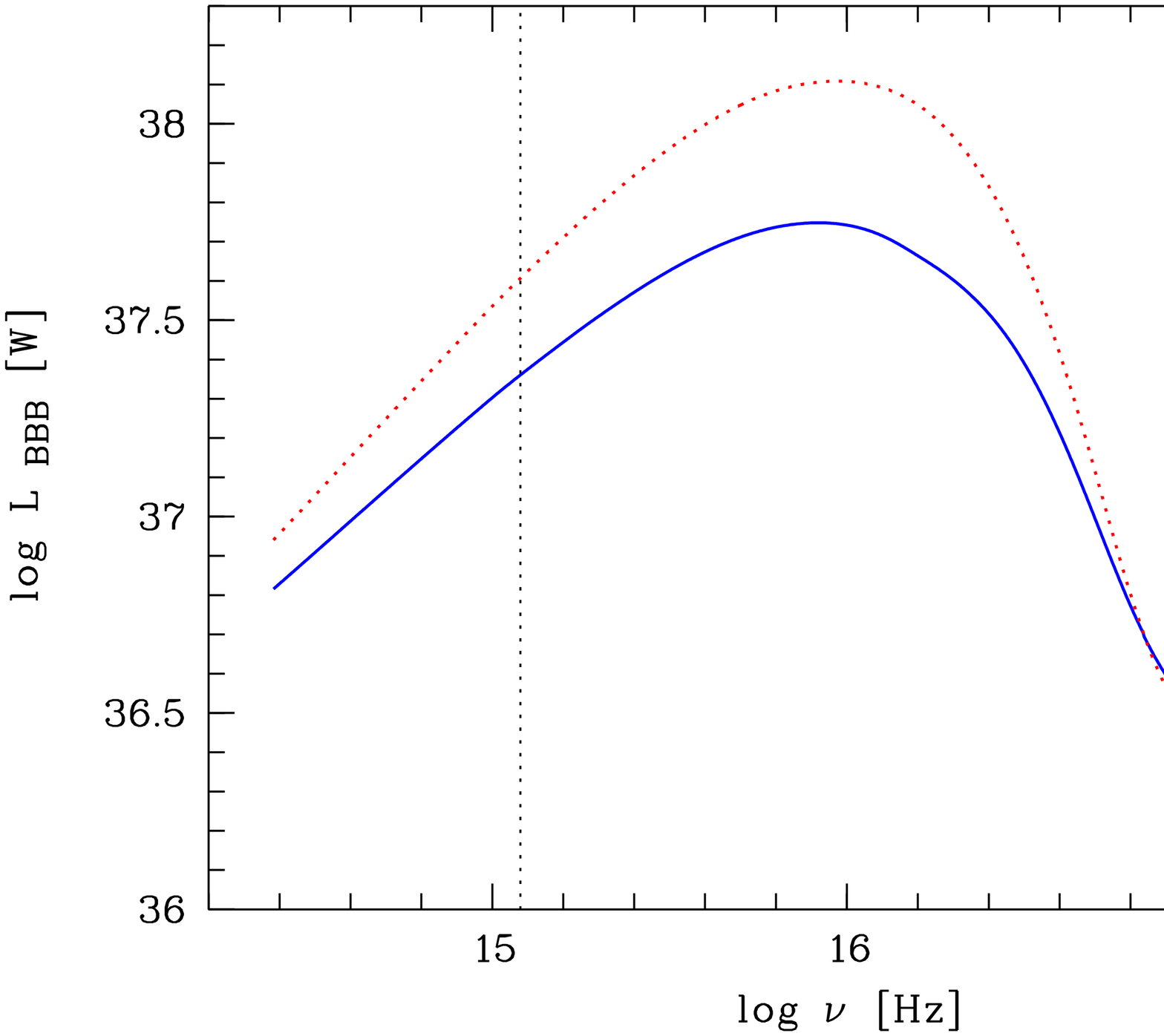}

\caption{\label{aver_sed} Average SEDs of NLS1s (solid blue line) and BLS1s (dotted red line). The left panel displays the SED uncorrected for intrinsic
reddening and the right panel shows the SED corrected for intrinsic reddening. The vertical dotted lines mark the frequencies at 2500\AA\ and 2 keV, which
are used to determine \aox.
}
\end{figure*}

\clearpage

\begin{figure}
\epsscale{0.75}
\plotone{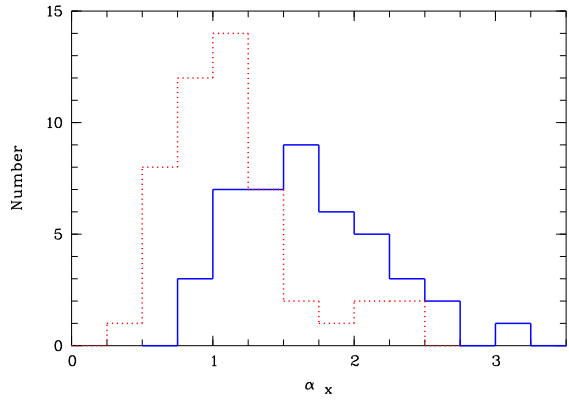}

\caption{\label{distr_ax} Distributions of the 
0.3-10.0 keV X-ray energy spectral slope \ax\  for NLS1s (solid blue line) and BLS1s (dotted red line).
}
\end{figure}

\begin{figure*}
\epsscale{1.6}
\plottwo{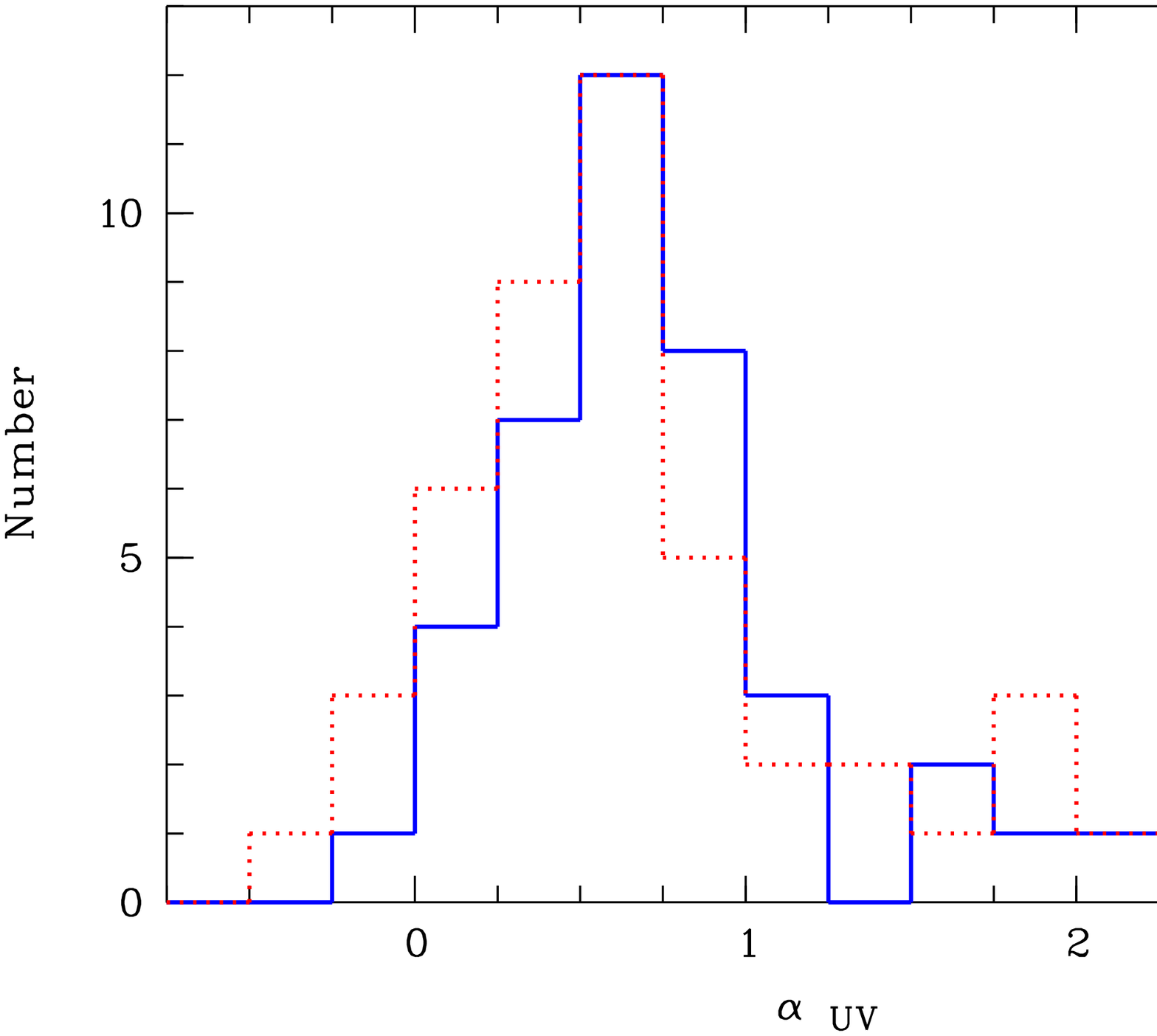}{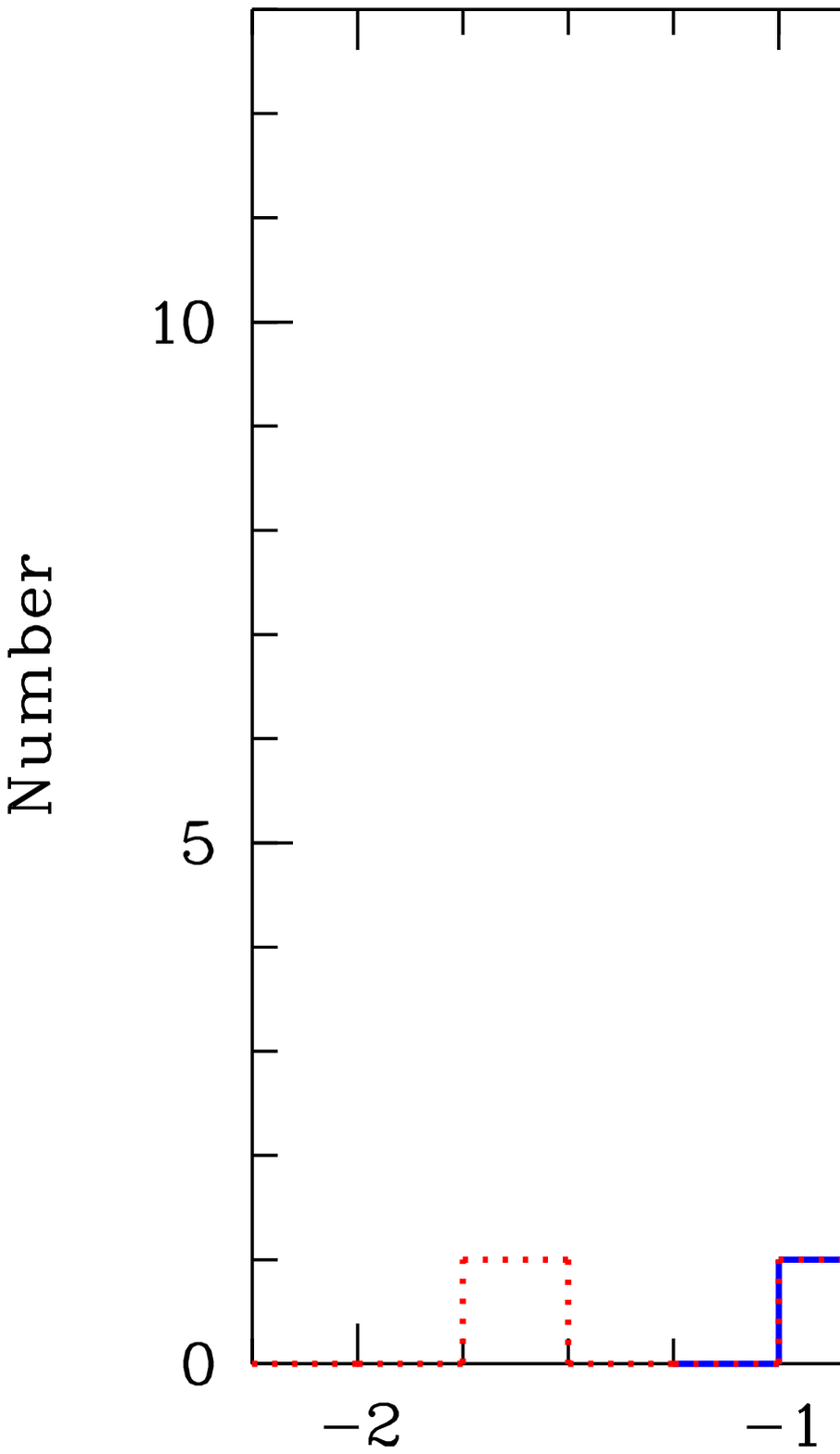}

\caption{\label{distr_auv} Distributions of the optical/UV 
spectral slope \auv\ uncorrected and corrected for intrinsic reddening (left and right panels, respectively). The coding for the lines is given in
Figure\,\ref{aver_sed}.
}
\end{figure*}

\begin{figure*}
\epsscale{1.6}
\plottwo{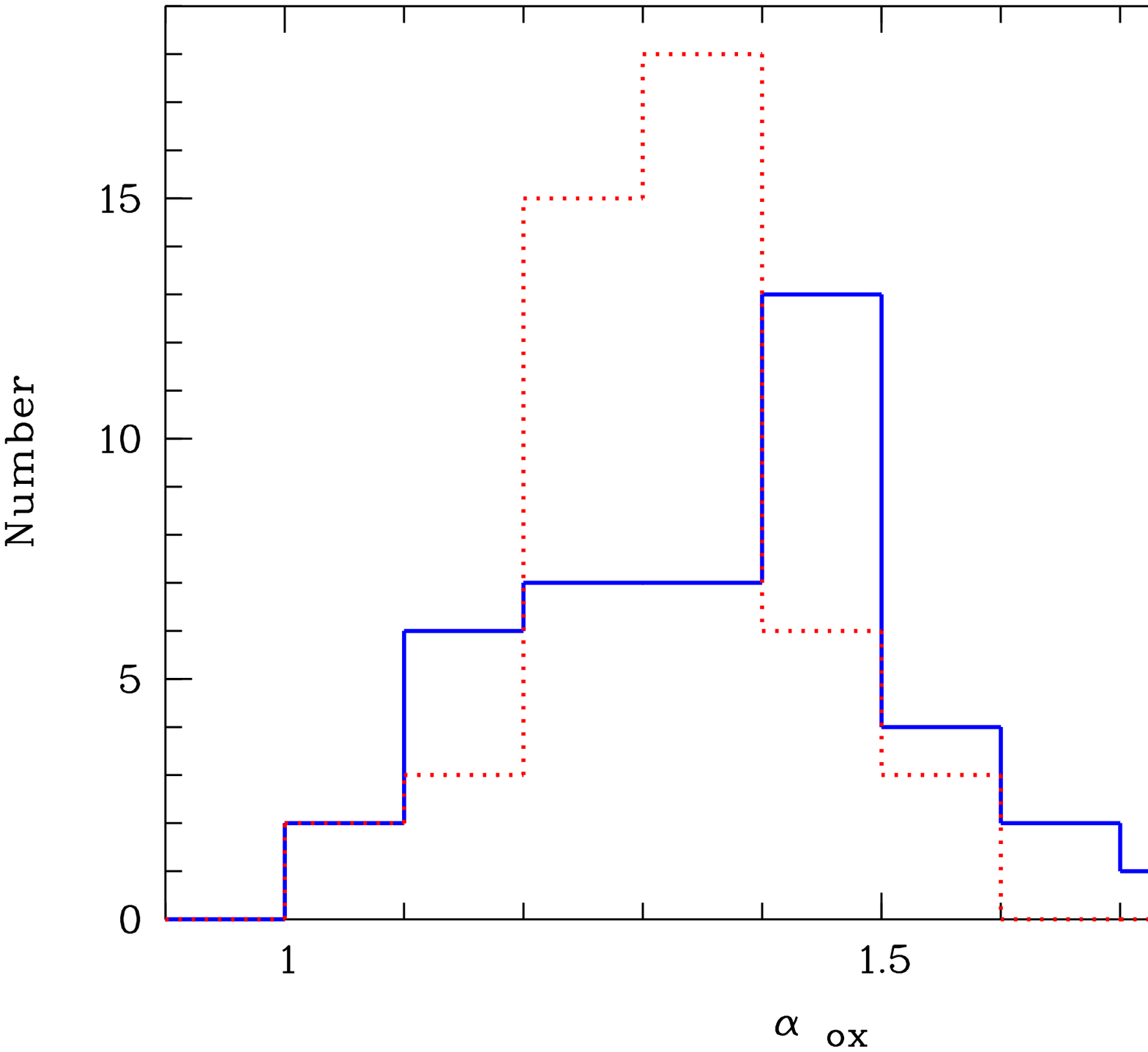}{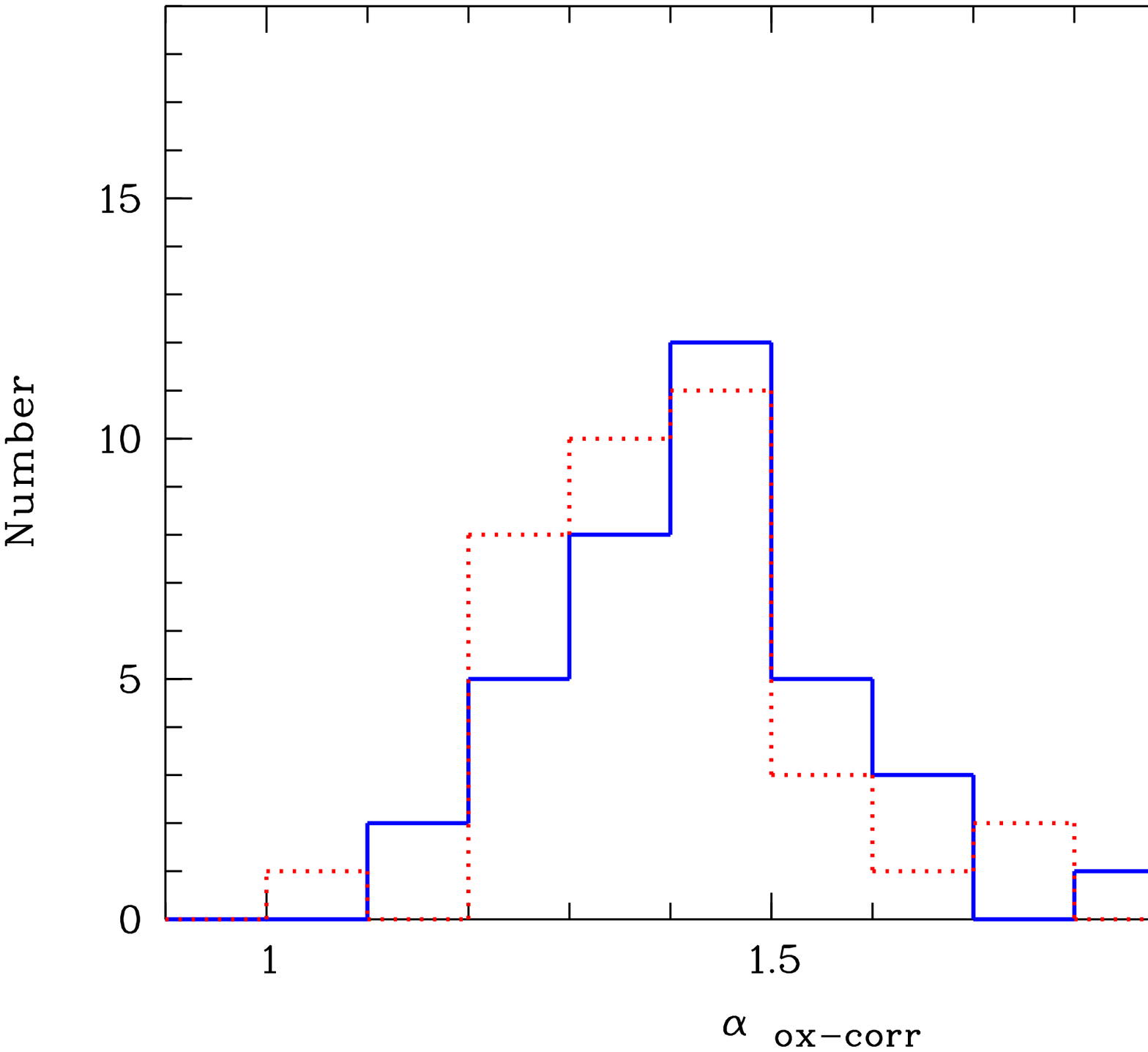}

\caption{\label{distr_aox} Distributions of the optical-to-X-ray 
spectral slope \aox\ uncorrected and corrected for intrinsic reddening (left and right panels, respectively). The coding for the lines is given in
Figure\,\ref{aver_sed}.
}
\end{figure*}

\begin{figure*}
\epsscale{1.6}
\plottwo{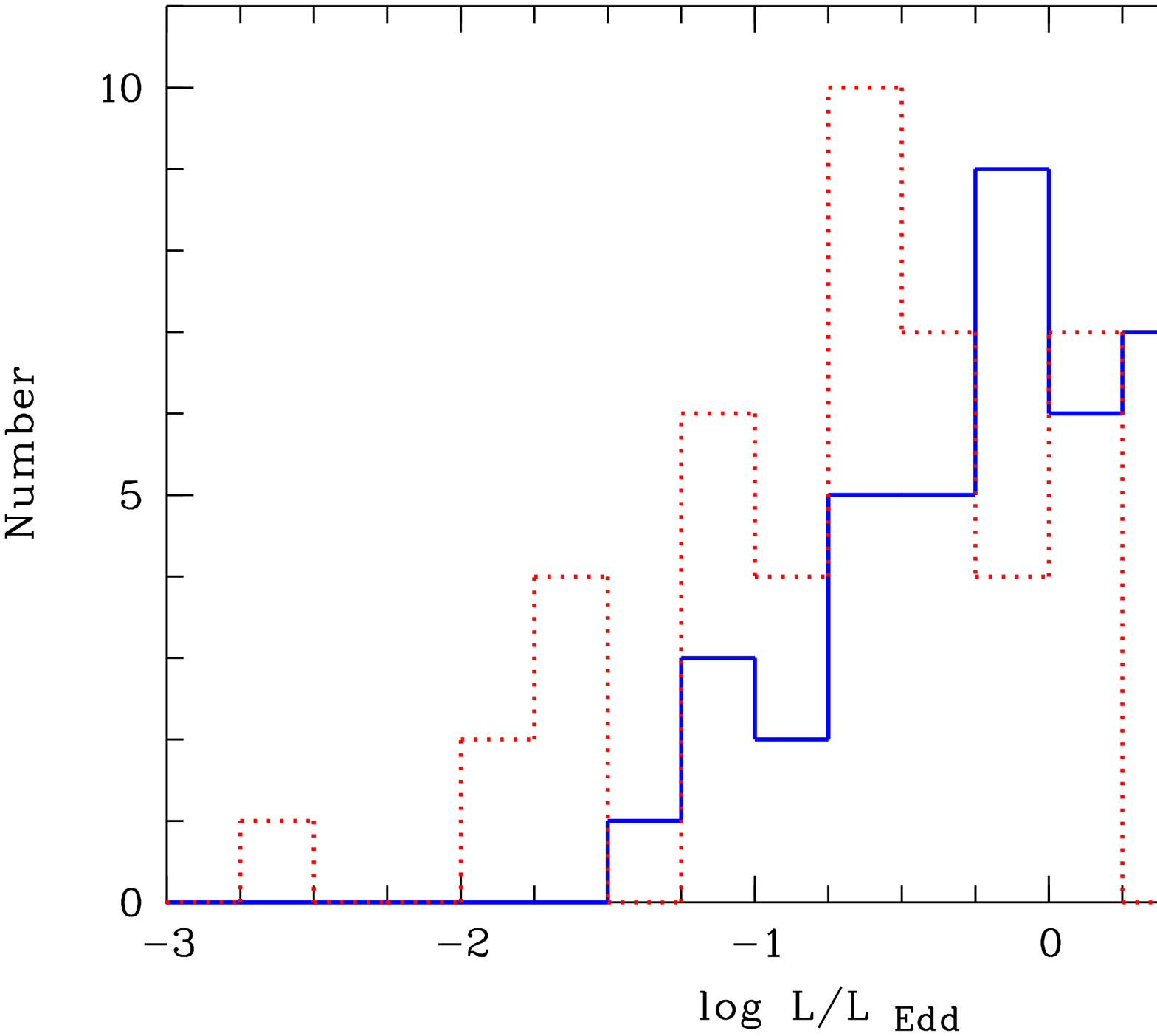}{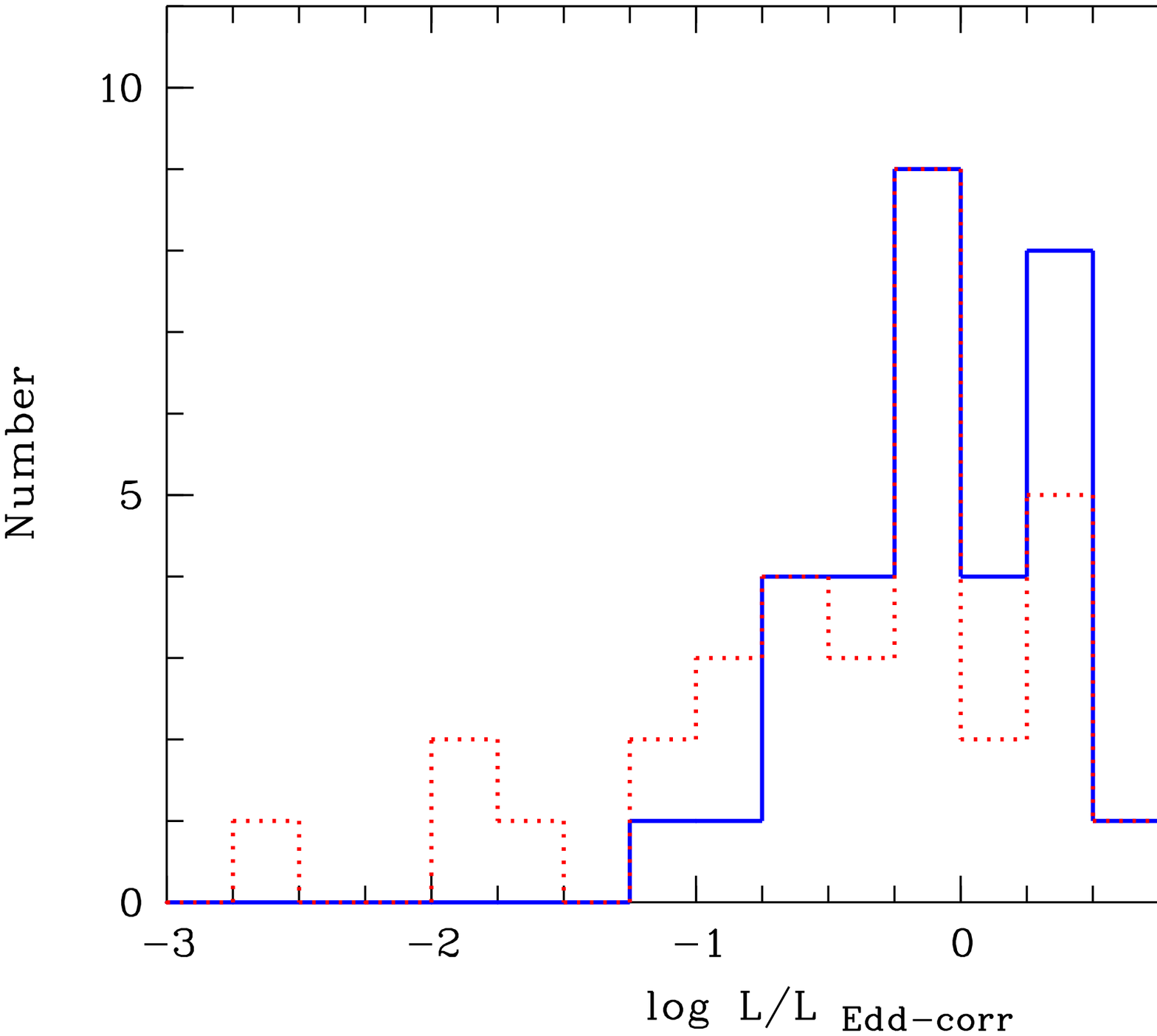}

\caption{\label{distr_l_ledd} Distributions of the Eddington ratios \lledd,
 uncorrected and corrected for intrinsic reddening (left and right panels, respectively). The coding for the lines is given in
Figure\,\ref{aver_sed}.
}
\end{figure*}

\begin{figure}
\epsscale{0.75}
\plotone{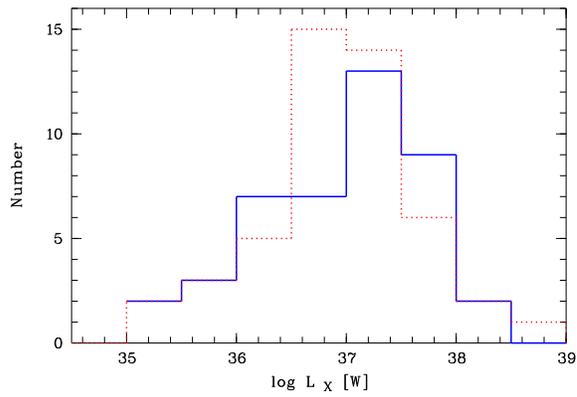}

\caption{\label{distr_lx} Distributions of the 
0.3-10.0 keV X-ray  luminosity log $L_{\rm X}$ for NLS1s (solid blue line) and BLS1s (dotted red line).
}
\end{figure}

\begin{figure*}
\epsscale{1.6}
\plottwo{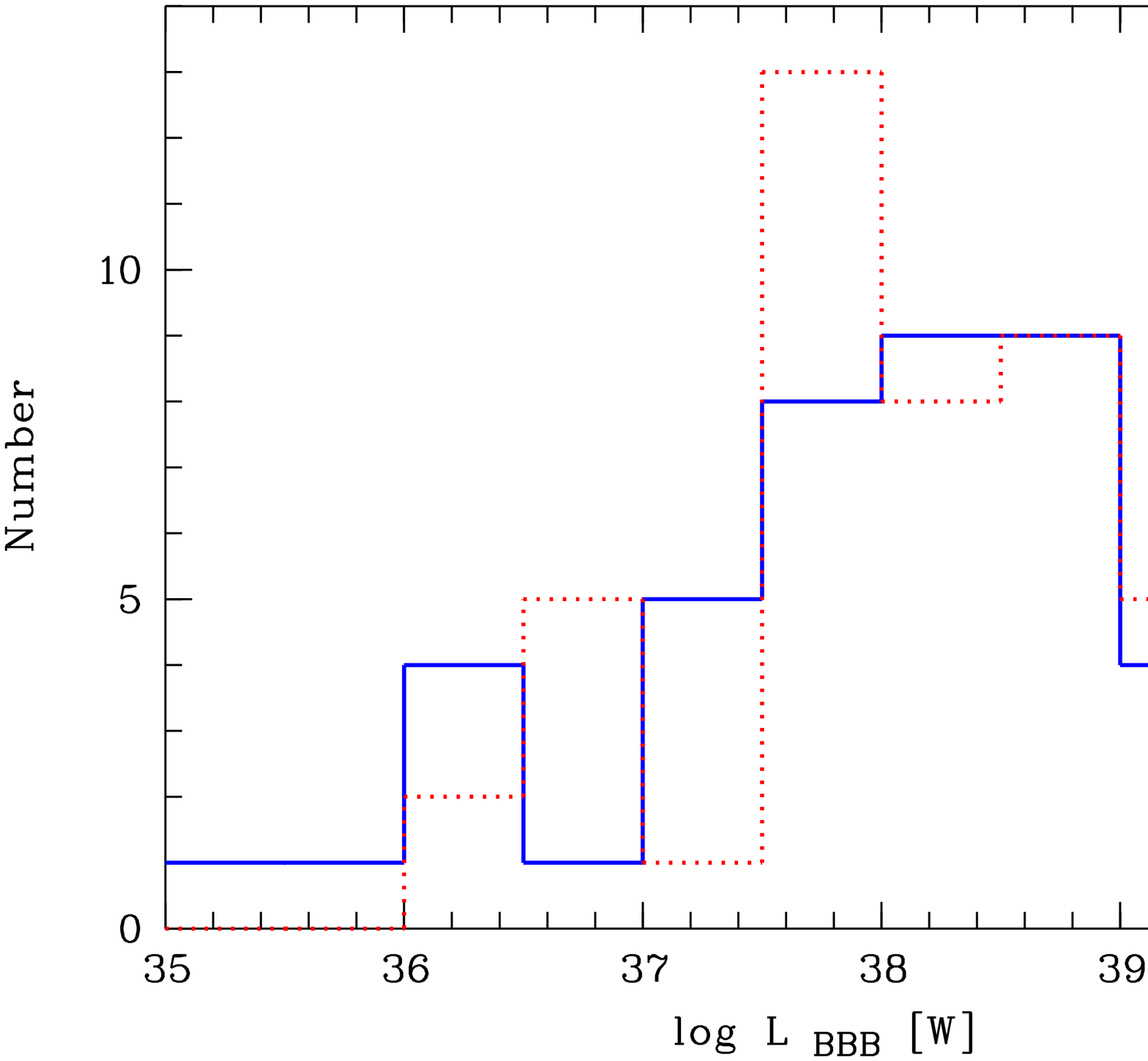}{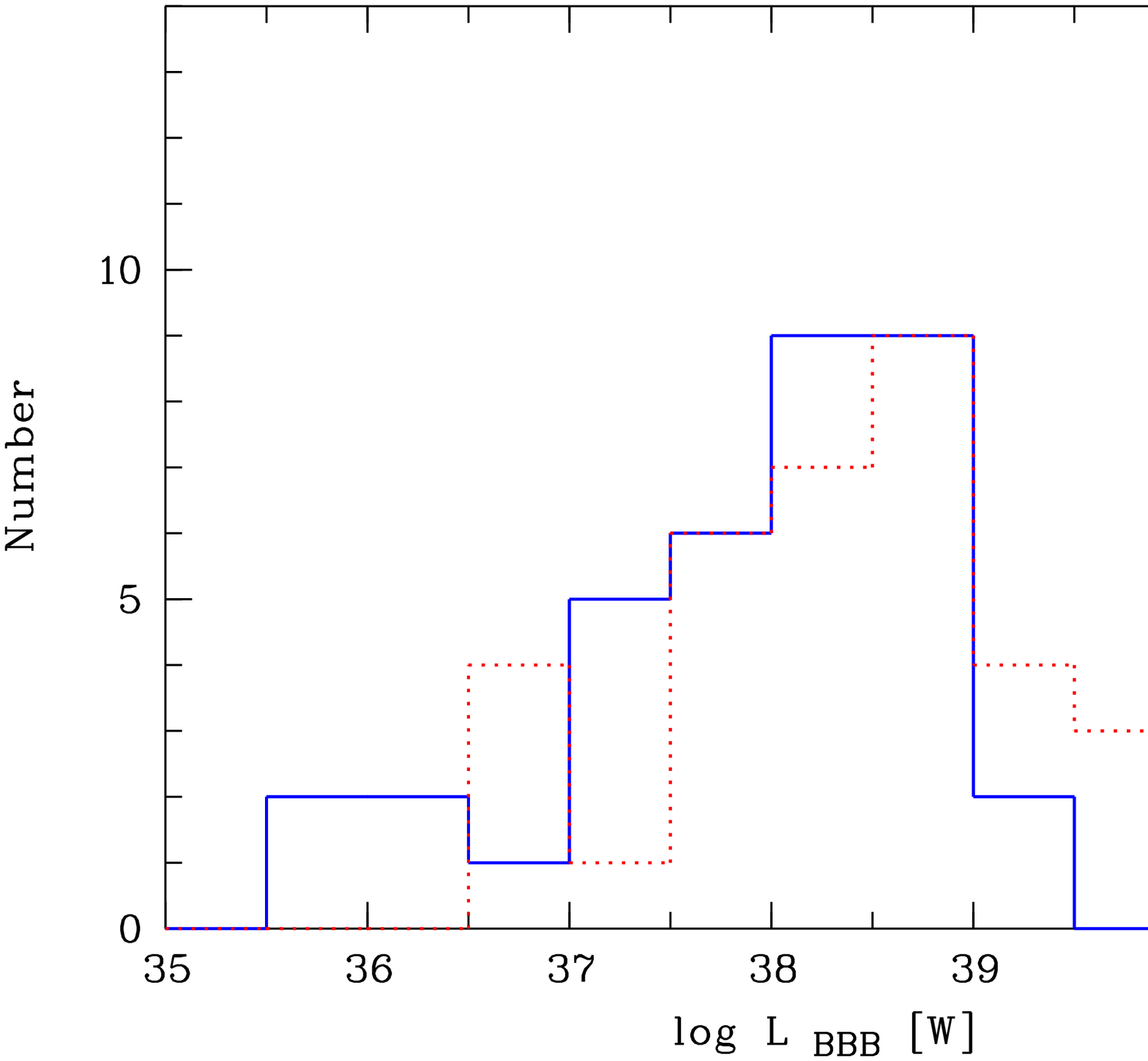}

\caption{\label{distr_lbbb} Distributions of the 
Big-Blue-Bump luminosity log $L_{\rm BBB}$ uncorrected and corrected for intrinsic reddening (left and right panels, respectively)
for NLS1s (solid blue line) and BLS1s (dotted red line).
}
\end{figure*}

\begin{figure*}
\epsscale{1.6}
\plottwo{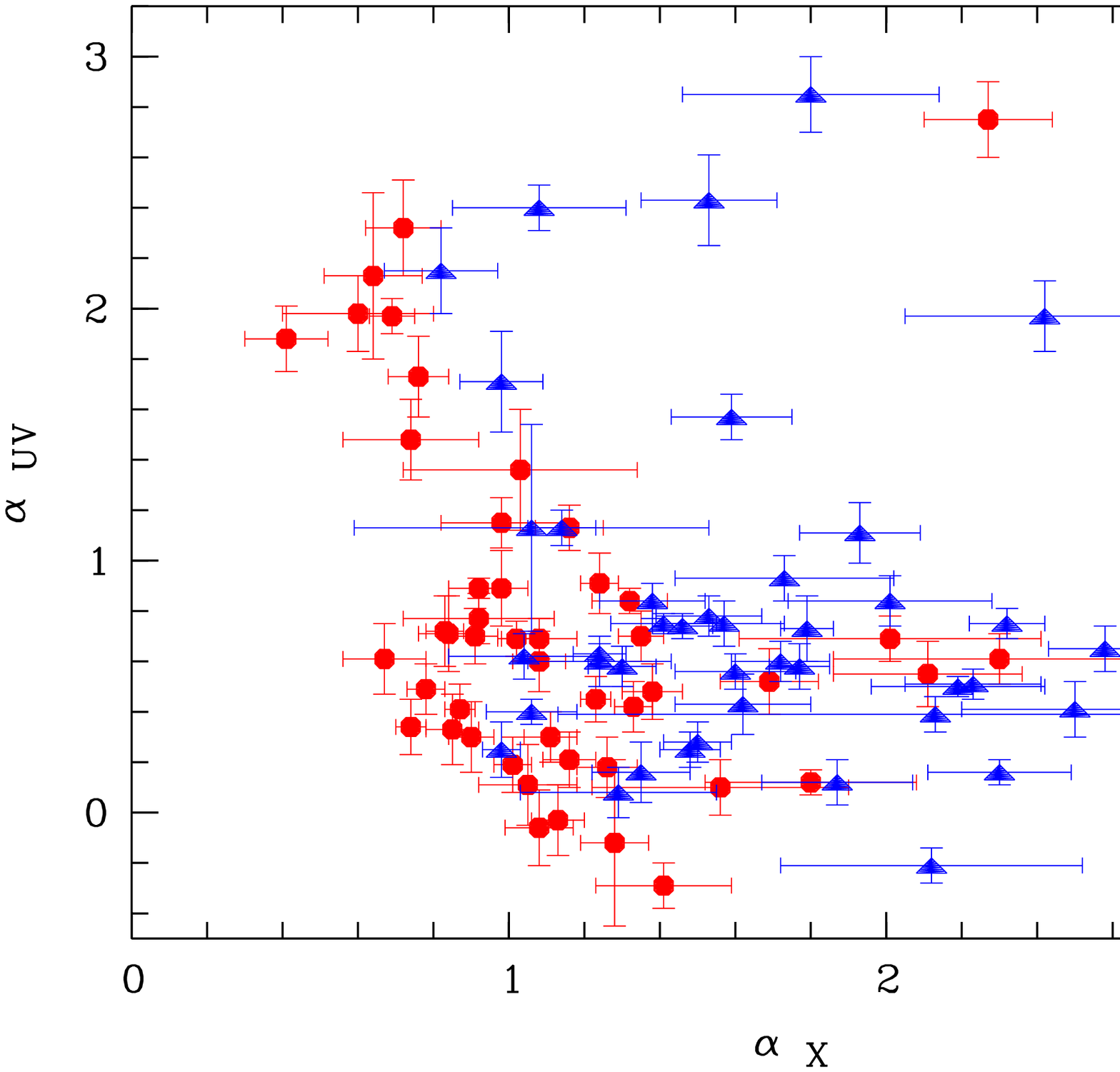}{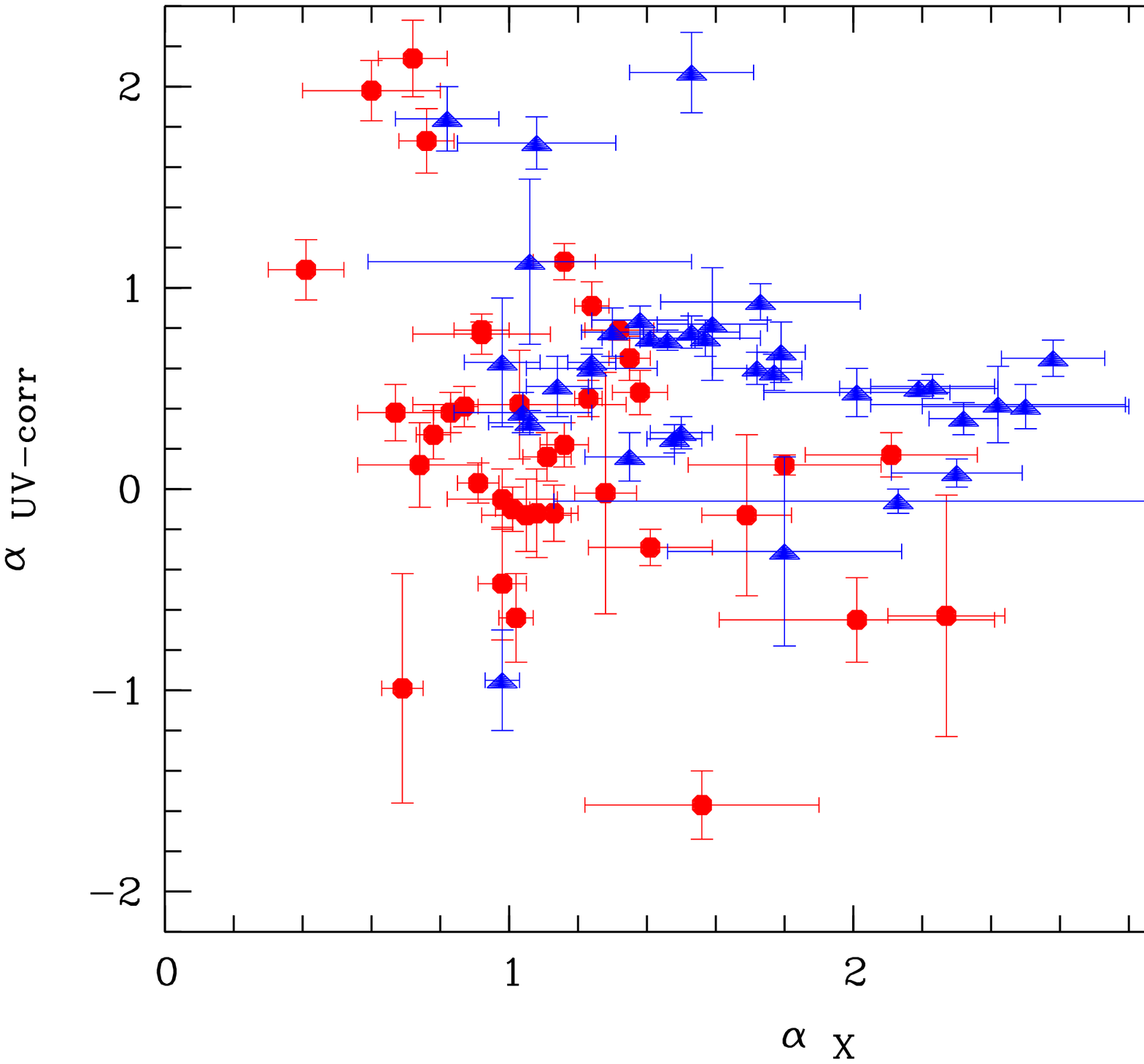}

\caption{\label{ax_auv} 0.3-10 keV X-ray energy spectral slope \ax\ and optical/UV spectral slope \auv. The left panel displays the \auv\ value
only corrected for Galactic reddening and the right panel the \auv\ value corrected for intrinsic and Galactic
reddening, determined from the Balmer decrement
given in Table\,\ref{obj_list}. NLS1s are displayed as blue triangles and BLS1s as red circles.  
}
\end{figure*}

\begin{figure*}
\epsscale{1.6}
\plottwo{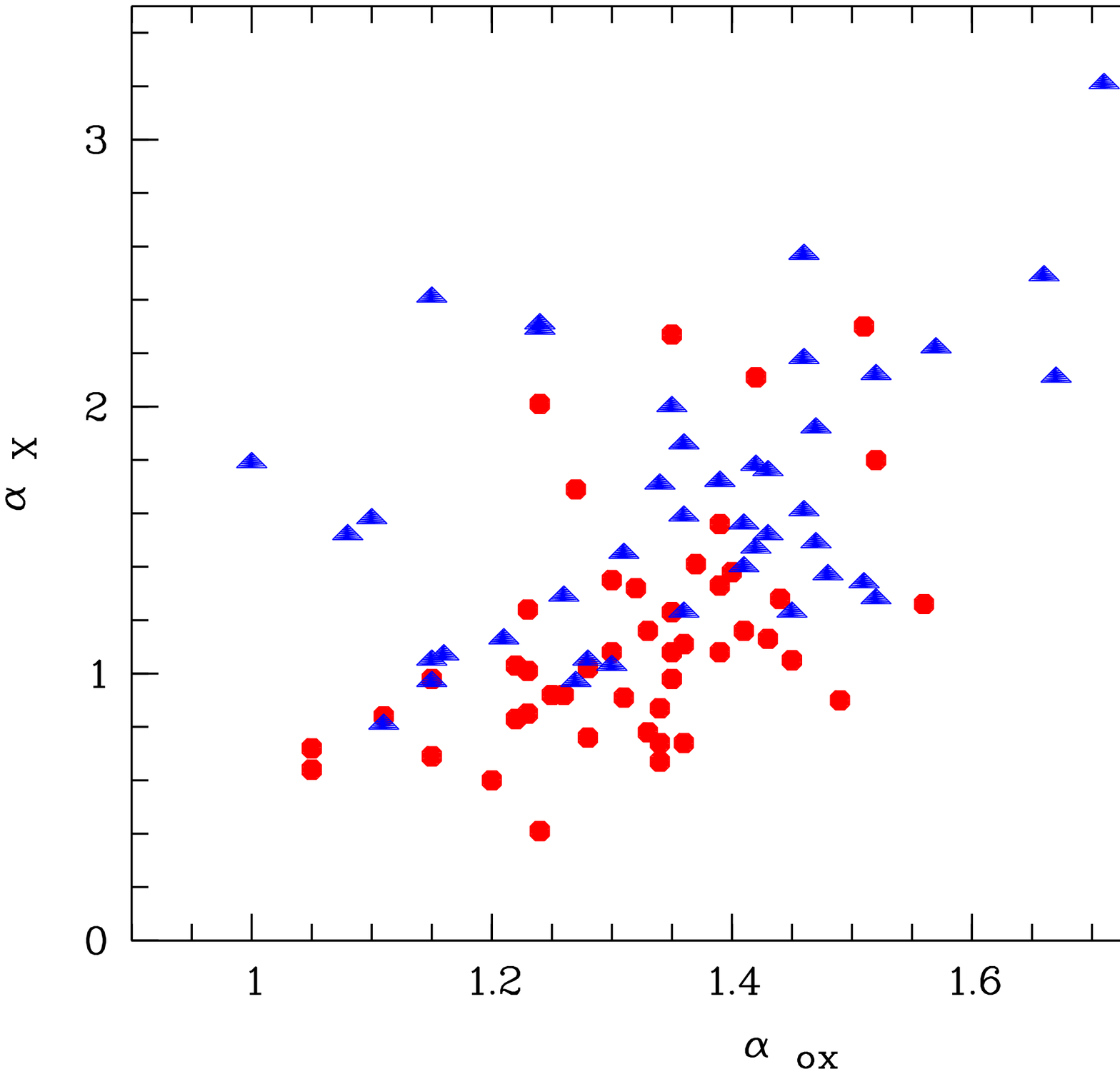}{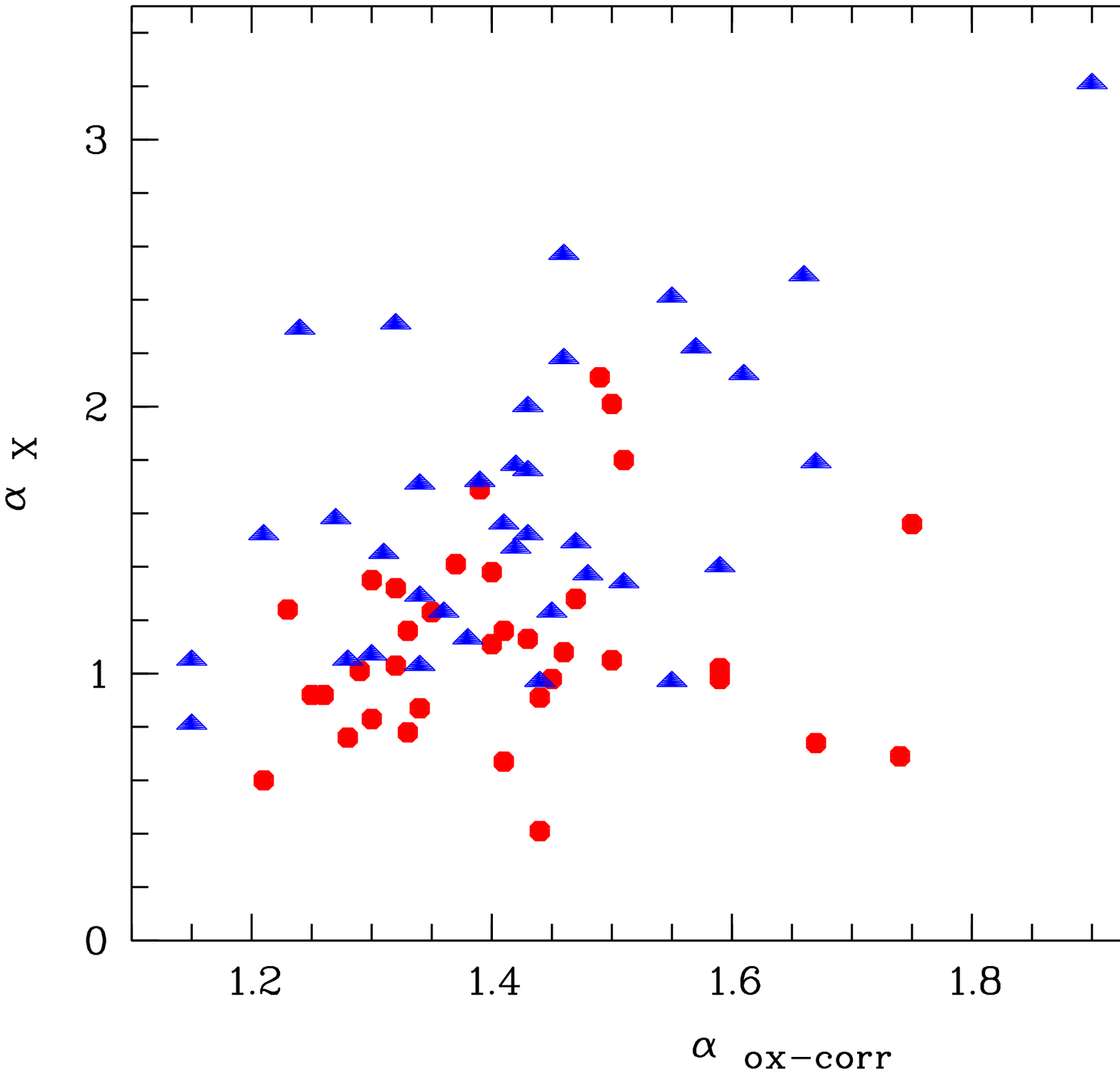}

\caption{\label{aox_ax} Optical-to-X-ray spectral slope \aox\ vs. X-ray spectral slope \ax. The left panel displays \aox\ 
only corrected for Galactic reddening and the right panel  corrected for intrinsic and galactic reddening, 
 NLS1s are displayed as blue triangles and BLS1s as red circles.  
}
\end{figure*}

\begin{figure*}
\epsscale{1.6}
\plottwo{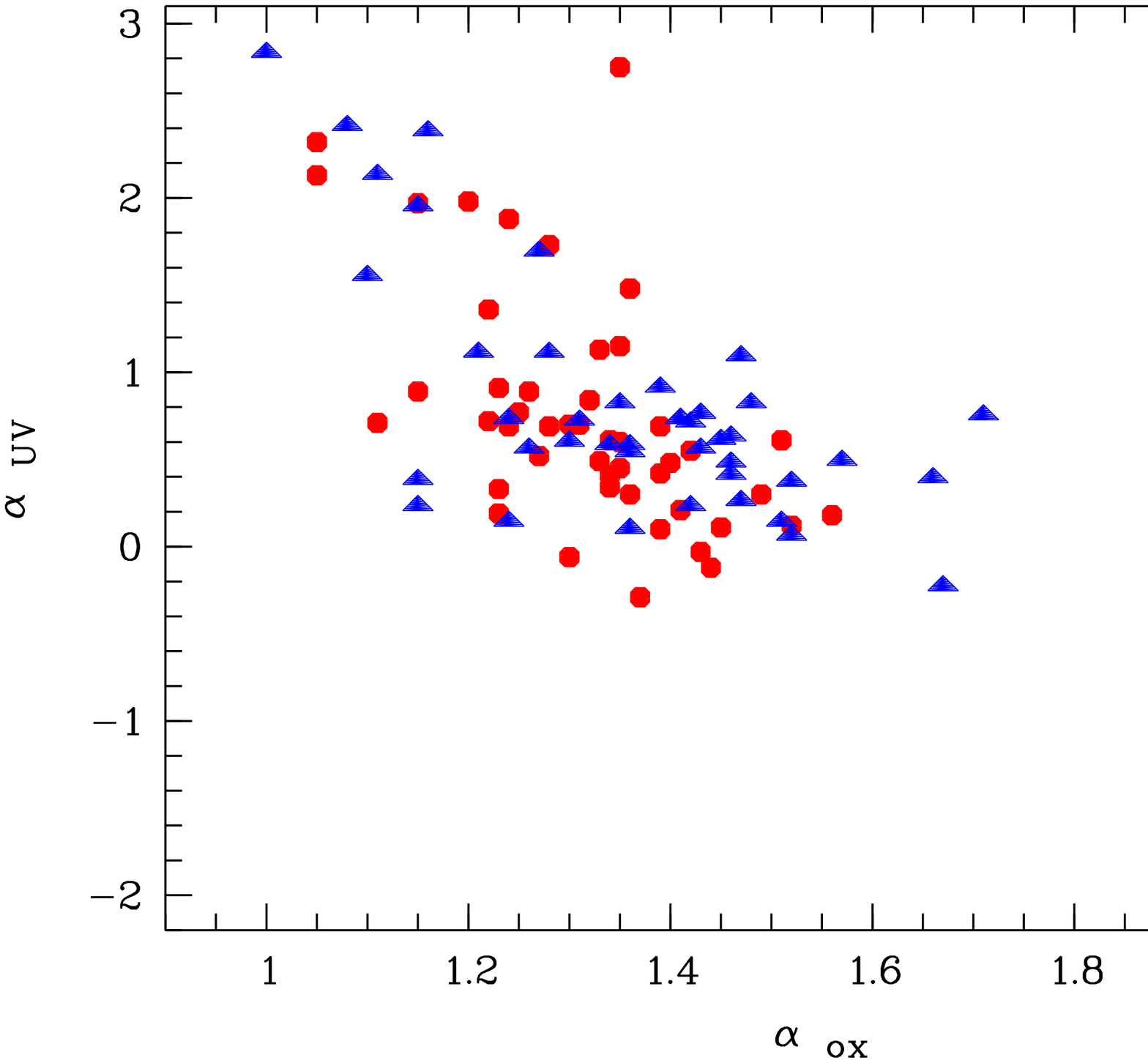}{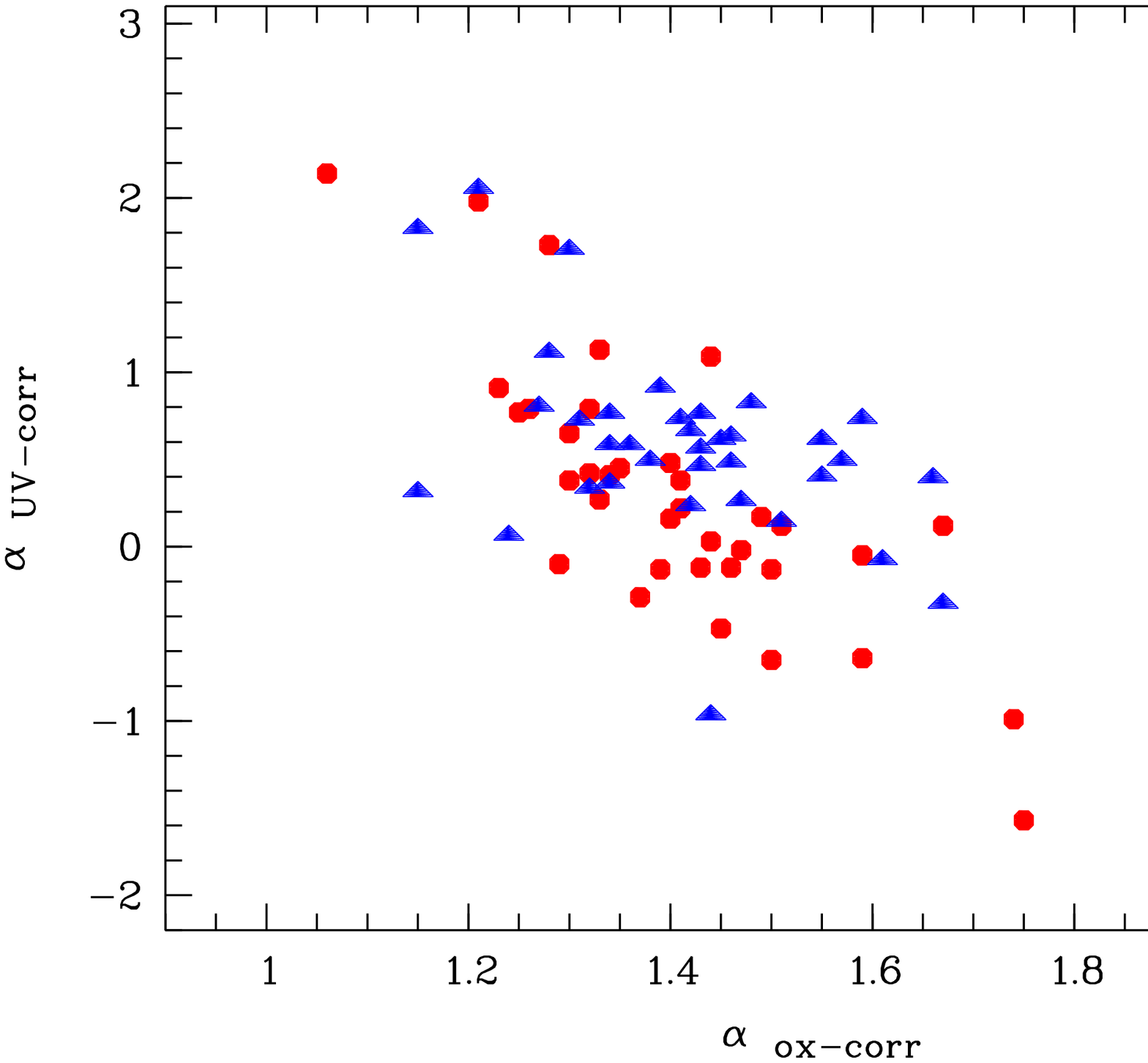}

\caption{\label{aox_auv} Optical-to-X-ray spectral slope \aox\ vs. UV/optical spectral slope \auv. The left panel displays the spectral slopes
only corrected for Galactic reddening and the right panel  corrected for intrinsic and Galactic reddening, 
 NLS1s are displayed as blue triangles and BLS1s as red circles.  
}
\end{figure*}

\begin{figure*}
\epsscale{1.5}
\plottwo{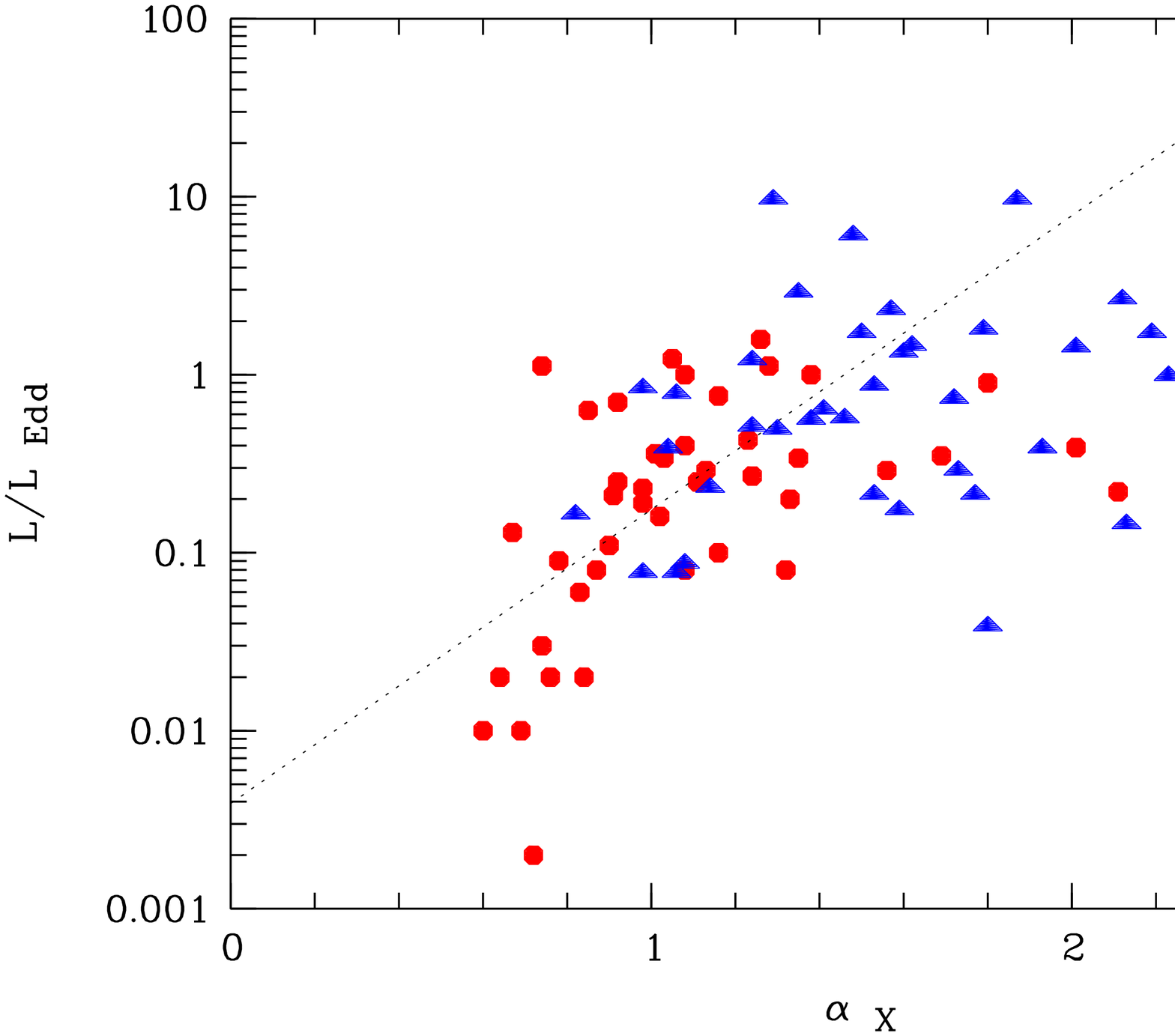}{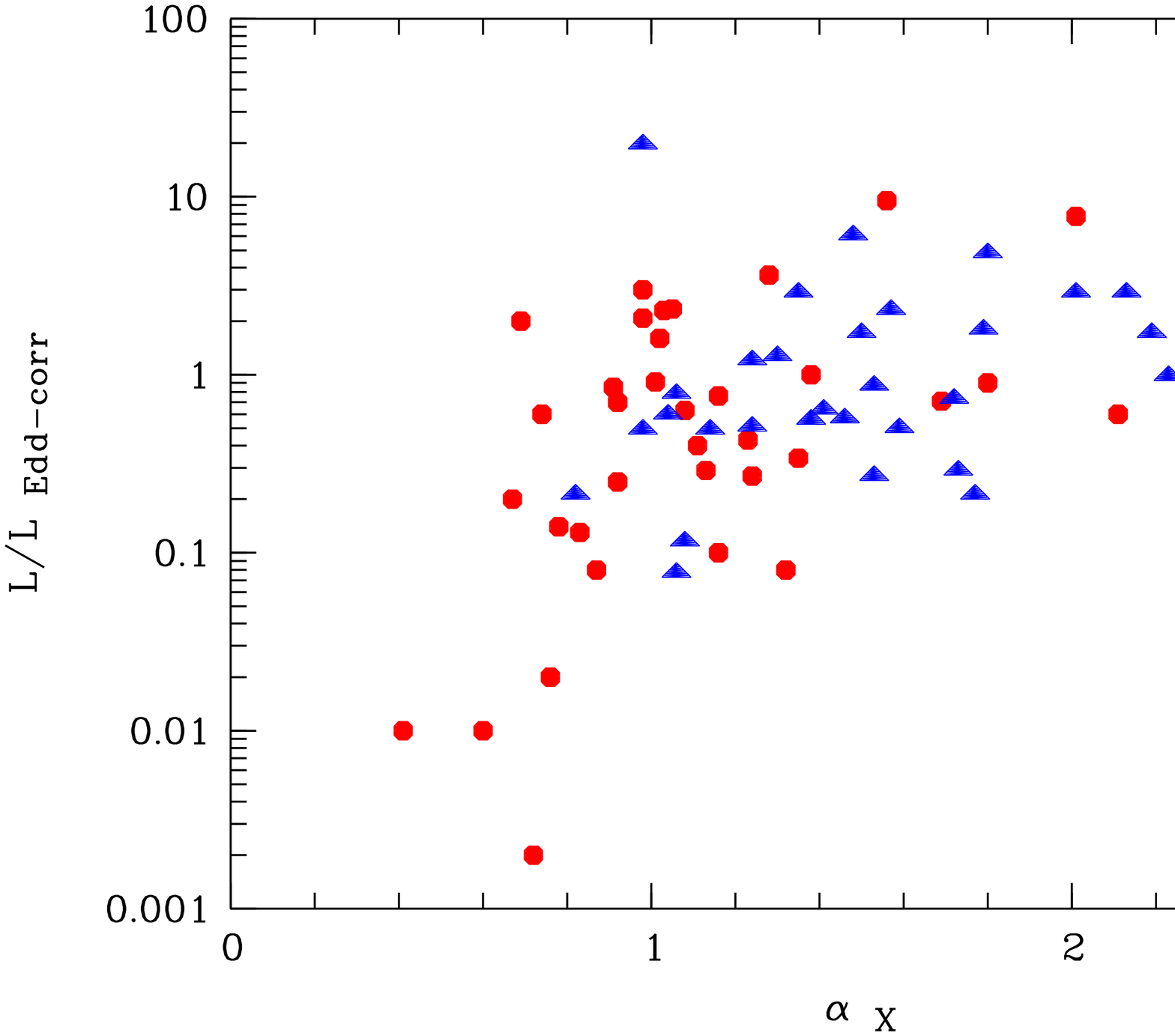}

\epsscale{1.5}
\plottwo{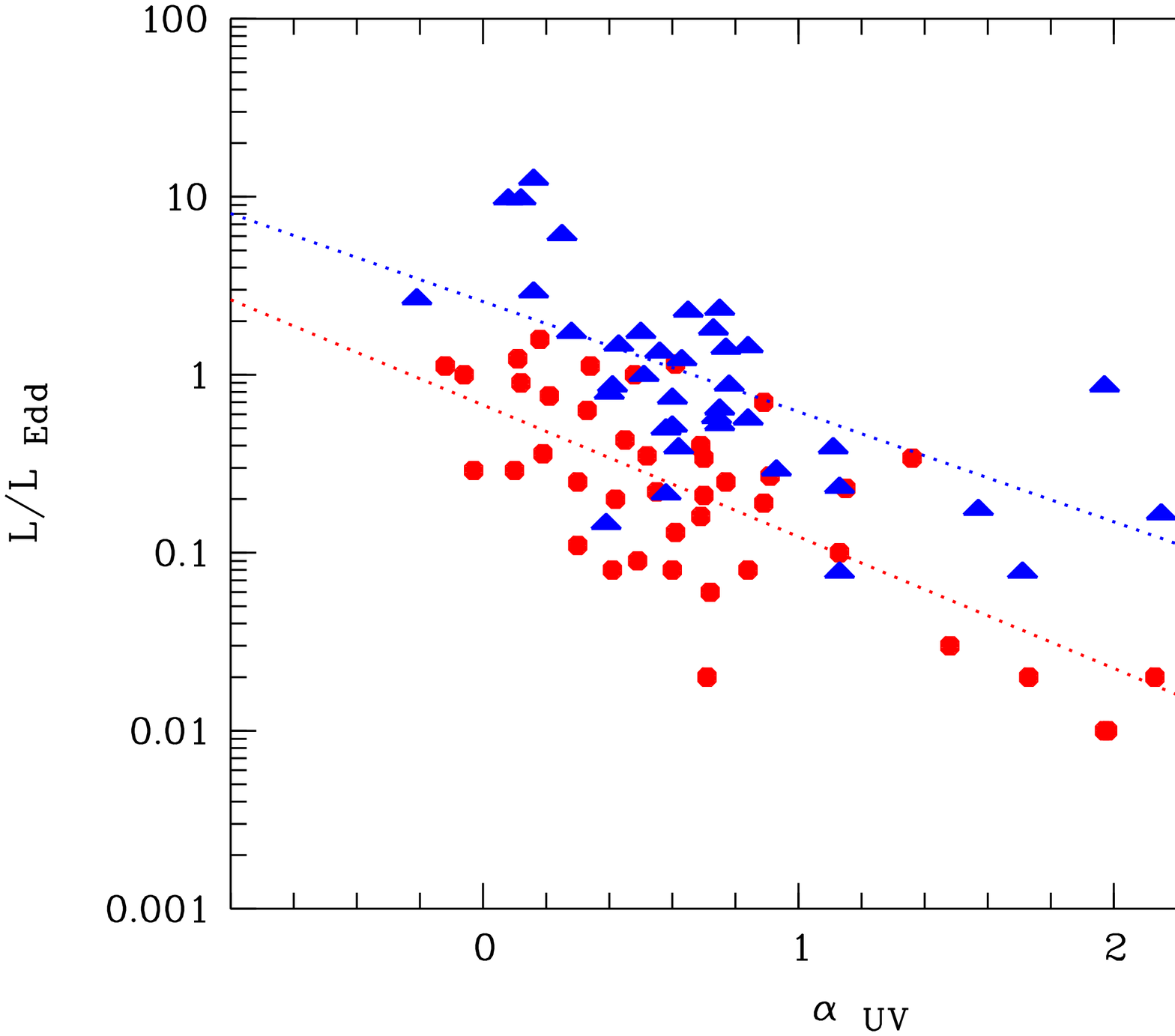}{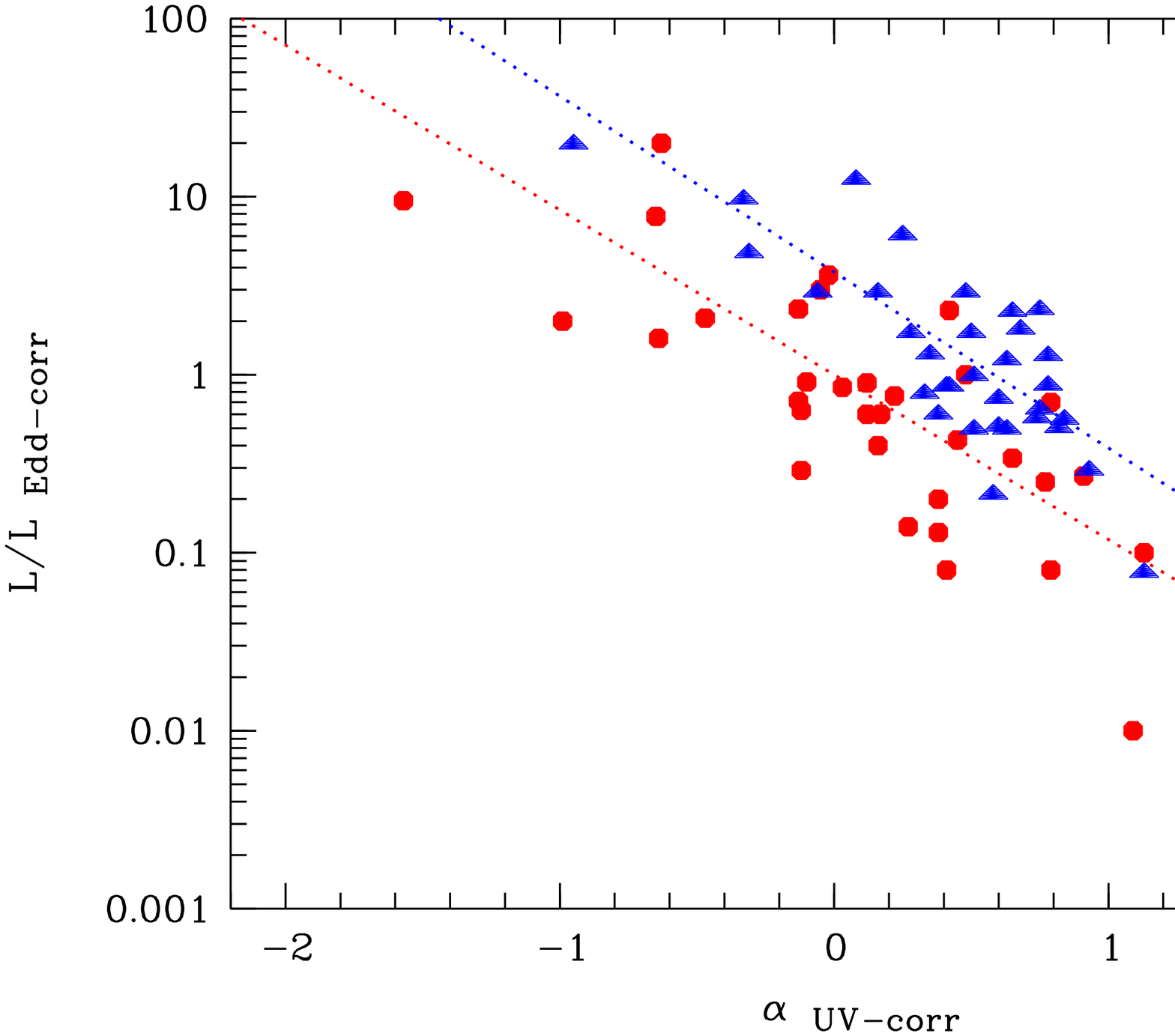}

\epsscale{1.5}
\plottwo{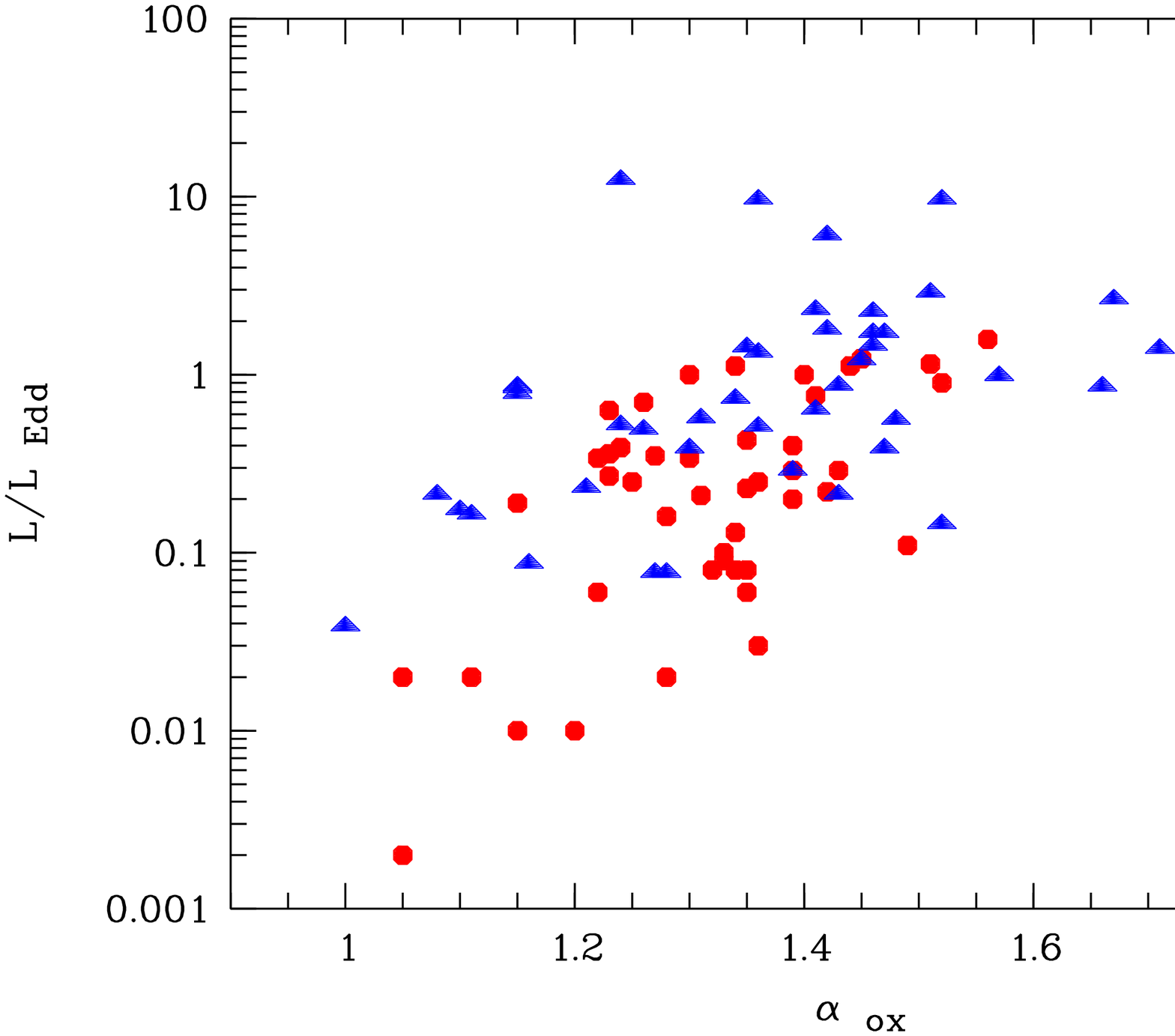}{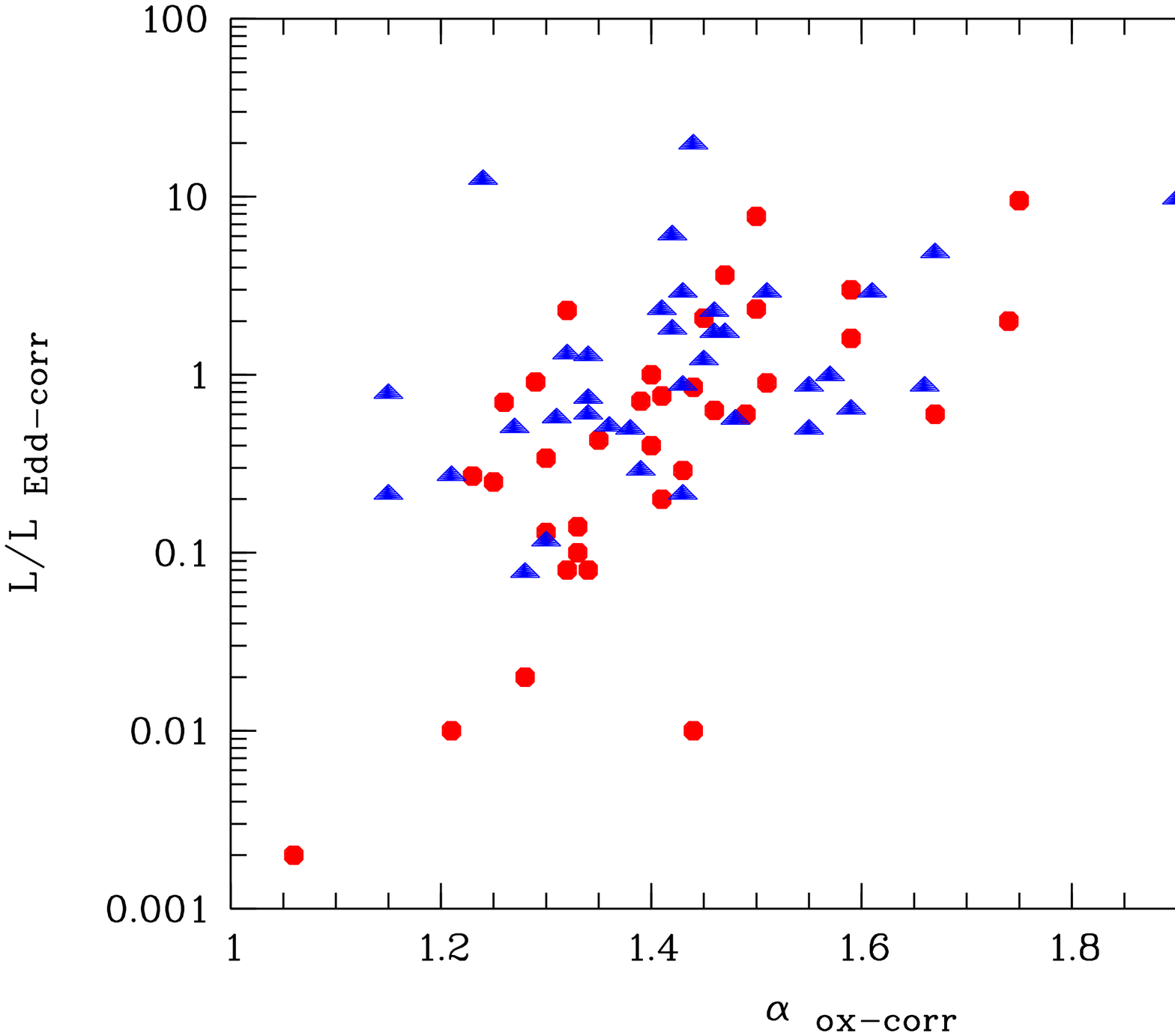}
\caption{\label{ax_auv_ledd} 0.2-2.0 keV X-ray energy spectral slope \ax, 
optical/UV spectral slope \auv, and optical-to-X-ray spectral slope \aox\
 vs. the Eddington ratio $L/L_{\rm Edd}$.
}
\end{figure*}

\begin{figure*}

\epsscale{1.5}
\plottwo{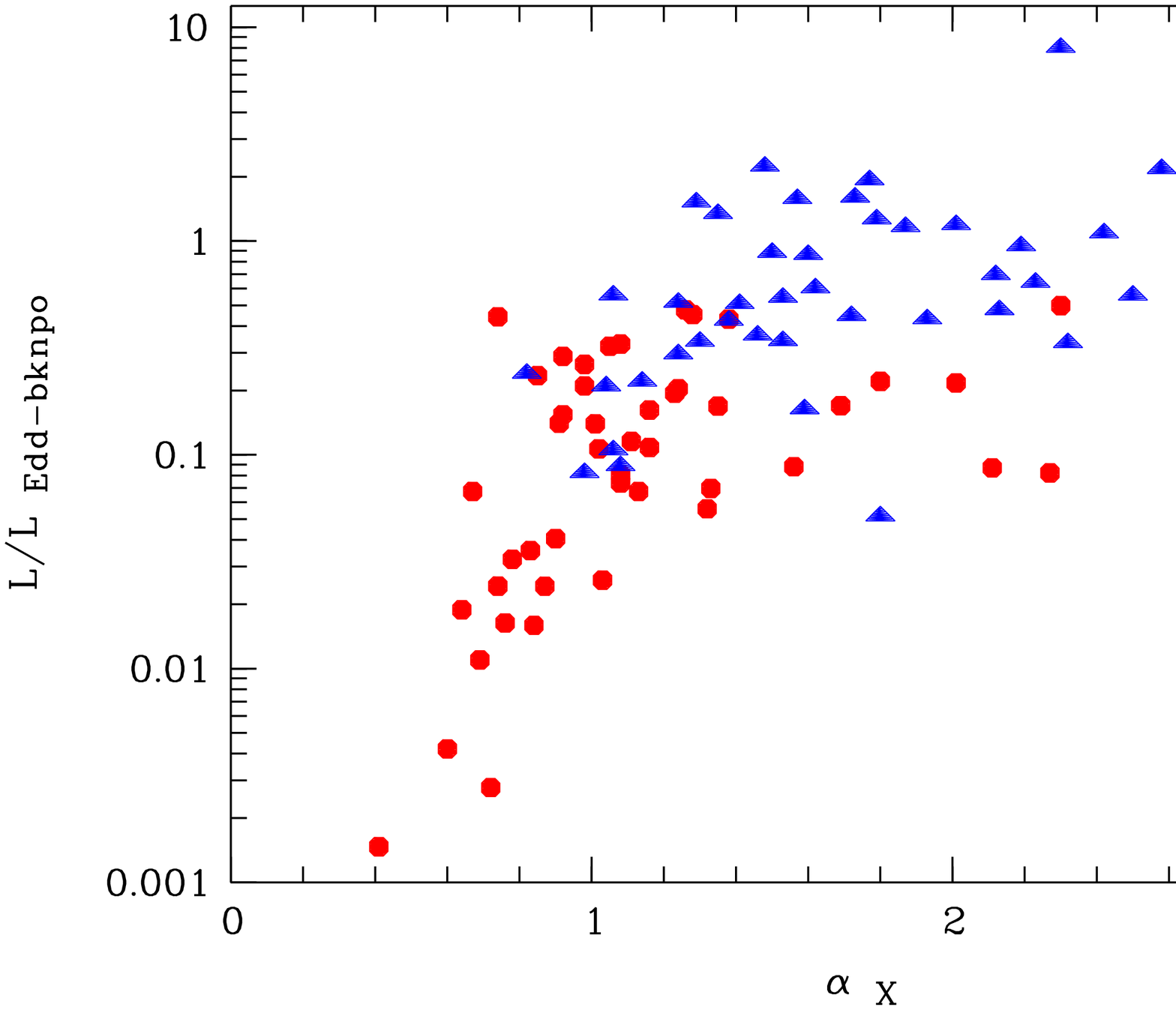}{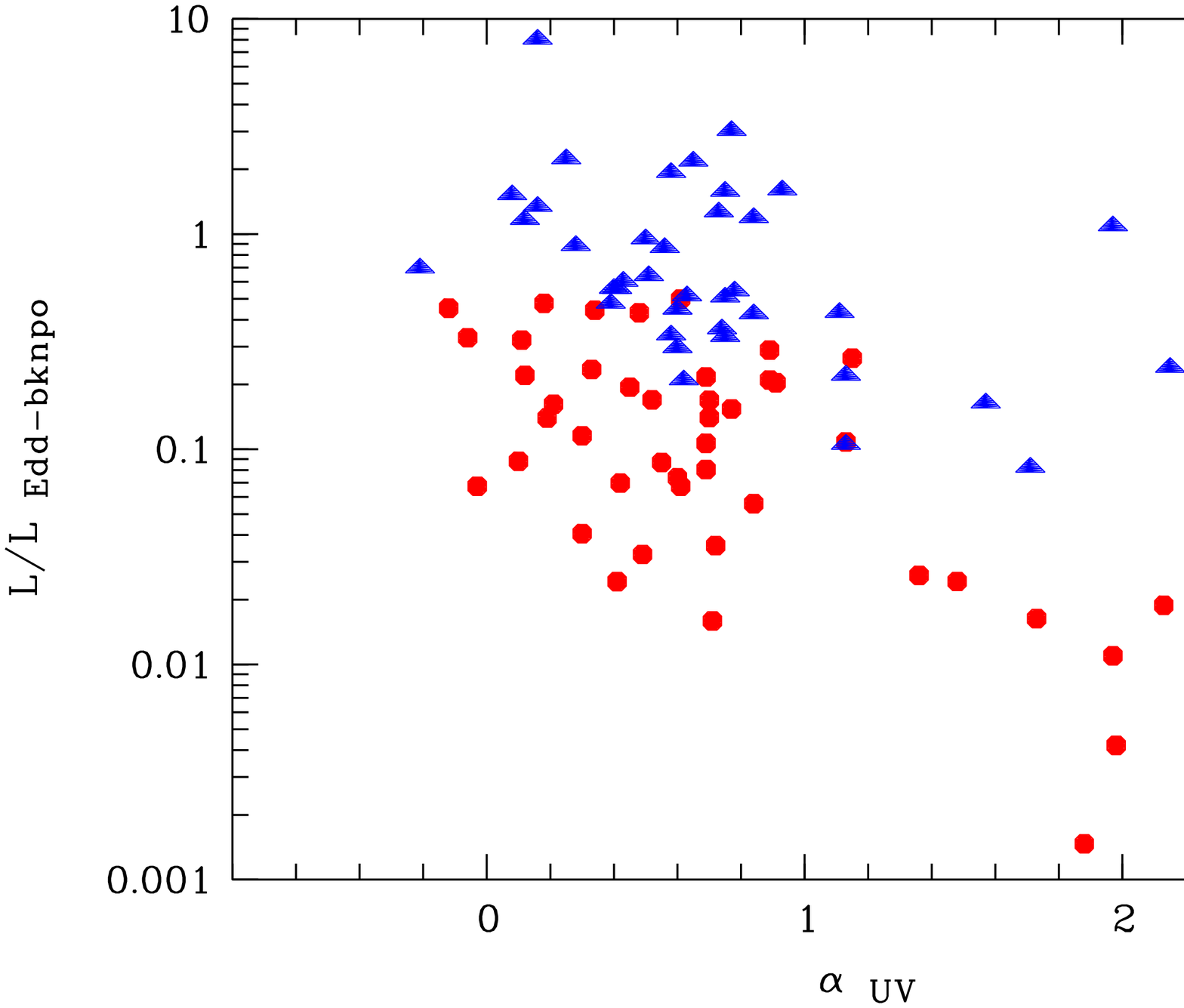}
\caption{\label{ax_auv_ledd_bknpo} 0.2-2.0 keV X-ray energy spectral slope \ax\ and  
optical/UV spectral slope \auv vs. the Eddington ratio $L/L_{\rm Edd-bknpo}$ using the double broken power law mode to fit the SED.
}
\end{figure*}

\begin{figure*}
\epsscale{1.6}
\plottwo{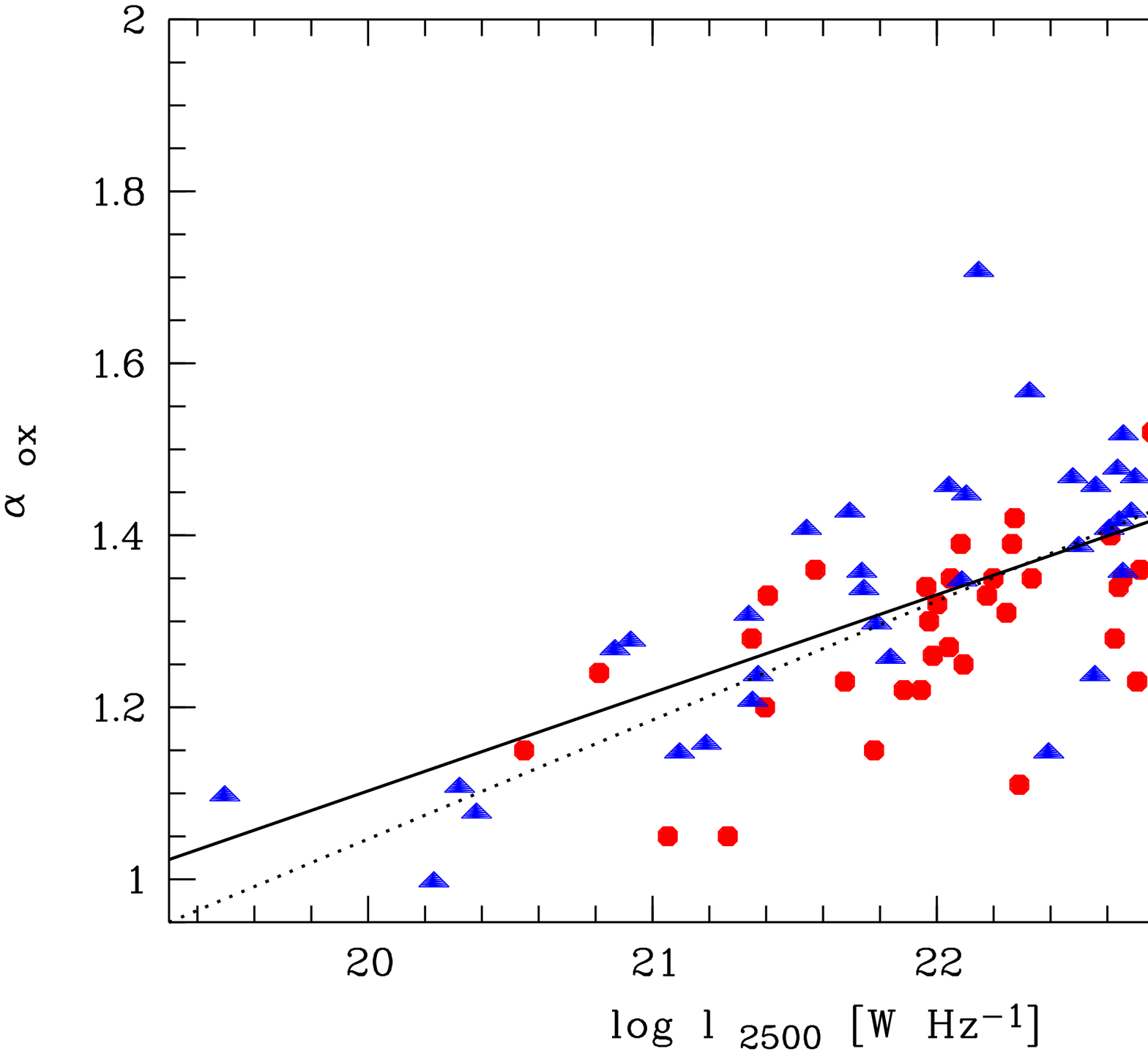}{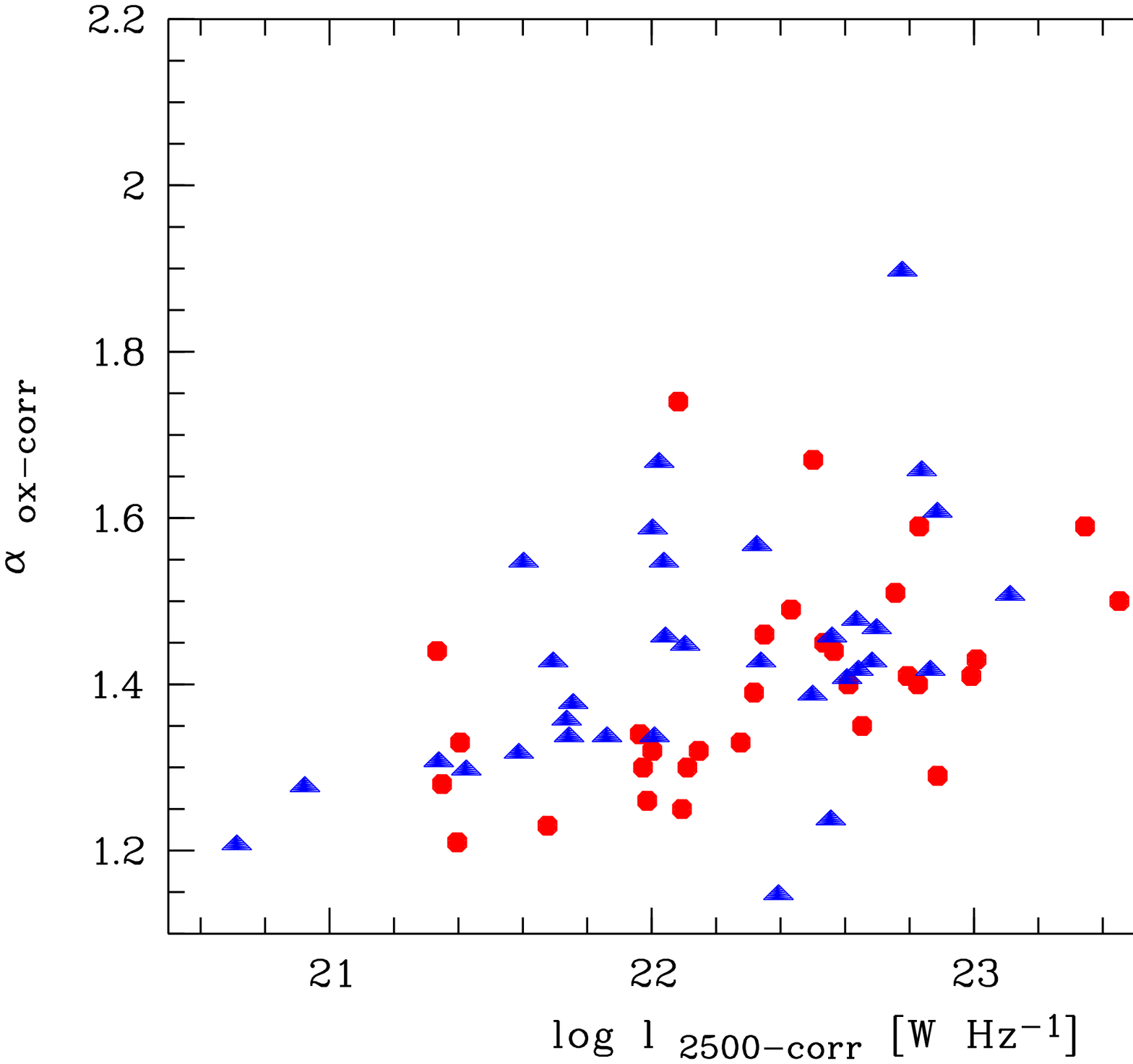}

\epsscale{1.4}
\plottwo{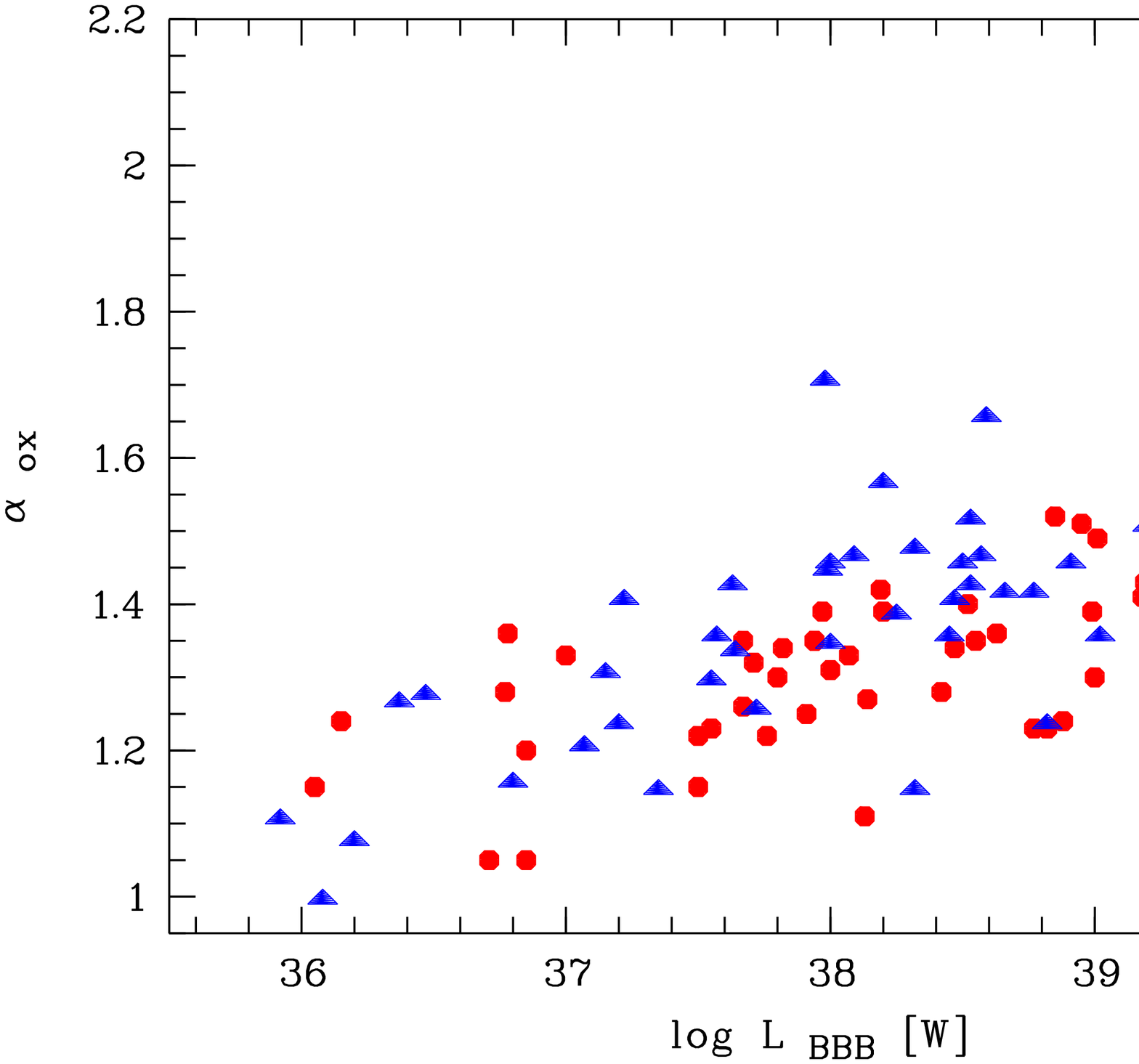}{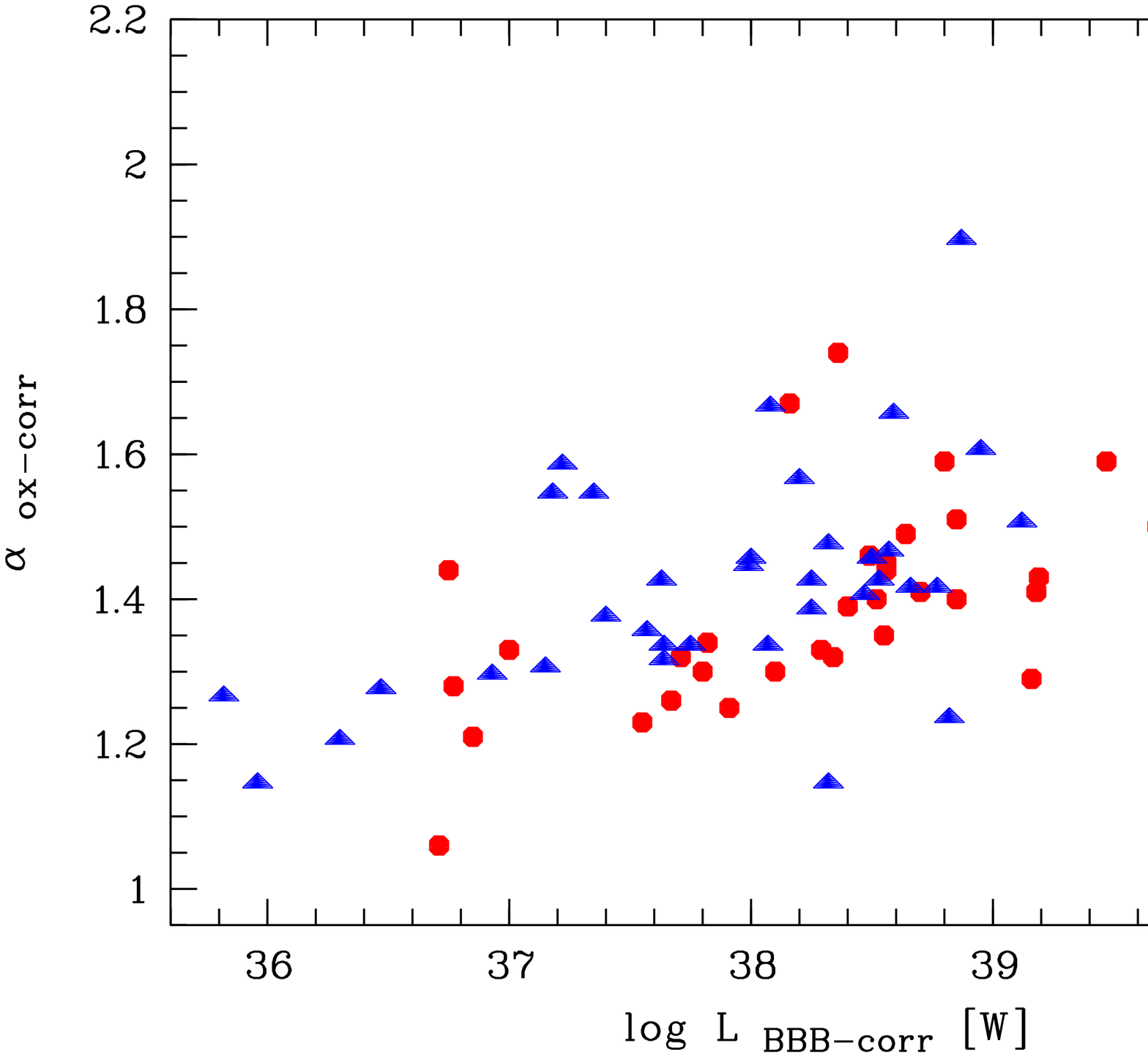}

\epsscale{1.4}
\plottwo{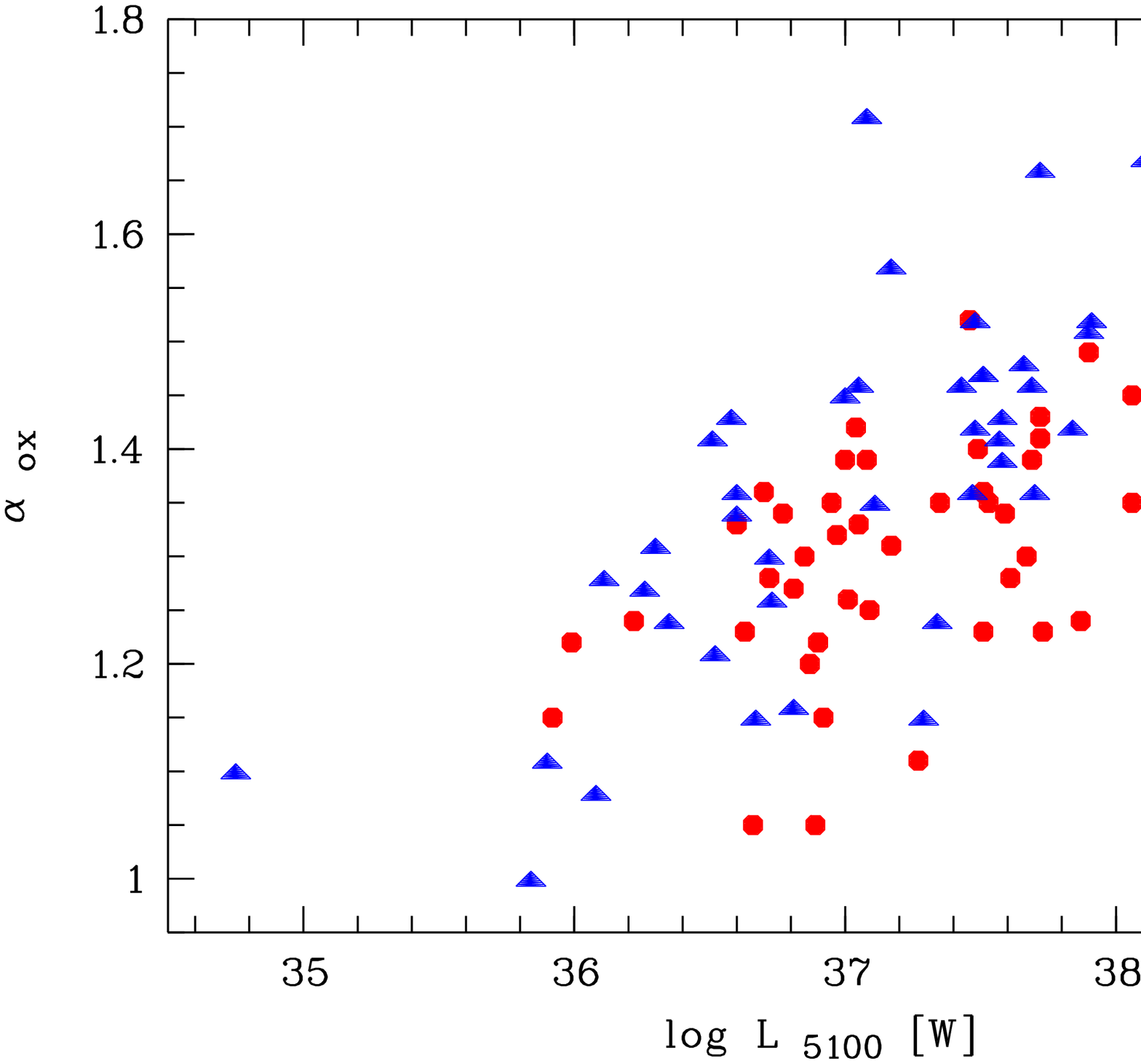}{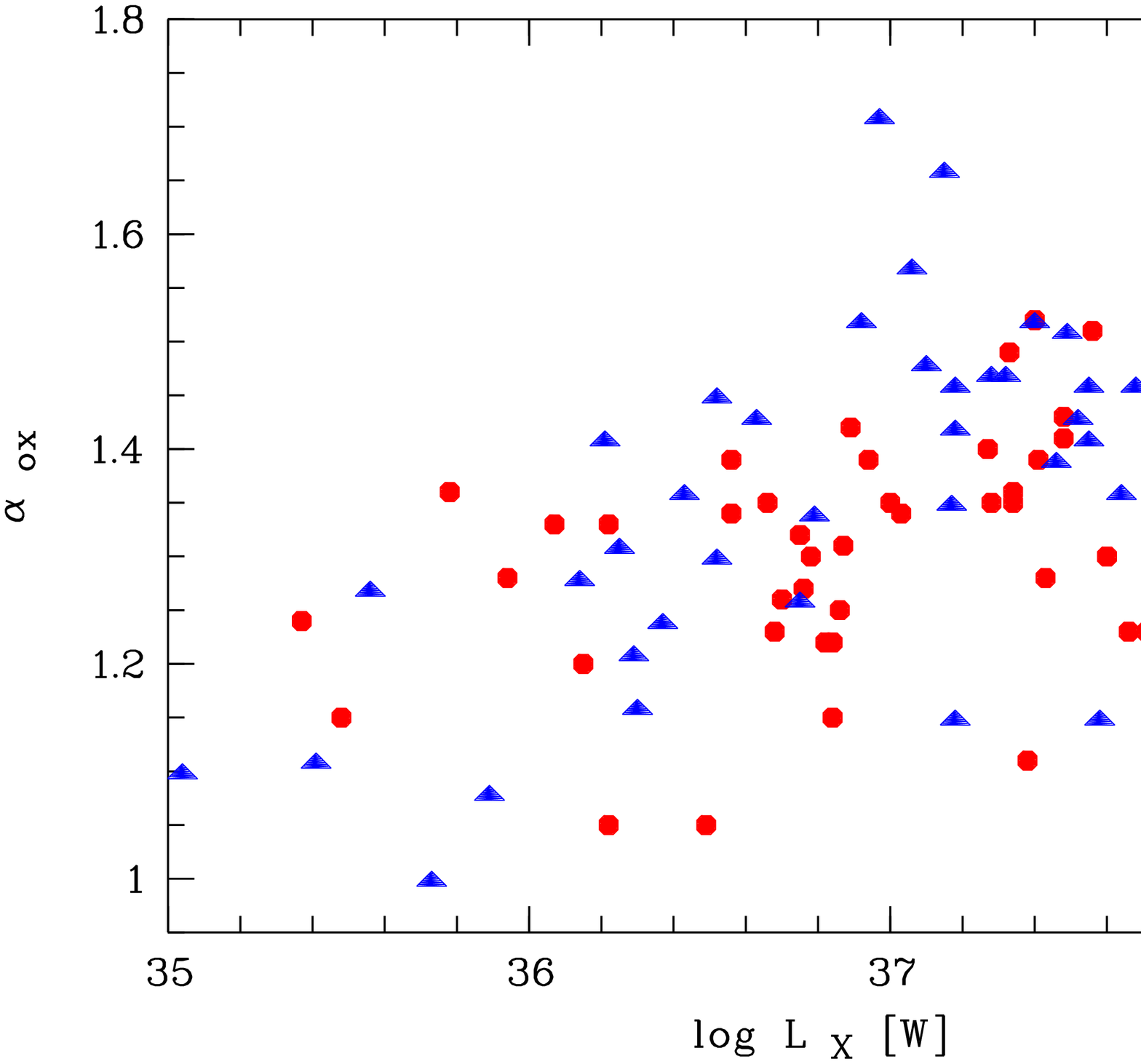}
\caption{\label{aox_l2500} Optical-to-X-ray spectral slope \aox\ vs. luminosity density at 2500\AA\ $l_{2500}$, the BBB luminosity $L_{\rm BBB}$,
luminosity at 5100\AA\ $L_{\rm 5100\AA}$, and rest-frame 0.2-2.0 keV X-ray luminosity $L_{\rm X}$. The solid line in the \aox-log $l_{2500\AA}$ plot
displays the relation between \aox and the luminosity density at 2500\AA\ as given in equation (1). The dashed line displays the same relation reported by
\citet{strateva05}
}
\end{figure*}

\begin{figure*}
\epsscale{1.6}

\plottwo{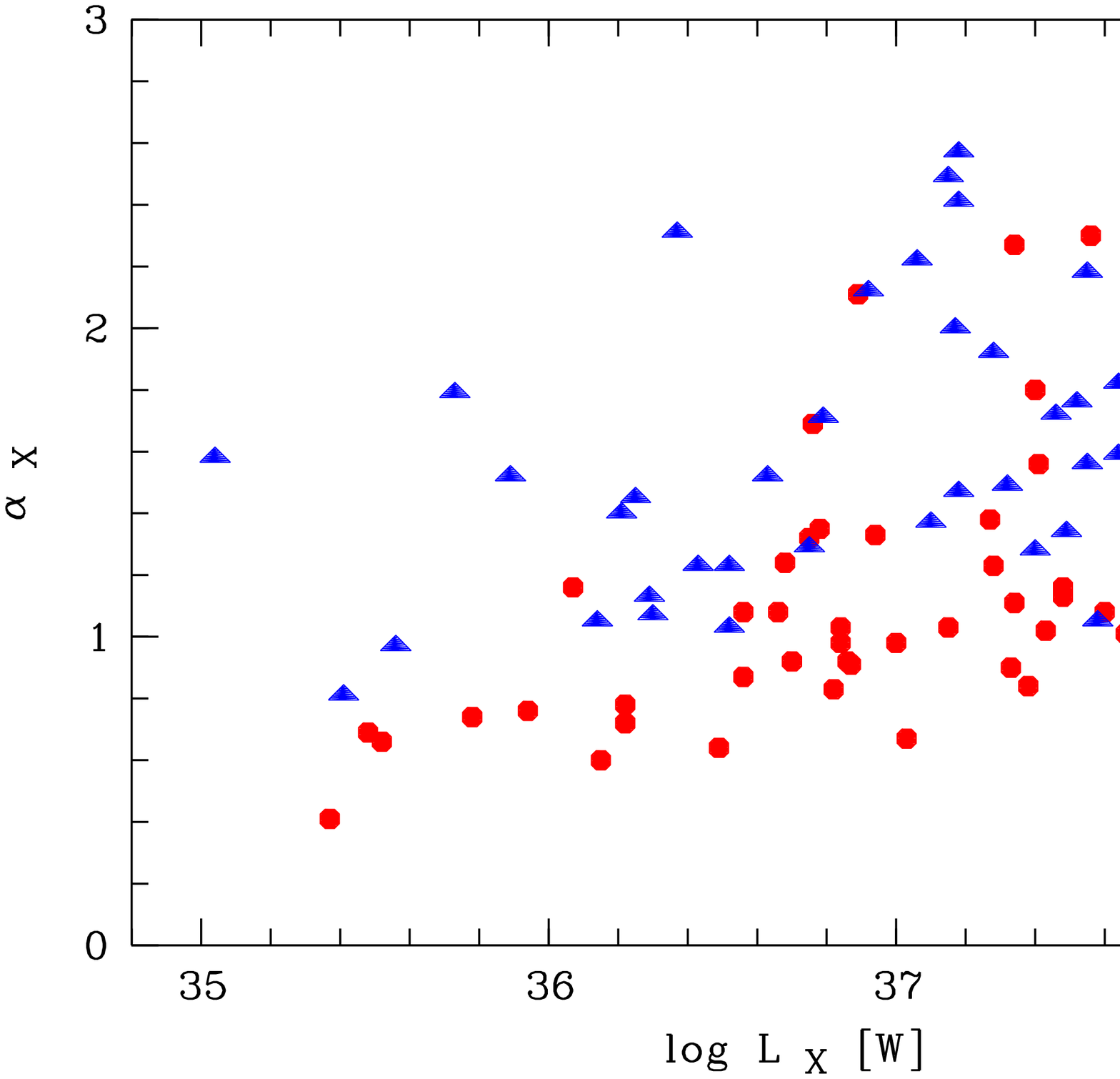}{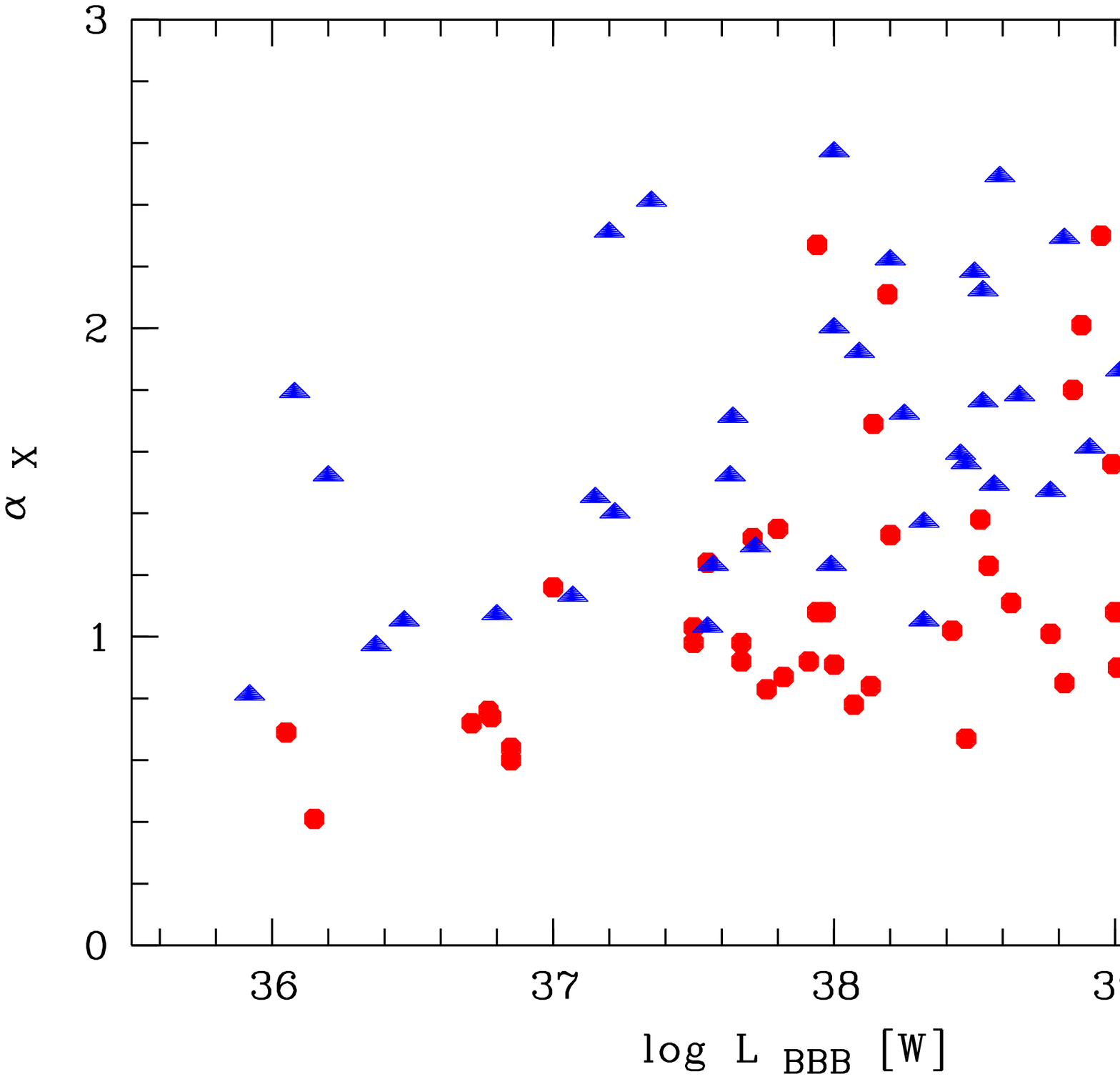}
\caption{\label{ax_l} X-ray spectral slope \ax\ vs. luminosities in the rest-frame 0.2-2.0 keV band and the BBB.   
}
\end{figure*}

\begin{figure*}
\epsscale{1.6}
\plottwo{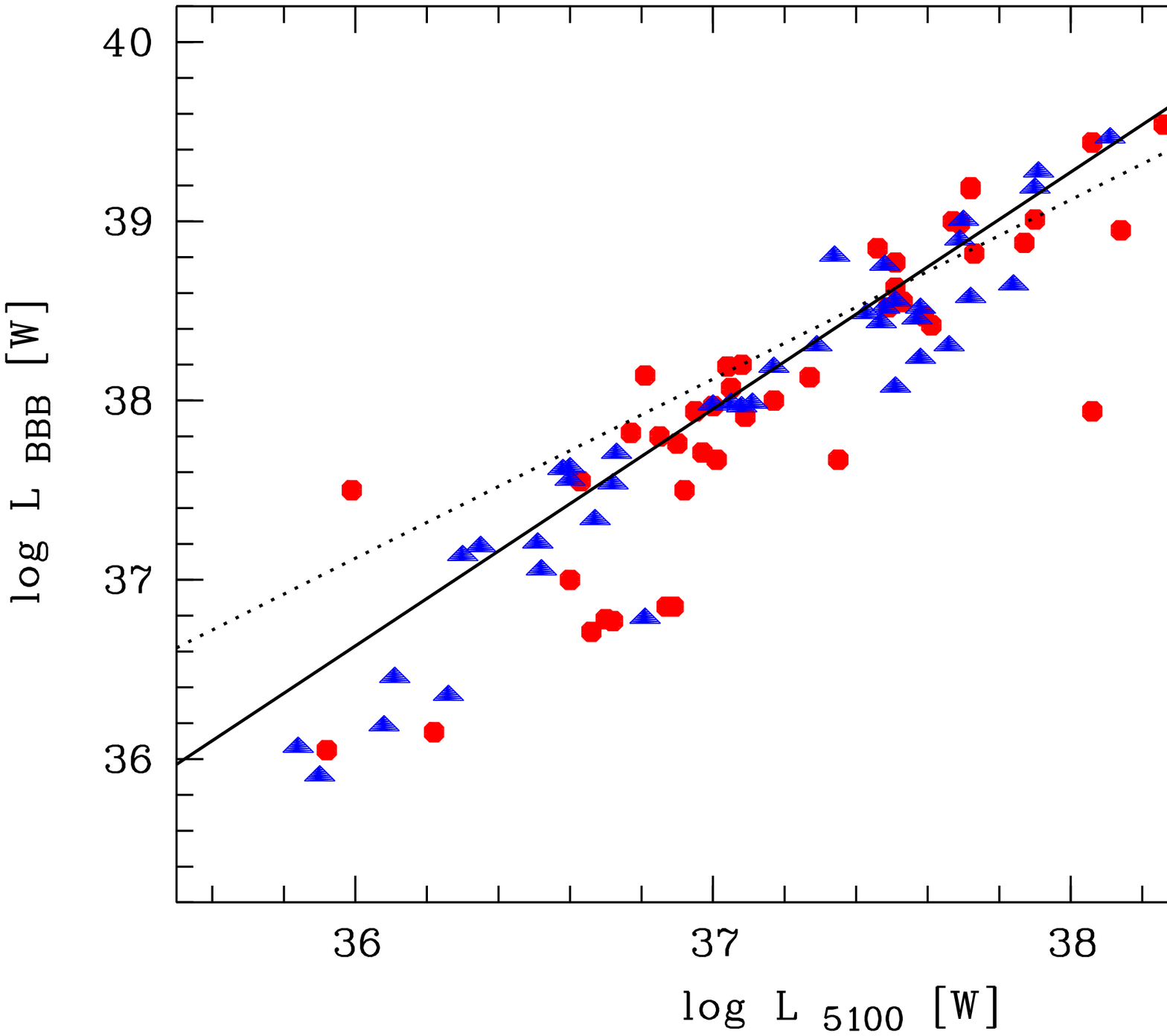}{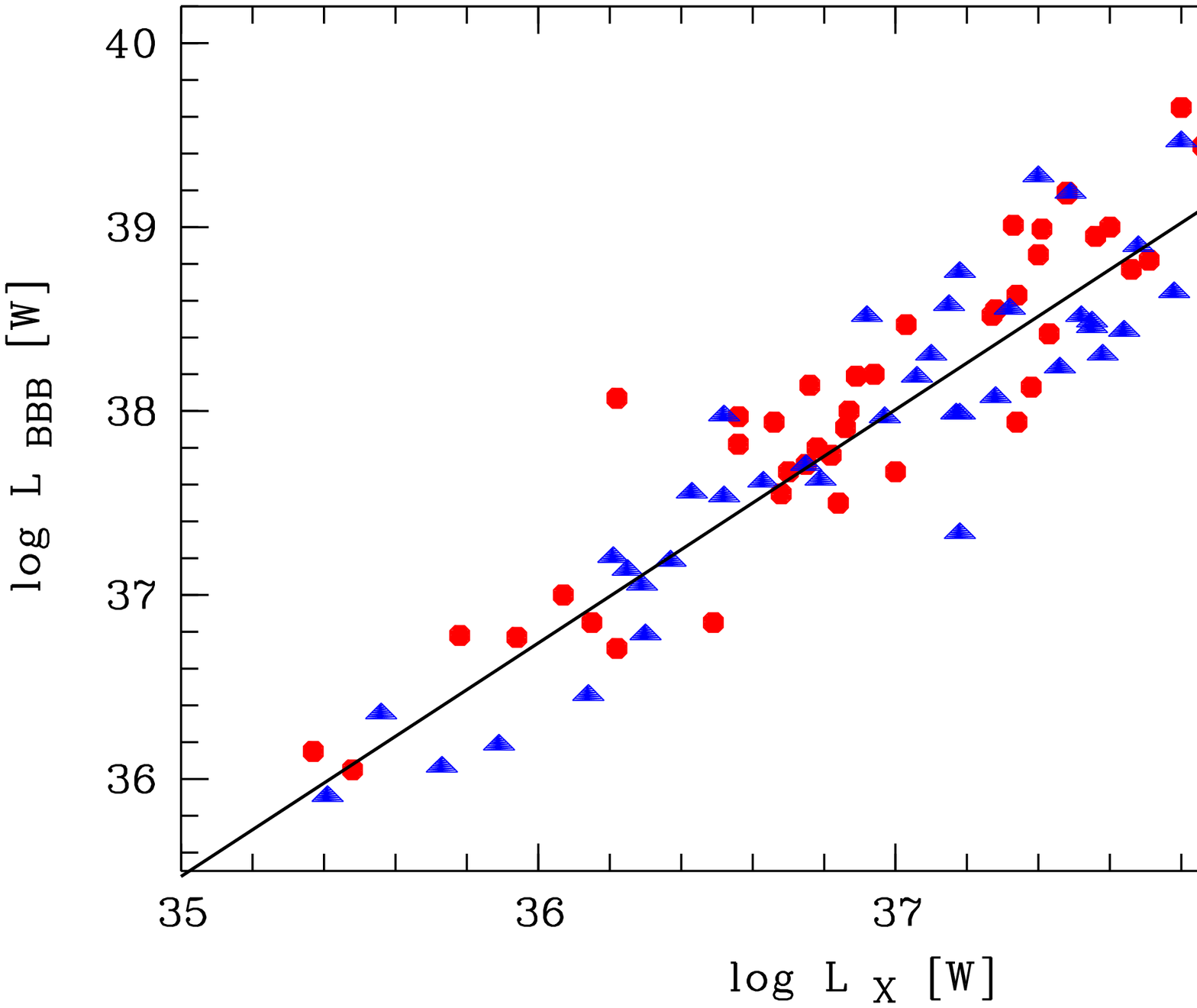}
\caption{\label{l5100_lbbb} Luminosity at 5100\AA\ and in the 0.2-2.0 keV band vs. the luminosity in the Big-Blue-Bump. 
The solid and dashed lines in the
left panel display our relation given in equation (12) and that given by \citet{elvis94}. The solid line in the right panel display equation (13).  
}
\end{figure*}

\begin{figure*}
\epsscale{1.6}
\plottwo{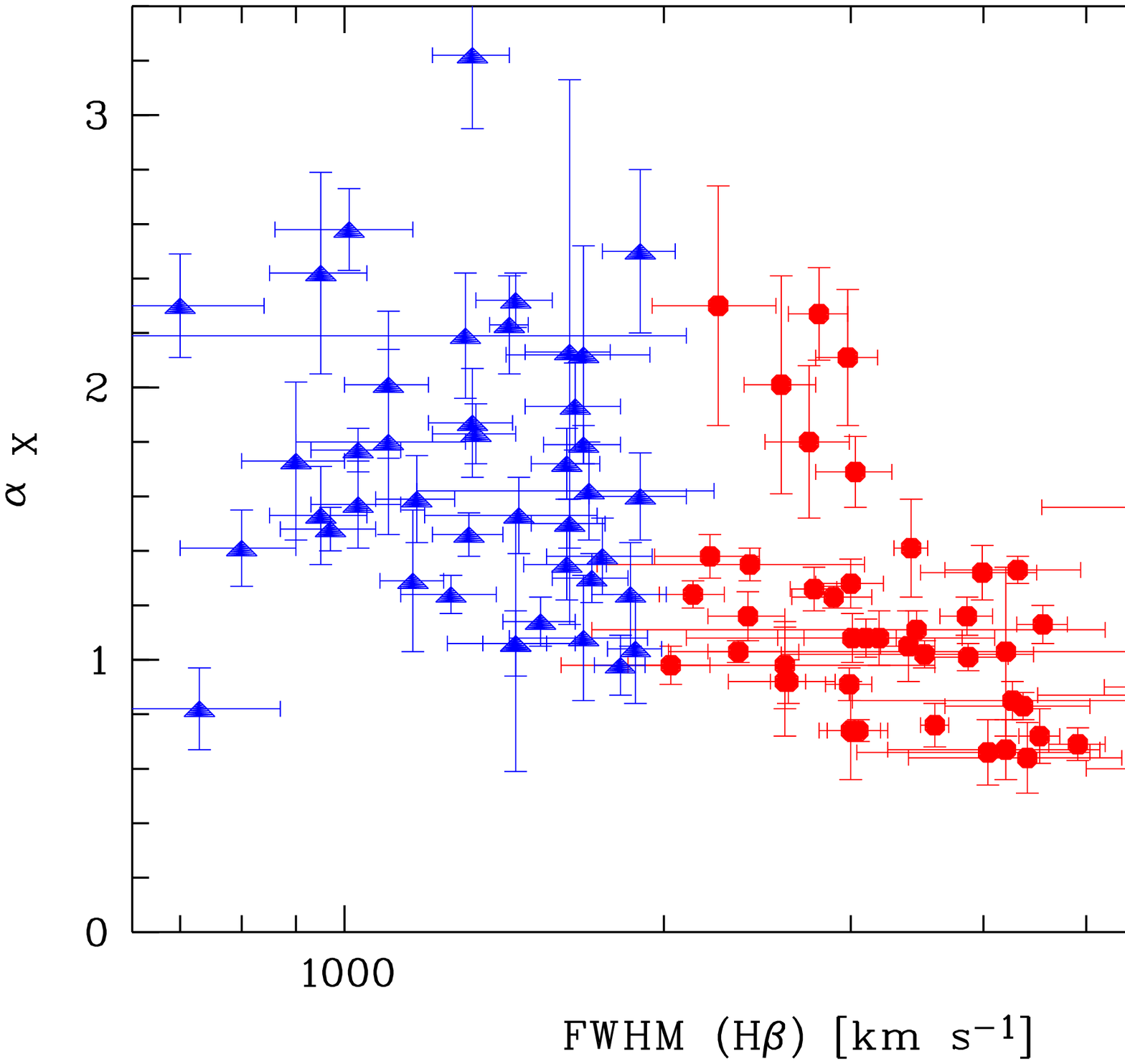}{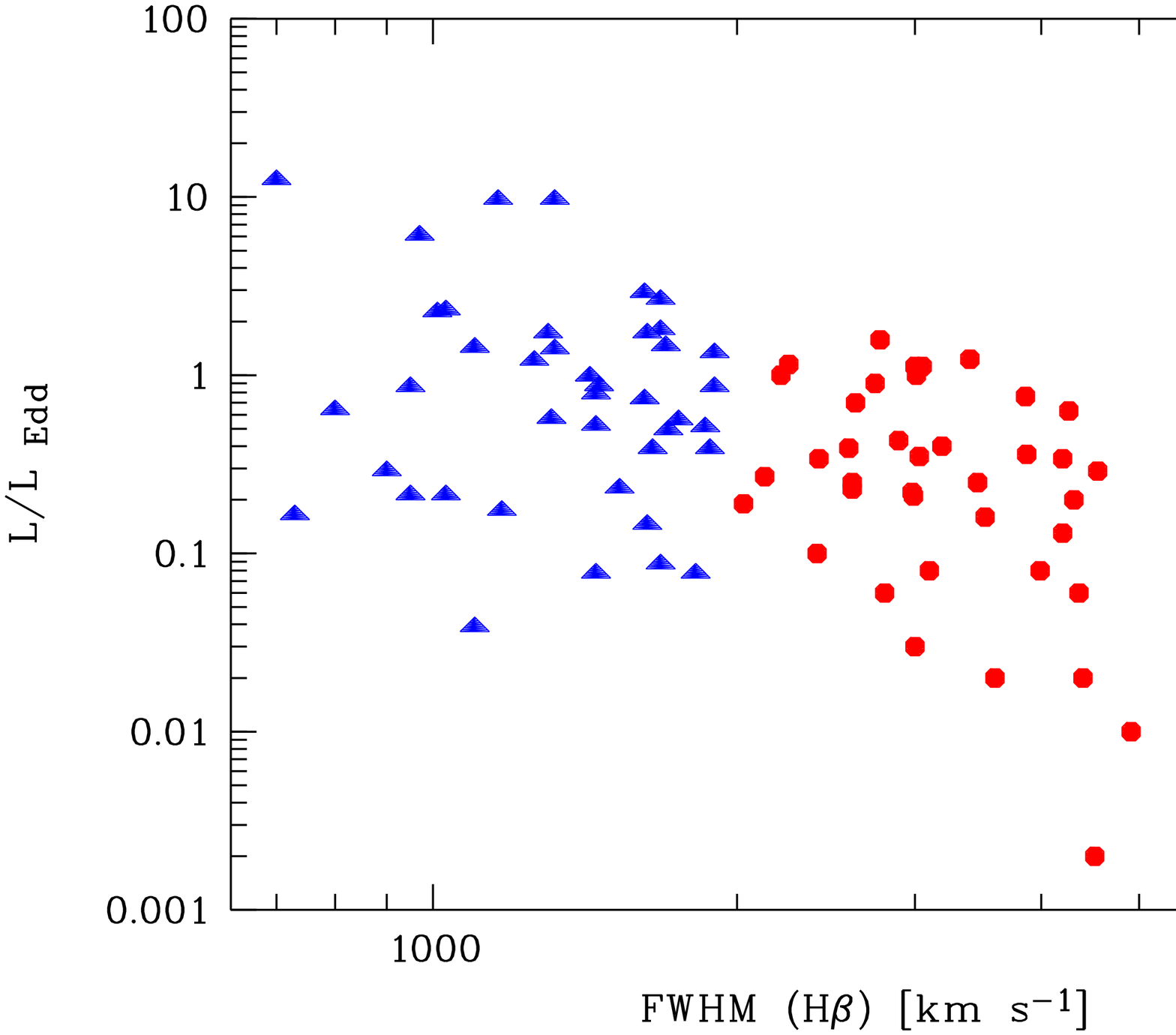}
\caption{\label{fwhb_ax} FWHM(H$\beta$) vs. 
0.2-2.0 keV X-ray energy spectral slope \ax\ (left) and \lledd\ (right). 
}
\end{figure*}

\begin{figure*}
\epsscale{1.5}
\plottwo{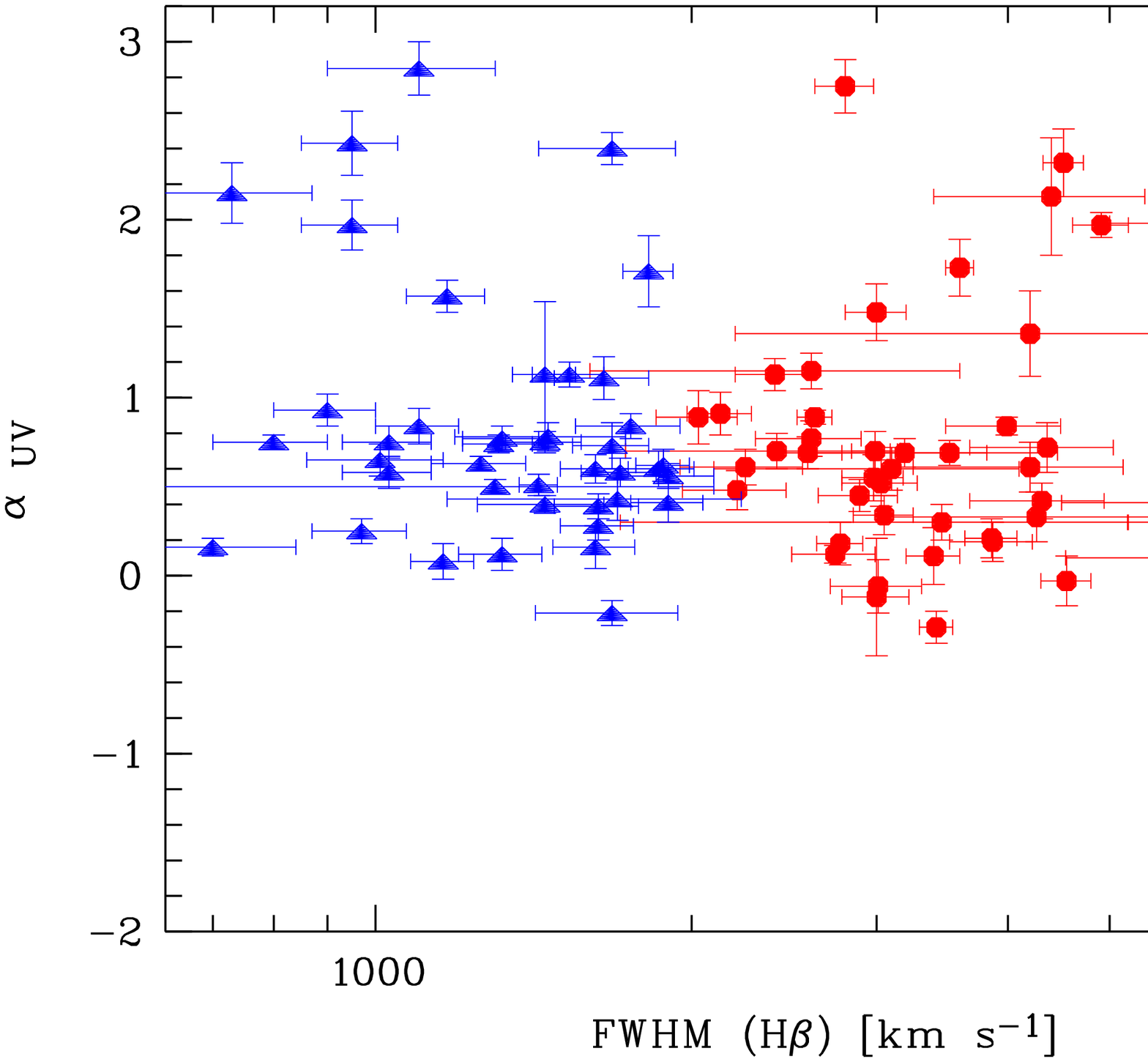}{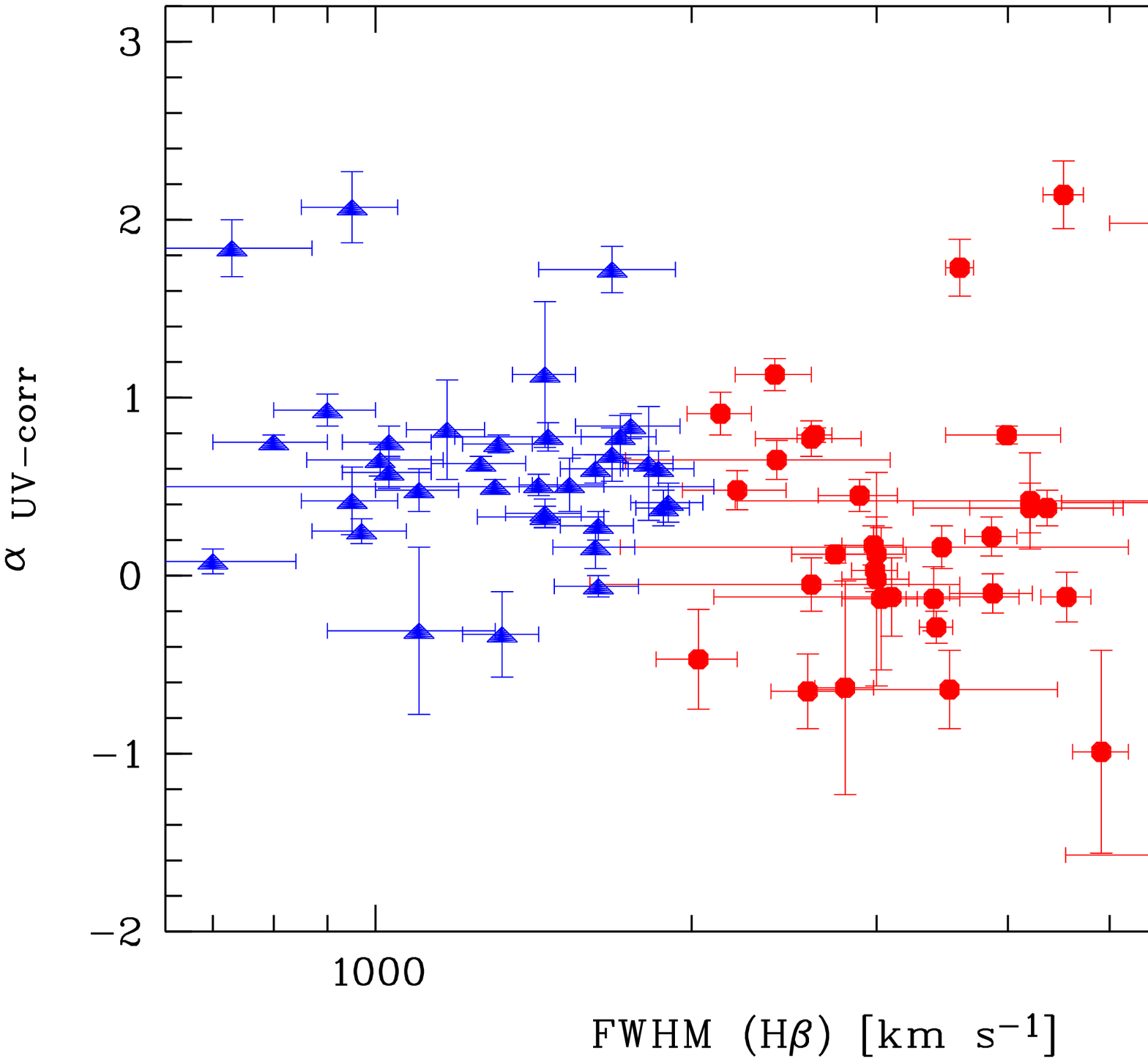}

\plottwo{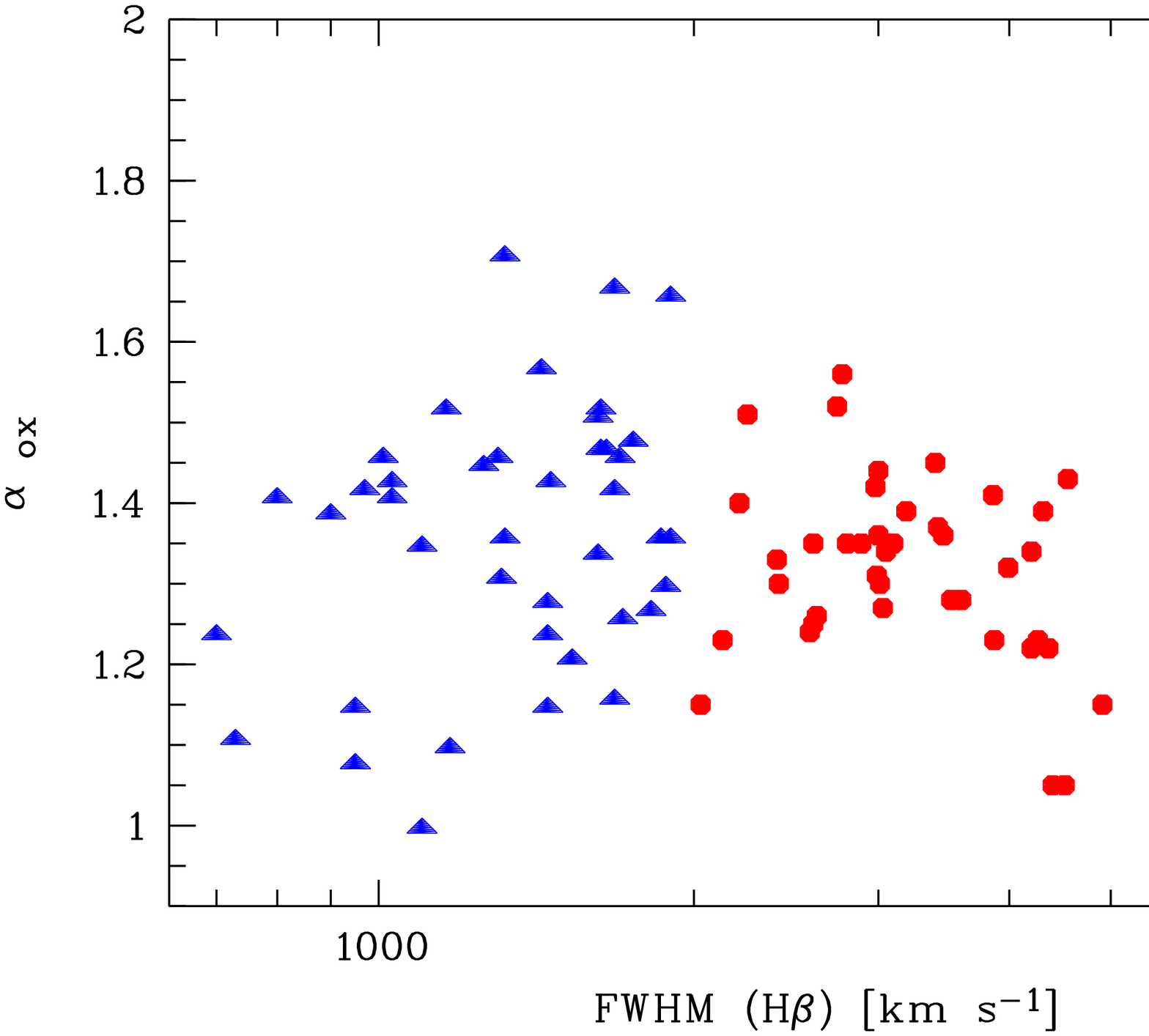}{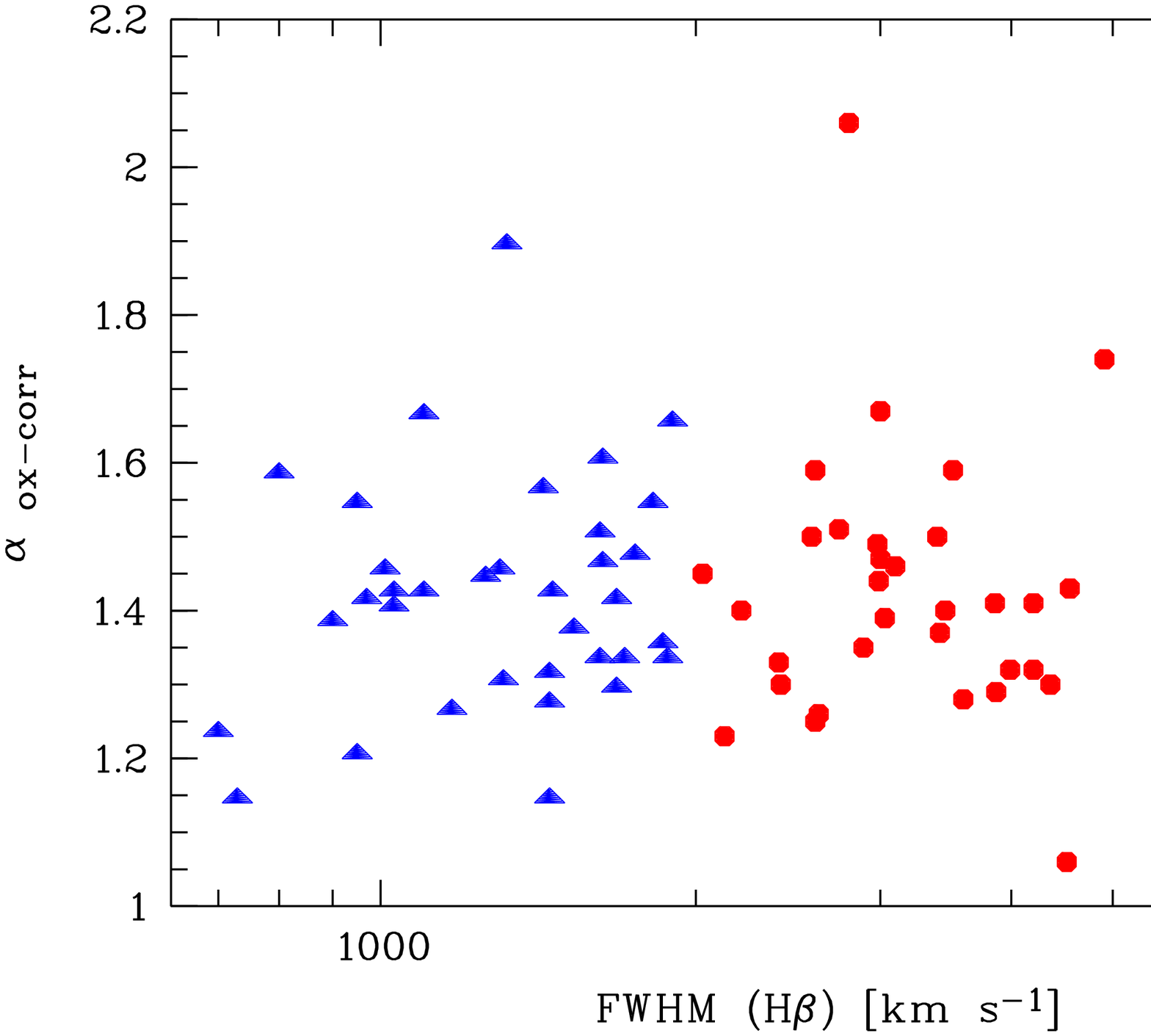}
\caption{\label{fwhb_auv_aox} FWHM(H$\beta$) vs. 
optical/UV spectral slope \auv\ and the optical-to-X-ray spectral slope \aox, both
uncorrected and corrected for intrinsic reddening. }
\end{figure*}

\clearpage

\begin{figure*}
\epsscale{1.6}
\plottwo{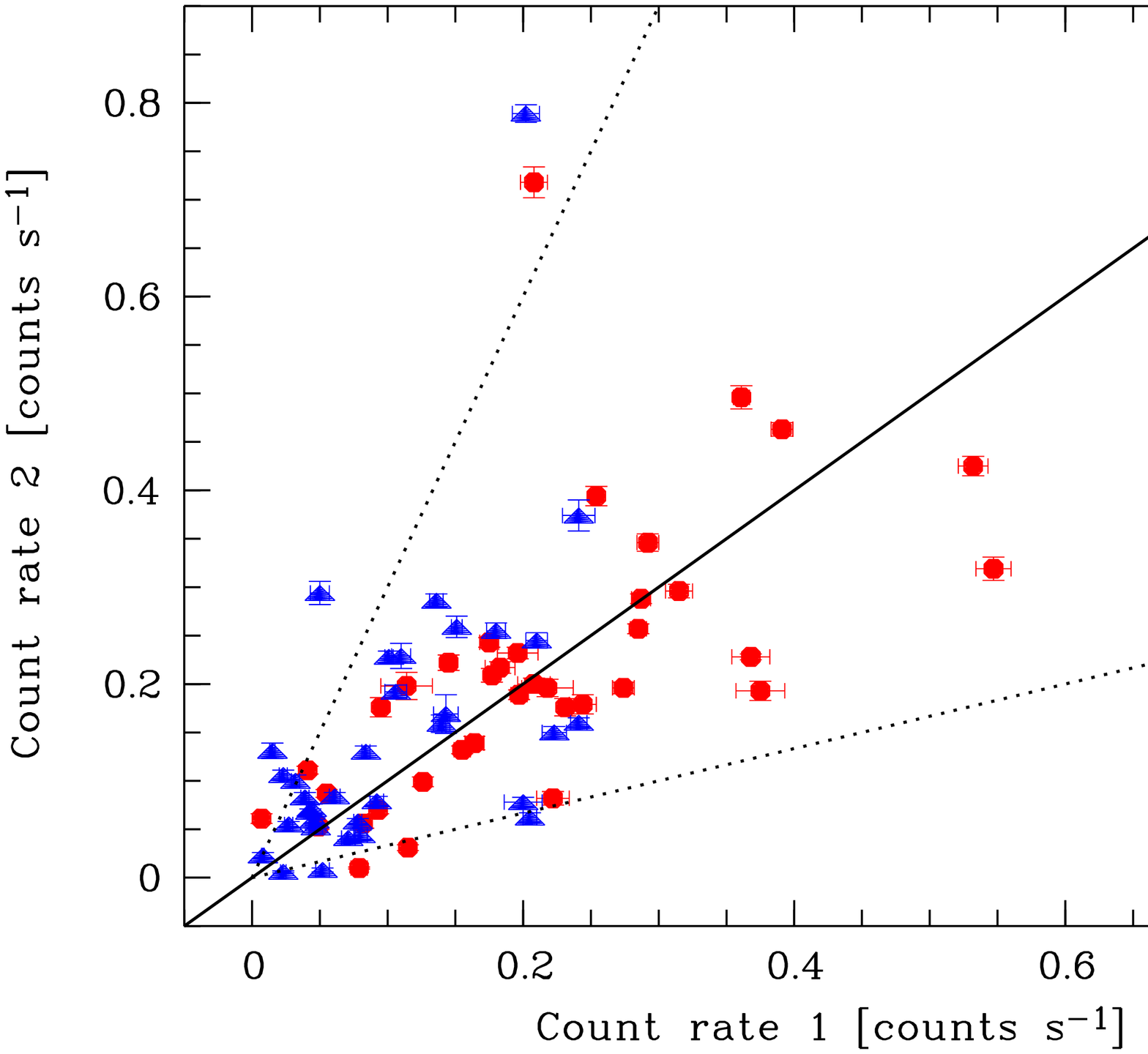}{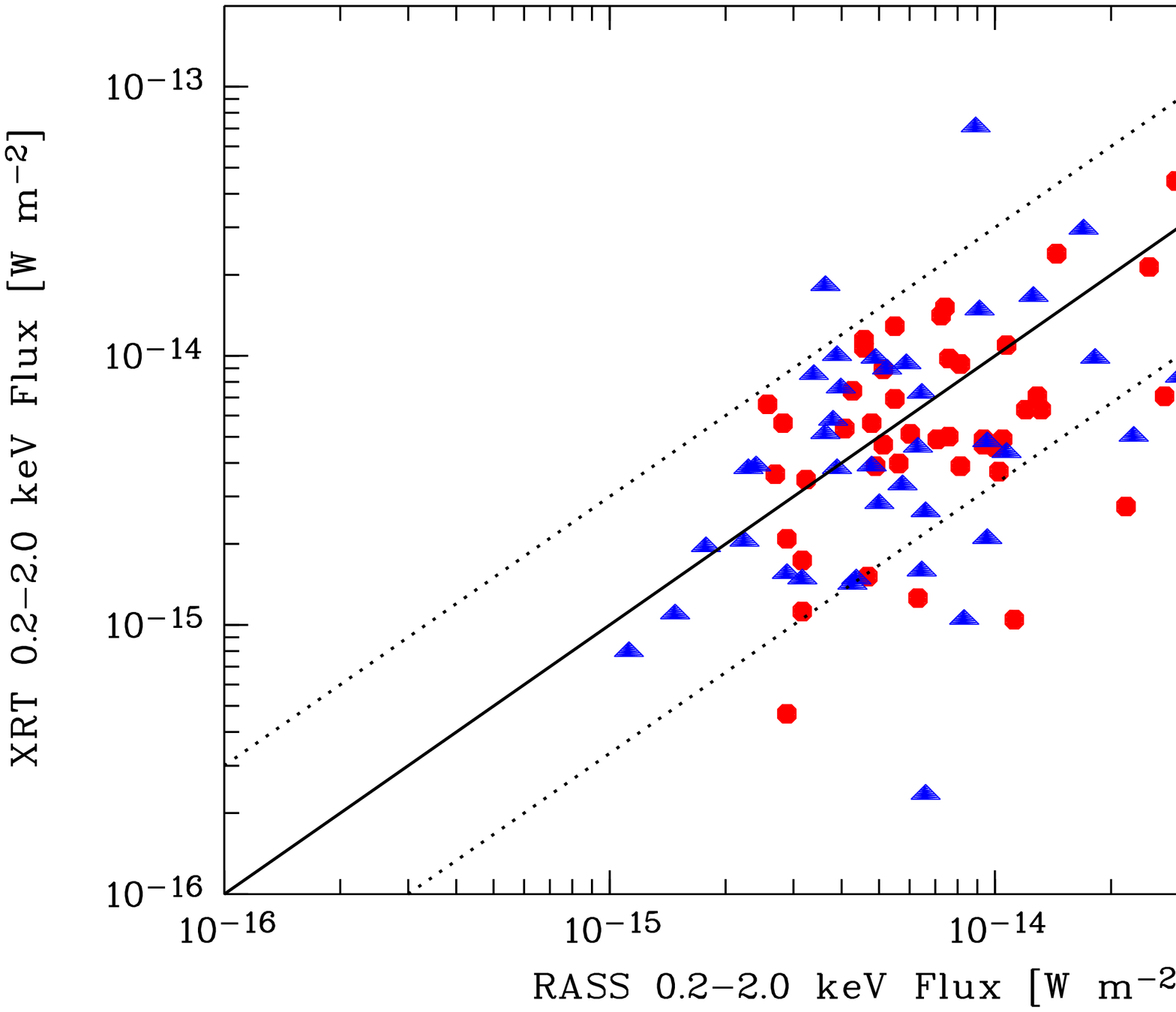}
\caption{\label{xray_cr_var} Short and long term X-ray flux variability in the 
soft X-ray selected AGN sample observed by \swift. The left panel shows the 
count rate variability of two \swift\ observation of the same AGN, the 
 right panel the 
rest frame 0.2-2.0 keV fluxes of the \swift\ XRT and ROSAT All-Sky Survey 
observations. 
}
\end{figure*}

\begin{figure*}
\epsscale{1.6}
\plottwo{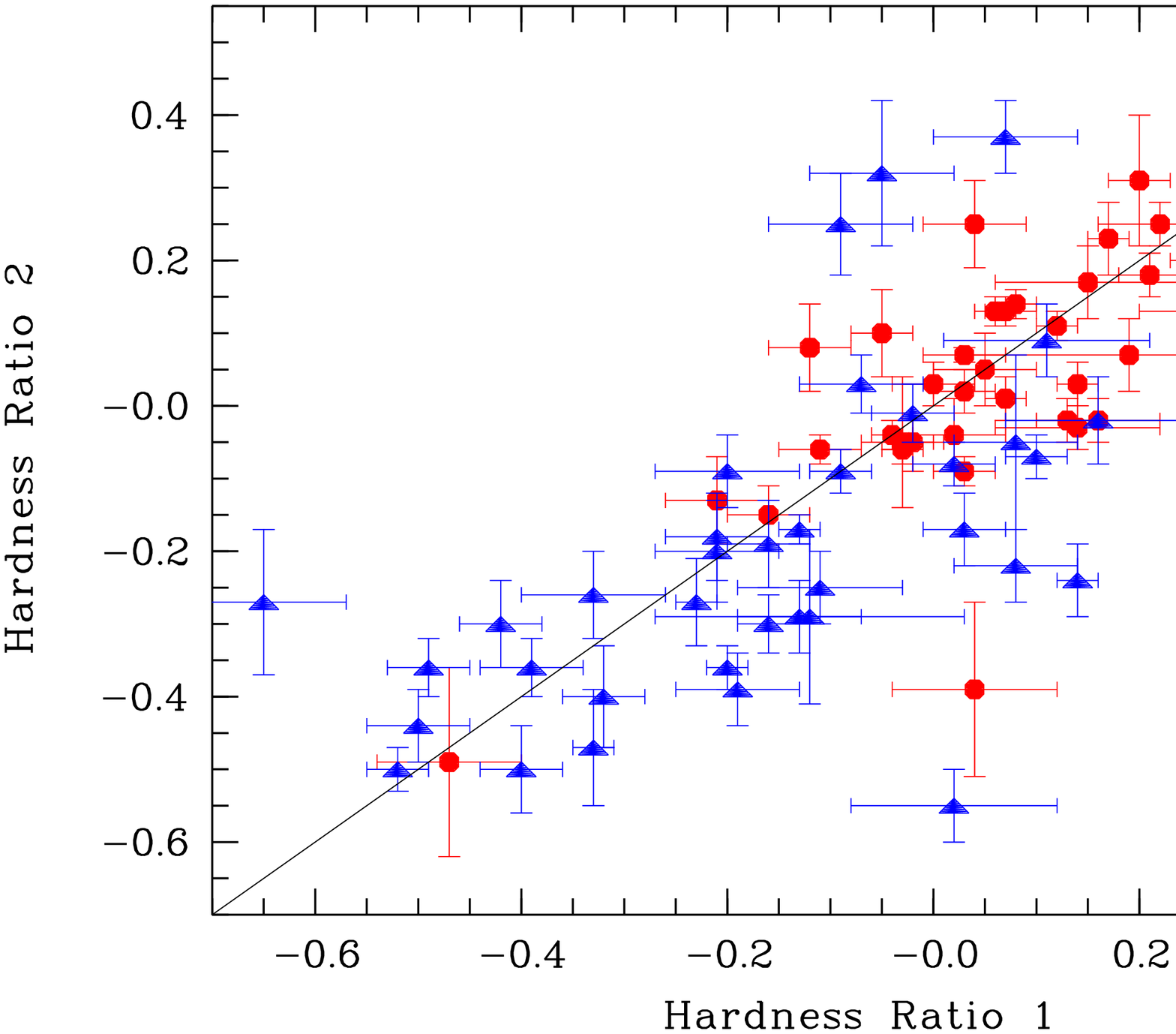}{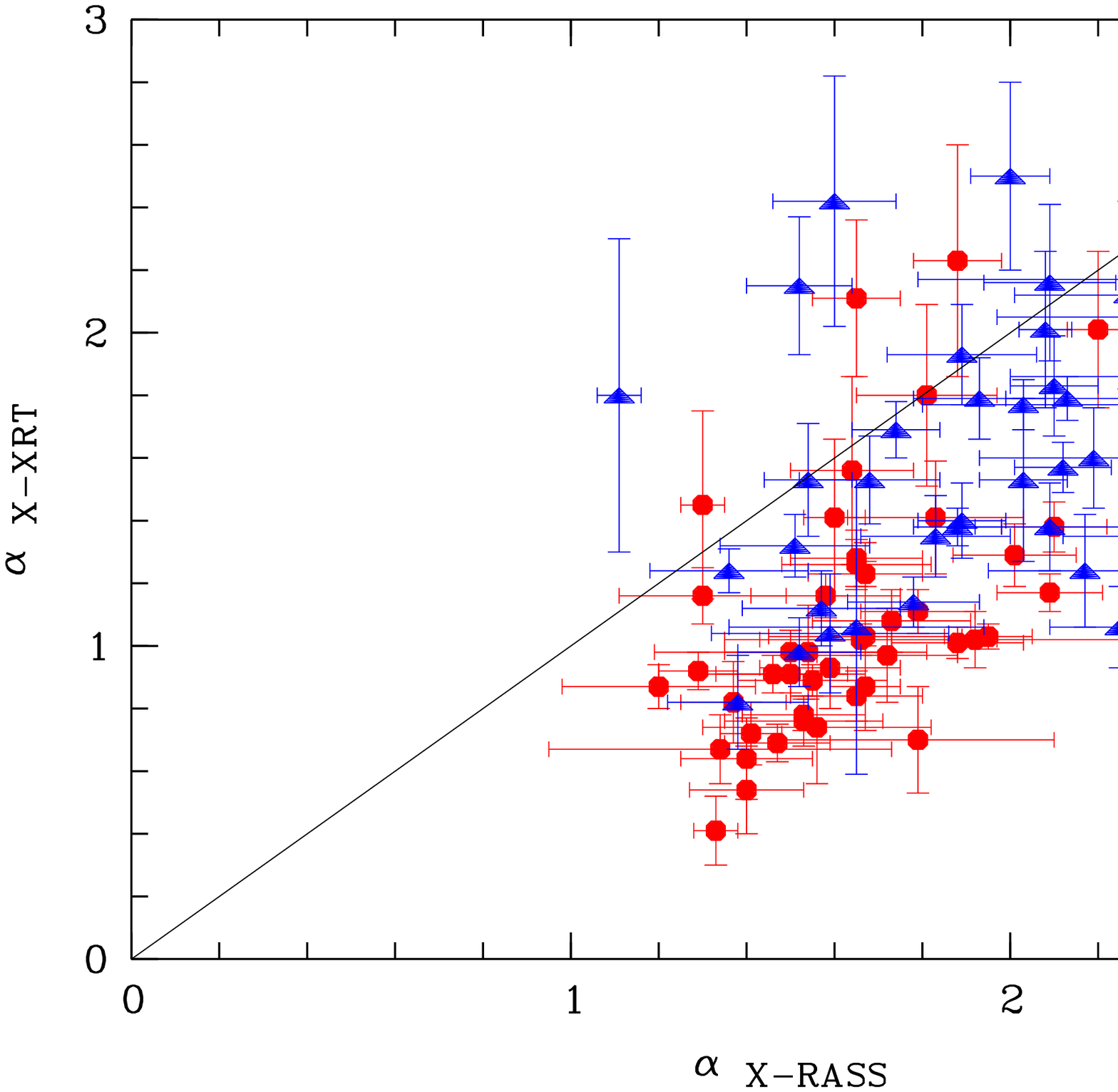}
\caption{\label{xray_hr_var} Short and long term X-ray spectral variability in the 
soft X-ray selected AGN sample observed by \swift. The left panel shows the 
Hardness ratio variability of two \swift\ observations of the same AGN, the 
 right panel the X-ray spectral slopes \ax\ 
 the \swift\ XRT and ROSAT All-Sky Survey 
observations. 
}
\end{figure*}

\clearpage

\begin{figure*}
\epsscale{1.6}
\plottwo{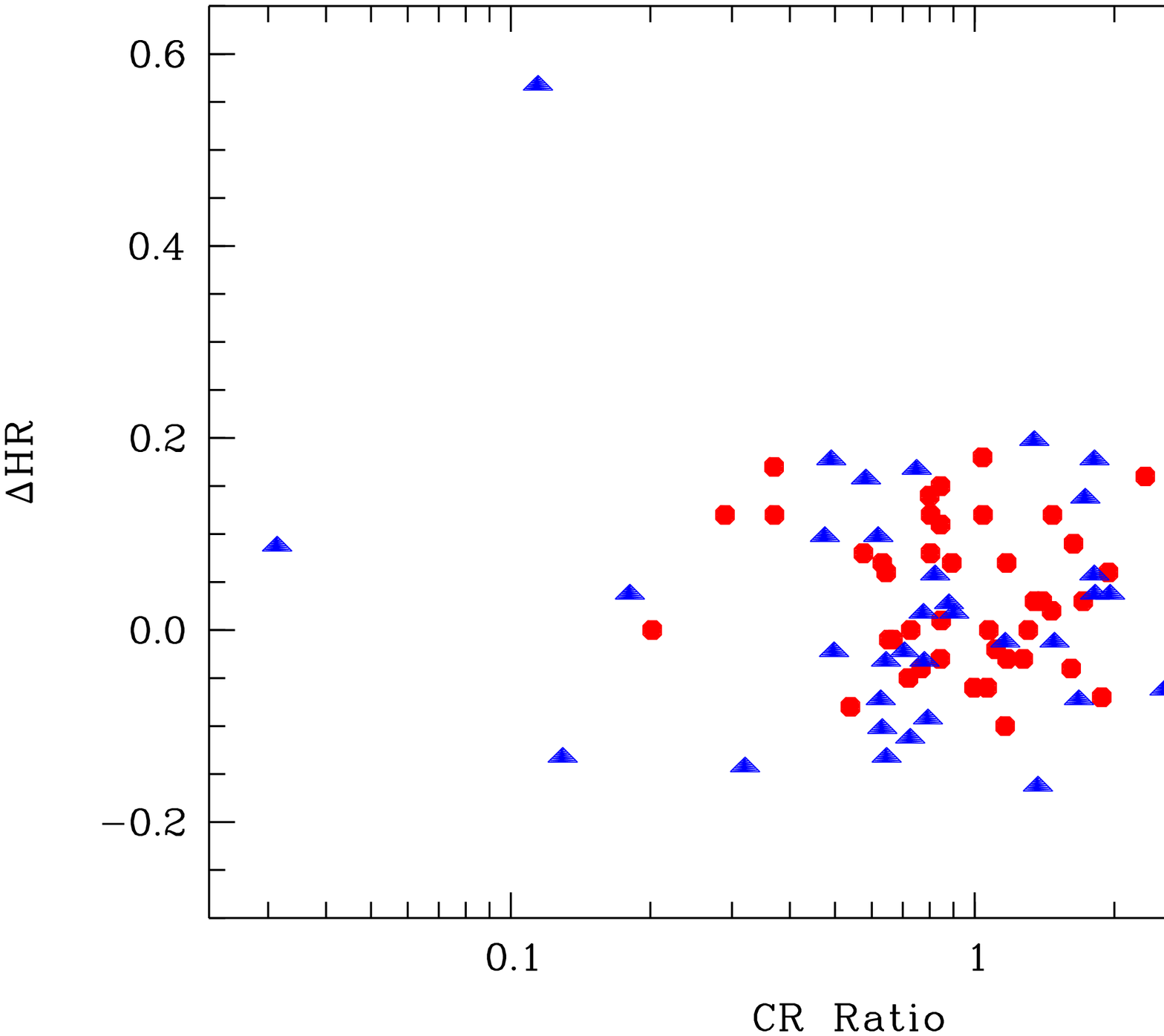}{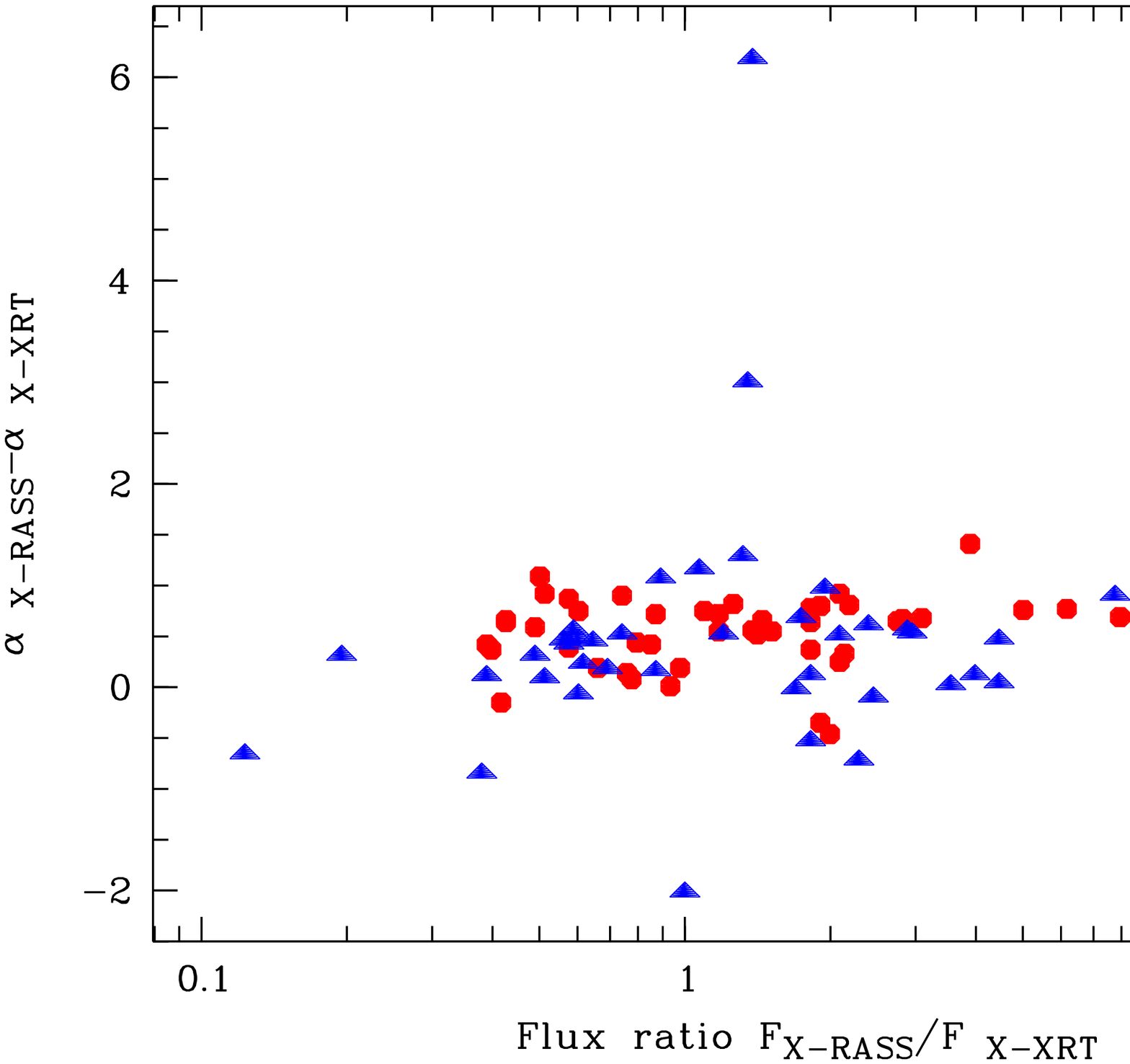}
\caption{\label{xray_var_cr_hr} Short term and long term X-ray variabilities. The left panel displays the Count rate ratio vs. the difference in the
hardness ratios in the \swift\ data, and the right panel shows the flux ratio between the RASS and the \swift\  observations vs. the differences in
the X-ray spectral slopes \ax.
}
\end{figure*}

\clearpage

\begin{figure*}
\epsscale{1.6}
\plottwo{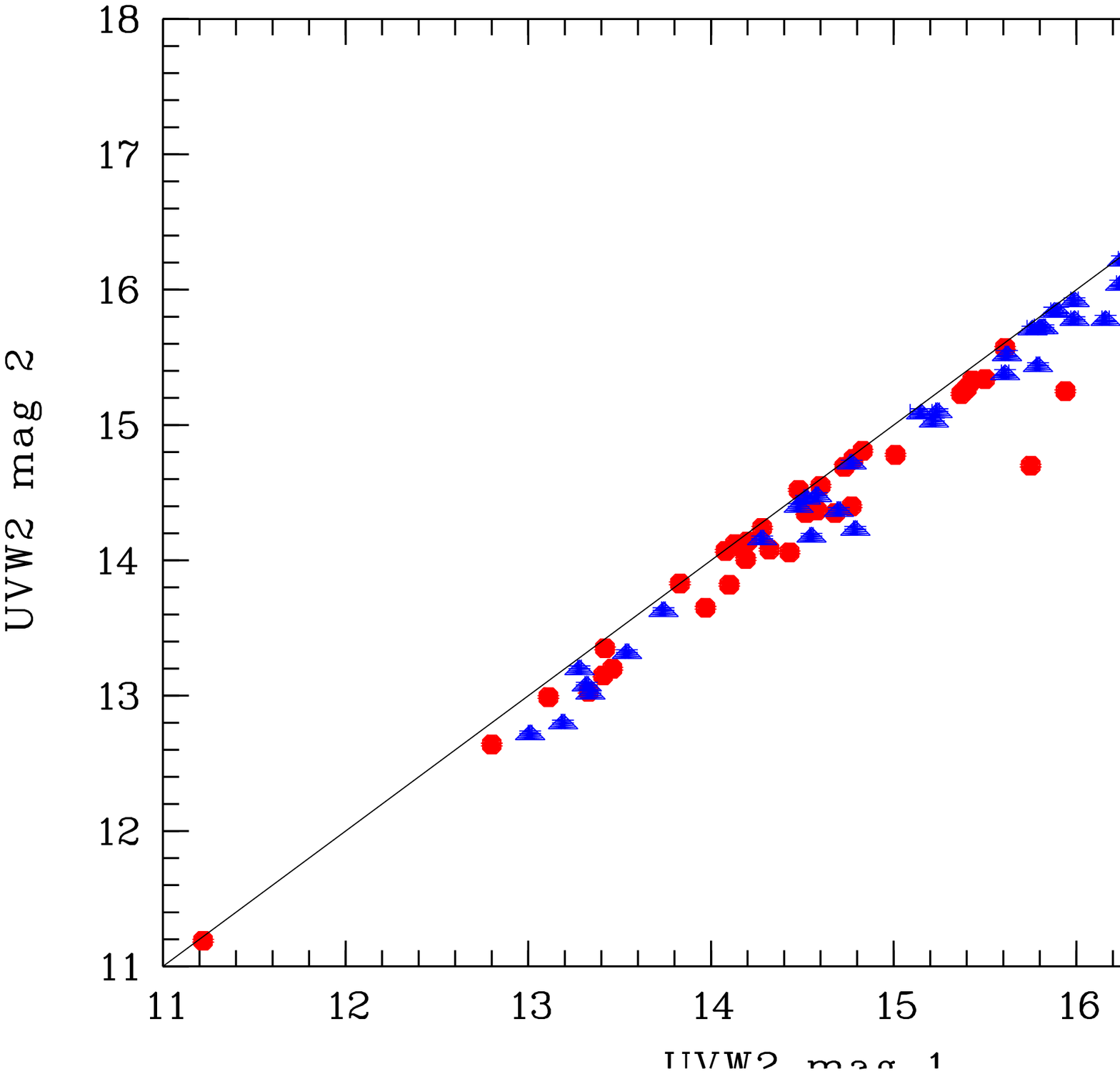}{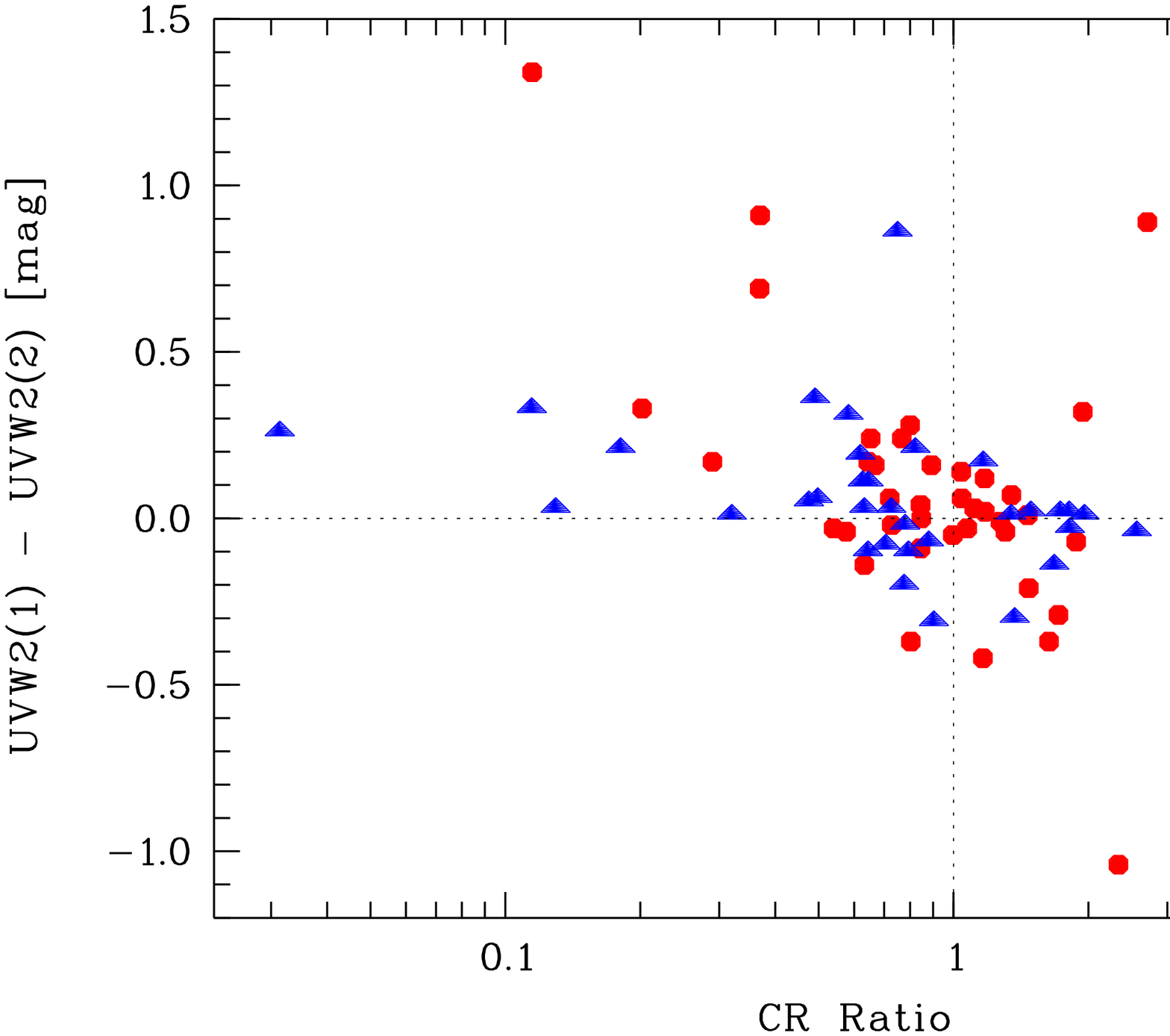}
\caption{\label{uvw2_var} UVOT W2 variability. The left panel displays the UVW2 magnitudes of the observations with the largest differences and the right
panel shows the difference between the X-ray flux ratio and the difference between the UVW2 magnitudes simultaneously taken with the X-ray data. 
}
\end{figure*}

\begin{figure*}
\epsscale{1.6}
\plottwo{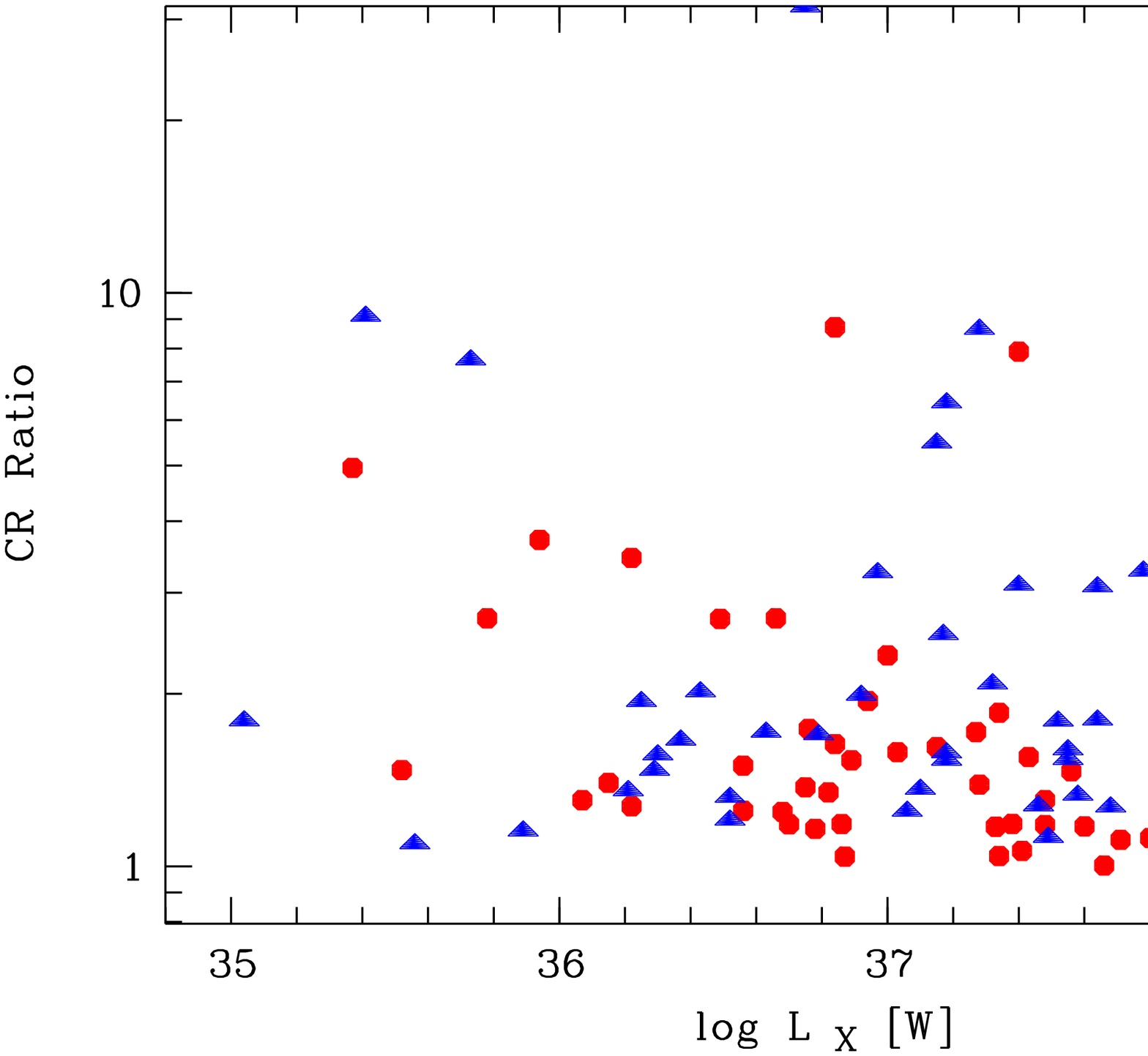}{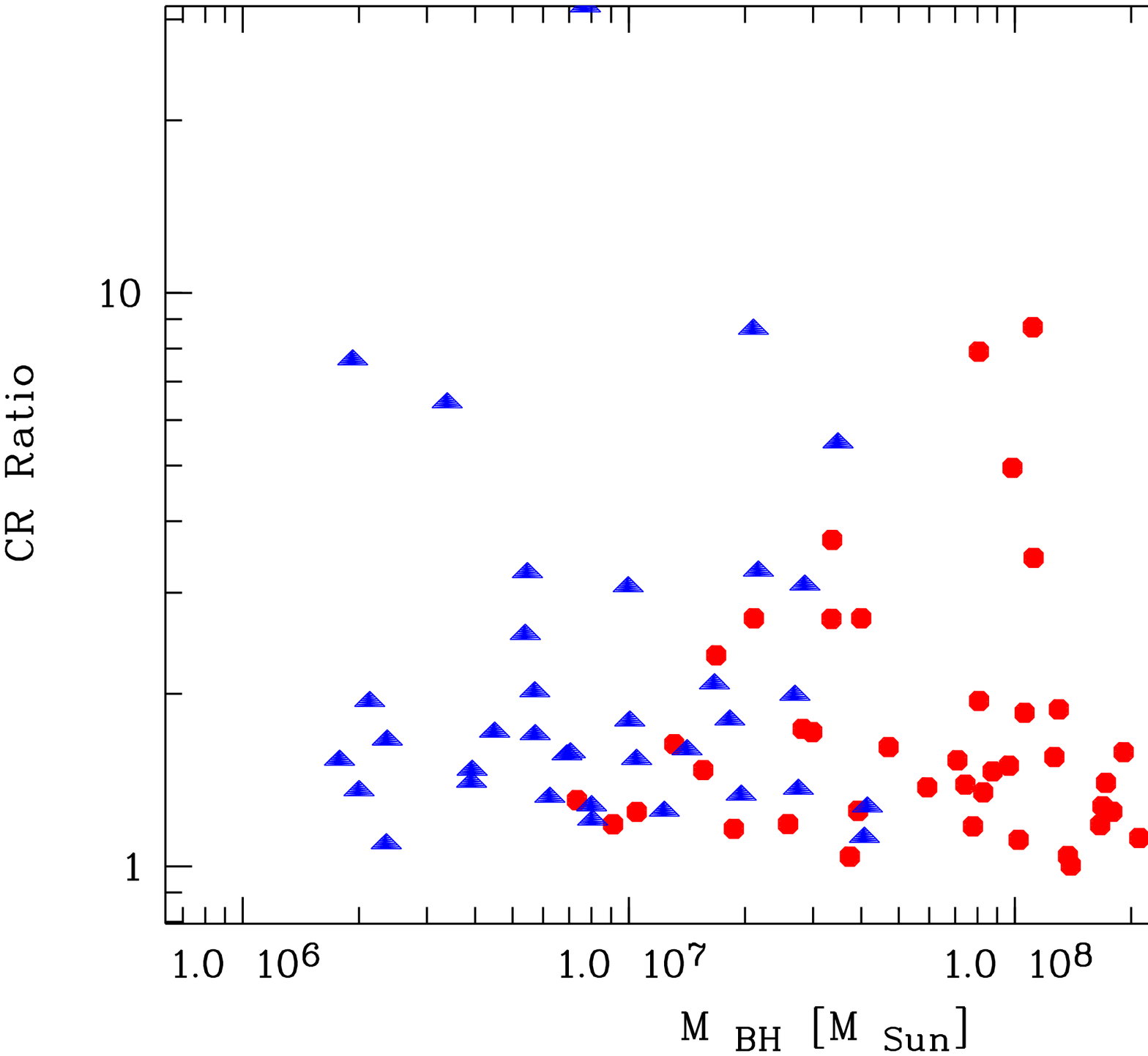}
\caption{\label{var_lx_mbh} Ratio of the \swift-XRT count rates of two epochs vs. 0.2-2.0 keV X-ray luminosity $L_{\rm X}$ (left panel) and the mass of
the central black hole (right panel). 
}
\end{figure*}

\clearpage

\begin{deluxetable}{rlcccccccrrcc}
\tabletypesize{\tiny}
\tablecaption{List of Soft X-ray selected AGN observed by \swift. 
\label{obj_list}}
\tablewidth{0pt}
\tablehead{
\colhead{\#} &
\colhead{Object} & \colhead{$\alpha_{2000}$} & 
\colhead{$\delta_{2000}$} & 
\colhead{z} & \colhead{$N_{\rm H, gal}$\tablenotemark{1}}
& \colhead{\ax\tablenotemark{2}}
& \colhead{log $F_{\rm X}$\tablenotemark{3}}
& \colhead{$E_{\rm B-V}$\tablenotemark{4}}
& \colhead{FWHM(H$\beta$)\tablenotemark{5}}
& \colhead{$M_{\rm BH}$\tablenotemark{6}}
& \colhead{H$\alpha$/H$\beta$}
& \colhead{$E_{\rm B-V~intr}$\tablenotemark{7}}
} 
\startdata
 1 & Mkn 335          & 00 06 19.5 &  +20 12 11 & 0.026 & 3.96 & 2.10\plm0.10 & --13.40 & 0.035 &   1710 &    7.7 & 3.30  & 0.070 \\
 2 & ESO 242-G008     & 00 25 00.2 & --45 29 34 & 0.059 & 1.48 & 1.56\plm0.26 & --14.31 & 0.010 &   3000 &   21.1 & 4.03  & 0.247 \\
 3 & Ton S 180        & 00 57 20.2 & --22 22 57 & 0.062 & 1.48 & 1.89\plm0.10 & --13.77 & 0.014 &    970 &    7.1 & 2.81  & 0.000 \\
 4 & QSO 0056-36      & 00 58 37.4 & --36 06 05 & 0.165 & 1.94 & 1.72\plm0.16 & --14.29 & 0.014 &   4550 &  448.4 & 3.10  & 0.012 \\
 5 & RX J0100.4--5113 & 01 00 27.1 & --51 13 54 & 0.062 & 2.42 & 1.73\plm0.18 & --13.99 & 0.015 &   3190 &   39.3 & \nodata & \nodata \\
 6 & RX J0105.6--1416 & 01 05 38.8 & --14 16 14 & 0.070 & 1.77 & 1.29\plm0.09 & --13.89 & 0.023 &   2600 &   25.7 & 2.75  & 0.000 \\
 7 & RX J0117.5--3826 & 01 17 30.6 & --38 26 30 & 0.225 & 2.08 & 2.09\plm0.15 & --14.18 & 0.013 &    900 &    8.0 & 2.65  & 0.000 \\
 8 & MS 0117-28       & 01 19 35.7 & --28 21 32 & 0.349 & 1.65 & 2.27\plm0.26 & --14.54 & 0.017 &   1680 &   90.4 & \nodata & \nodata \\
\enddata

\tablenotetext{1}{Galactic column density from \citet{dic90} in units of
$10^{20}$ cm$^{-2}$.}
\tablenotetext{2}{X-ray energy spectral \ax\  in the 0.2-2.0 keV band during the
RASS as given in \citet{gru01}.} 
\tablenotetext{3}{Rest frame 0.2-2.0 keV X-ray flux during the RASS as given in \citet{gru01} in units of W m$^{-2}$.}
\tablenotetext{4}{Galactic reddening $E_{\rm B-V}$ from \citet{schlegel98}.}
\tablenotetext{5}{FWHM(H$\beta$) is given in units of km s$^{-1}$ and the values were taken from \citet{gru04a}. We separated NLS1s and BLS1s by the 
common criterion  of 2000 km s$^{-1}$}
\tablenotetext{6}{Black hole mass $M_{\rm BH}$ in units of $10^6$\msun determined by the method described in \citet{kaspi00} using FWHM(H$\beta$) and 
$L_{\rm 5100\AA}$ as given in \citet{gru04a}.}
\tablenotetext{7}{Intrinsic reddening $E_{\rm B-V}-intr$ derived from the Balmer decrement.}
\tablenotetext{8}{Taken from \citet{lira99}.}
\end{deluxetable}

\begin{deluxetable}{rlccccrrrrrrr}
\tabletypesize{\tiny}
\tablecaption{Observation summary of the  \swift-XRT and UVOT observations. 
\label{obslog}}
\tablewidth{0pt}
\tablehead{
\colhead{\#} &
\colhead{Object} & 
\colhead{Target ID} &
\colhead{segment} & 
\colhead{$T_{\rm start}$} & 
\colhead{$T_{\rm stop}$} 
& \colhead{$T_{\rm XRT}$\tablenotemark{1}}
& \colhead{$T_{\rm V}$\tablenotemark{1}}
& \colhead{$T_{\rm B}$\tablenotemark{1}}
& \colhead{$T_{\rm U}$\tablenotemark{1}}
& \colhead{$T_{\rm W1}$\tablenotemark{1}}
& \colhead{$T_{\rm M2}$\tablenotemark{1}}
& \colhead{$T_{\rm W2}$\tablenotemark{1}}
} 
\startdata
 1 & Mkn 335\tablenotemark{2}   &  90006  & 003 & 2008-06-15 00:48  & 2008-06-15 02:45 &  2323 &  200 &  200 &  200 &  401 &  447 & 803 \\
 2 & ESO 242-G008     & 35756 & 001 & 2006-08-16 14:13 & 2006-08-16 20:52 &  3886 &  324 &  324 &  324 &  644 &   821 & 1293 \\
   &                  &       & 002 & 2006-08-17 00:03 & 2006-08-17 16:13 &  4375 &  470 &  470 &  470 &  939 &  1266 & 1881 \\
   &                  &       & 003 & 2006-11-05 09:47 & 2006-11-05 18:16 &  7703 &  632 &  632 &  632 & 1264 &  1776 & 2536 \\
 3 & Ton S 180        & 35757 & 001 & 2006-09-12 02:33 & 2006-09-12 13:51 &   862 &   92 &   92 &   92 &  185 &    38 &  372 \\
   &                  &       & 003 & 2007-05-16 02:15 & 2007-05-16 16:54 &  4994 &  506 &  506 &  506 & 1020 &  1370 & 2035 \\
   &                  &       & 004 & 2007-09-27 00:54 & 2007-09-27 04:20 &  2248 & \nodata & \nodata & \nodata & \nodata & \nodata & \nodata \\
   &                  &       & 005 & 2007-12-30 21:22 & 2007-12-31 23:16 & 10953 &  900 &  900 &  900 & 1787 & 2346 & 3585 \\
\enddata

\tablenotetext{1}{XRT and UVOT exposure times are given in s.}
\tablenotetext{2}{\swift\ observed Mkn 335 several times as a monitoring project \citet{grupe07b, grupe08}. Here we used one of the observations during the more typical high-state.}
\tablenotetext{3}{The Narrow Line Seyfert 1 galaxy RX J0148.3--2758 has been observed by \swift\ multiple times and the results of the
observations obtained in 2005 have already been published in \citep{grupe06}. Therefore here we only list the new, previously unpublished
observations. }
\tablenotetext{4}{PG 1211+143 was monitored by \swift\ for more than a month \citep{bachev09}. Here we picked only one of the observations during high state 
}
\tablenotetext{5}{Mkn 766 has been observed through a monitoring campaign by \swift (Grupe et al. in prep). 
 Here we only present the first segment of data as a representative observation.}
\tablenotetext{6}{3C 273 has been observed by \swift\ many times as a calibration target as well as a science target. Here we only picked
two recent observations in 2009 which were observed in 5 UVOT filters in the $5^{'}\times 5^{'}$ UVOT hardware window.}
\tablenotetext{7}{UVOT observations done with the grisms.}
\tablenotetext{8}{MCG+08-23-067 was  previously observed in 2005. However, these observations were all performed with the UV grism} 
\tablenotetext{9}{The data of
NGC 5548 are from an observing campaign in 2007 with simultaneous observations with Suzaku \citep{krongold09}.}
\tablenotetext{10}{Mkn 684 has been observed a few times before the listed observations. However, for these observation the
UVOT was using the grisms.}
\tablenotetext{11}{Due to a bright star 3$^{'}$ away from the position of RX J2248.6--5109 UVOT could not observe the field in order to avoid damaging
the instrument.}
\end{deluxetable}

\clearpage

\begin{deluxetable}{rlcrrrrrr}
\tabletypesize{\tiny}
\tablecaption{Results of the  \swift-UVOT photometry 
\label{uvot_res}}
\tablewidth{0pt}
\tablehead{
\colhead{\#} &
\colhead{Object} & \colhead{segment(s)}  
& \colhead{ V\tablenotemark{1}}
& \colhead{ B\tablenotemark{1}}
& \colhead{ U\tablenotemark{1}}
& \colhead{ UVW1\tablenotemark{1}}
& \colhead{ UVM2\tablenotemark{1}}
& \colhead{ UVW2\tablenotemark{1}}
} 
\startdata
 1 & Mkn 335          & 003 & 14.20\plm0.01 & 14.45\plm0.01 & 13.27\plm0.01 & 13.11\plm0.01 & 12.97\plm0.01 & 13.01\plm0.01 \\
 2 & ESO 242-G008     & 001+002 & 16.09\plm0.02 & 16.70\plm0.02 & 16.01\plm0.01 & 15.97\plm0.01 & 15.97\plm0.02 & 15.94\plm0.01 \\
   &                  & 003 & 15.93\plm0.02 & 16.44\plm0.02 & 15.54\plm0.01 & 15.42\plm0.01 & 15.32\plm0.01 & 15.25\plm0.01 \\
 3 & Ton S 180        & 001 & 14.46\plm0.02 & 14.58\plm0.01 & 13.39\plm0.01 & 13.09\plm0.01 & 13.12\plm0.03 & 12.85\plm0.01 \\
   &                  & 003 & 14.56\plm0.01 & 14.70\plm0.01 & 13.49\plm0.01 & 13.22\plm0.01 & 13.01\plm0.01 & 13.02\plm0.01 \\
   &                  & 005 & 14.29\plm0.01 & 14.44\plm0.00 & 13.23\plm0.00 & 12.94\plm0.00 & 12.71\plm0.00 & 12.73\plm0.00 \\
 4 & QSO 0056-36      & 001 & 15.76\plm0.01 & \nodata       & \nodata       & \nodata       & 14.26\plm0.04 & 14.10\plm0.01 \\
   &                  & 002 & 15.55\plm0.05 & 15.65\plm0.02 & 14.30\plm0.02 & 14.05\plm0.01 & 13.82\plm0.01 & 13.80\plm0.01 \\
\enddata

\tablenotetext{1}{ UVOT magnitudes corrected for Galactic reddening using the $E_{\rm (B-V)}$ values by \citet{schlegel98} as given in
Table\,\ref{obj_list}.}

\end{deluxetable}

\begin{deluxetable}{rlccccccccc}
\tabletypesize{\tiny}
\tablecaption{Spectral analysis of the \swift\ XRT data.  \label{xrt_res}}
\tablewidth{0pt}
\tablehead{
\colhead{\#} 
& \colhead{Object} 
& \colhead{segment(s)}  
& \colhead{Model\tablenotemark{1}} 
& \colhead{$\alpha_{\rm X,soft}$} 
& \colhead{$E_{\rm break}$\tablenotemark{2}}
& \colhead{$\alpha_{\rm X,hard}$}
& \colhead{log $F_{\rm 0.2-2.0 keV}$\tablenotemark{3}}
& \colhead{$\chi^2/\nu$}
& \colhead{CR\tablenotemark{4}} 
& \colhead{HR\tablenotemark{5}}
} 
\startdata
 1 & Mkn 335          & 003     & (1)  & 1.31\plm0.09           & --- & ---  & --13.37 & 34/37 & 1.237\plm0.030 &
 +0.03\plm0.02 \\          
 2 & ESO 242-G008     & 001-002 & (1)  & 0.74$^{+0.18}_{-0.17}$ & --- & ---  & --14.41 & 14/12 & 0.041\plm0.003 & +0.14\plm0.08 \\
   &                  & 003 & (1)  & 1.17$^{+0.11}_{-0.10}$ & --- & --- &  --14.41 & 37.0/34 & 0.111\plm0.004 & --0.03\plm0.03 \\
   &                  & 003 & (2)  & 1.46$^{+0.13}_{-0.14}$  & 1.35$^{+0.24}_{-0.23}$ & 0.75$^{+0.17}_{-0.15}$ &  --14.41 & 24.5/32 \\
 3 & Ton S 180        & 001 & (1)  & 1.56\plm0.17	    & --- & --- &  --13.72 & 17/14 & 0.495\plm0.026 & --0.29\plm0.05  \\
   &                  & 003 & (1)  & 1.48\plm0.06	    & --- & --- &  --13.66 & 63/65 & 0.543\plm0.011 & --0.23\plm0.02 \\
   &                  & 004 & (1)  & 1.40\plm0.12	    & --- & --- &  --14.07 & 20/16 & 0.202\plm0.010 & --0.13\plm0.06 \\
   &                  & 005 & (1)  & 1.58\plm0.04	    & --- & --- &  --13.52 & 122/125 & 0.789\plm0.009 & --0.22\plm0.01 \\
\enddata

\tablenotetext{1}{Spectral models used are: (1): absorbed power law, (2): absorbed broken power law, (3) absorbed blackbody
plus power law}
\tablenotetext{2}{The break energy $E_{\rm break}$ is given in units of keV.}
\tablenotetext{3}{the unabsorbed rest-frame 0.2-2.0 X-ray flux log $F_{\rm 0.2-2.0 keV}$ is given in units of W m$^{-2}$}
\tablenotetext{4}{CR is the XRT count rate given in units of counts s$^{-1}$.}
\tablenotetext{5}{The hardness ratio is defined as HR=(H-S)/(H+S) where S and H are the backgroud subtracted numbers of counts in the 
0.3-1 and 1-10 keV energy bands, respectively} 
\end{deluxetable}

\begin{deluxetable}{rlcccccccccccccc}
\rotate
\tabletypesize{\tiny}
\tablecaption{Properties of the Spectral energy Distribution
\label{sed_res}}
\tablewidth{0pt}
\tablehead{
\colhead{\#} &
\colhead{Object} & \colhead{segment(s)}  
& \colhead{ \ax\tablenotemark{1}}
& \colhead{ \auv\tablenotemark{2}}
& \colhead{ \aox\tablenotemark{4}}
& \colhead{ $\alpha_{\rm UV-corr}$\tablenotemark{3}}
& \colhead{ $\alpha_{\rm ox-corr}$\tablenotemark{5}}
& \colhead{ log \lx\tablenotemark{6}}
& \colhead{ log \lbbb\tablenotemark{6}}
& \colhead{ $L/L_{\rm Edd}$}
& \colhead{ log \lbbb\tablenotemark{6}\tablenotemark{7}}
& \colhead{ log 5100\tablenotemark{6}}
& \colhead{ log \lbbb-corr\tablenotemark{6}}
& \colhead{ $L/L_{\rm Edd}$-corr}
& \colhead{ log 5100-corr\tablenotemark{6}}
} 
\startdata
 1 & Mkn 335         & 003 &  1.30\plm0.09  & $+$0.58\plm0.08 & 1.26 & $+$0.78\plm0.12 & 1.34 & 36.75 & 37.72 & 0.51 &  37.53  & 36.73 & 38.07 & 1.32 & 36.83 \\
 2 & ESO 242-G008    & 001+002 & 0.74\plm0.18 & +1.48\plm0.16  & 1.36 & $+$0.12\plm0.21 & 1.67 & 35.78 & 36.78 & 0.03 &  36.81  & 36.70 & 38.16 & 0.60 & 36.90 \\
   &                 & 003 &  1.17\plm0.10  & $+$0.98\plm0.15 & 1.29 & $+$0.40\plm0.19 & 1.66 & 36.44 & 37.24 & 0.07 &  37.22  & 36.73 & 38.59 & 2.14 & 36.77 \\
 3 & Ton S 180        & 005 &  1.48\plm0.08 & $+$0.25\plm0.07 & 1.42 & $+$0.25\plm0.07 & 1.42 & 37.18 & 38.77 & 6.30 &  38.31  & 37.48 & 38.77 & 6.30 & 37.48 \\
 4 & QSO 0056-36      & 003 &  1.13\plm0.07 & $-$0.03\plm0.14 & 1.43 & $-$0.12\plm0.14 & 1.43 & 37.48 & 39.19 & 0.29 &  38.58  & 37.72 & 39.19 & 0.29 & 37.72 \\ 
 5 & RX J0100.4--5113 & 001 &  1.08\plm0.10 & $+$0.69\plm0.08 & 1.39 & ---             & ---  & 36.56 & 37.85 & 0.30 &  37.60  & 37.00 & ---  & ---  & ---   \\
 6 & RX J0105.6--1416 & 001 &  0.92\plm0.20 & $+$0.77\plm0.10 & 1.25 & $+$0.77\plm0.10 & 1.25 & 36.86 & 37.91 & 0.25 &  37.70  & 37.09 & 37.91 & 0.25 & 37.09 \\
 7 & RX J0117.5--3826 & 001 &  1.73\plm0.29 & $+$0.93\plm0.09 & 1.39 & $+$0.93\plm0.09 & 1.39 & 37.46 & 38.25 & 0.30 &  38.22  & 37.58 & 38.25 & 0.30 & 37.58 \\
\enddata

\tablenotetext{1}{X-ray slope in the 0.2-2.0 keV energy range }
\tablenotetext{2}{Optical/UV slope corrected for Galactic reddening}
\tablenotetext{3}{Optical/UV slope corrected for Galactic and intrinsic reddening}
\tablenotetext{4}{UV-to-X-ray spectra slope \aox, as defined by \citet{tananbaum79}}
\tablenotetext{5}{UV-to-X-ray spectra slope \aox, as defined by \citet{tananbaum79} corrected for Galactic and intrinsic reddening}
\tablenotetext{6}{Luminosities are given in units of W.}
\tablenotetext{7}{Using the broken power law model as shown in Figure 1.}

\end{deluxetable}

\begin{deluxetable}{lccccccccc}
\tablecaption{Mean, standard deviation, and median of the parameters of the whole sample, NLS1s (43) and BLS1s(49)
\label{mean}}
\tablewidth{0pt}
\tablehead{
& \multicolumn{3}{c}{all}
& \multicolumn{3}{c}{NLS1s}
& \multicolumn{3}{c}{BLS1s} \\
 \colhead{Property}
& \colhead{mean} & \colhead{sd\tablenotemark{1}} & \colhead{median} 
& \colhead{mean} & \colhead{sd\tablenotemark{1}} & \colhead{median}
& \colhead{mean} & \colhead{sd\tablenotemark{1}} & \colhead{median}
} 
\startdata
\ax   & +1.38 & 0.55 & +1.26 &                  +1.68 & 0.51 & +1.59 &    +1.11 & 0.42 & +1.03 \\
\auv  & +0.81 & 0.65 & +0.65 &                  +0.85 & 0.66 & +0.65 &    +0.79 & 0.67 & +0.61 \\
\aox  & +1.34 & 0.14 & +1.35 &                   +1.36 & 0.16 & +1.39 &   +1.32 & 0.11 & +1.33 \\
$\alpha_{\rm UV-corr}$ & +0.44 & 0.67 & +0.42 & +0.61 & 0.51 & +0.60 &    +0.28 & 0.78 & +0.17 \\ 
$\alpha_{\rm ox-corr}$ & +1.42 & 0.17 & +1.41 &  +1.42 & 0.15 &  +1.42 &  +1.42 & 0.18 & +1.40 \\
log $L_{\rm X}$ [W] & 36.94 & 0.73 & 37.00 &         36.93 & 0.75 & 37.10 &   36.96 & 0.71 & 36.89 \\ 
log $L_{\rm BBB-expo}$\tablenotemark{2} [W] & 38.00 & 0.96 & 38.07 &       37.86 & 0.99 & 38.01 &   38.11 & 0.92 & 38.07 \\
log $L_{\rm BBB-bknpo}$\tablenotemark{3} [W] & 37.80 & 0.80 & 37.92 &       37.72 & 0.986 & 38.07 &   37.89 & 0.75 & 37.82 \\
\enddata

\tablenotetext{1}{Standard deviation of the sample mean}

\end{deluxetable}

\begin{deluxetable}{rccccccccccccc}
\rotate
\tabletypesize{\tiny}
\tablecaption{Correlation analysis\tablenotemark{1}
\label{correlations}}
\tablewidth{0pt}
\tablehead{
 \colhead{ }
& \colhead{ \ax}
& \colhead{ \auv}
& \colhead{ \aox}
& \colhead{ $\alpha_{\rm UV-corr}$}
& \colhead{ $\alpha_{\rm ox-corr}$}
& \colhead{ log \lx}
& \colhead{ log \lbbb}
& \colhead{ $L/L_{\rm Edd}$}
& \colhead{ log 5100}
& \colhead{ log \lbbb-corr}
& \colhead{ $L/L_{\rm Edd}$-corr}
& \colhead{ log 5100-corr}
& \colhead{ FWHM(H$\beta$)}
} 
\startdata
& $r_l$,(\#) & --0.15 (87)& +0.47 (87)& --0.24 (71)& +0.43 (71) & +0.34 (90)& +0.23 (87)& 0.51 (86)& 0.28 (87)& 0.26 (71)& 0.57 (71)& 0.24 (71)& --0.58 (90)\\
 \rb{ \ax} & P & 0.1606    & $<$0.0001 & 0.0408      & 0.0002    & 0.0009    & 0.0301    & $<$0.0001& 0.0079   & 0.0263   & $<$0.0001& 0.0408   & $<$0.0001 \\
& --0.19, --1.8 (87) & $r_{\rm s}$, $T_{\rm s}$ (\#) & --0.63 (87) & & +0.03 (71) & --0.67 (87) & --0.82 (87) & --0.63 (86) & --0.77 (87) & --0.54 (71) &
--0.31 (71) & --0.43 (71) & --0.04 (87)   \\
 \rb{ \auv} & 0.0769 & P & $<$0.0001 & \rb{---} & 0.8011 & $<$0.0001  & $<$0.0001 & $<$0.0001 & $<$0.0001 & $<$0.0001 & 0.0189 & 0.0001 & 0.7098 \\
& +0.48, +5.0 (87) & --0.57, -6.3 (87) & & --0.27 (71) & & +0.46 (87) & +0.64 (87) & +0.55 (86) & +0.64 (87) & 0.49 (71) & +0.33 (71) & 0.53 (71) & --0.19 (87) \\
 \rb{ \aox} & $<$0.0001 & $<$0.0001 & \rb{1} & 0.0235 & \rb{---} & $<$0.0001 & $<$0.0001 & $<$0.0001 & $<$0.0001 & $<$0.0001 & 0.0044 & $<$0.0001 & 0.0745 \\
& --0.14, --1.2 (71) & & --0.21, -1.8 (71) & & --0.68 (67) & --0.43 (71) & --0.48 (71) & --0.20 (70) & --0.45 (71) & --0.81 (71) & --0.72 (71) & --0.55
(71) & $-$0.08 (71) \\
 \rb{ $\alpha_{\rm UV-corr}$} & 0.2419 & \rb{---} & 0.0844 & \rb{1} & $<$0.0001 & 0.0001 & $<$0.0001 & 0.0898 & $<$0.0001 & $<$0.0001 & $<$0.0001 & $<$0.0001 &
 0.5011 \\
\enddata

\tablenotetext{1}{The part above the diagonal in the  gives the linear correlation coefficient $r_{\rm l}$ and 
the corresponding probability P and number of sources involved. The lower part of the table lists $R_{\rm S}$, $T_{\rm S}$, number of sources, and the
probability.}

\end{deluxetable}

\begin{appendix}
\section{Spectral Energy Distribution Plots}

Here we display the SEDs for each AGN with the power law model with exponential cutoff (Model A) for the reddening uncorrected and corrected
 UVOT data and the double broken power law model (Model B). Note that the SEDs with intrinsic reddening corrected UVOT data are only shown for those
 AGN for which the Balmer decrement was measured.

\def\eps@scaling{0.6}%

\begin{figure*}
\epsscale{0.60}
\plotthree{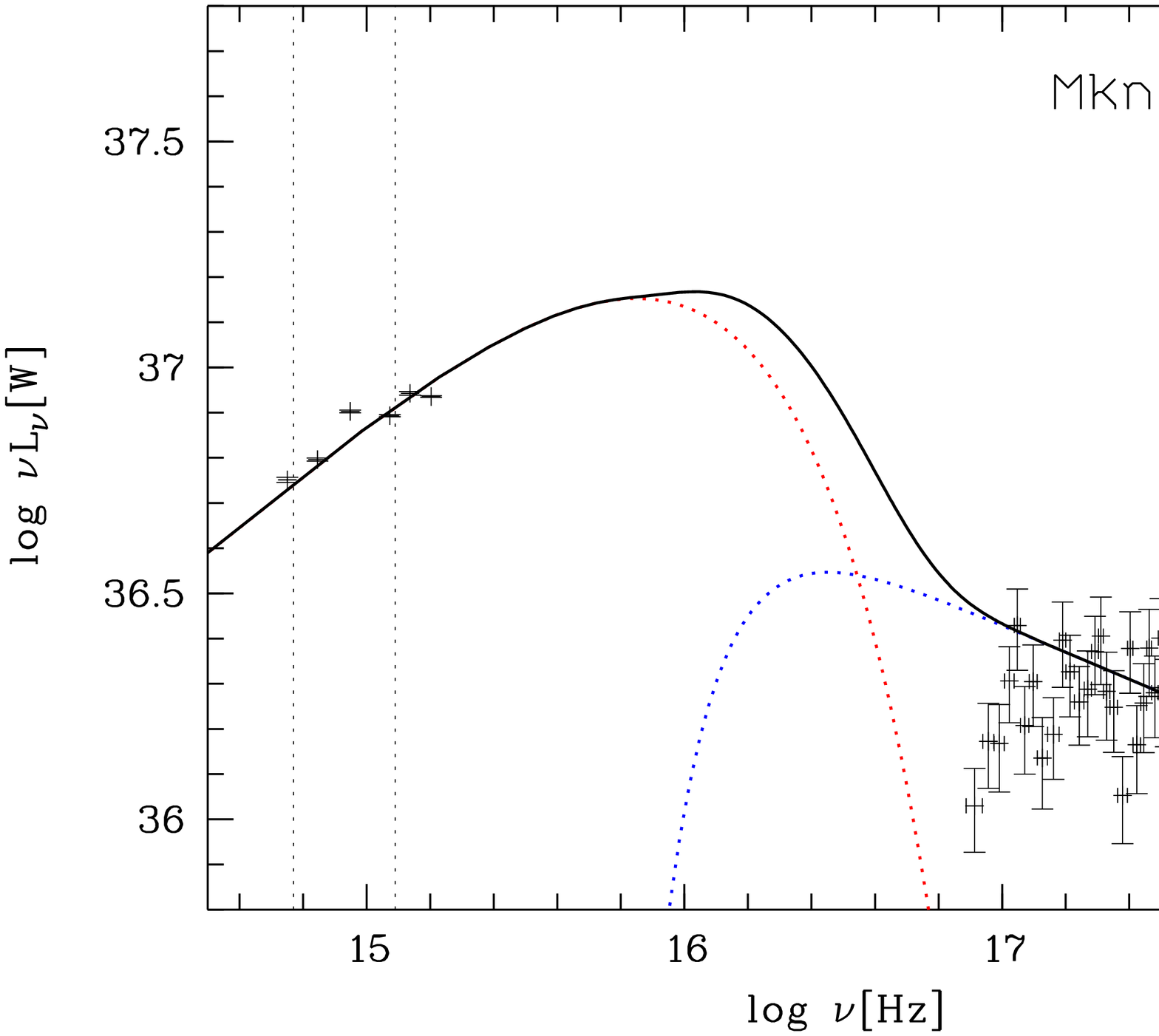}{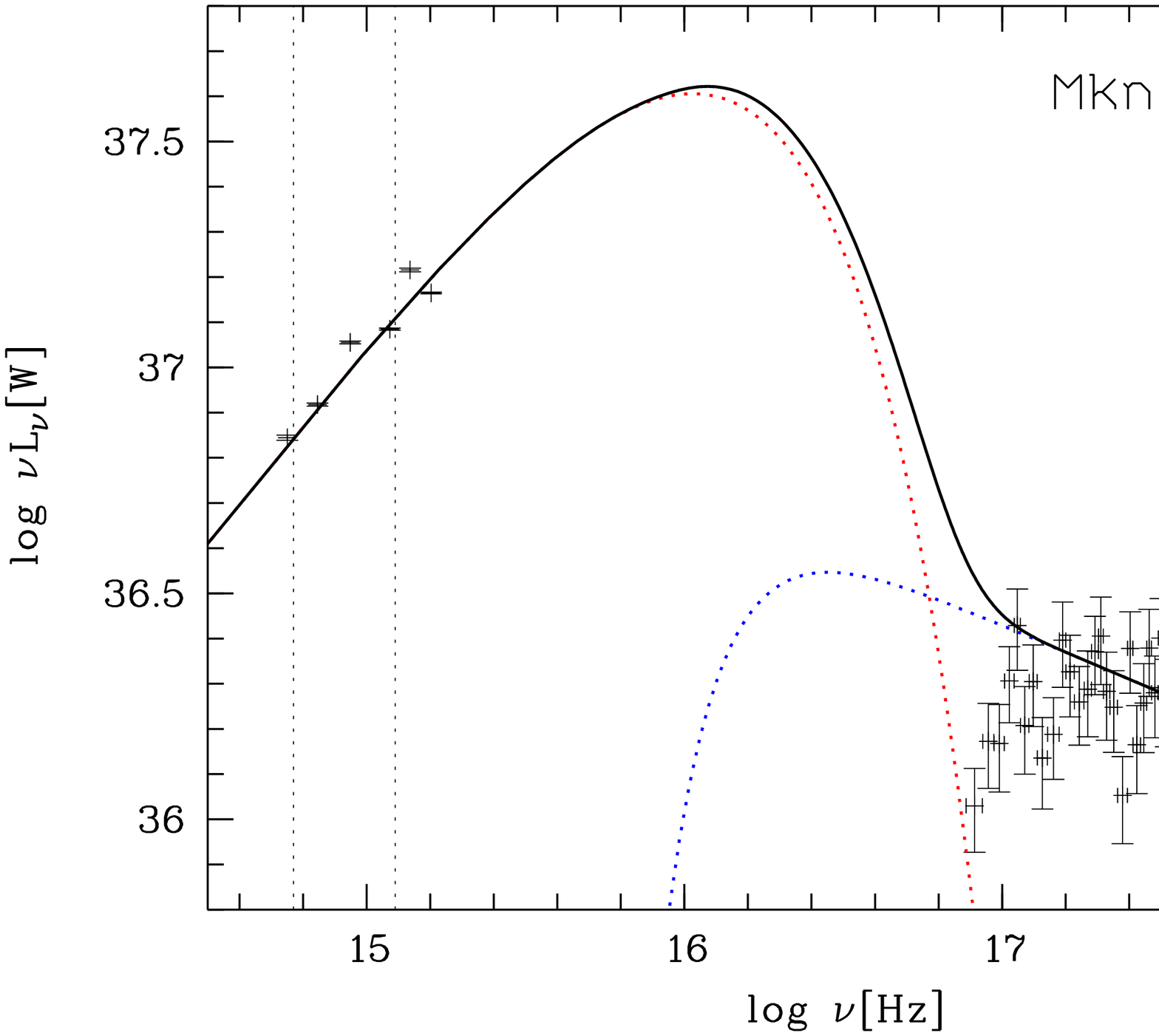}{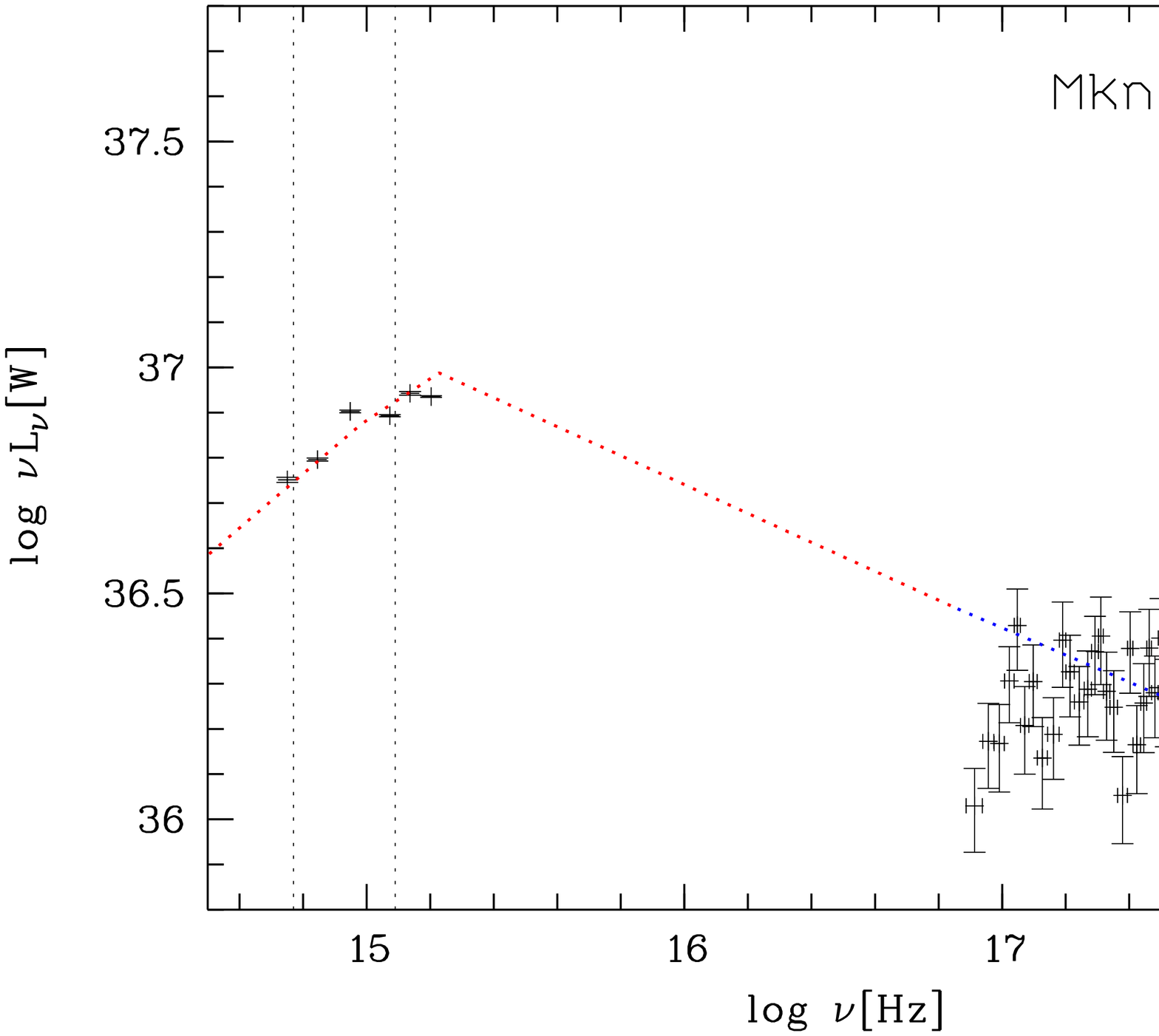}

\plotthree{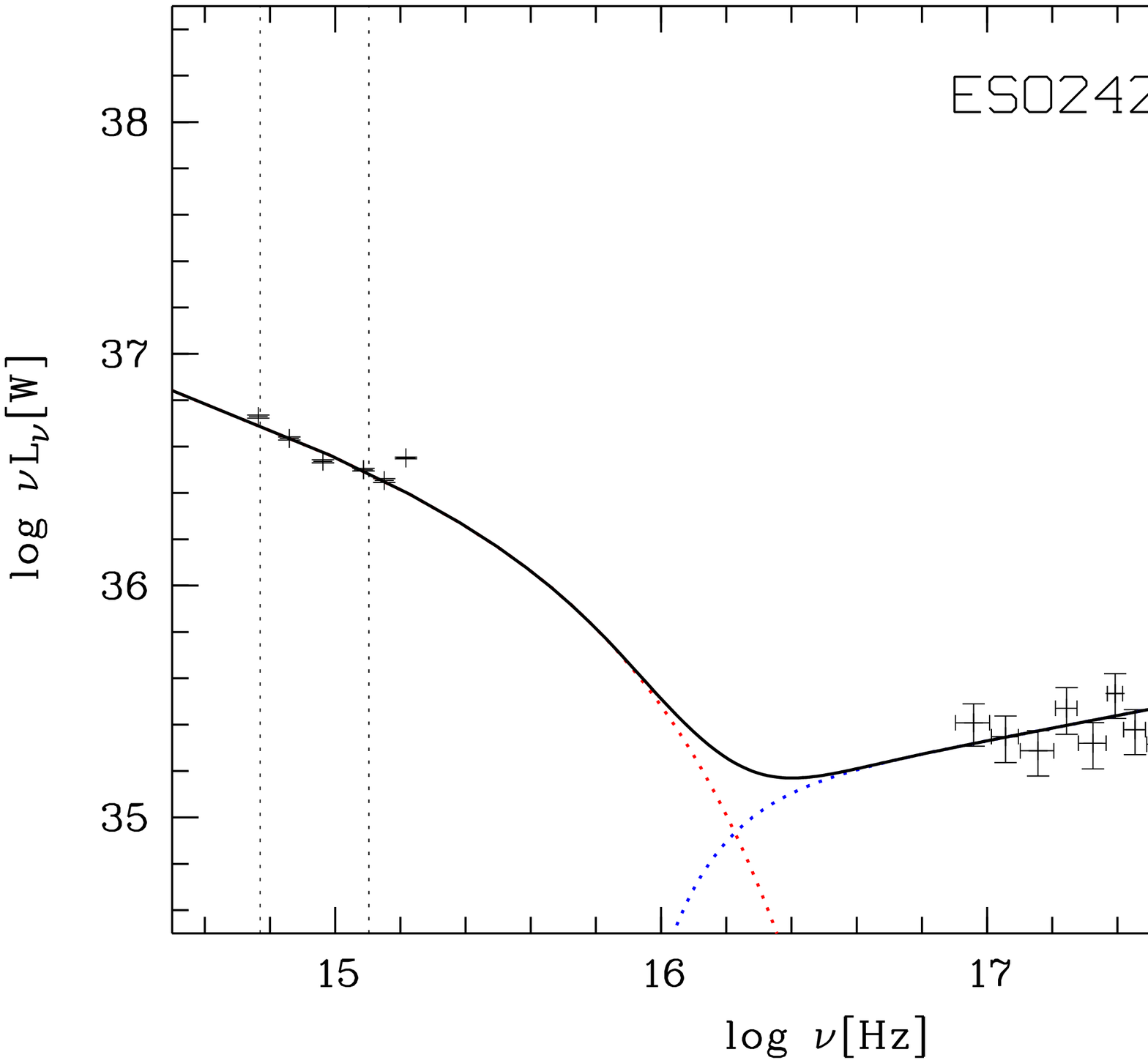}{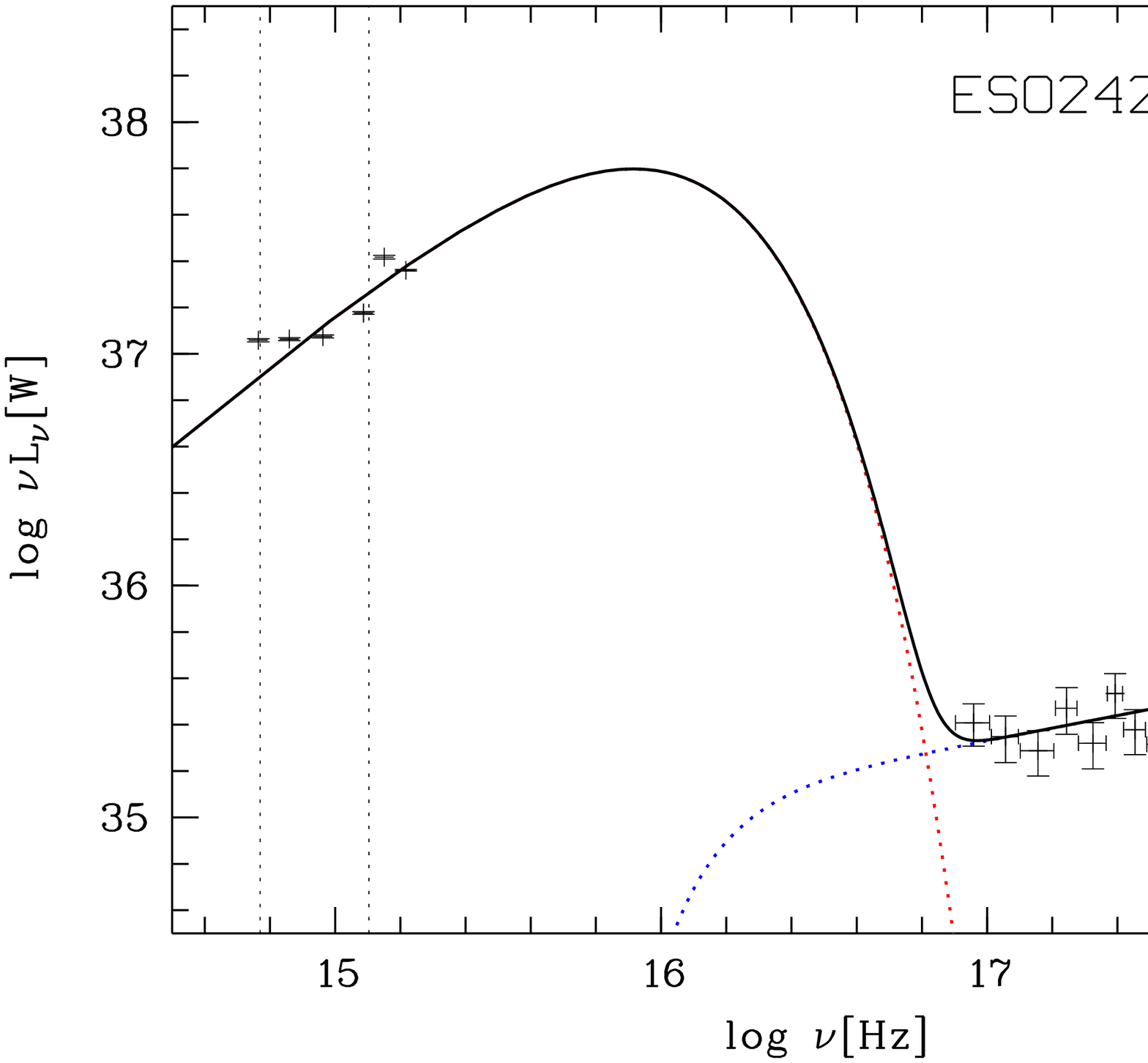}{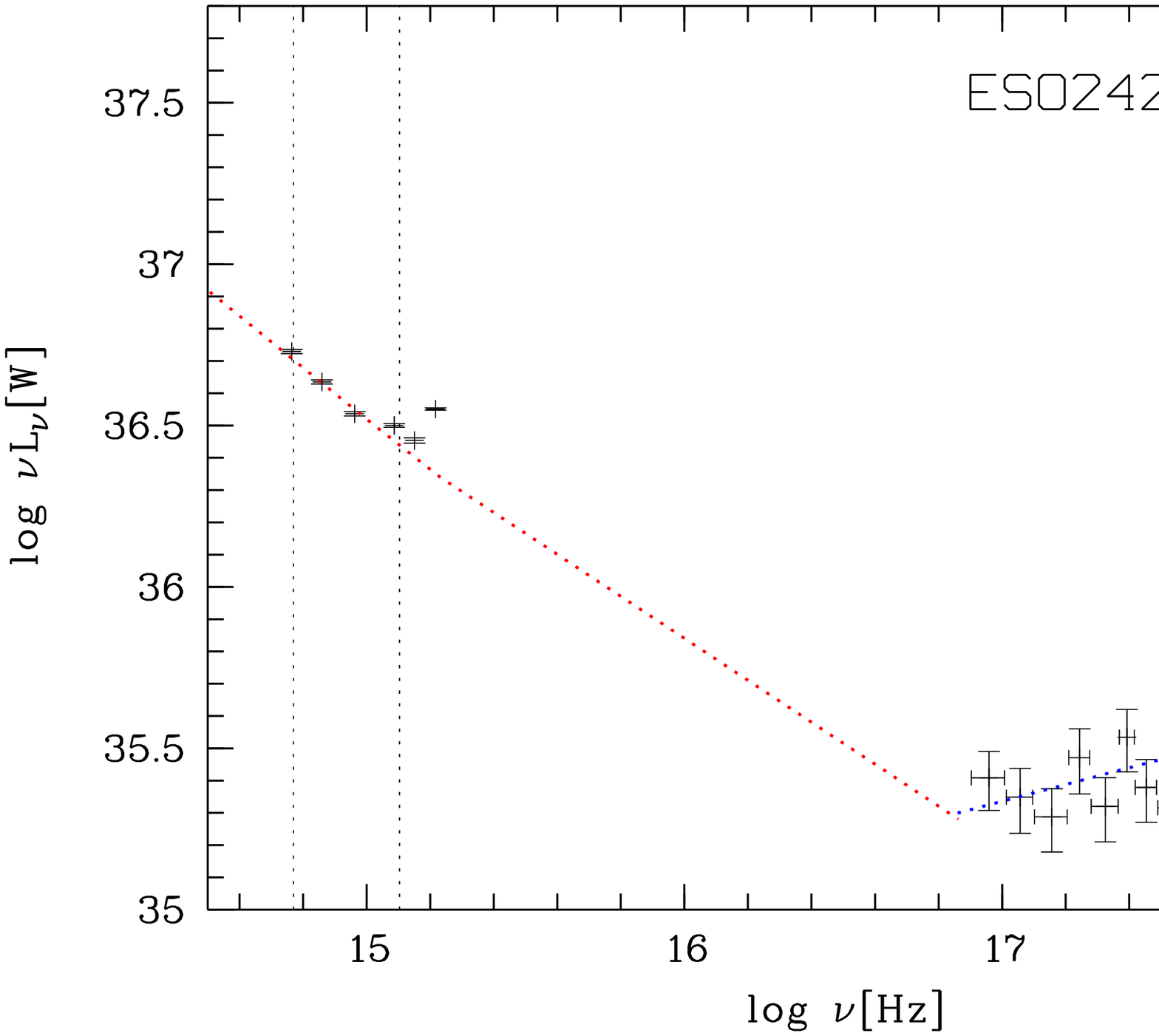}

\plotthree{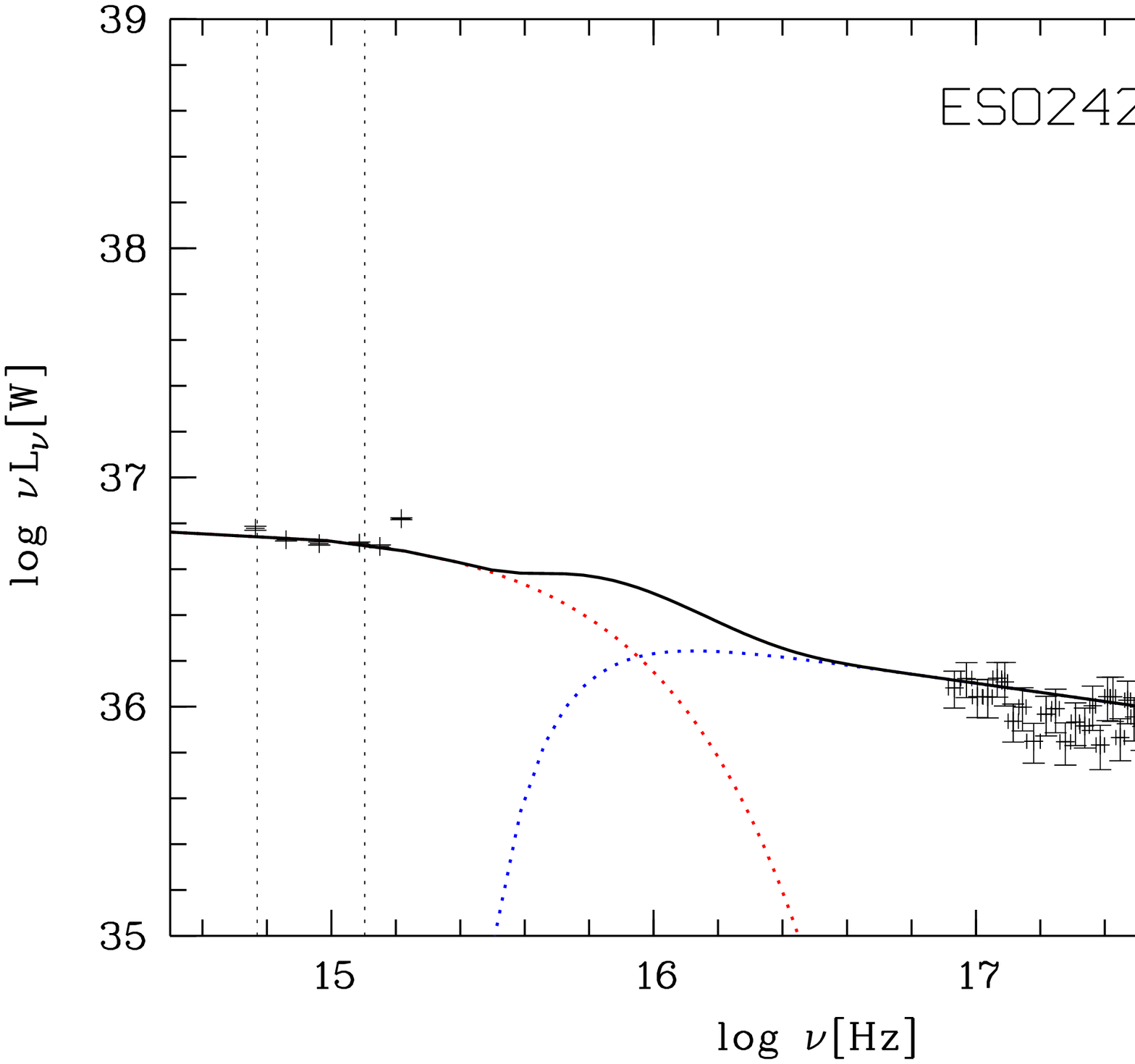}{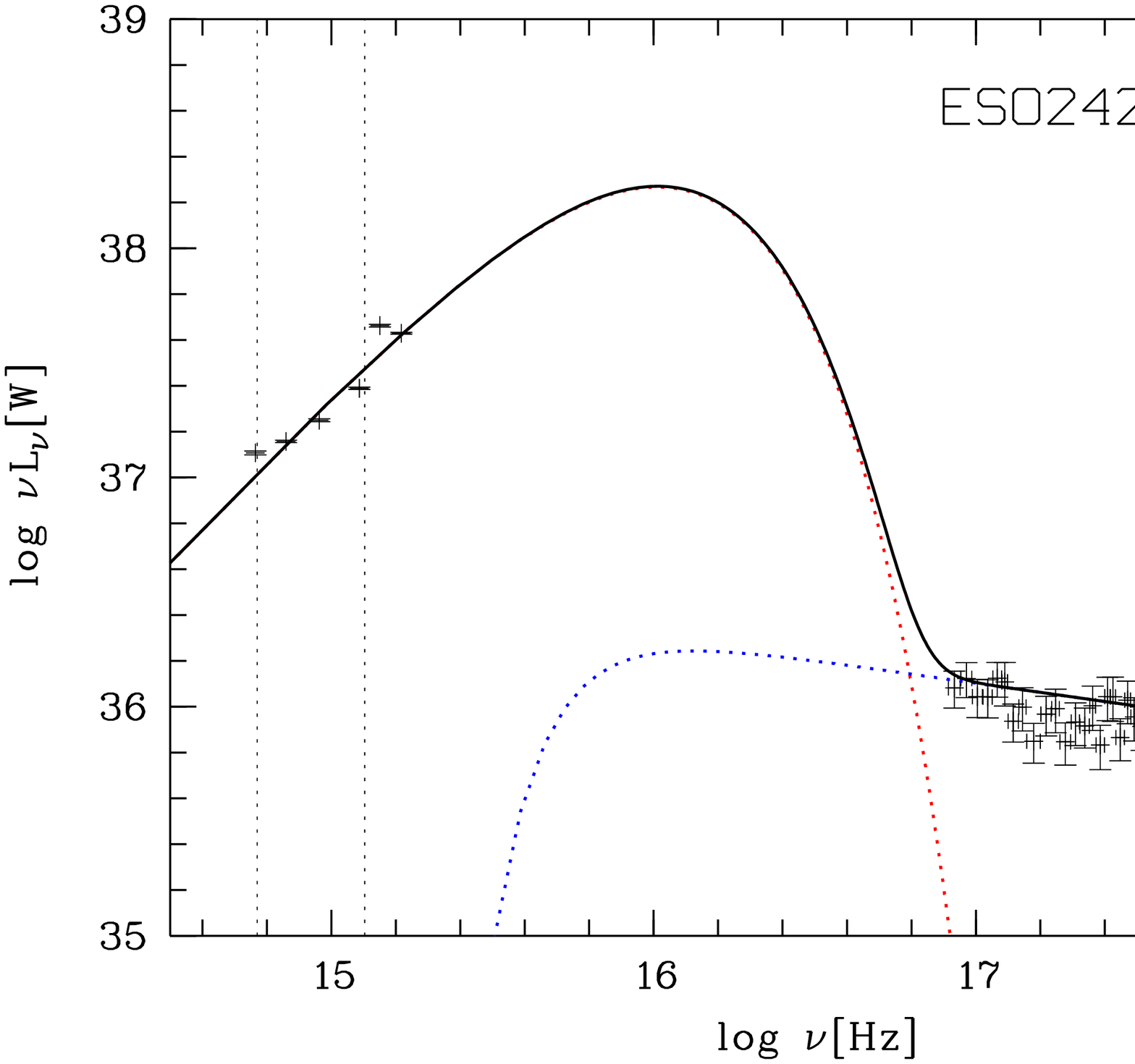}{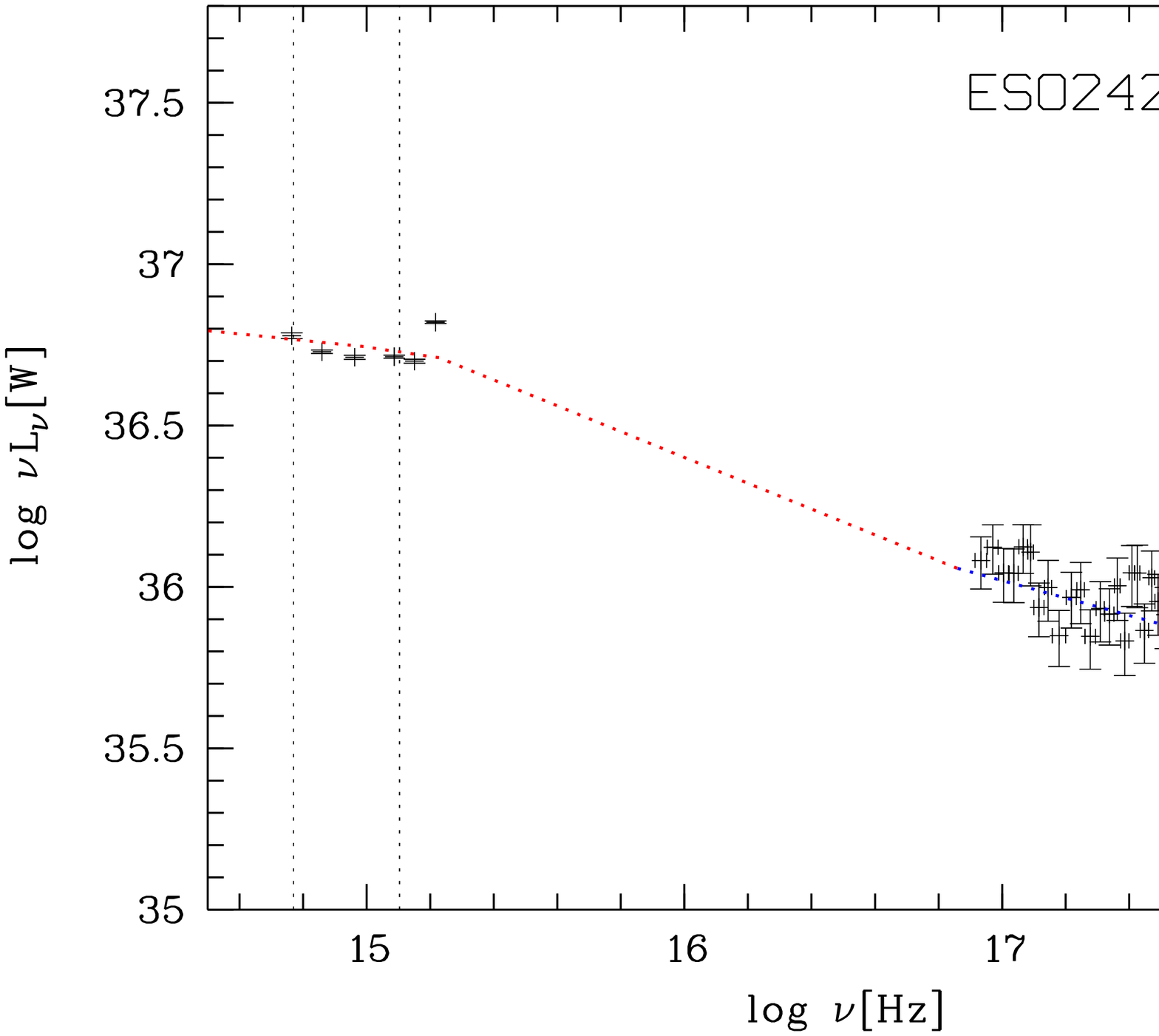}

\plotthree{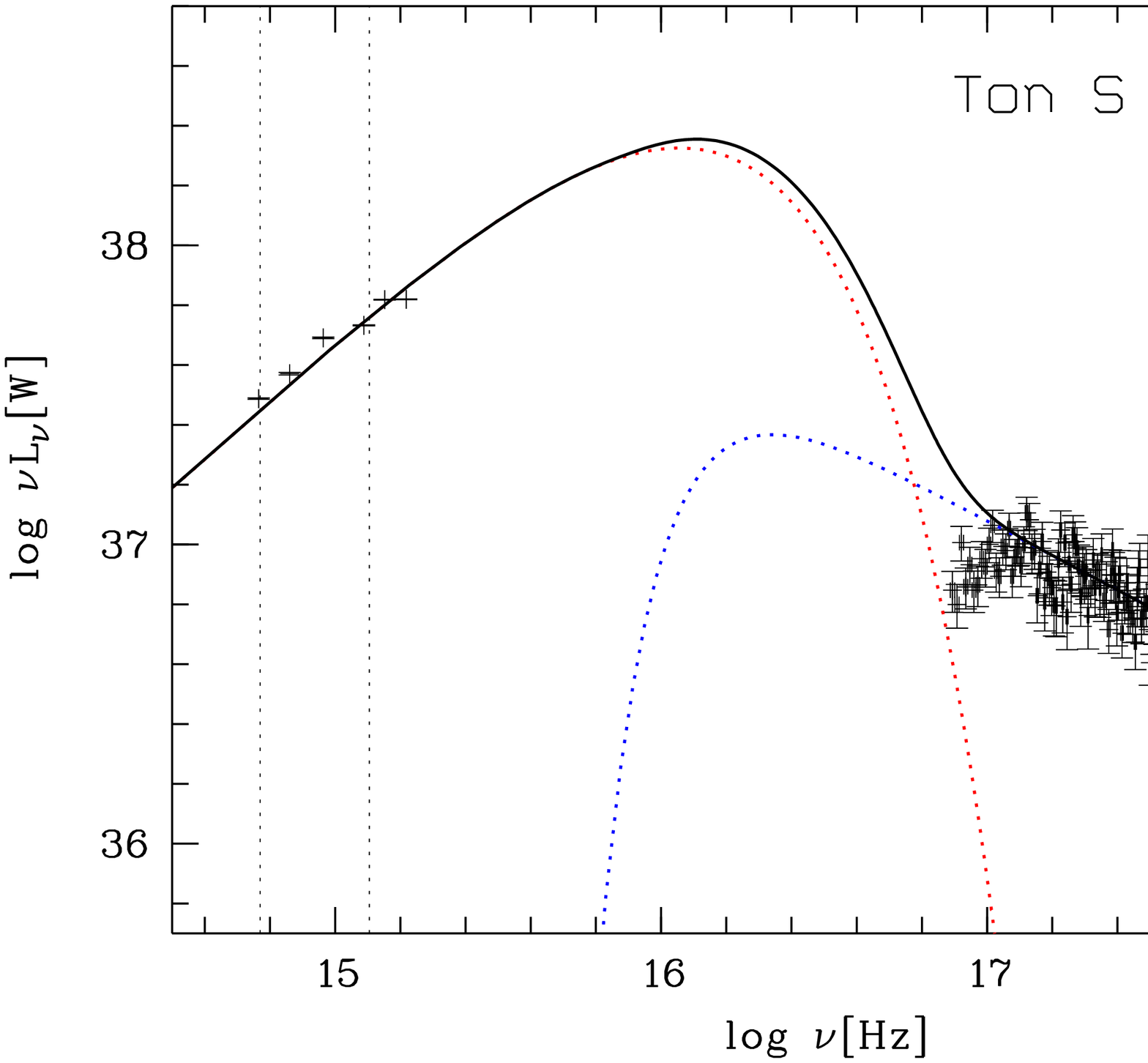}{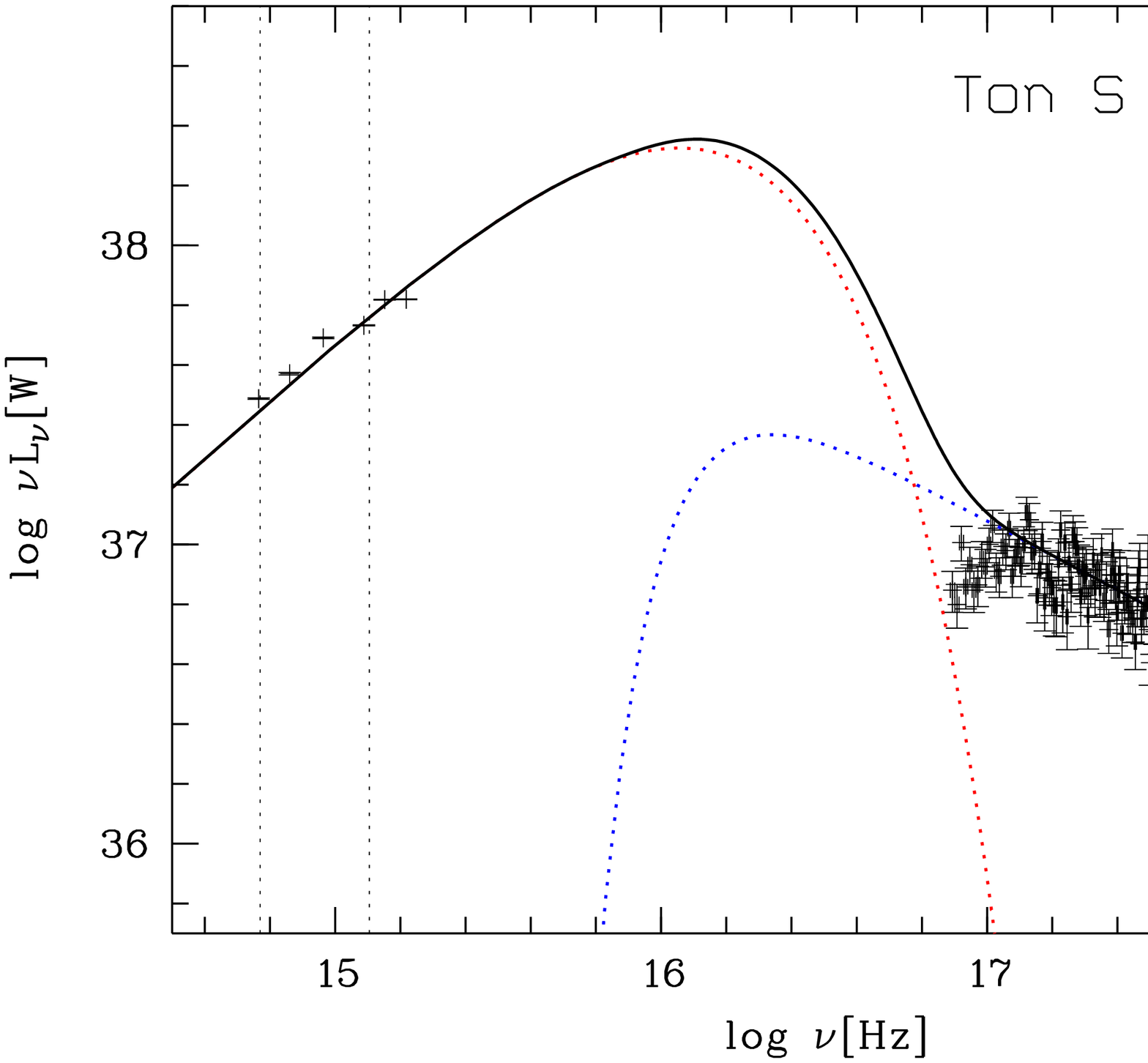}{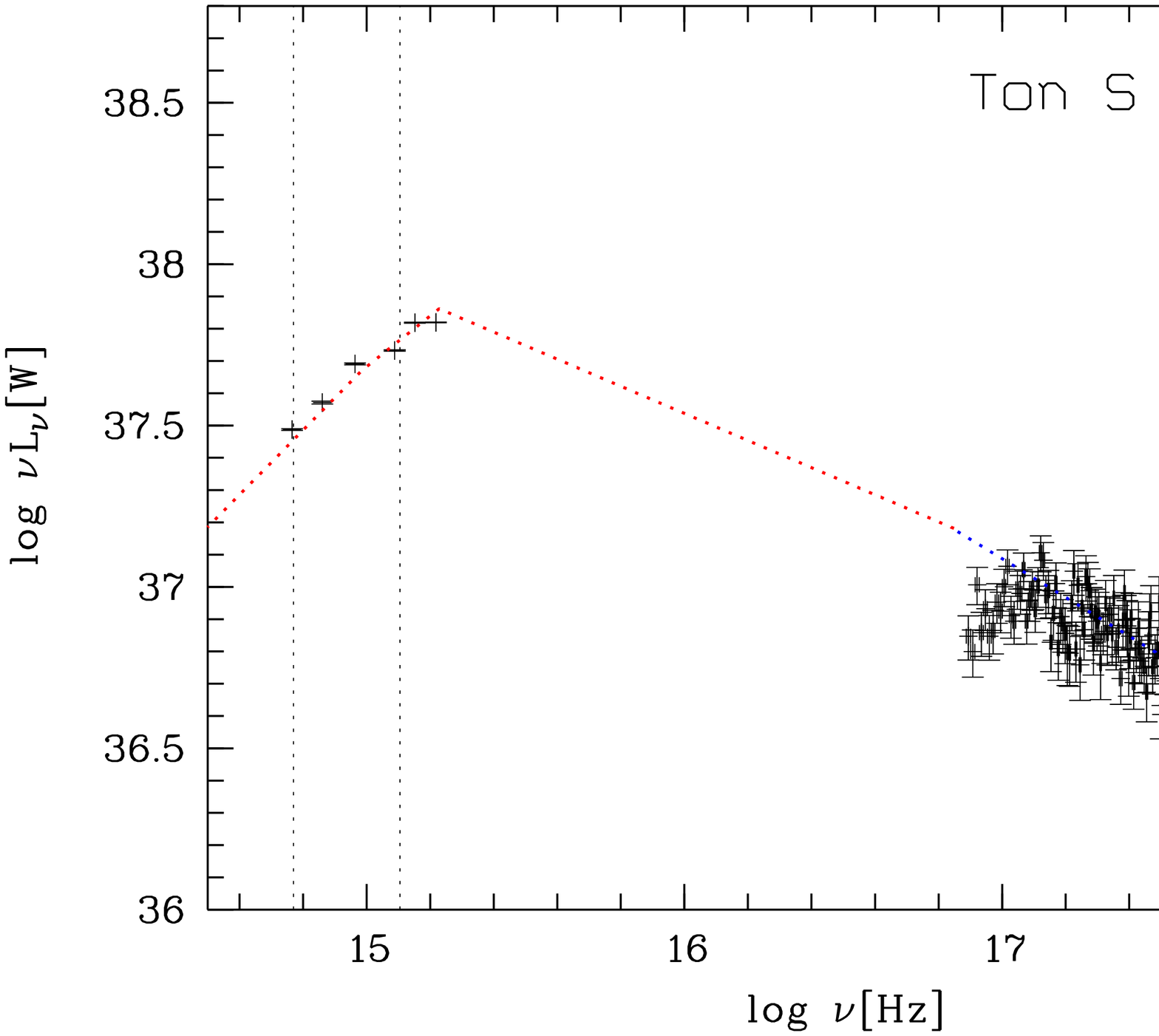}

\caption{Spectral Energy Distributions of the AGN of our sample. The first column shows the UVOT and XRT data fitted by an power law with exponential cut off and an absorbed
power law (Model A) to the intrinsic reddening uncorrected UVOT data, the second column displays the same model to the reddening corrected UVOT data, and the third column
shows the data fitted by a double broken power law (Model B).
}
\end{figure*}

\begin{figure*}
\epsscale{0.60}
\plotthree{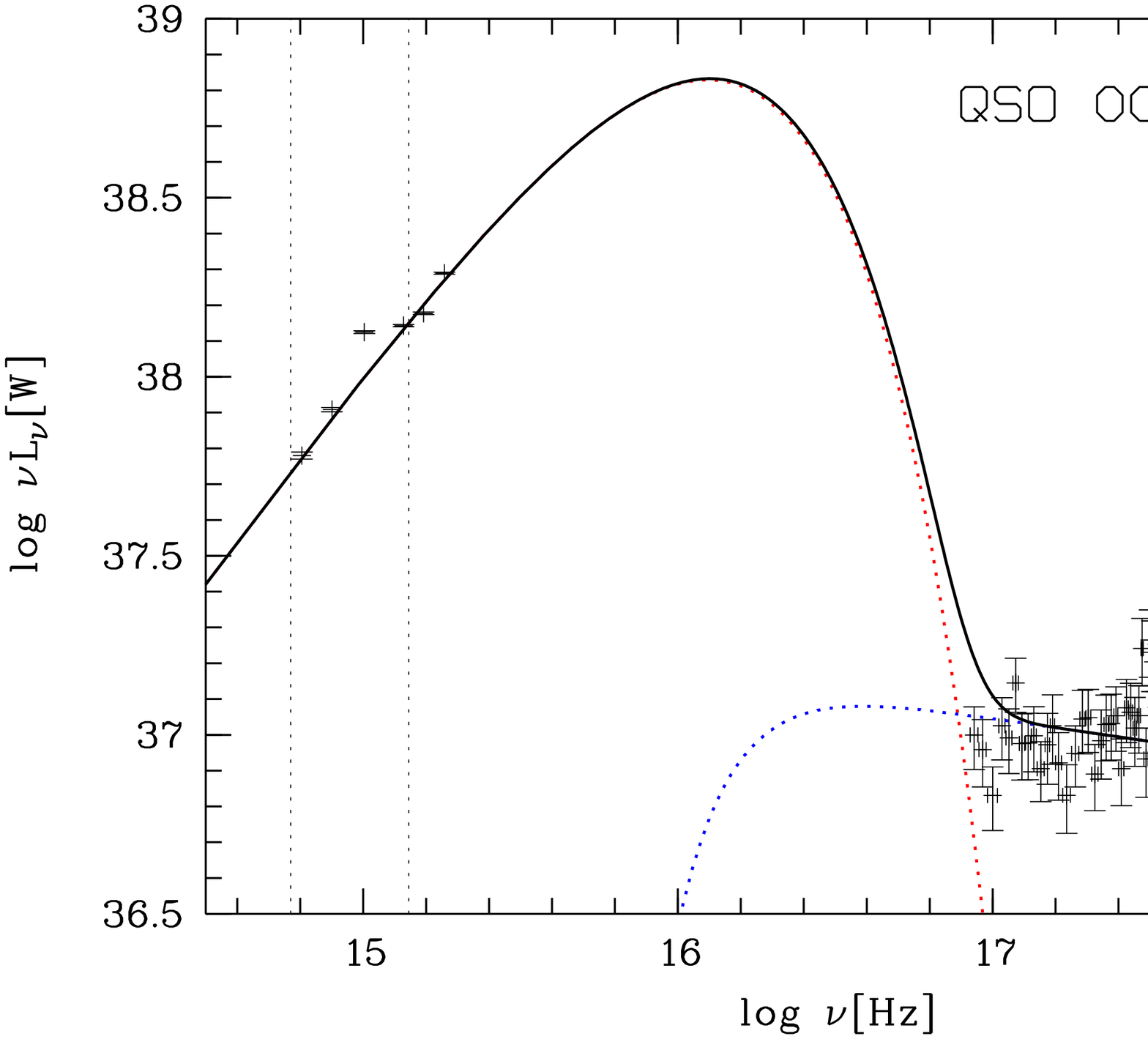}{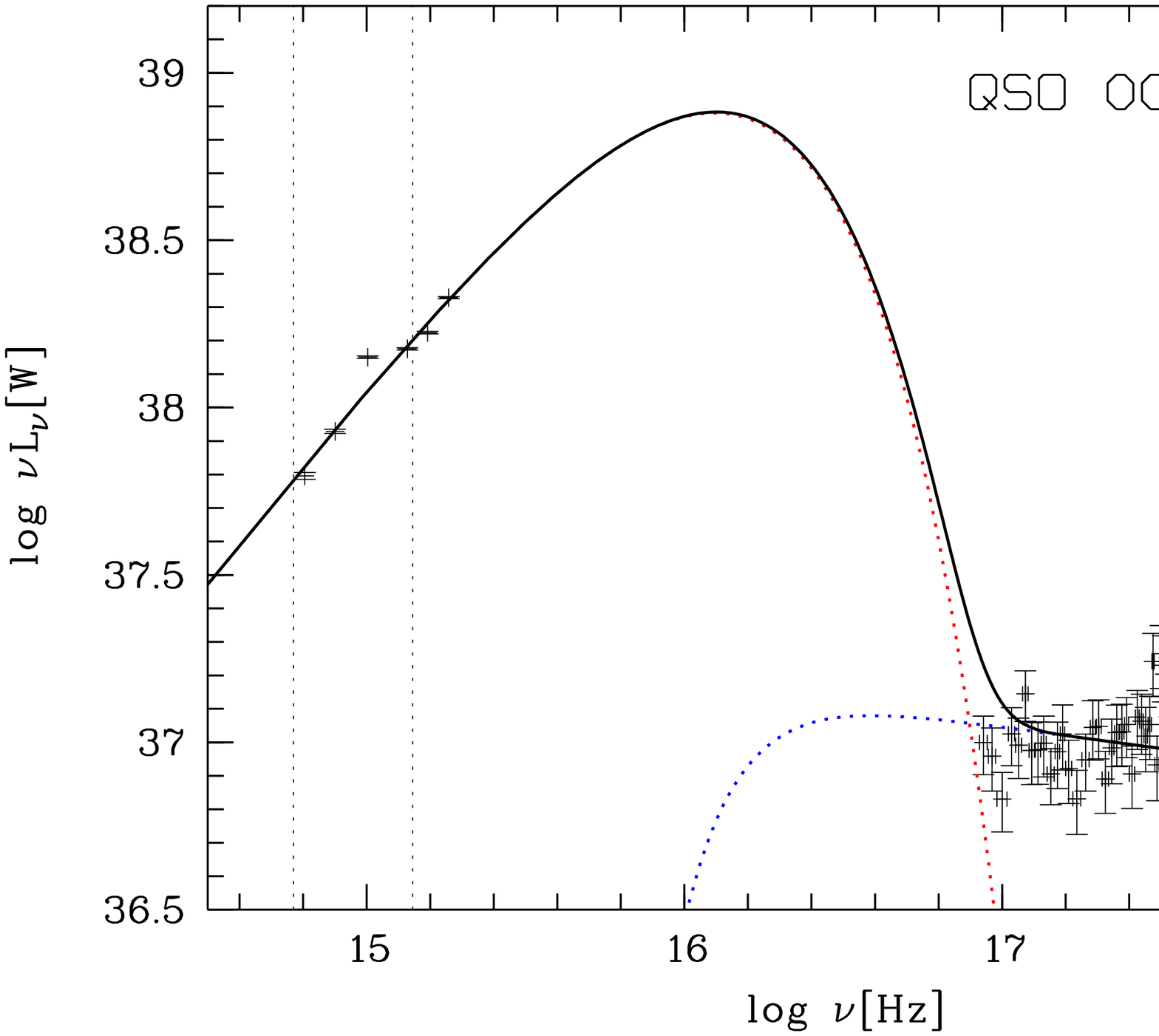}{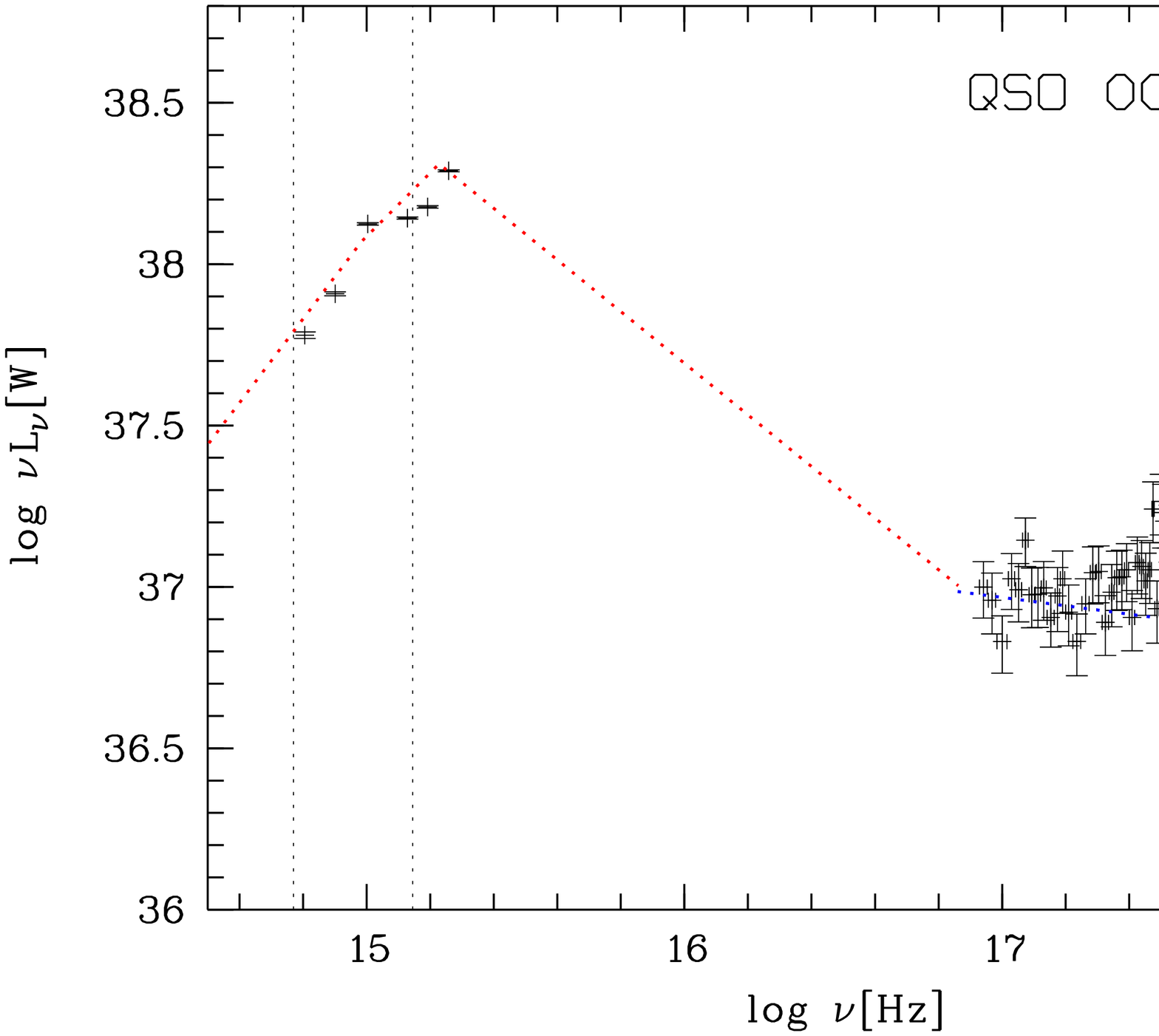}

\plotthree{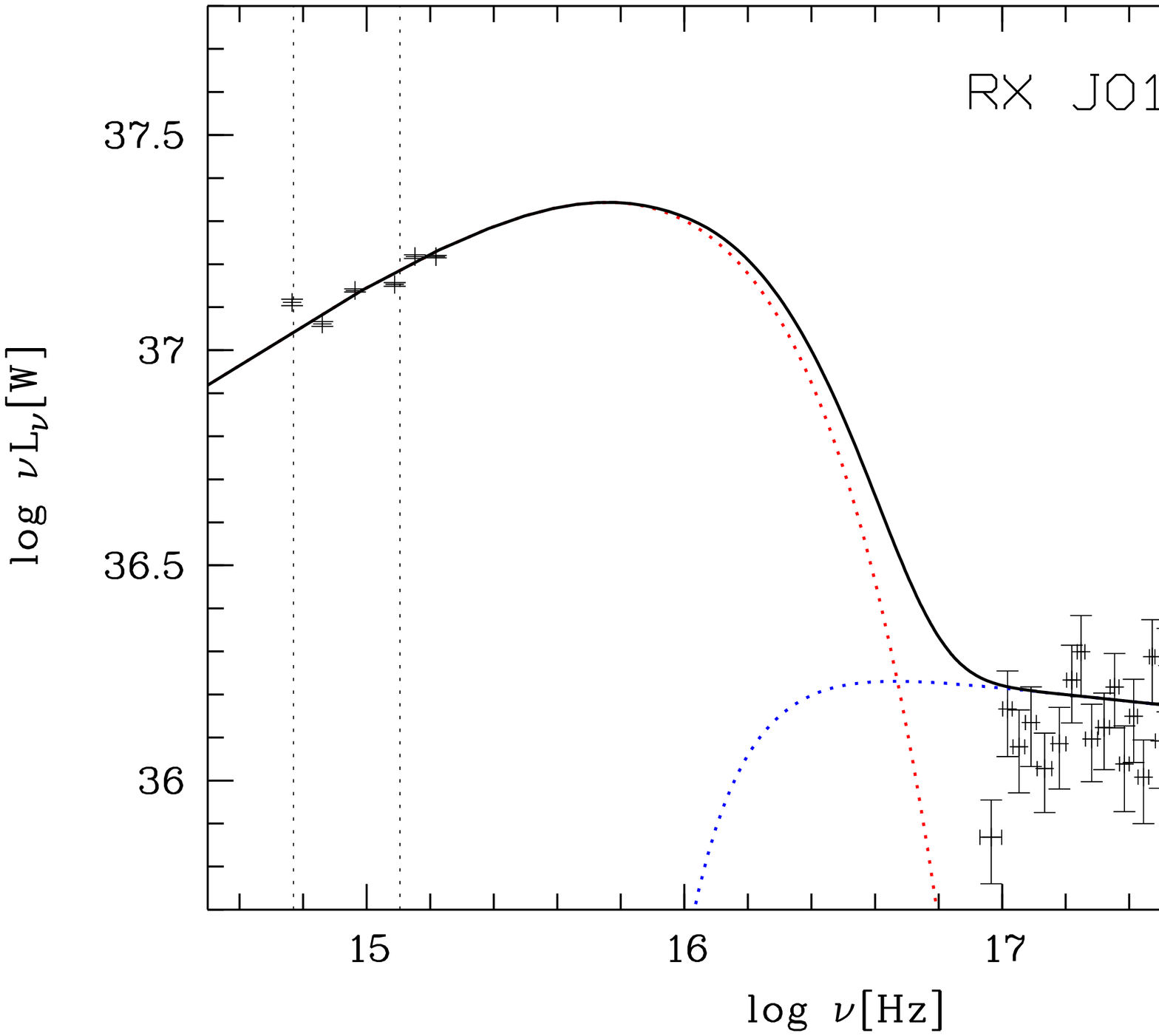}{}{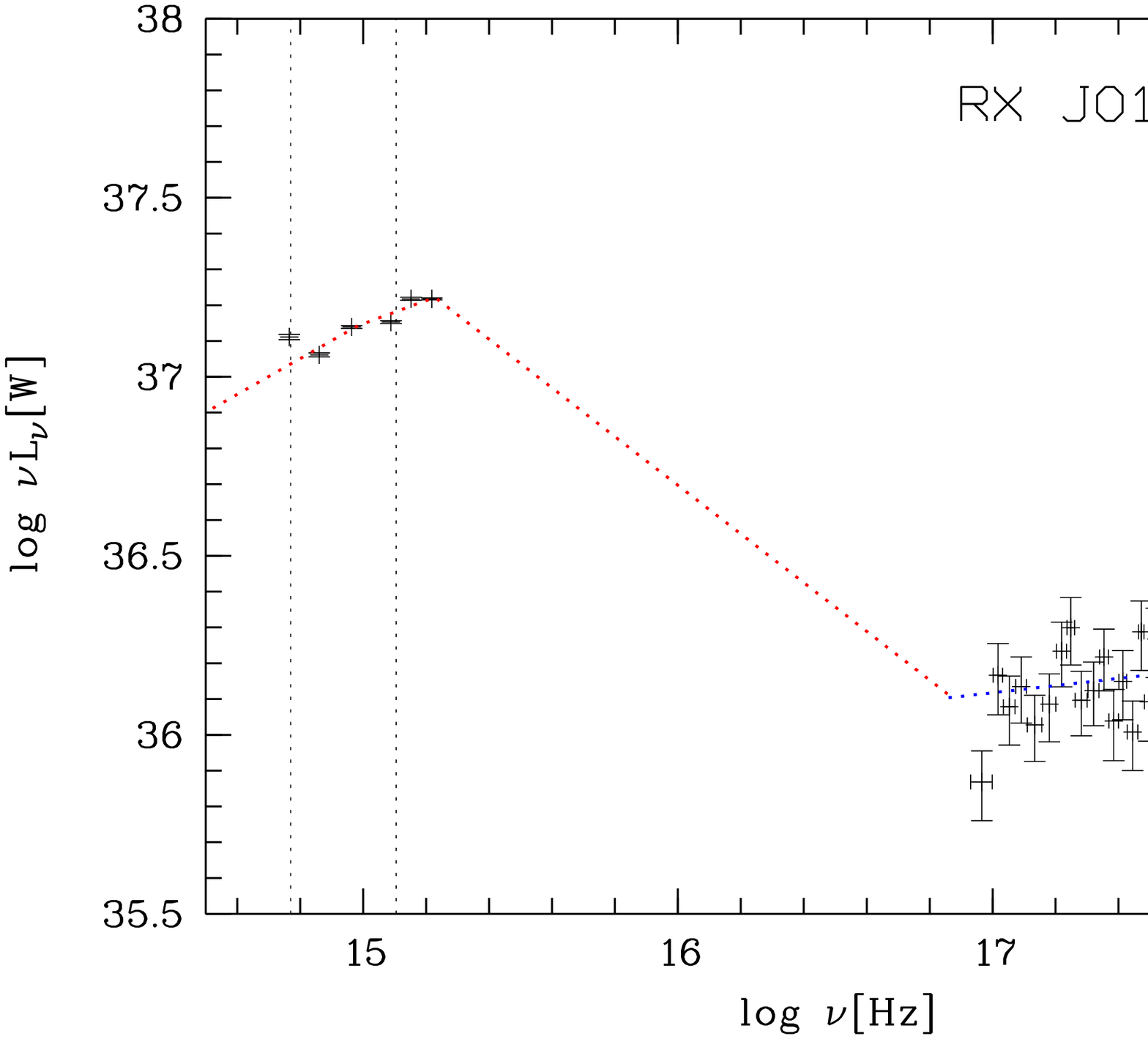}

\plotthree{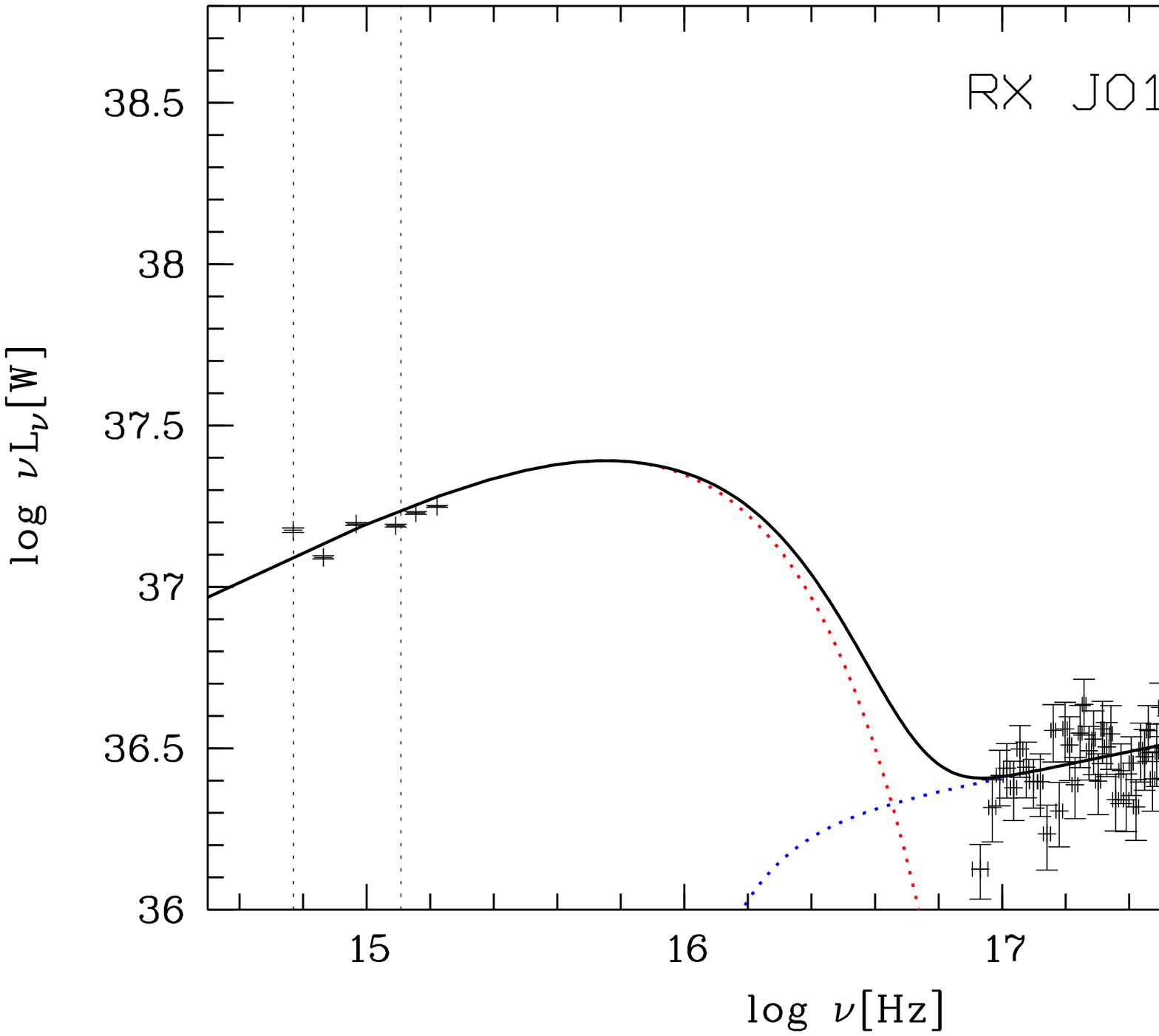}{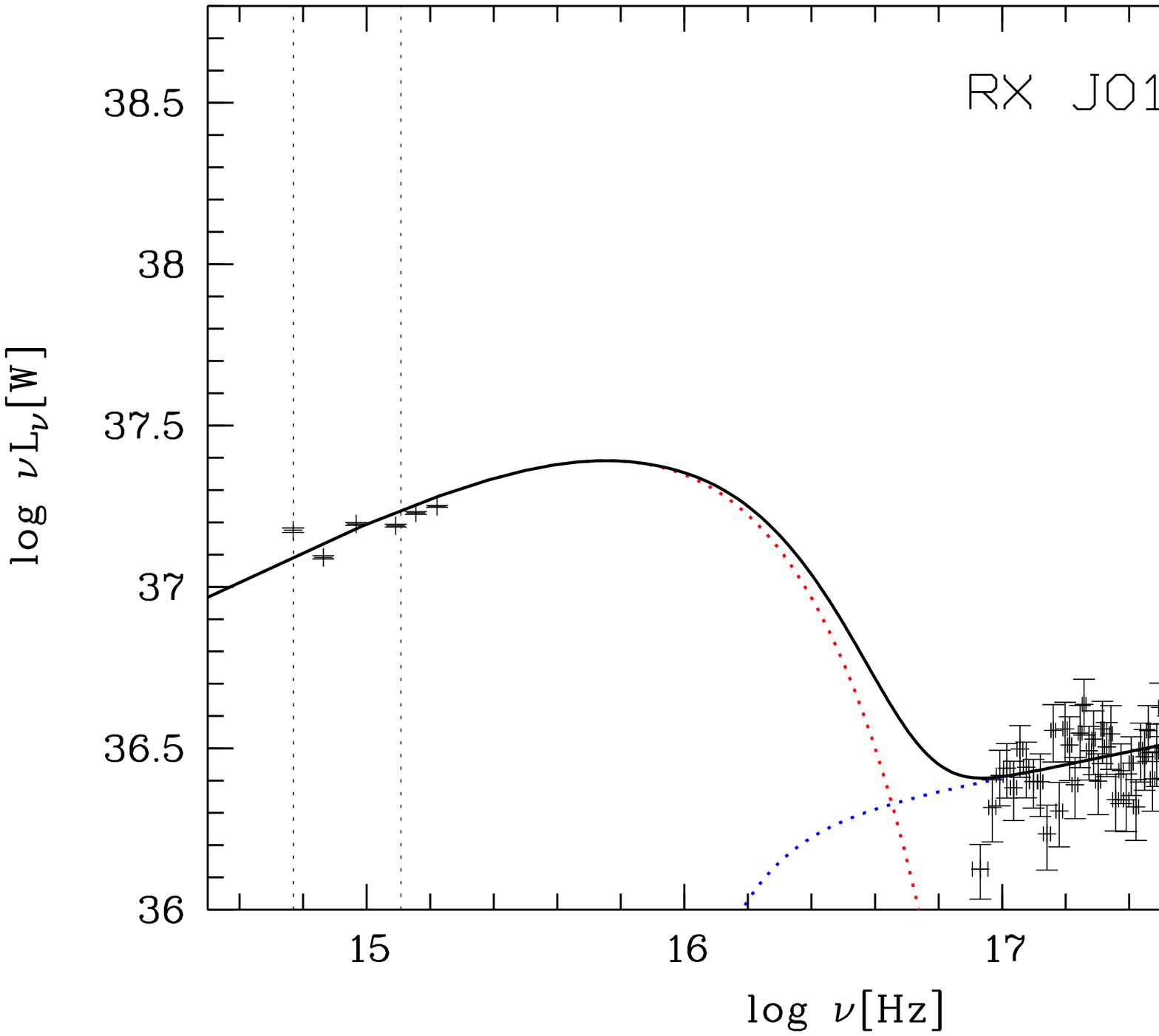}{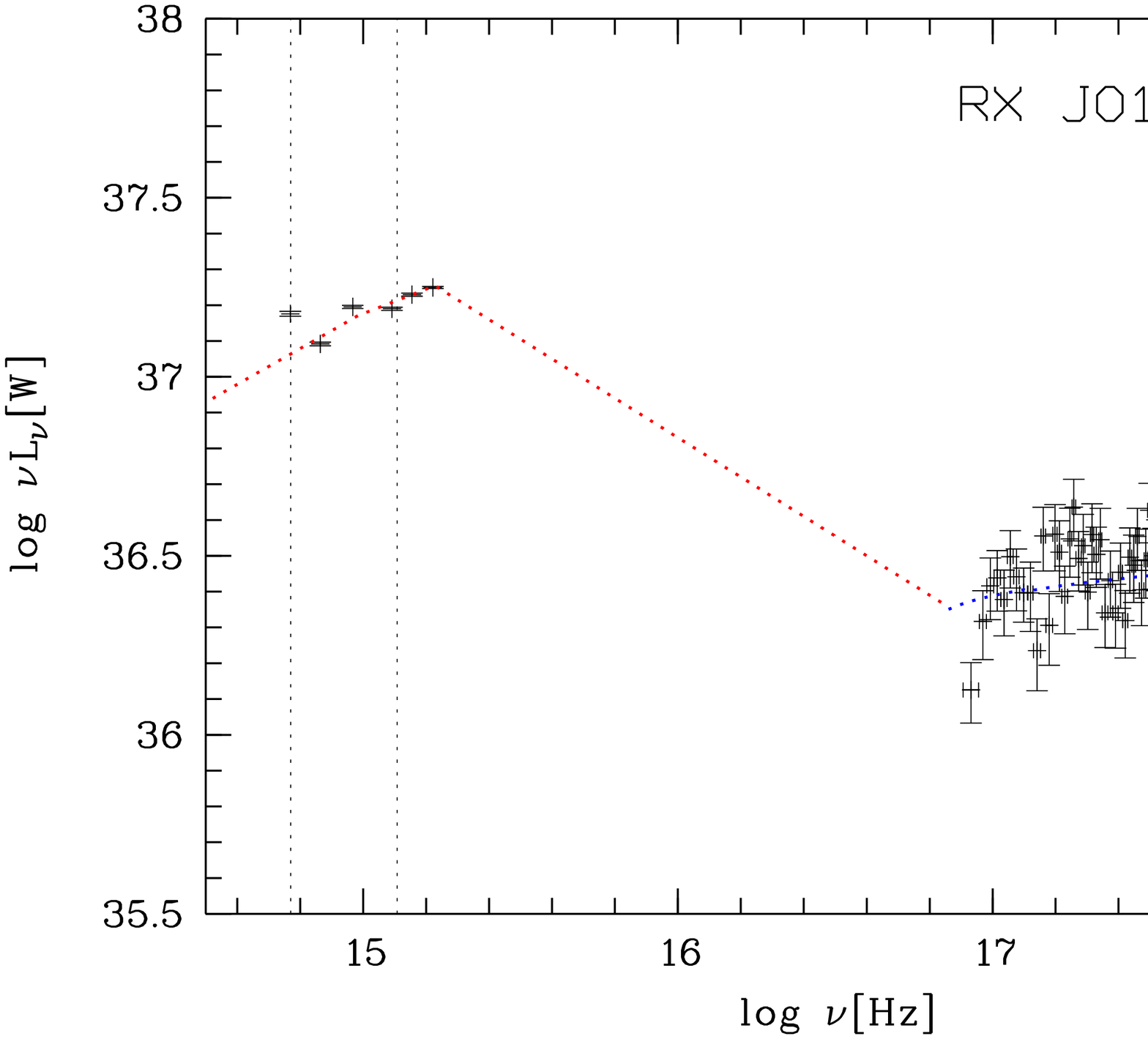}

\plotthree{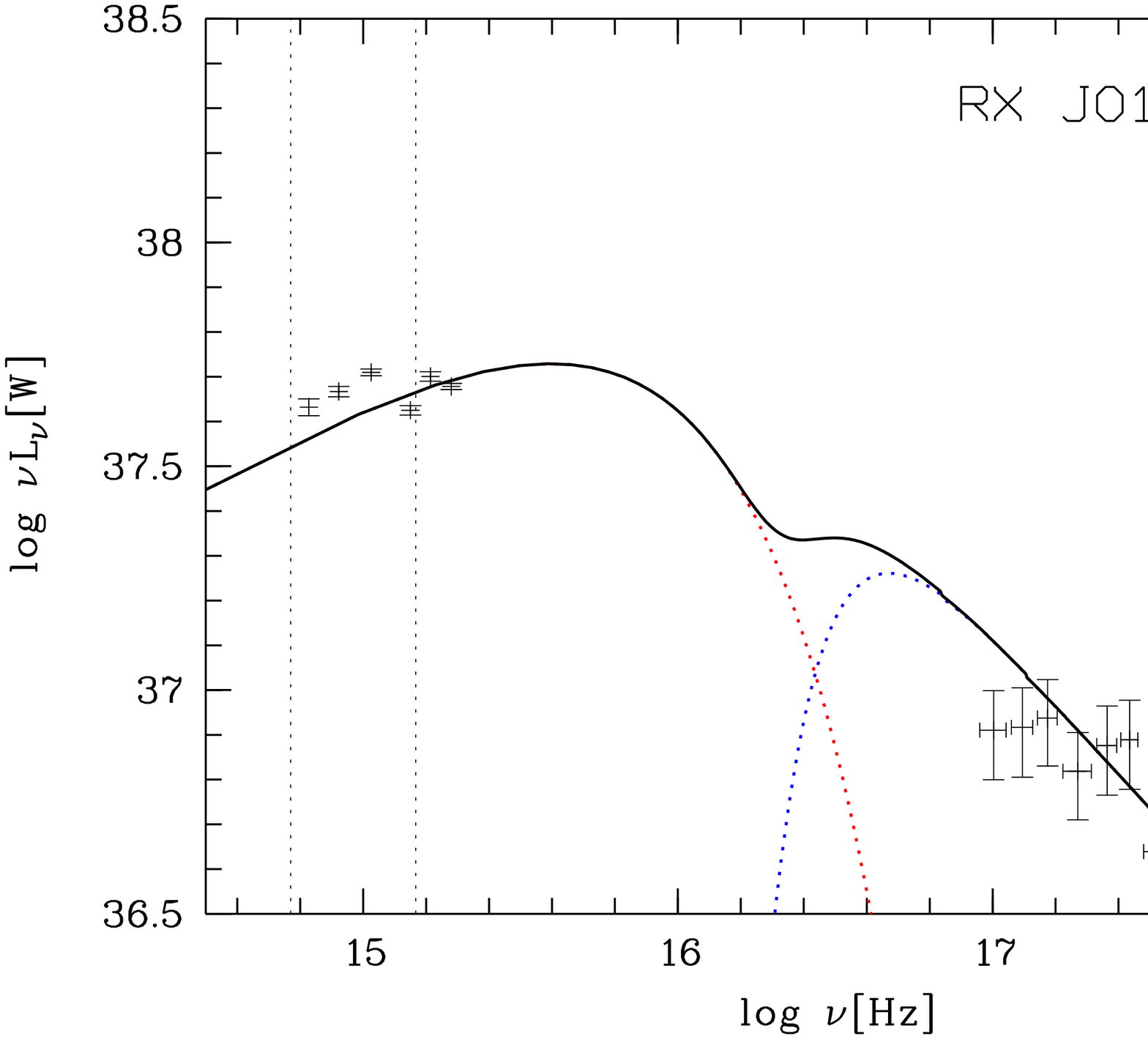}{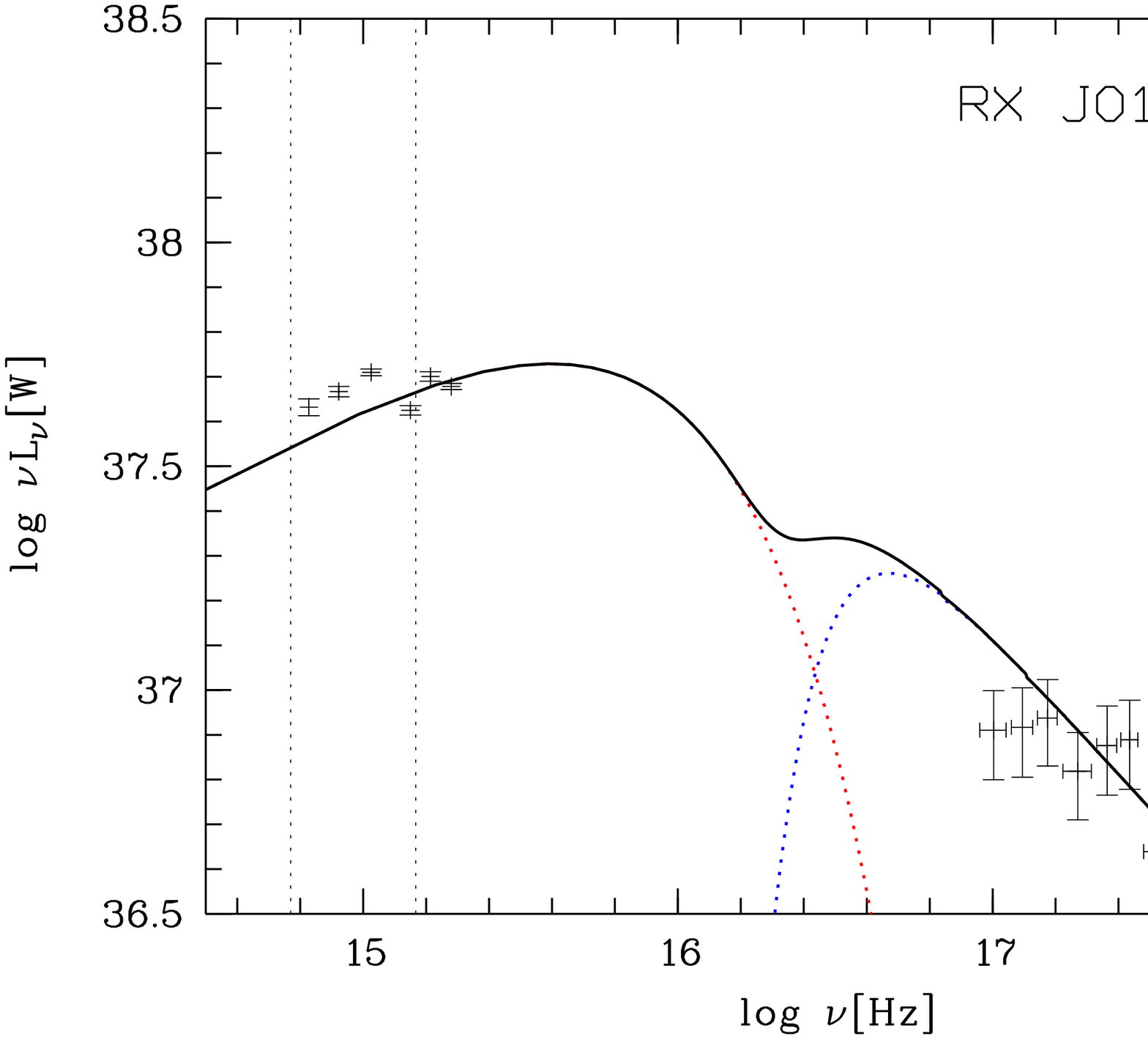}{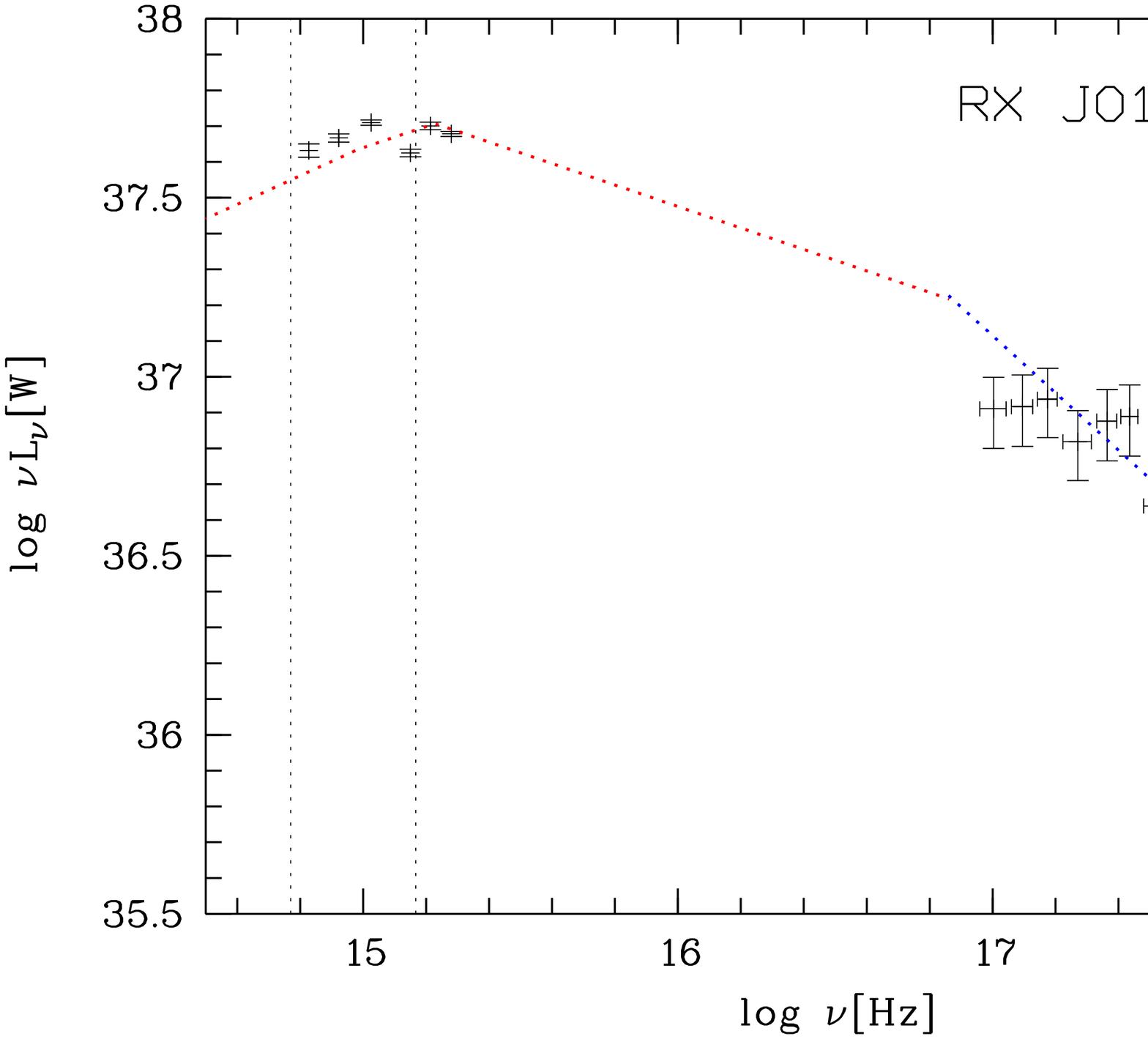}

\plotthree{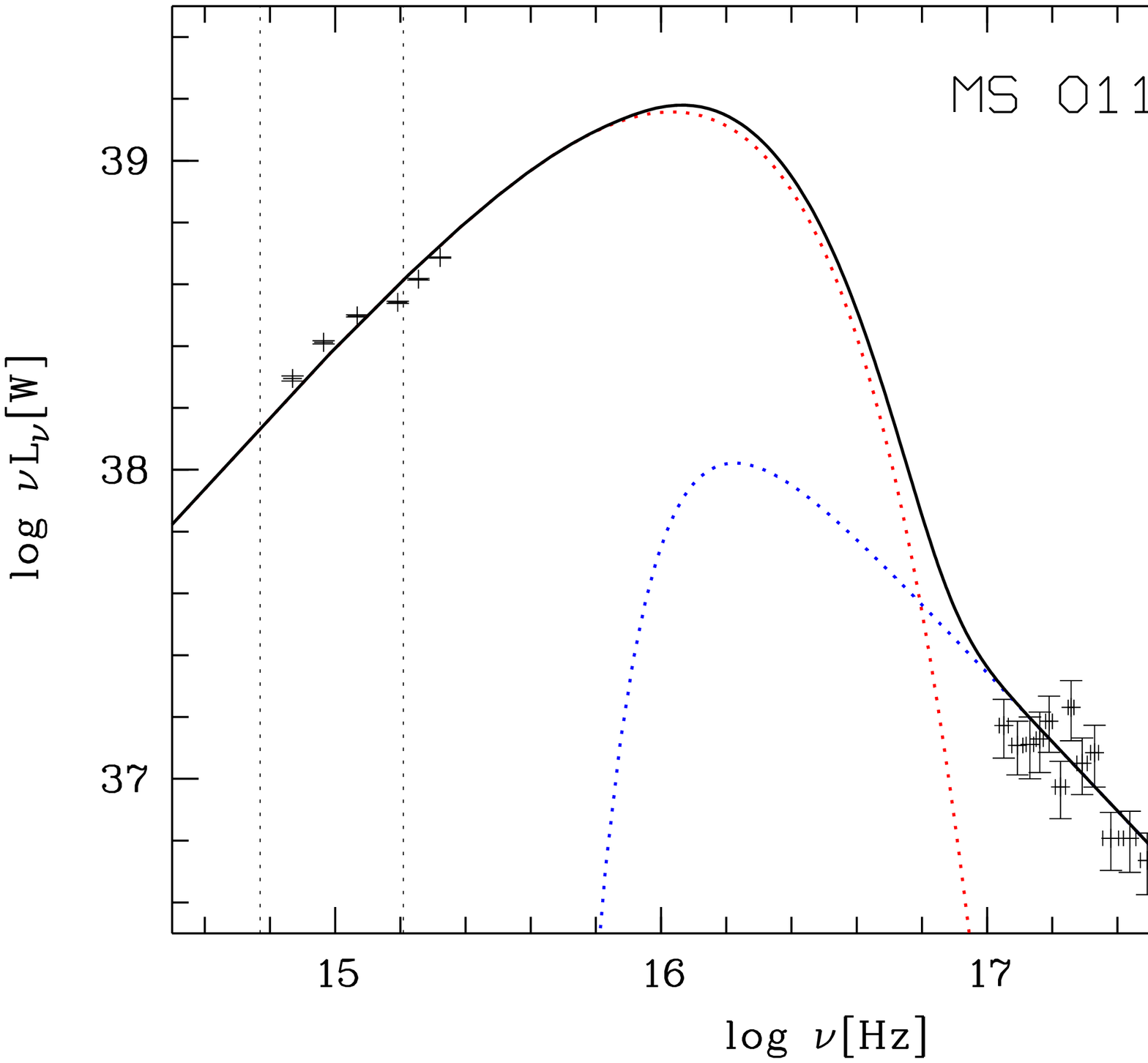}{f25_t.ps}{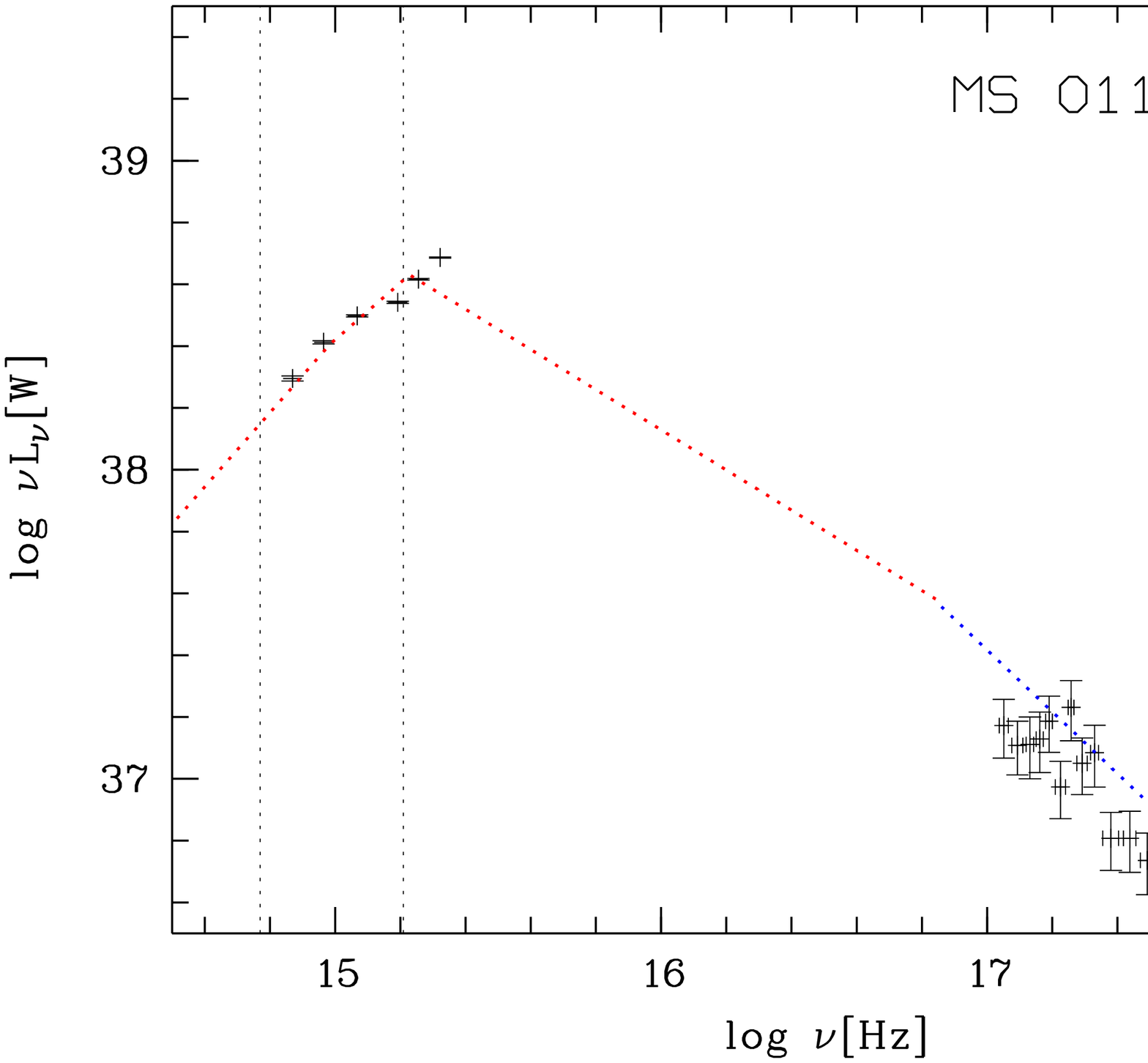}

\end{figure*}

\begin{figure*}
\epsscale{0.60}
\plotthree{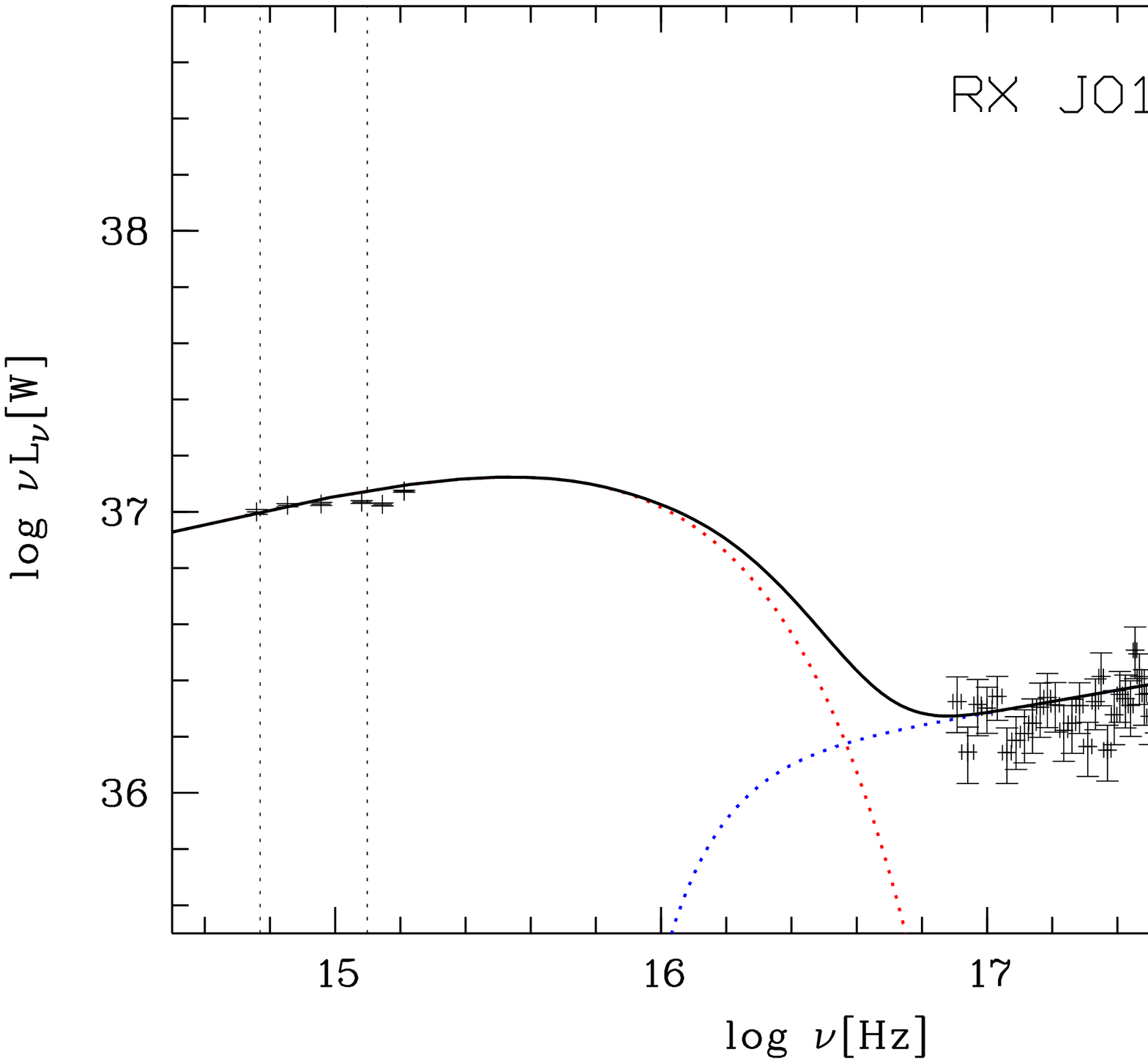}{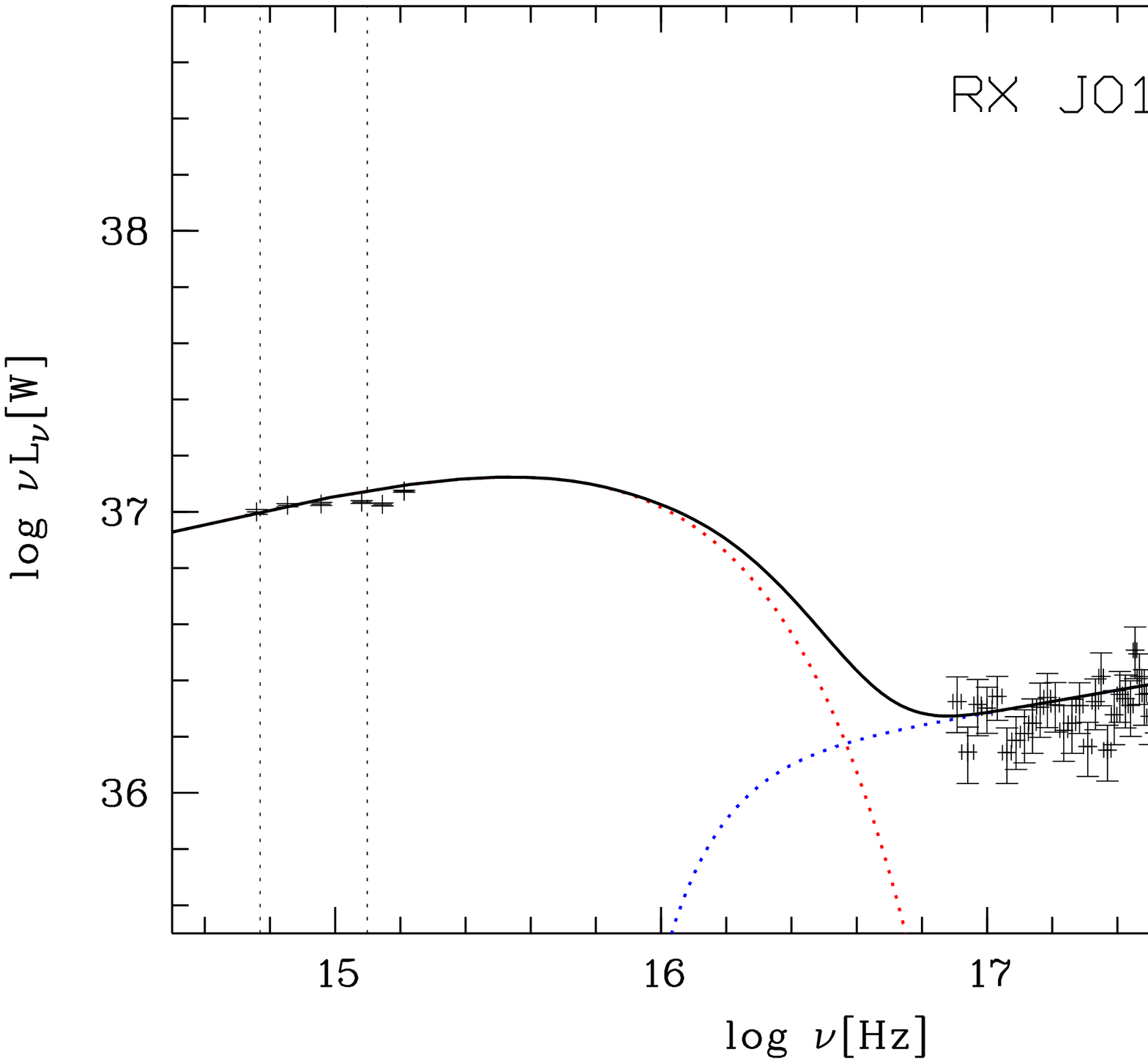}{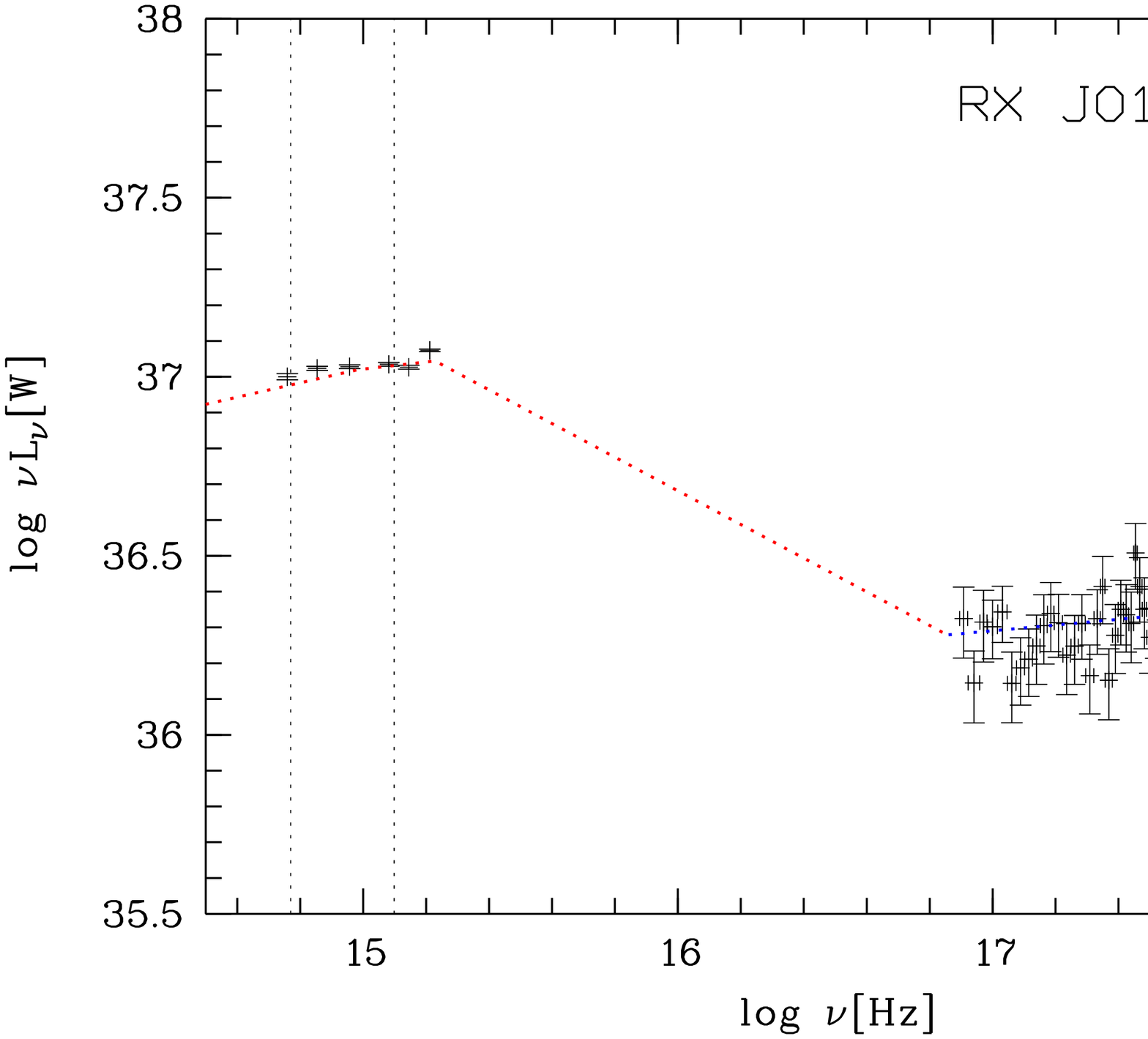}

\plotthree{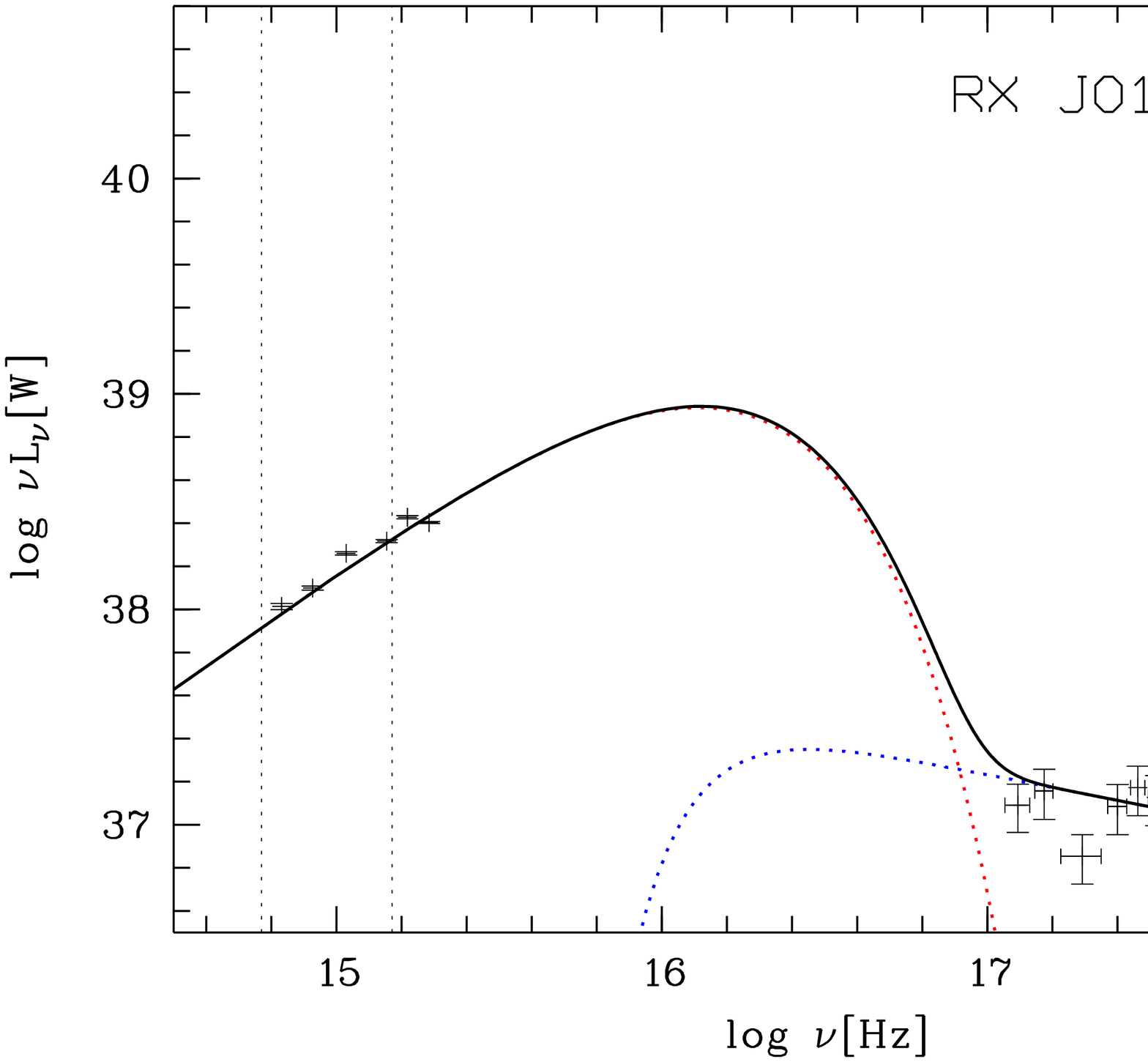}{f25_t.ps}{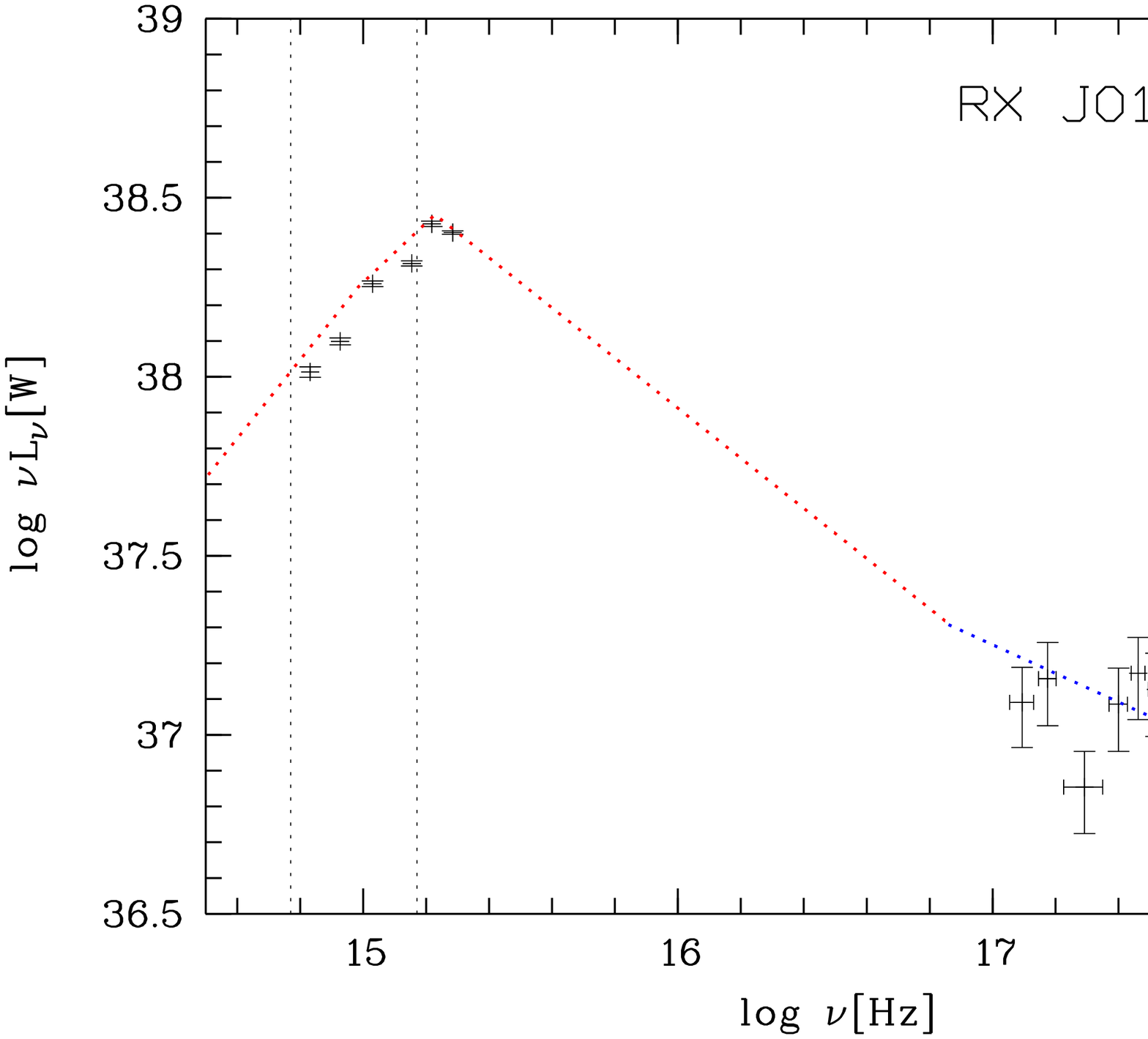}

\plotthree{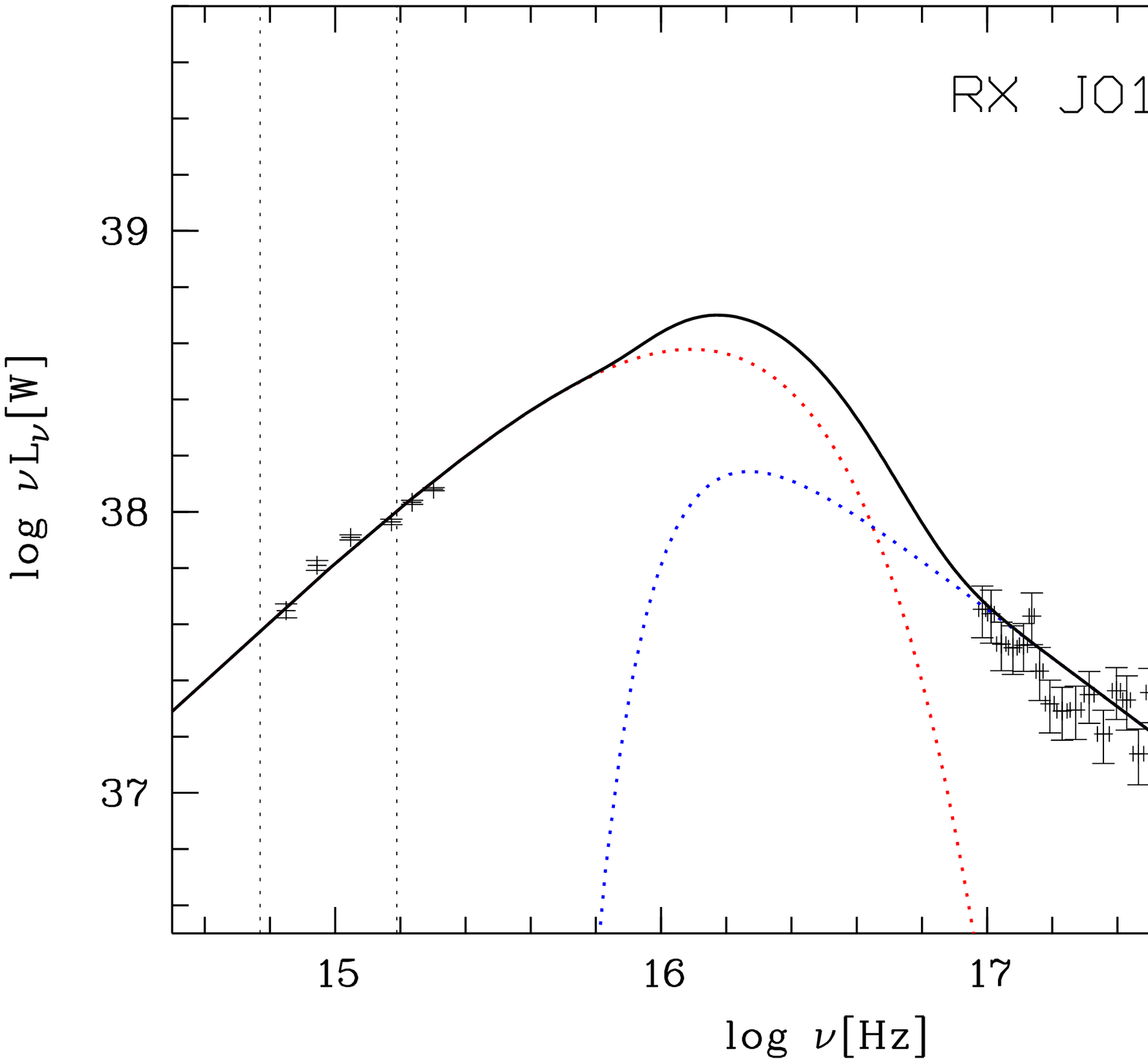}{f25_t.ps}{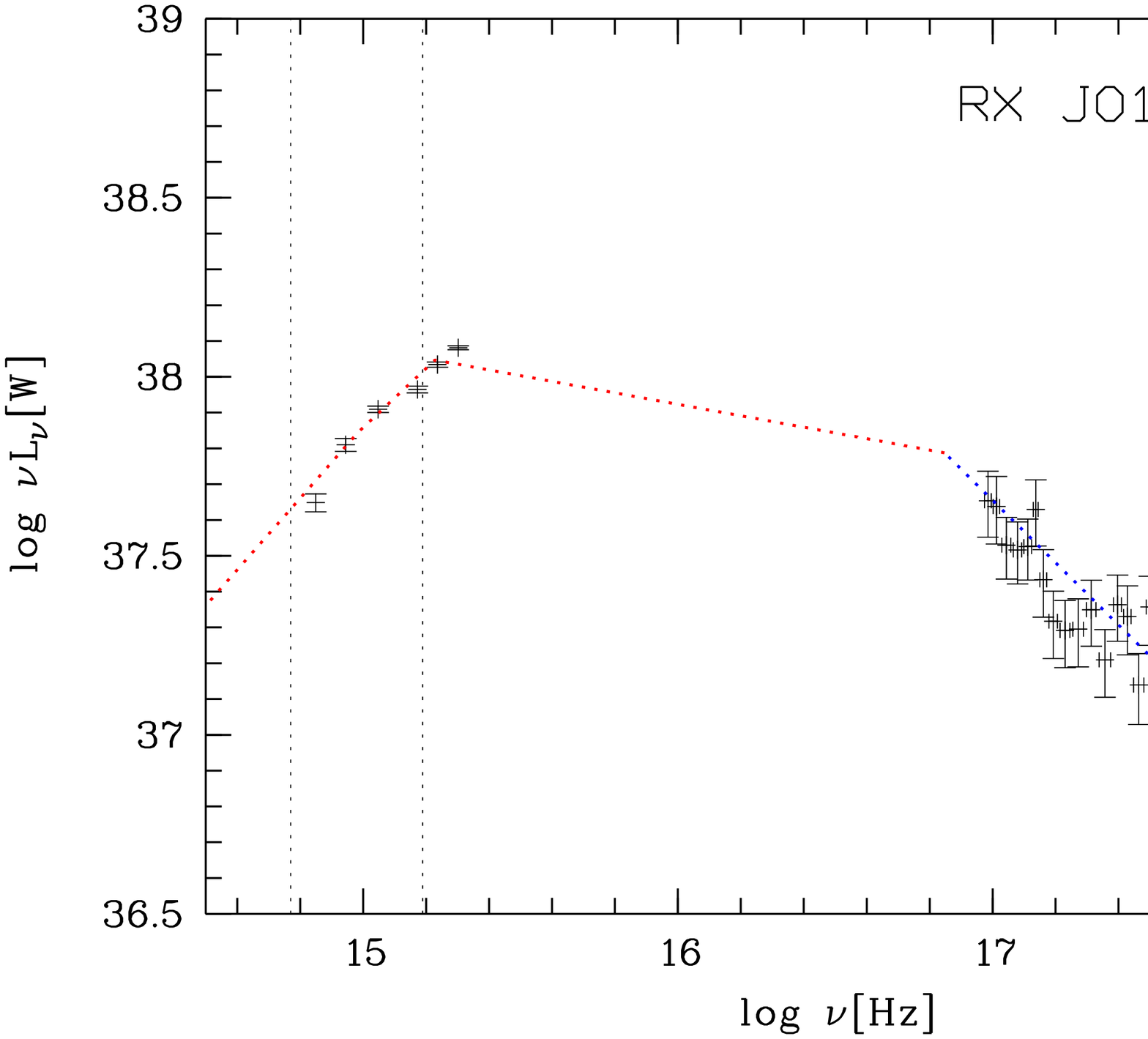}

\plotthree{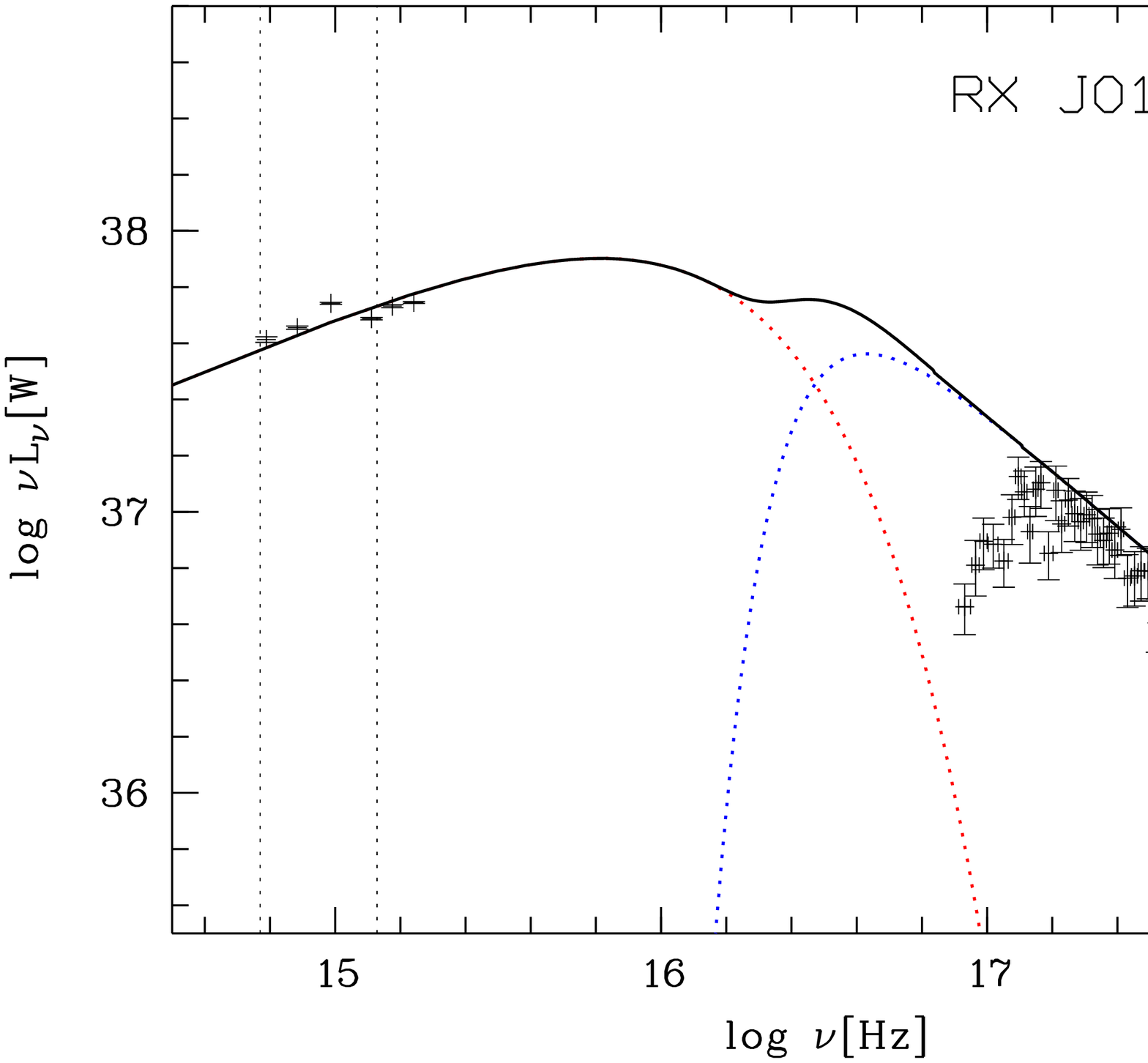}{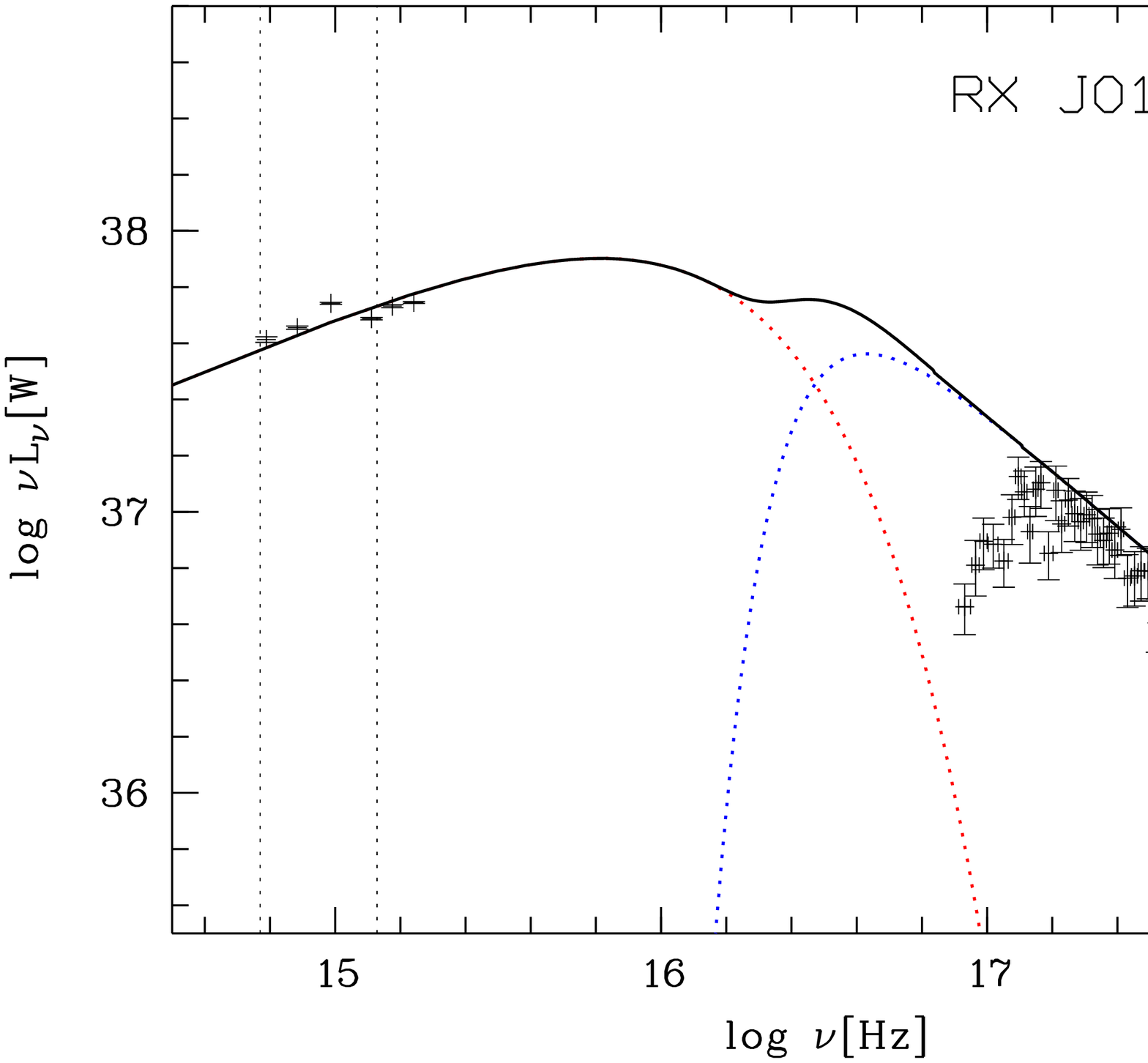}{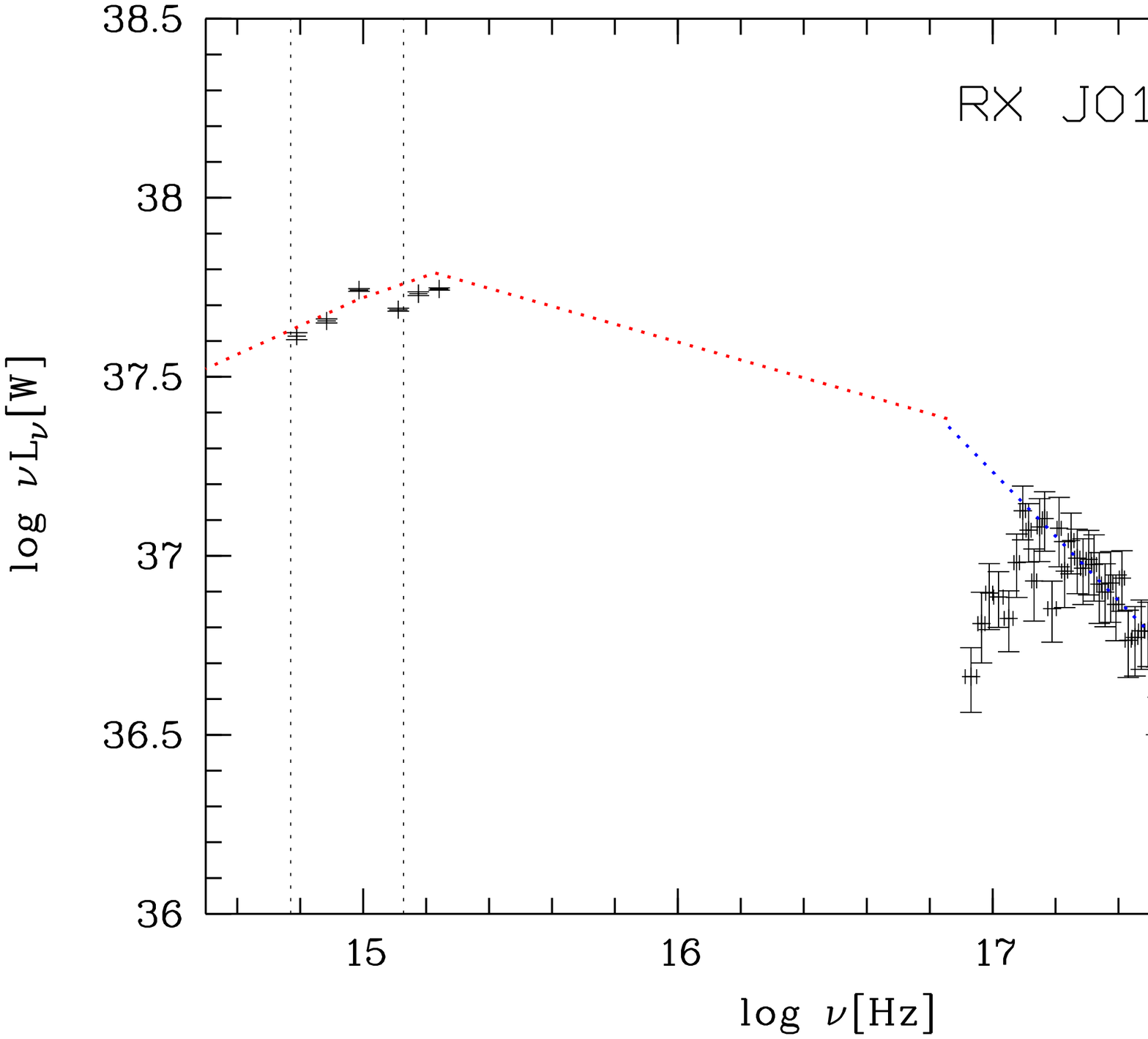}

\plotthree{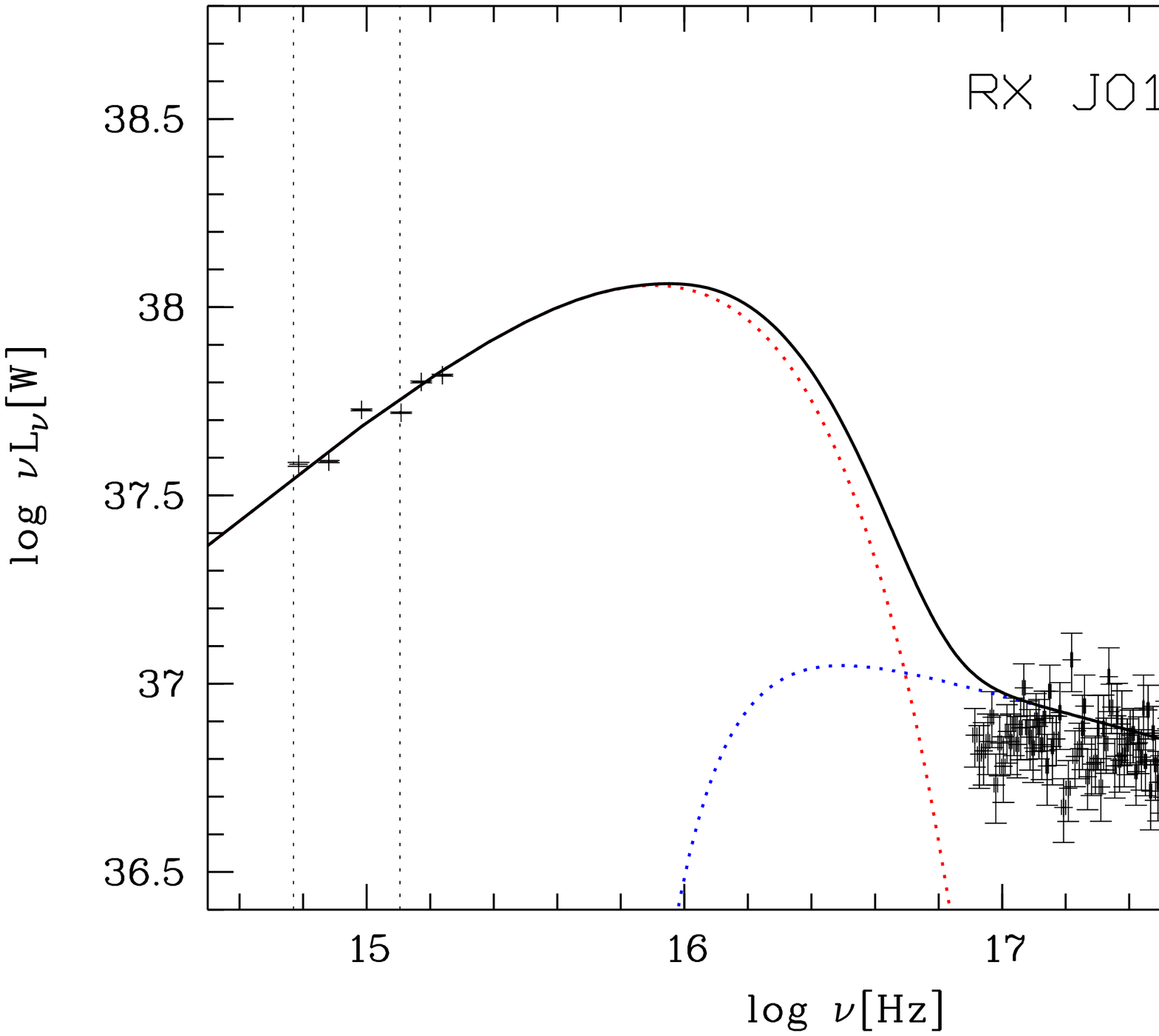}{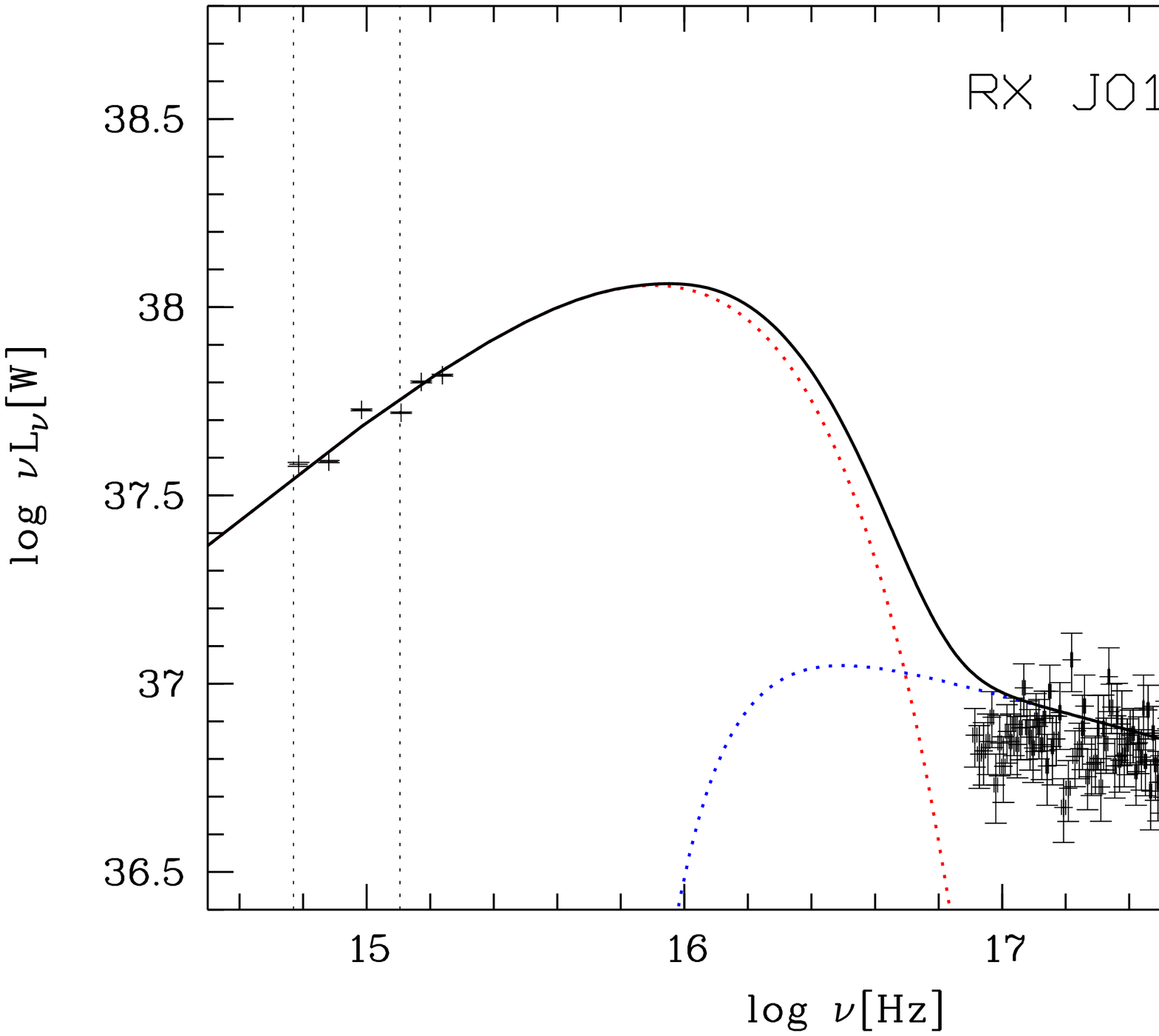}{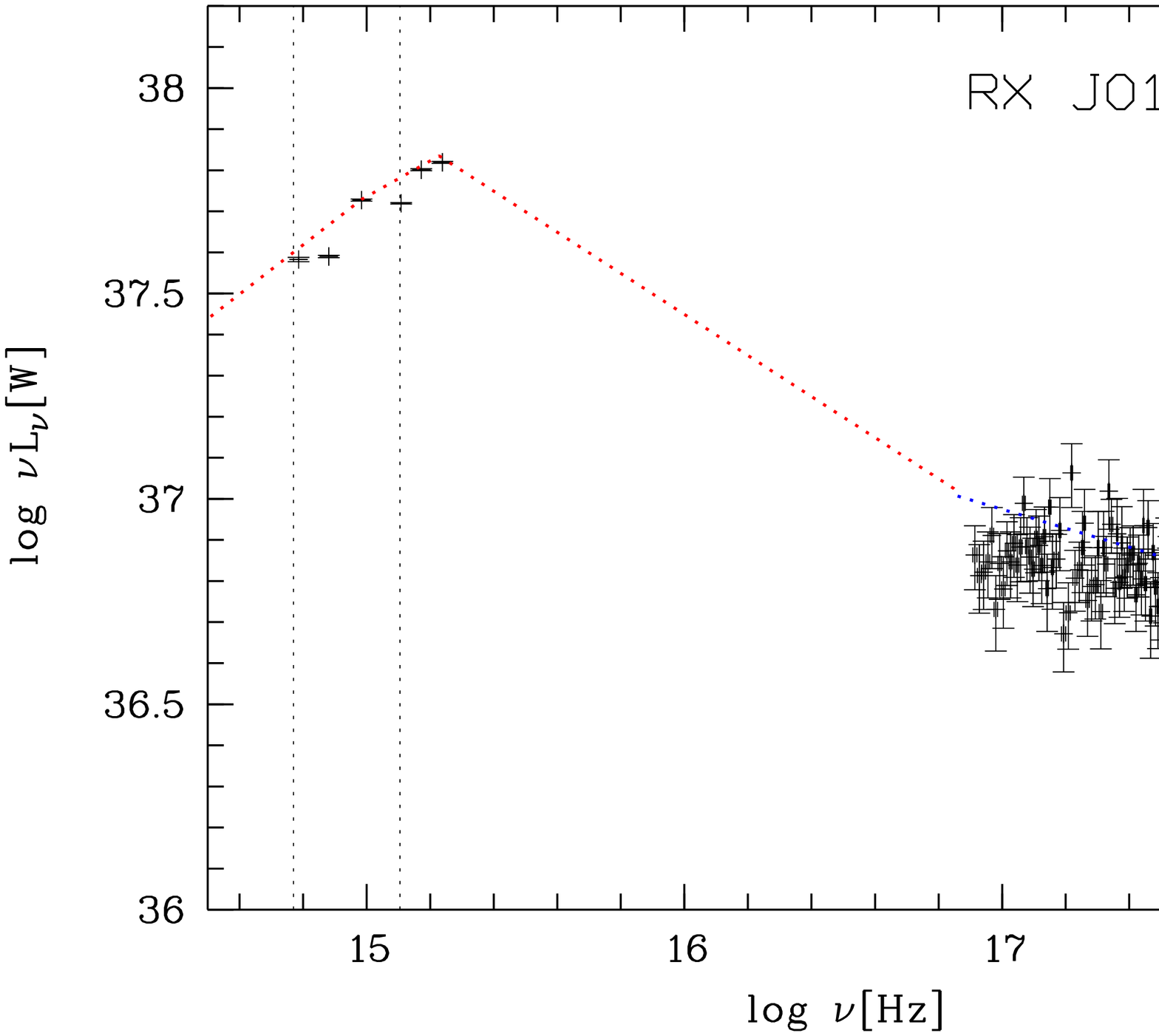}

\end{figure*}

\begin{figure*}
\epsscale{0.60}
\plotthree{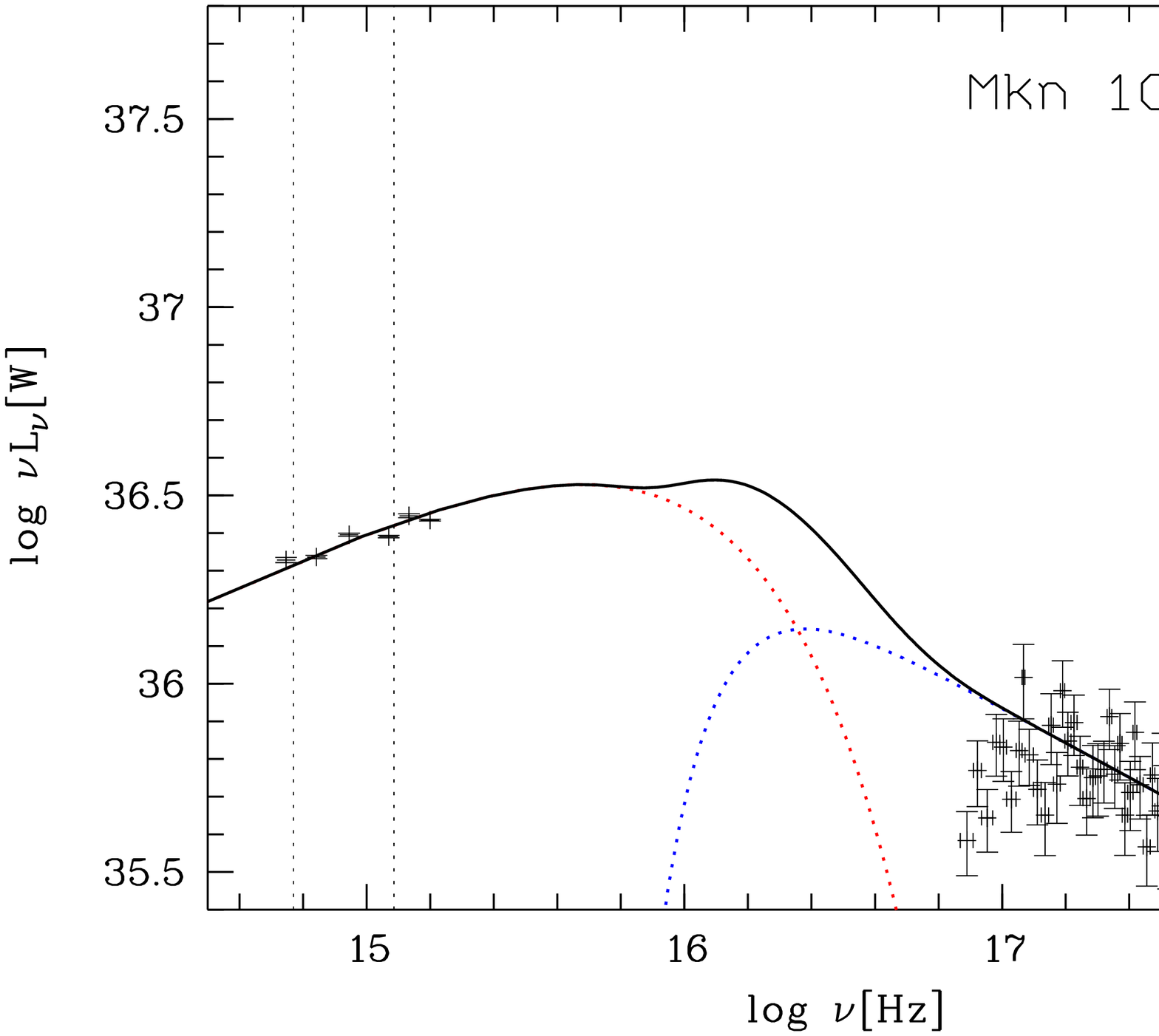}{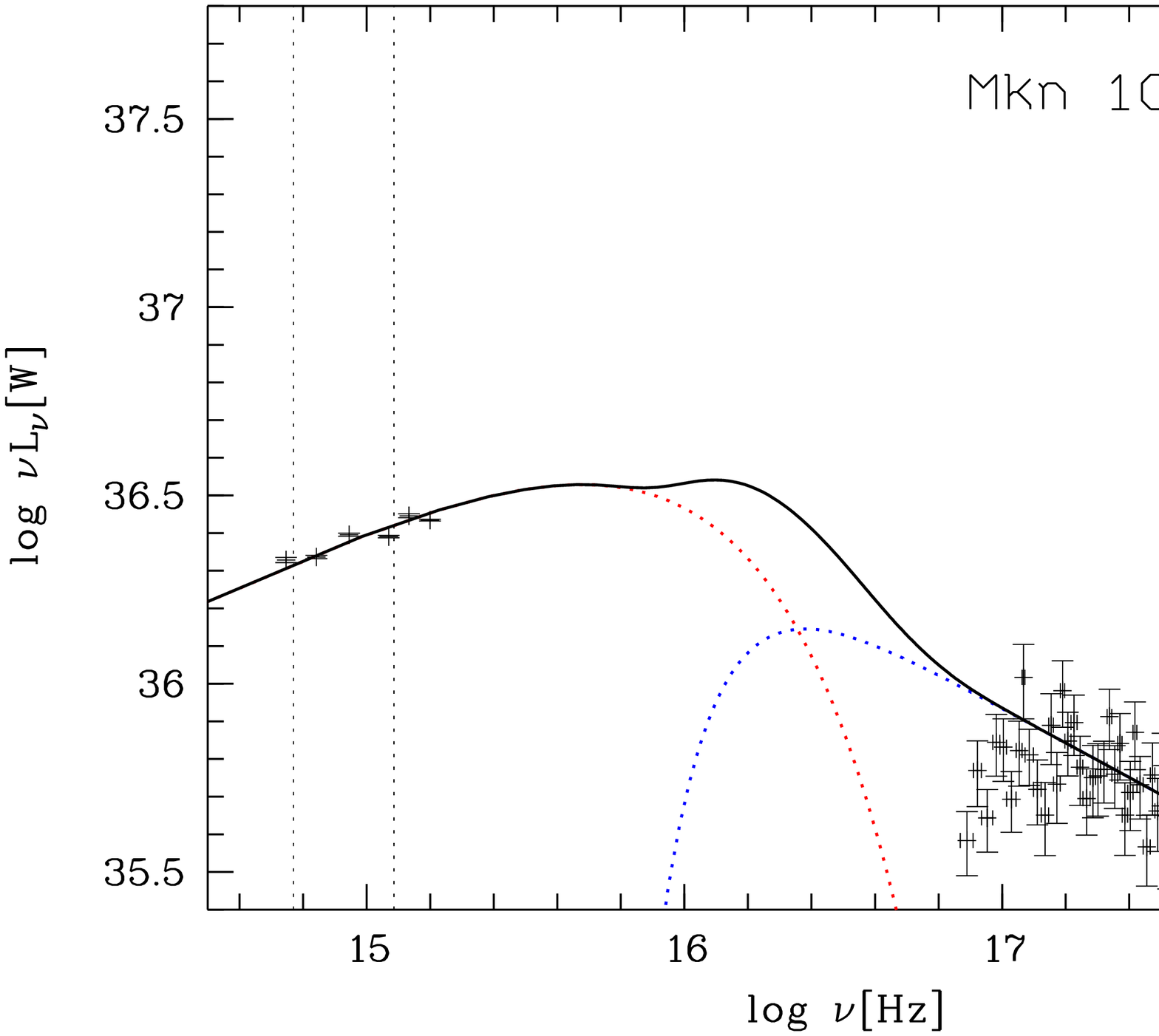}{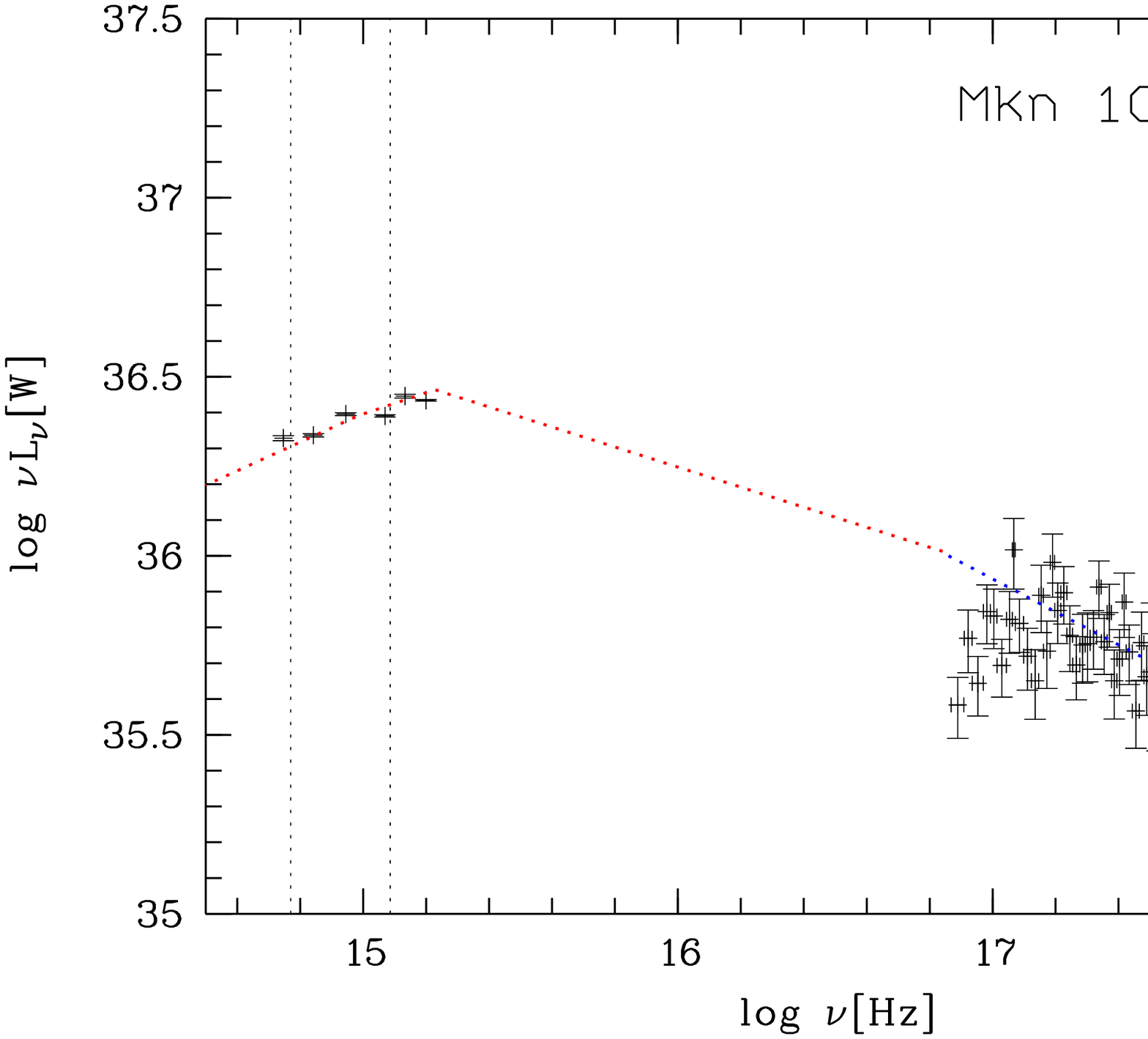}

\plotthree{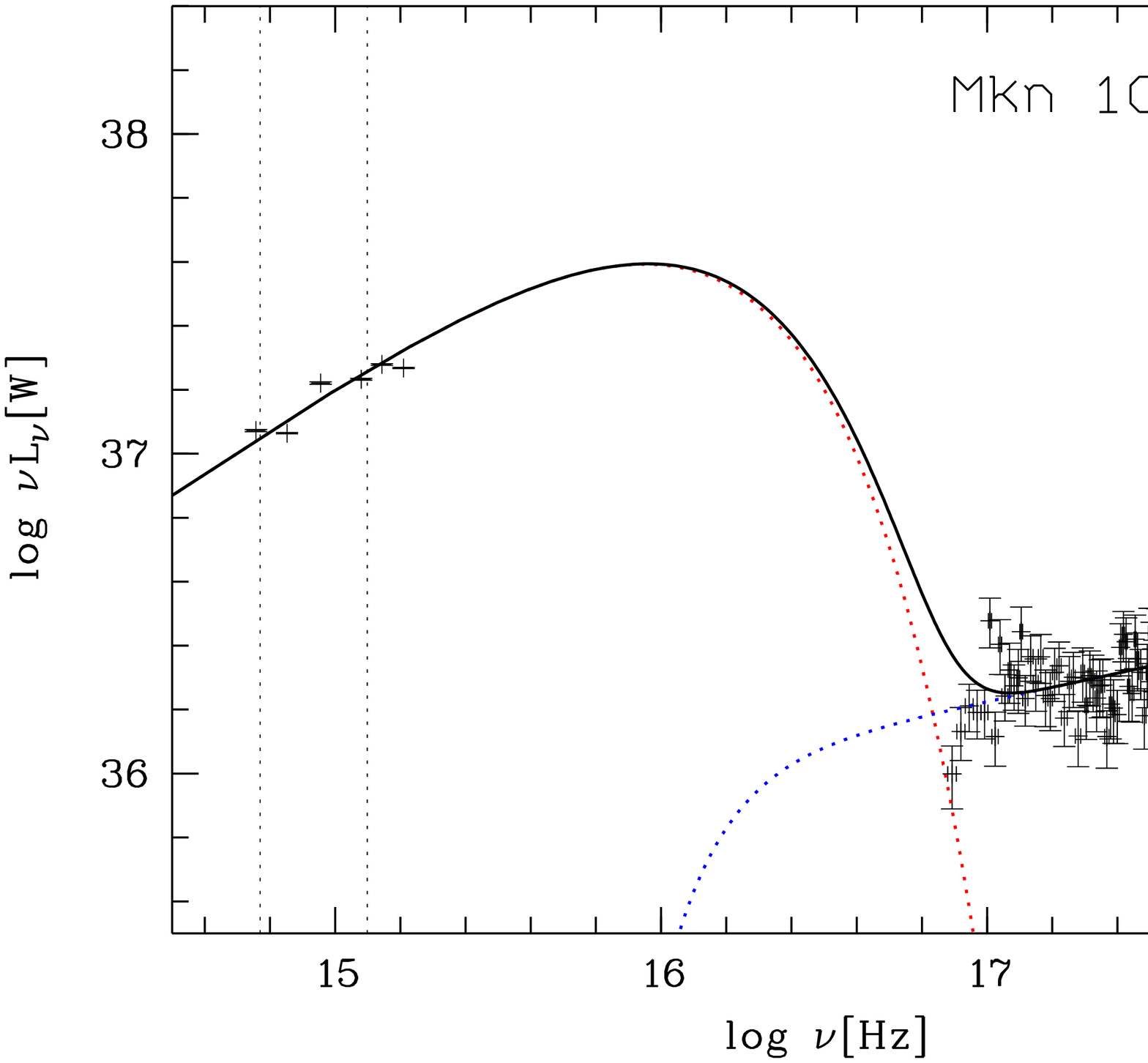}{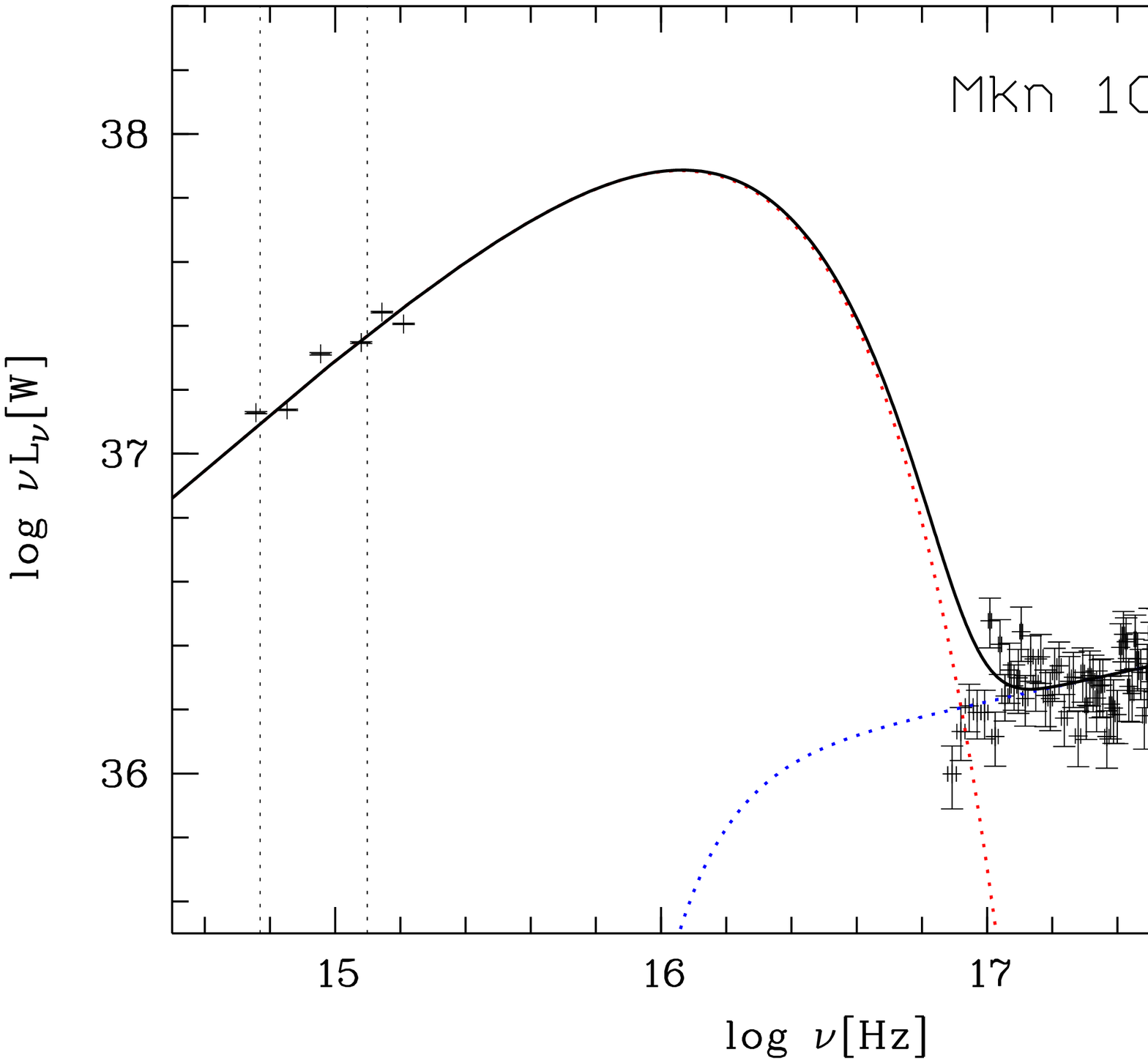}{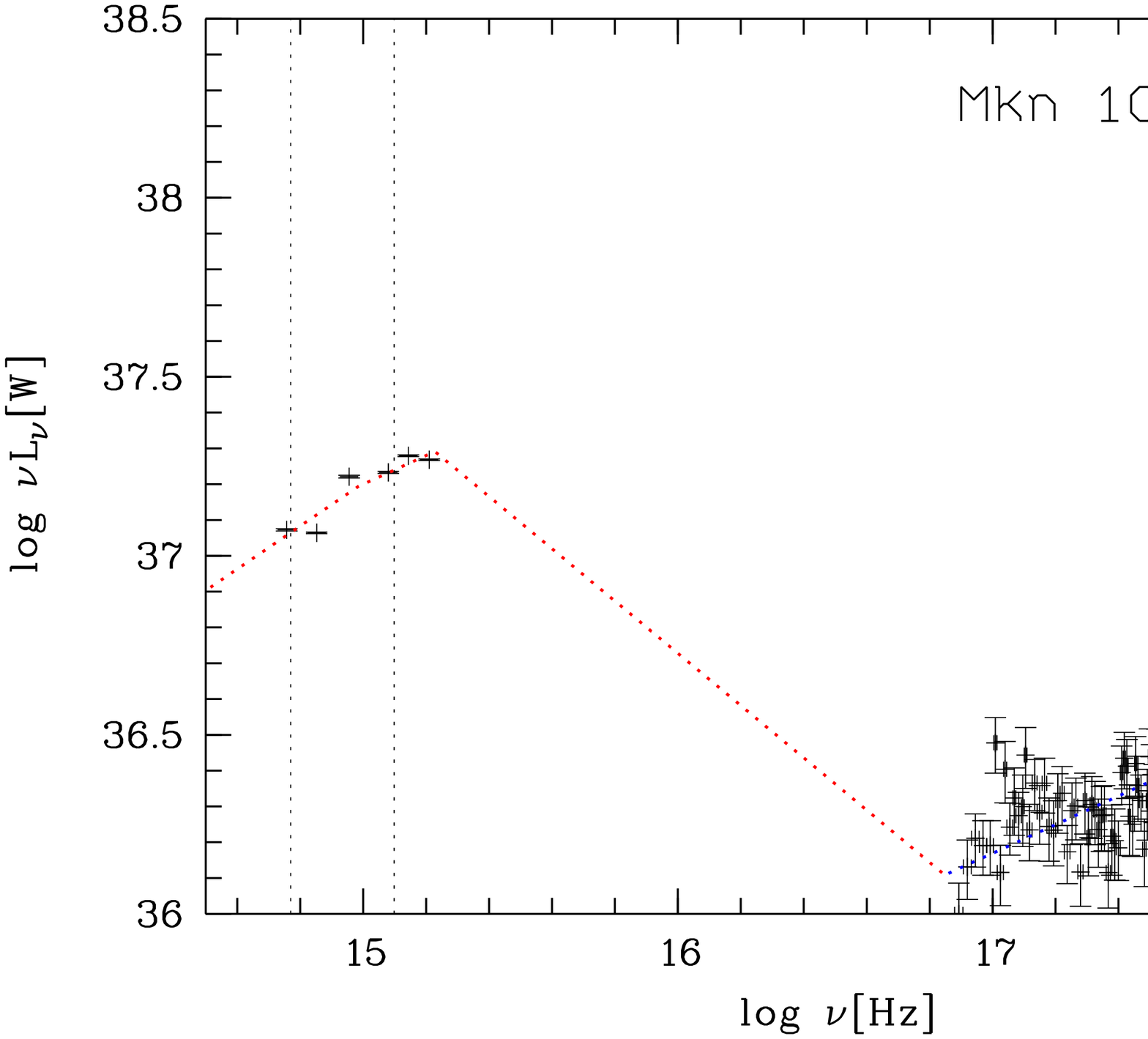}

\plotthree{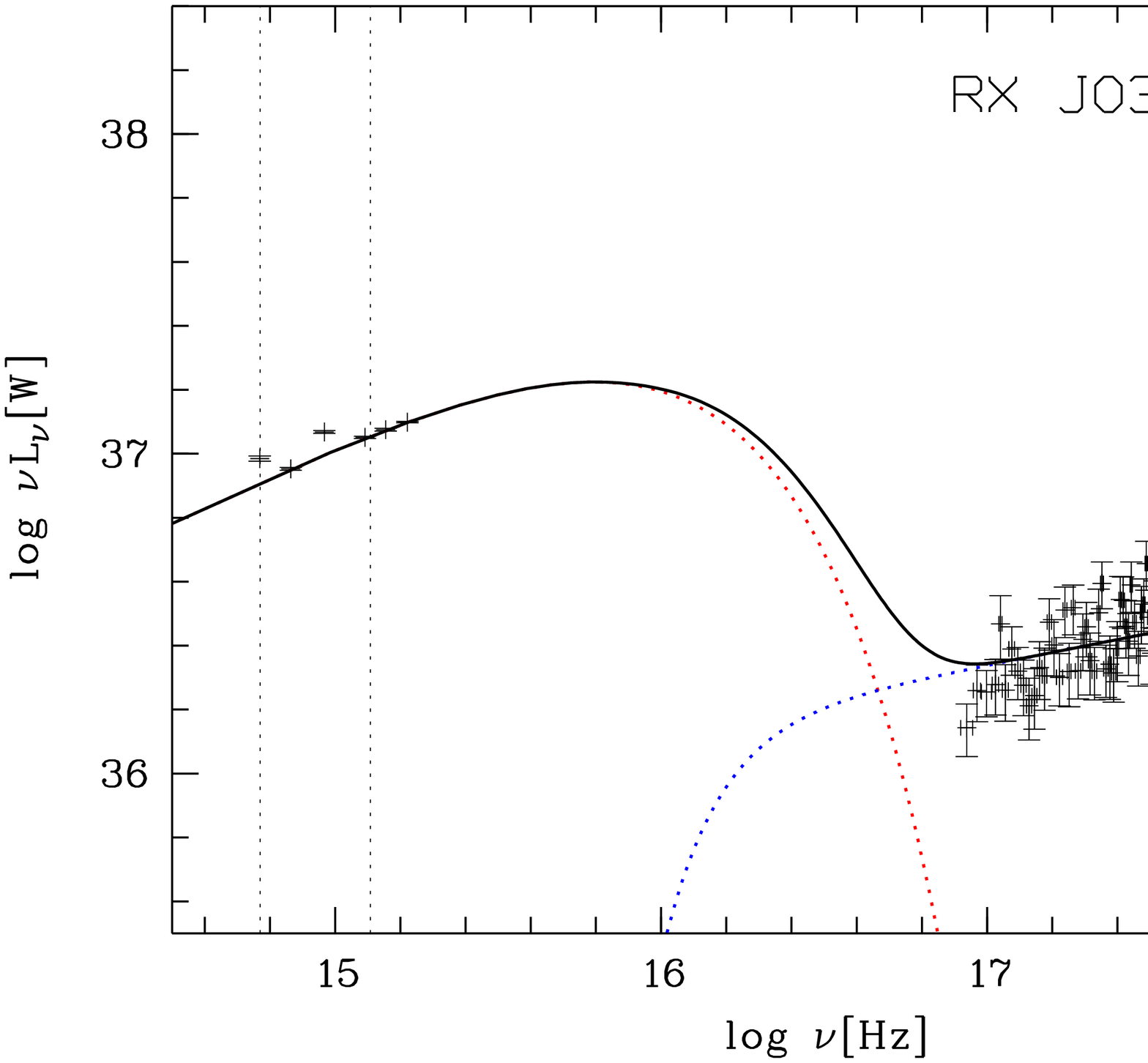}{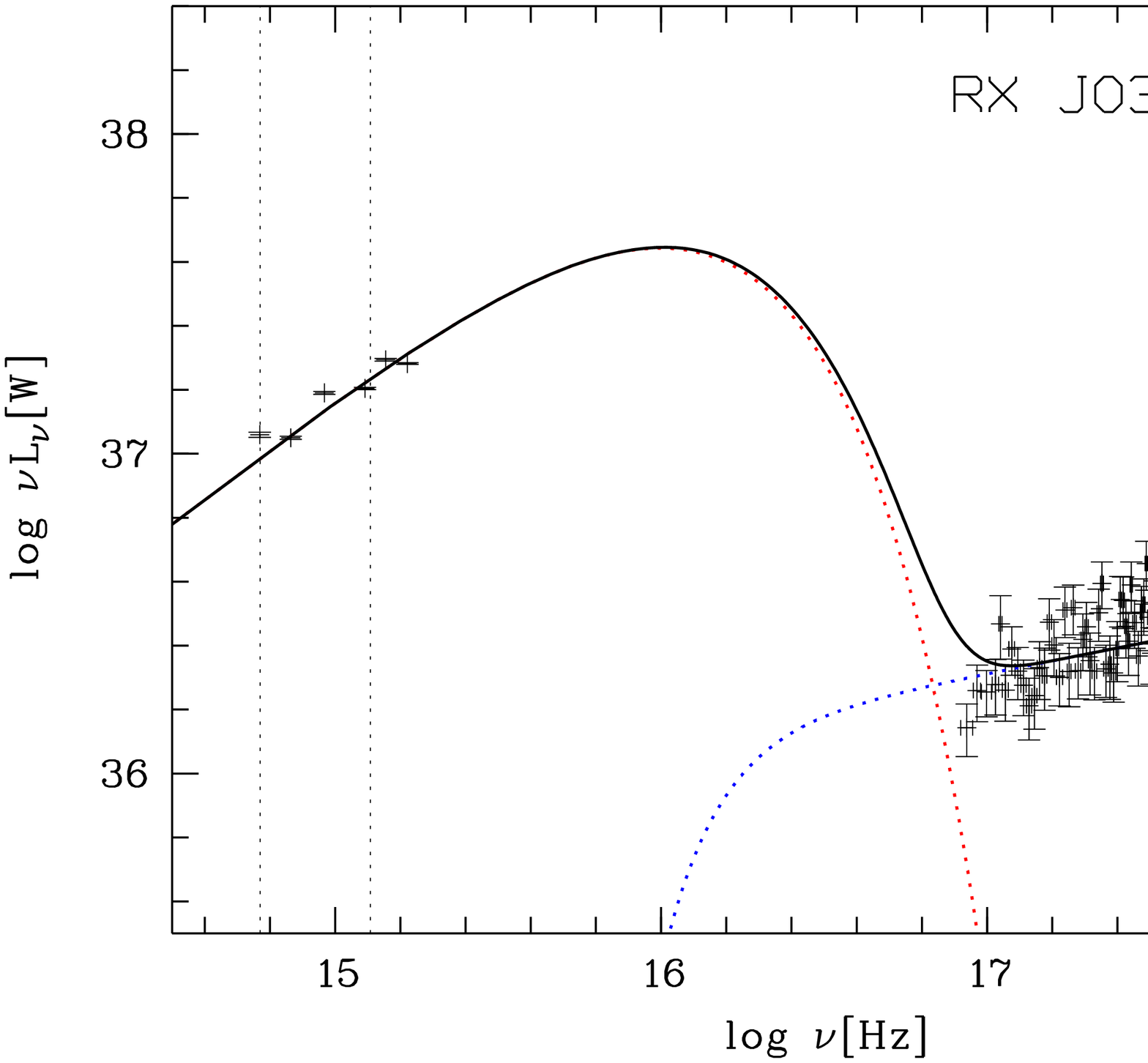}{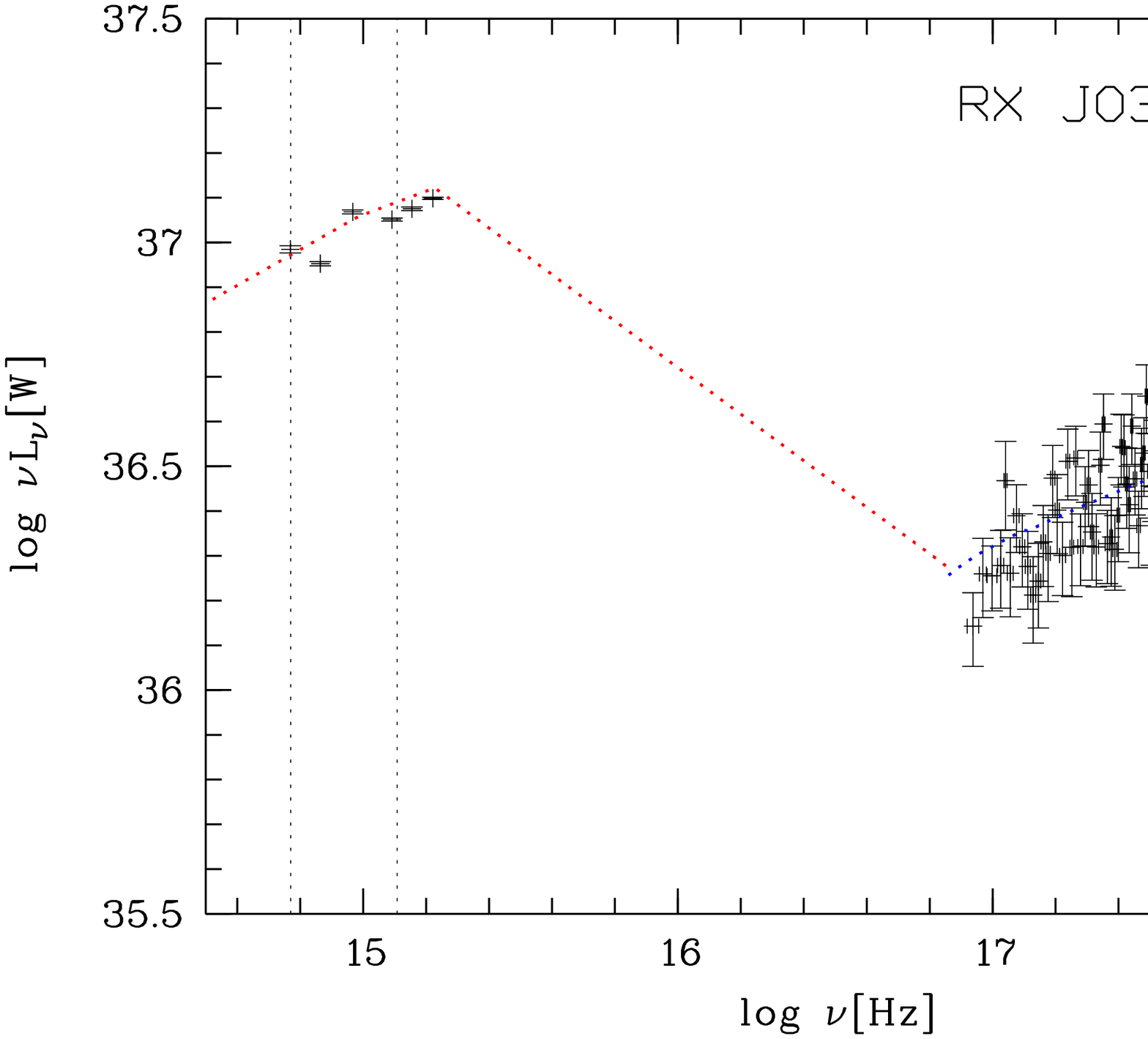}

\plotthree{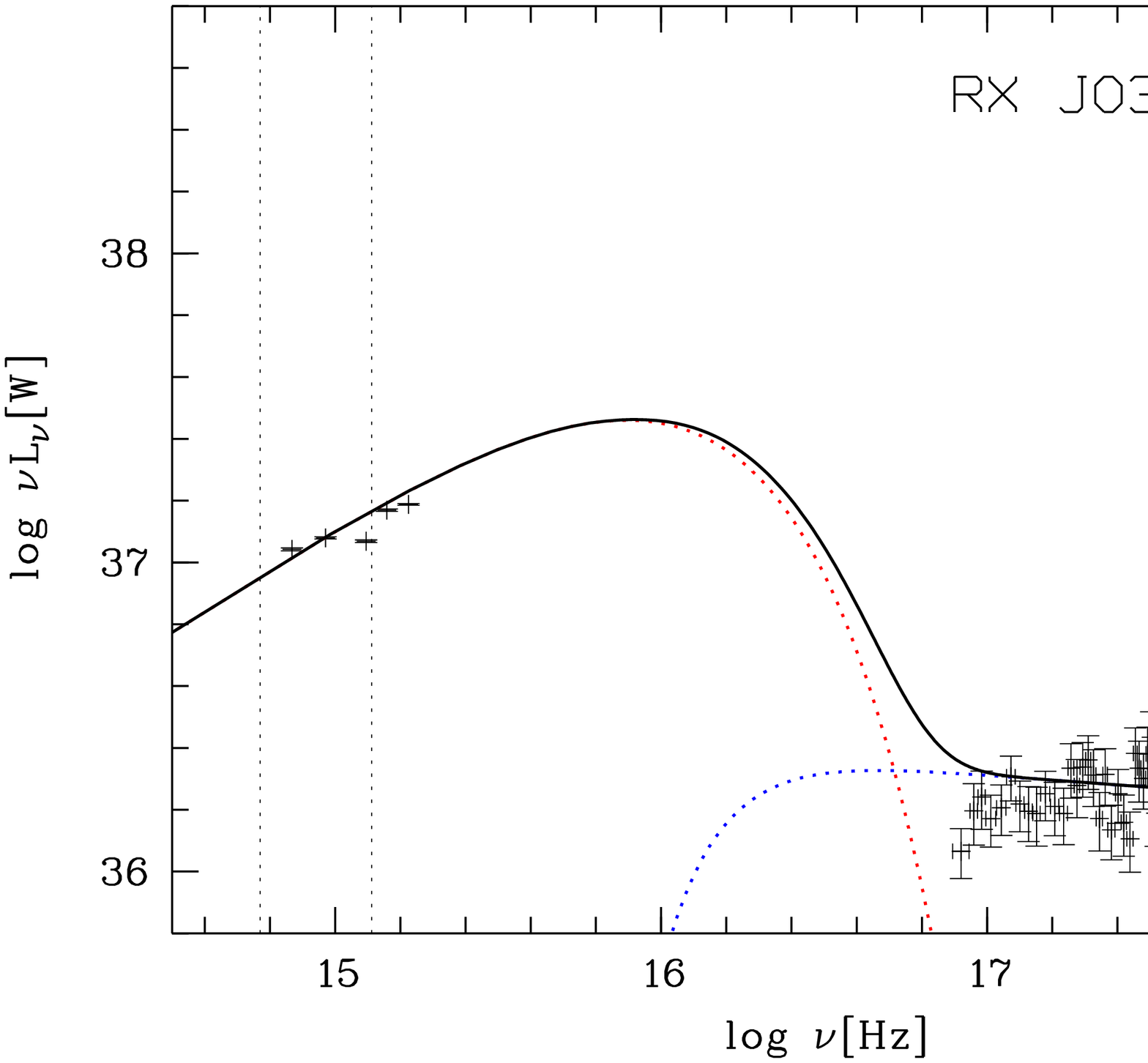}{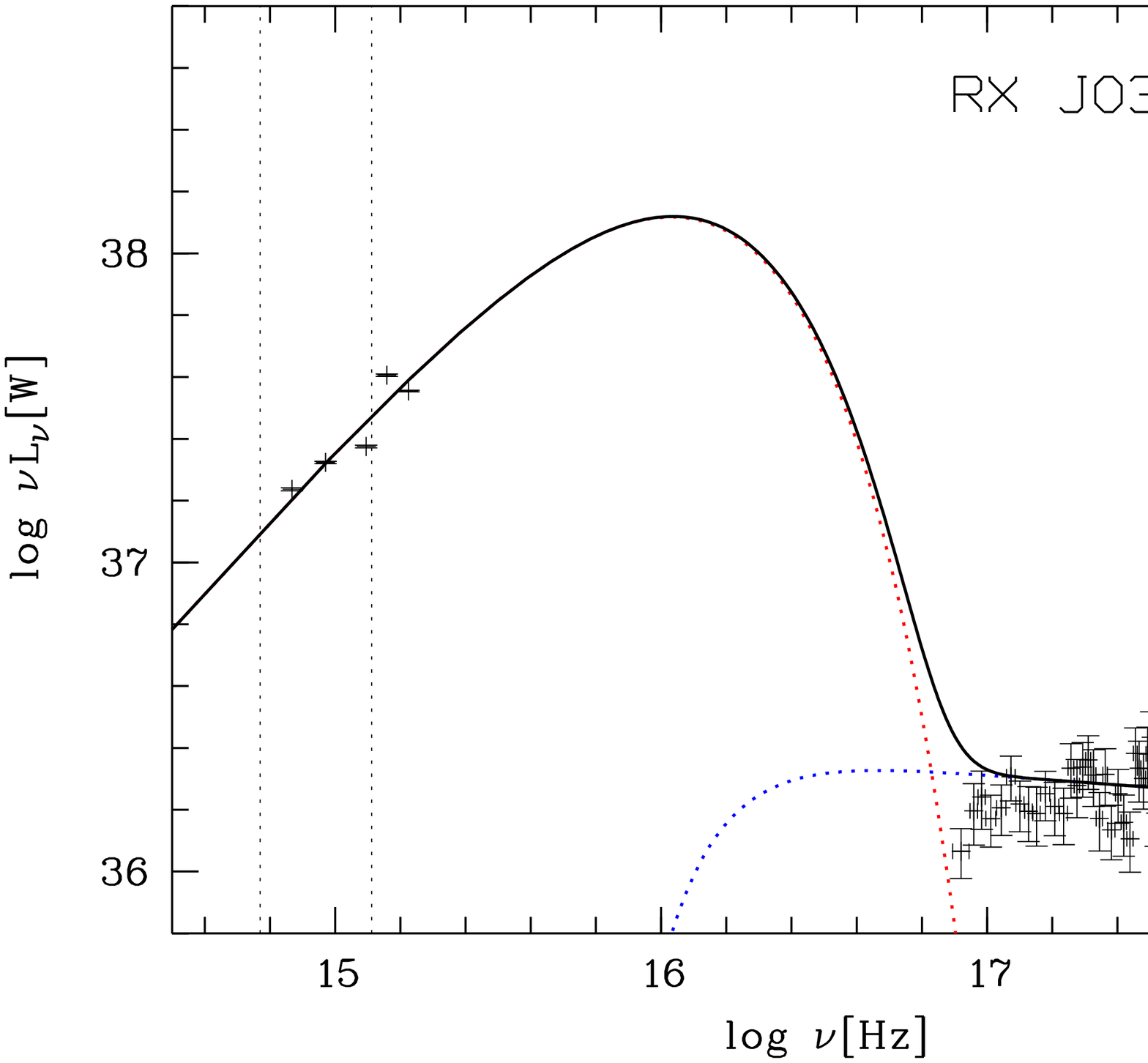}{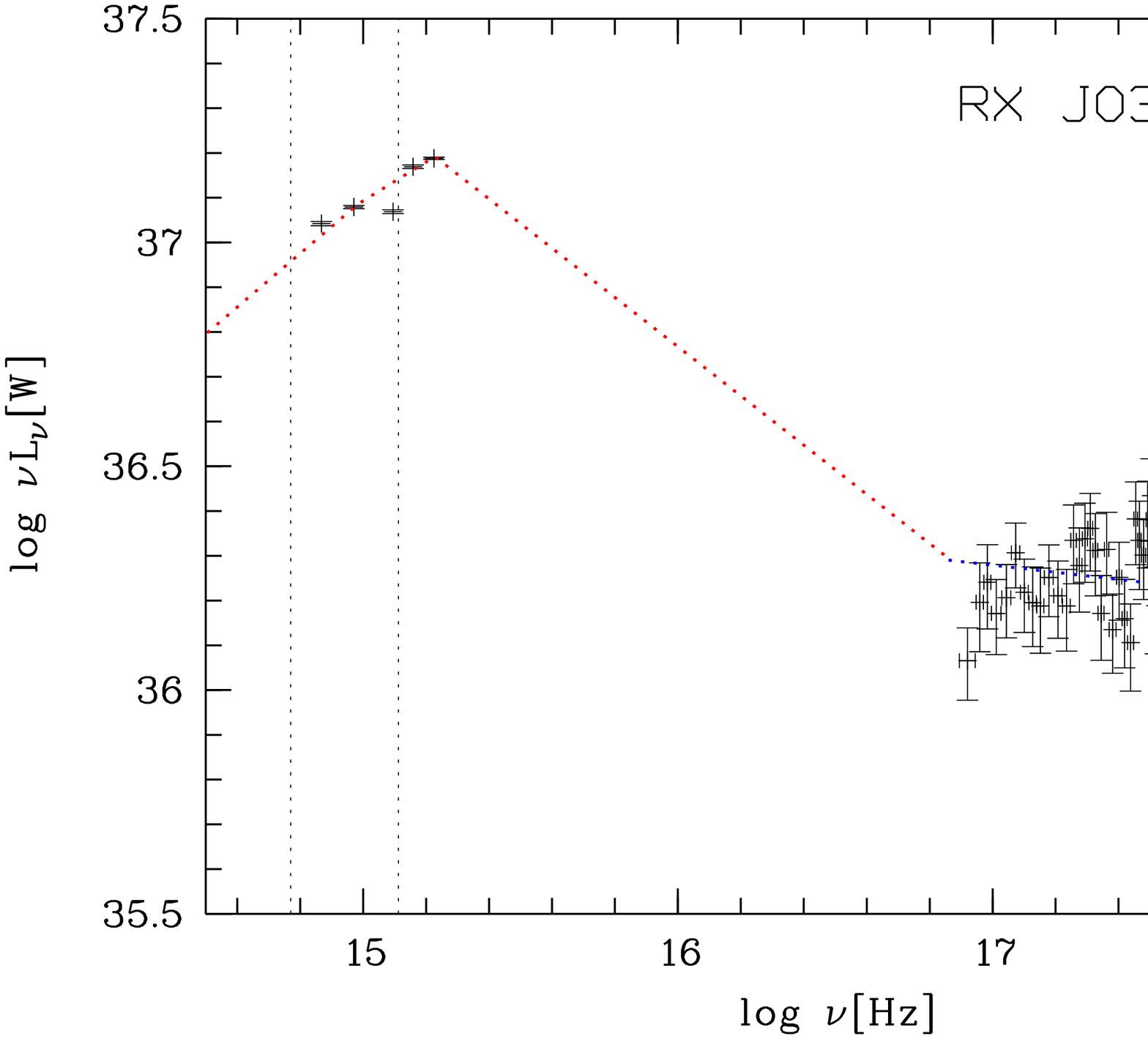}

\plotthree{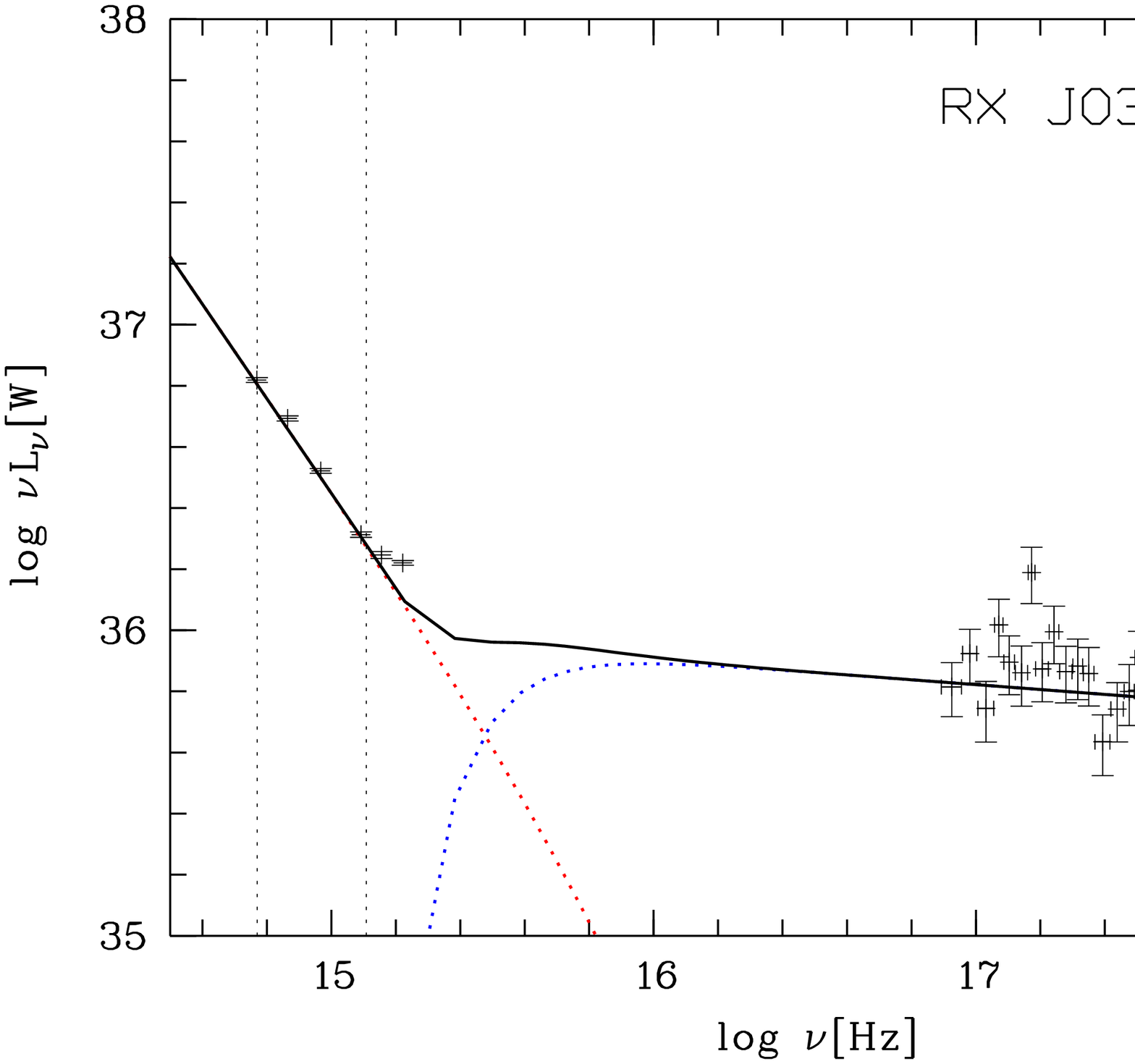}{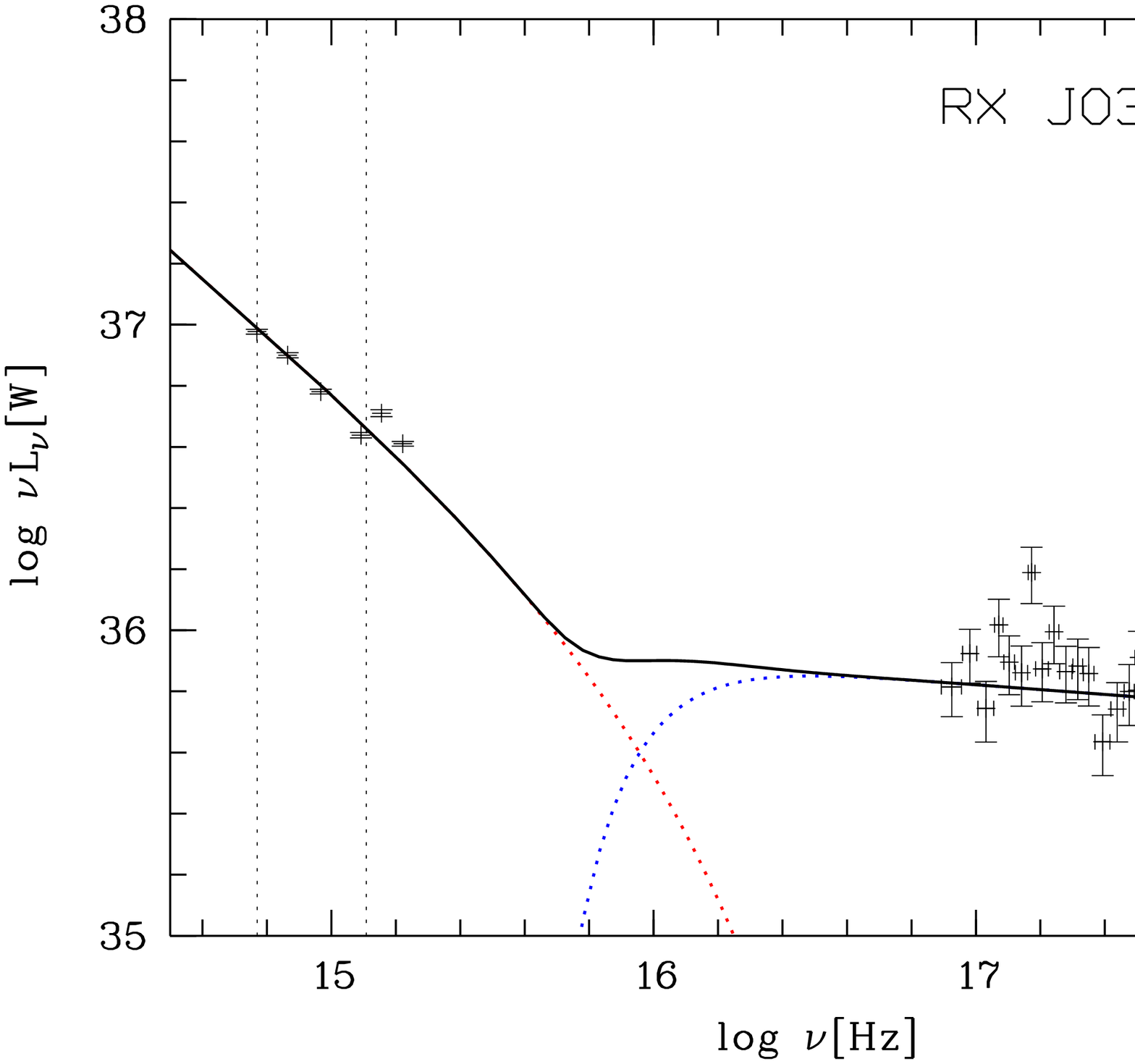}{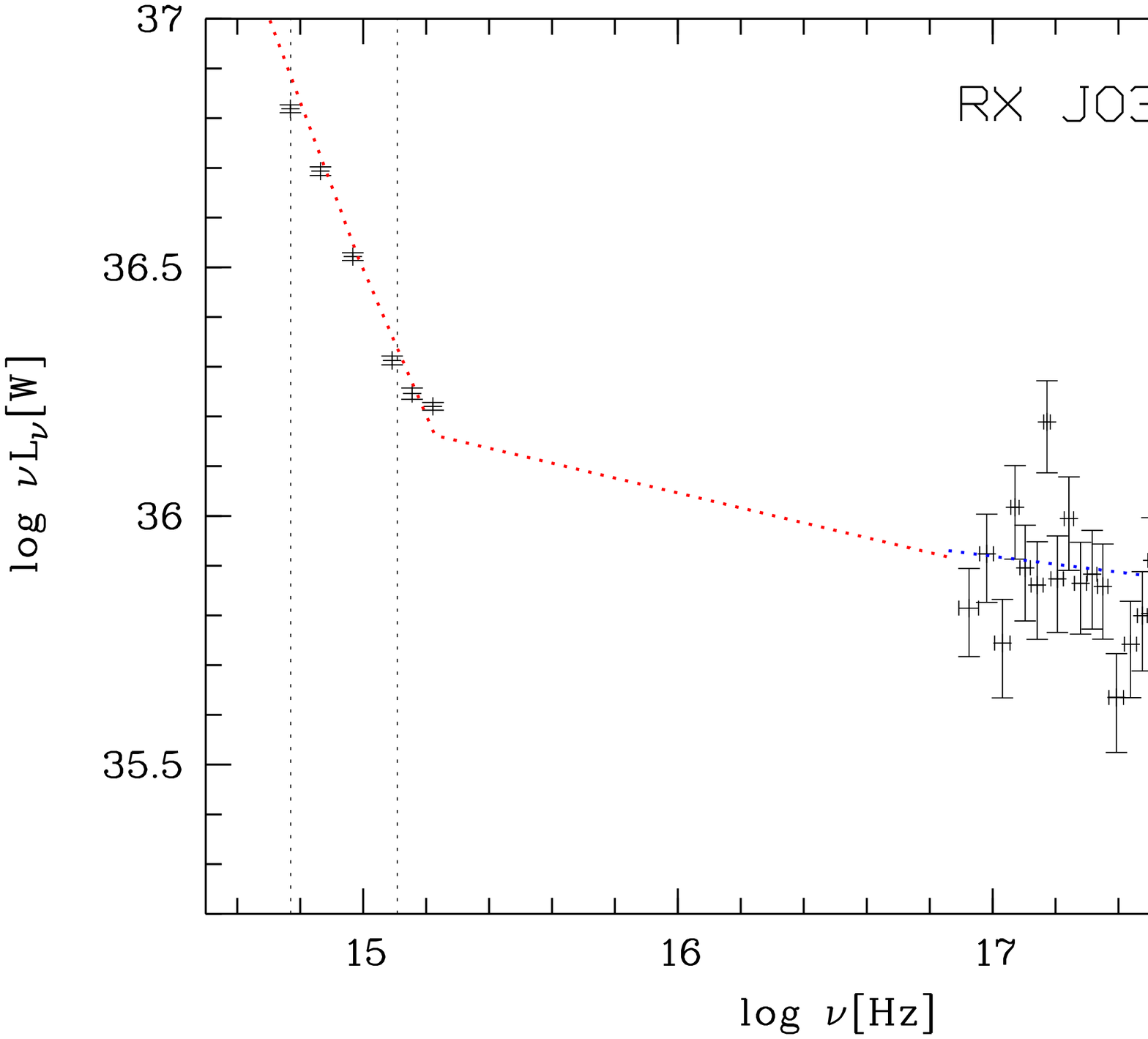}

\end{figure*}

\begin{figure*}
\epsscale{0.60}
\plotthree{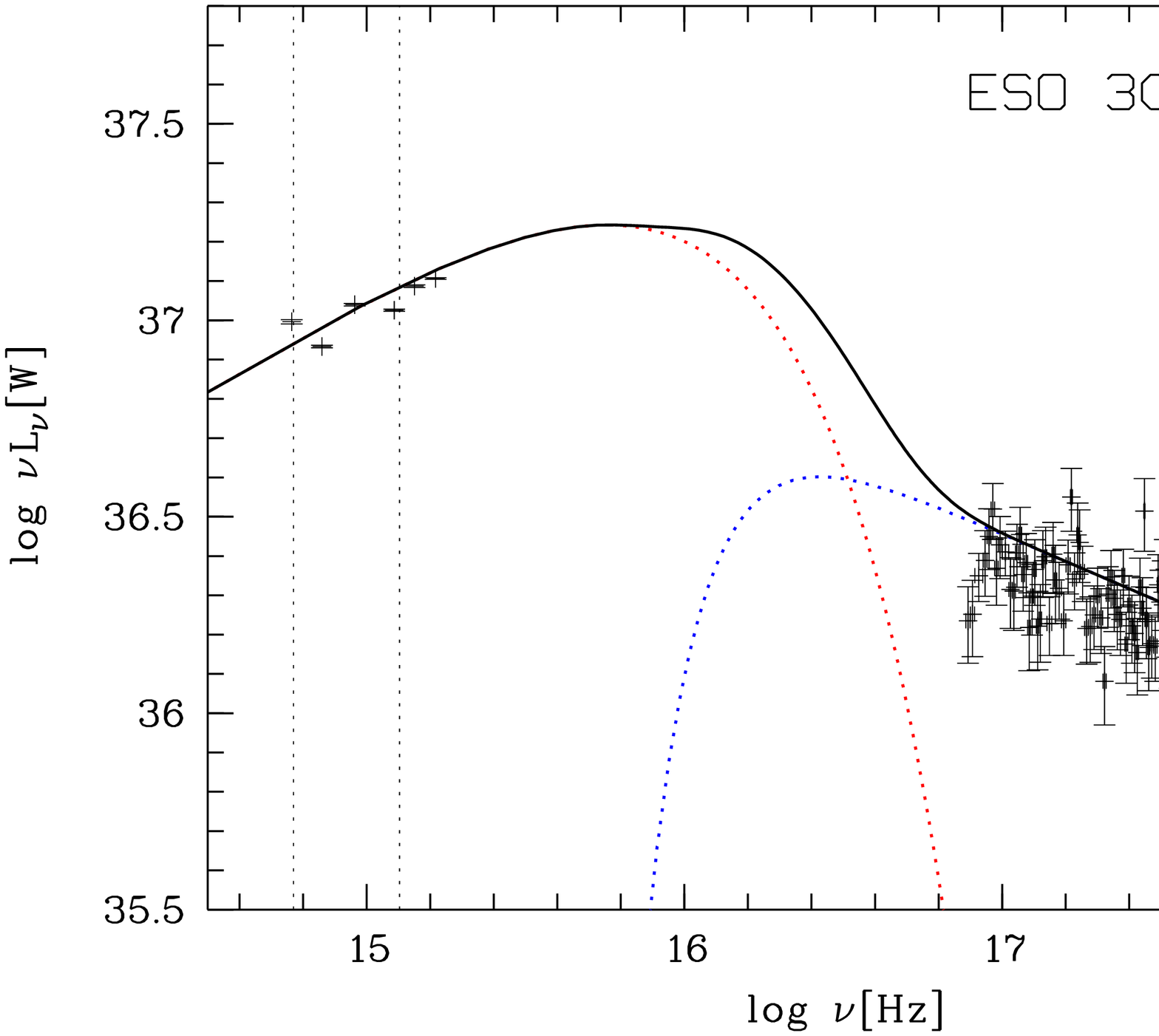}{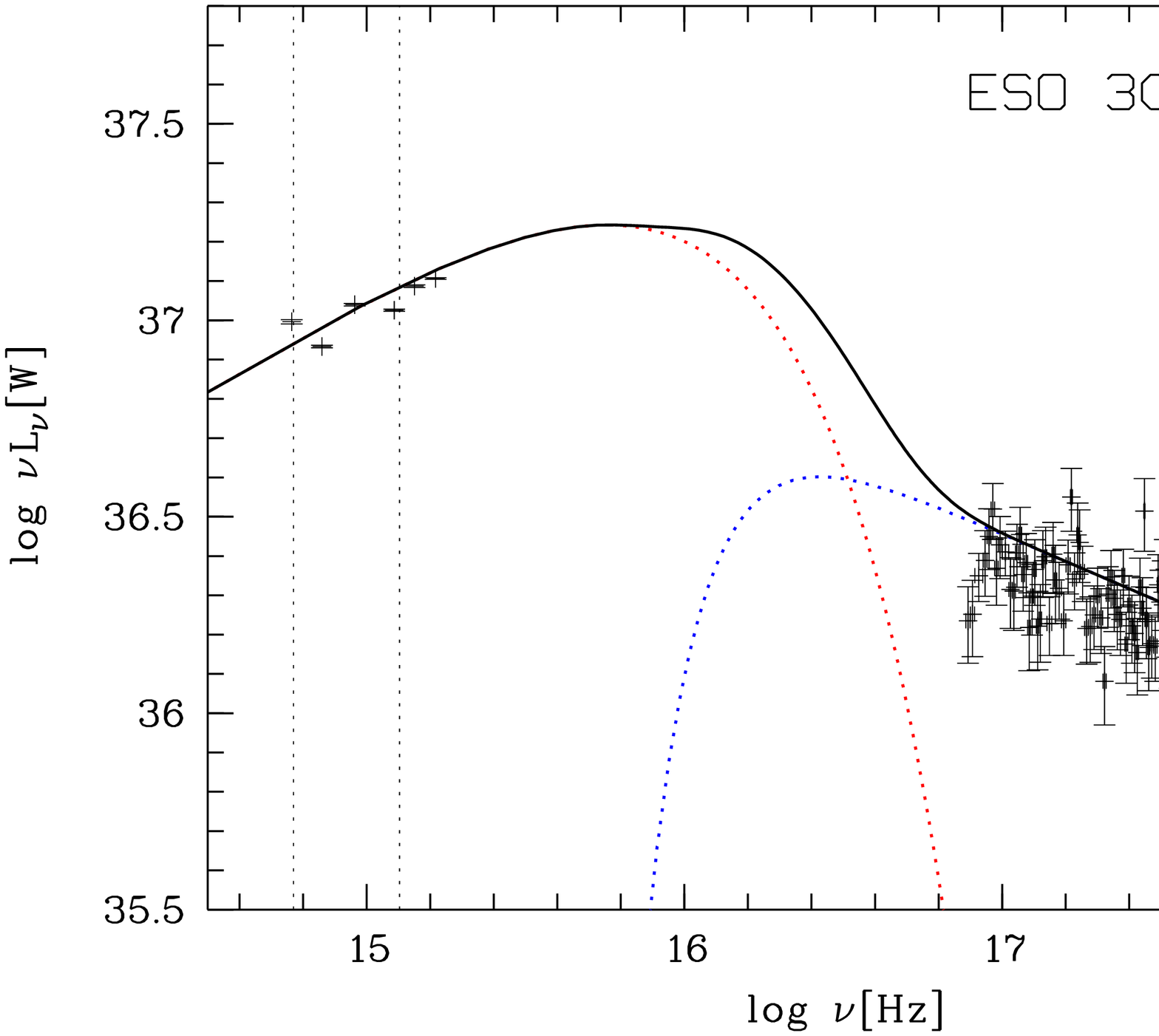}{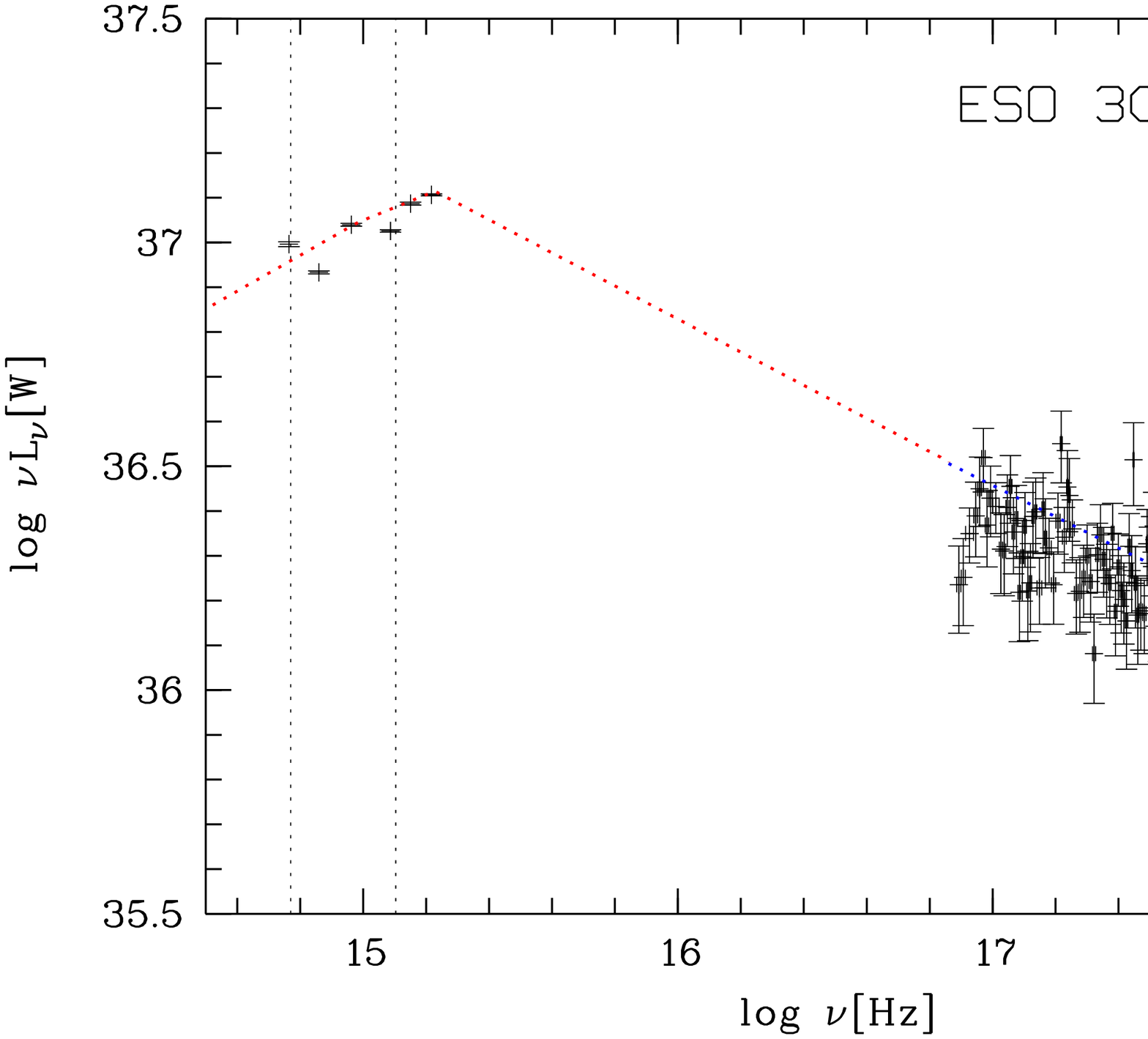}

\plotthree{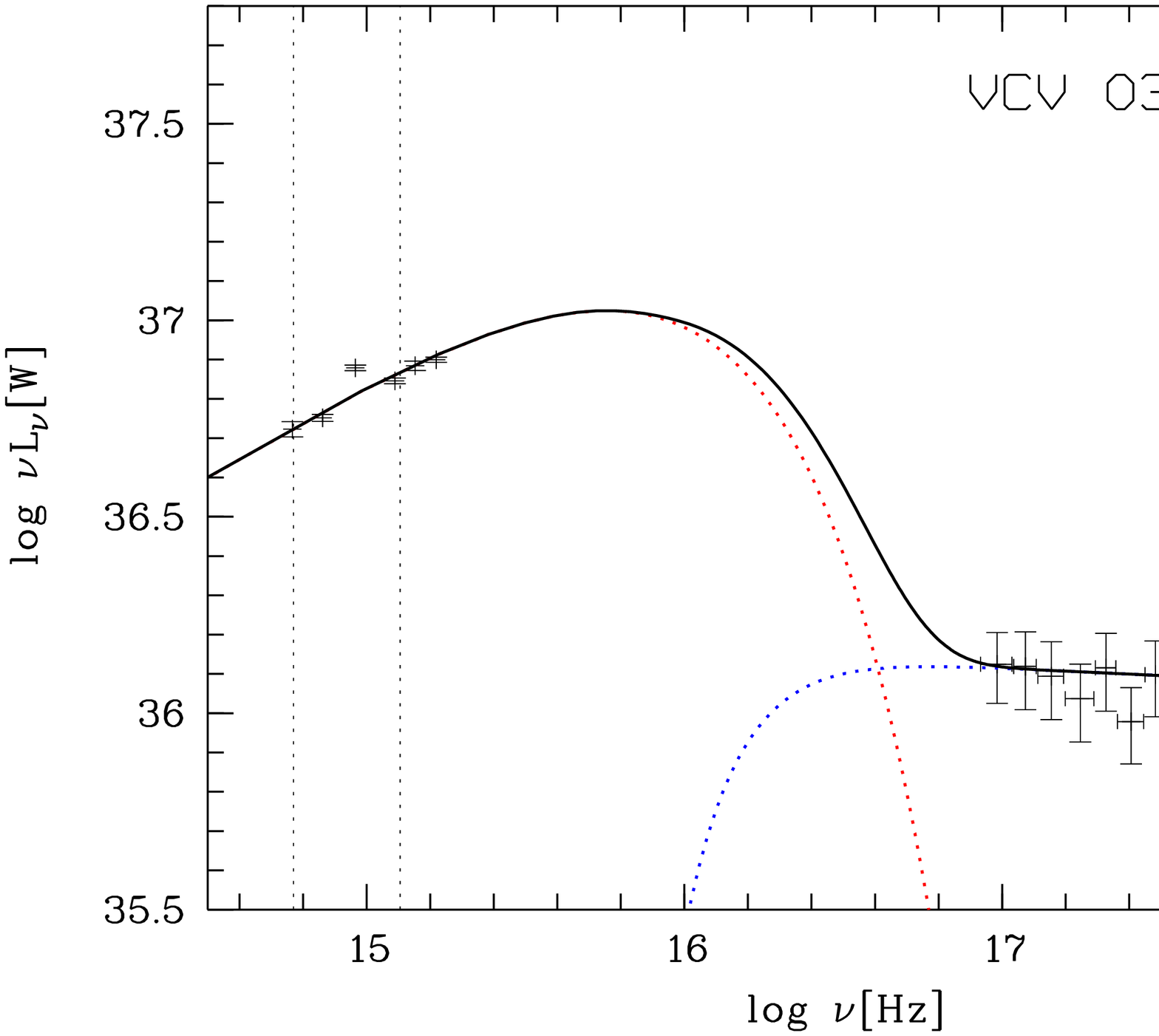}{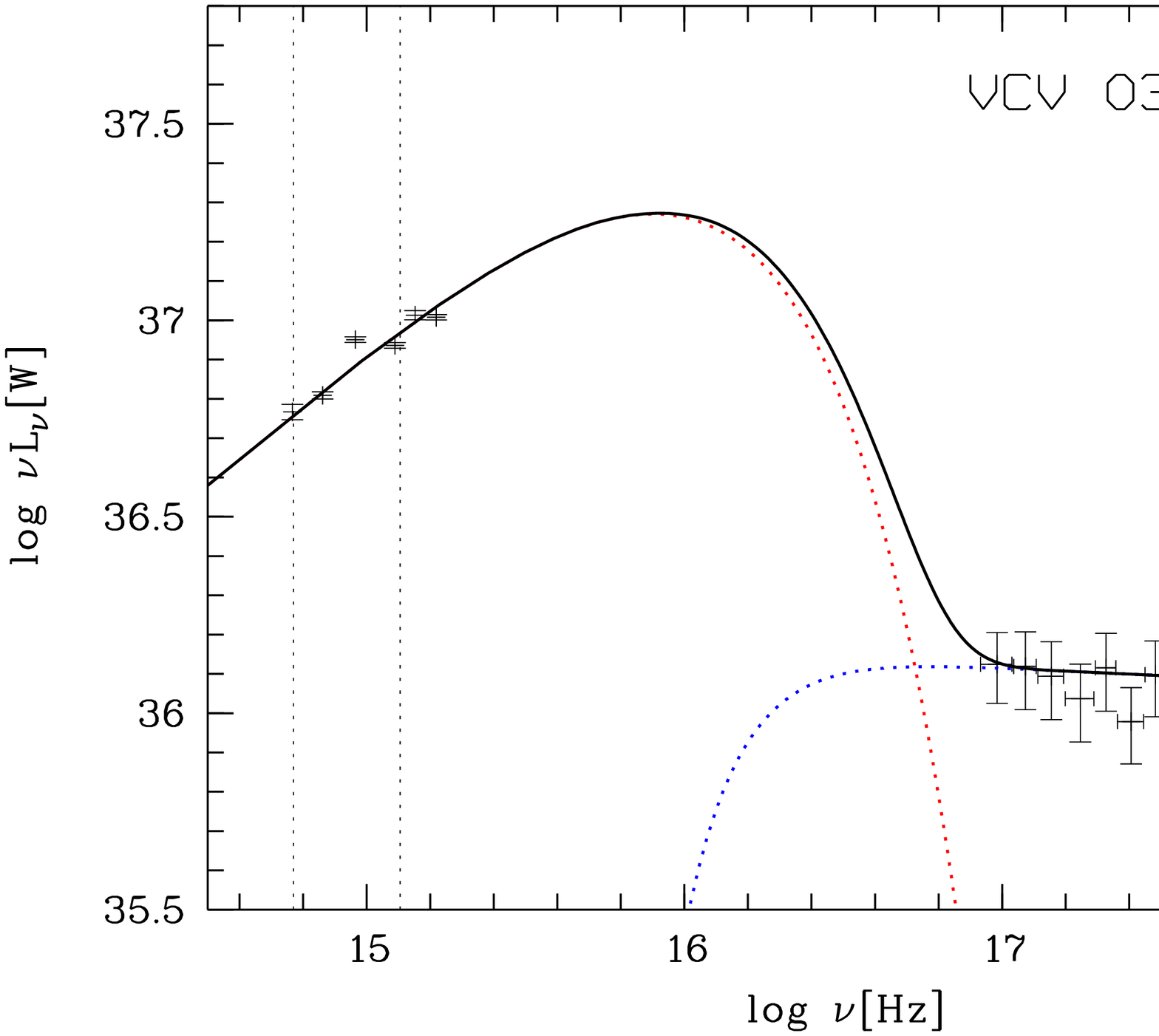}{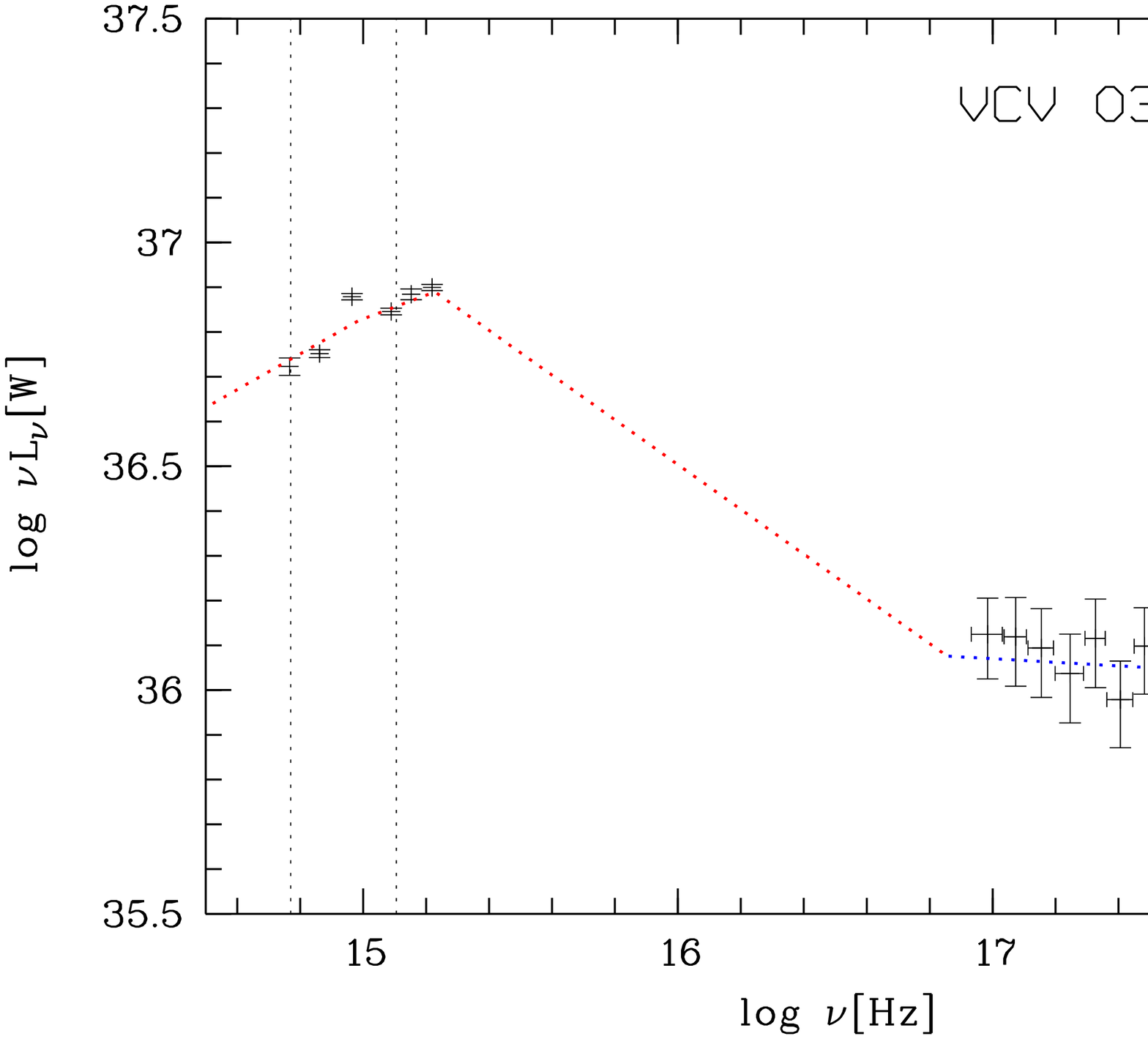}

\plotthree{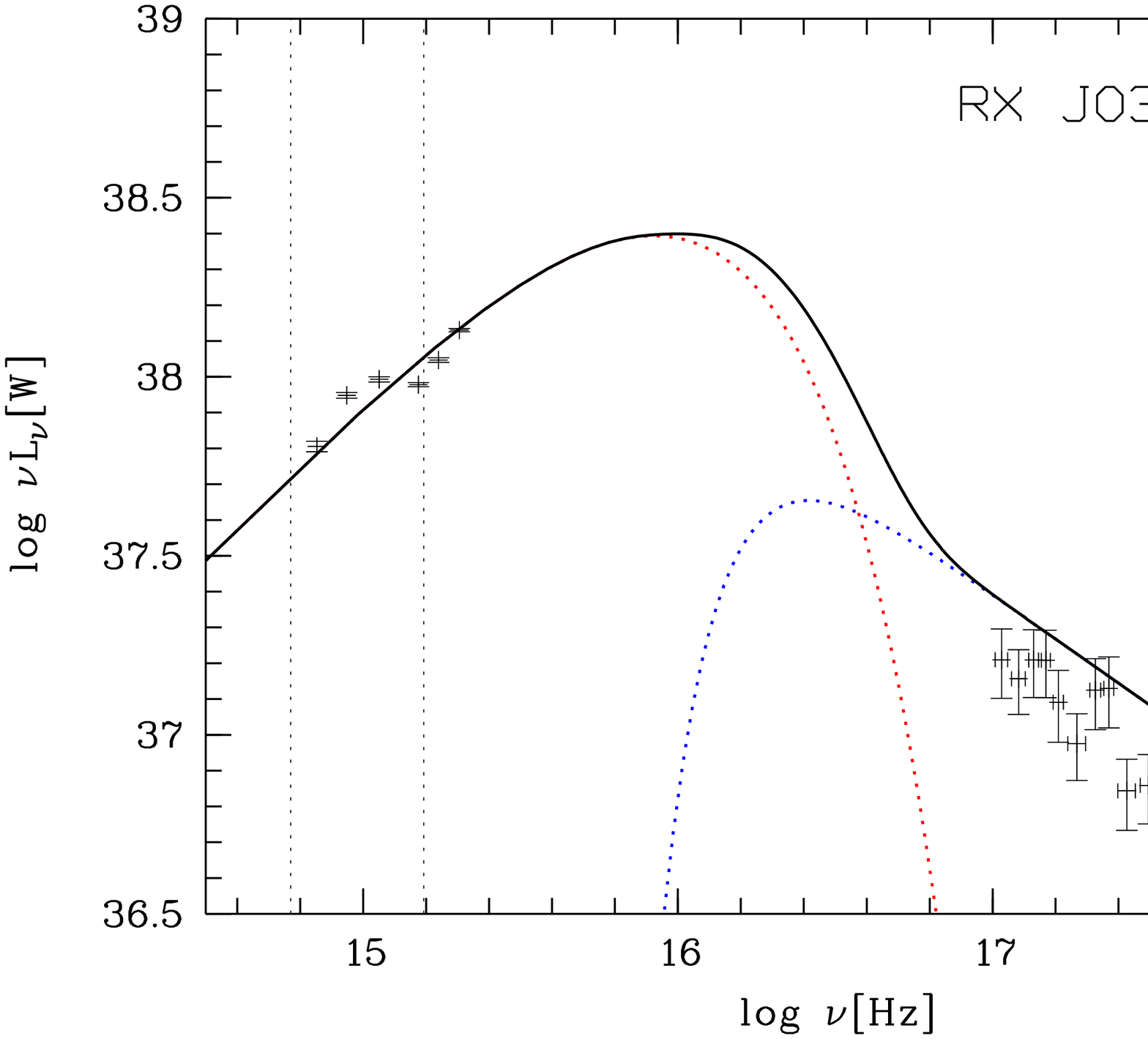}{f25_t.ps}{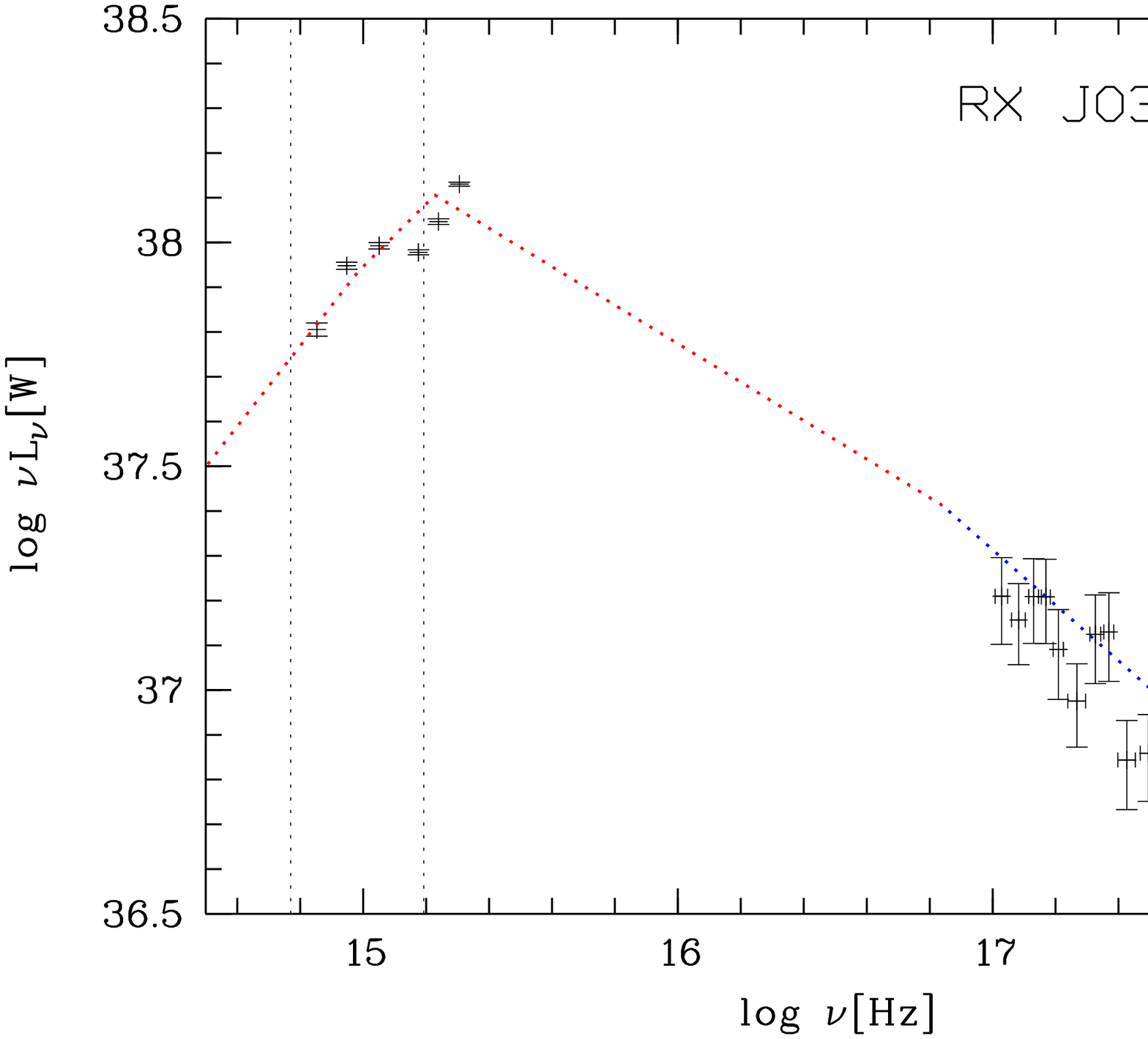}

\plotthree{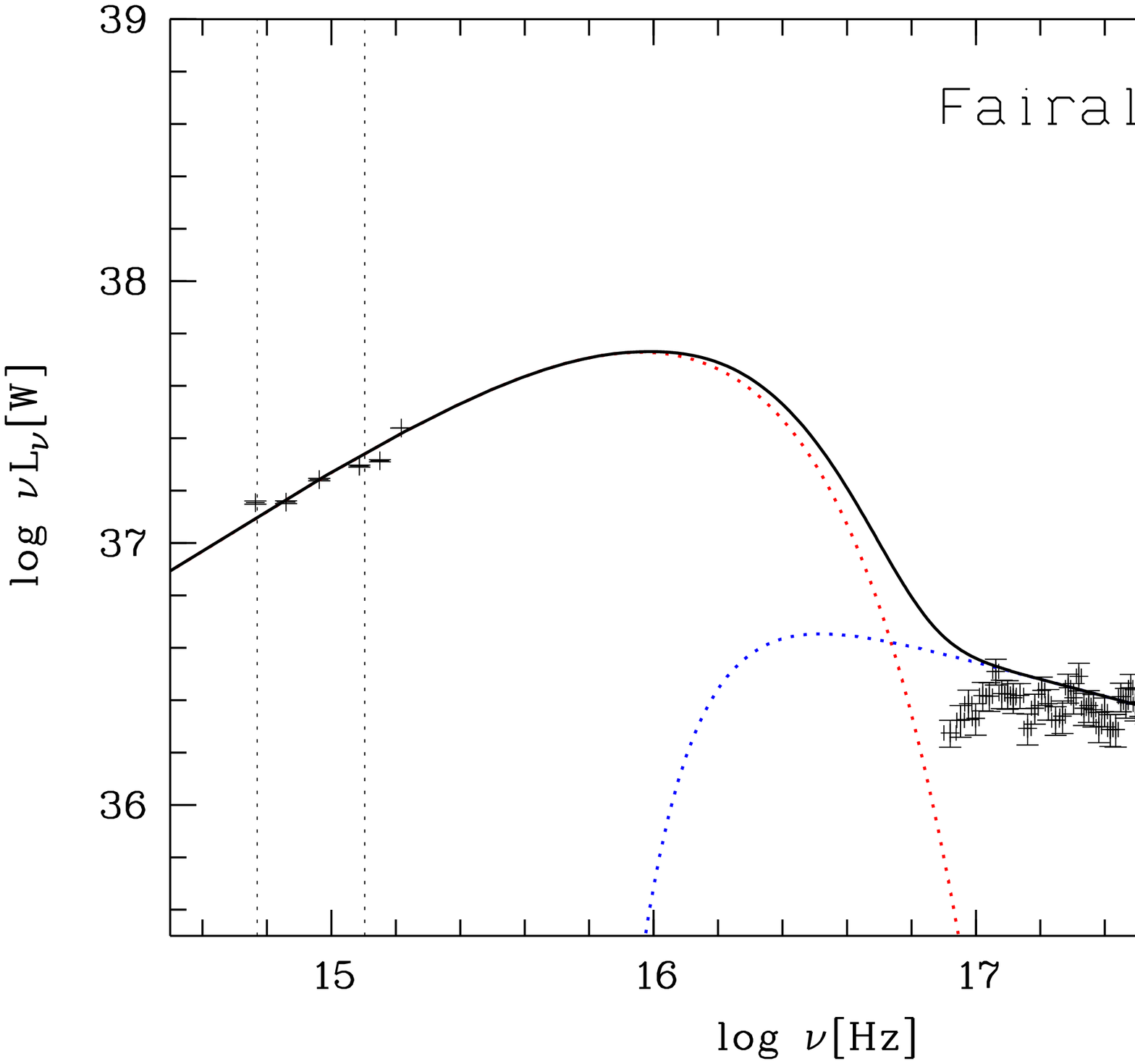}{f25_t.ps}{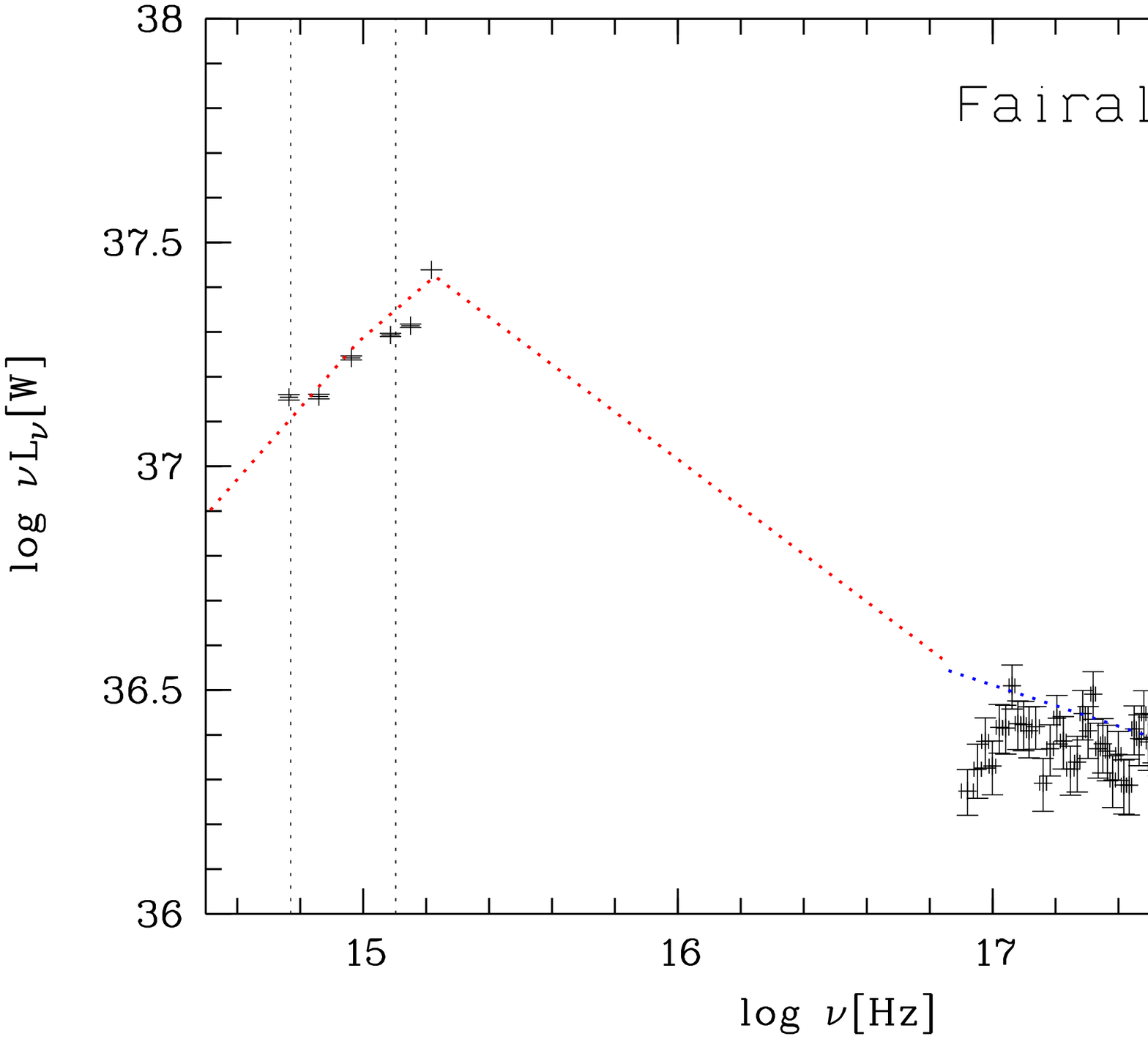}

\plotthree{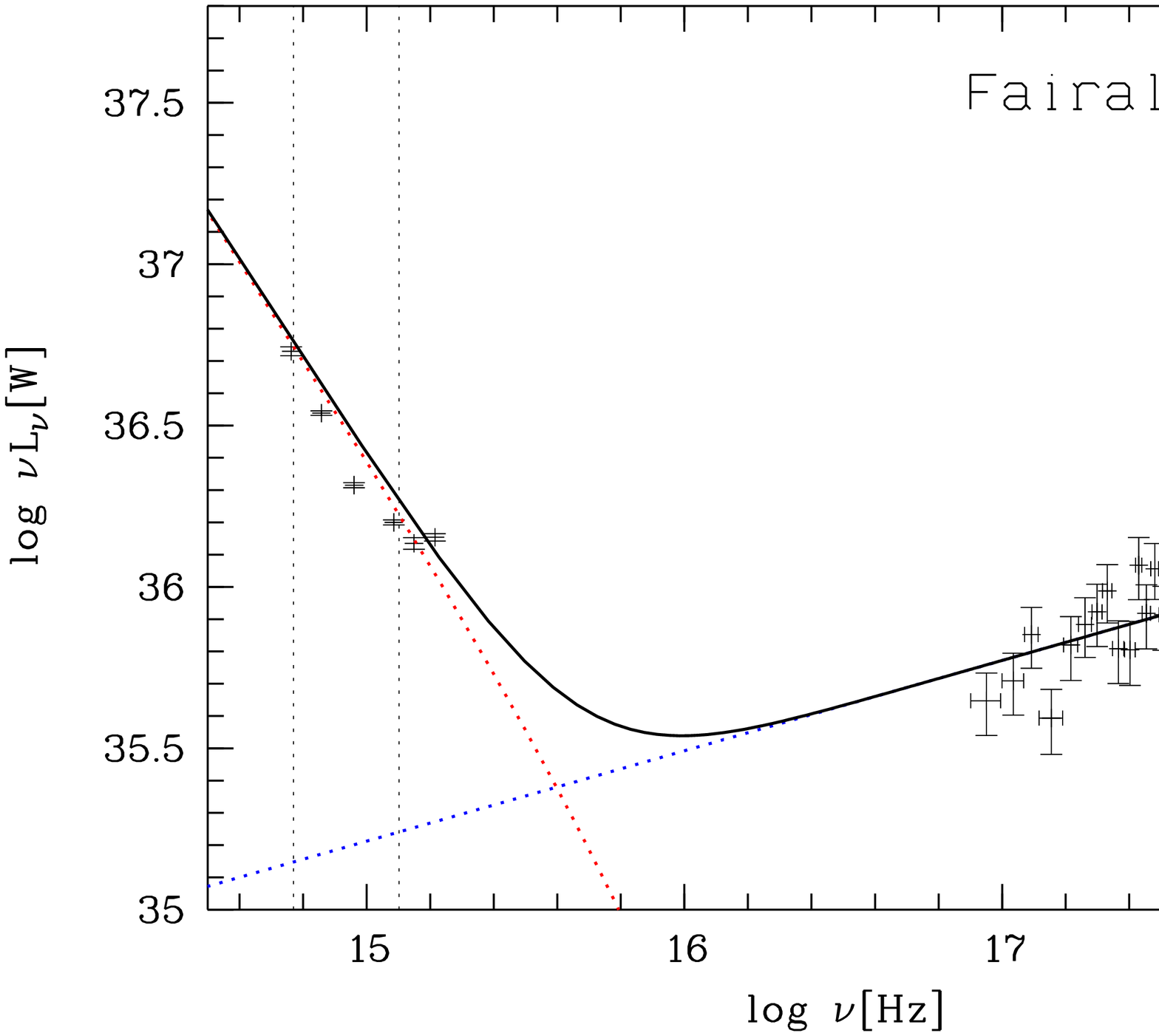}{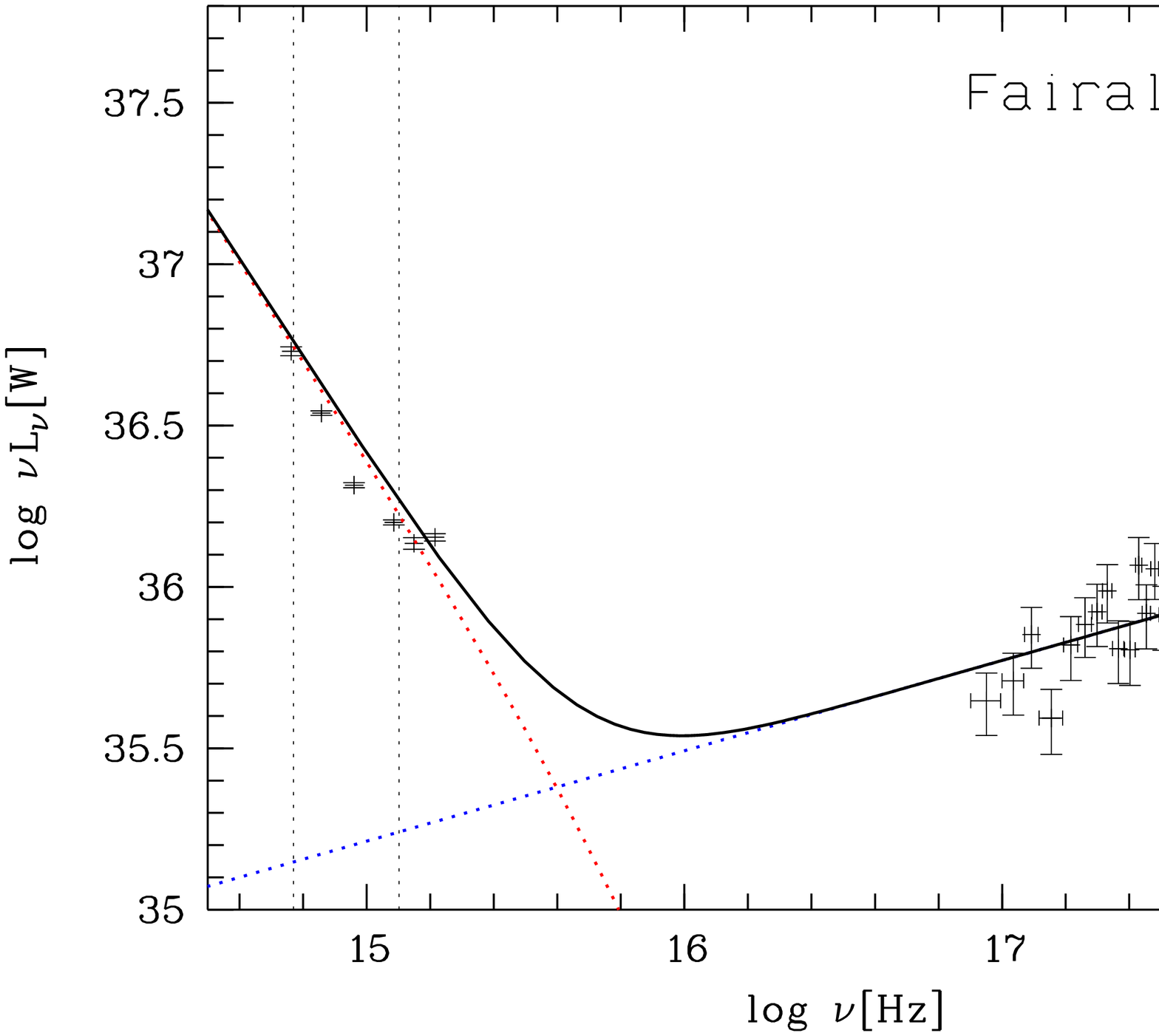}{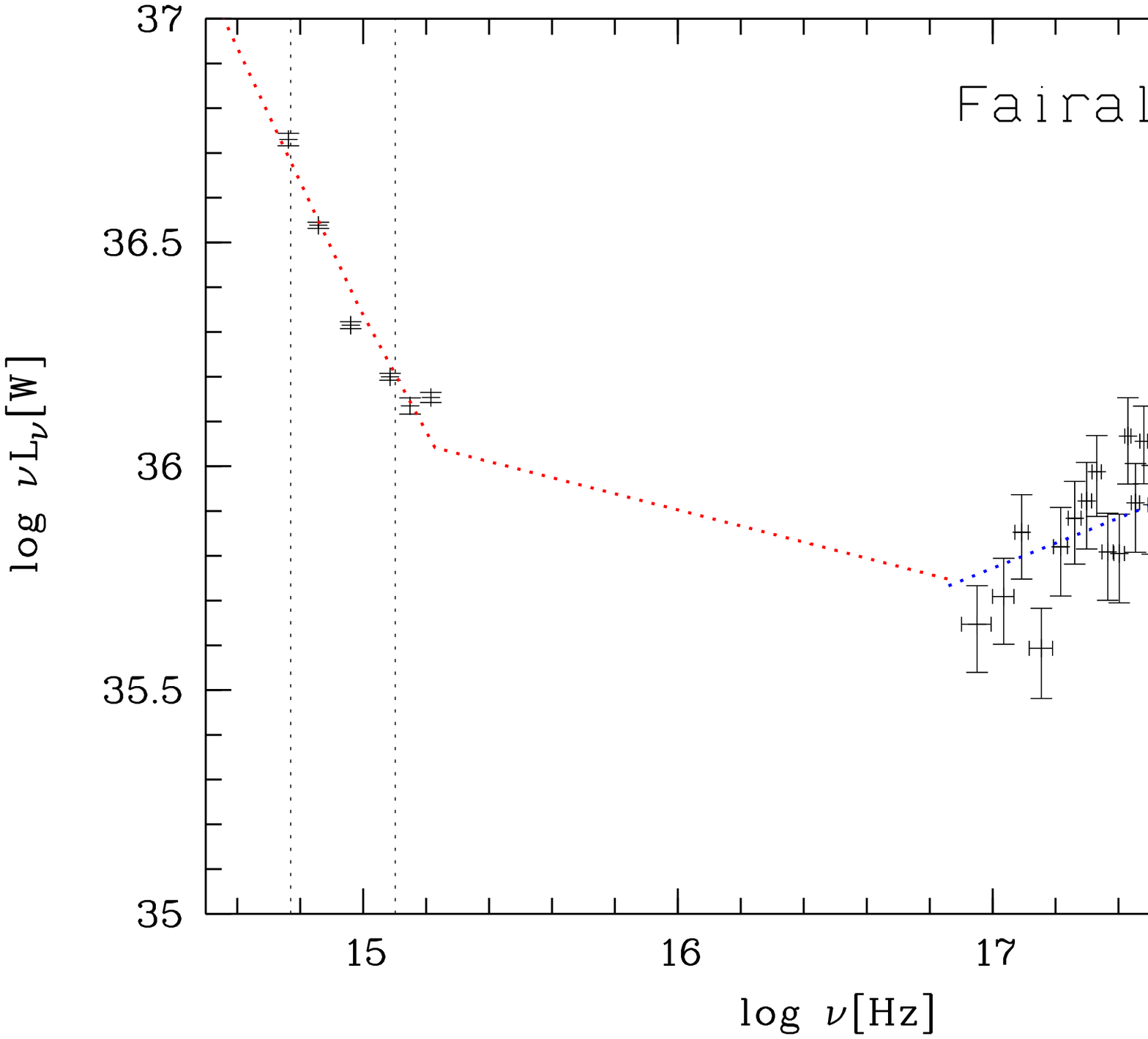}

\end{figure*}

\begin{figure*}
\epsscale{0.60}
\plotthree{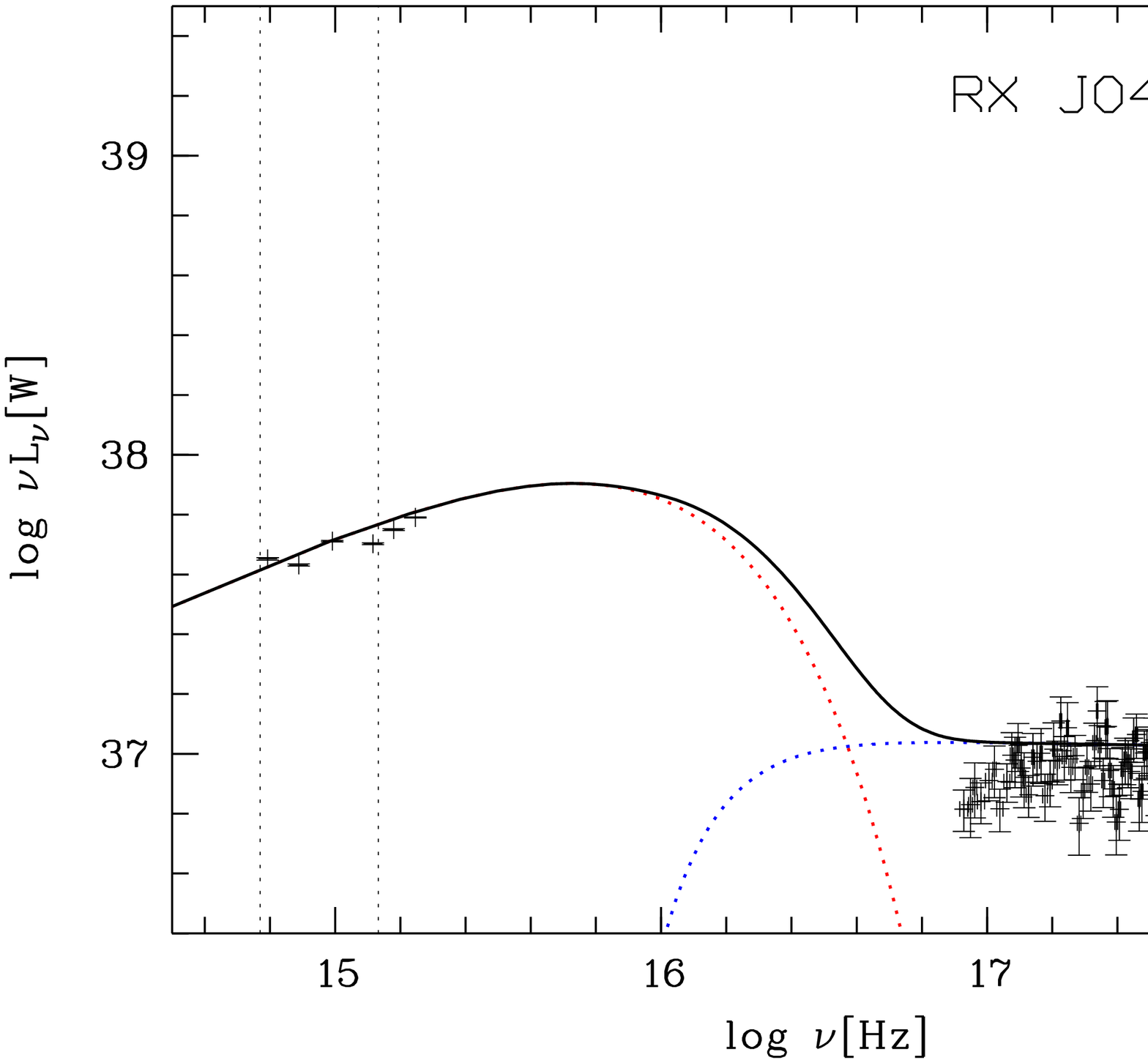}{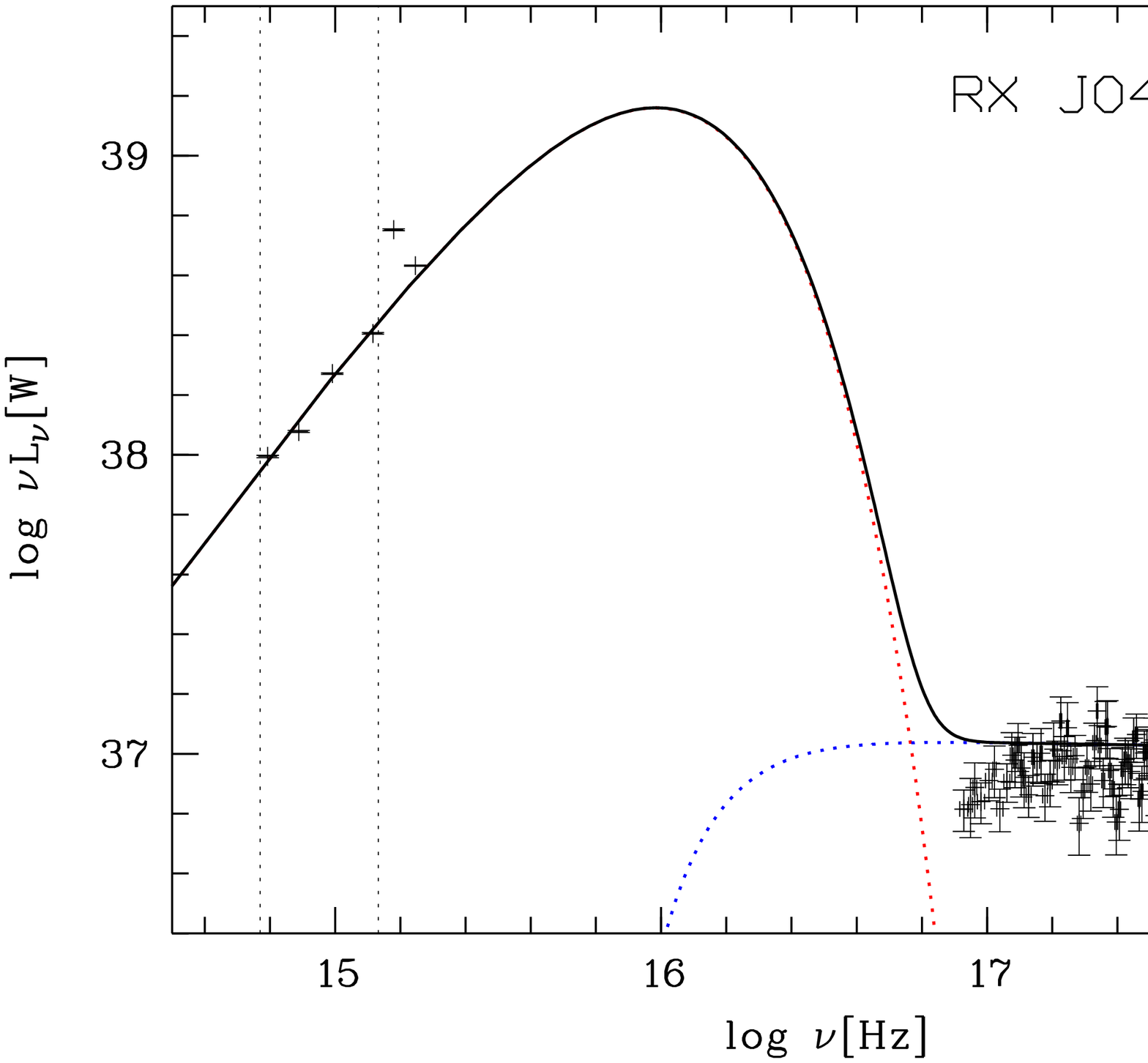}{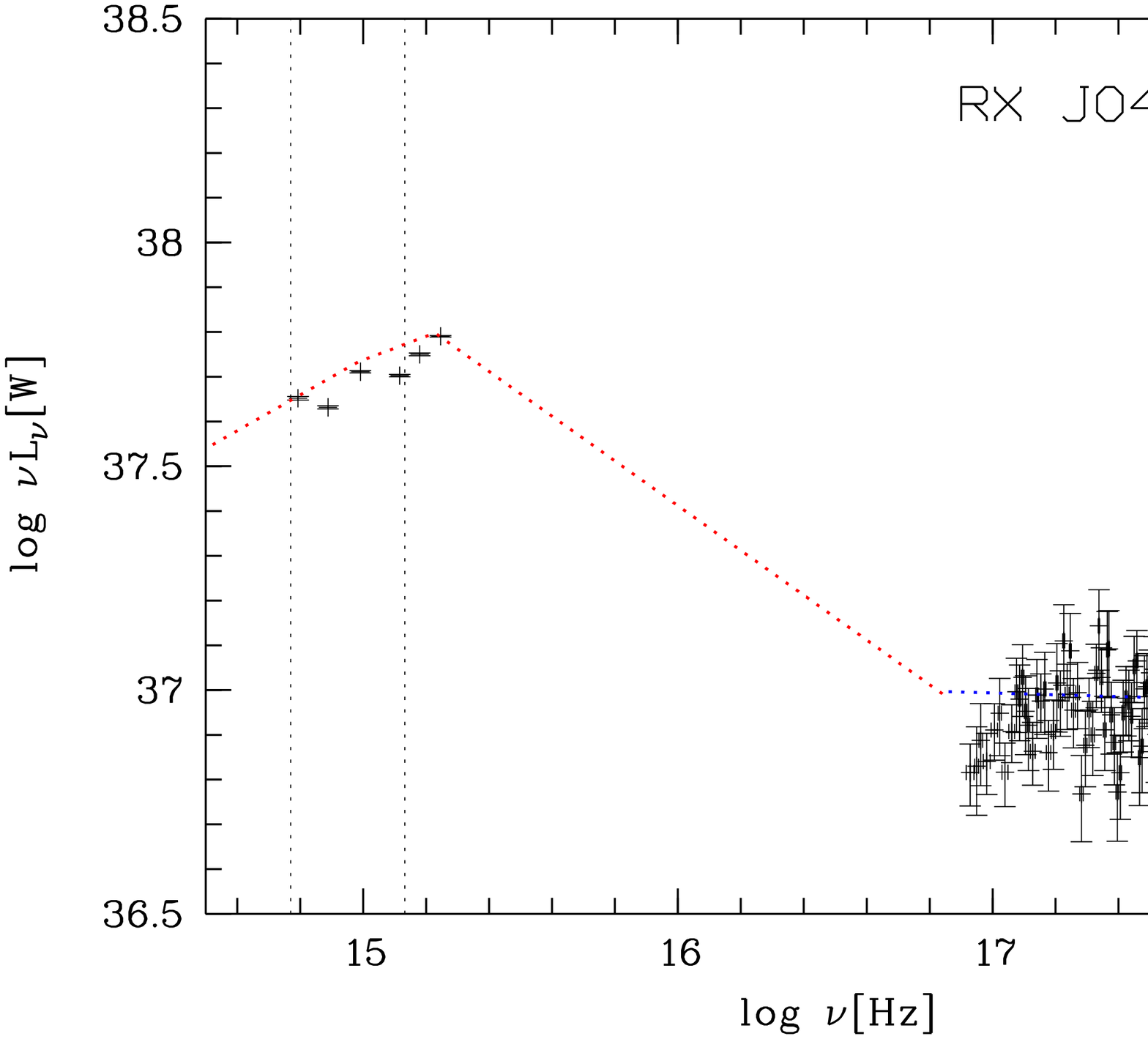}

\plotthree{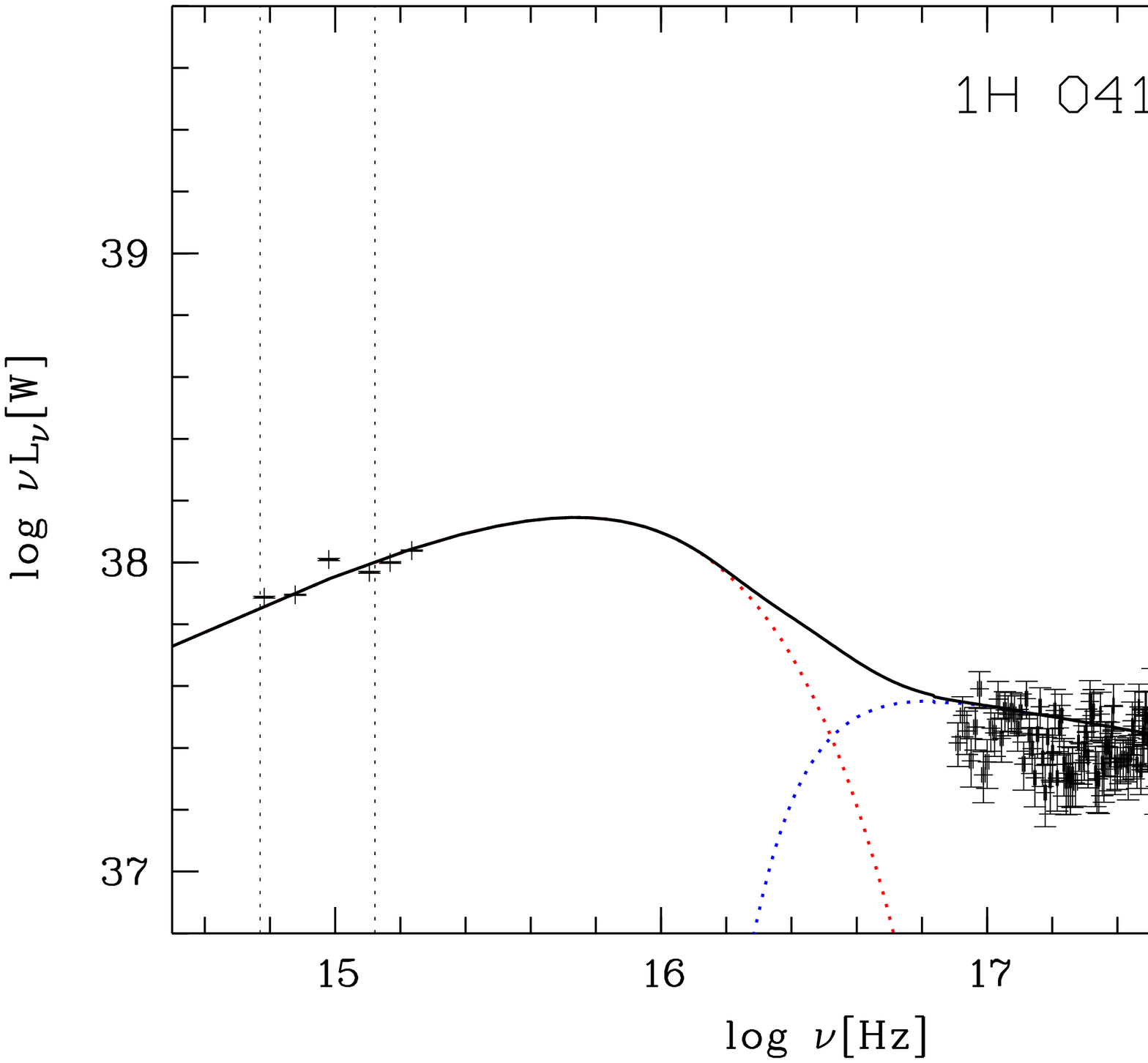}{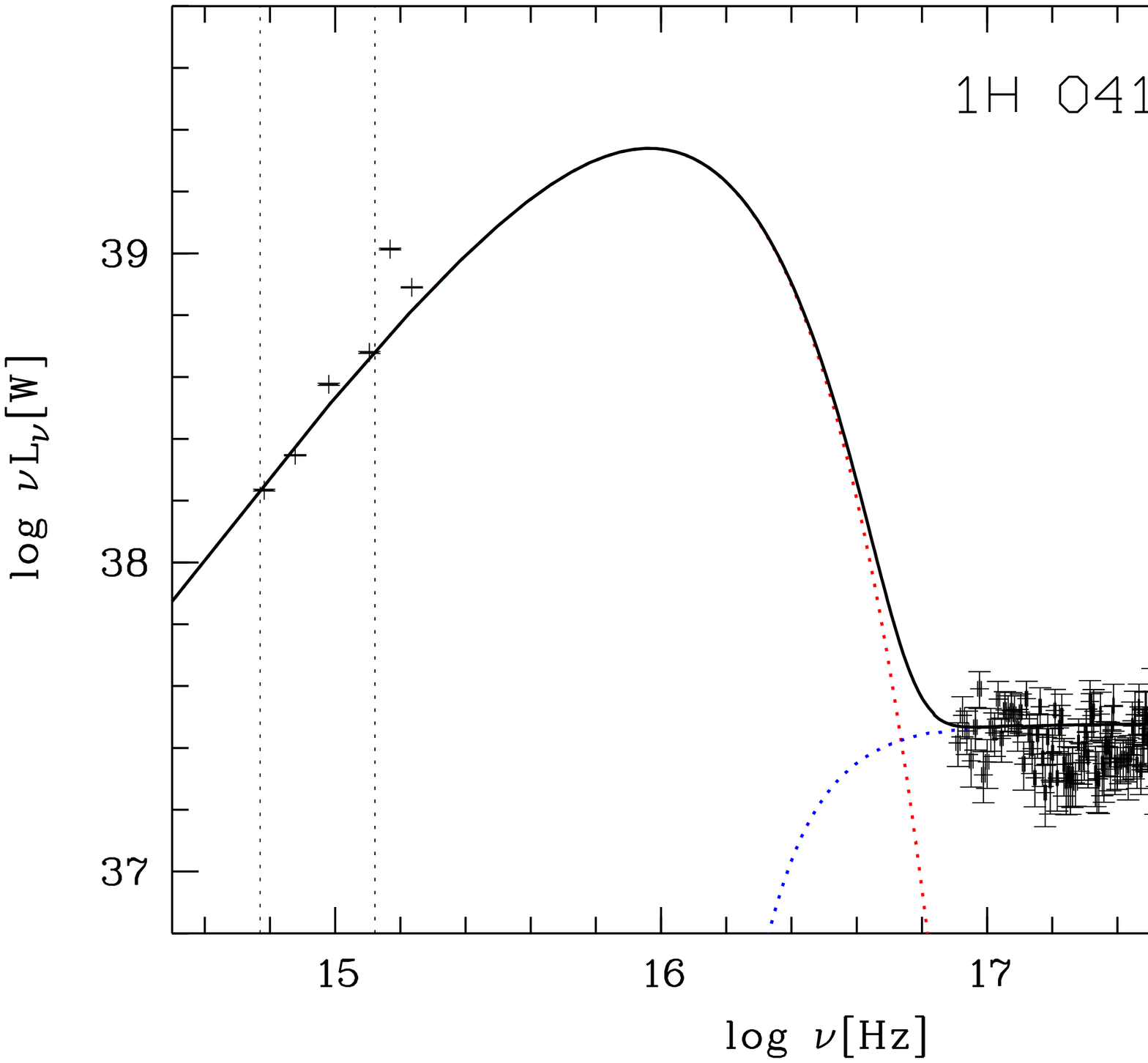}{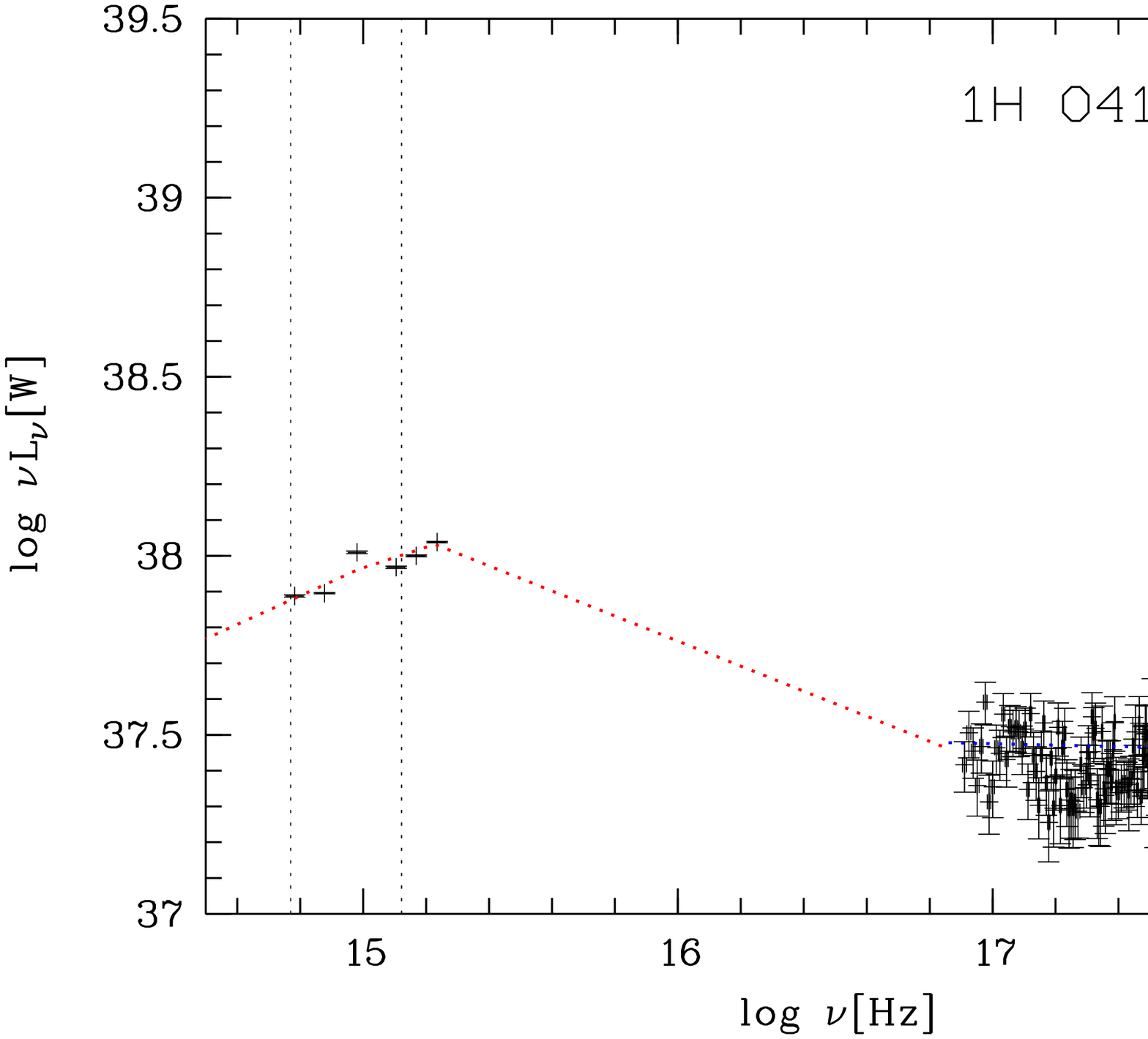}

\plotthree{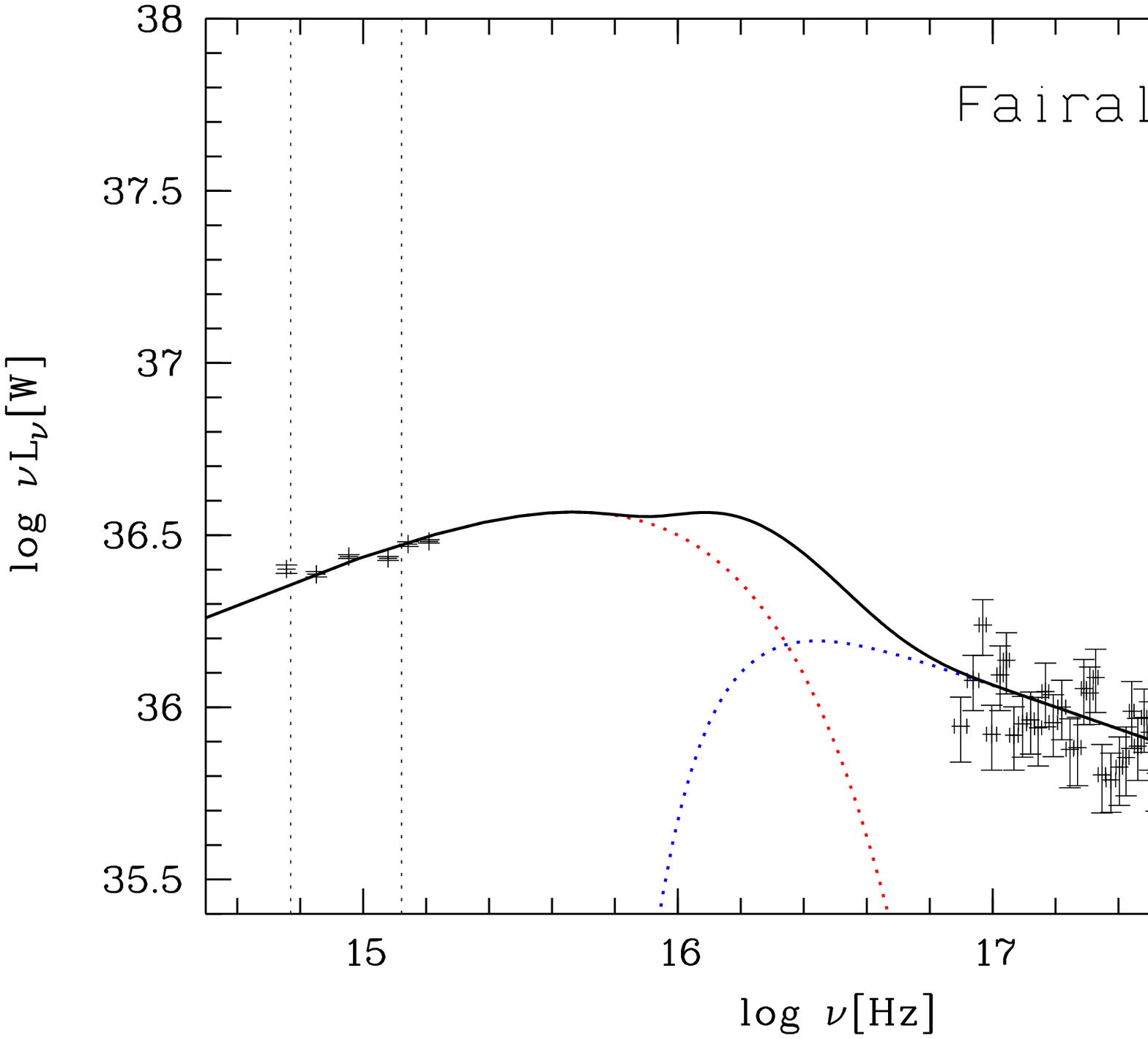}{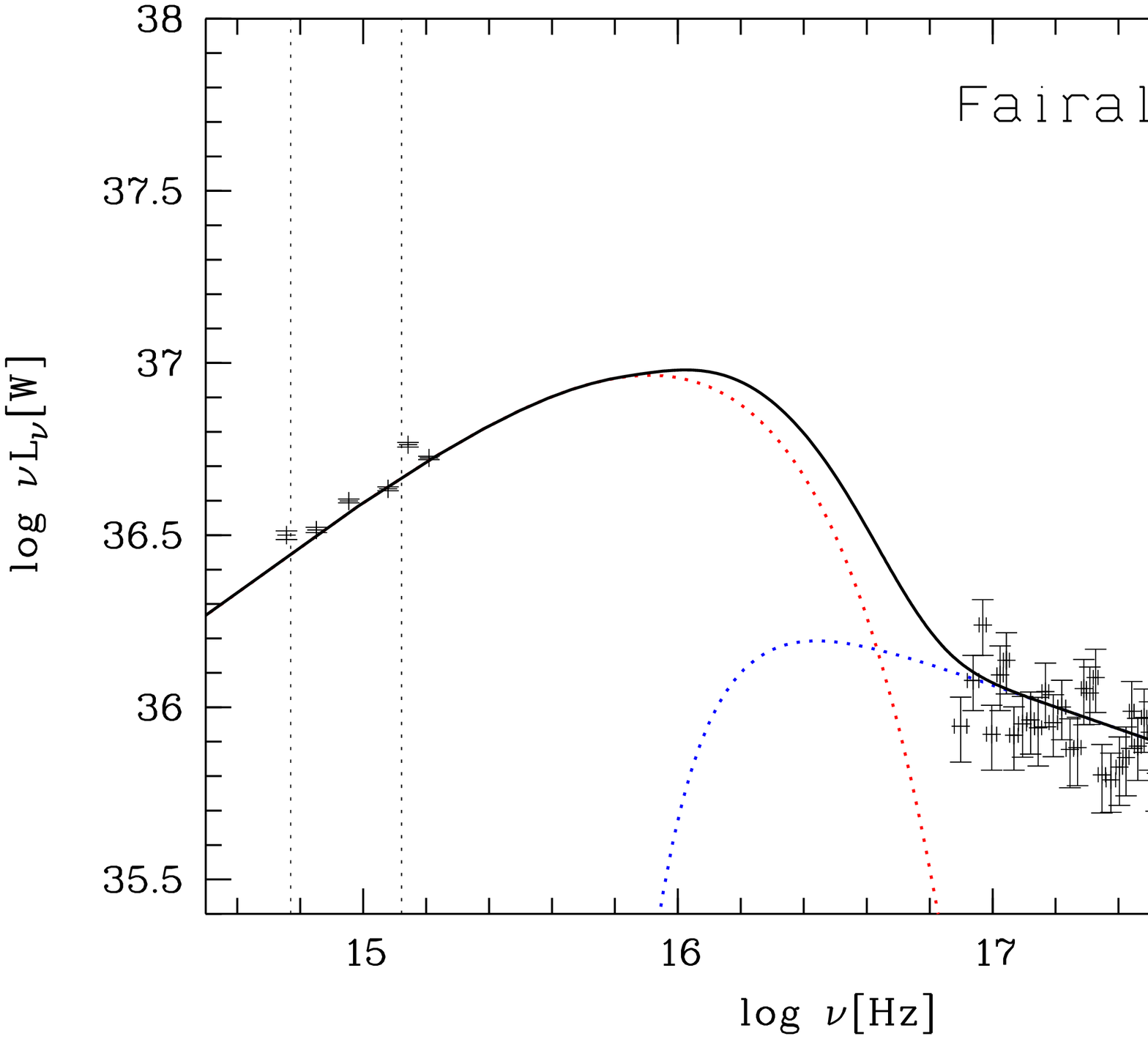}{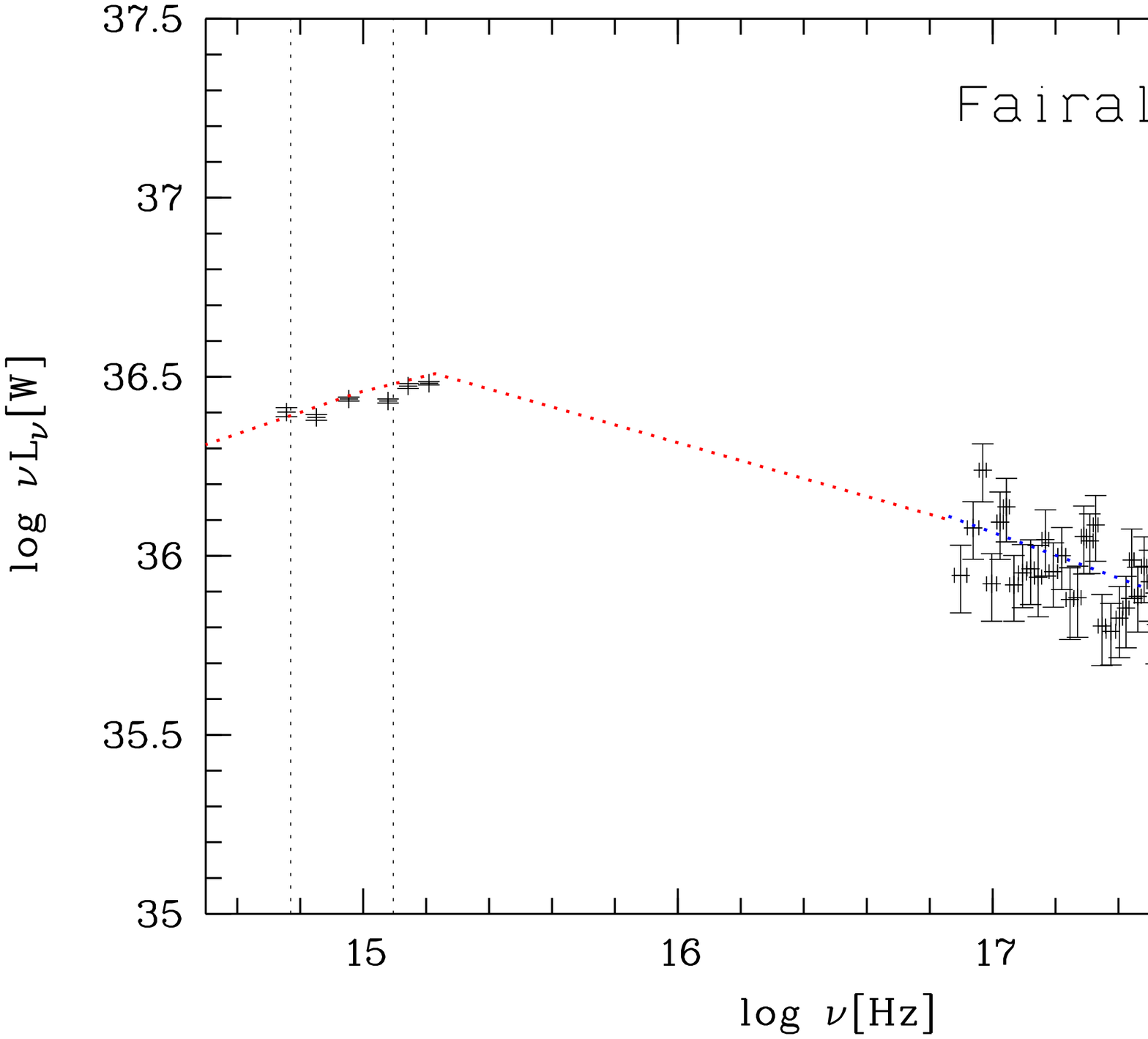}

\plotthree{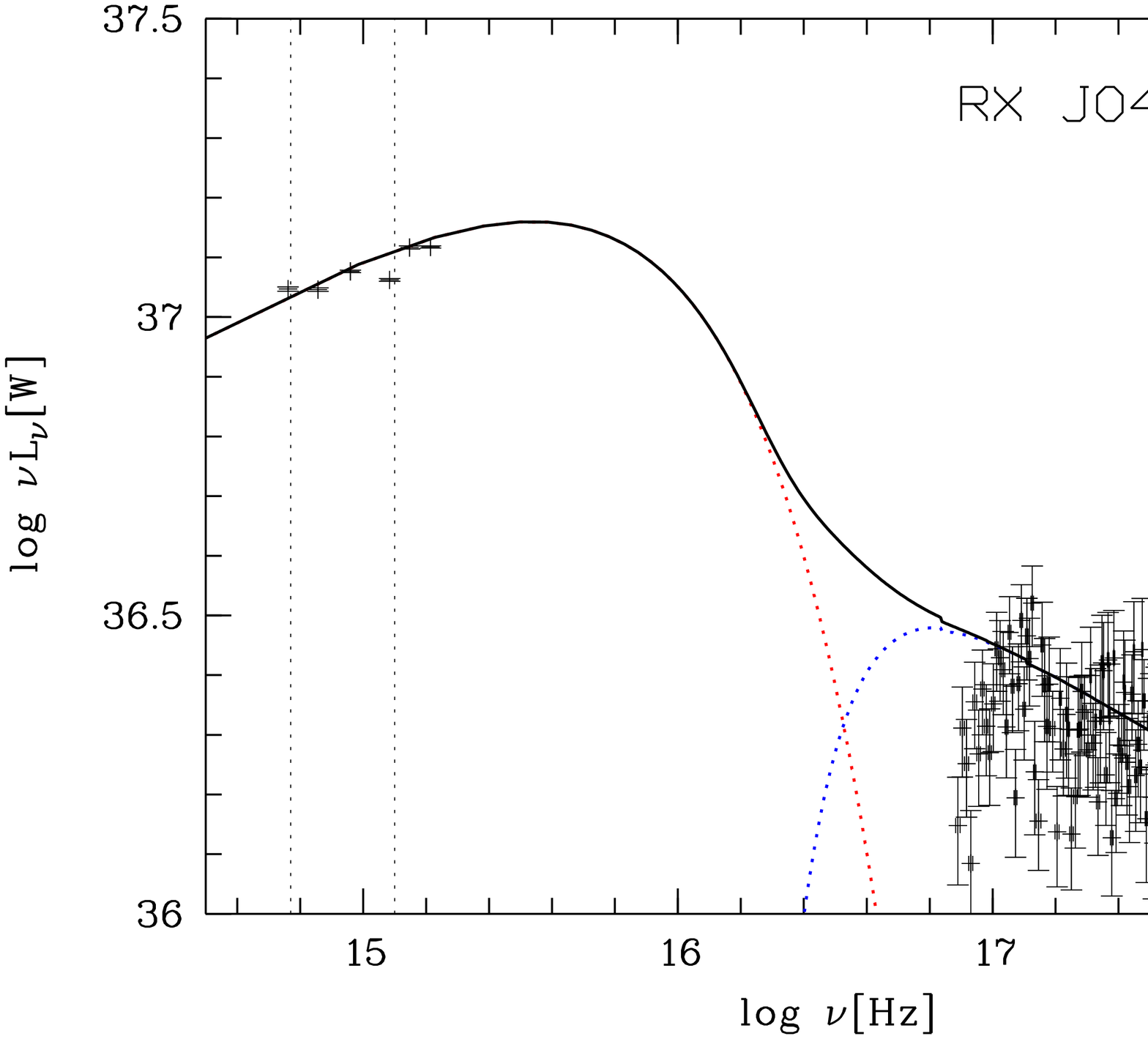}{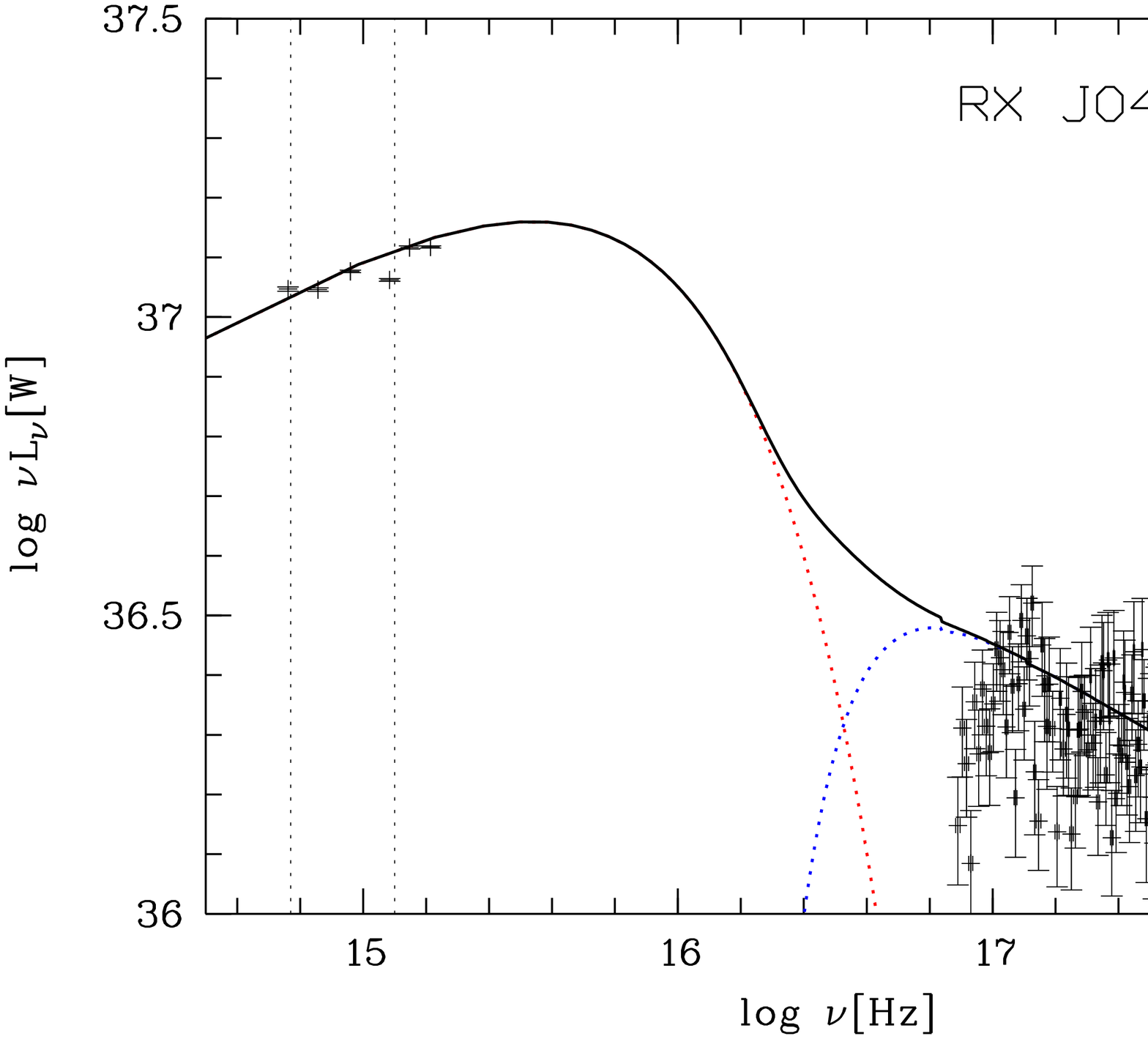}{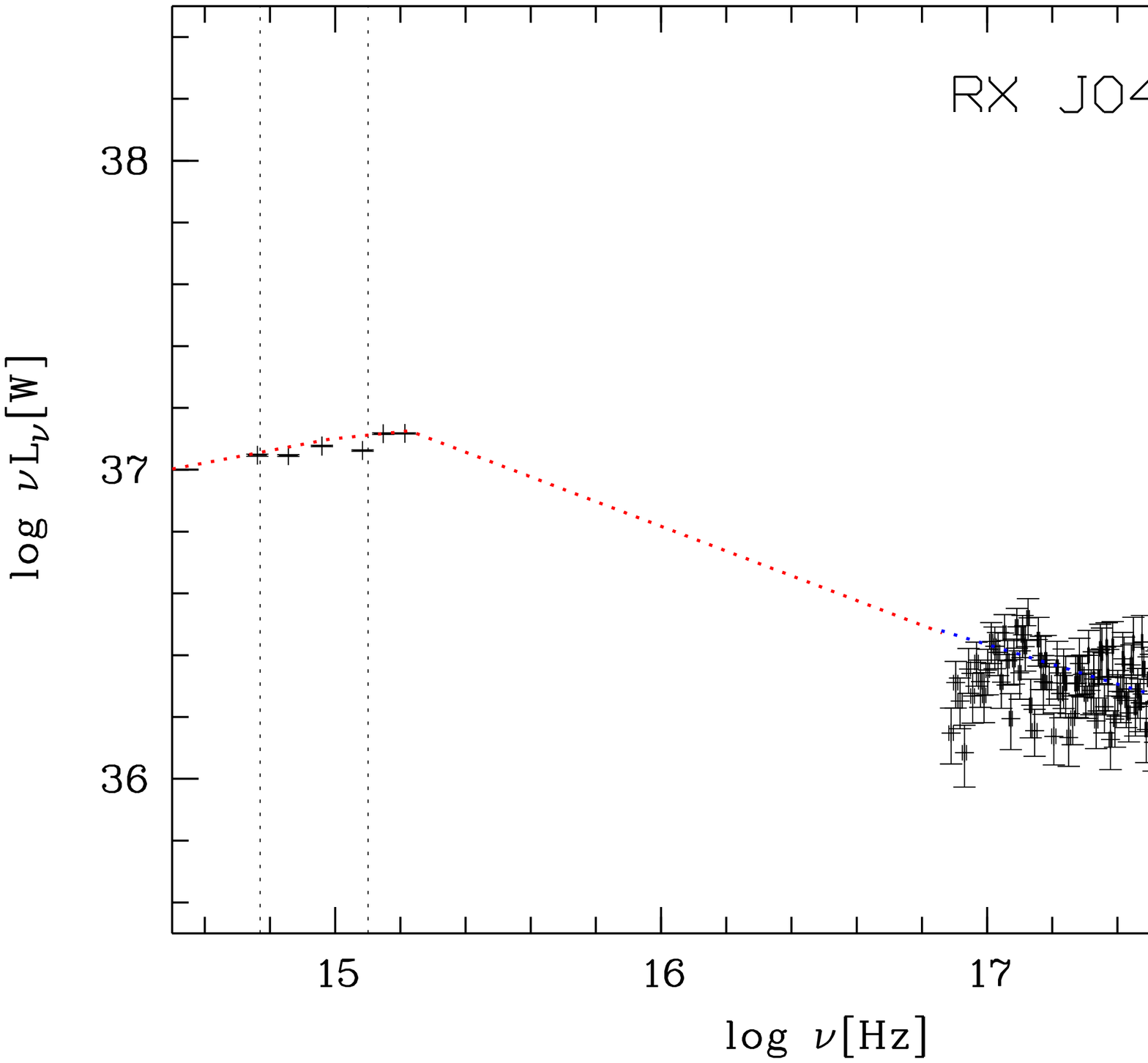}

\plotthree{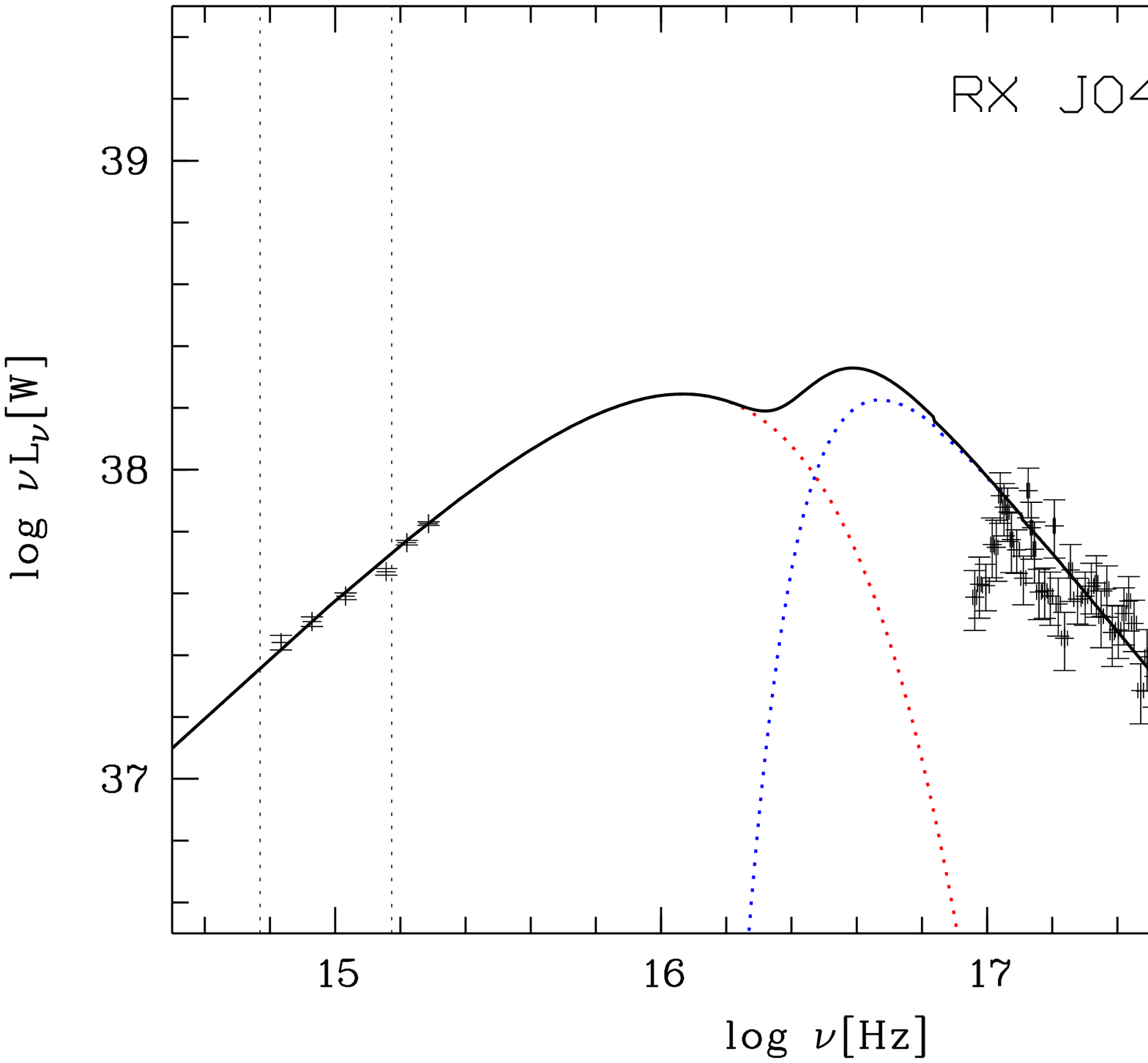}{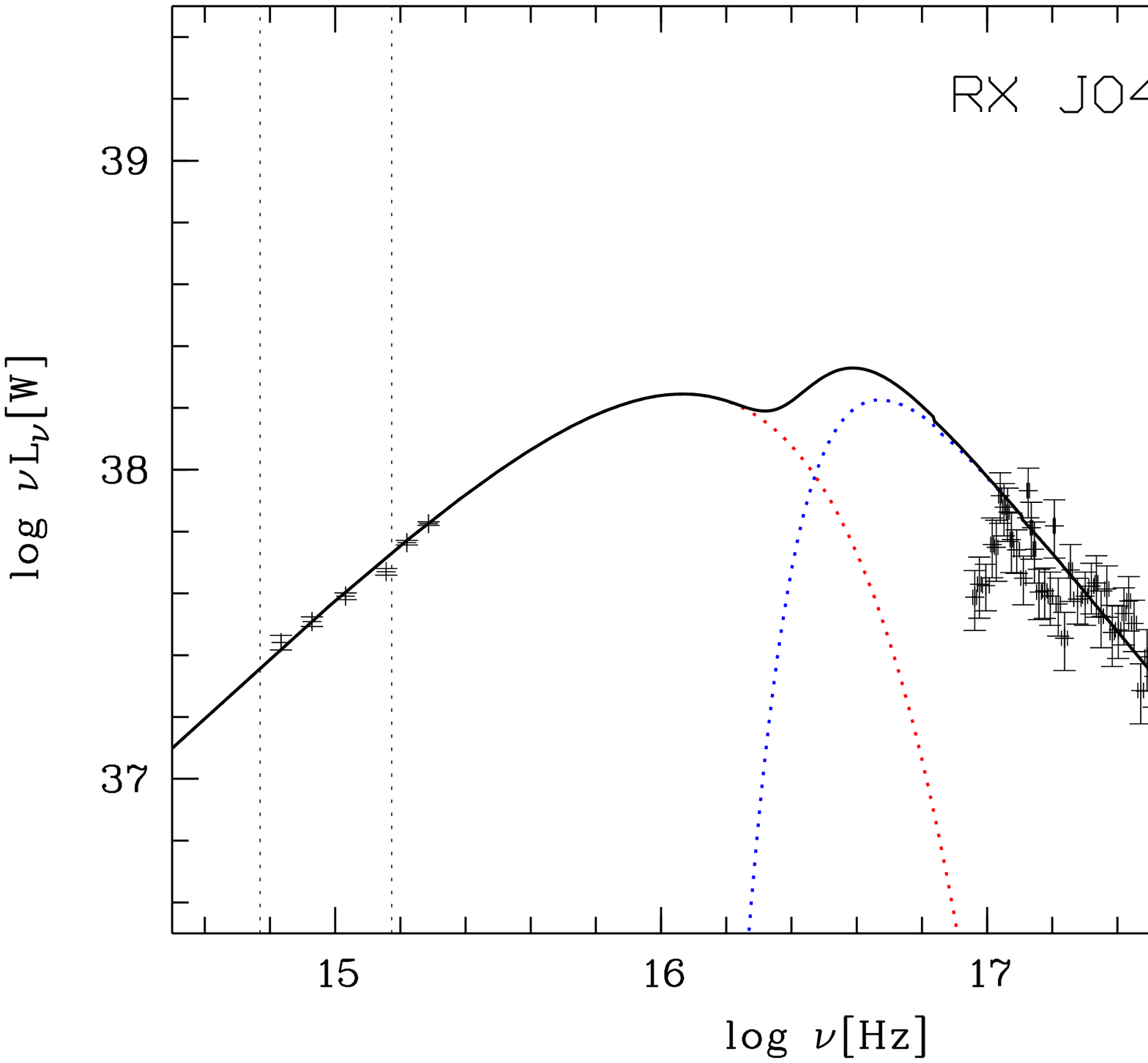}{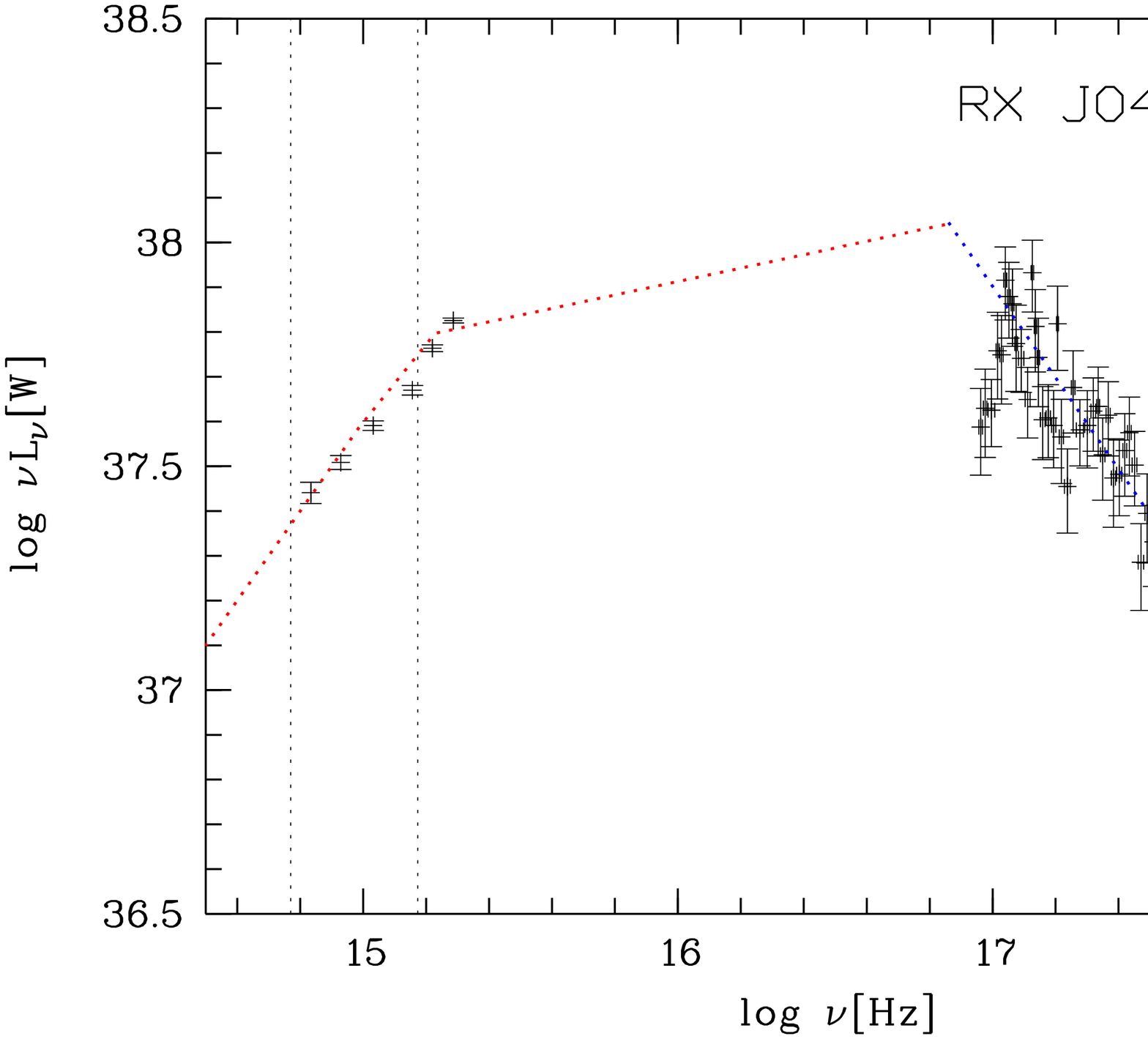}

\end{figure*}

\begin{figure*}
\epsscale{0.60}
\plotthree{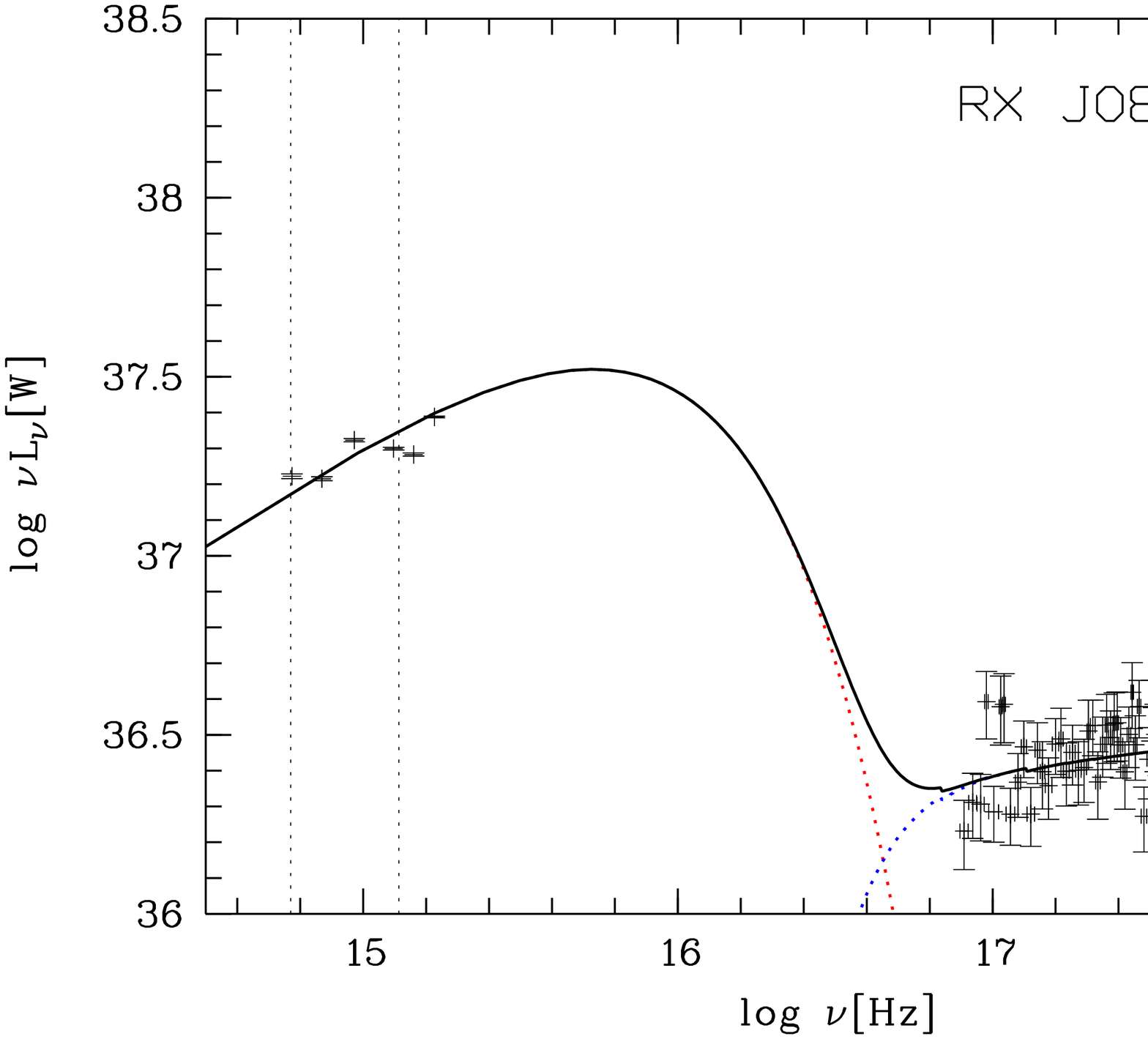}{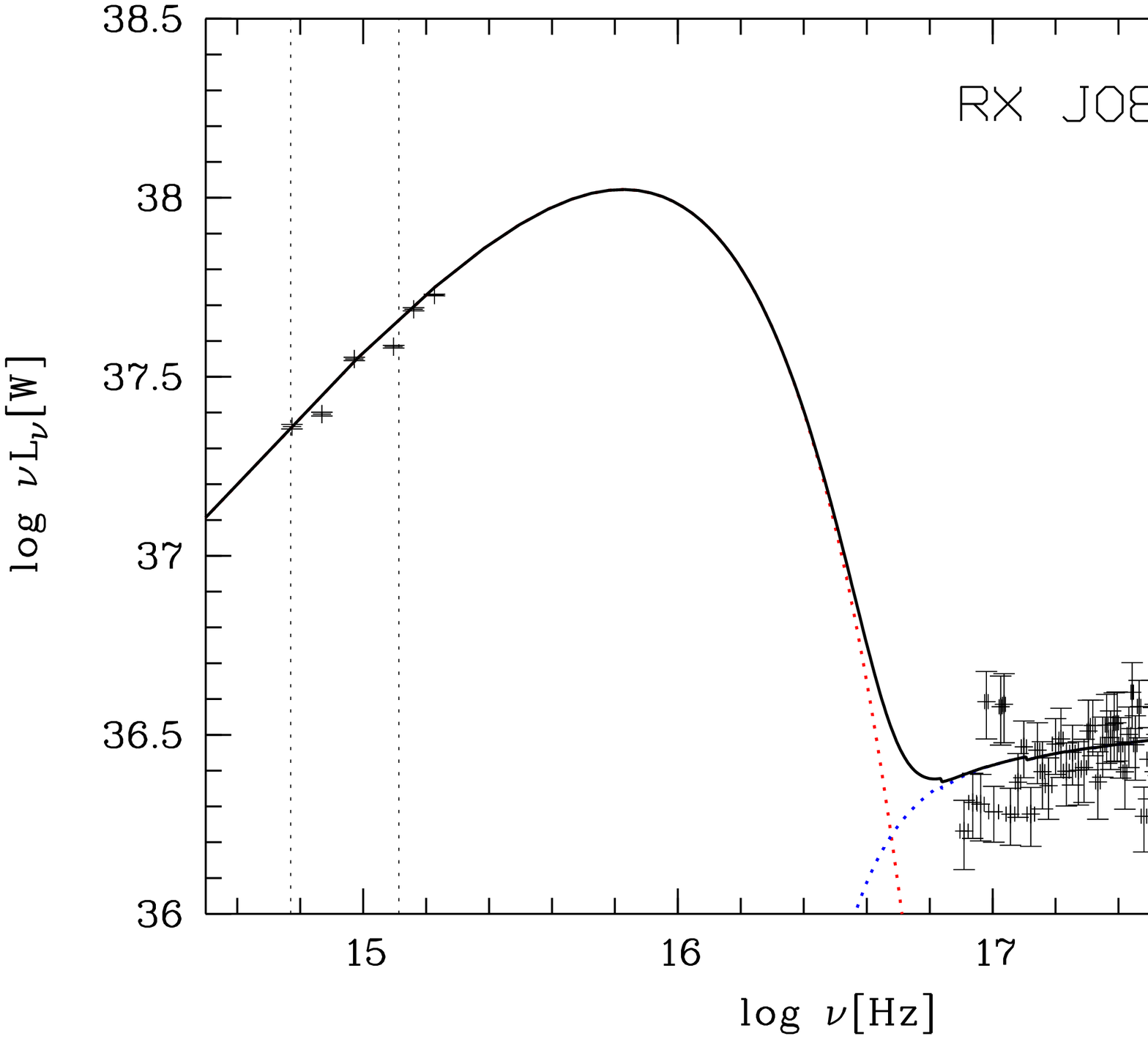}{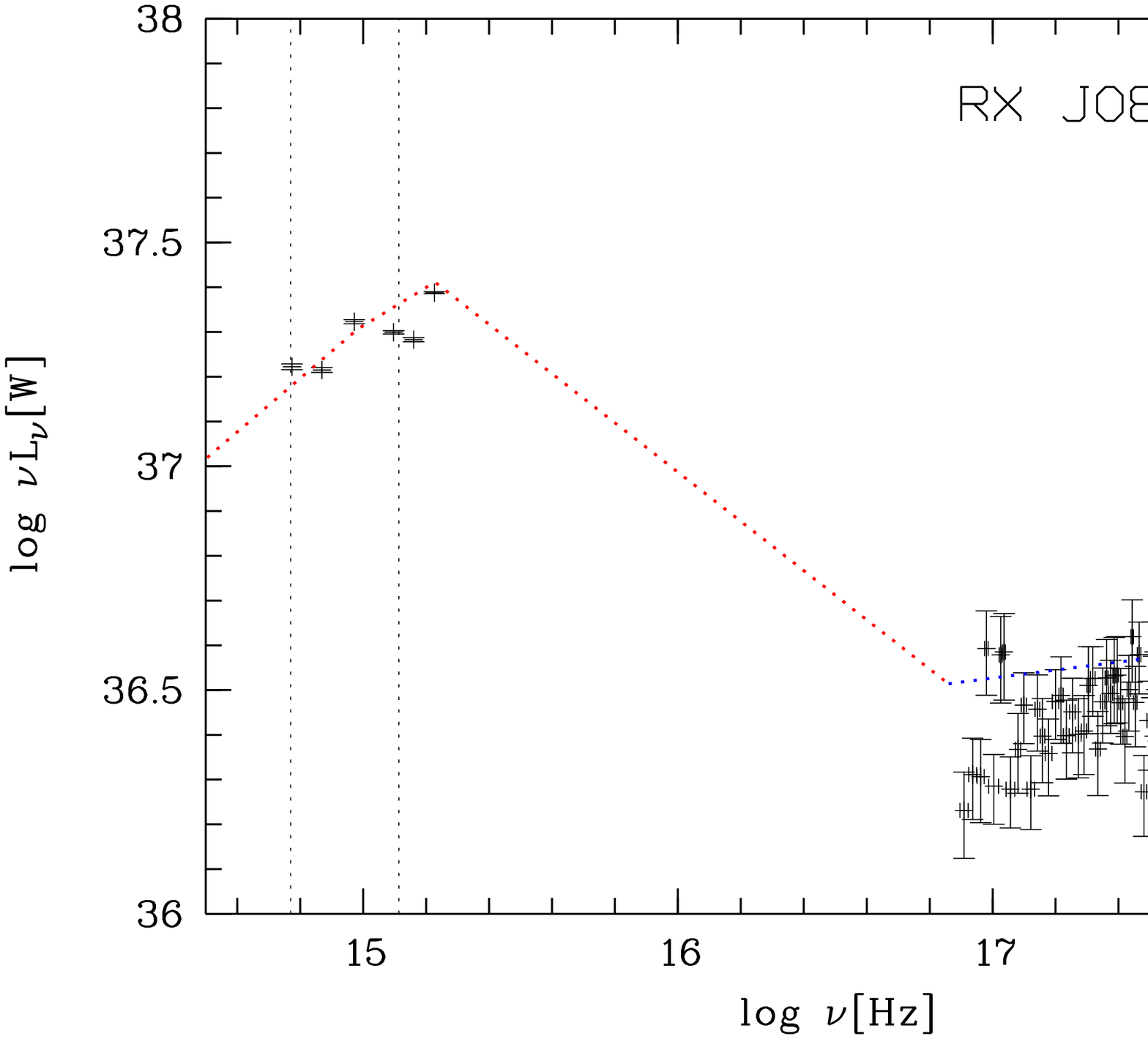}

\plotthree{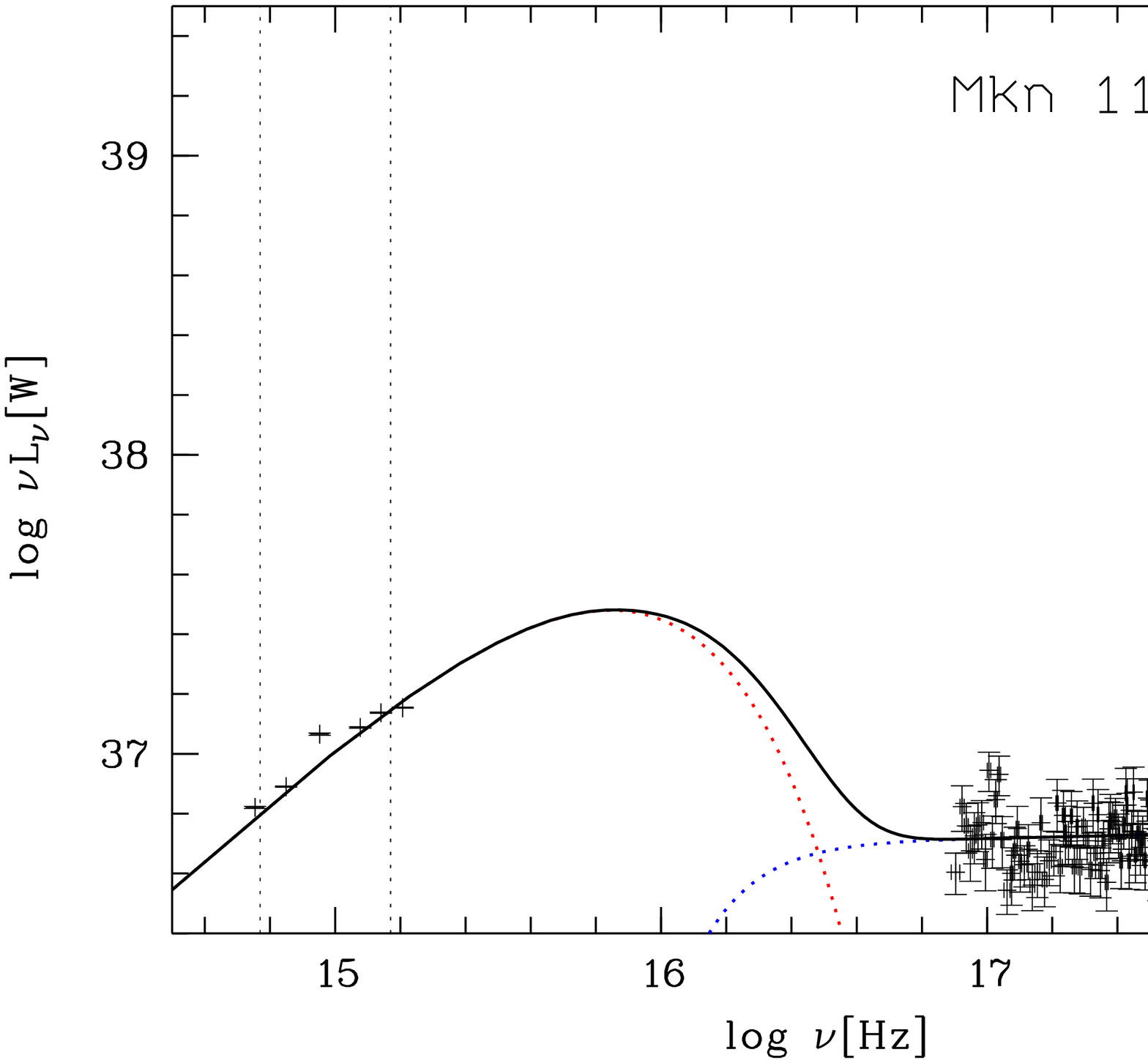}{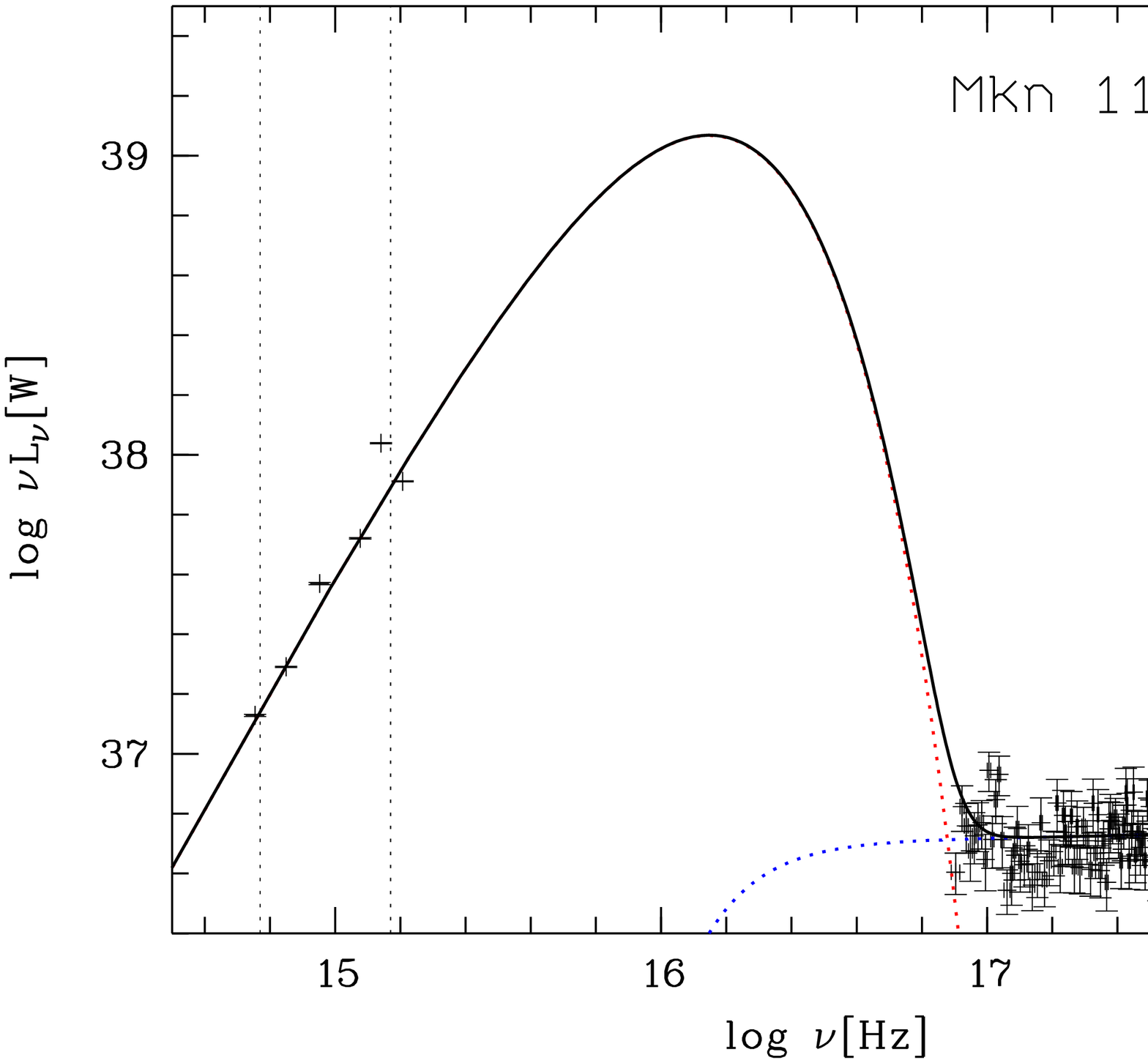}{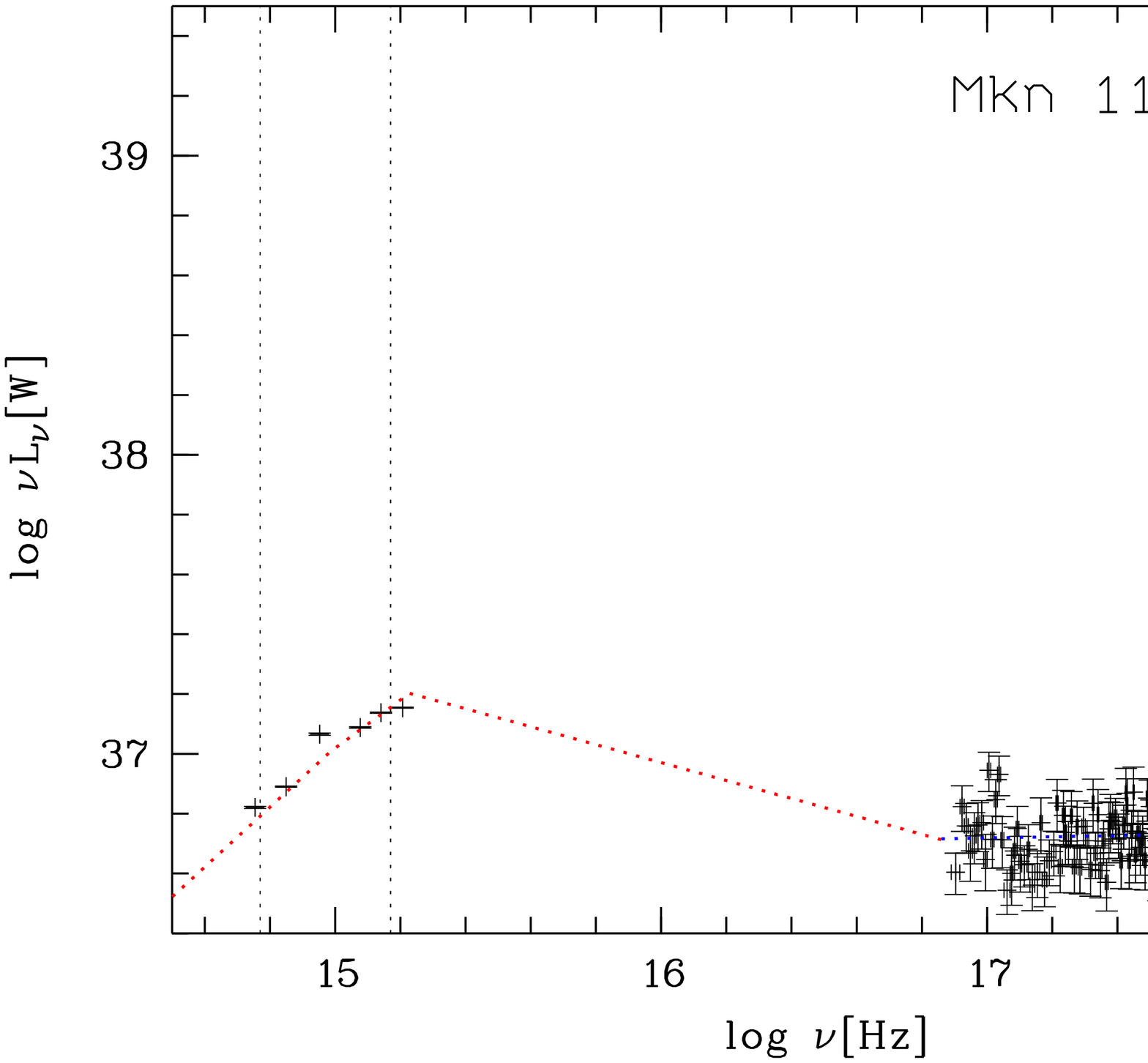}

\plotthree{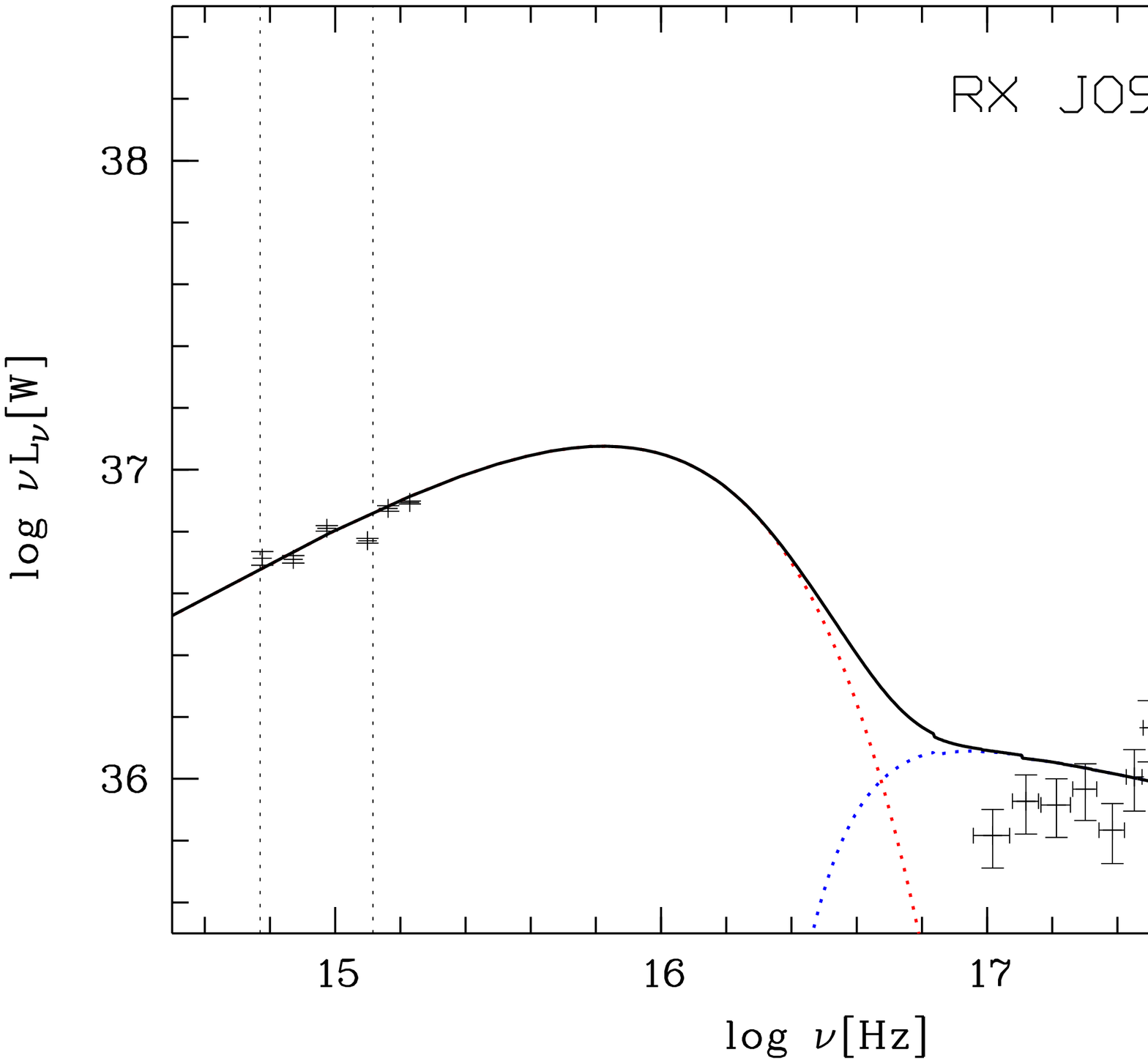}{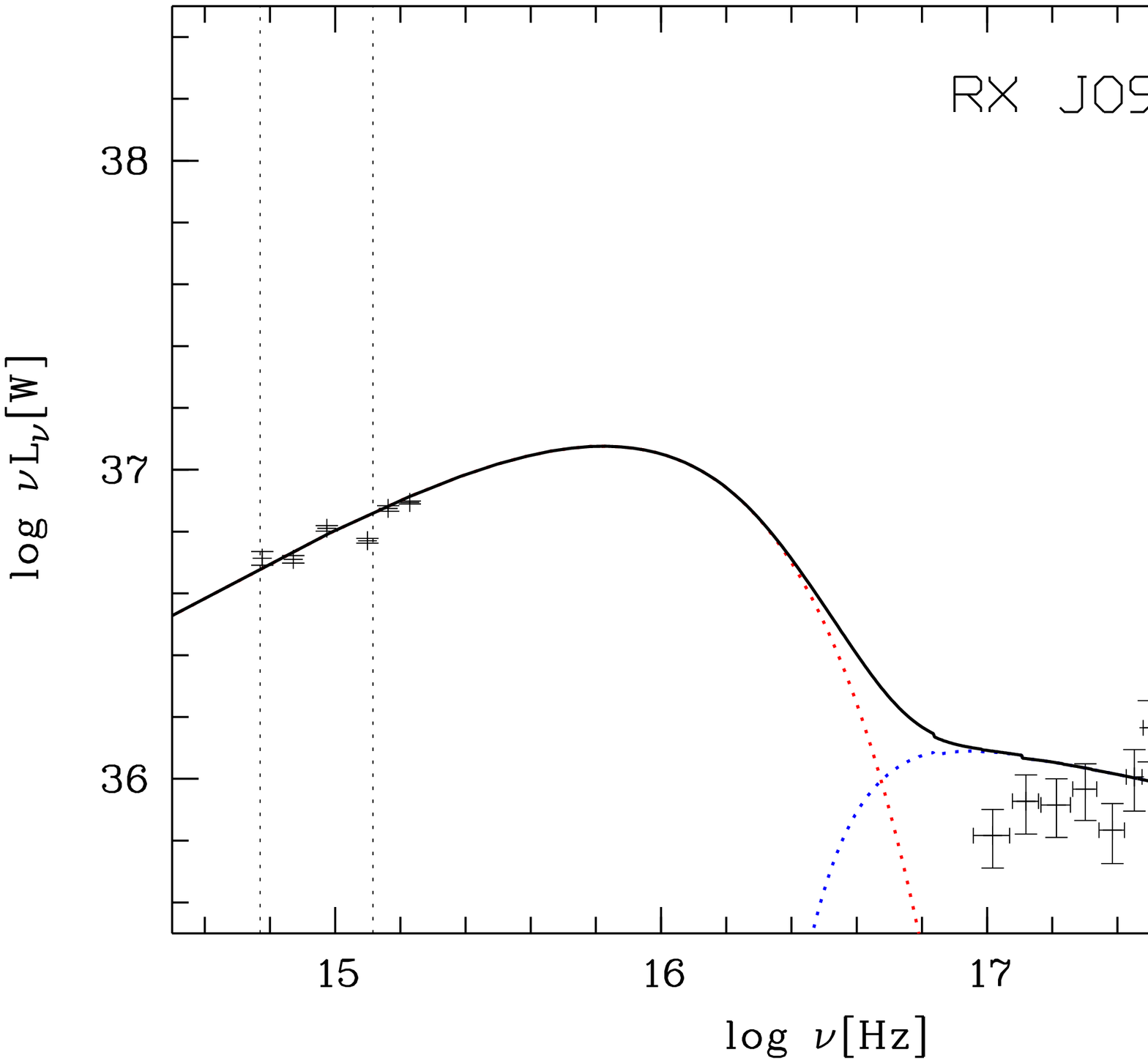}{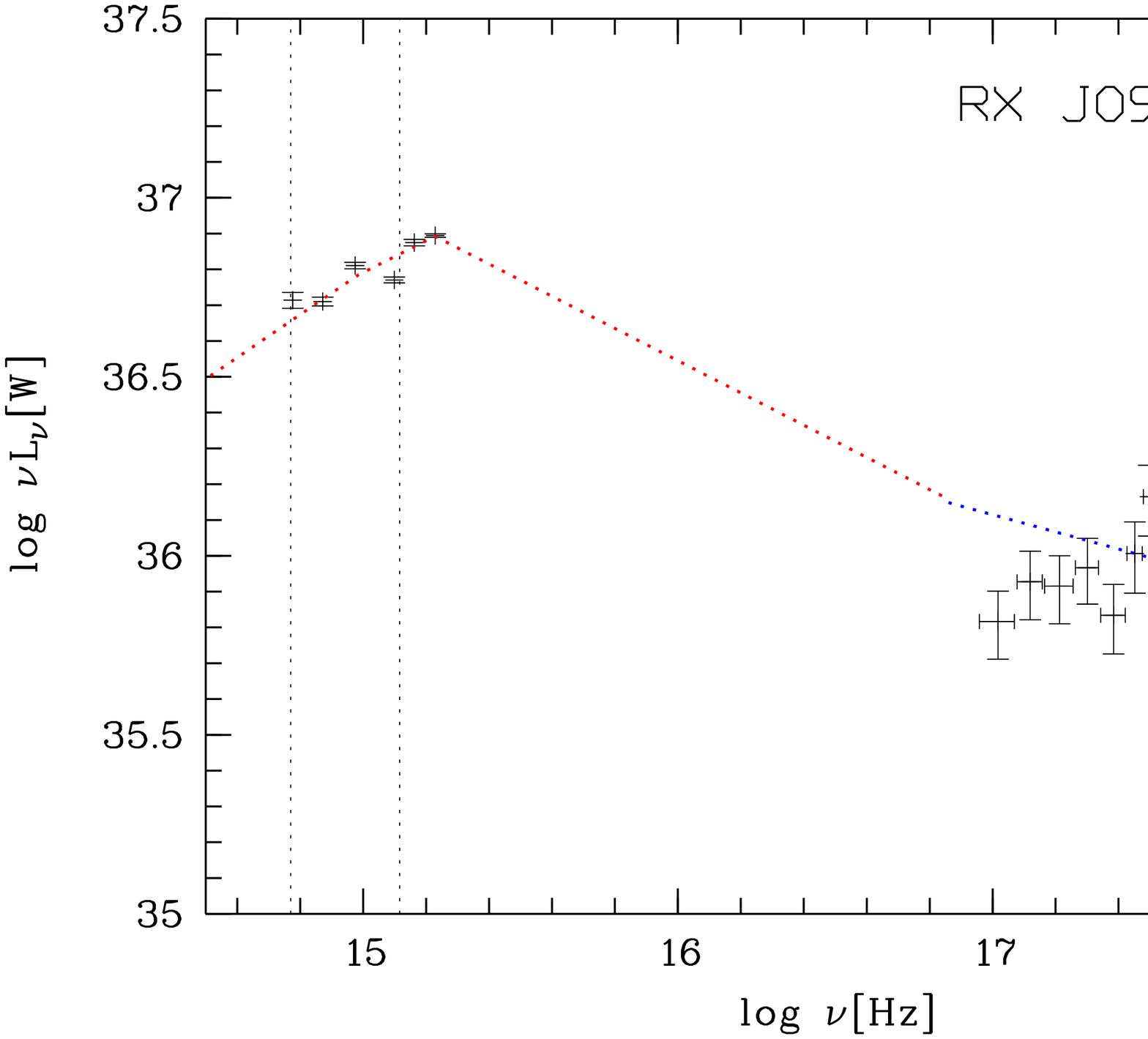}

\plotthree{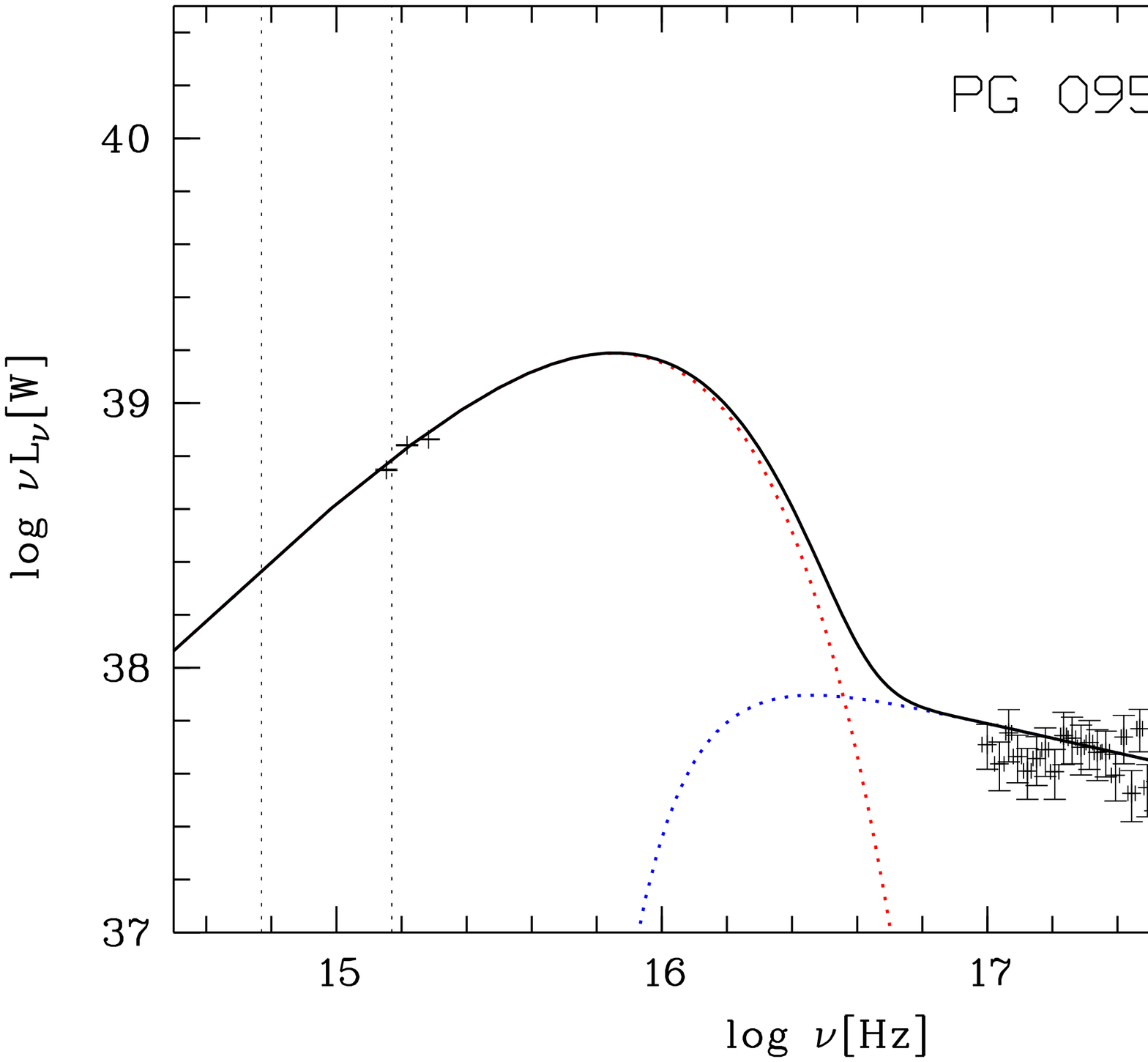}{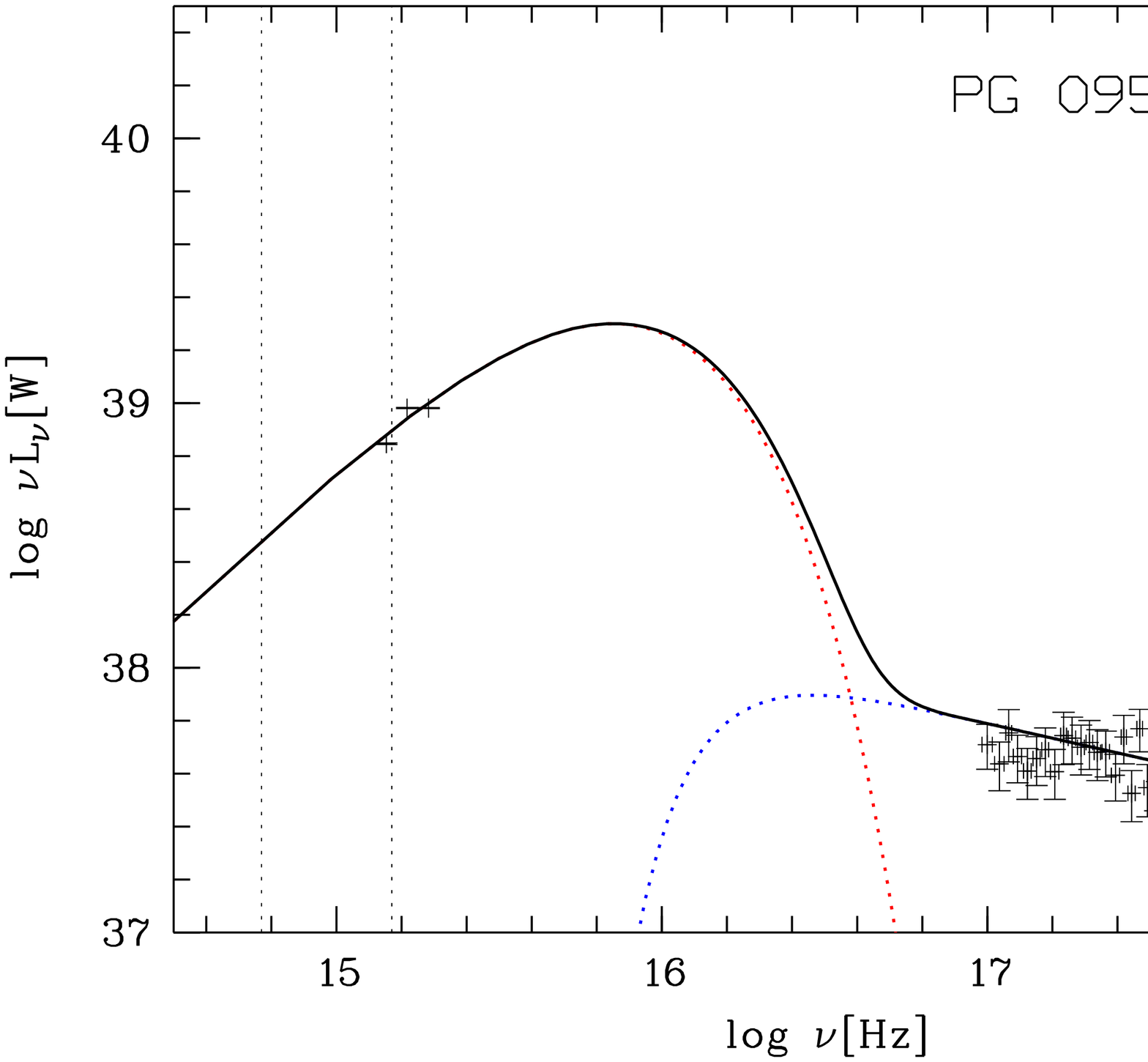}{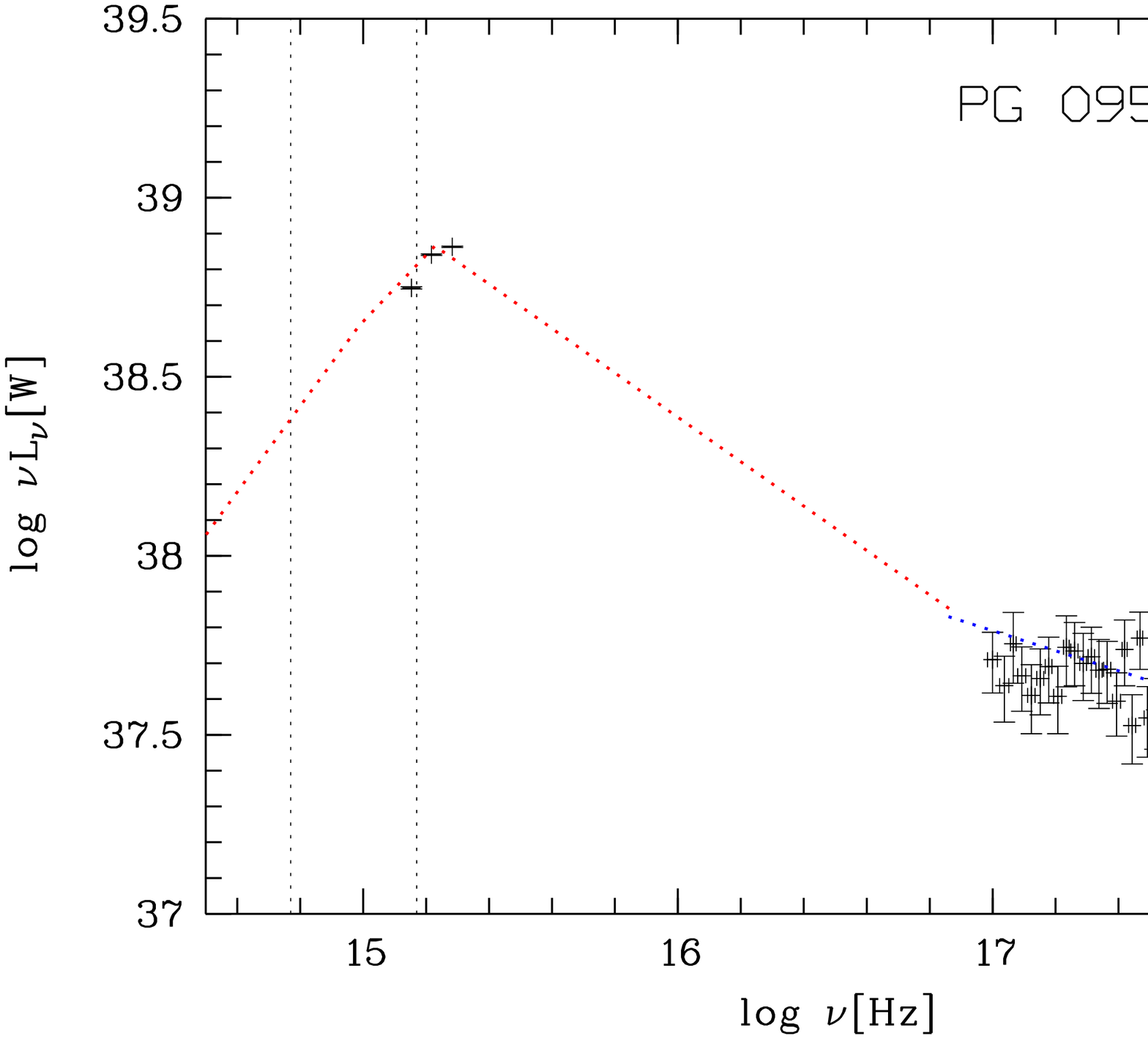}

\plotthree{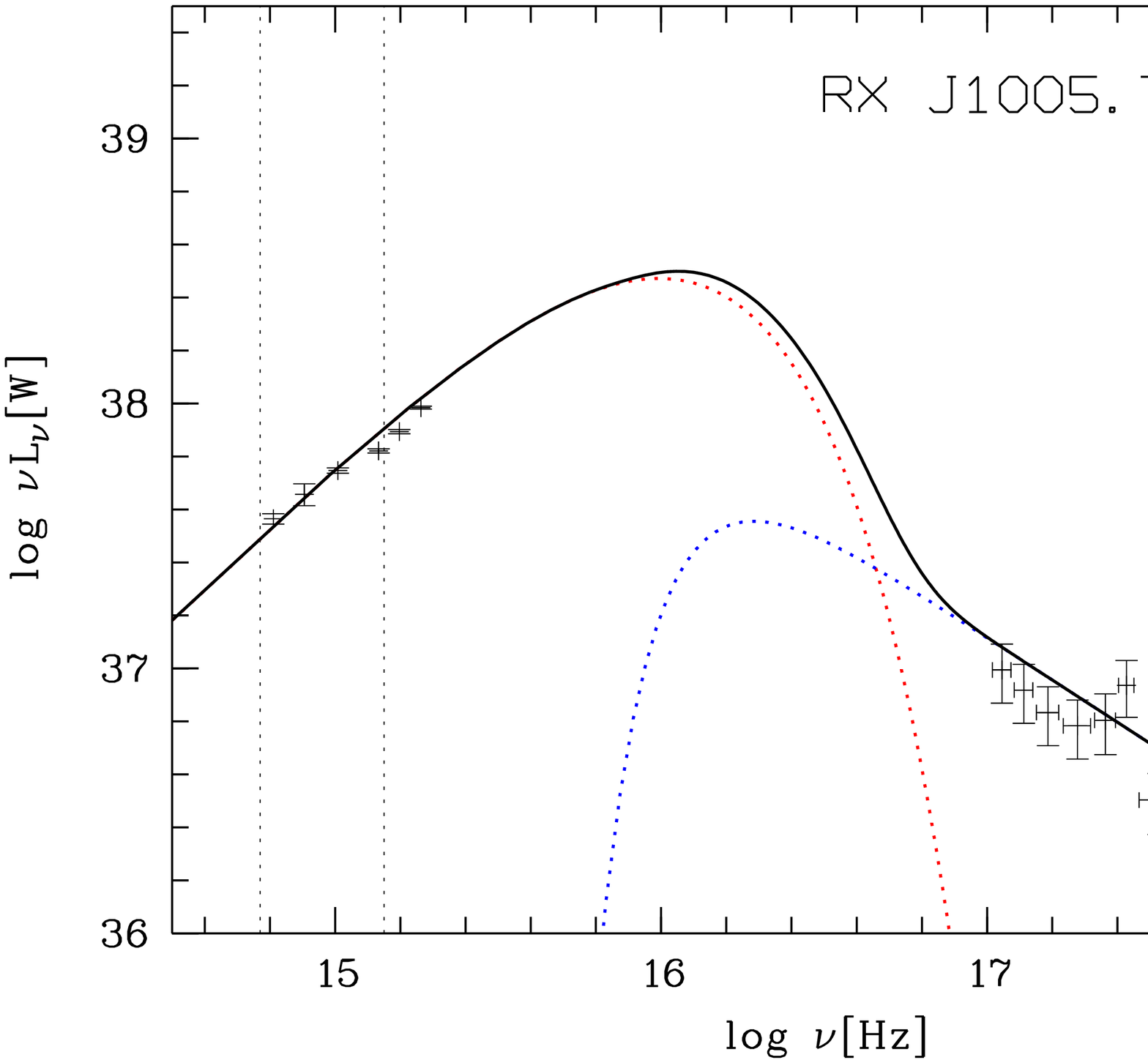}{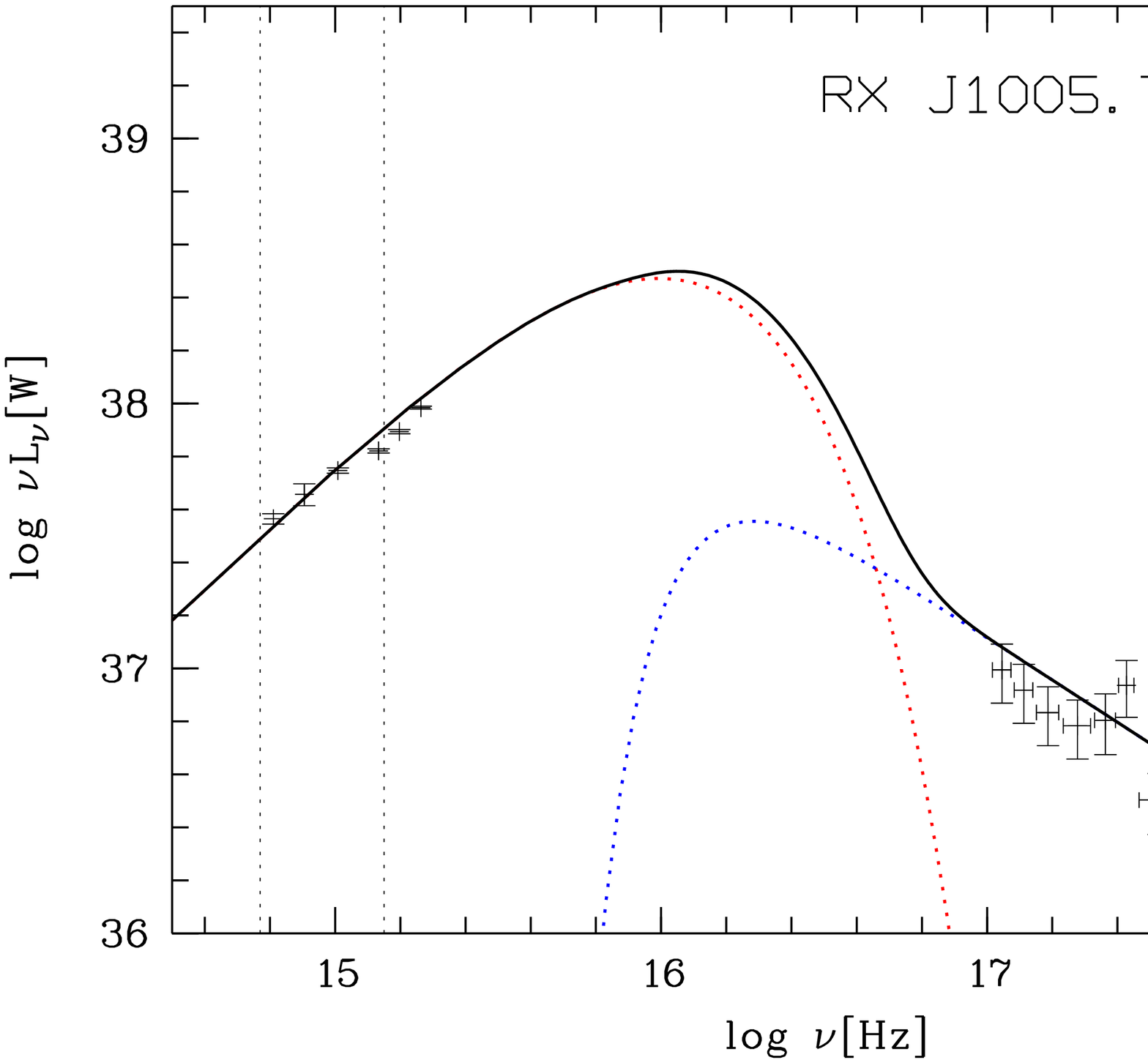}{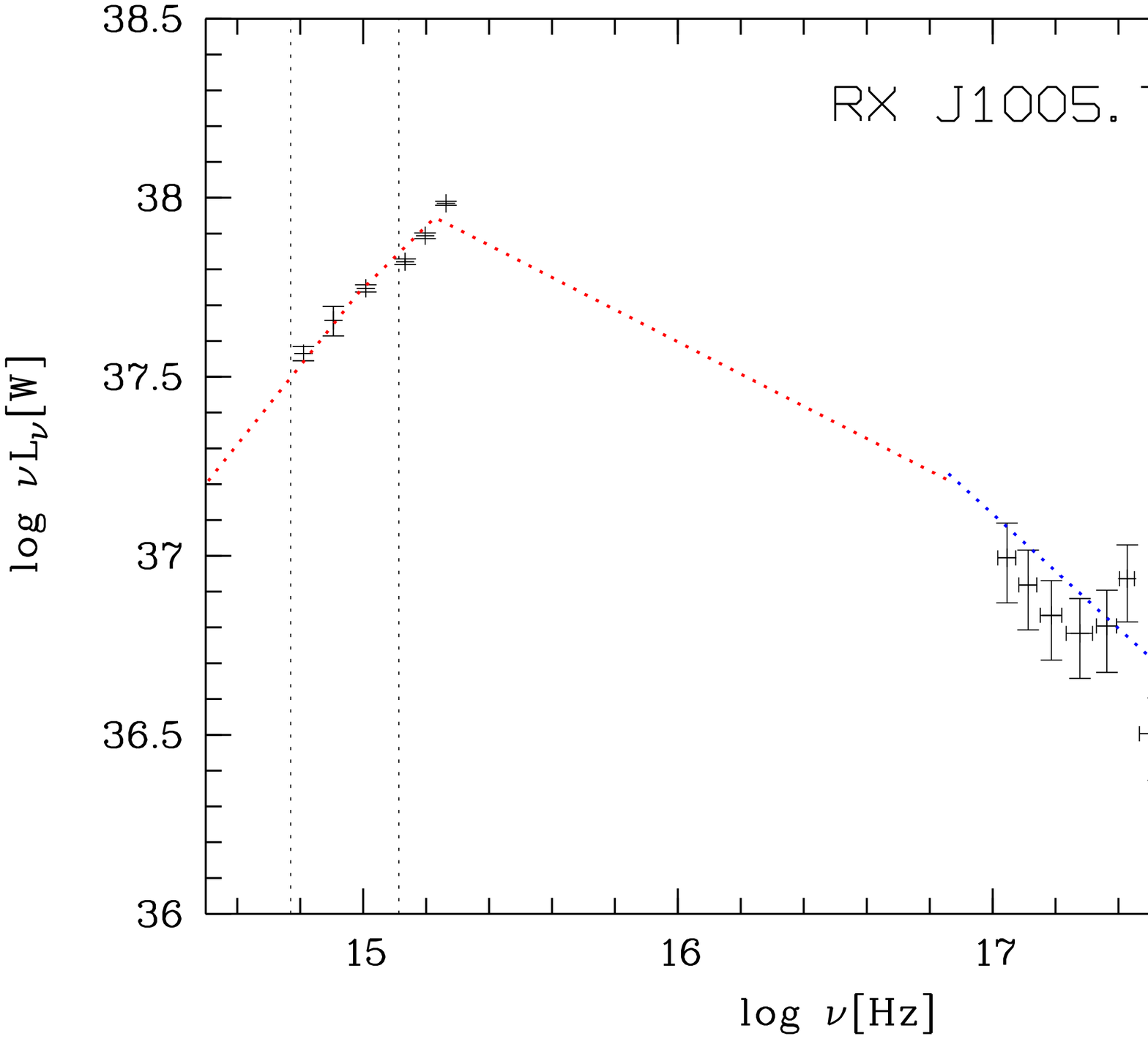}

\end{figure*}

\begin{figure*}
\epsscale{0.60}
\plotthree{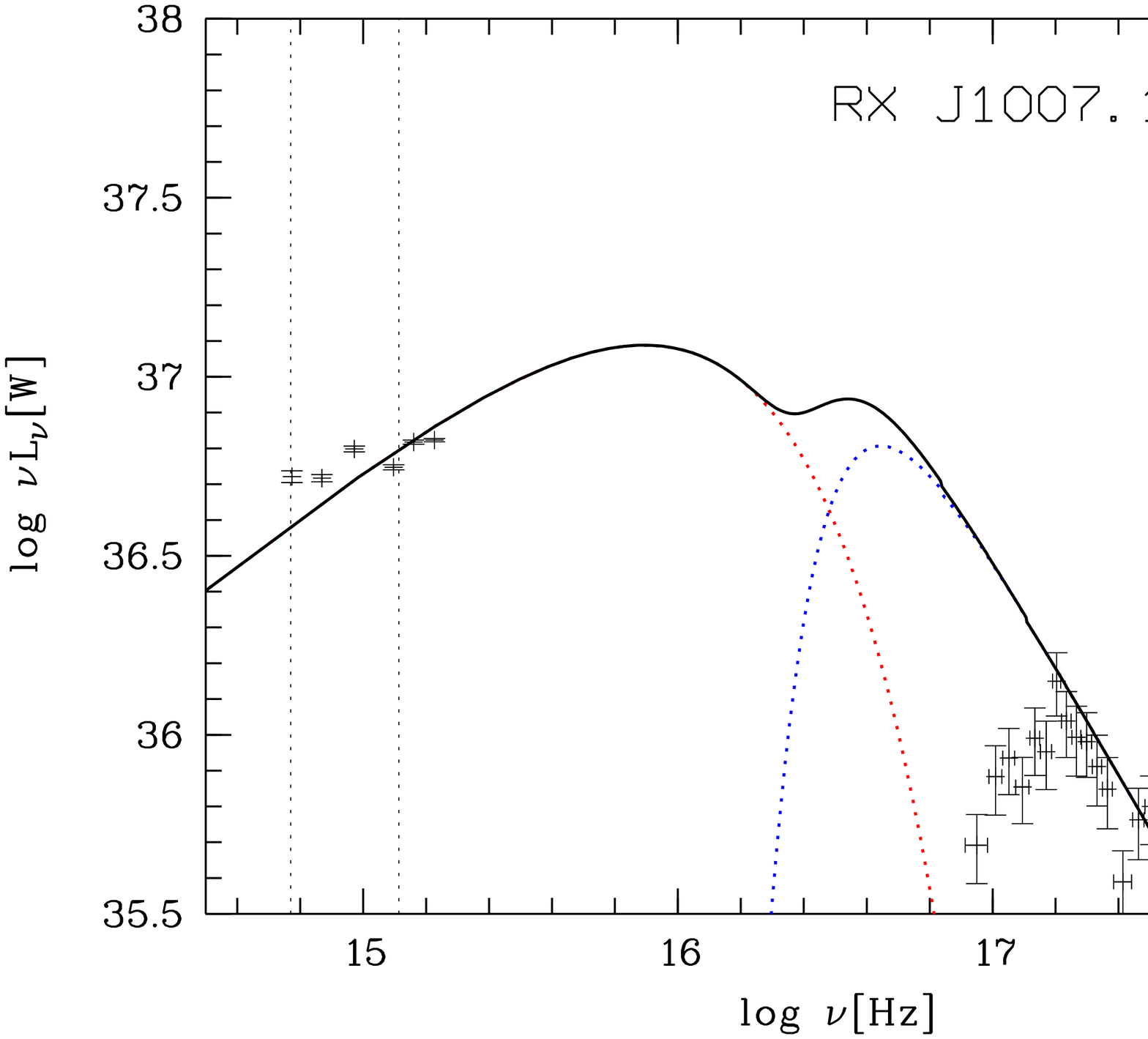}{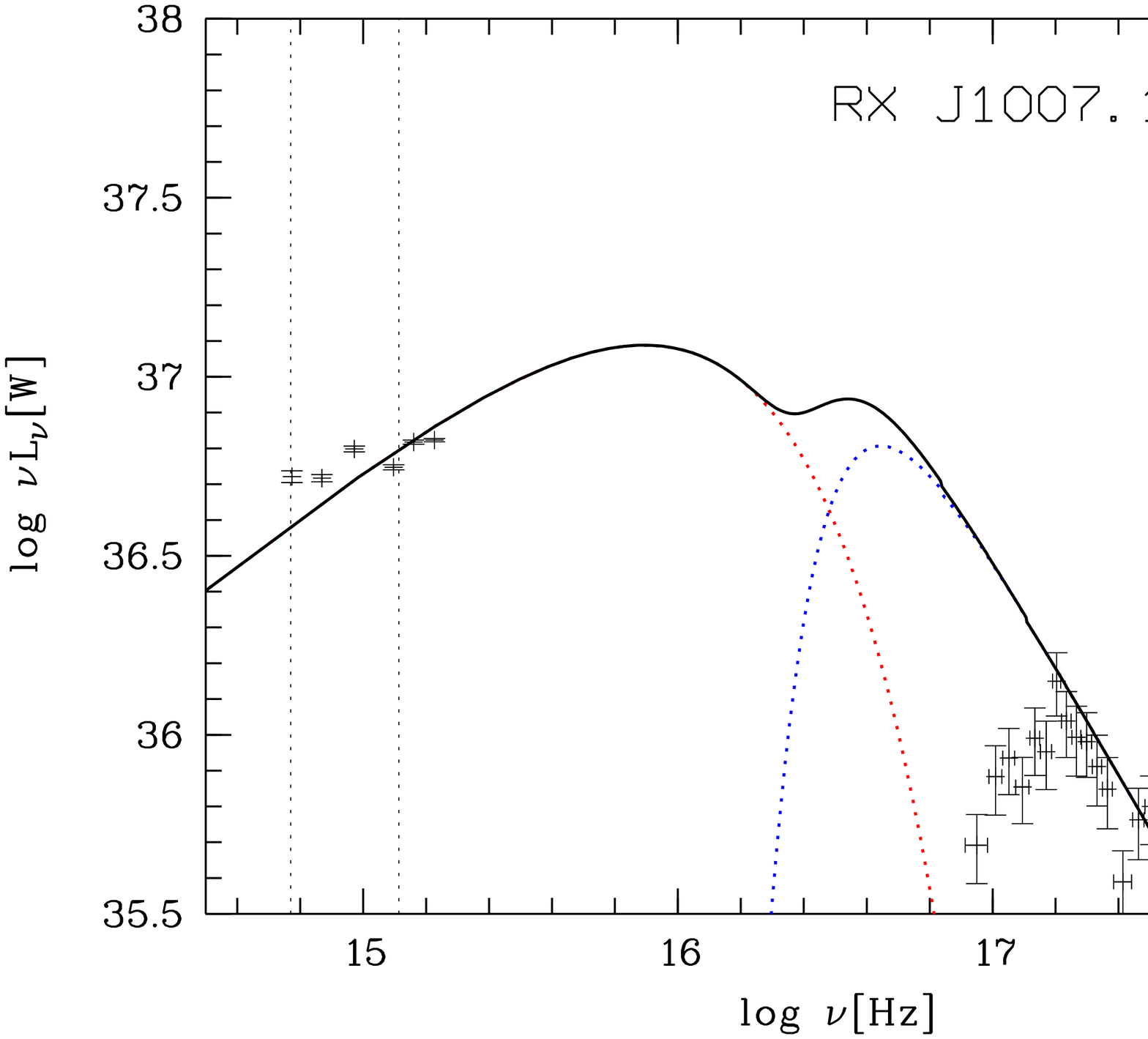}{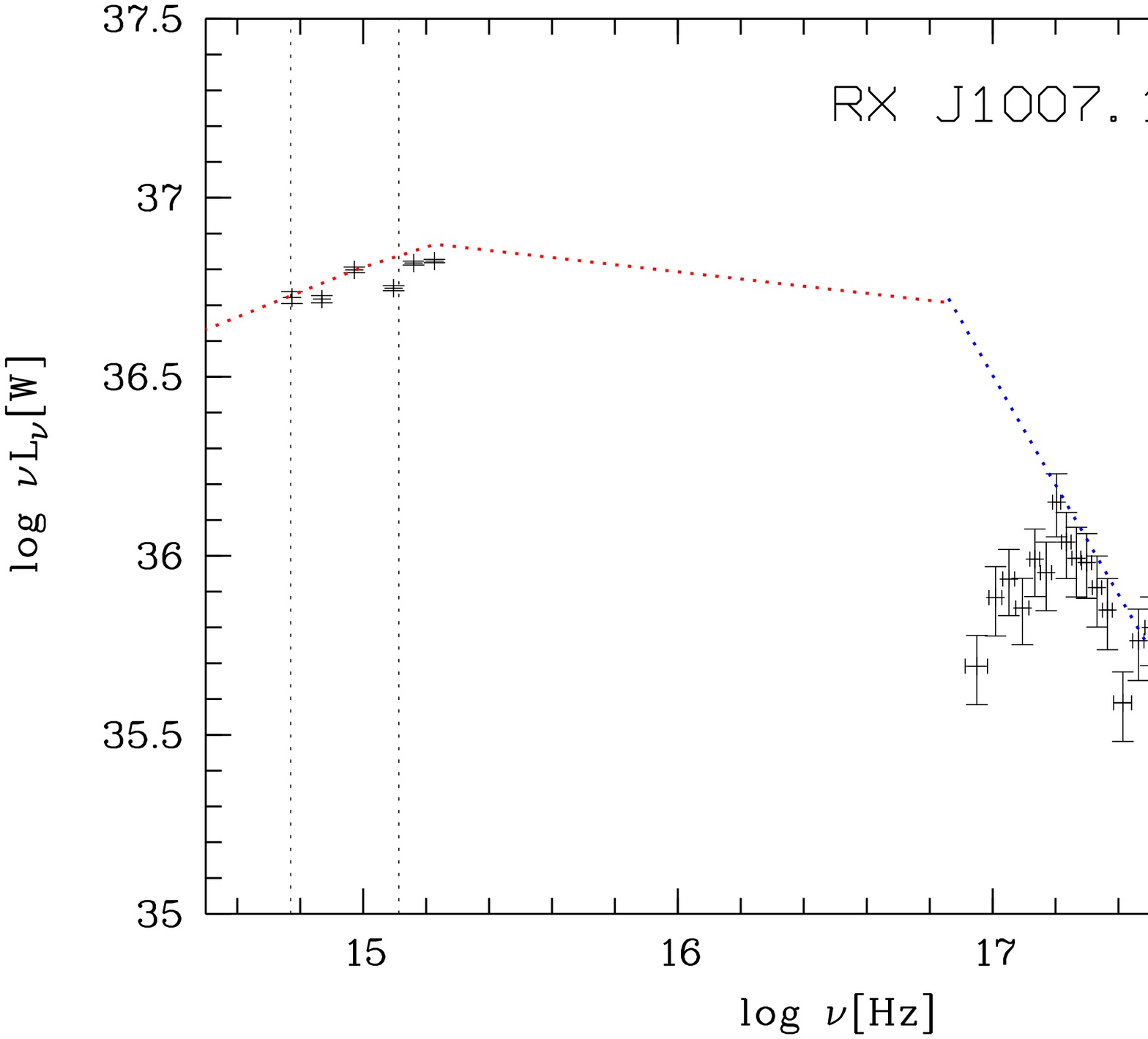}

\plotthree{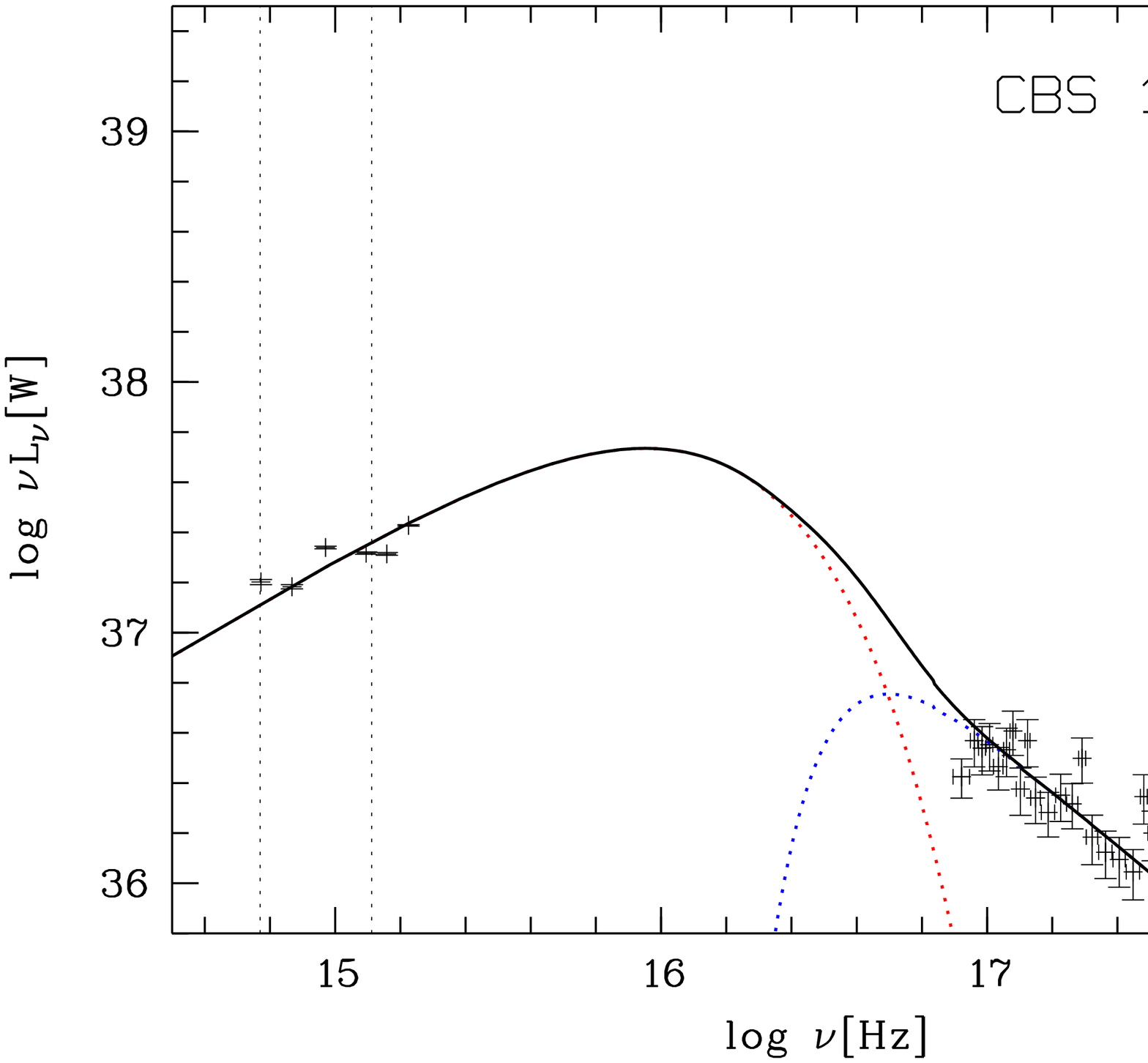}{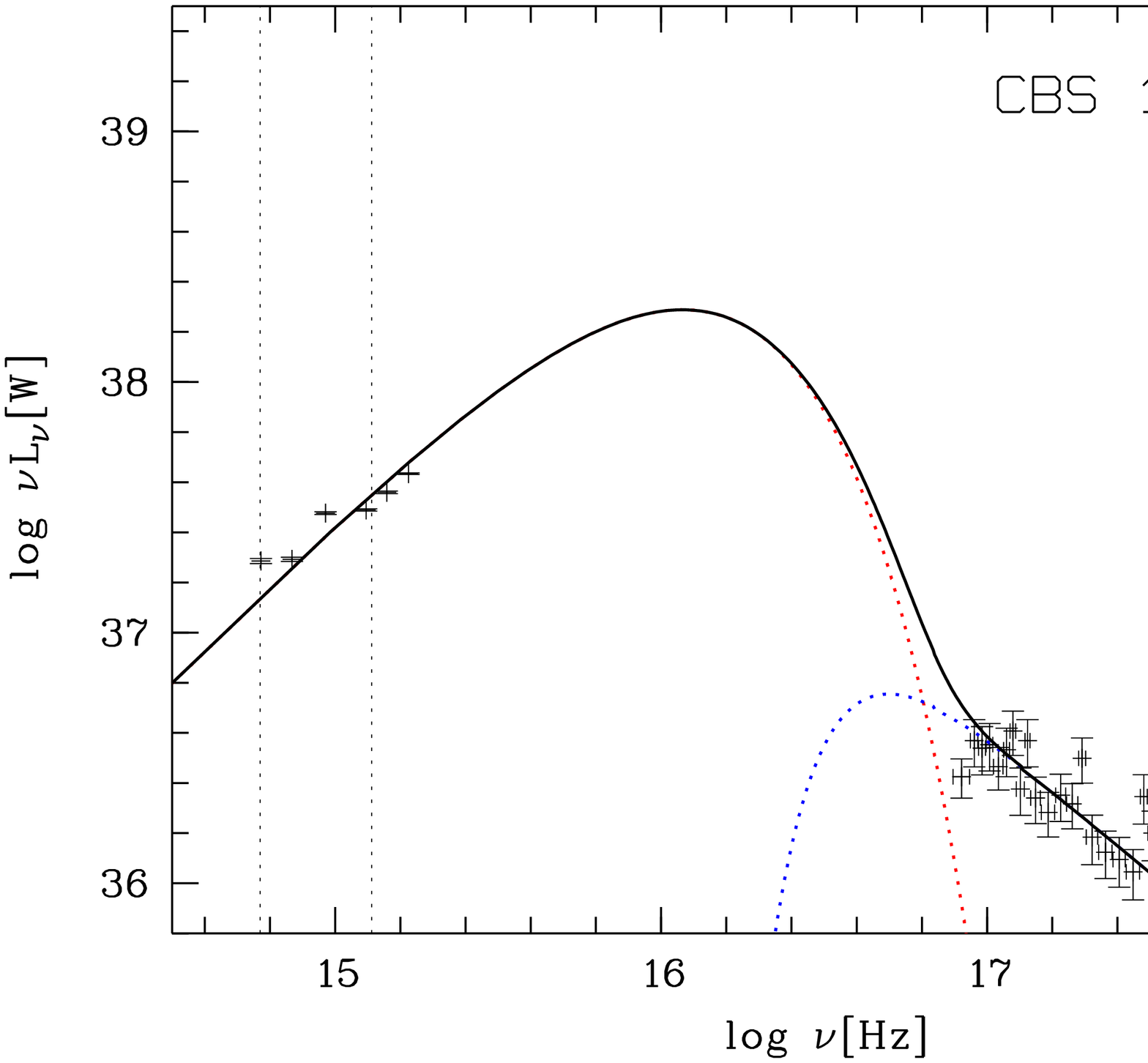}{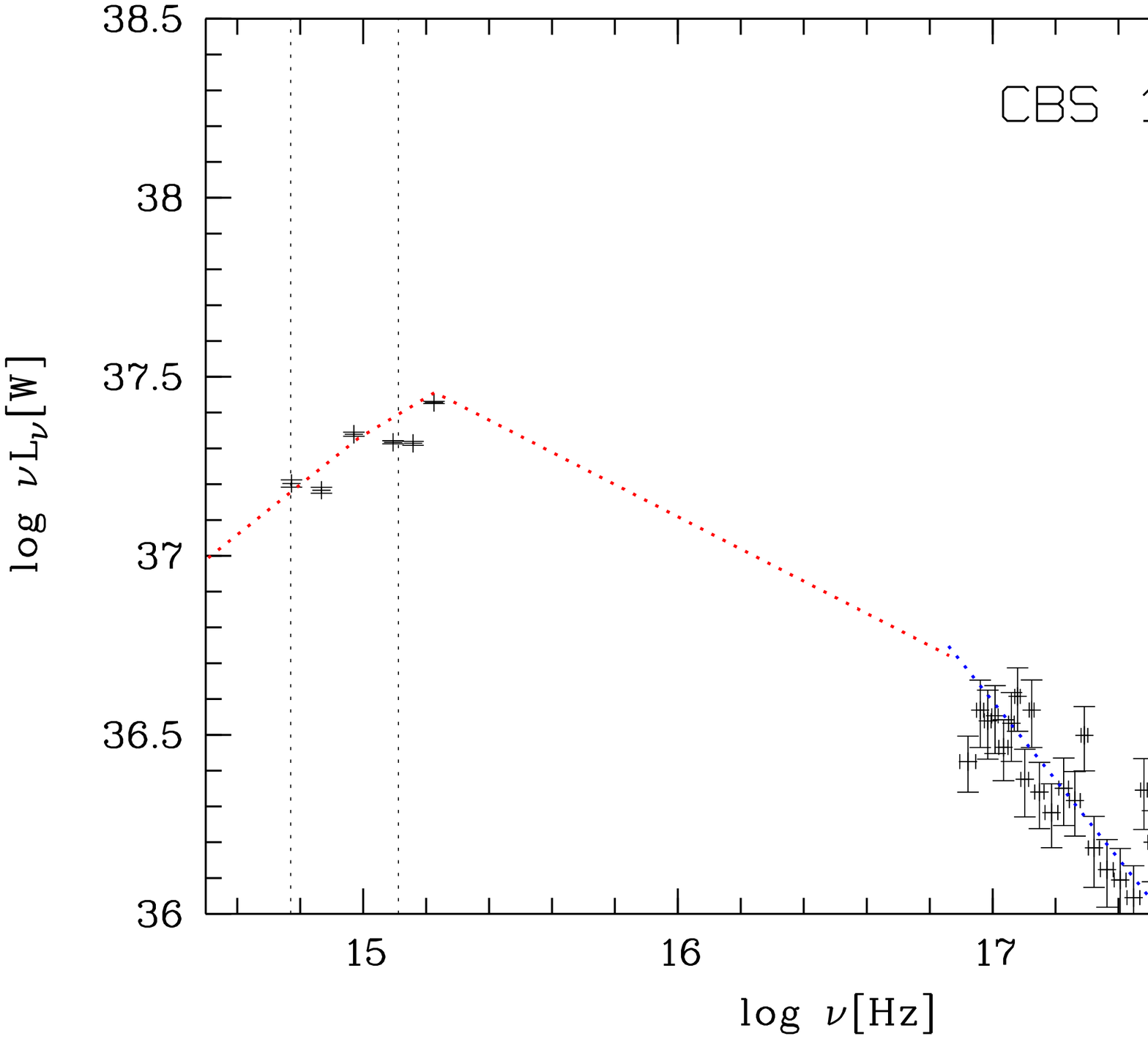}

\plotthree{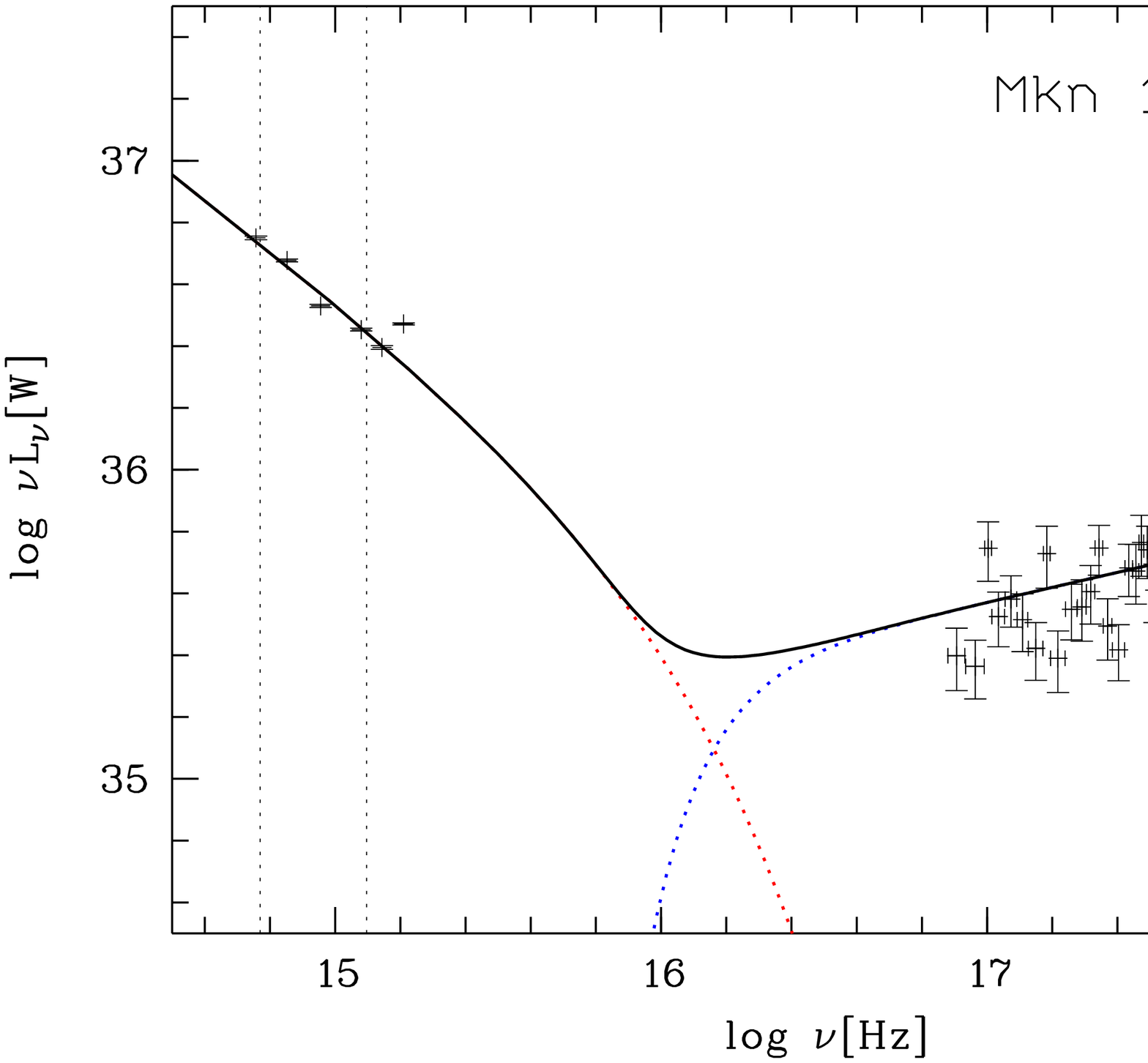}{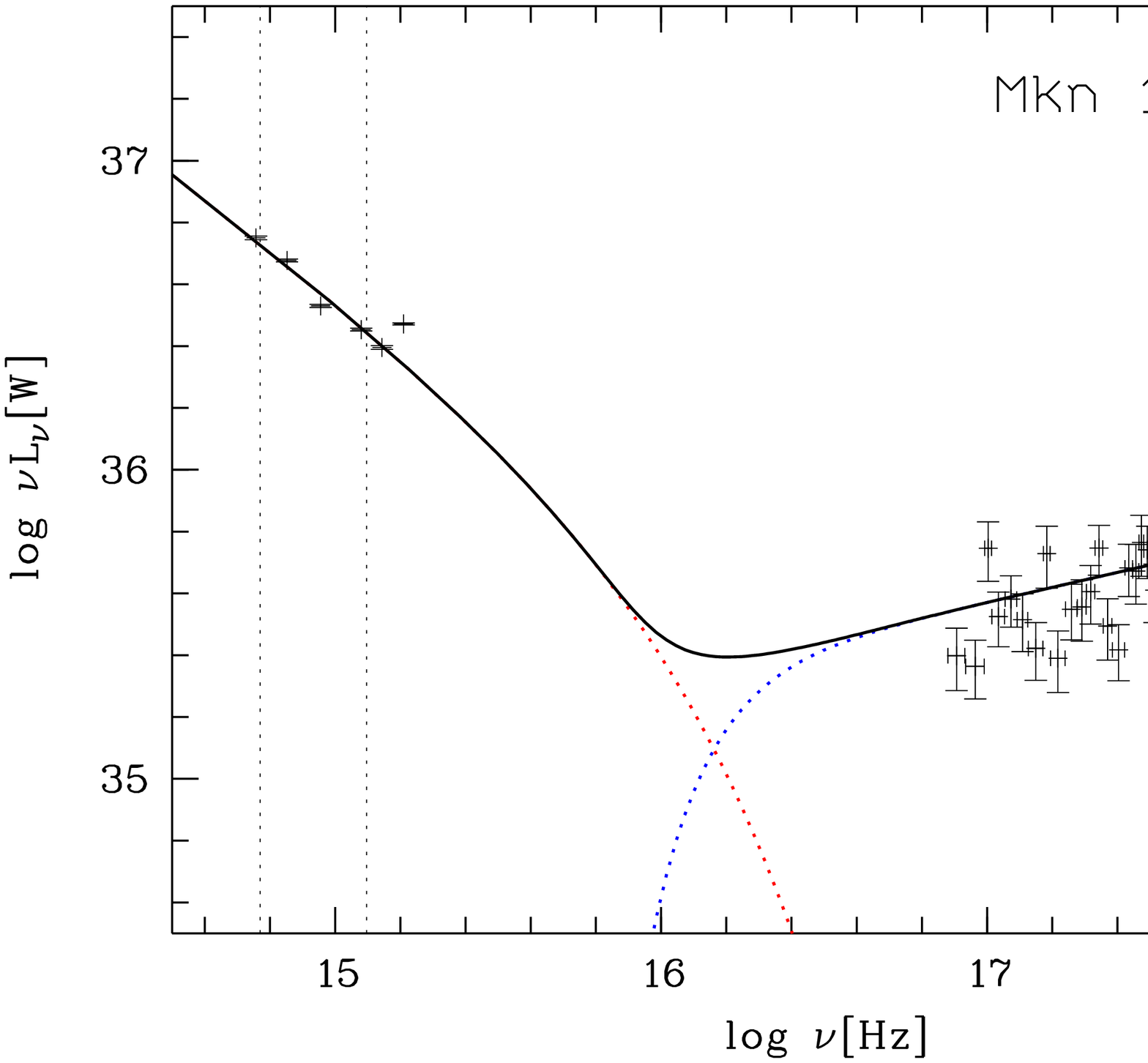}{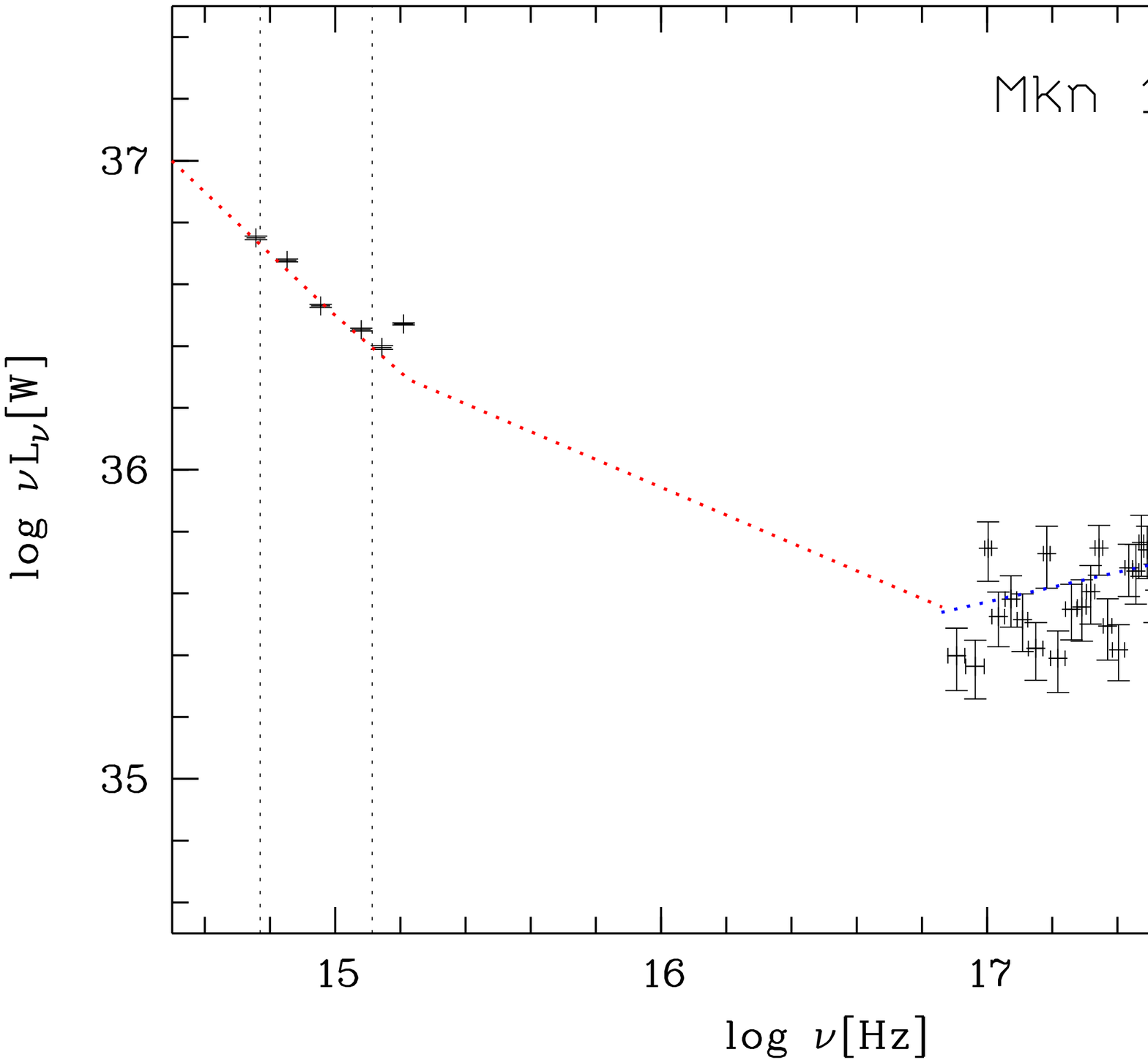}

\plotthree{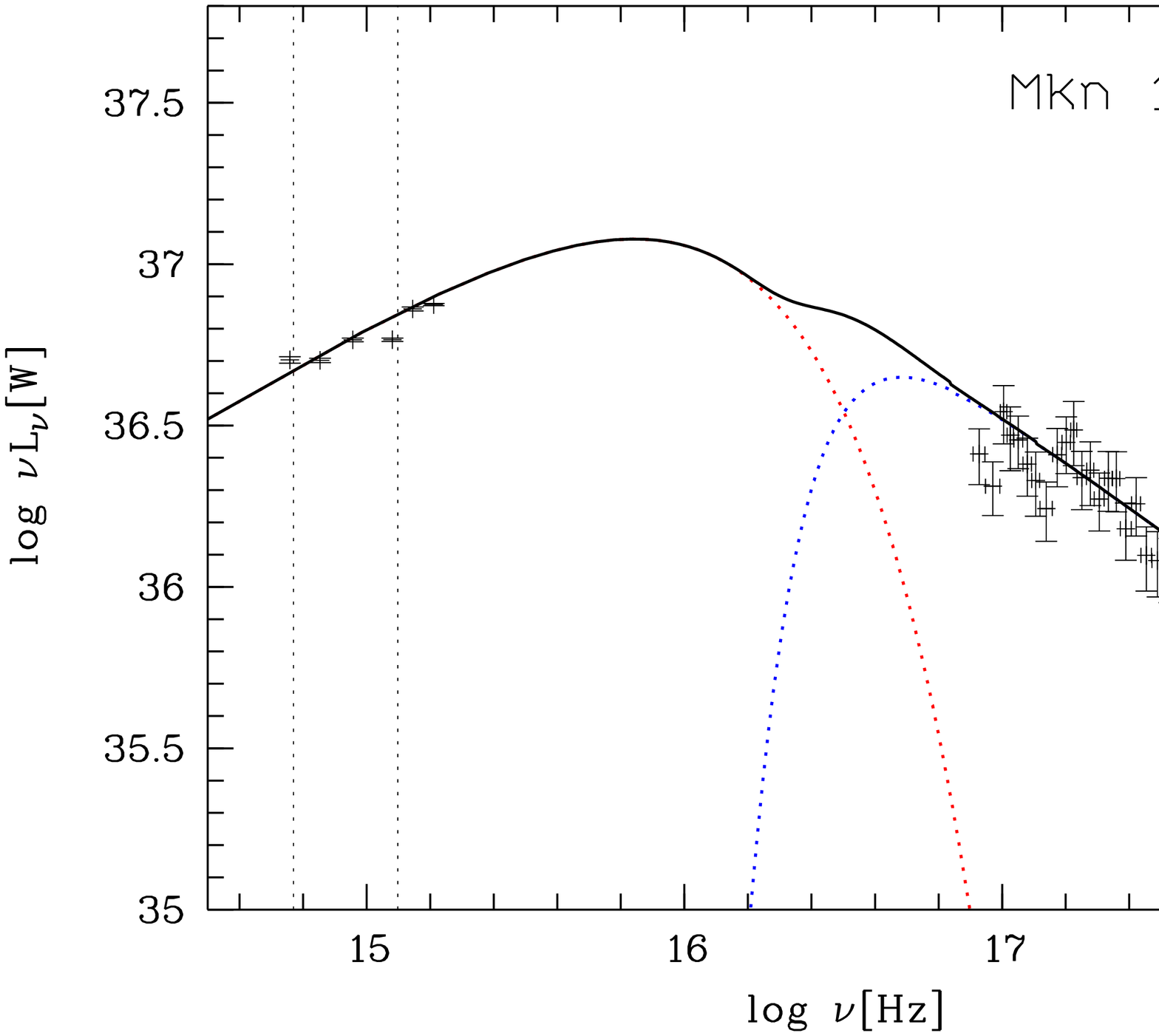}{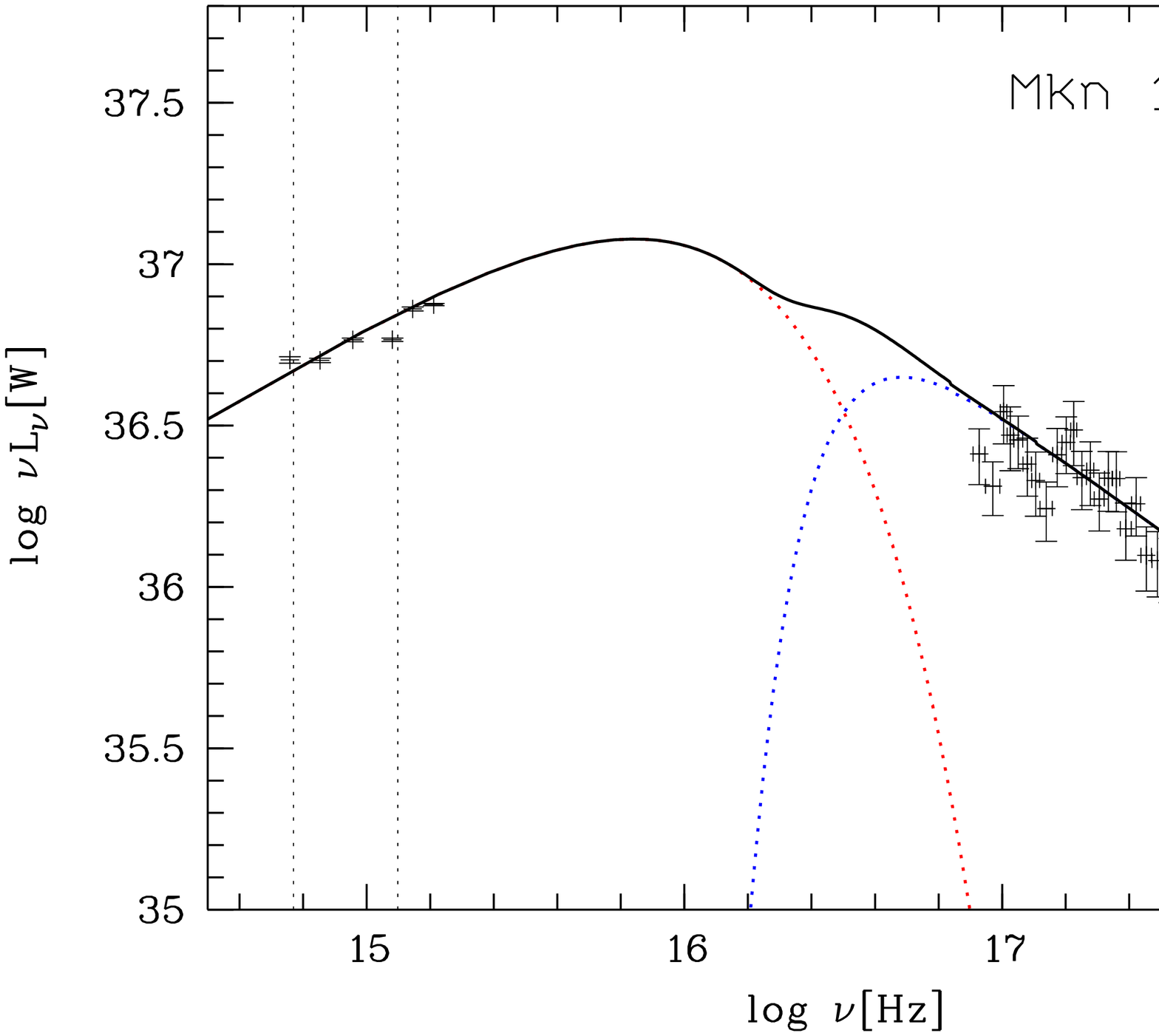}{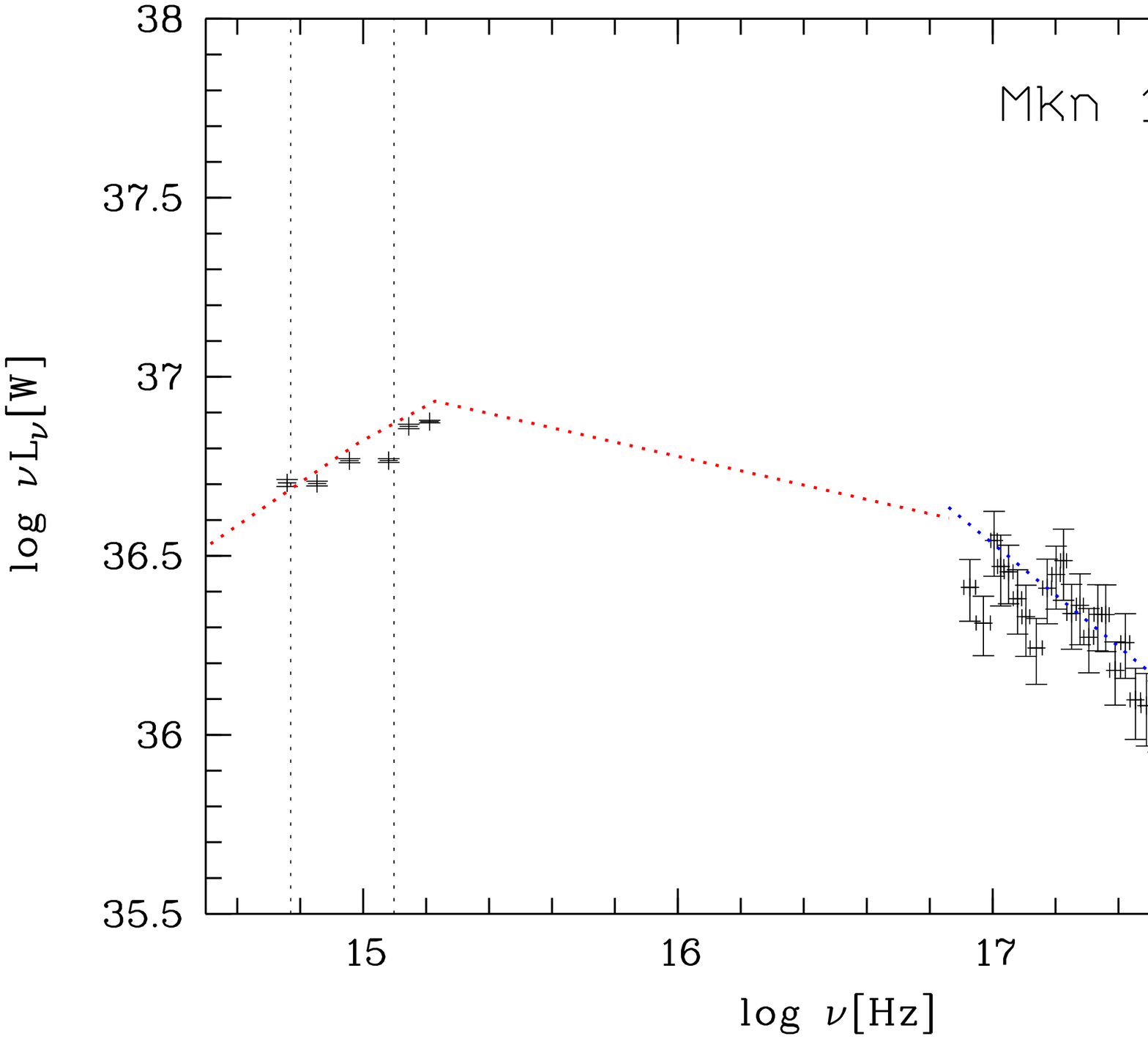}

\plotthree{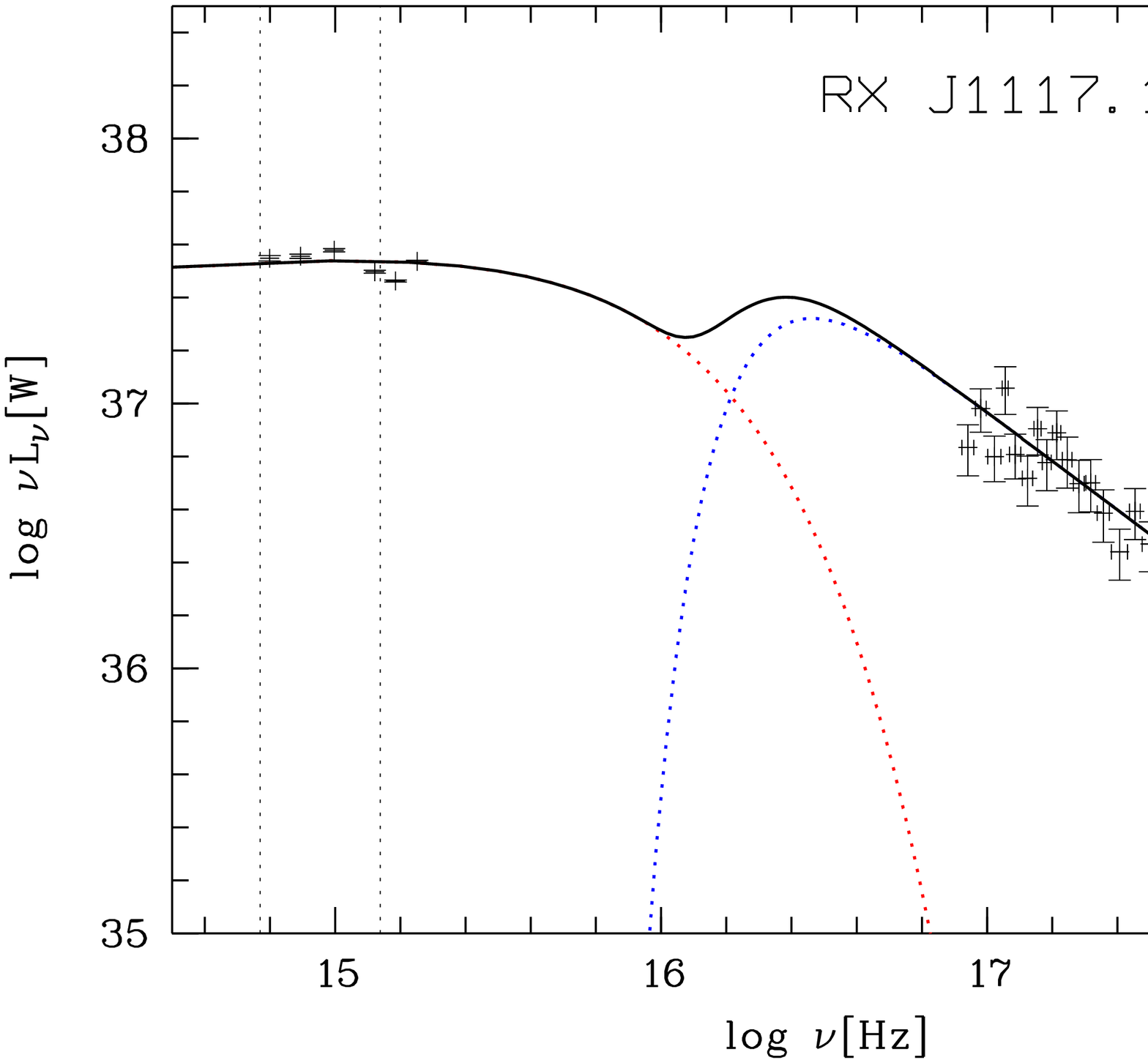}{f25_t.ps}{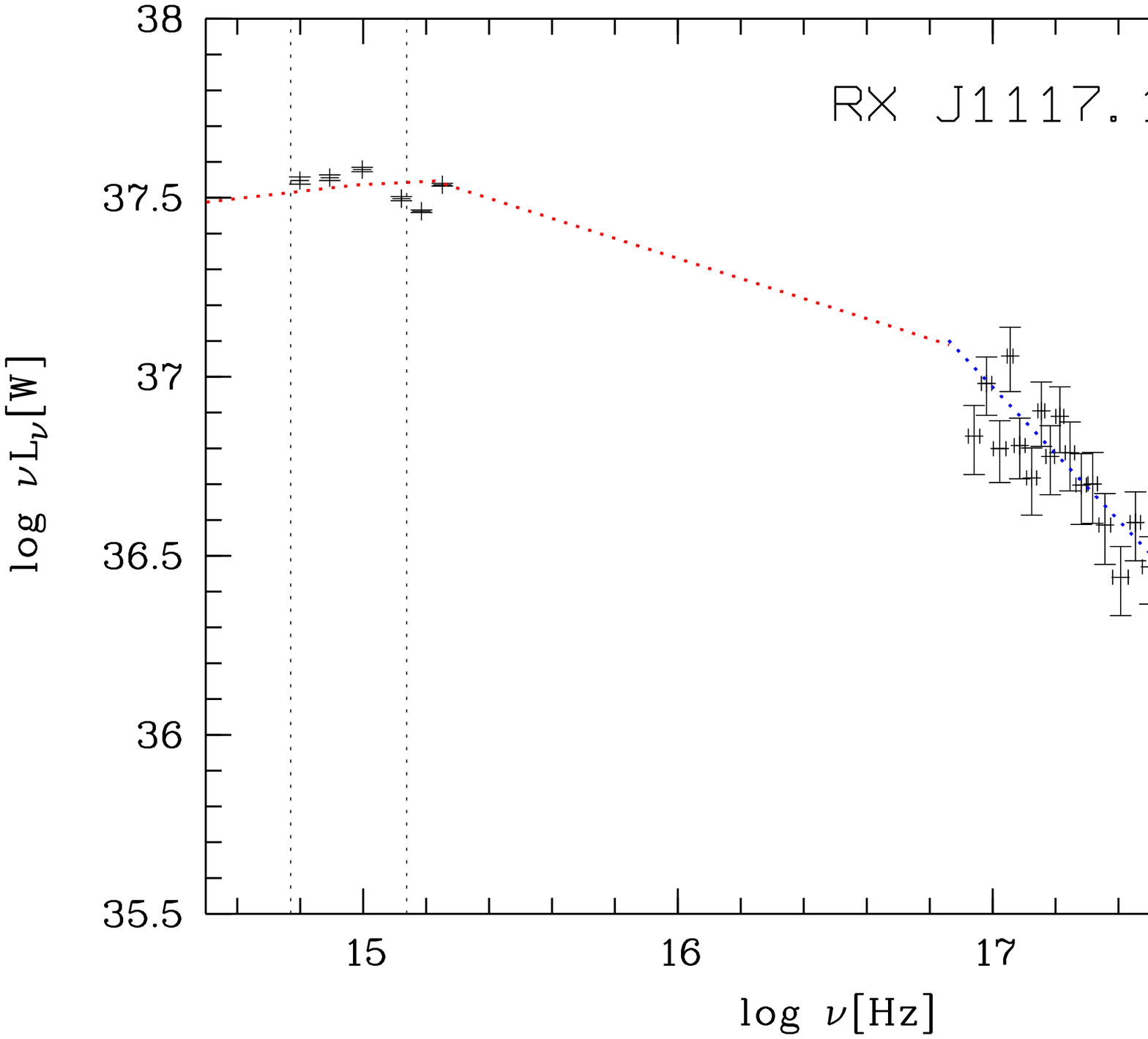}

\end{figure*}

\begin{figure*}
\epsscale{0.60}
\plotthree{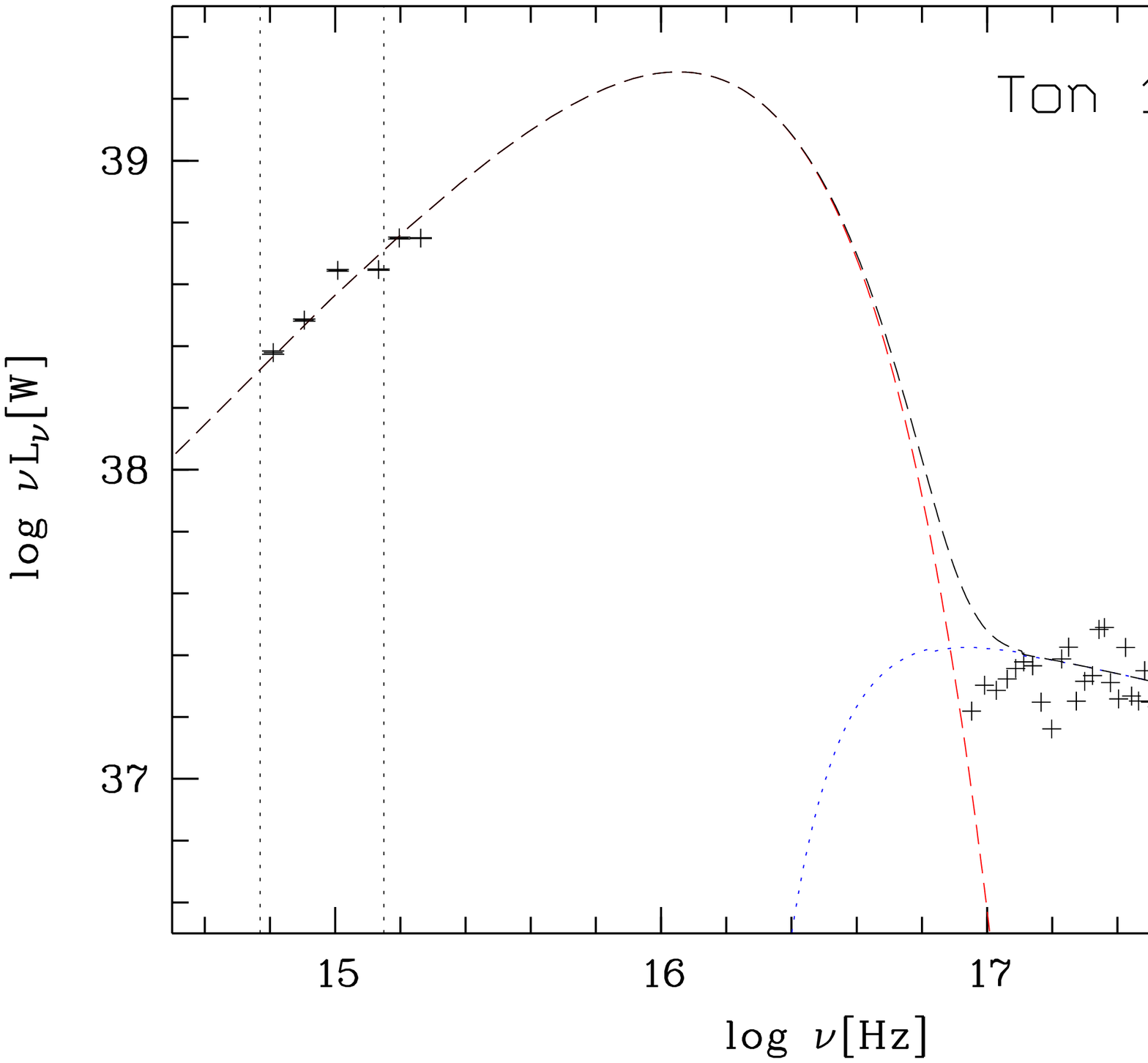}{f25_t.ps}{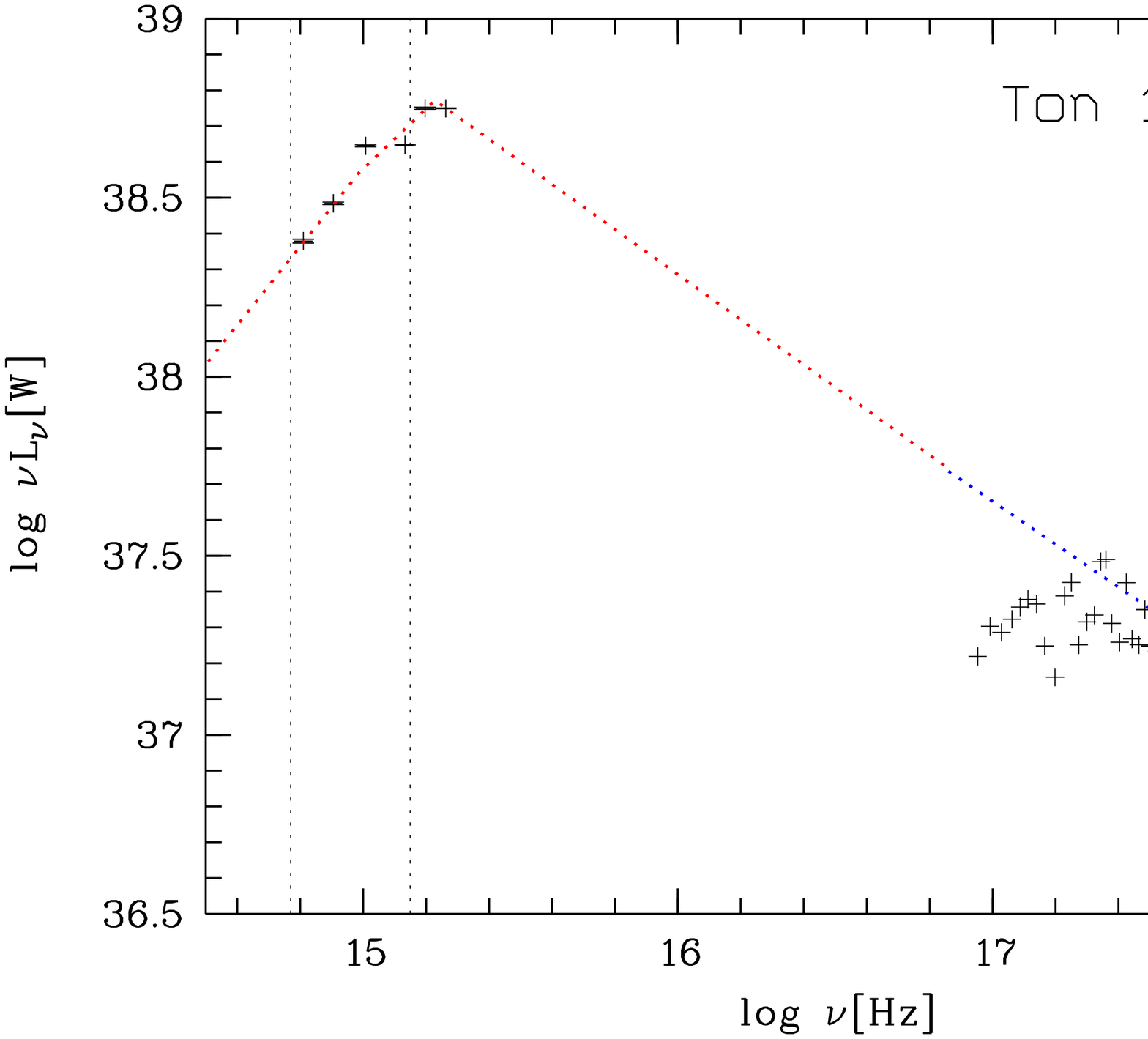}

\plotthree{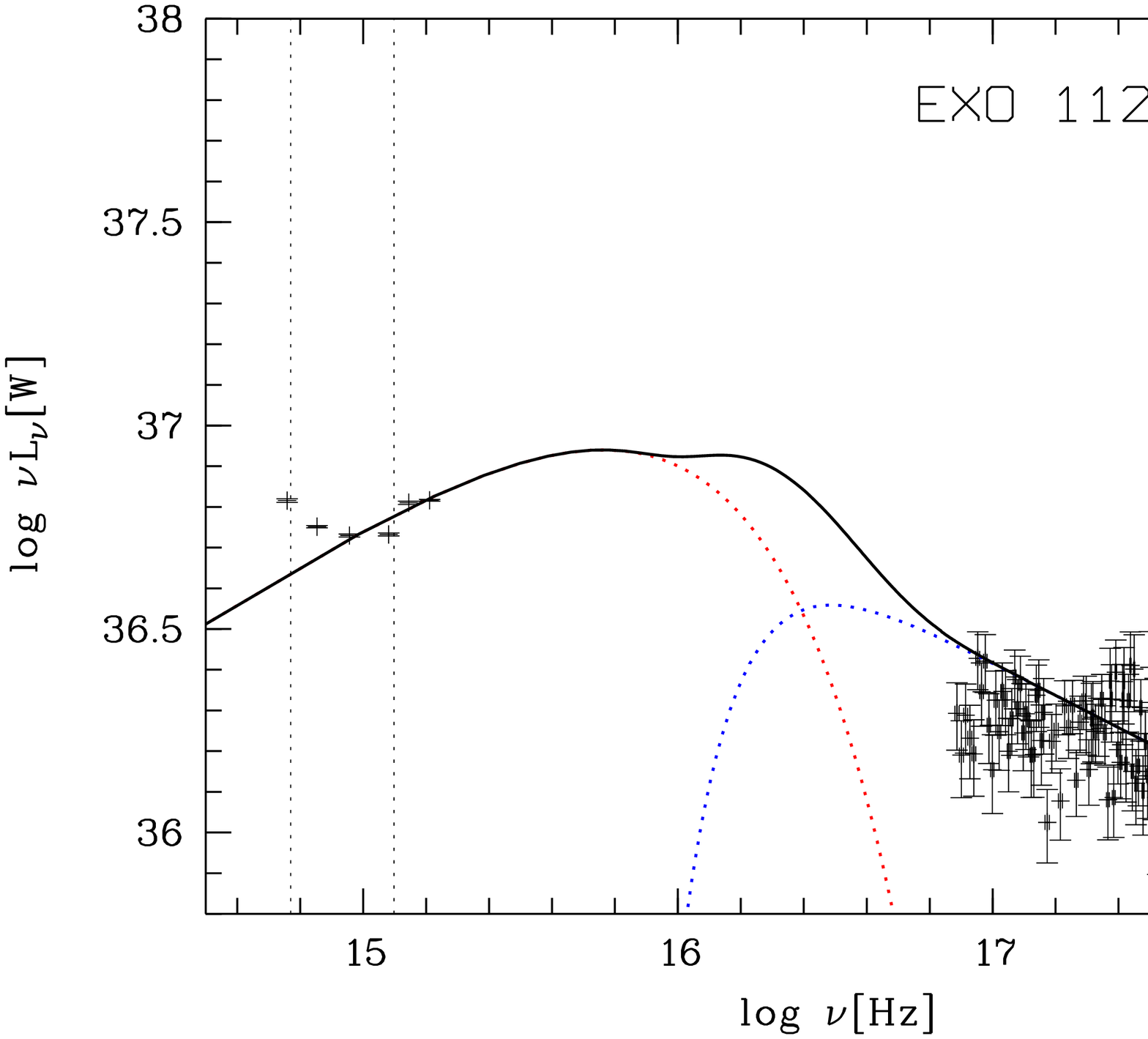}{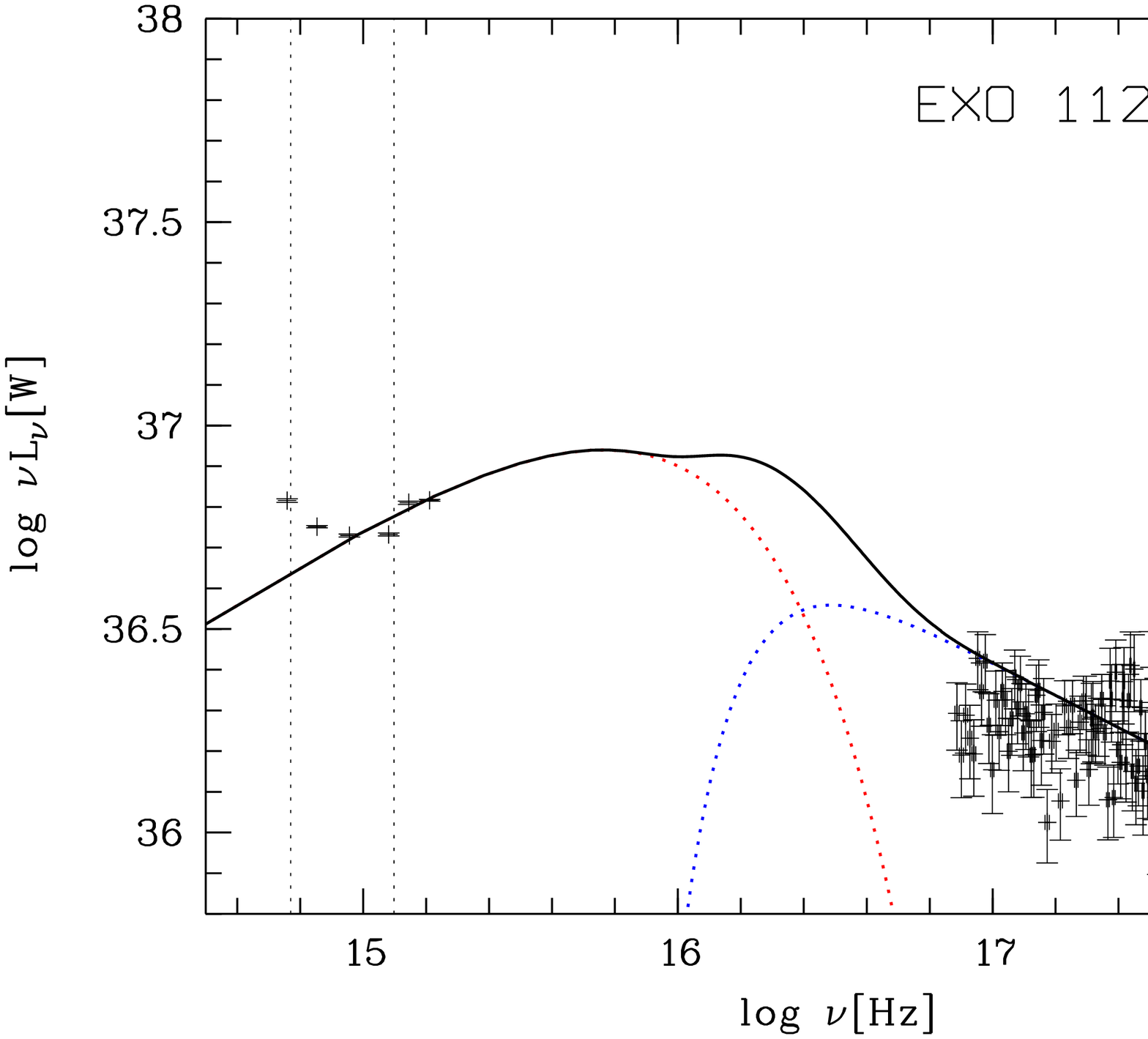}{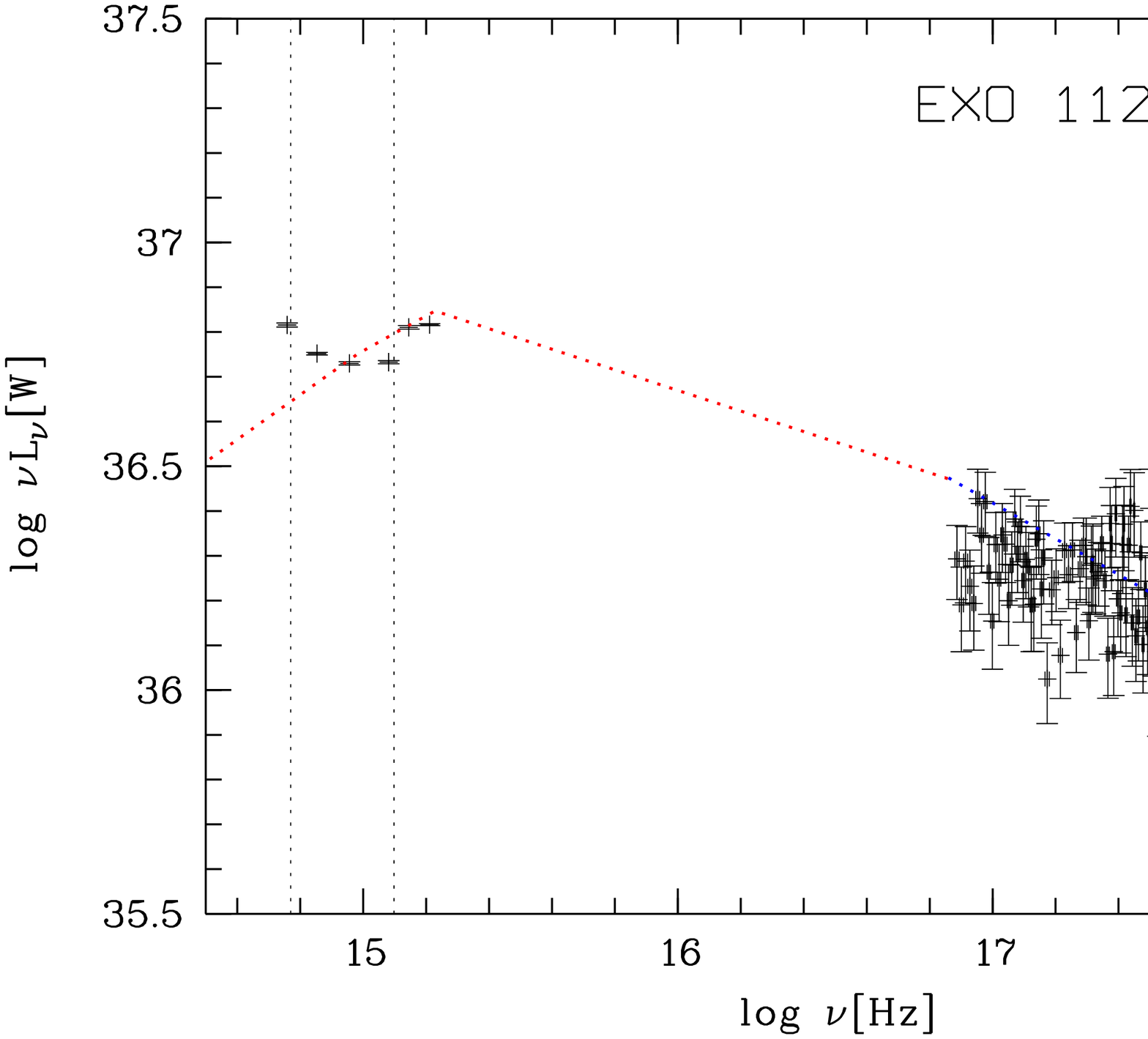}

\plotthree{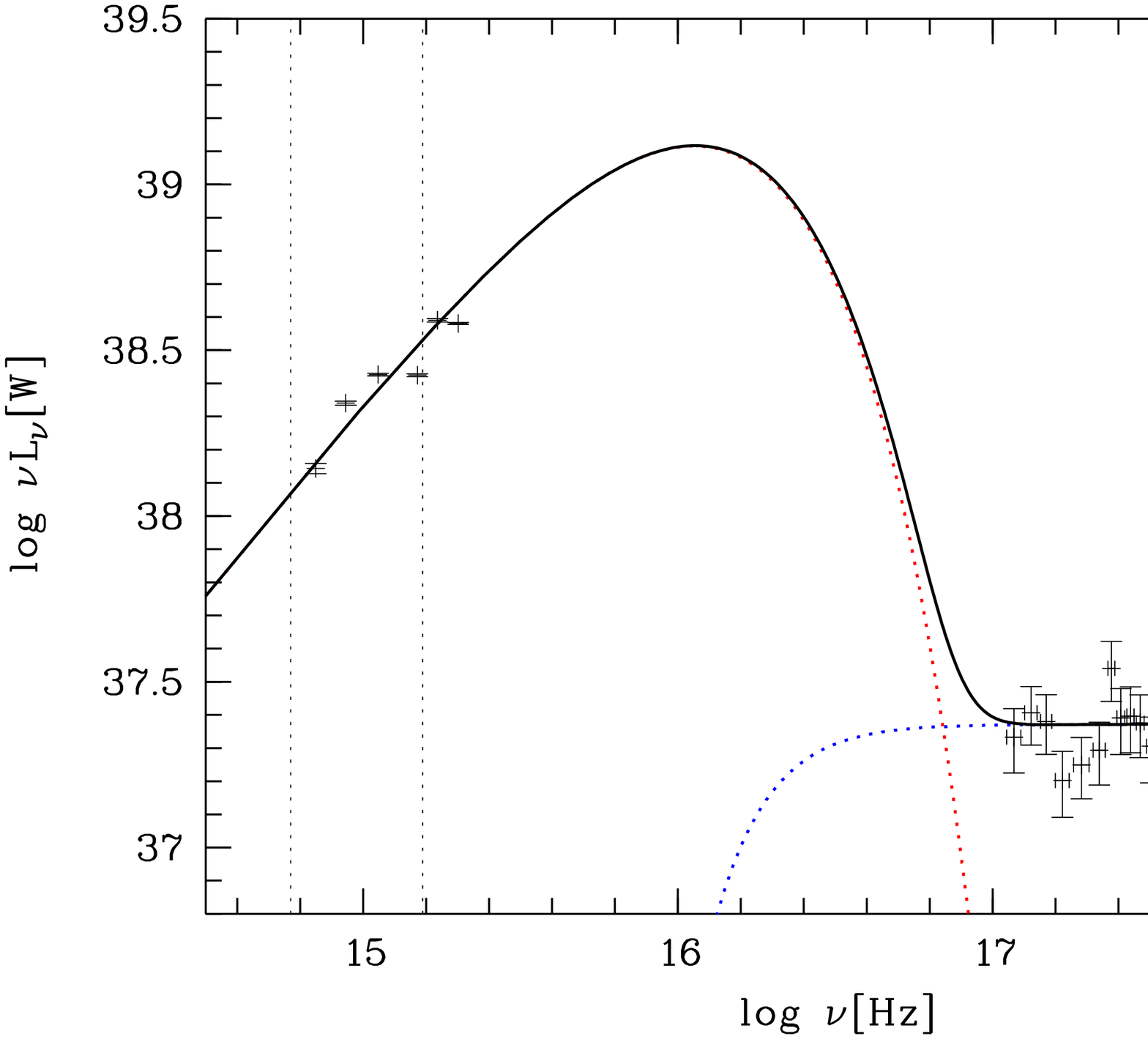}{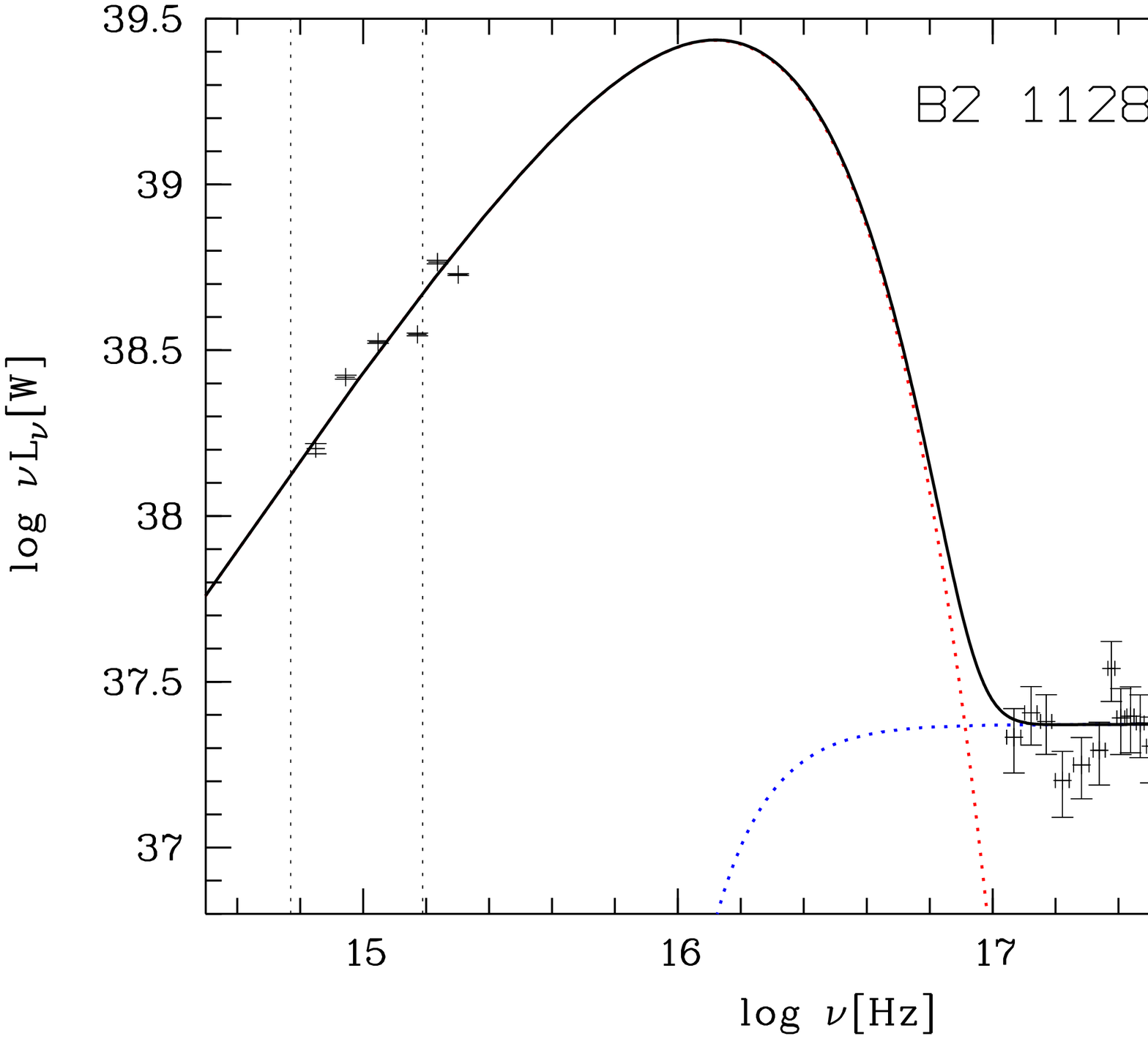}{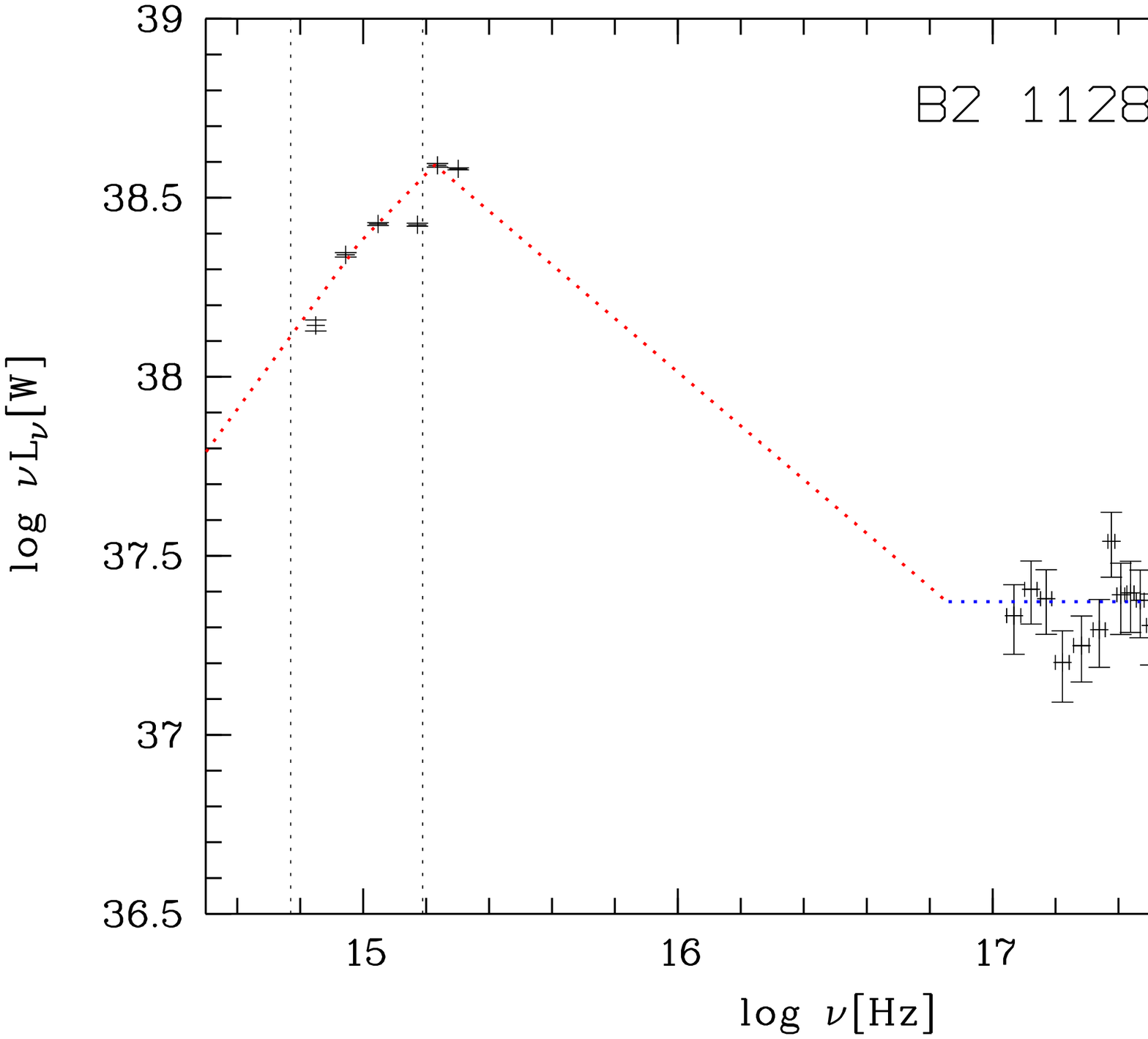}

\plotthree{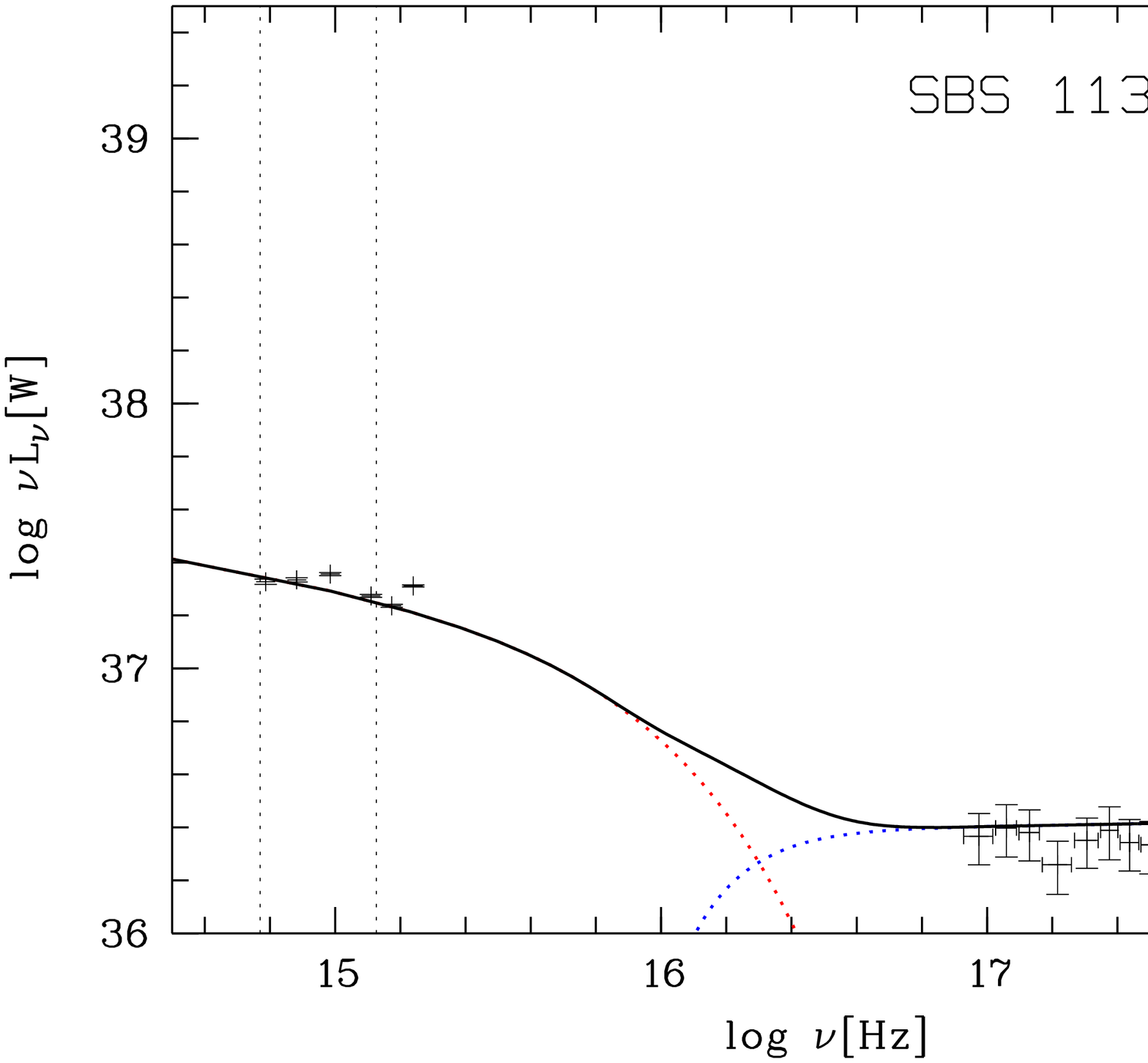}{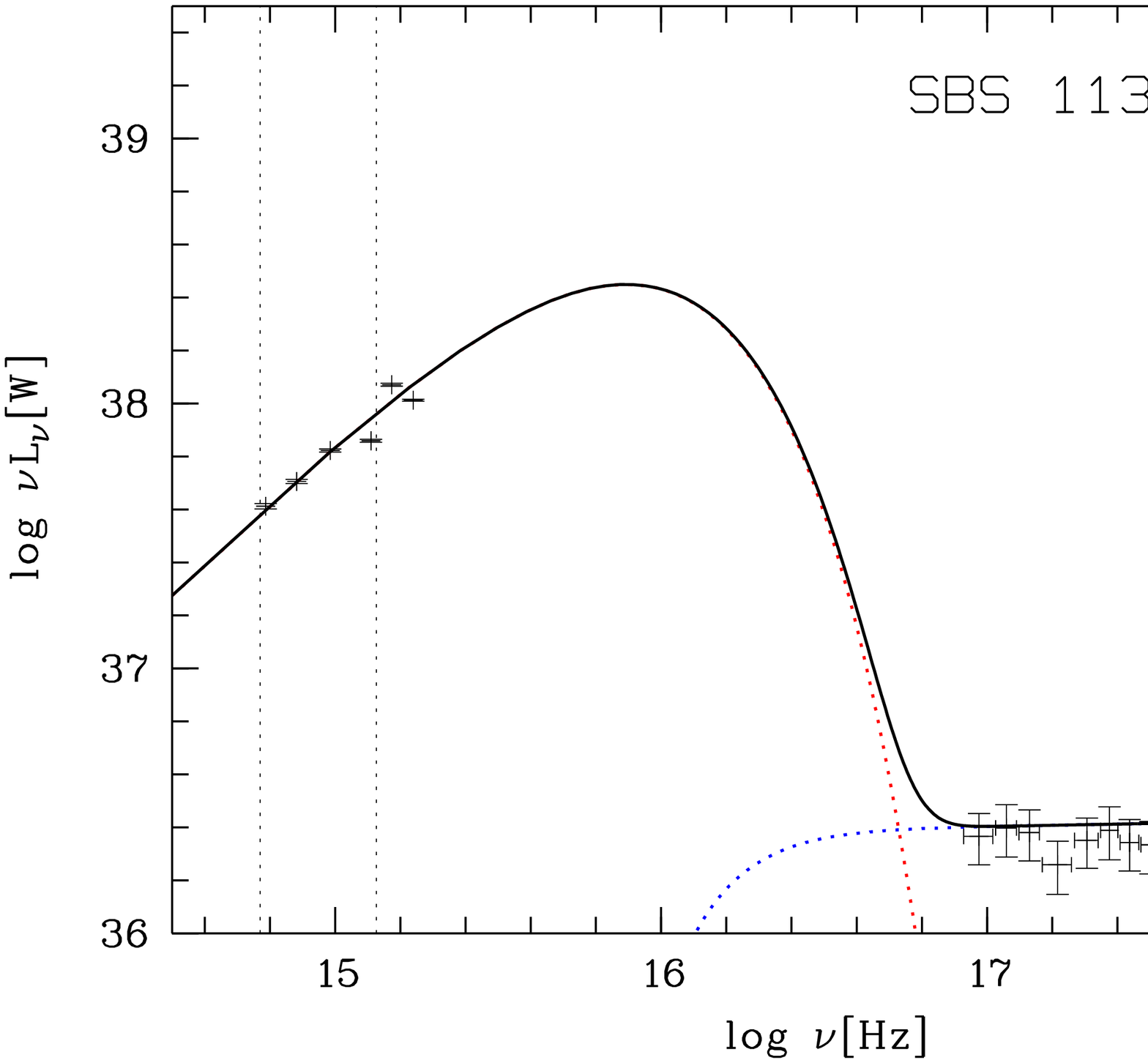}{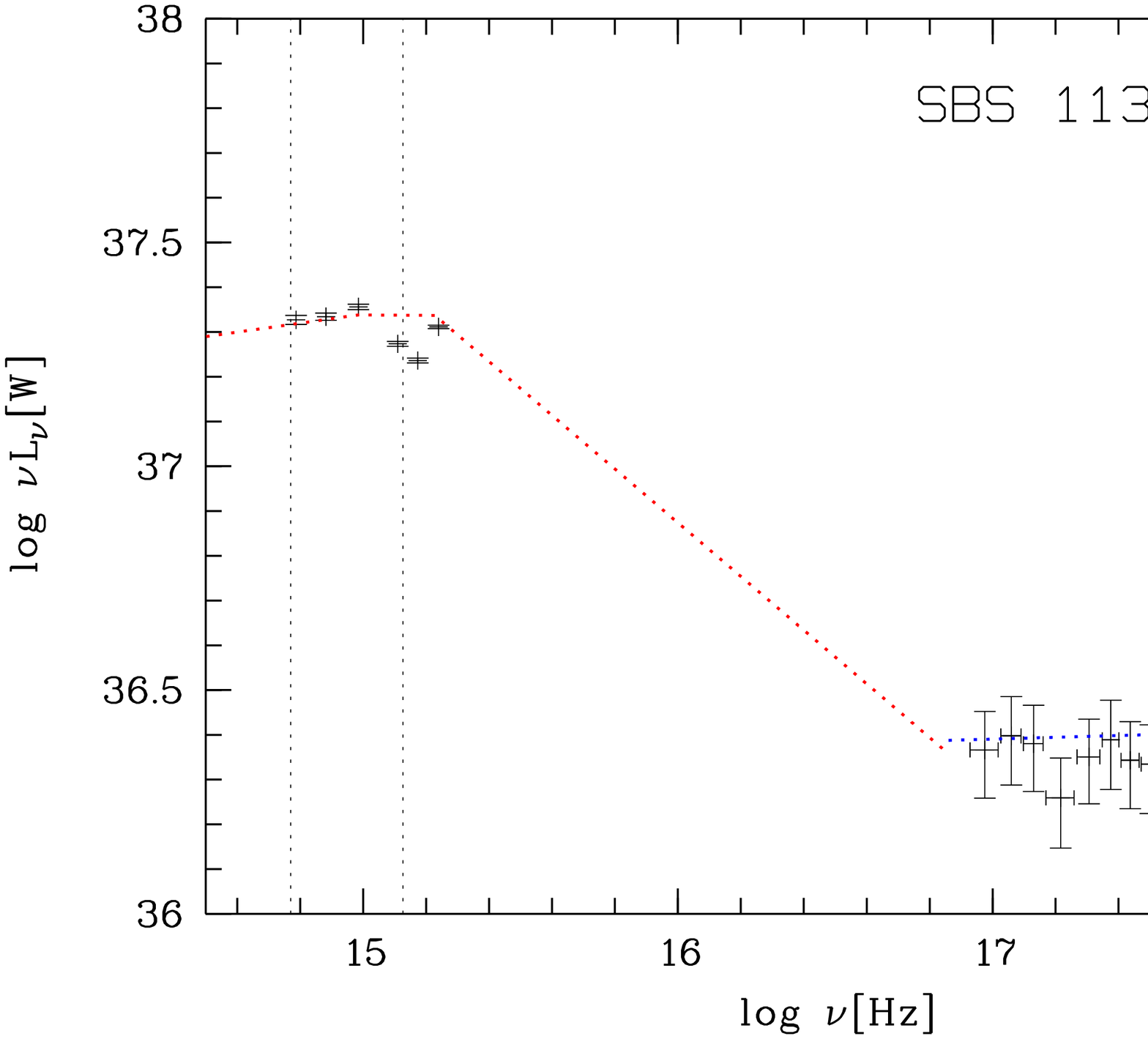}

\plotthree{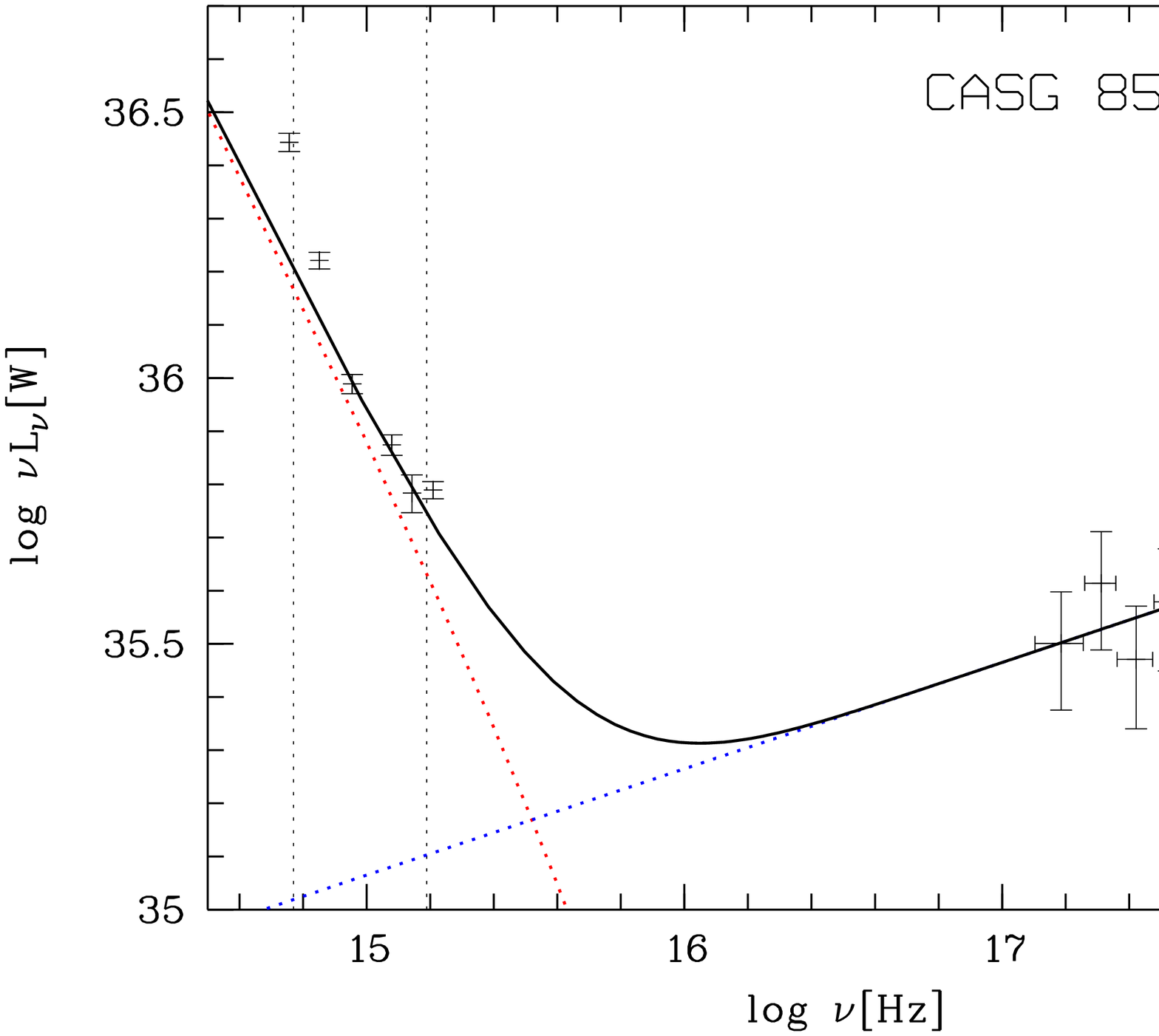}{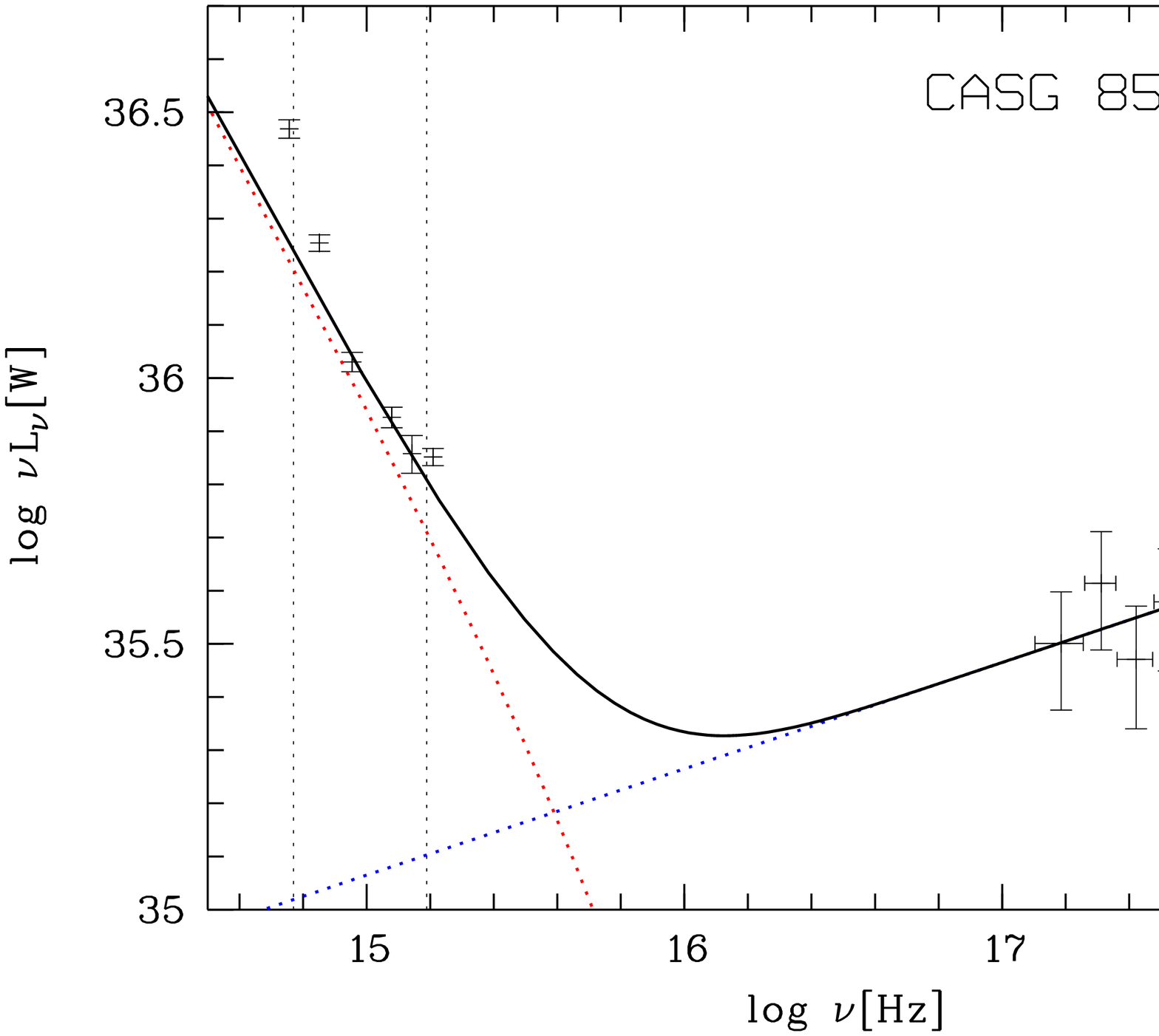}{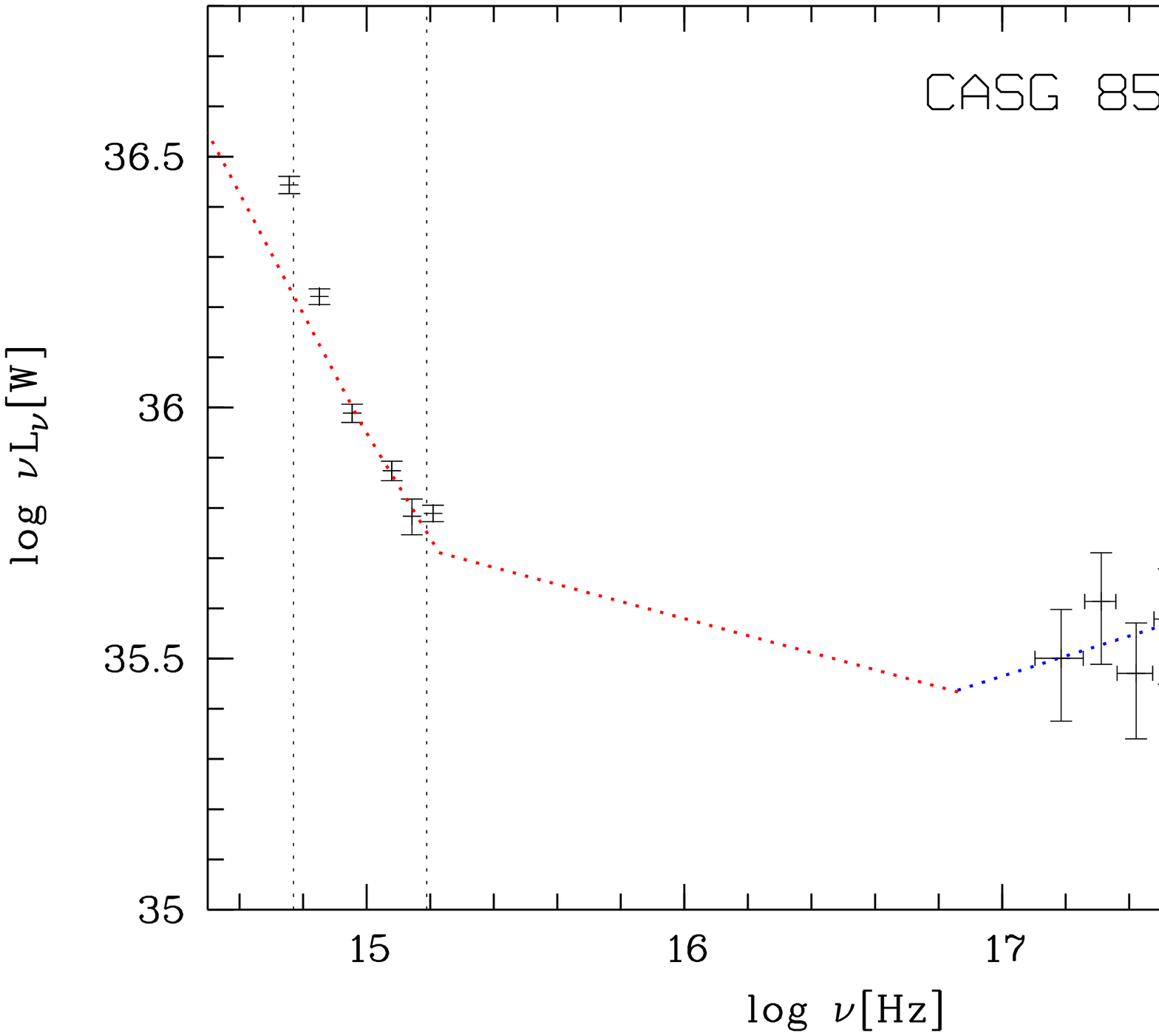}

\end{figure*}

\begin{figure*}
\epsscale{0.60}
\plotthree{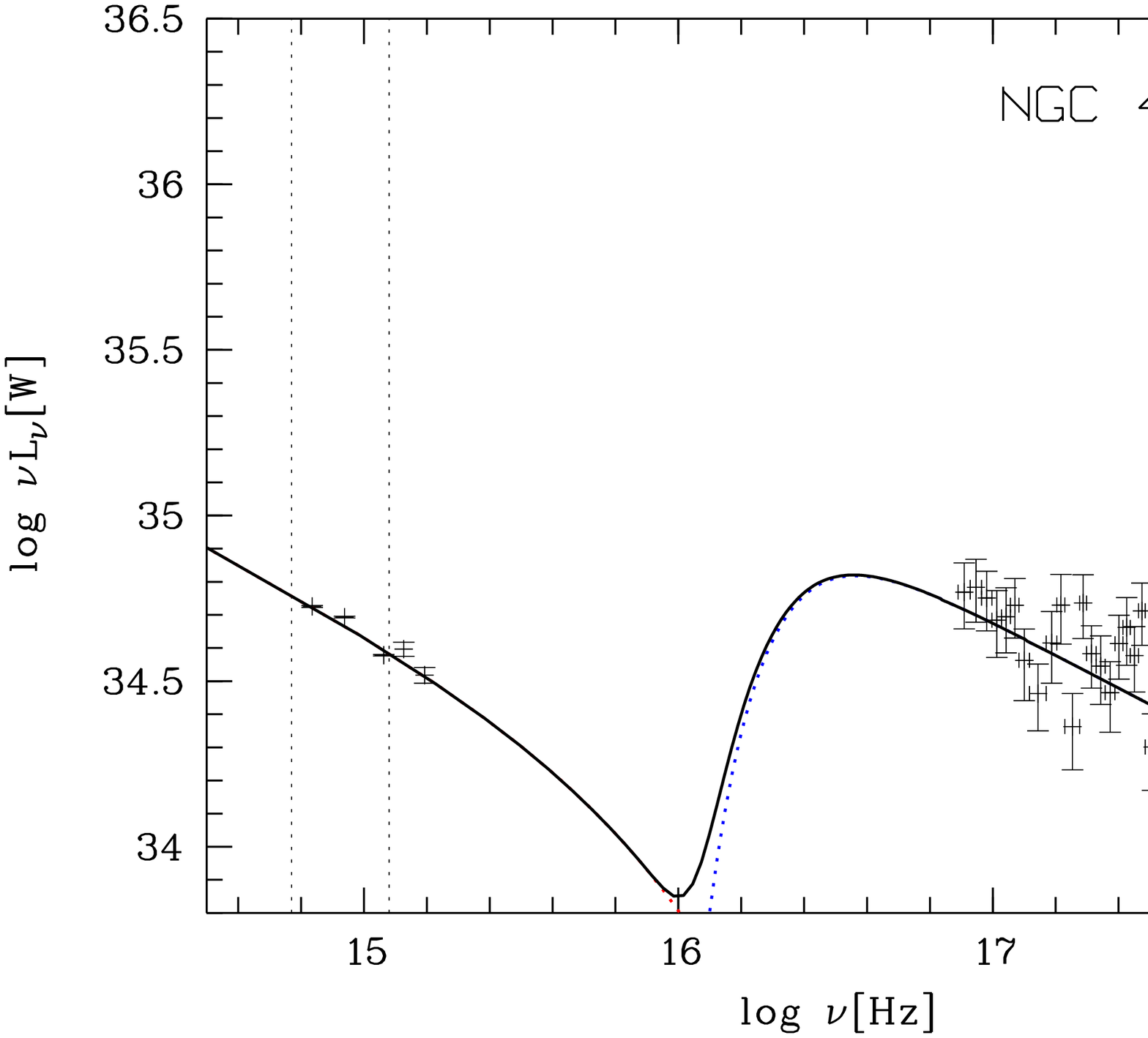}{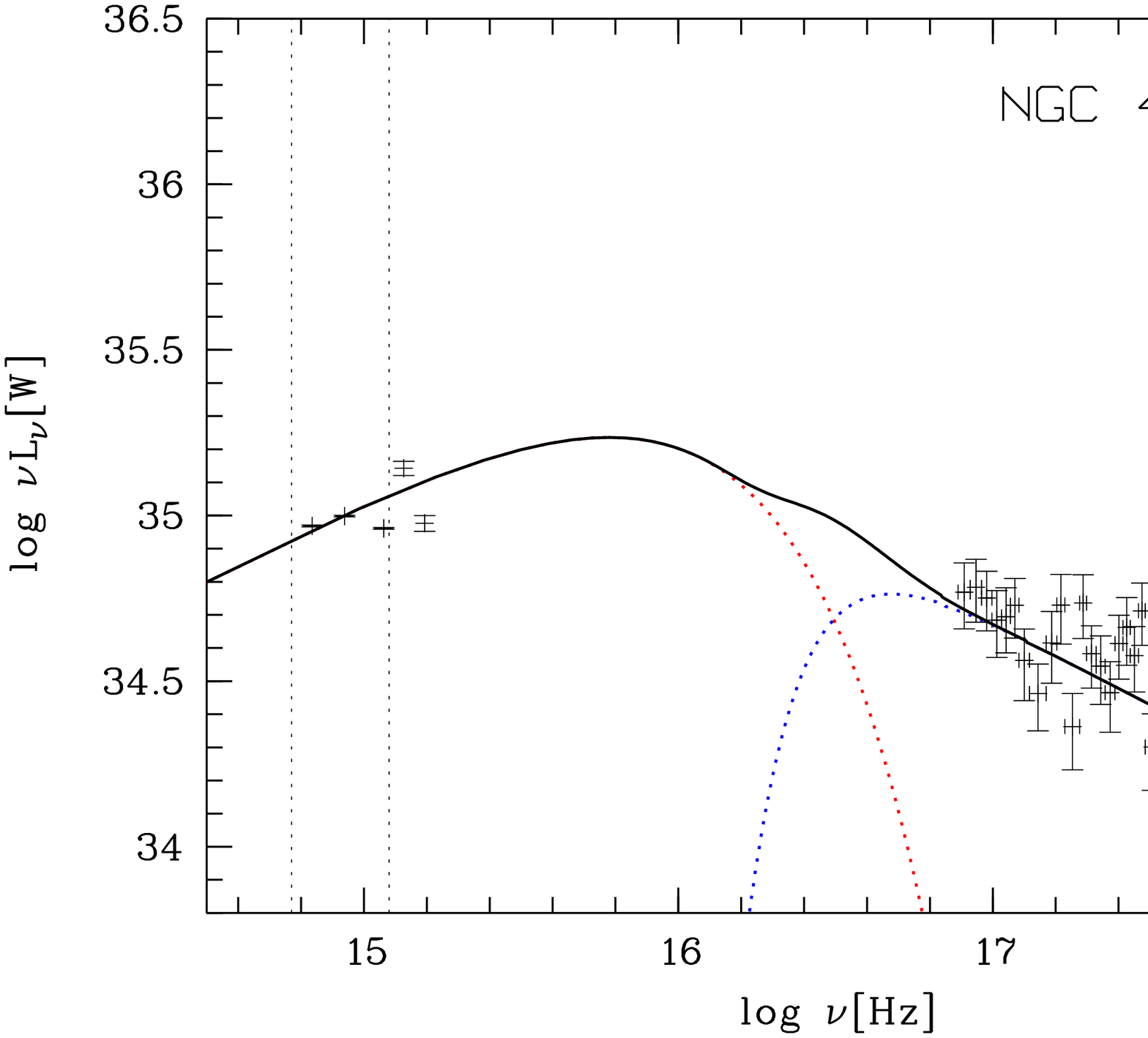}{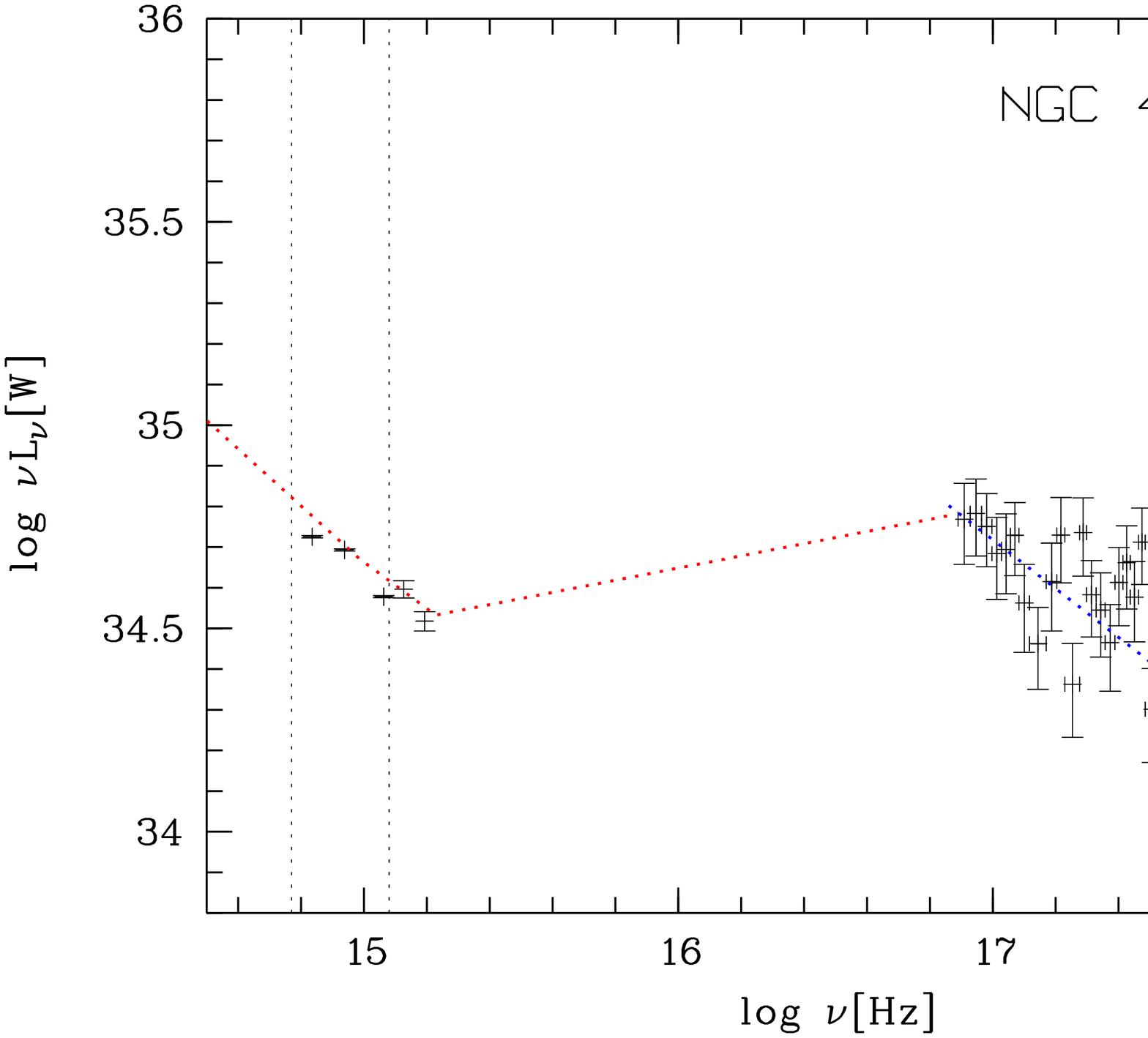}

\plotthree{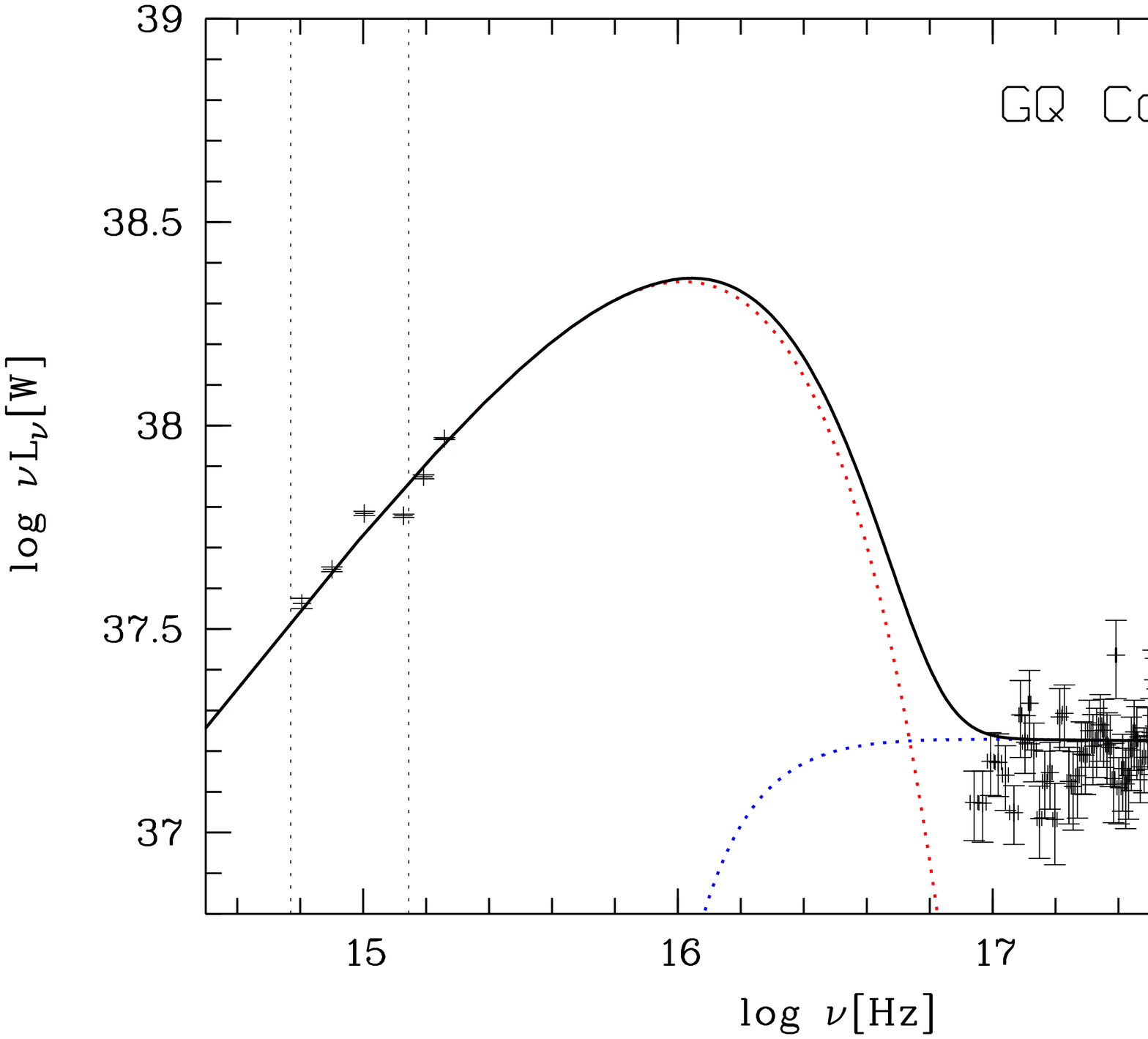}{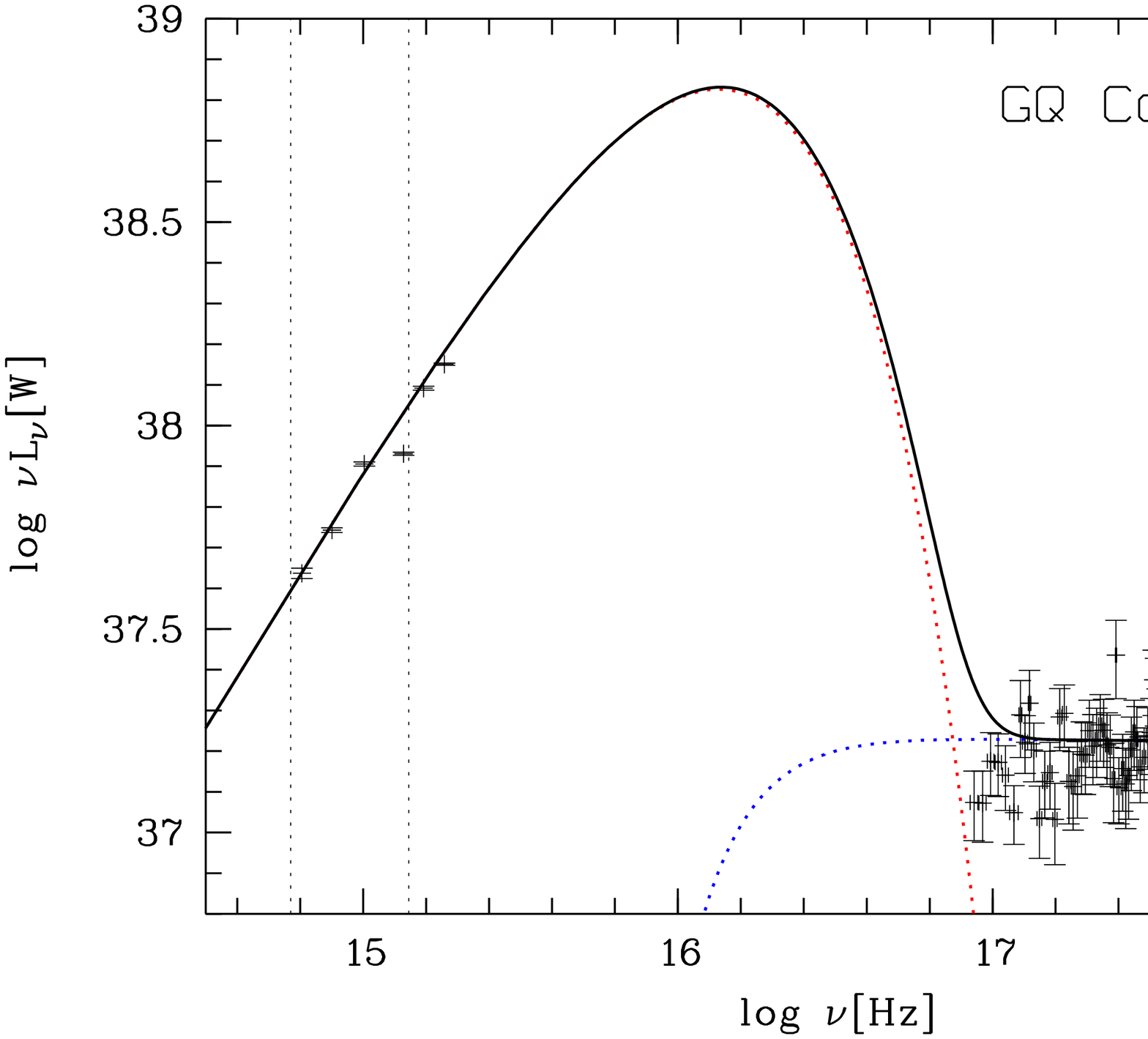}{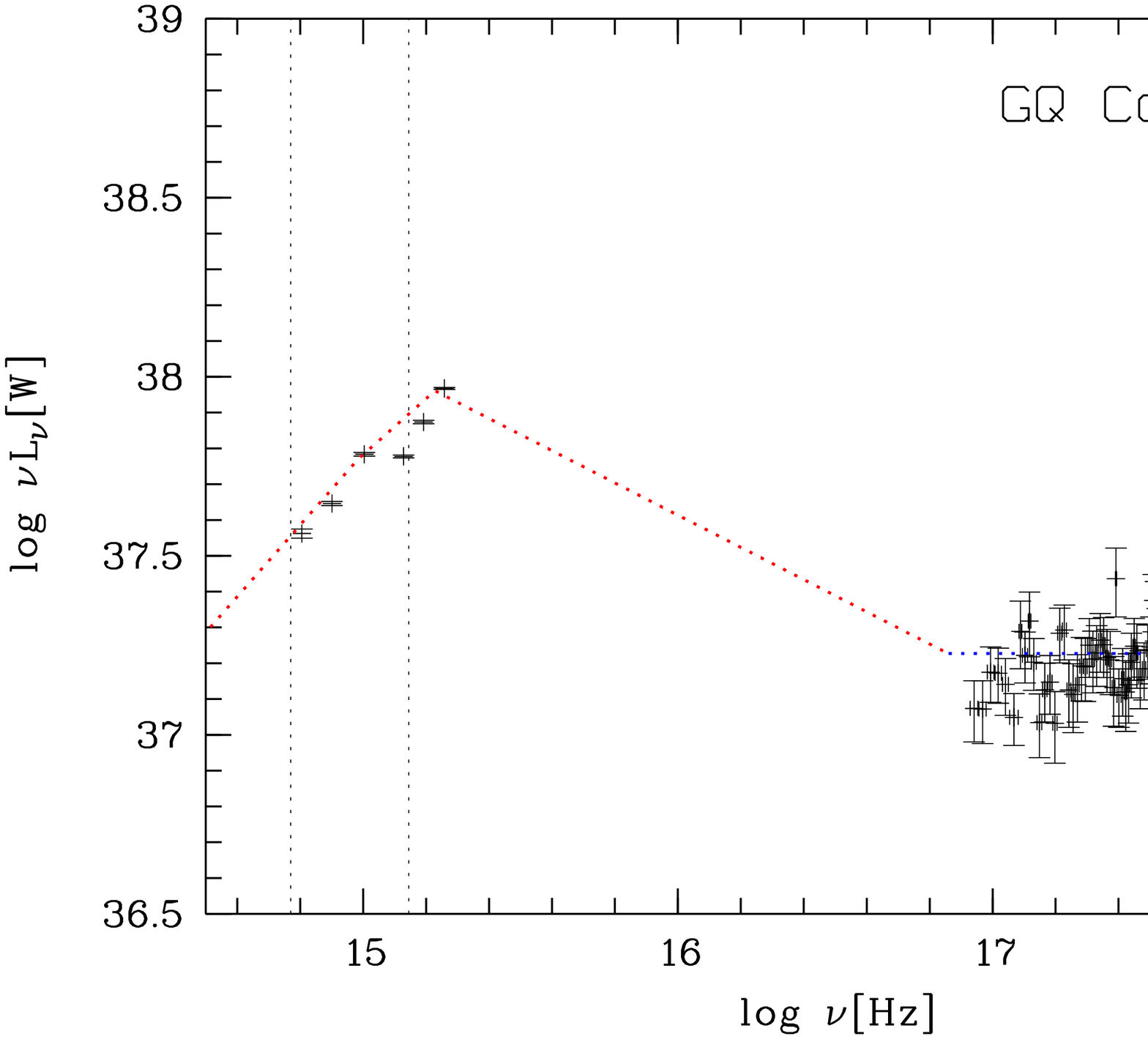}

\plotthree{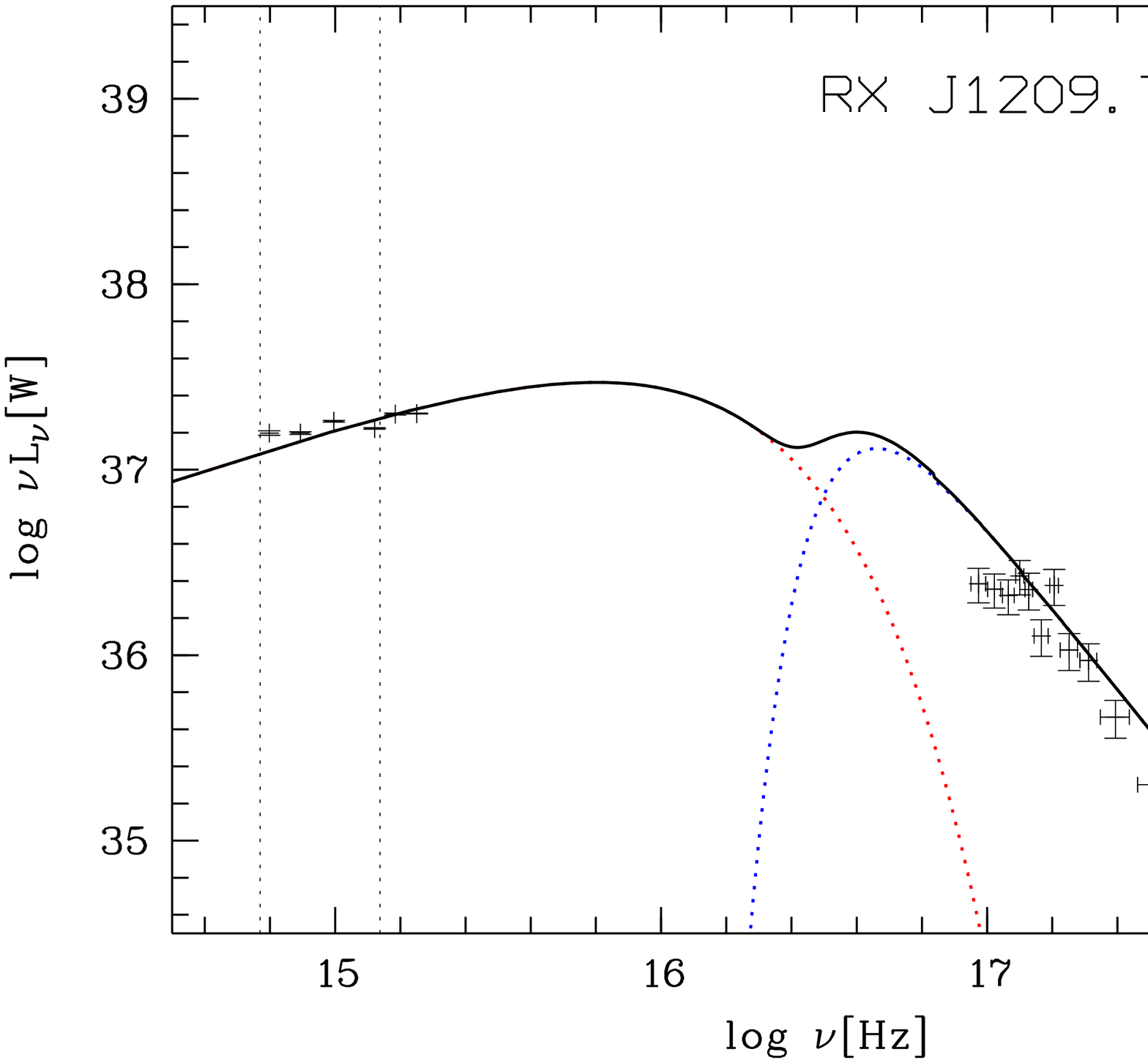}{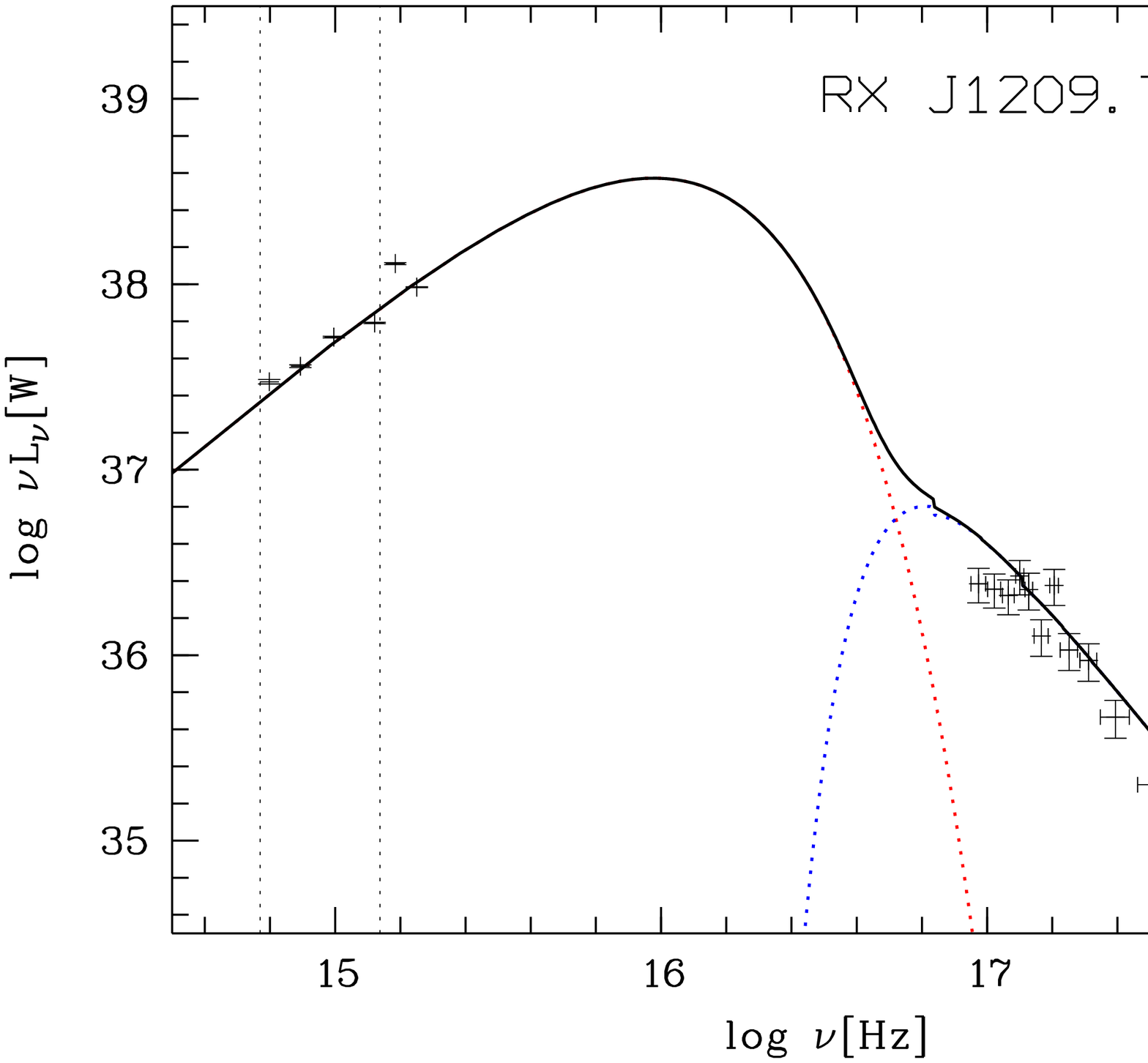}{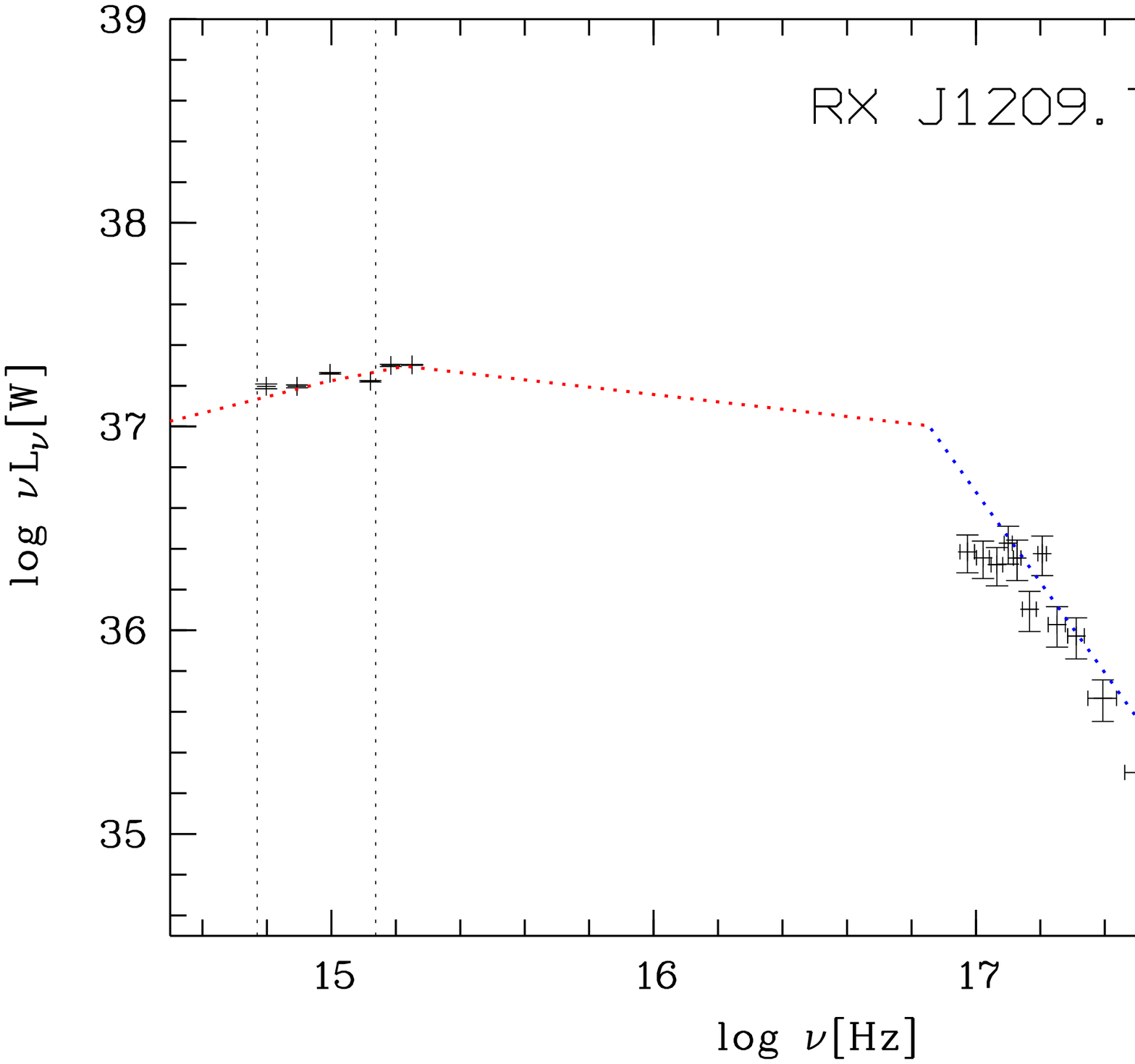}

\plotthree{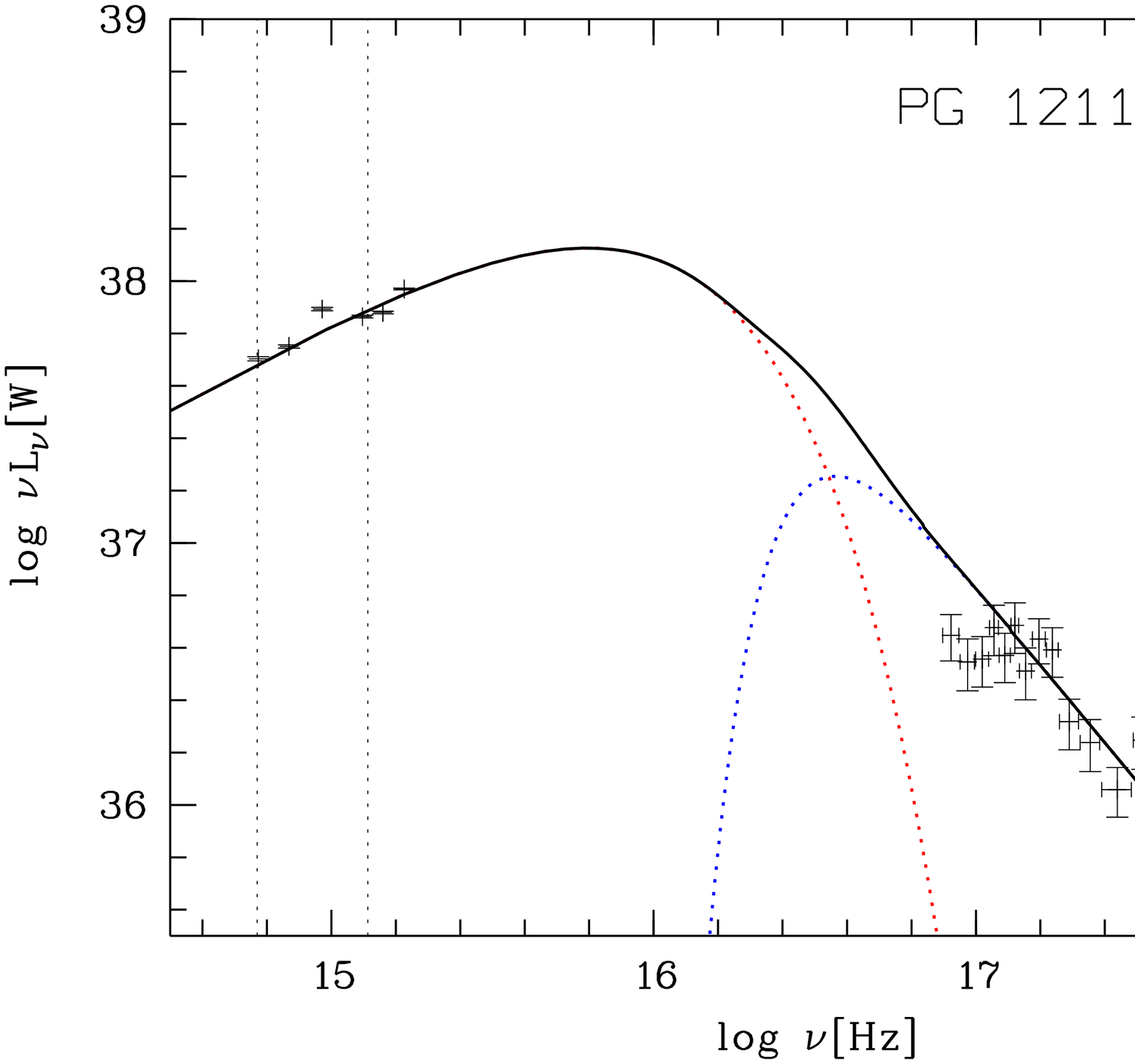}{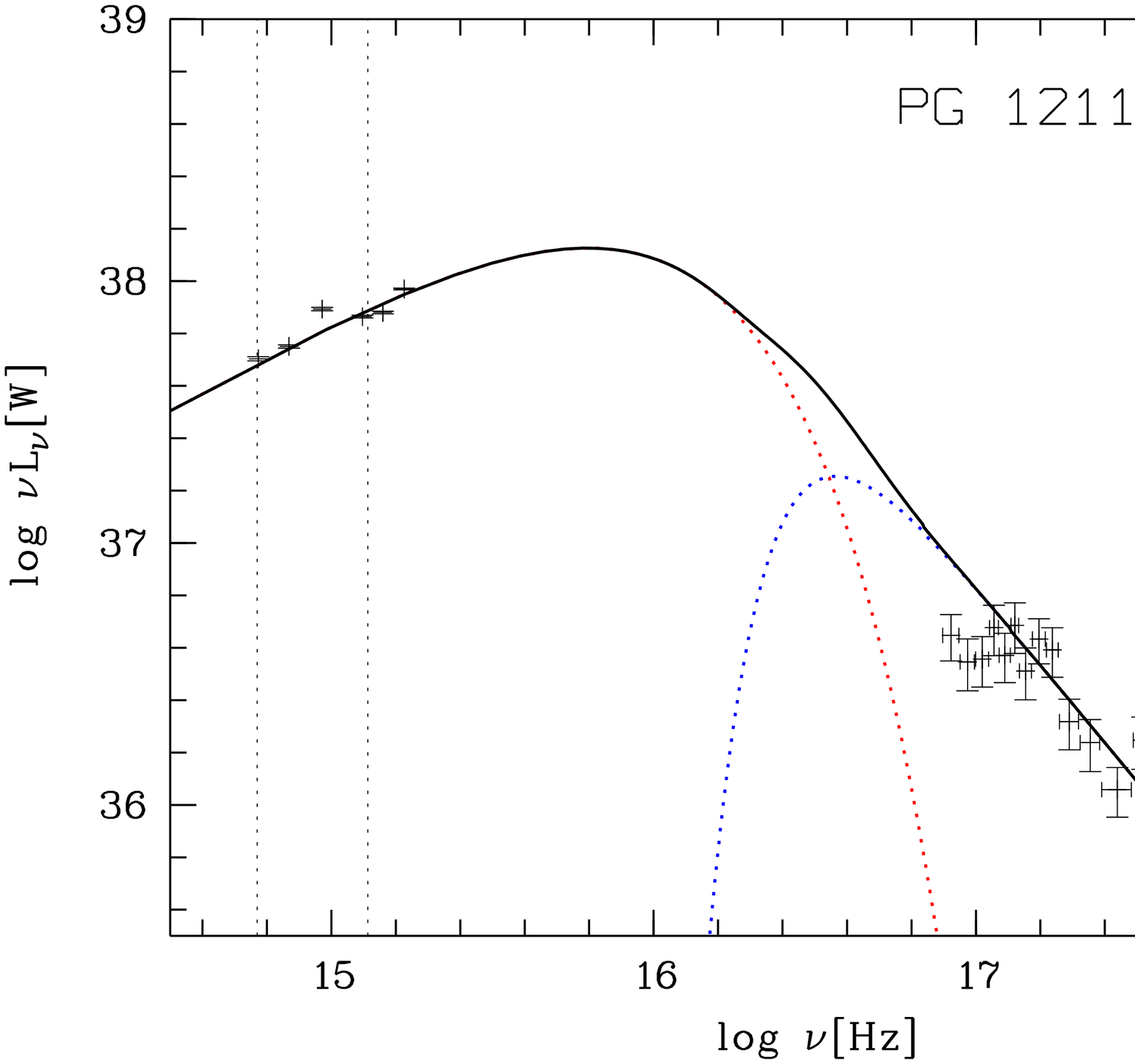}{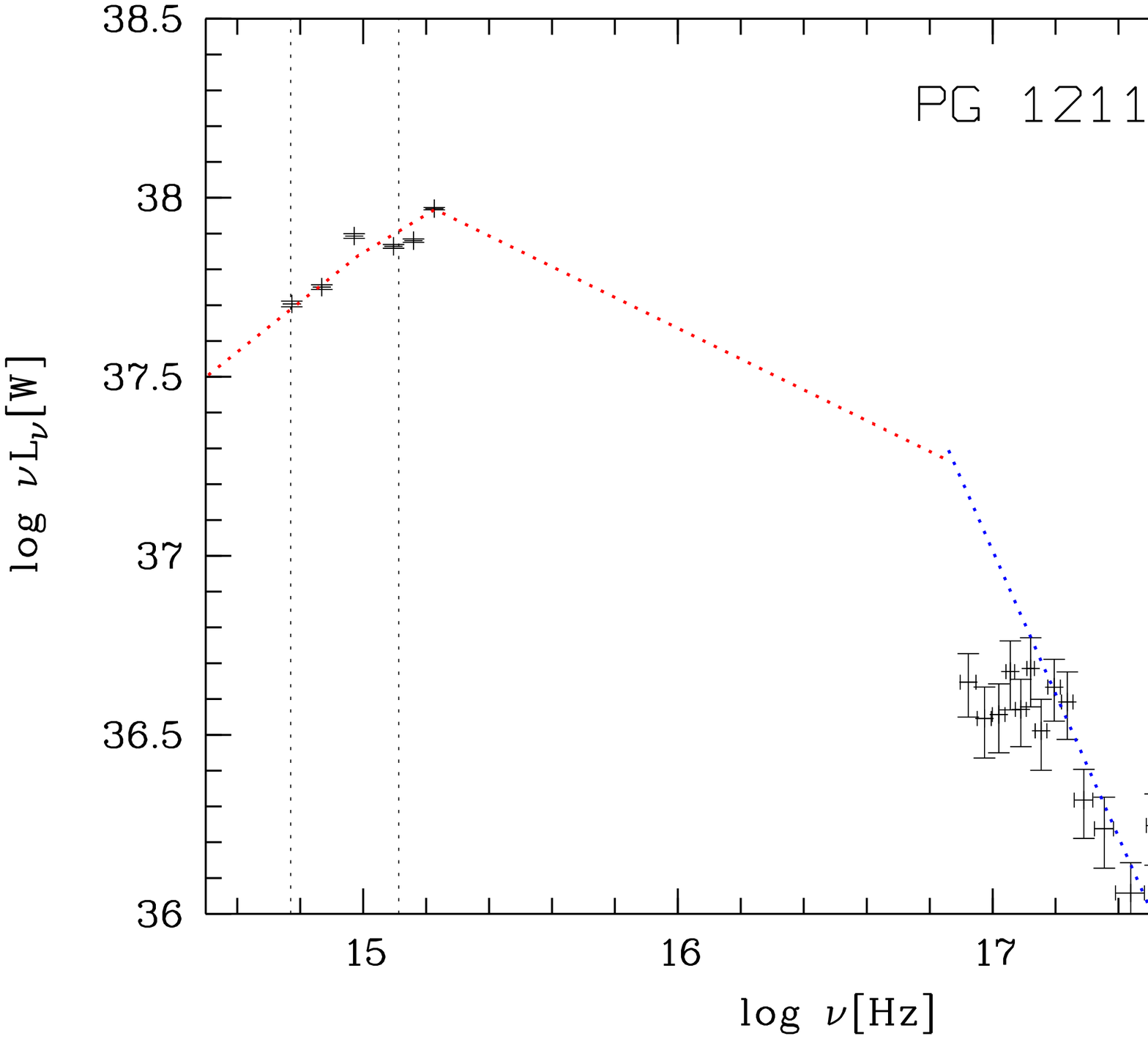}

\plotthree{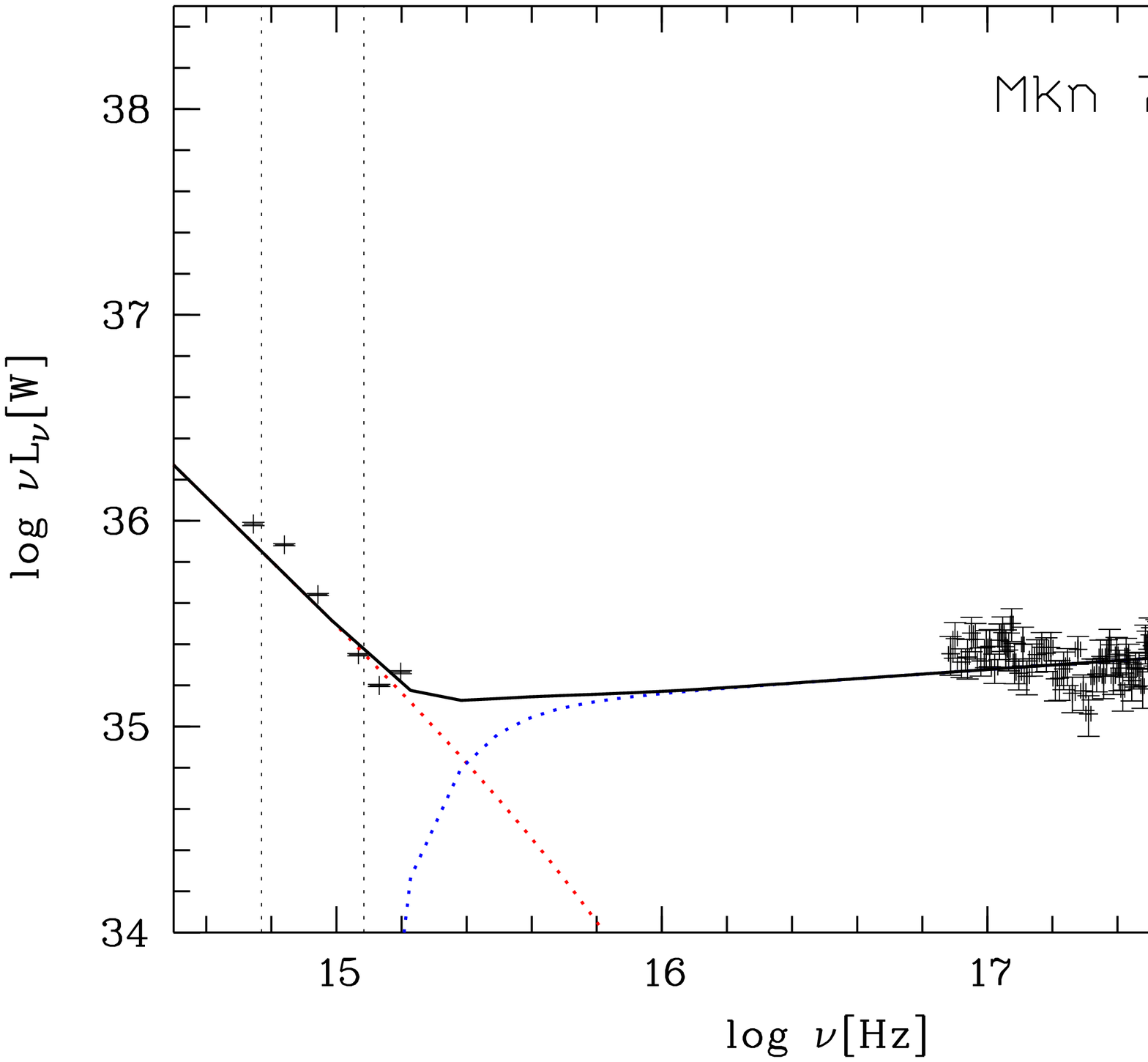}{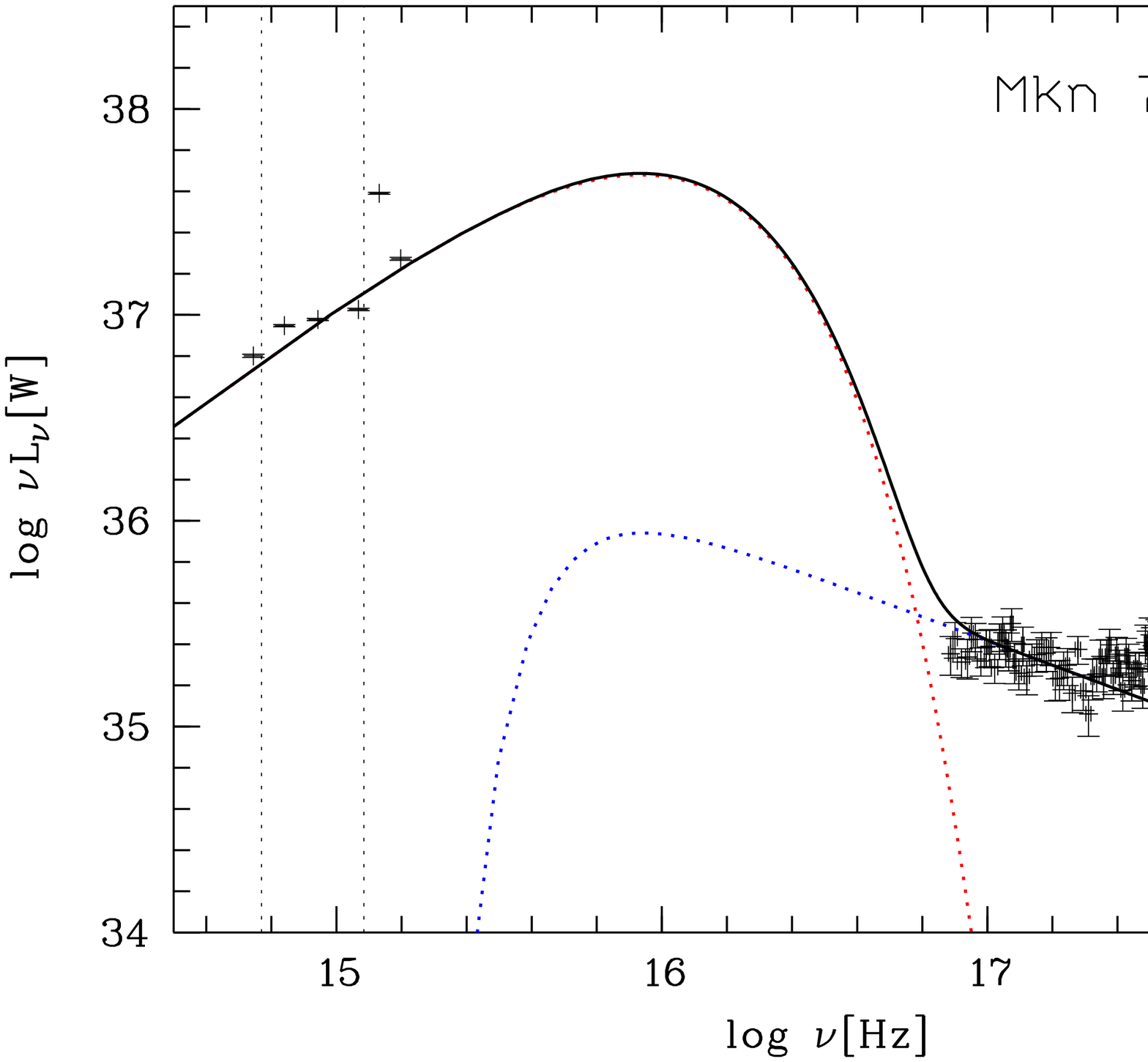}{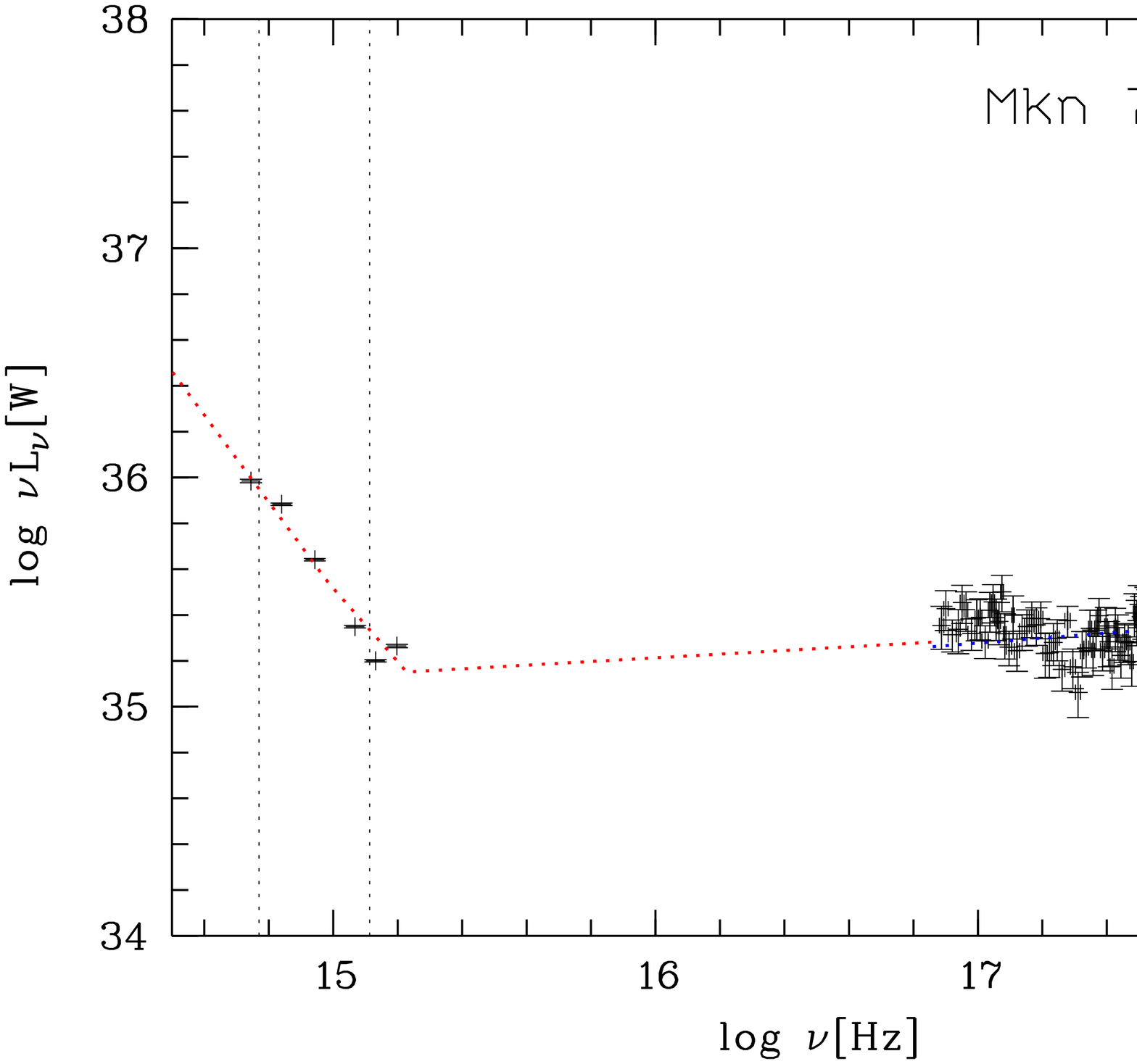}

\end{figure*}

\begin{figure*}
\epsscale{0.60}
\plotthree{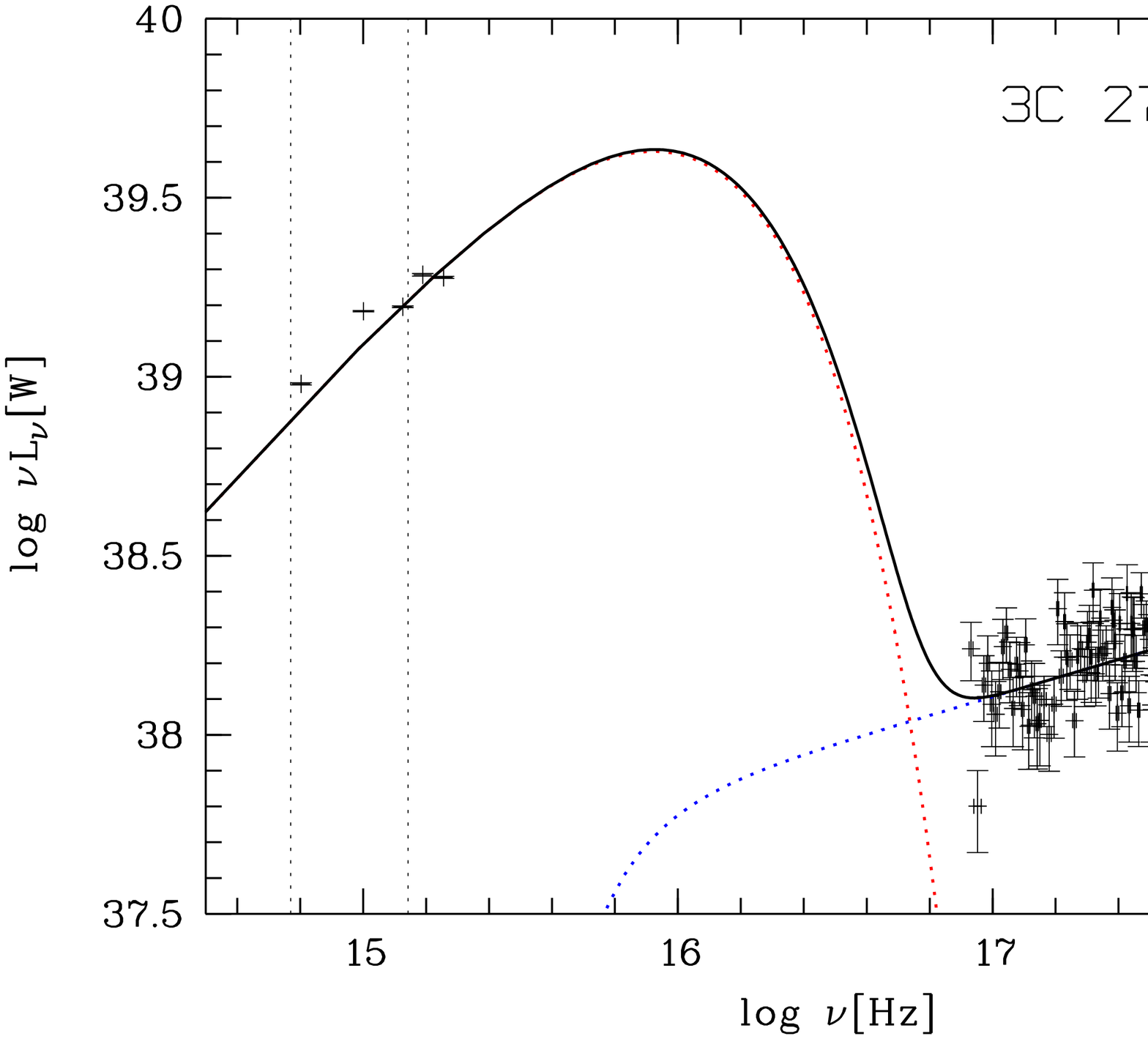}{f25_t.ps}{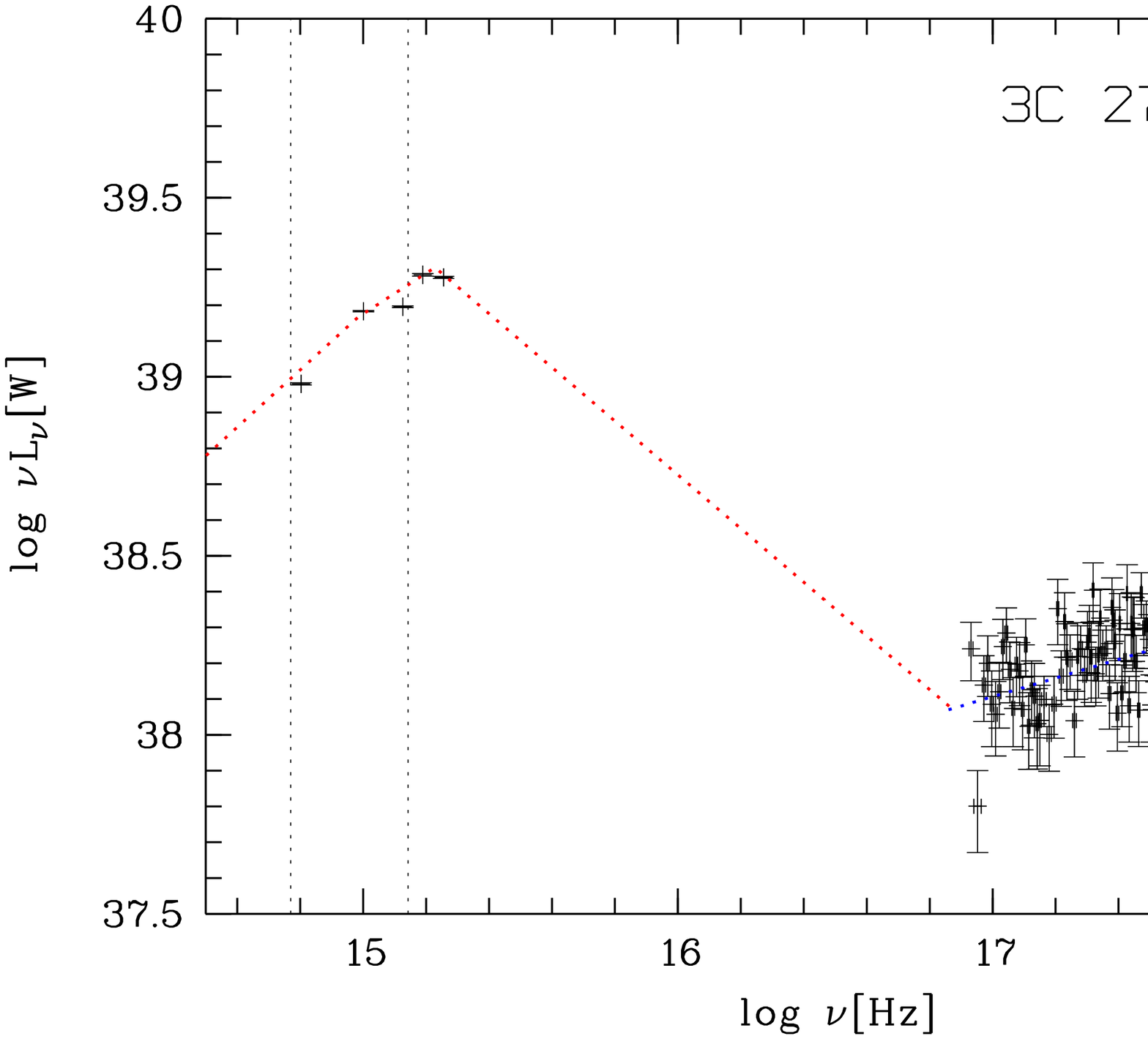}

\plotthree{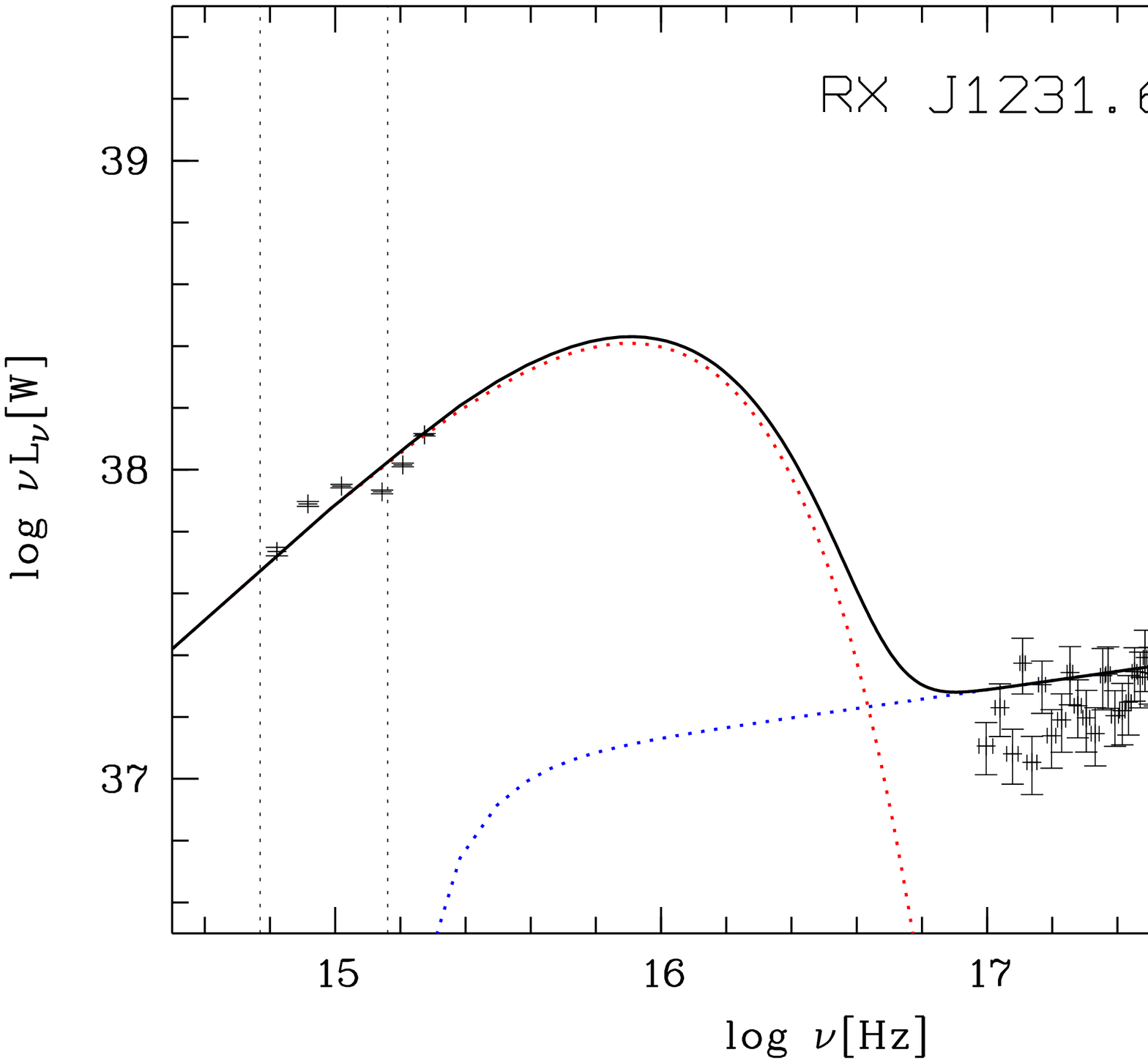}{f25_t.ps}{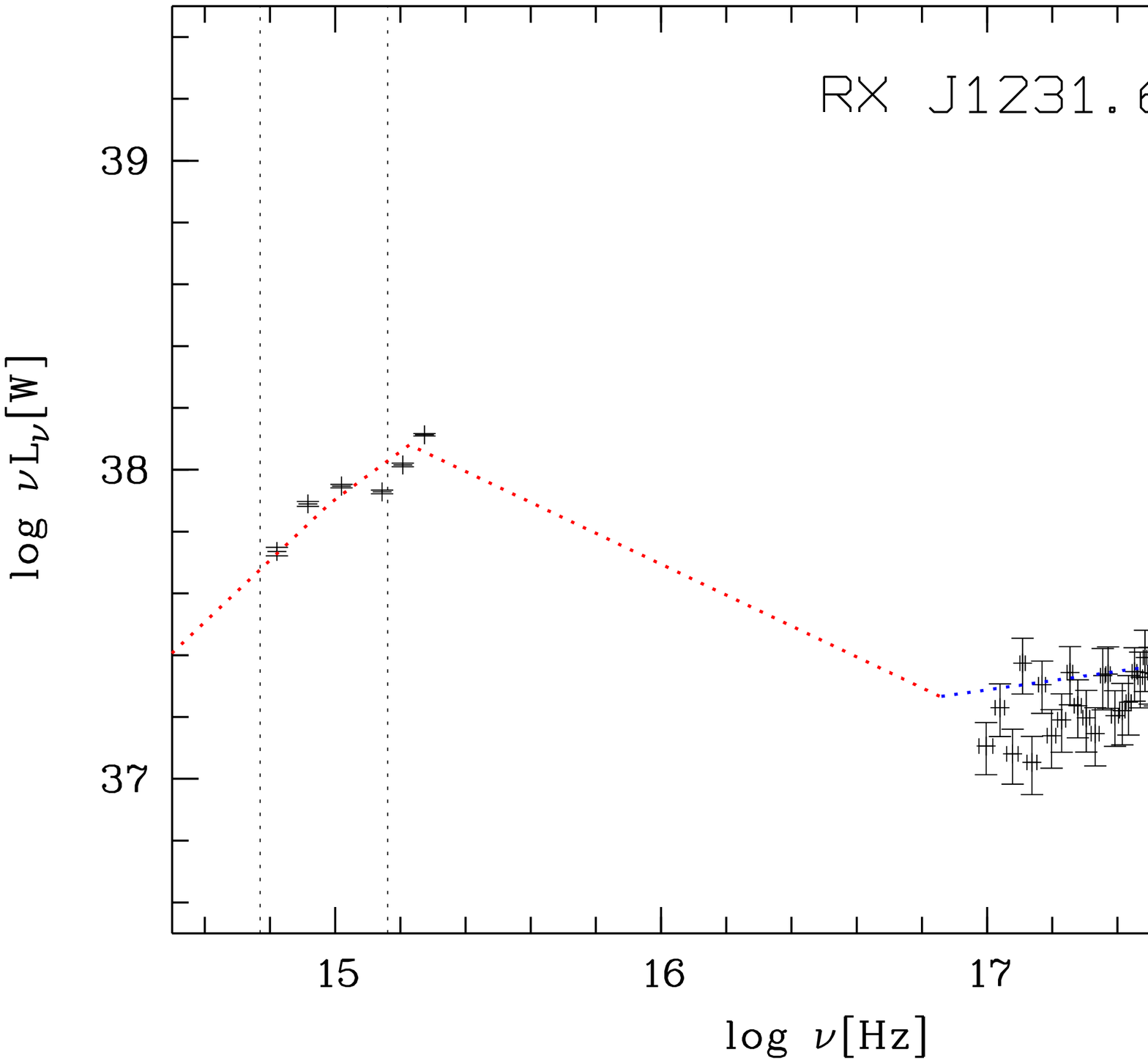}

\plotthree{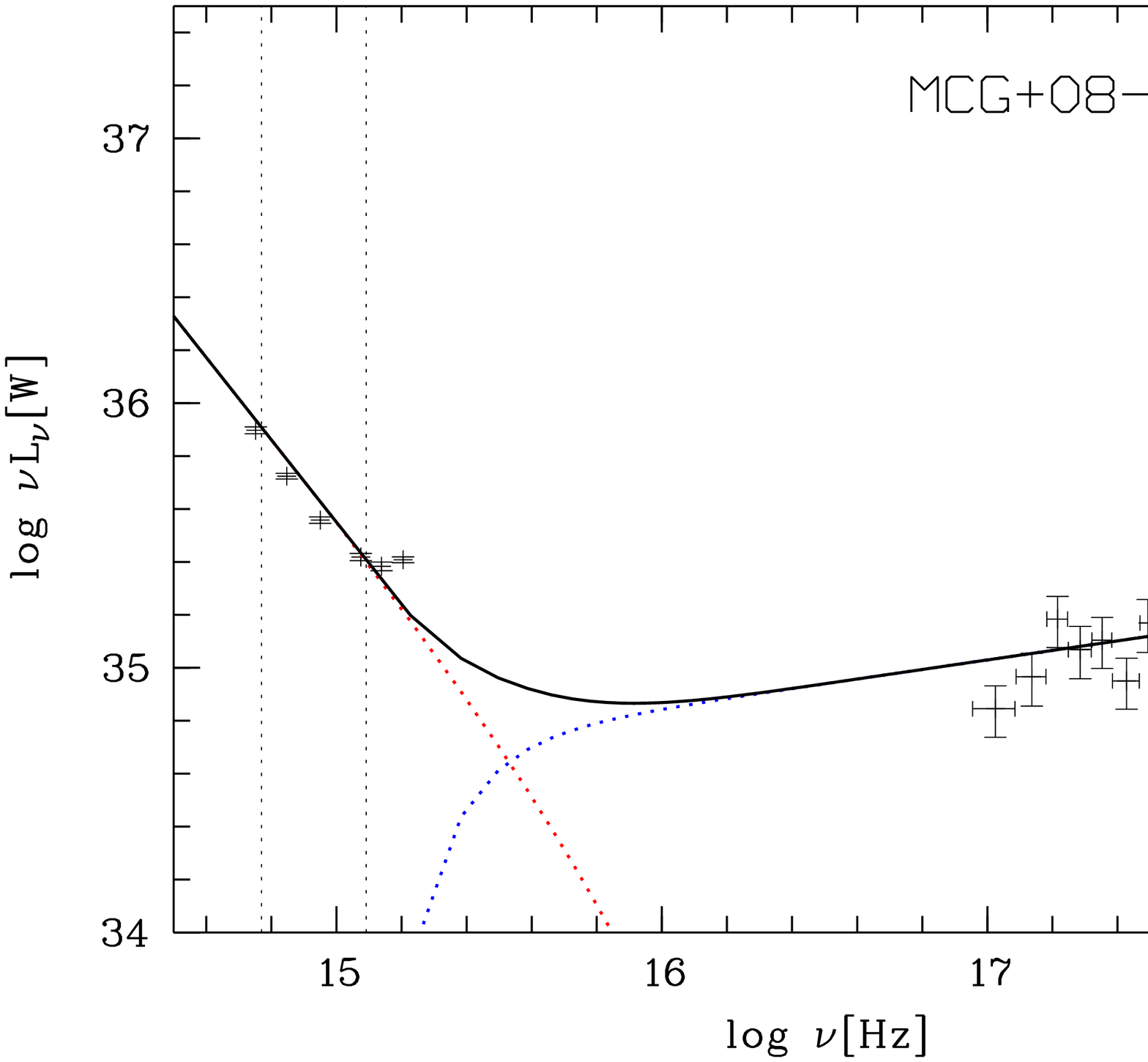}{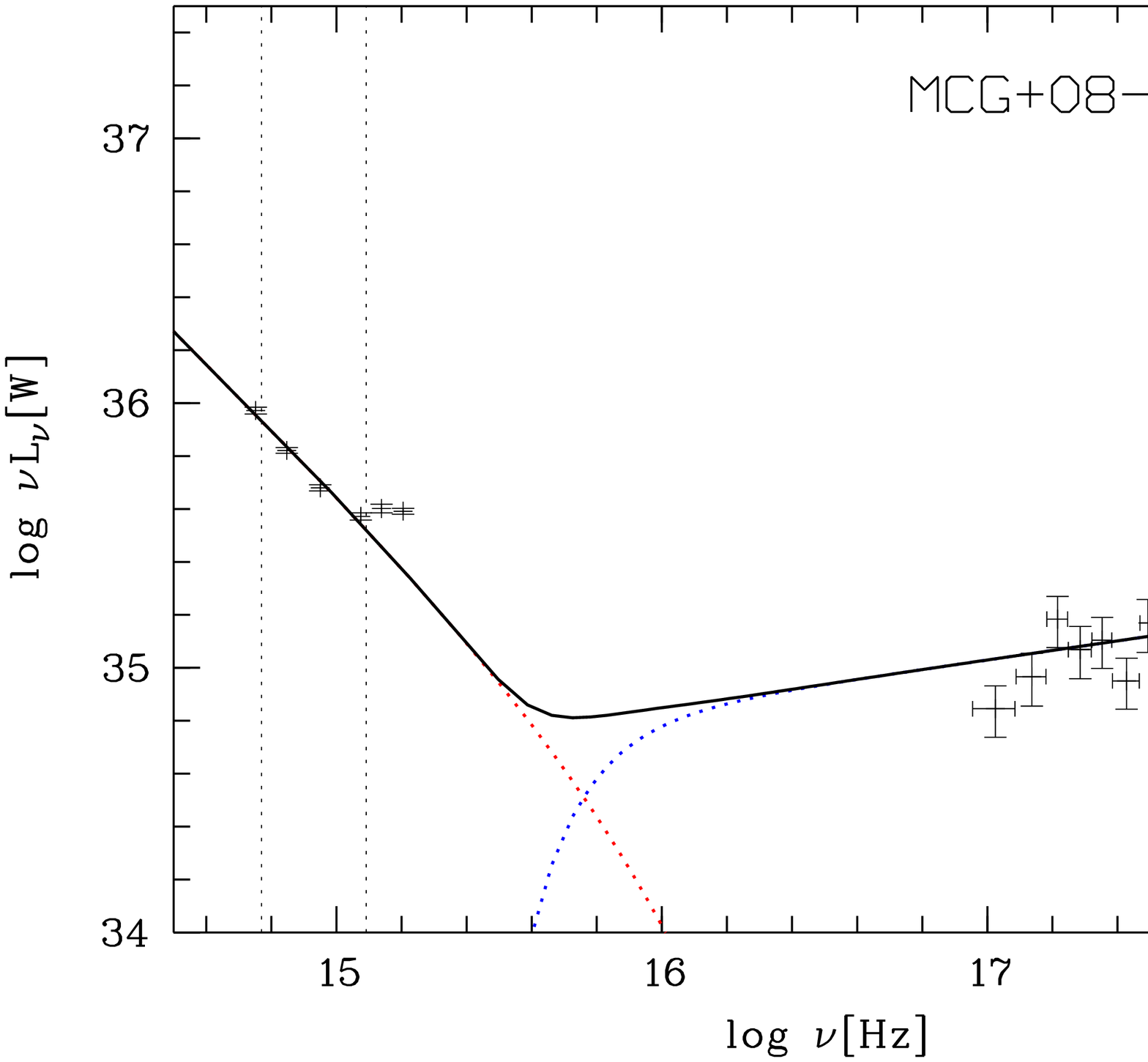}{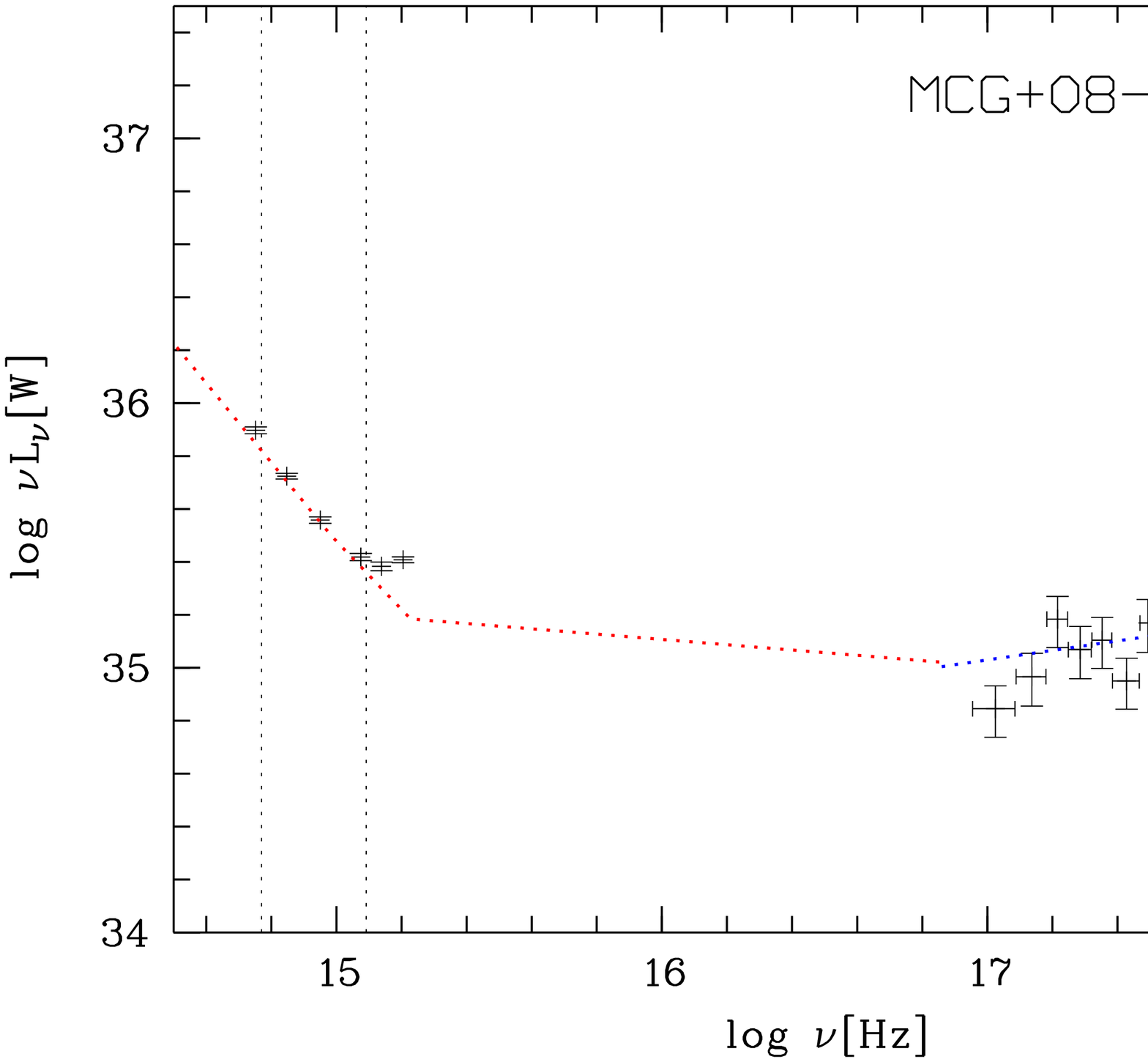}

\plotthree{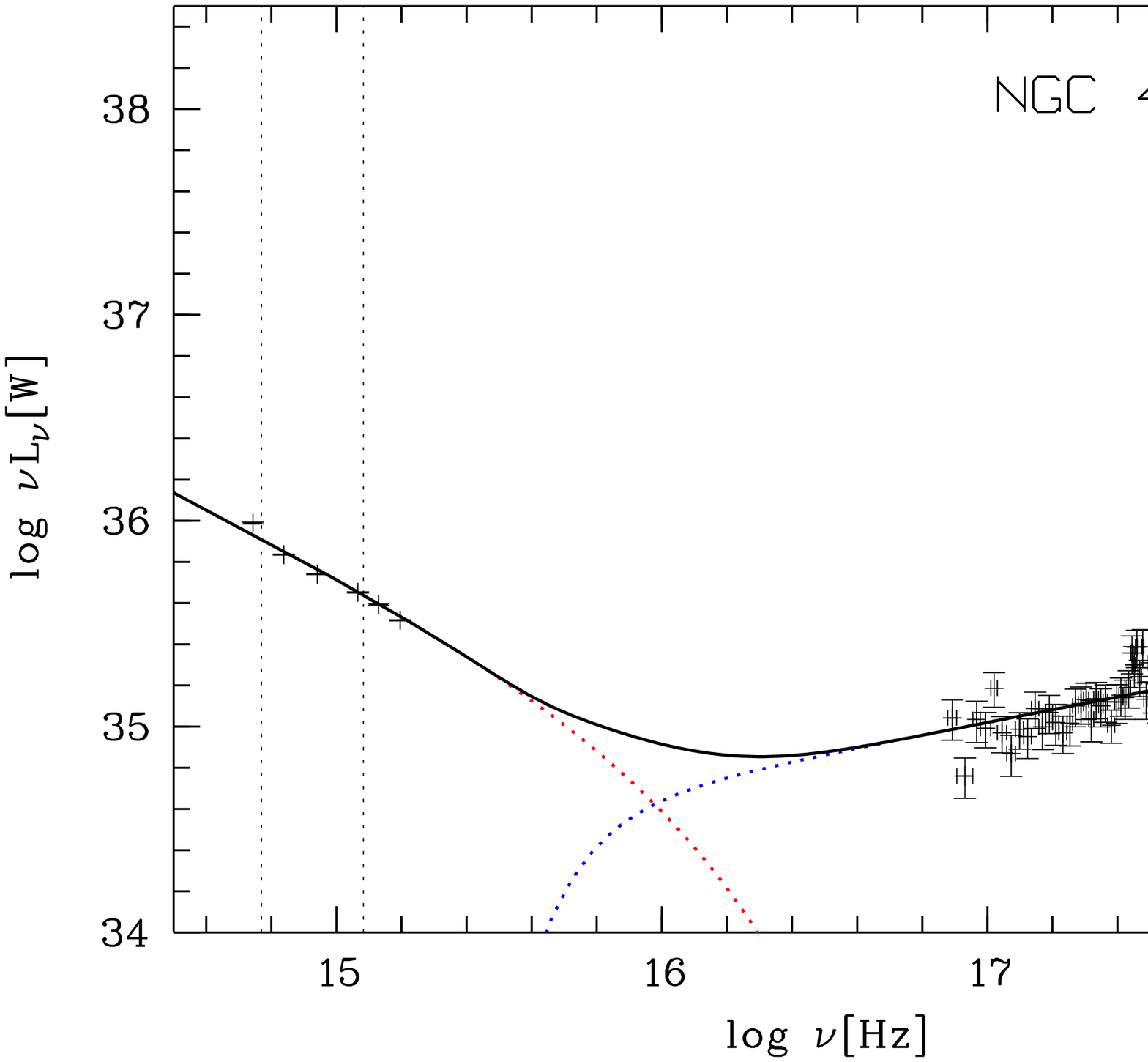}{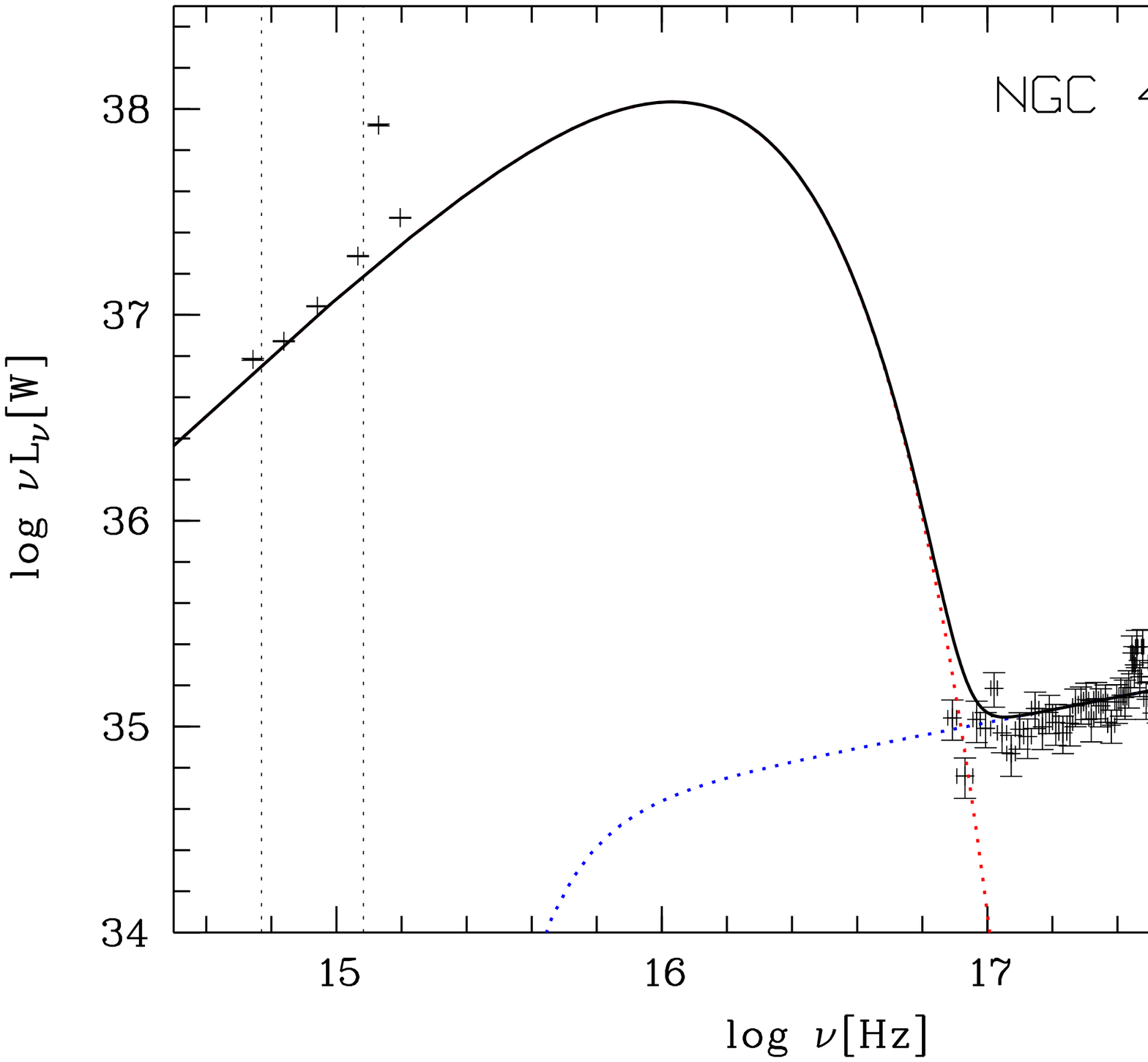}{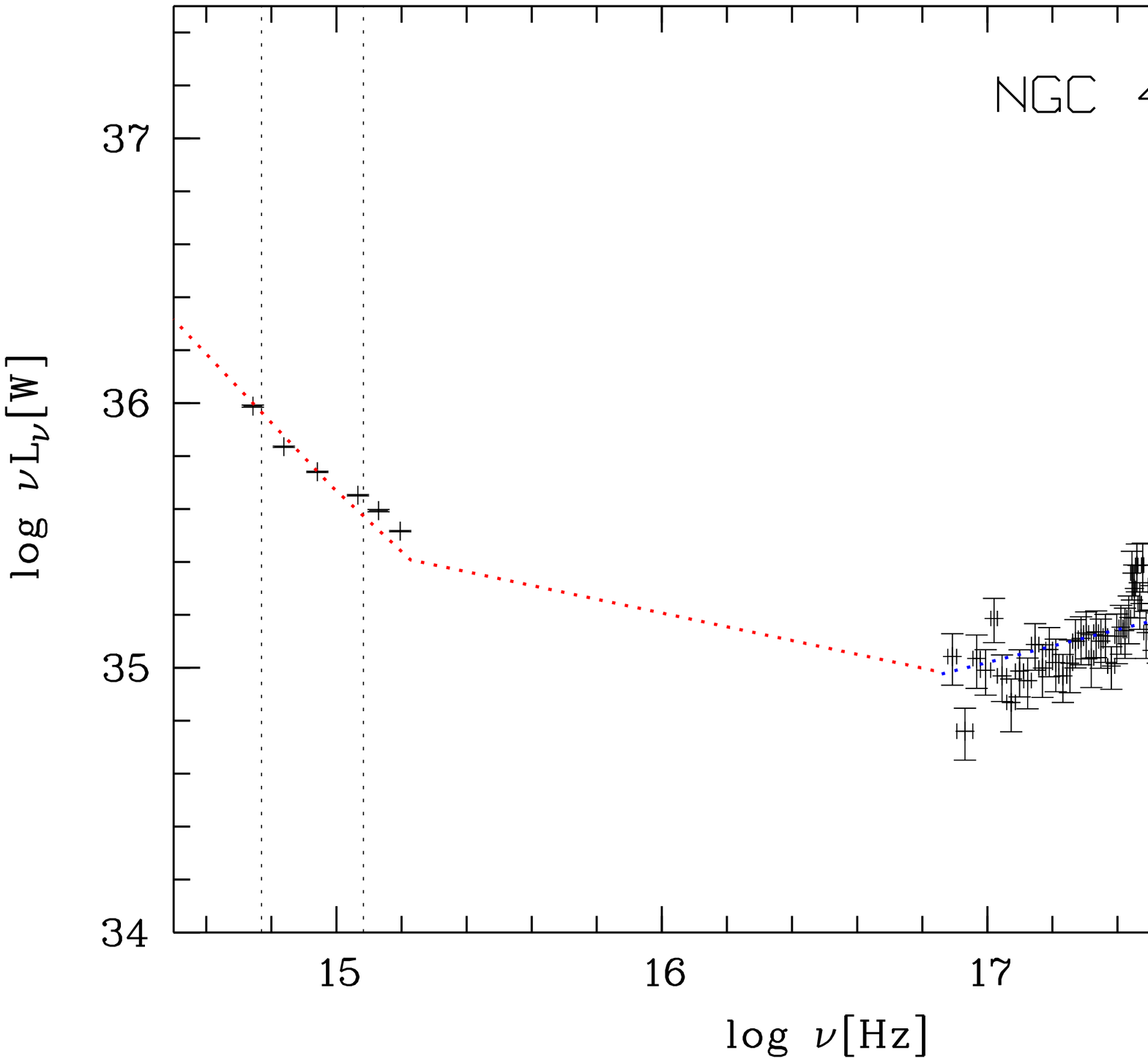}

\plotthree{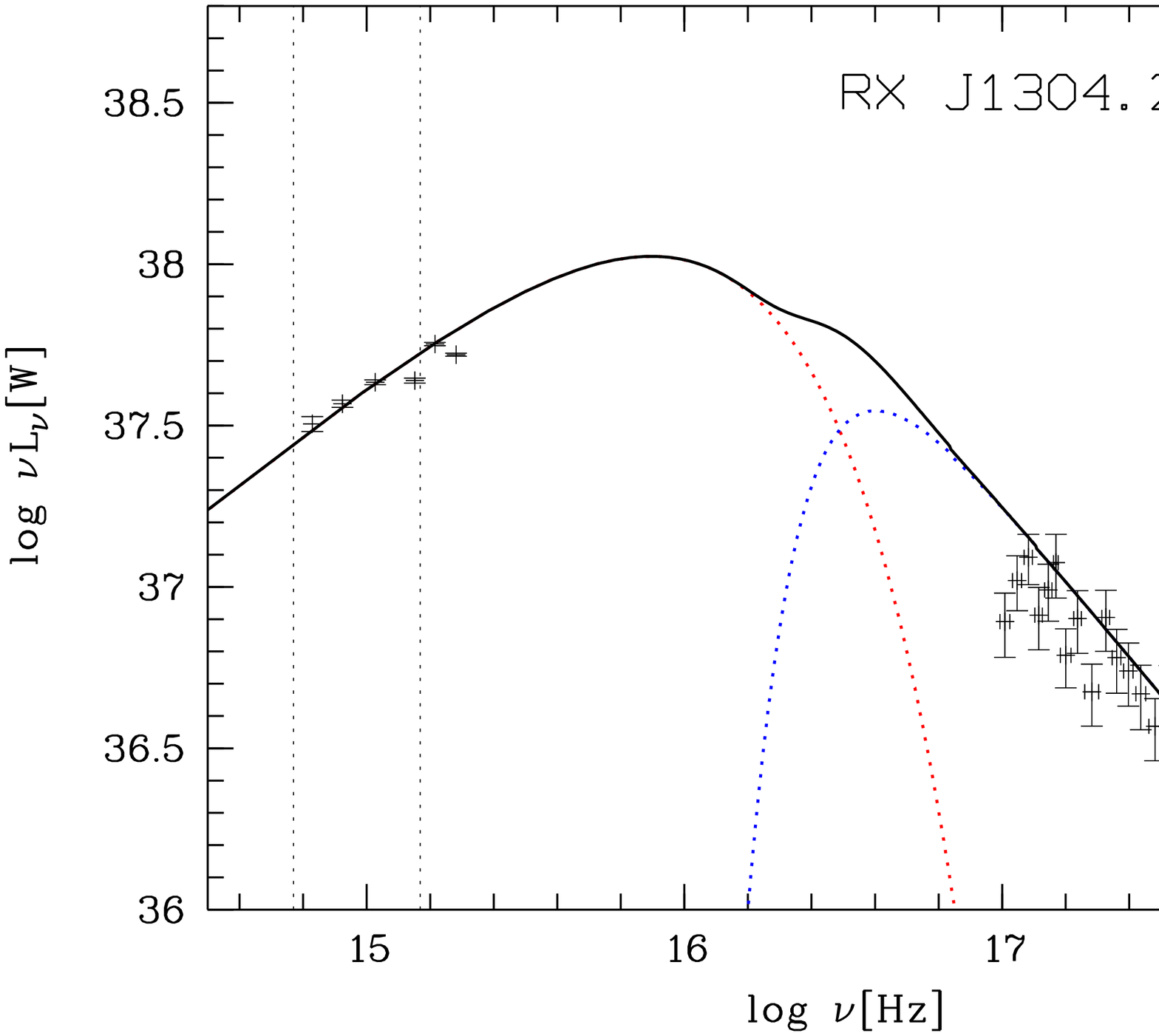}{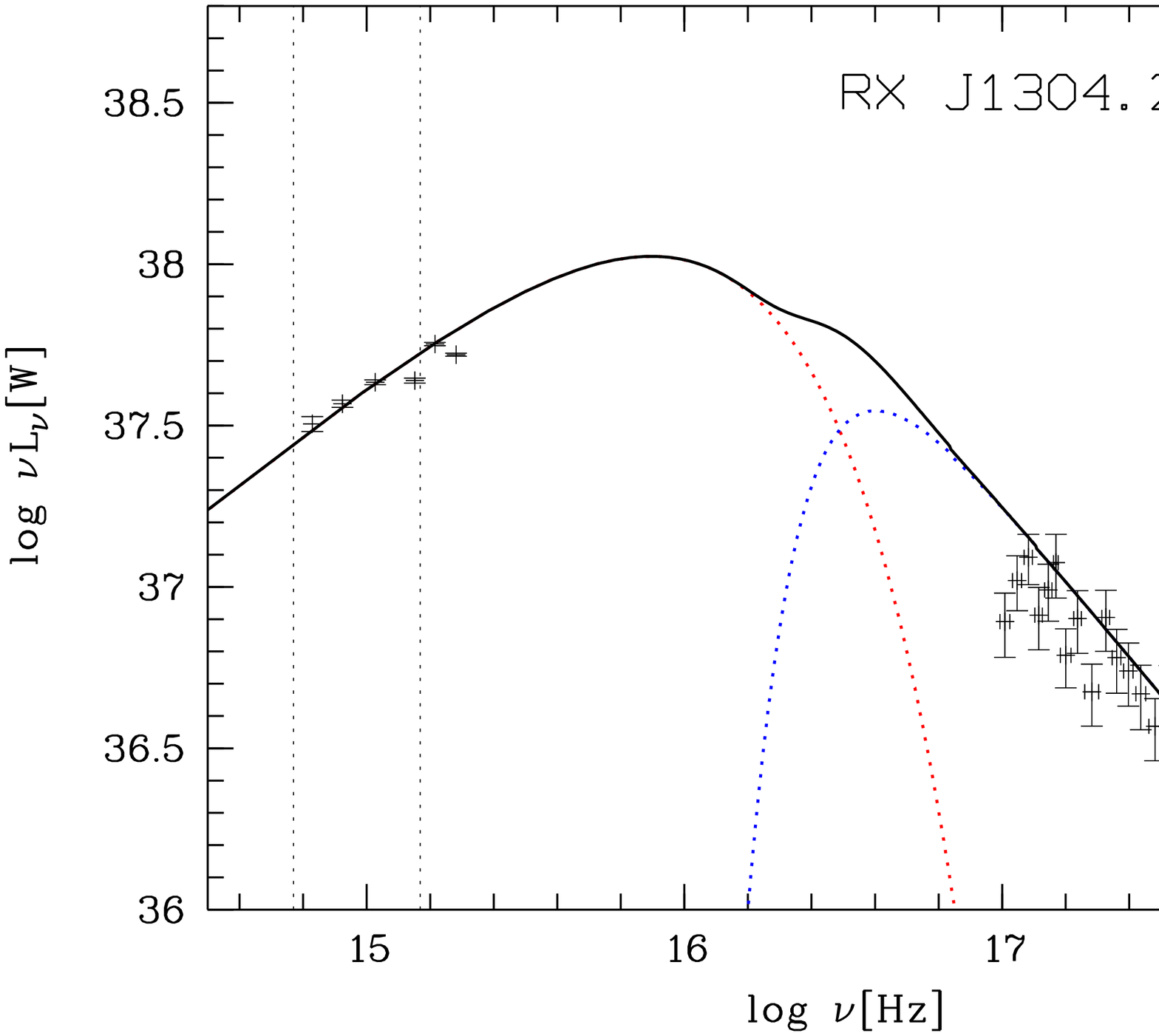}{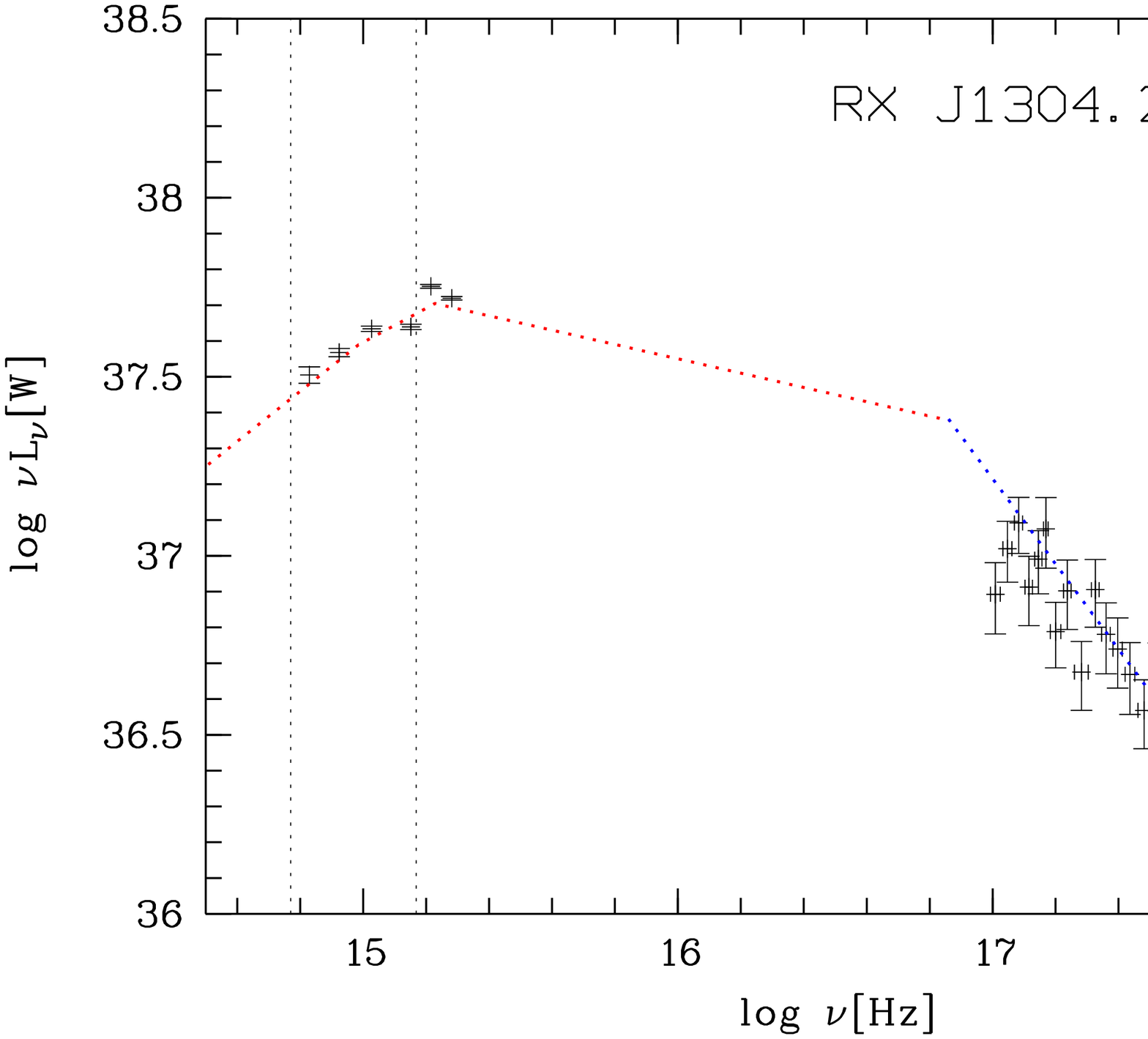}
\end{figure*}

\begin{figure*}
\epsscale{0.60}
\plotthree{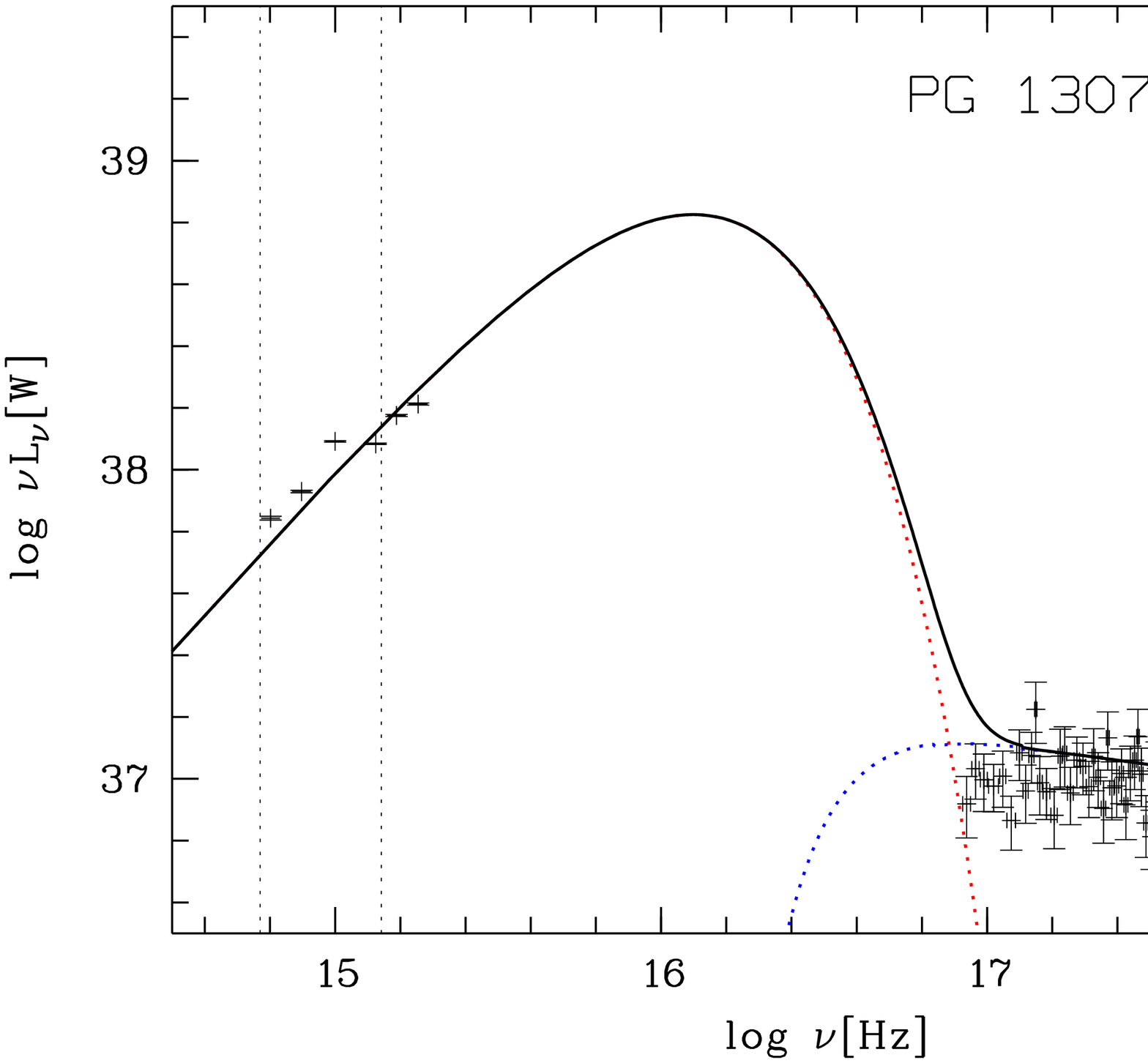}{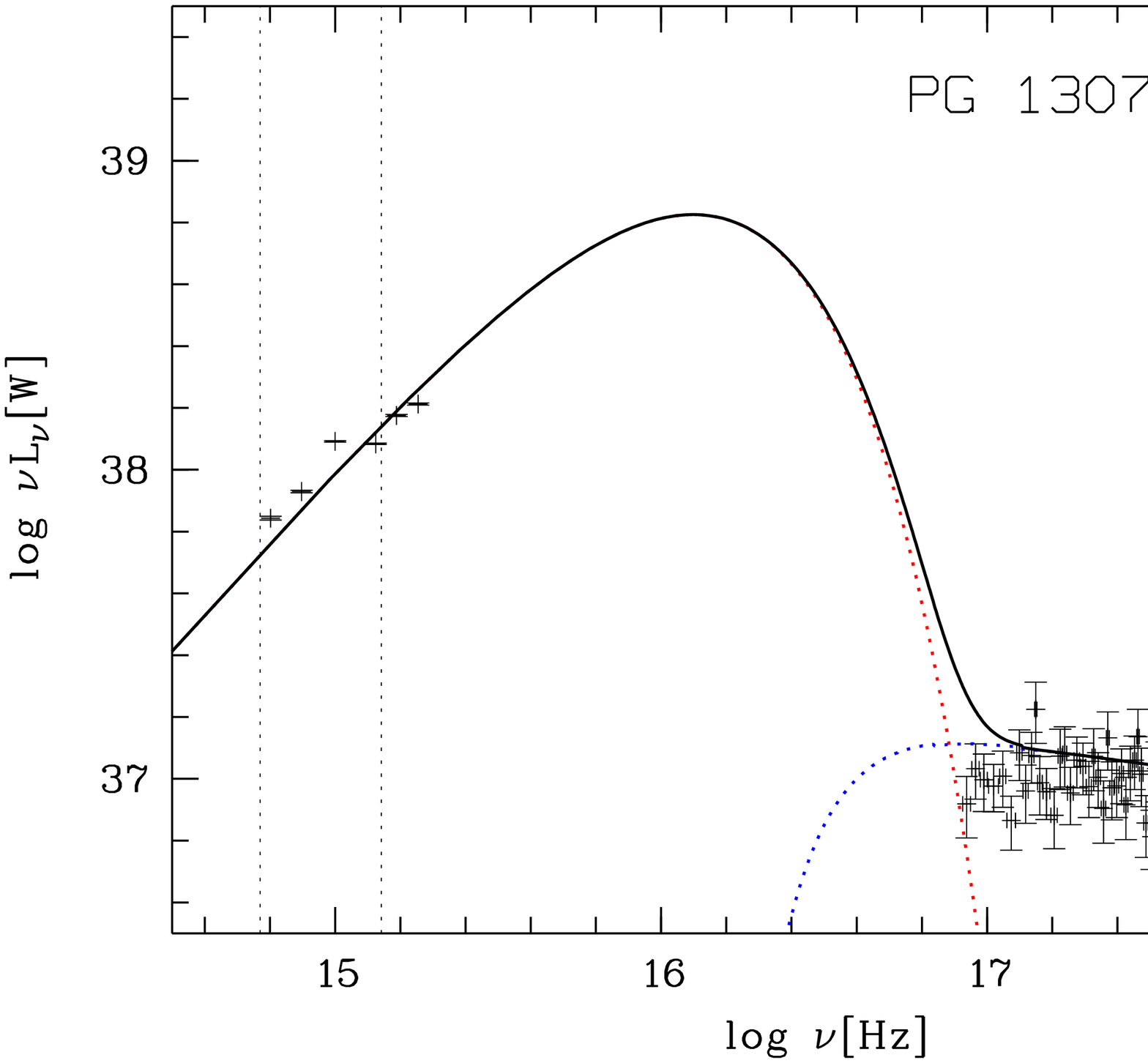}{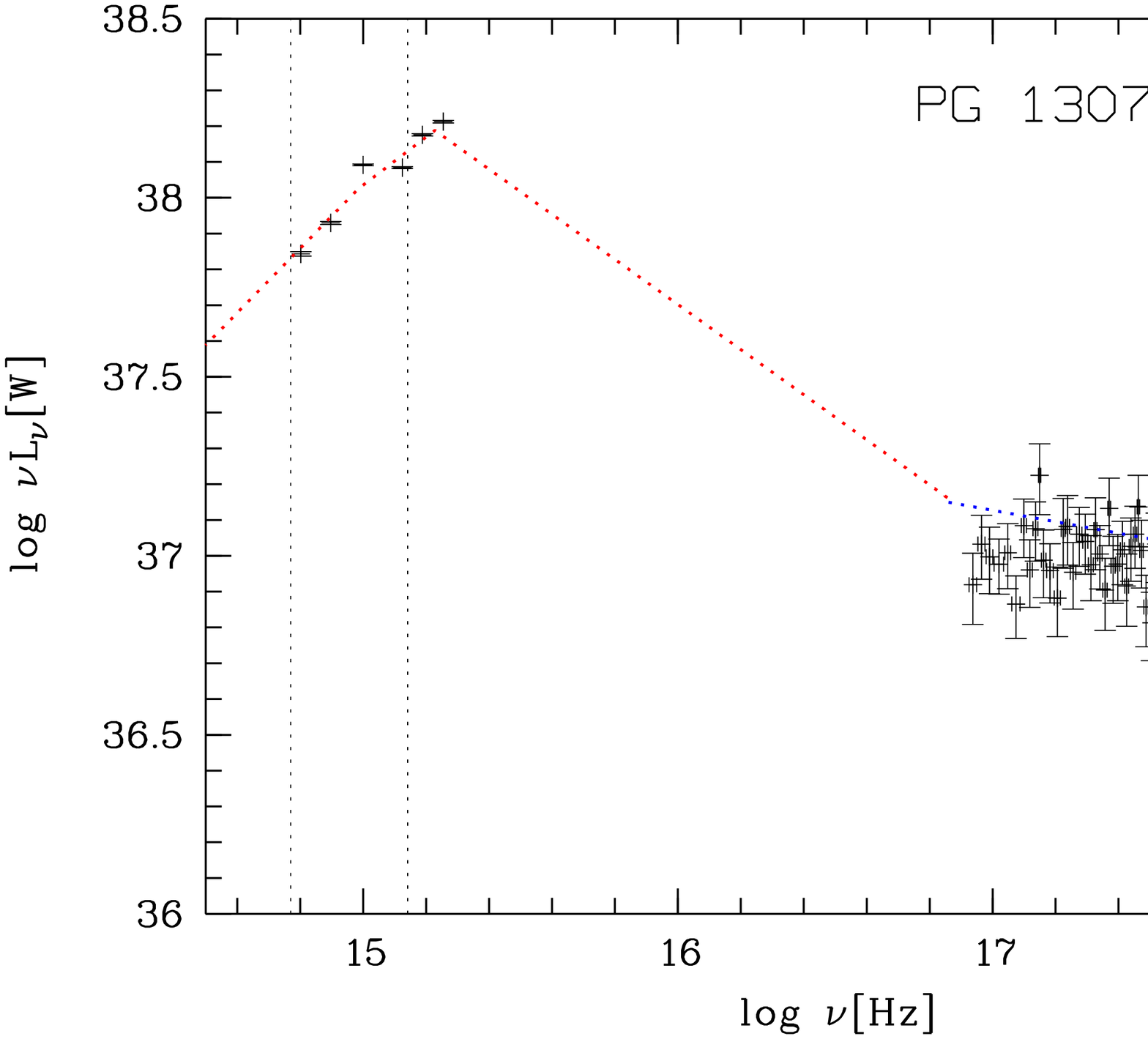}

\plotthree{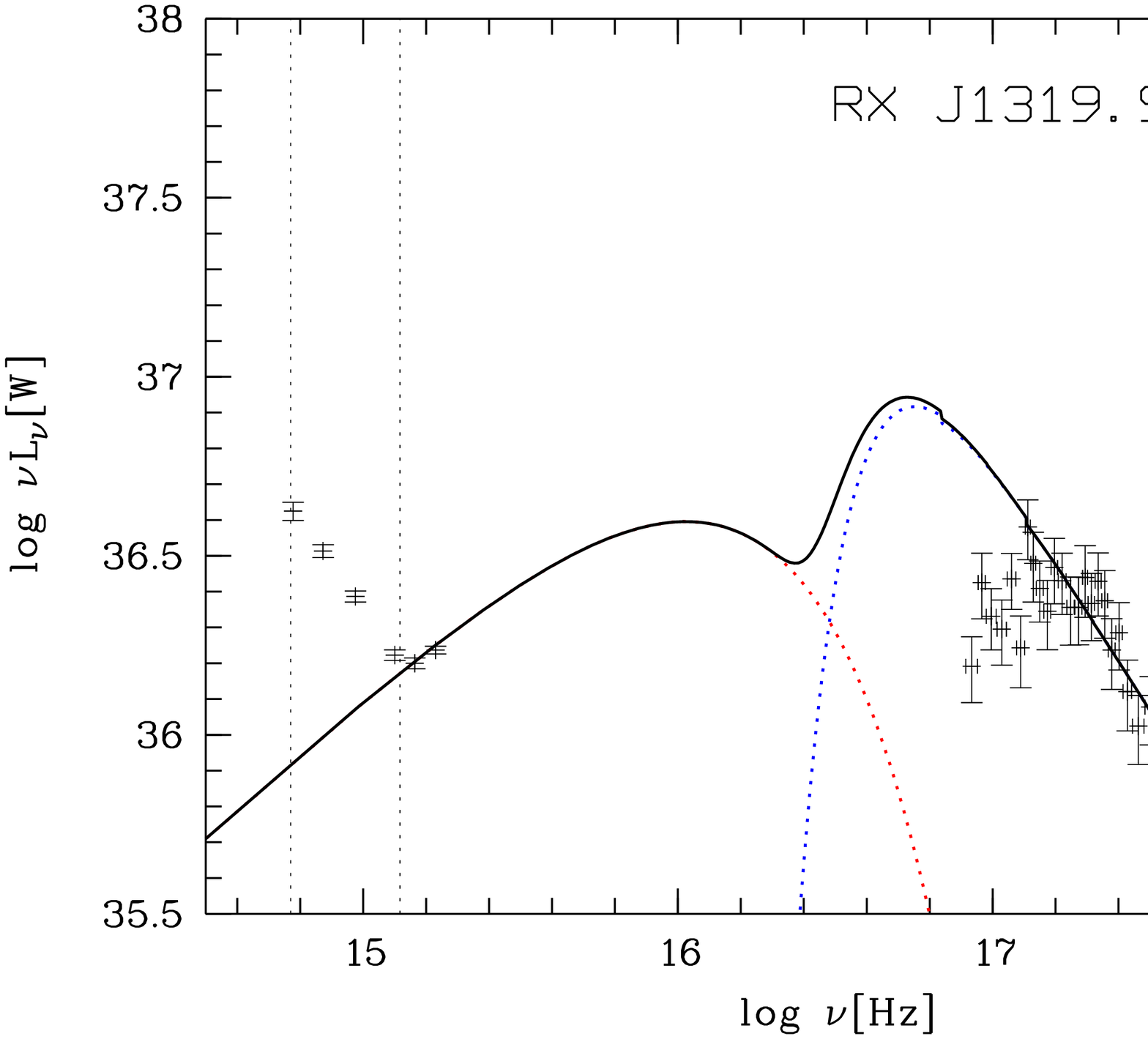}{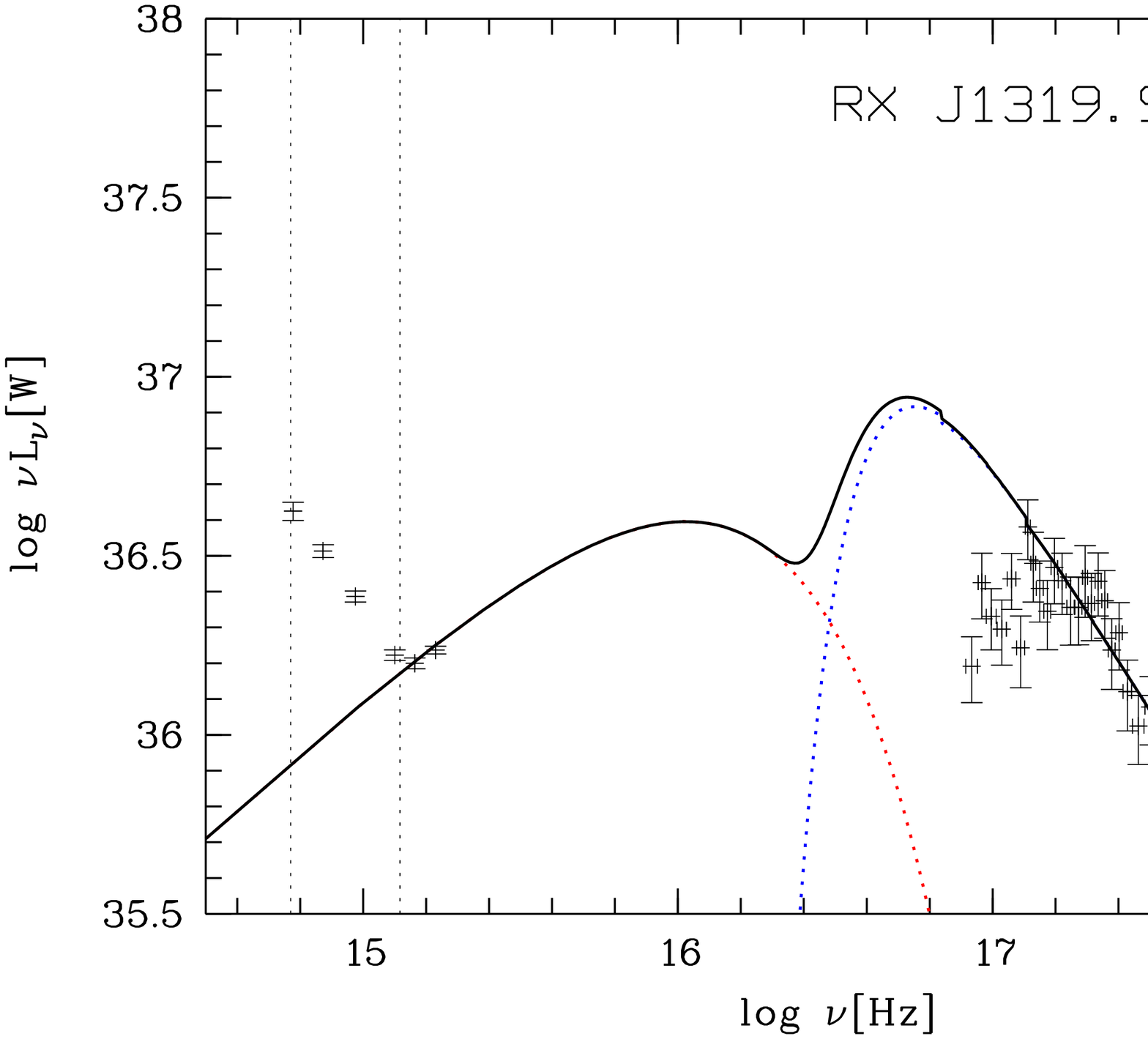}{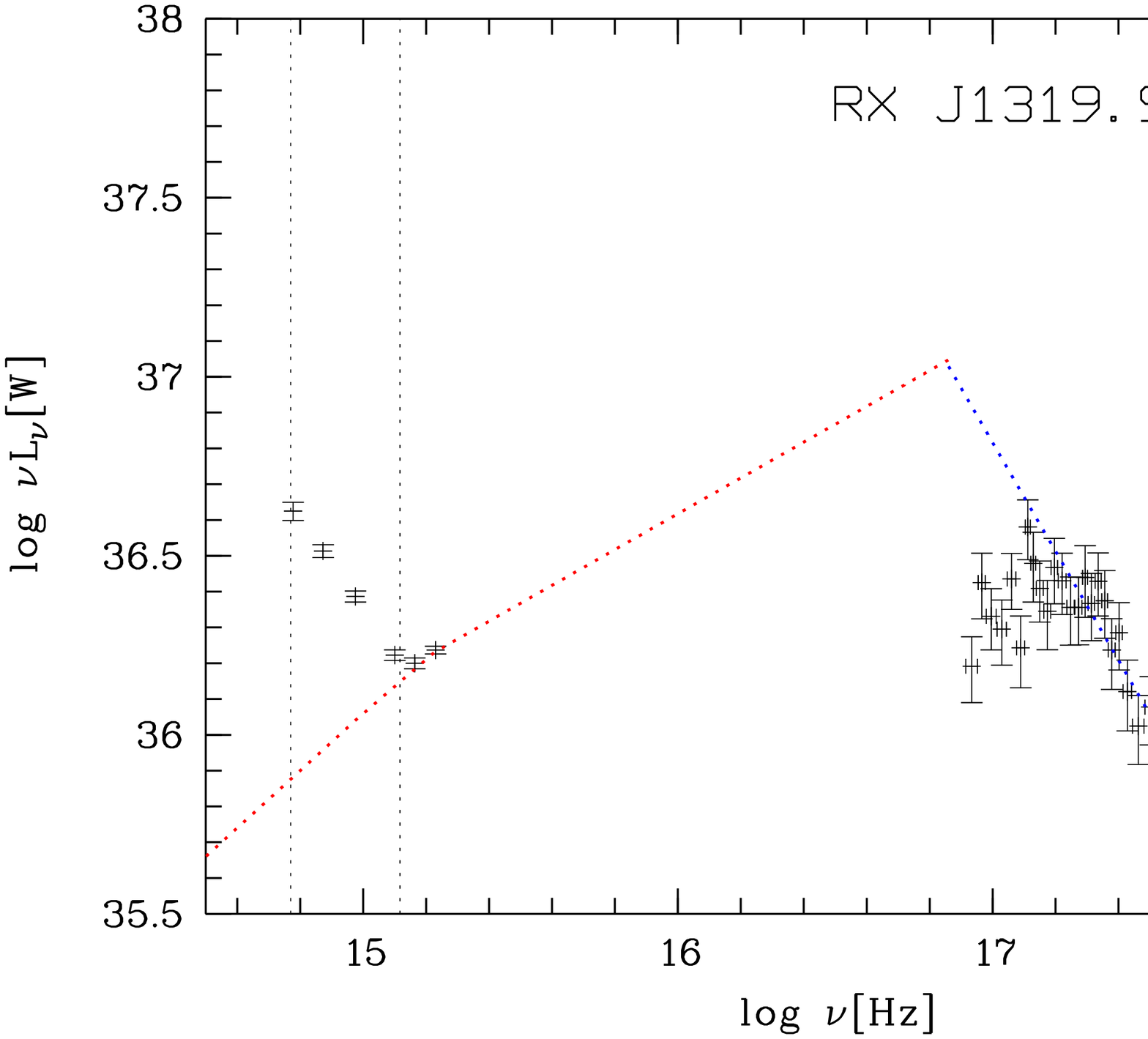}

\plotthree{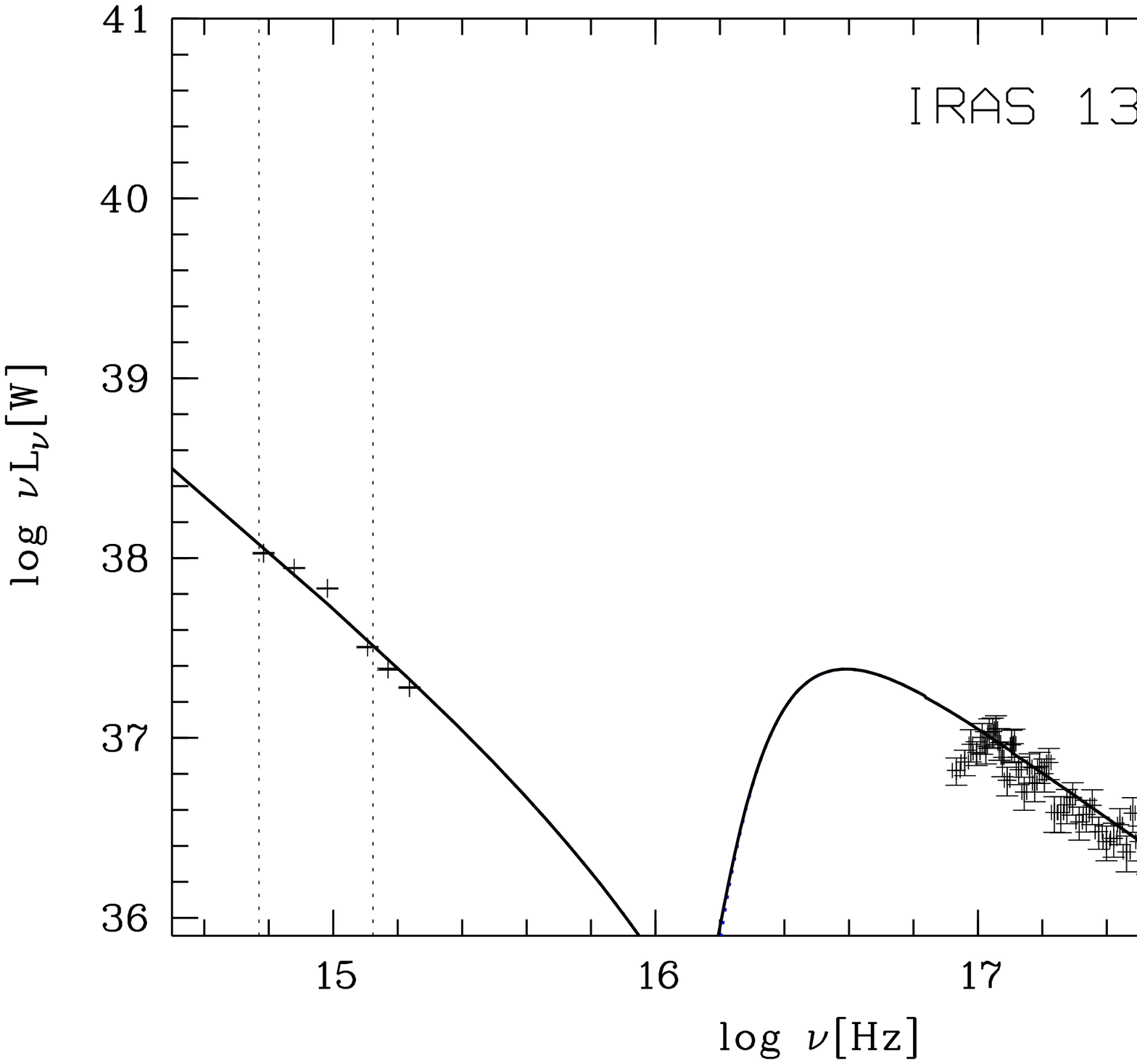}{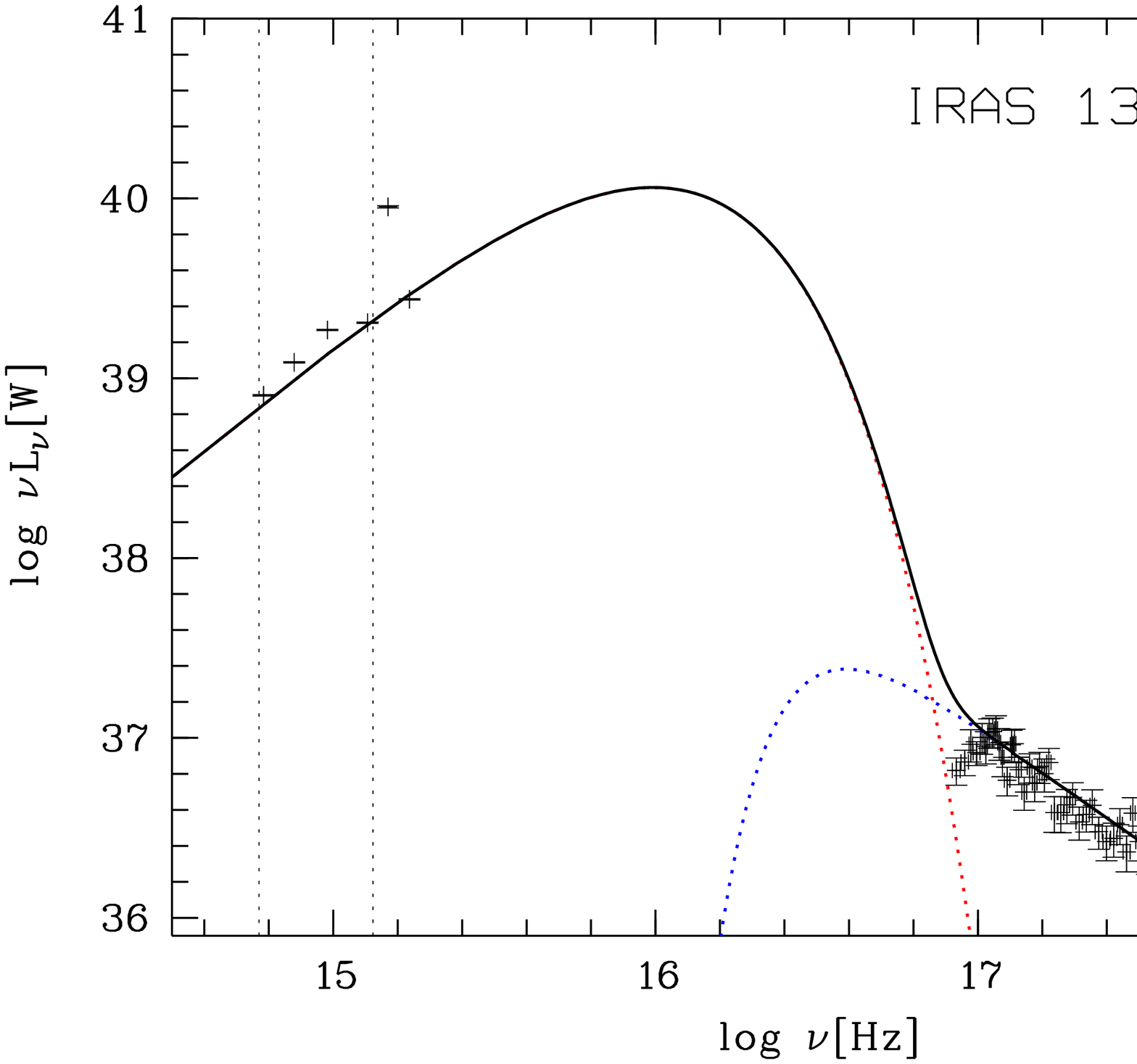}{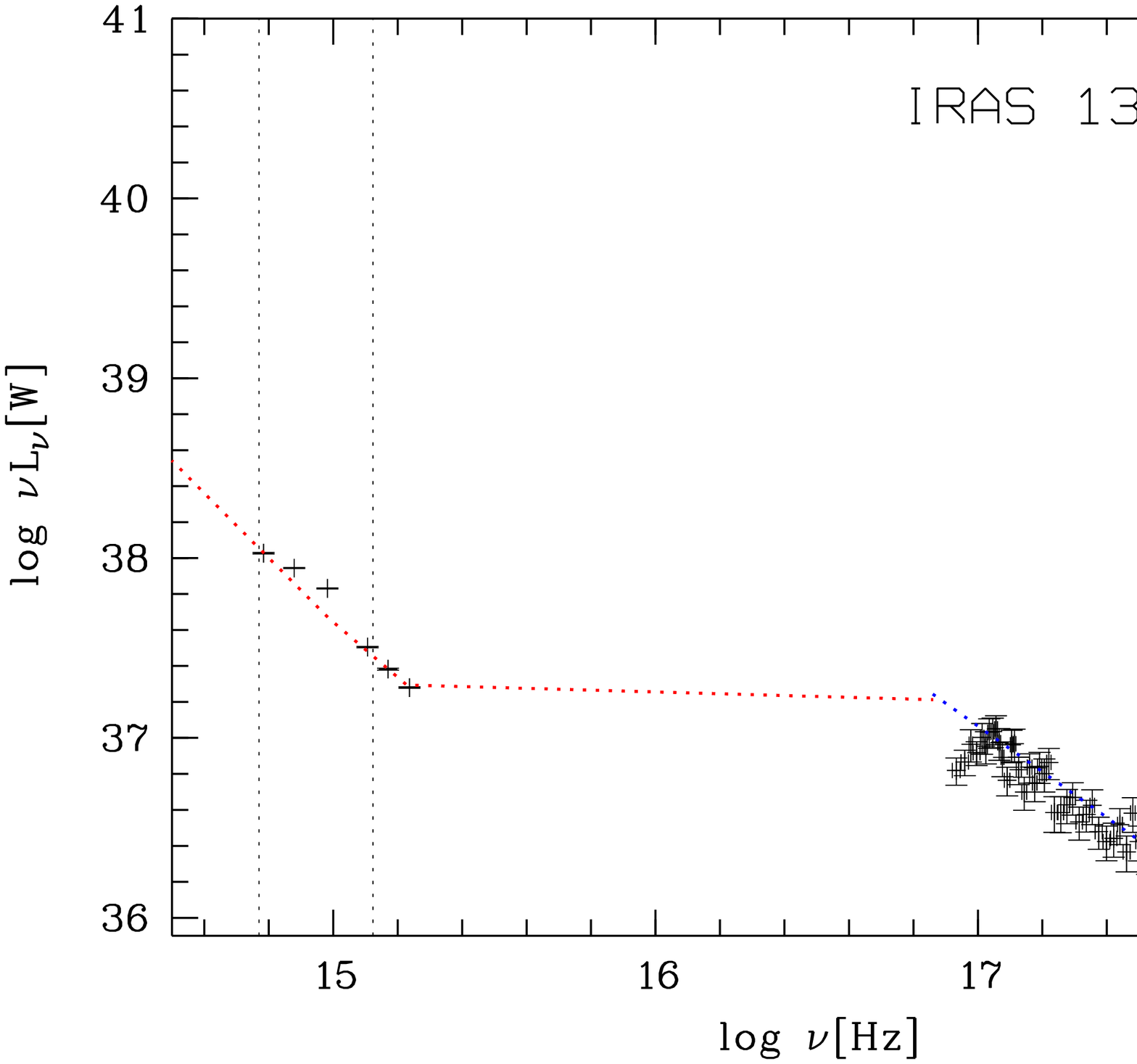}

\plotthree{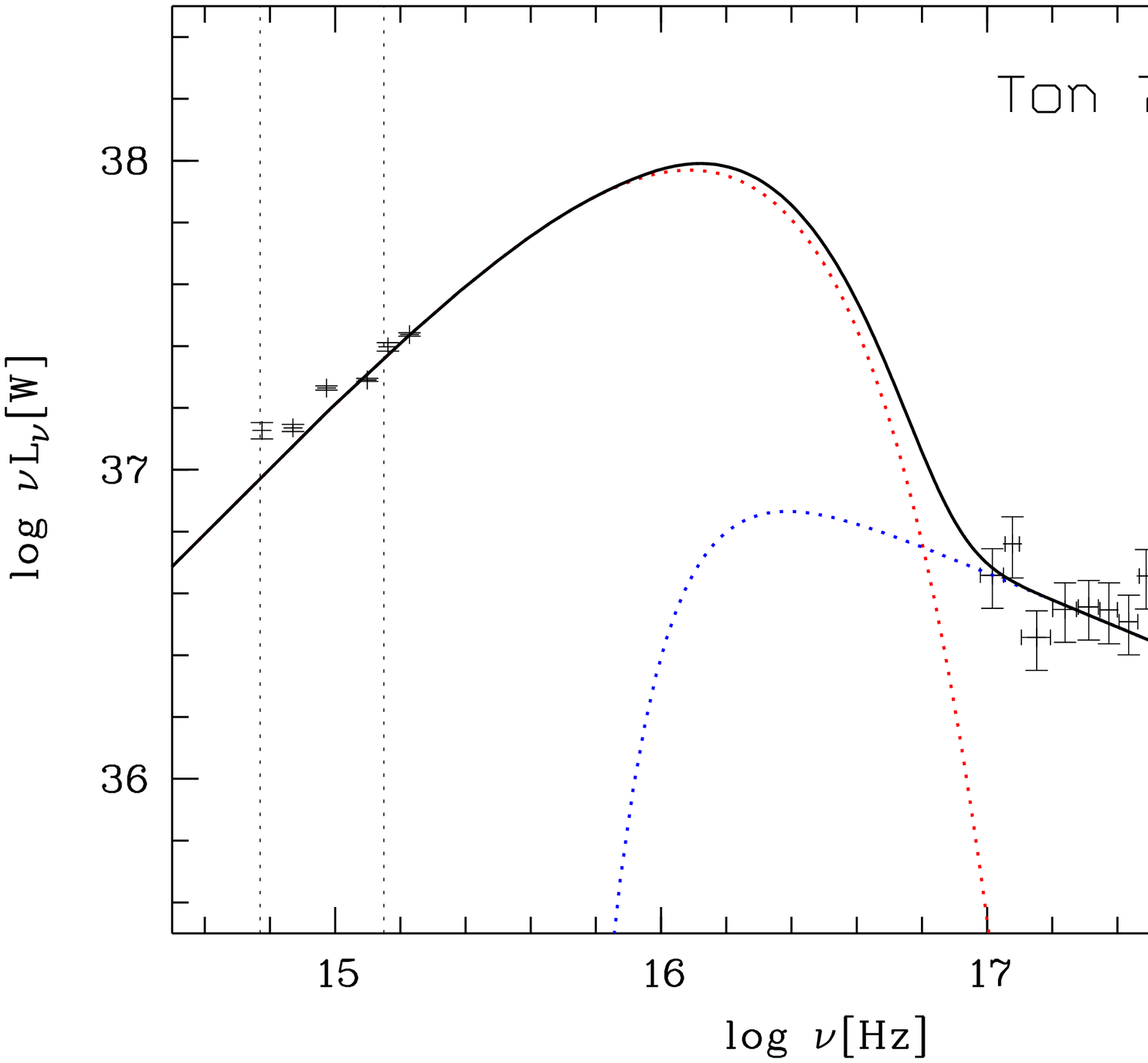}{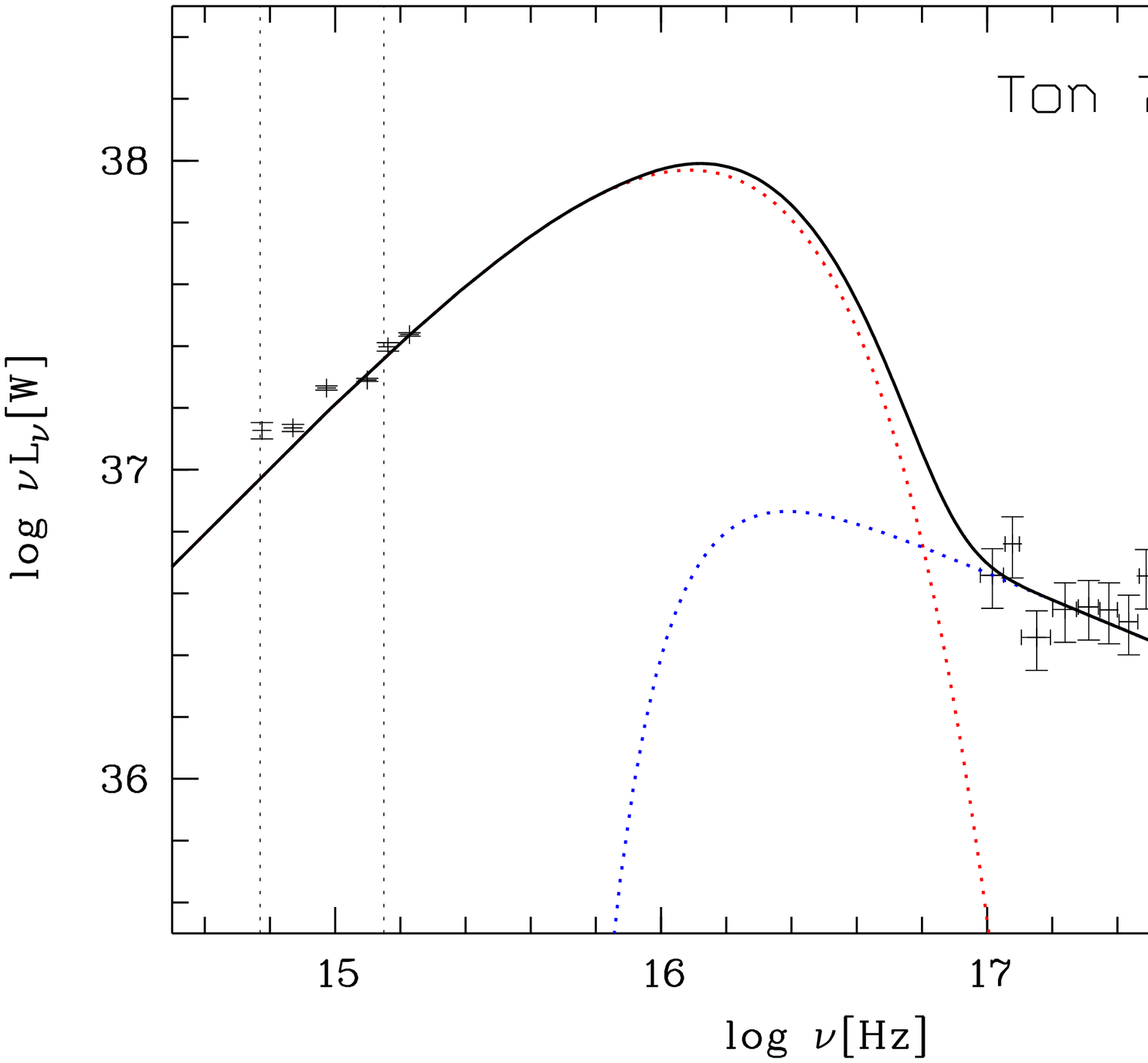}{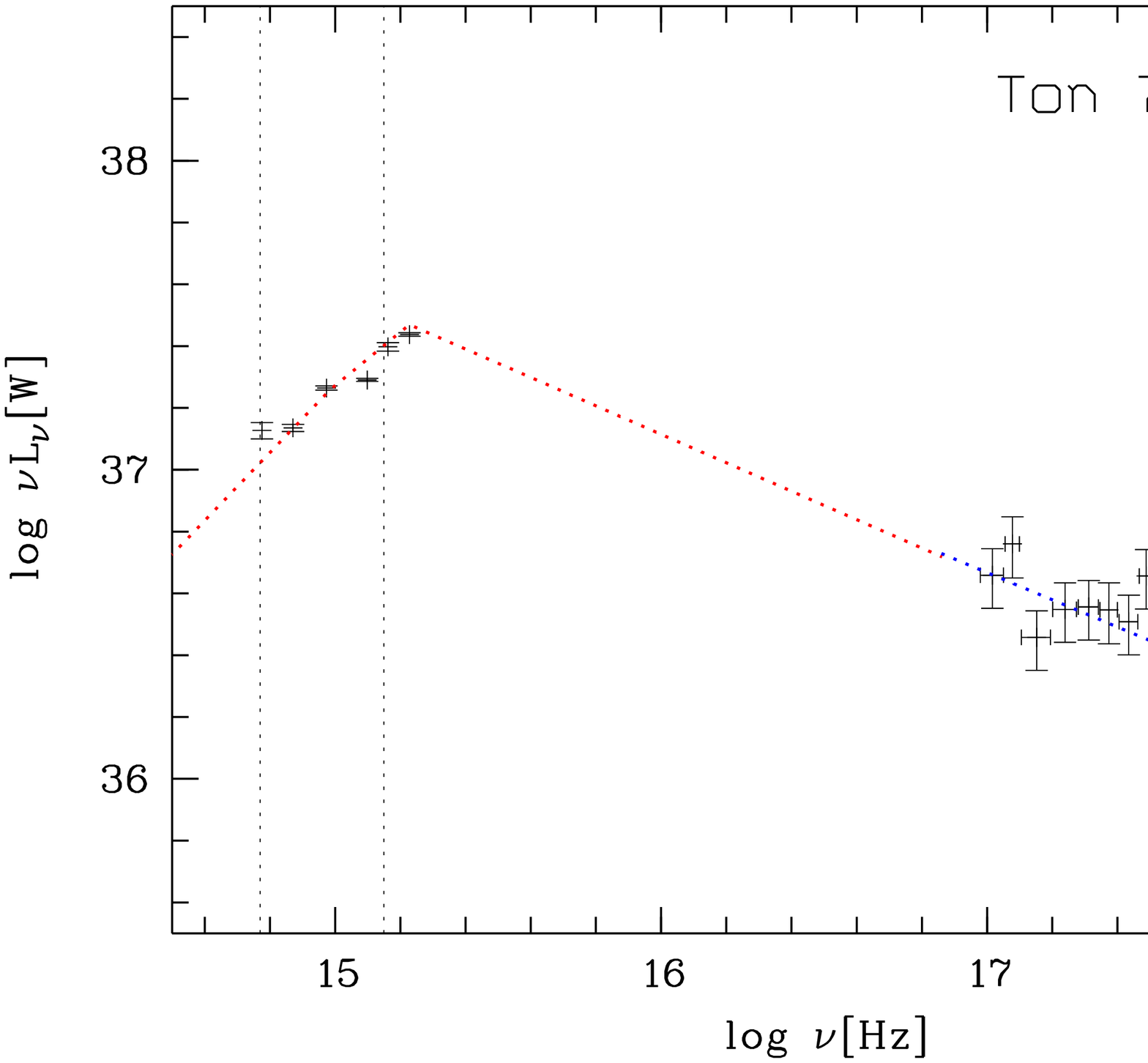}

\plotthree{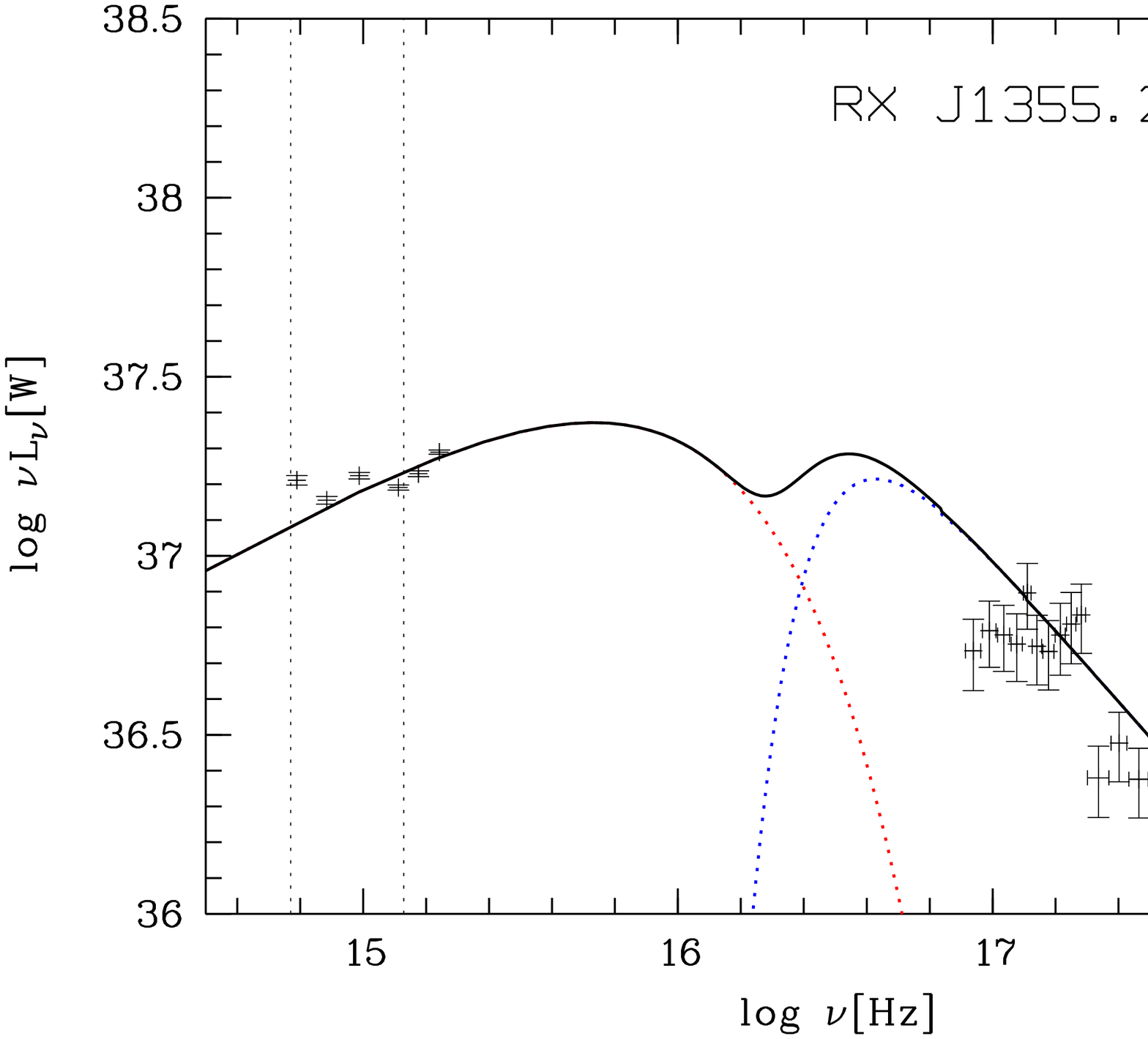}{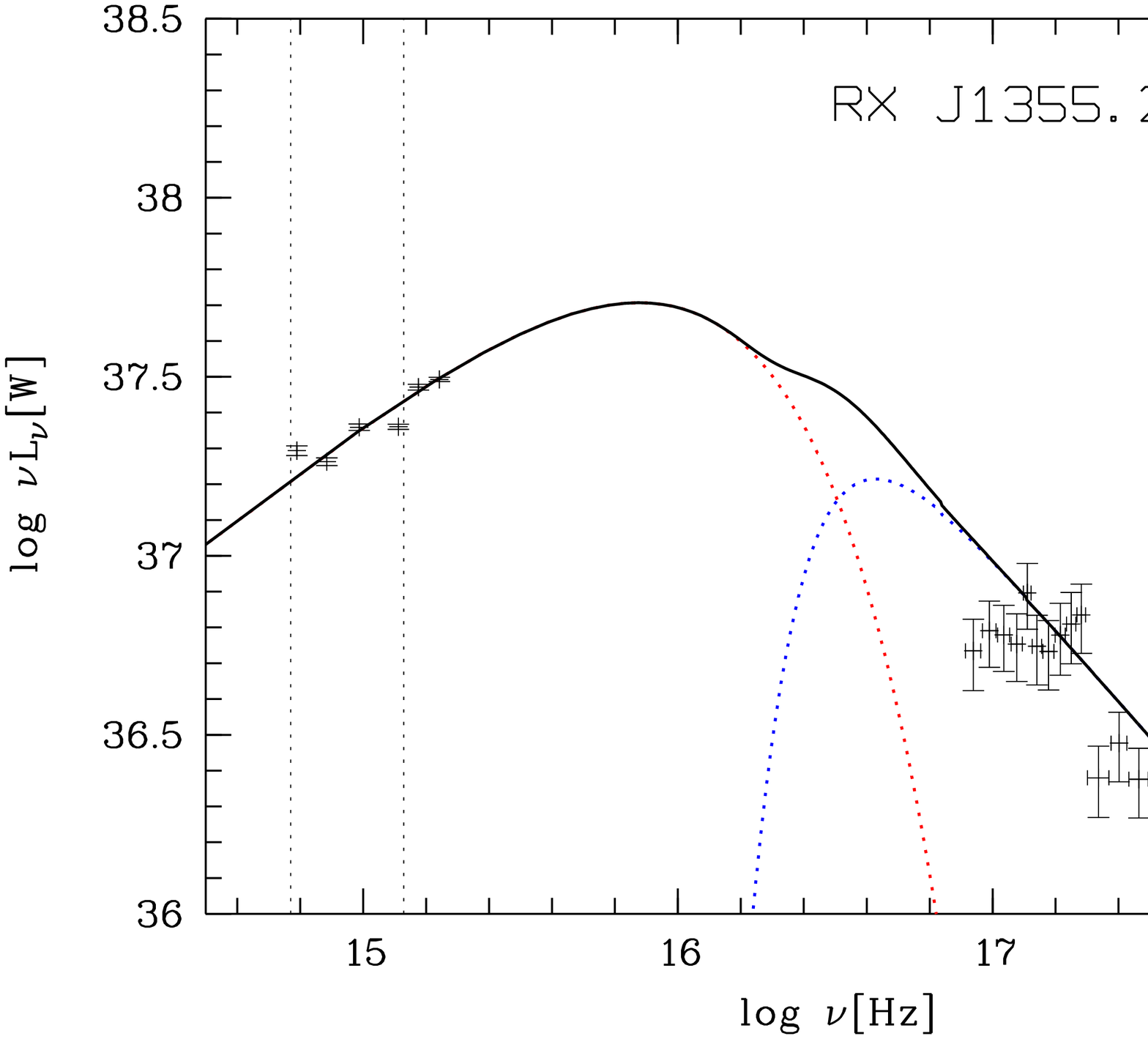}{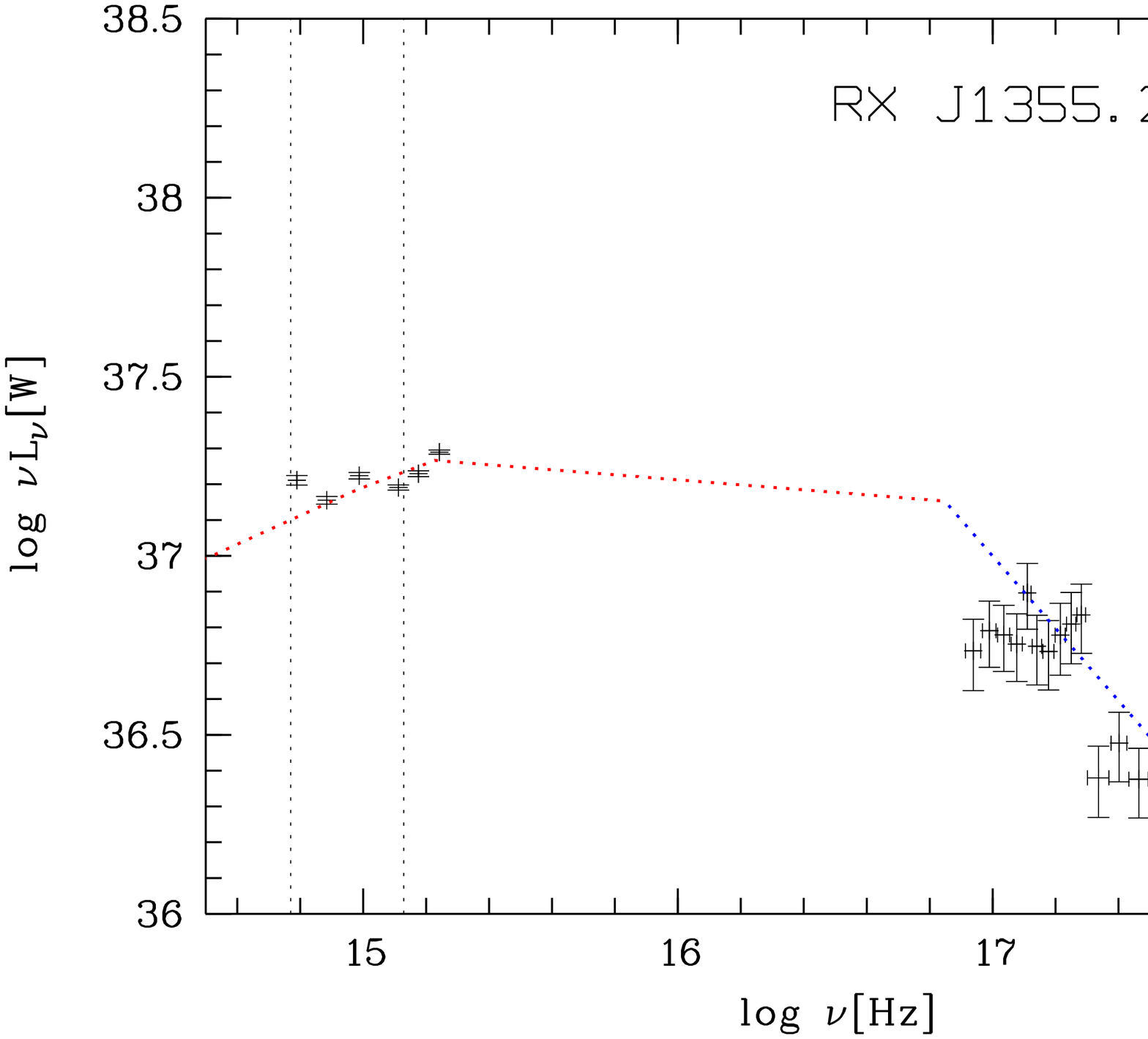}
\end{figure*}

\begin{figure*}
\epsscale{0.60}
\plotthree{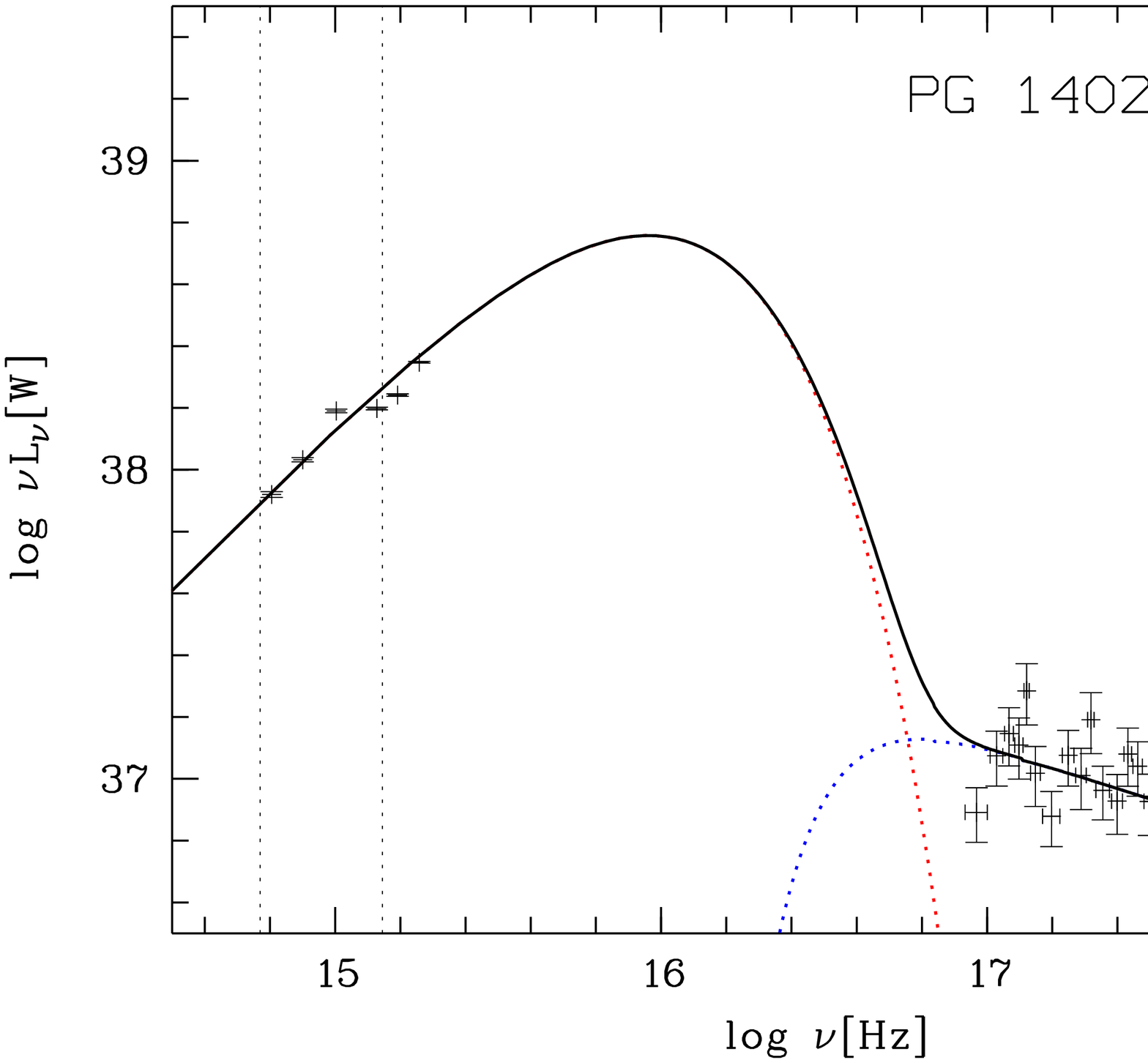}{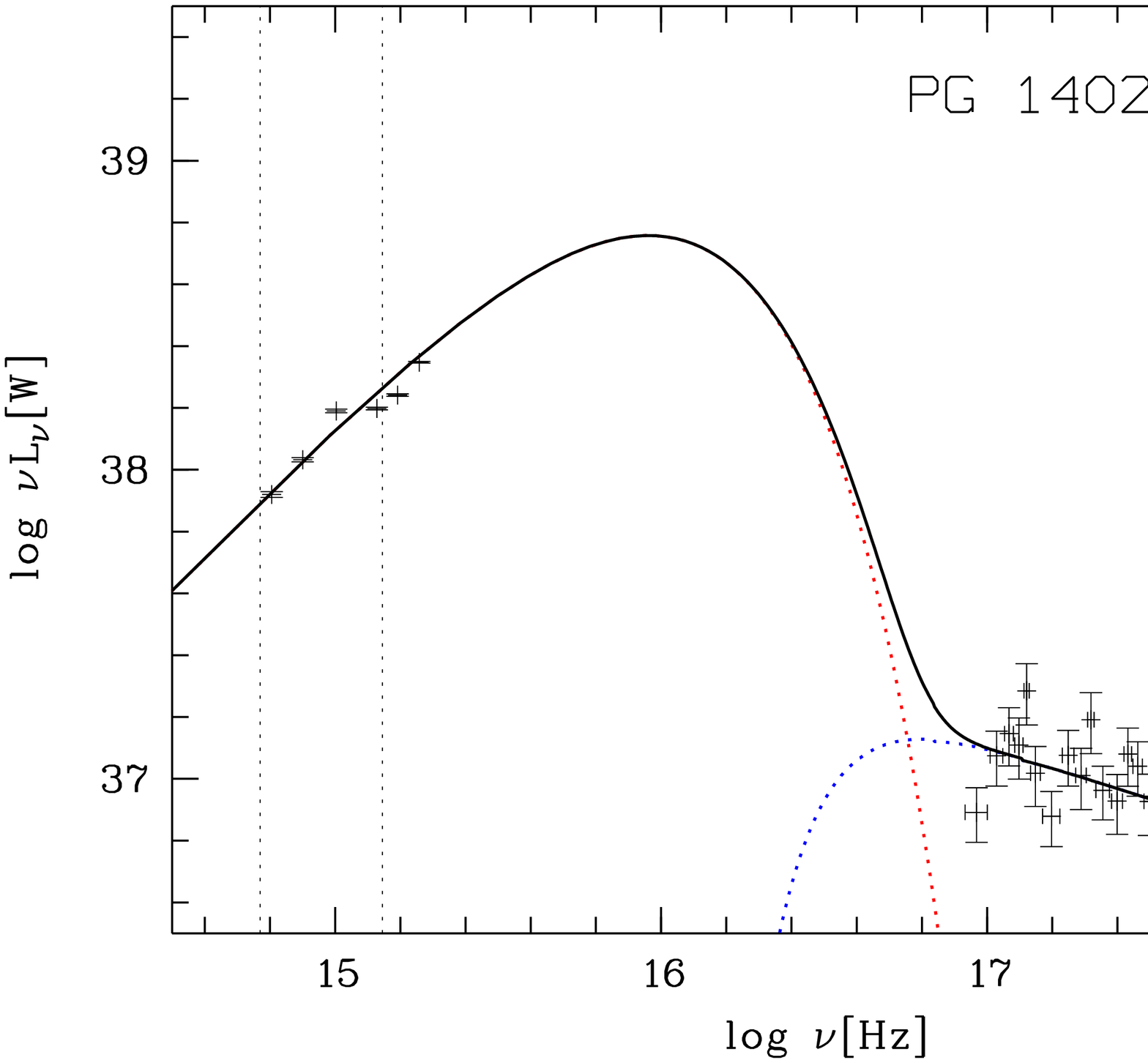}{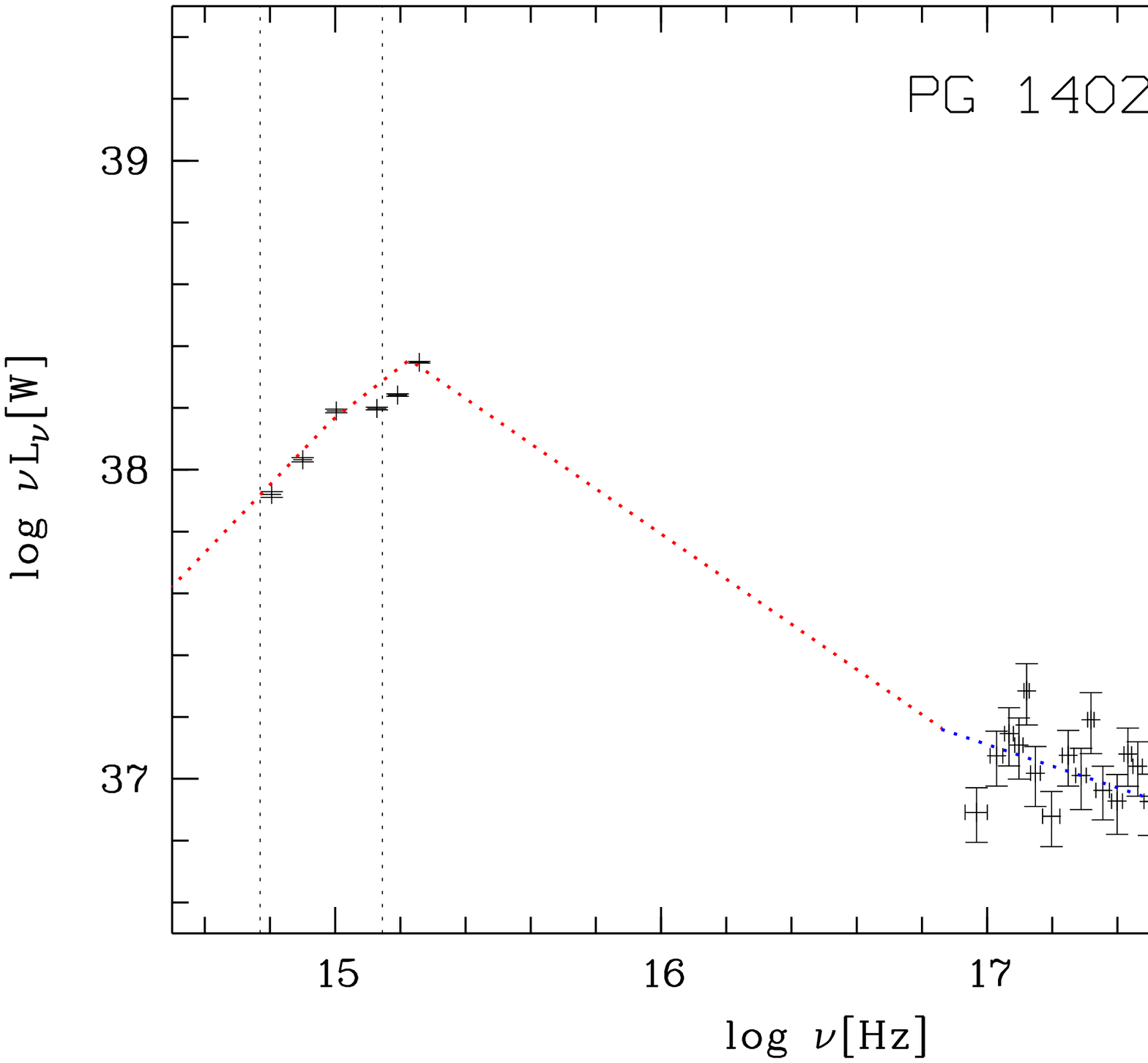}

\plotthree{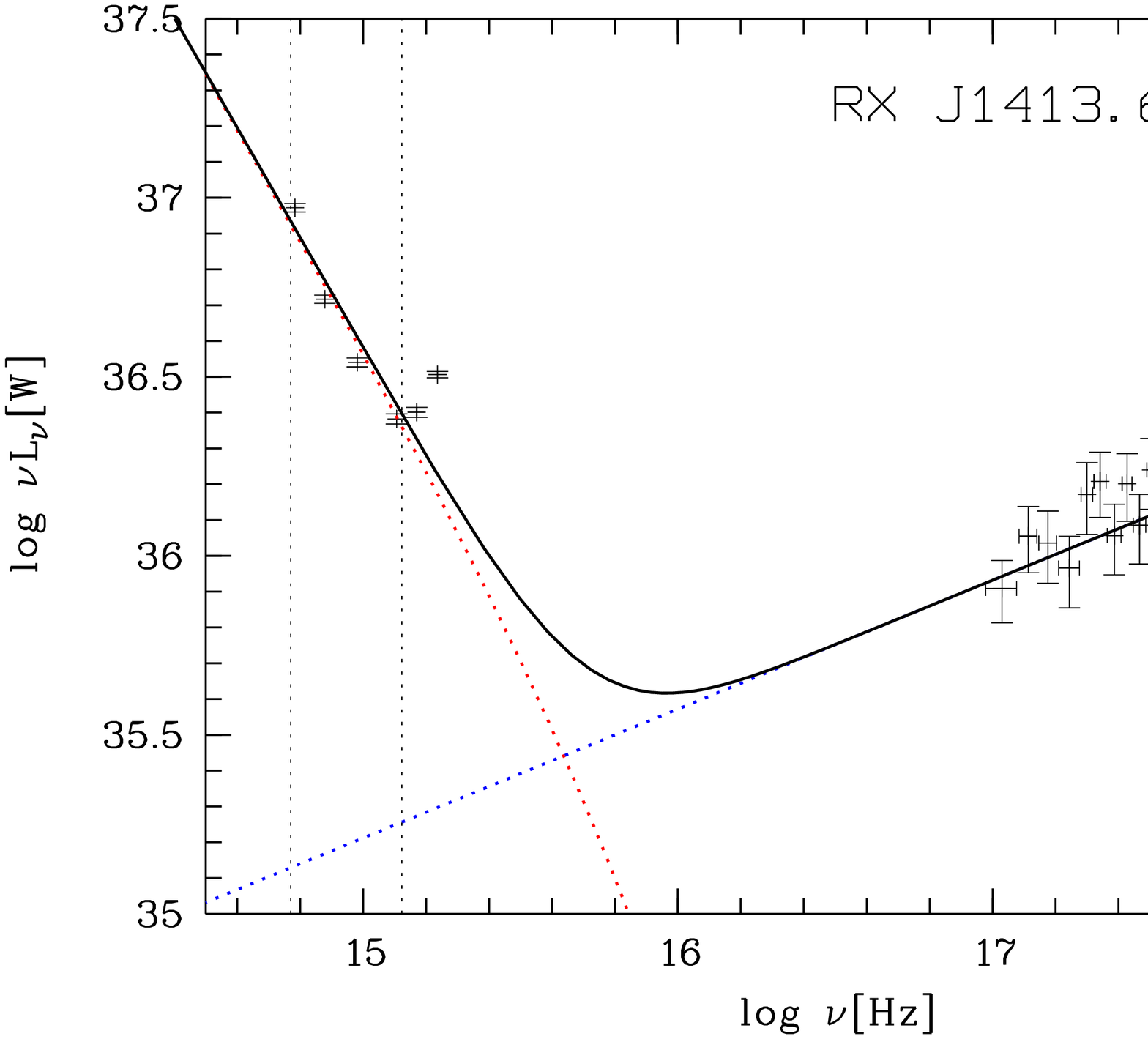}{f25_t.ps}{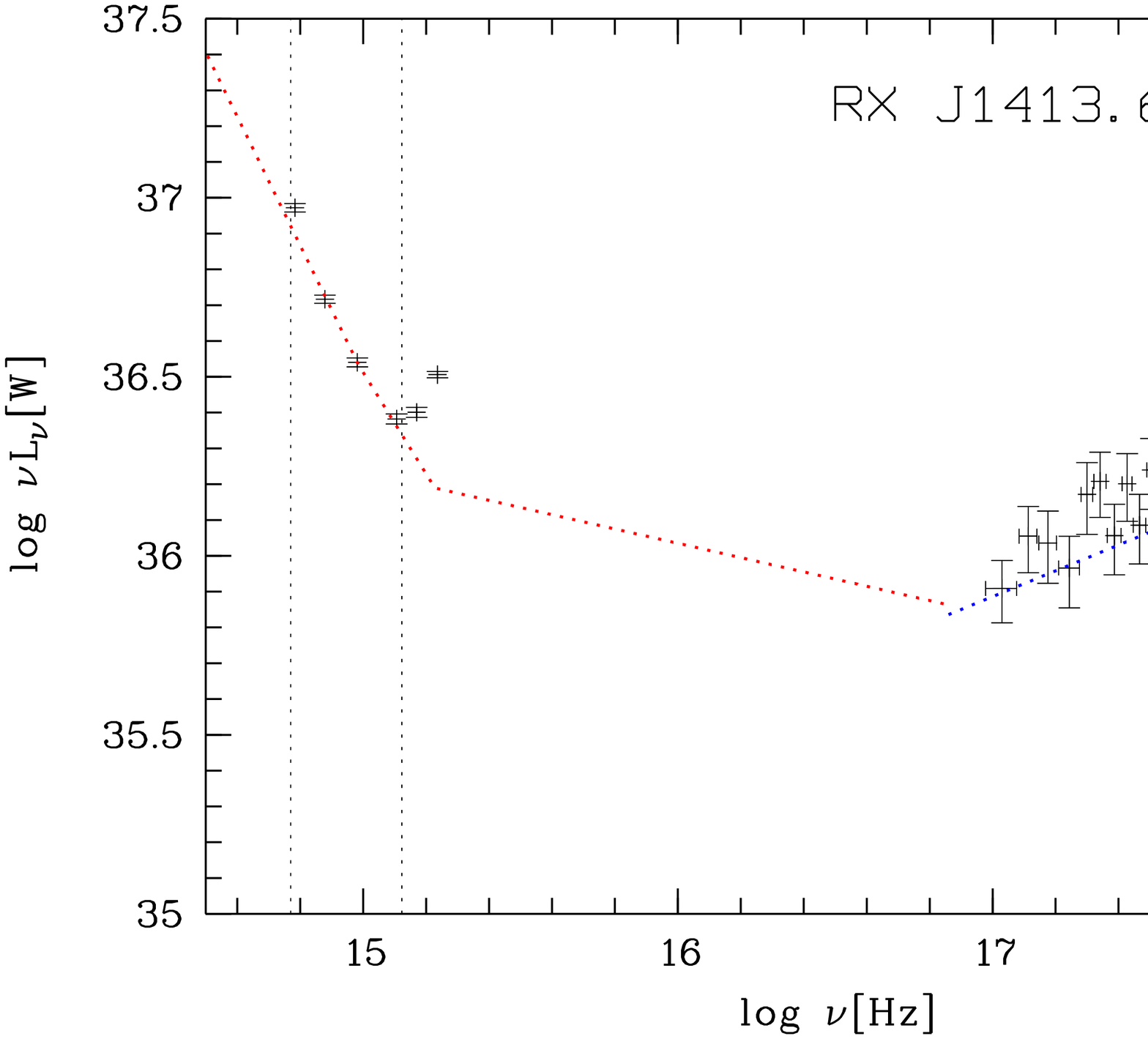}

\plotthree{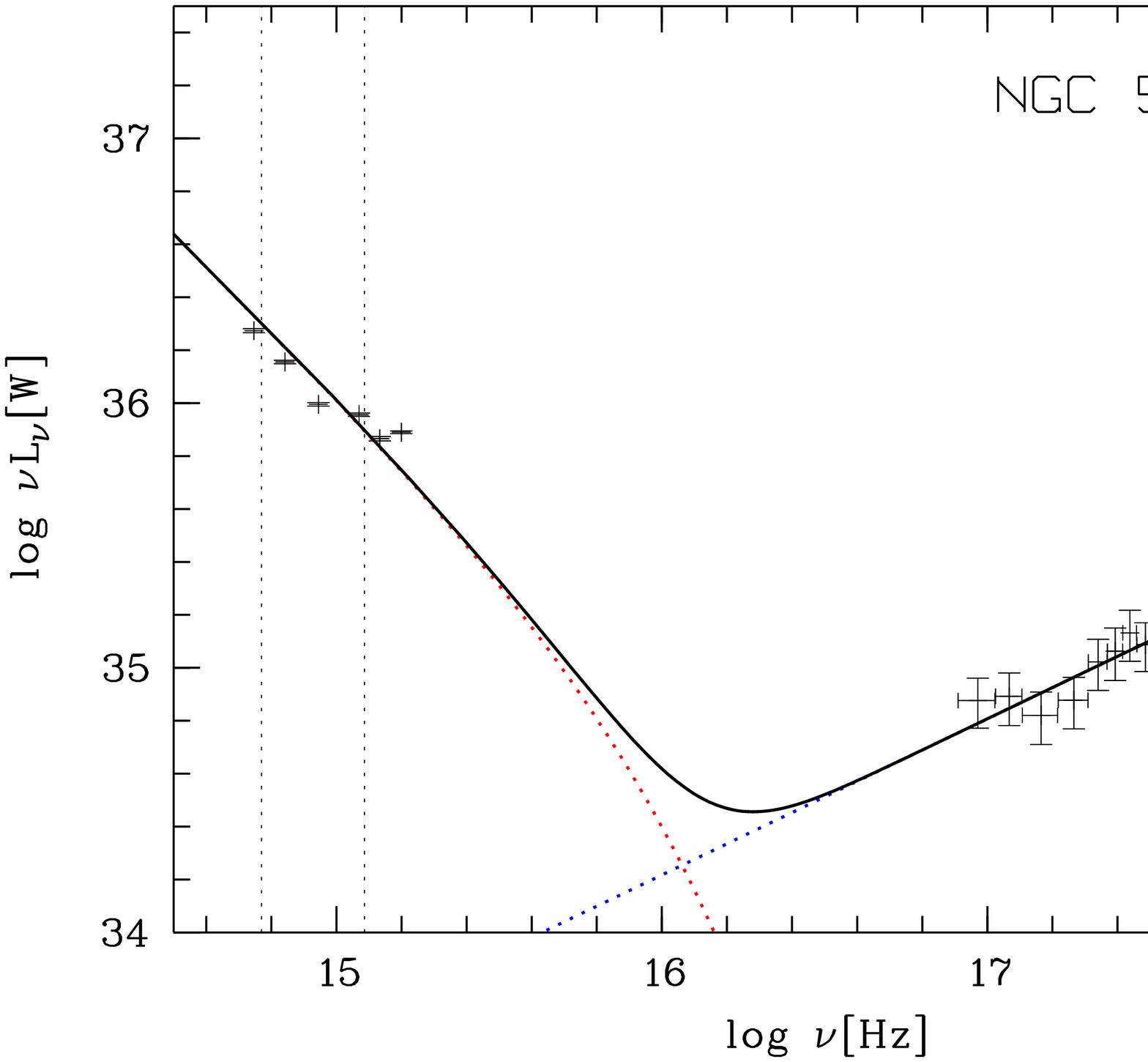}{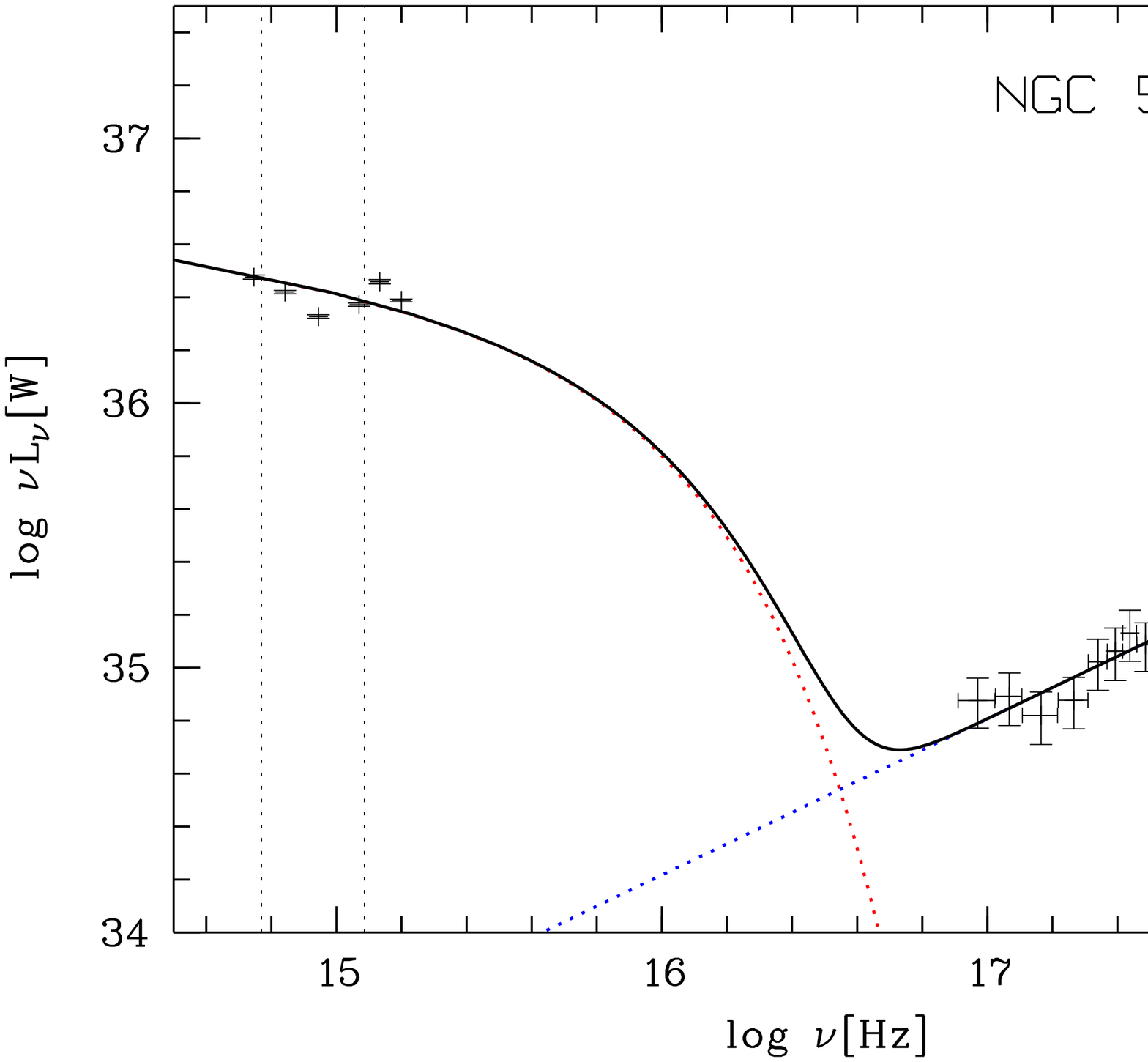}{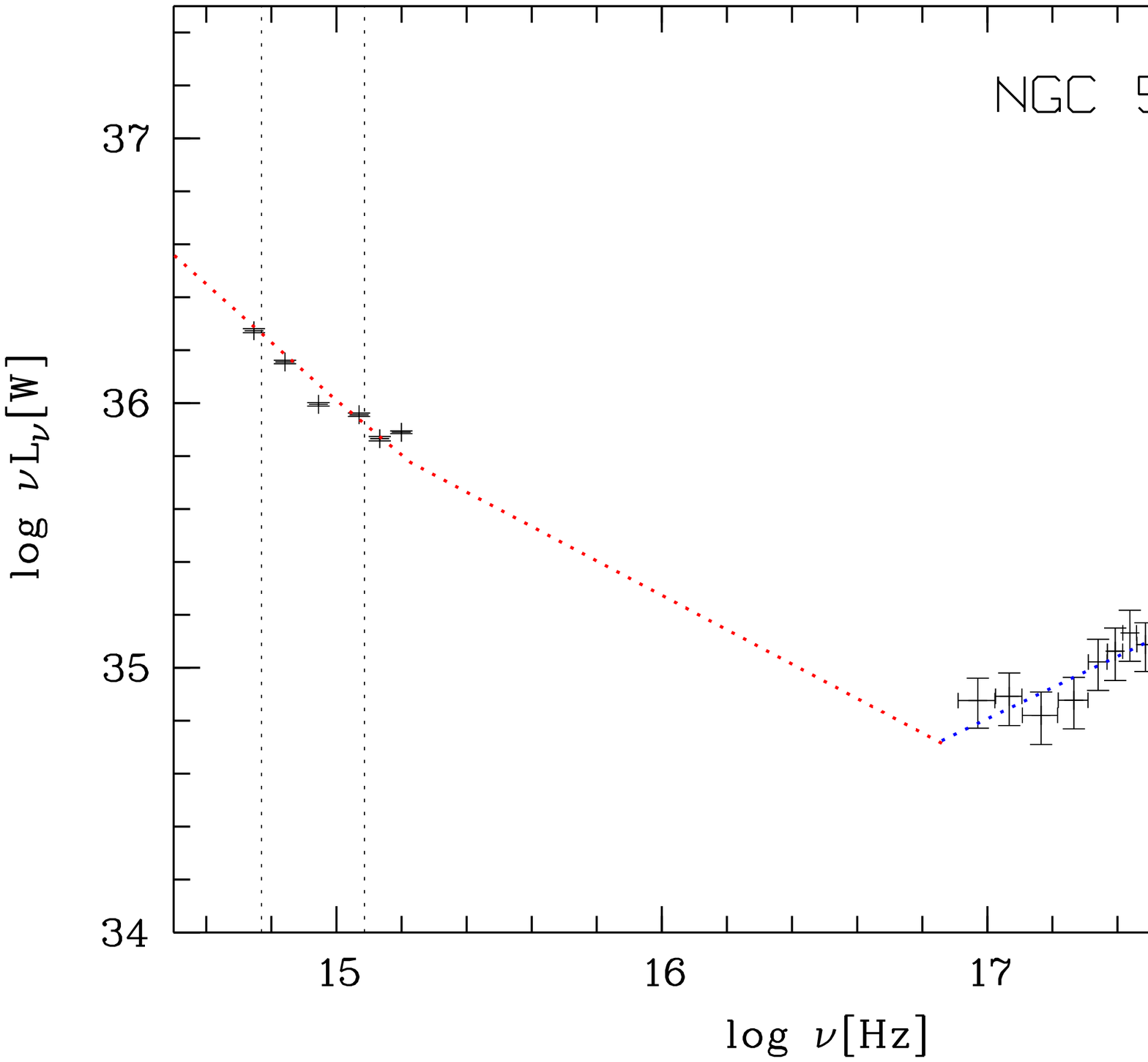}

\plotthree{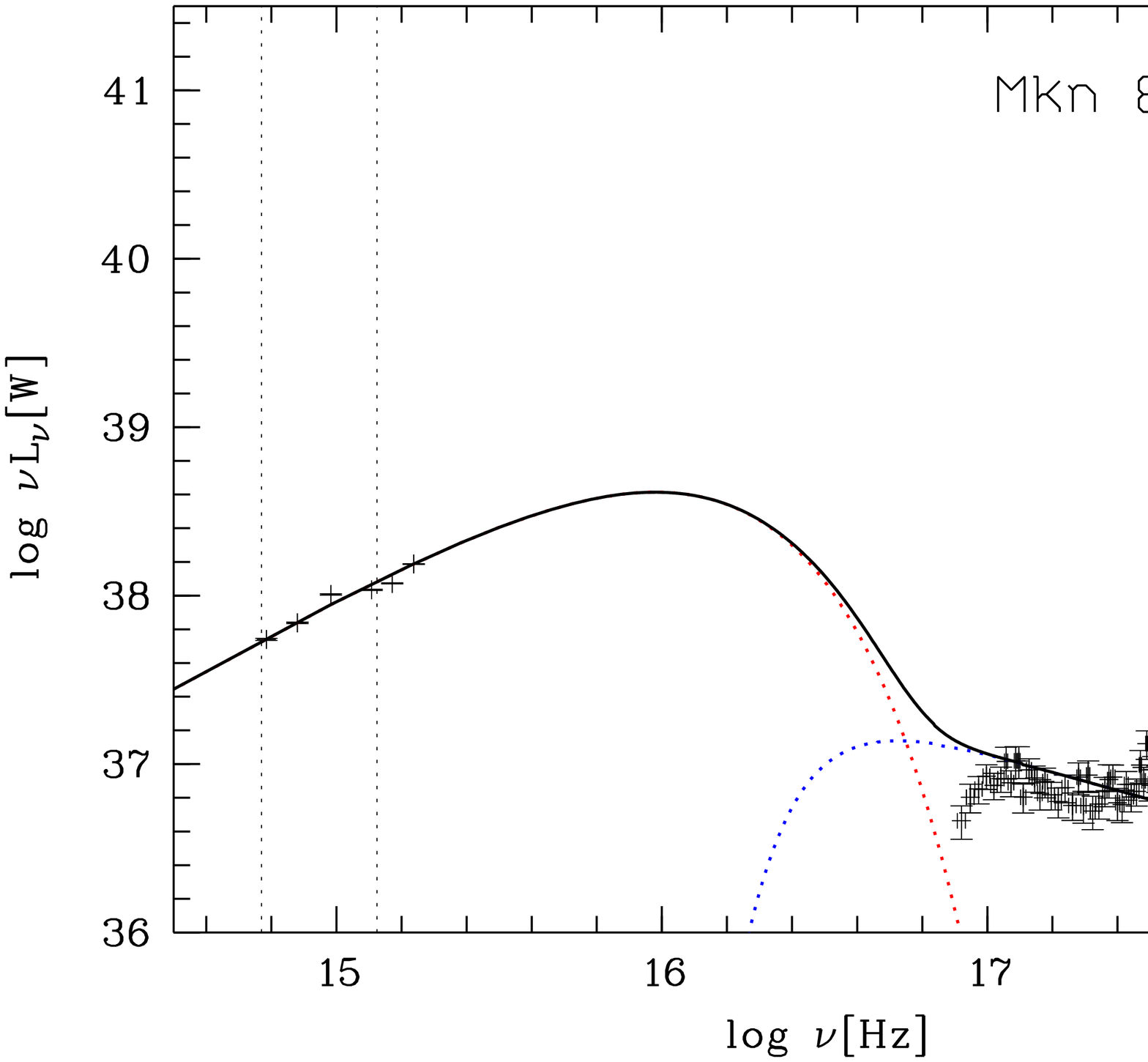}{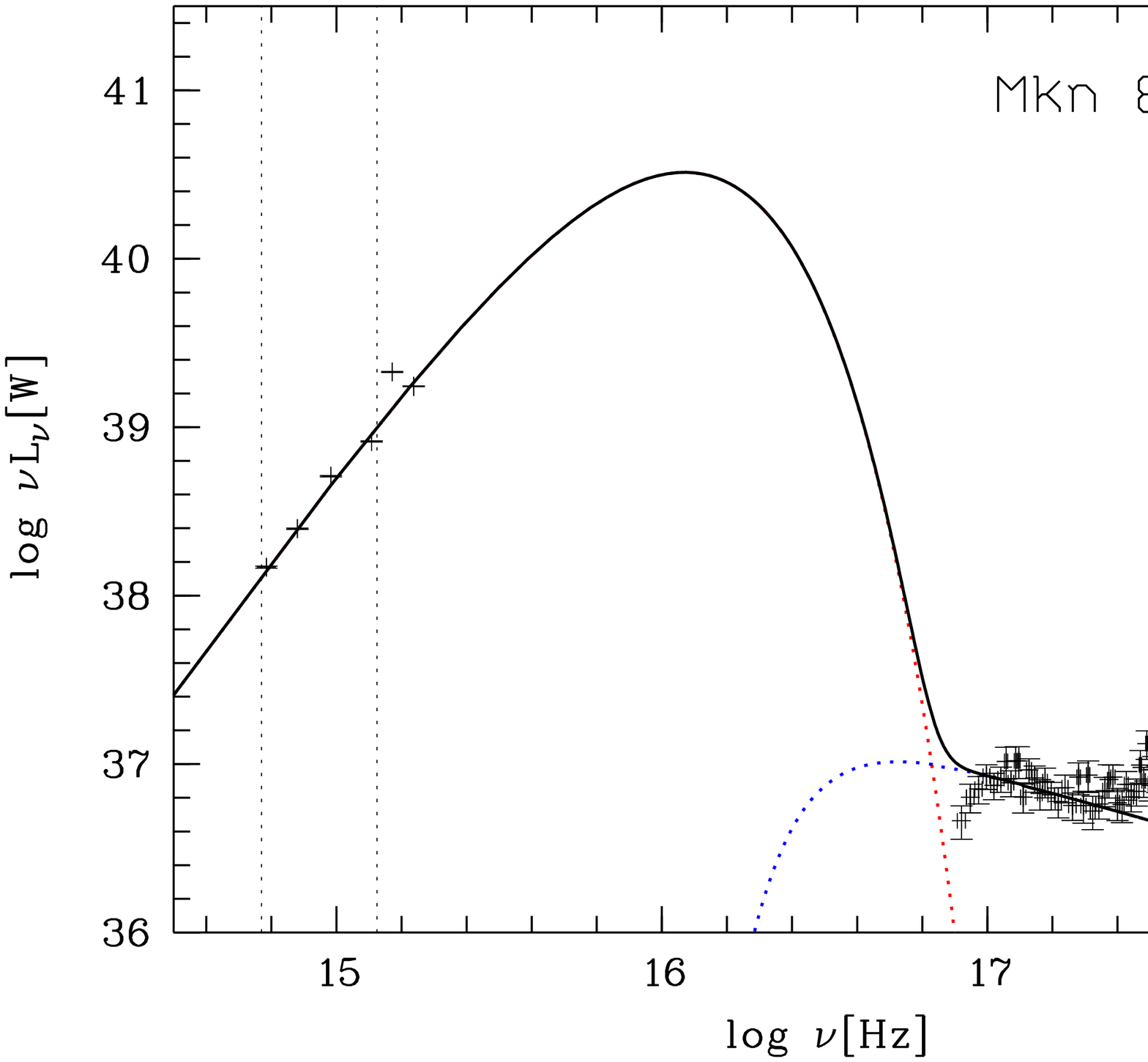}{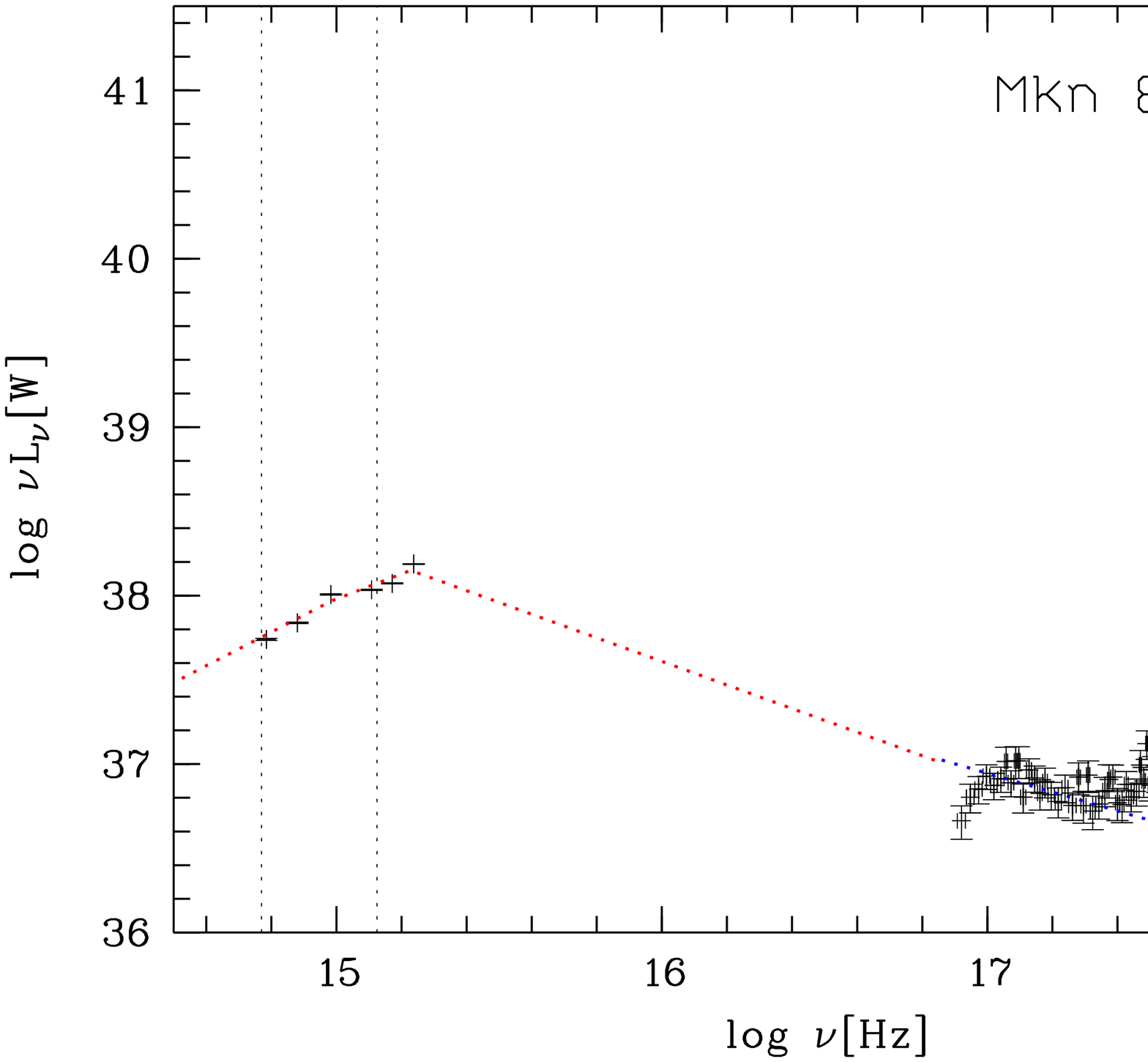}

\plotthree{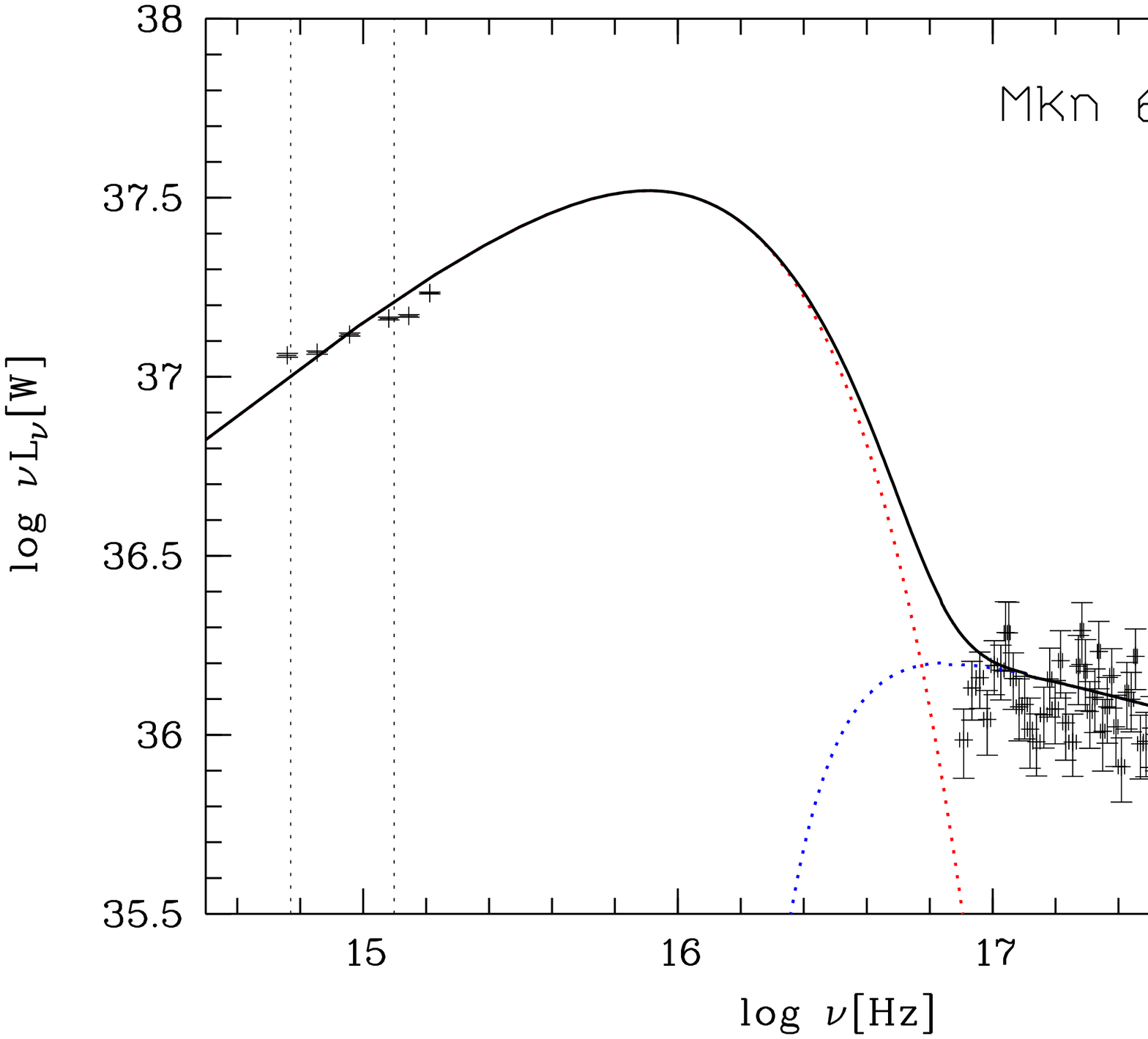}{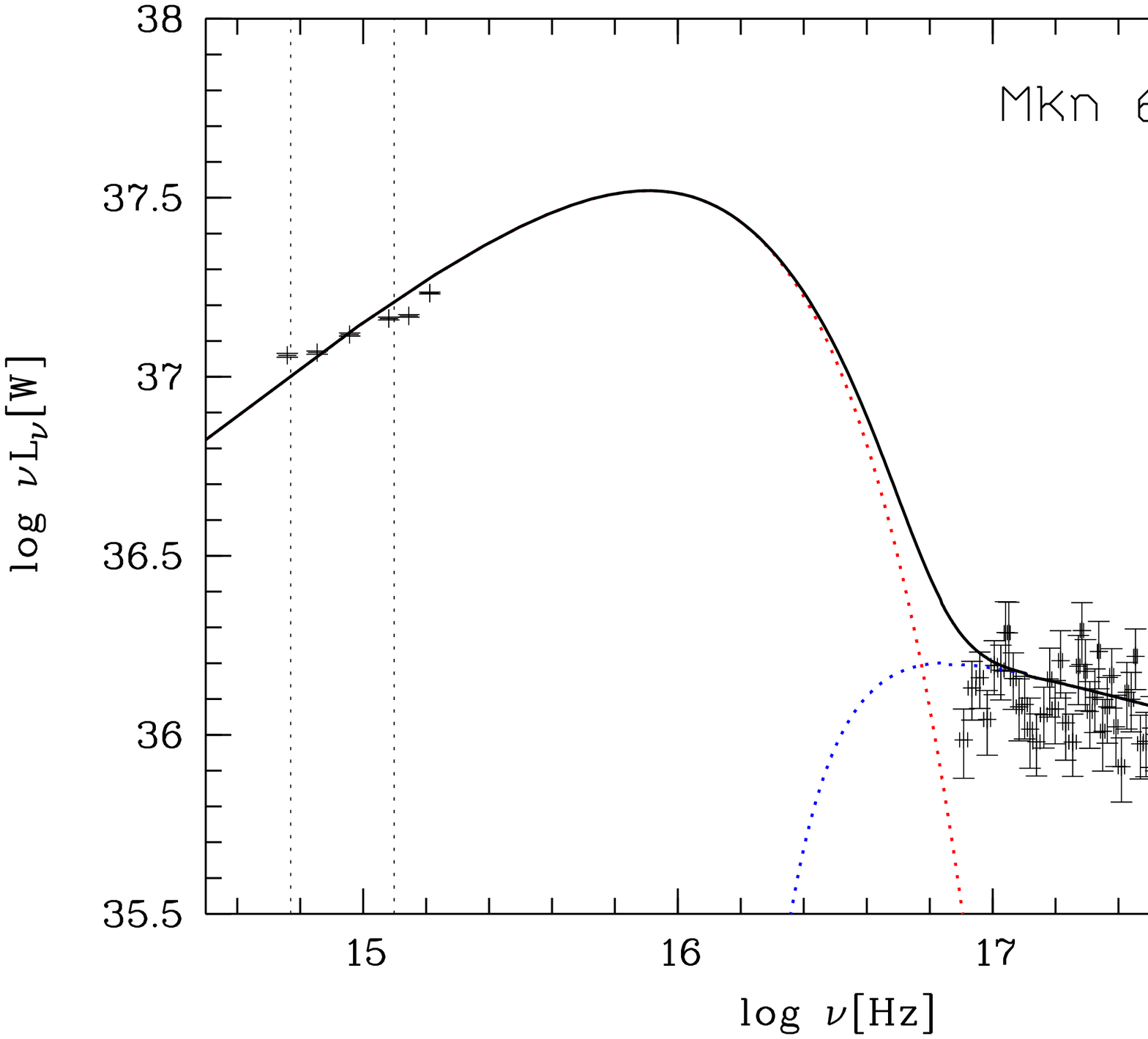}{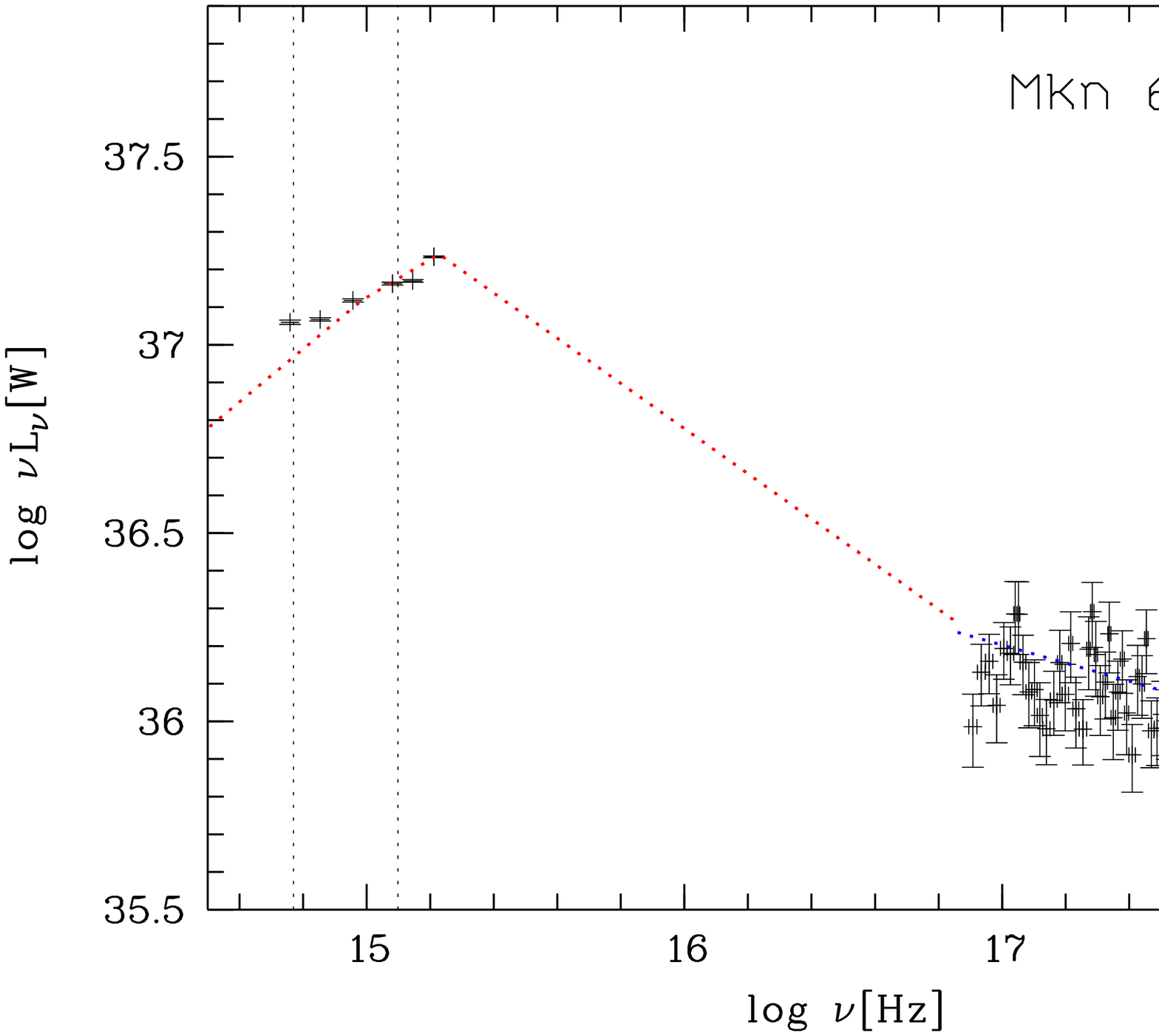}
\end{figure*}

\begin{figure*}
\epsscale{0.60}
\plotthree{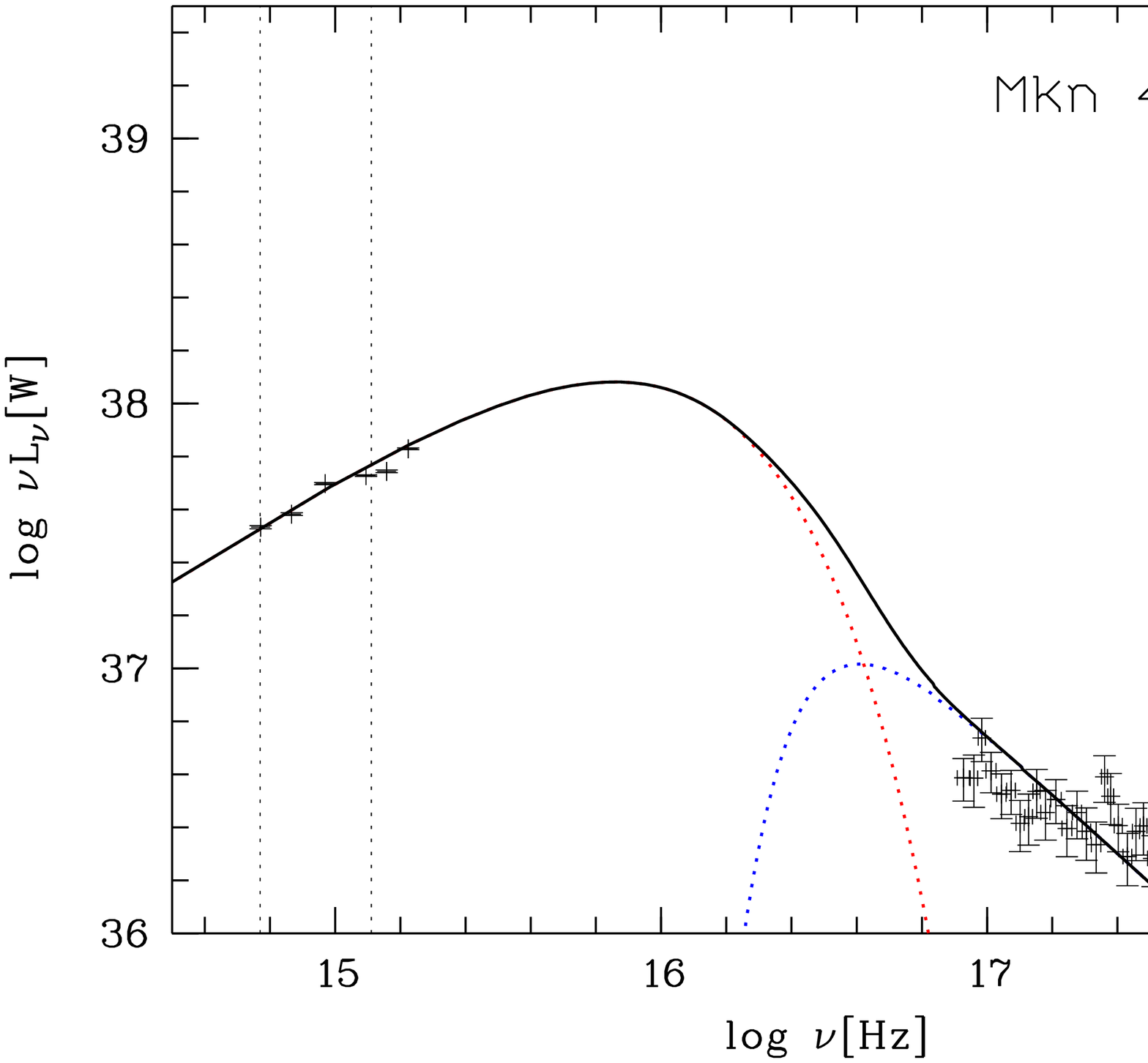}{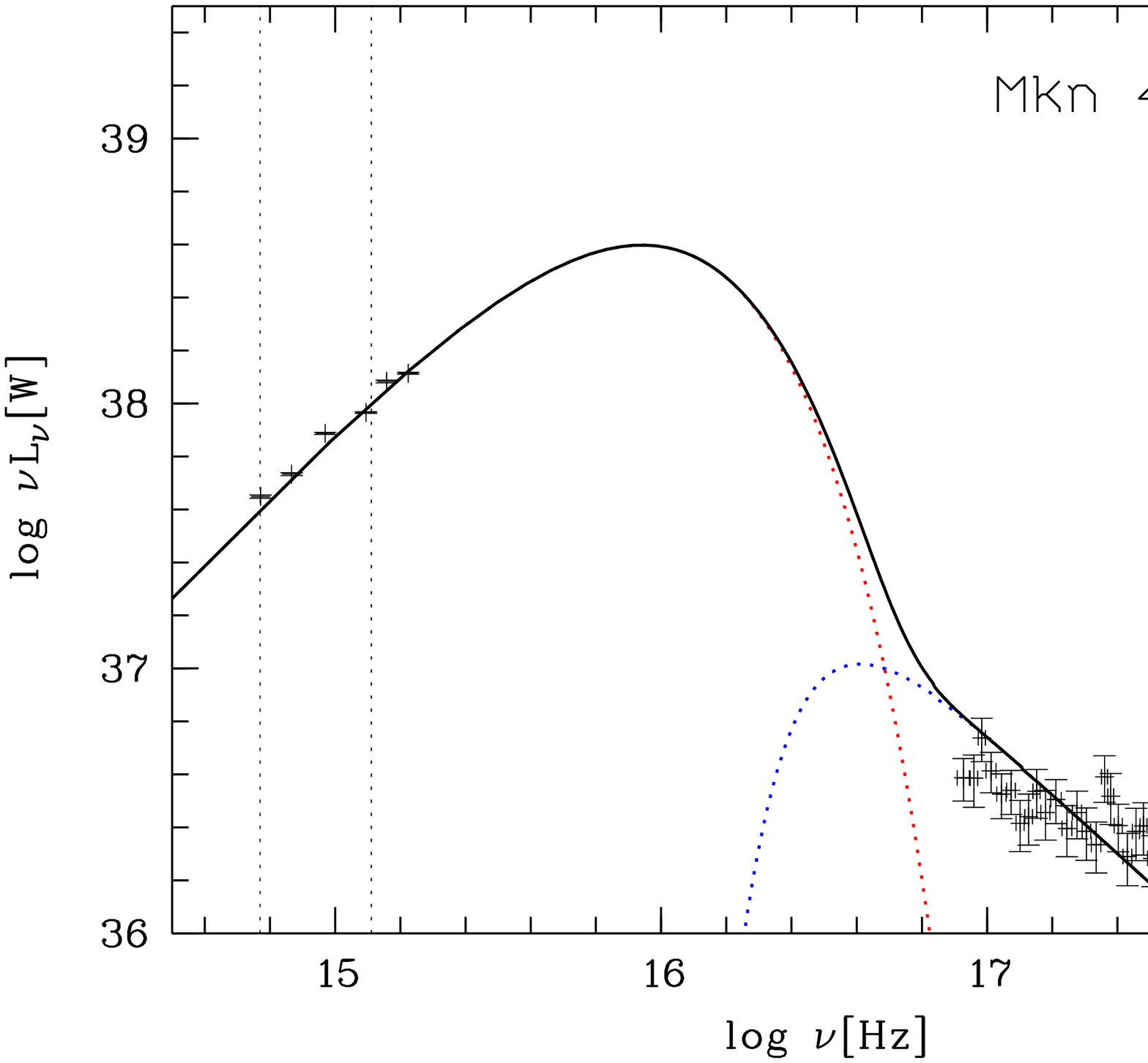}{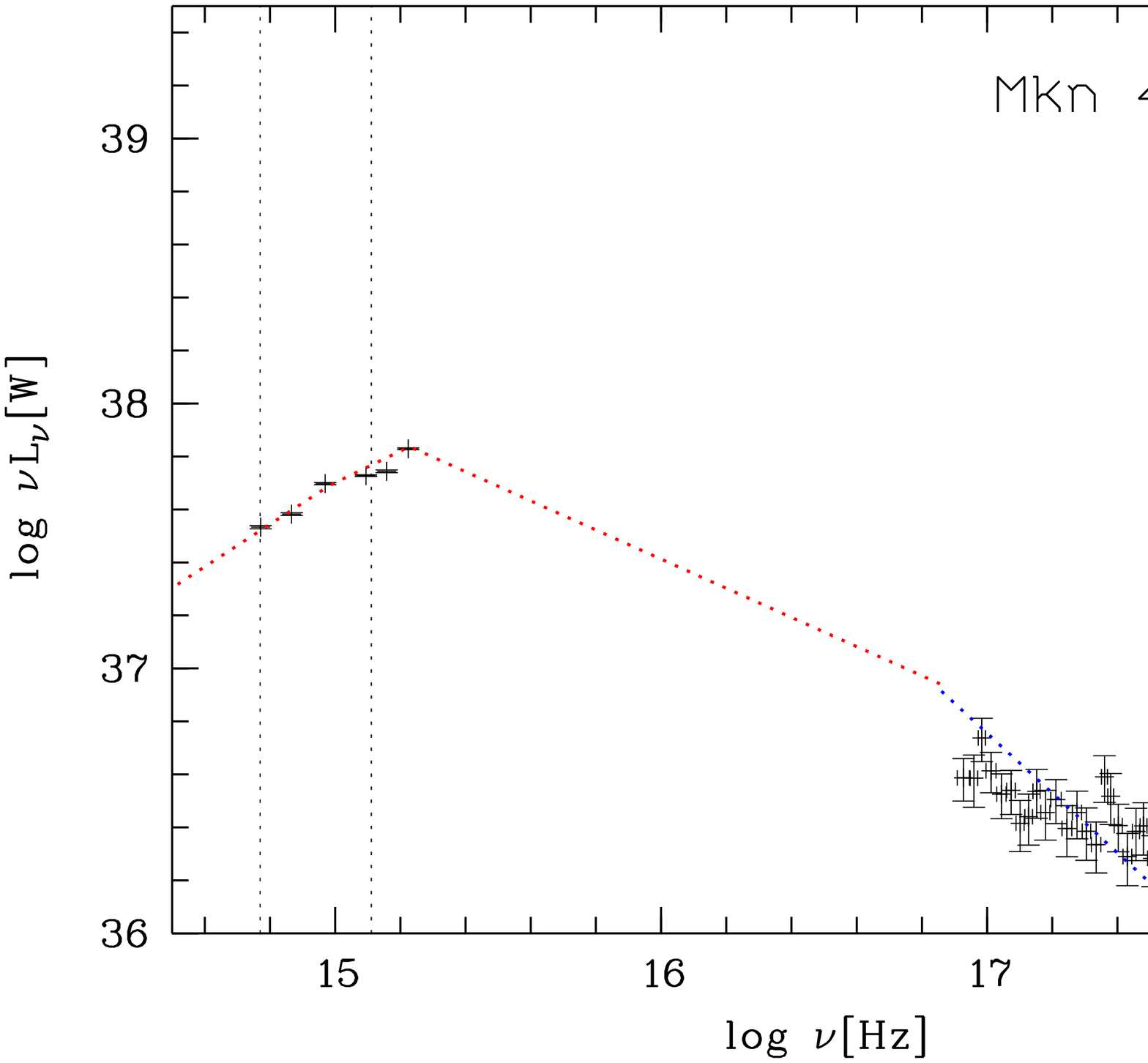}

\plotthree{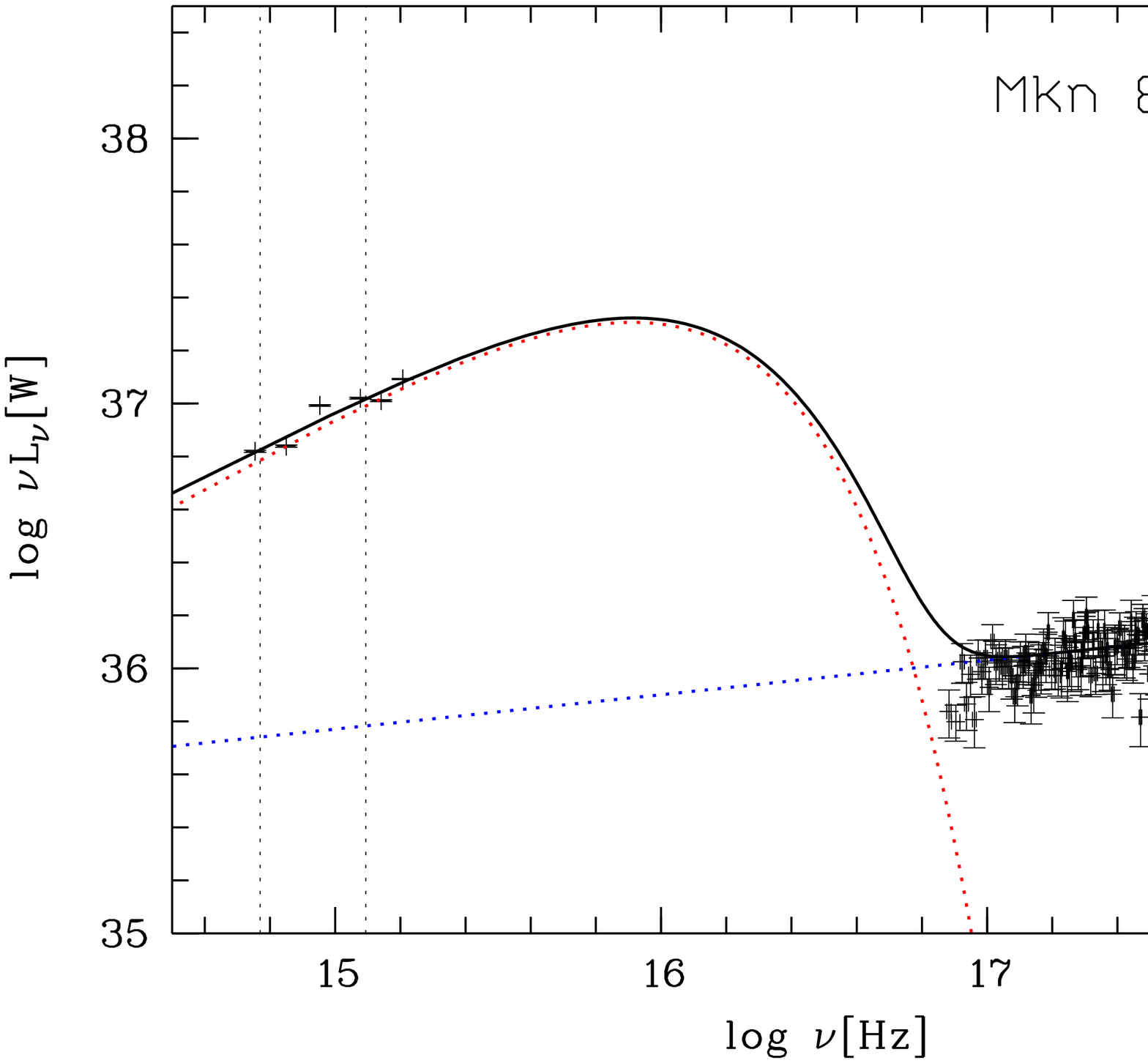}{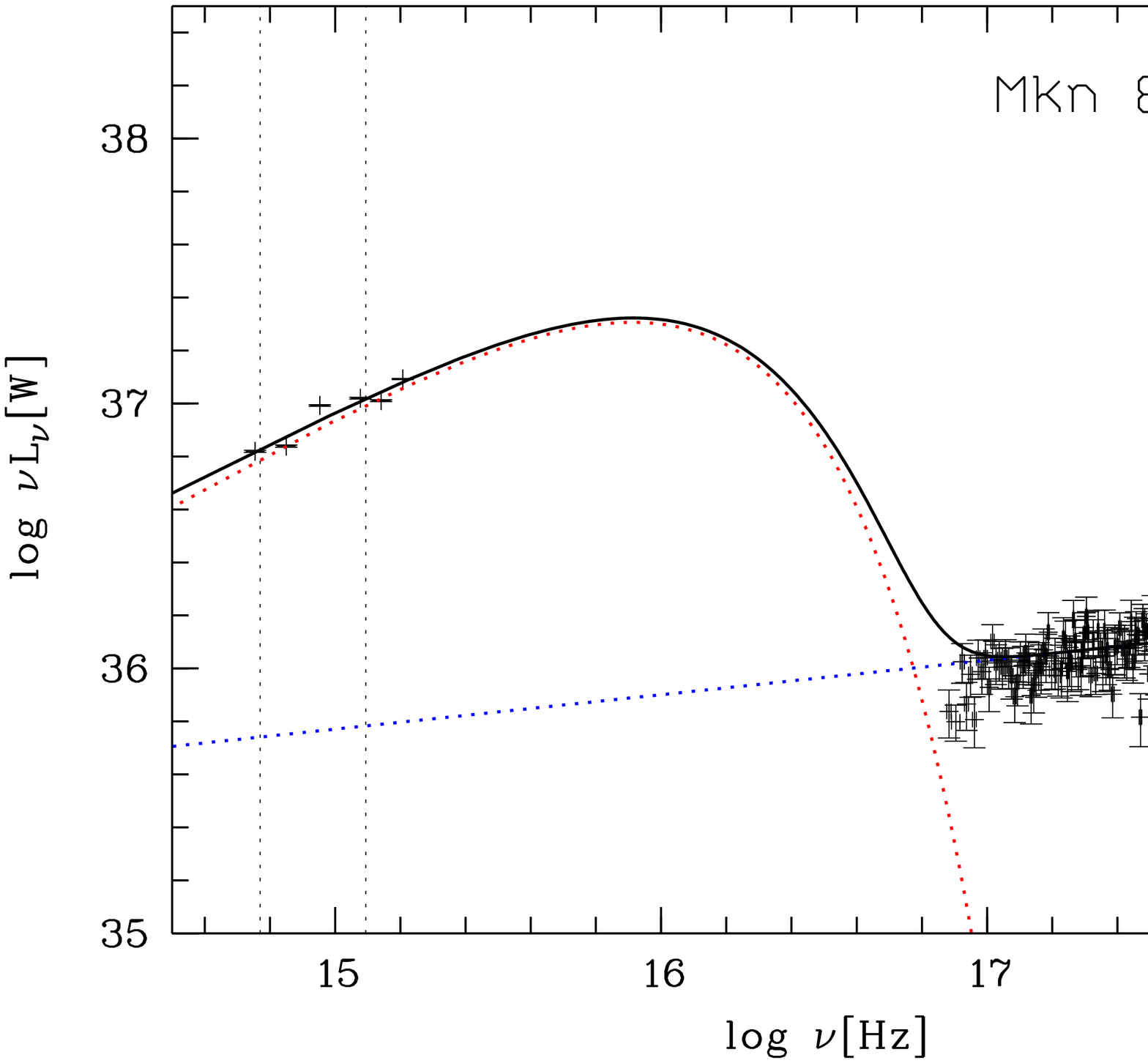}{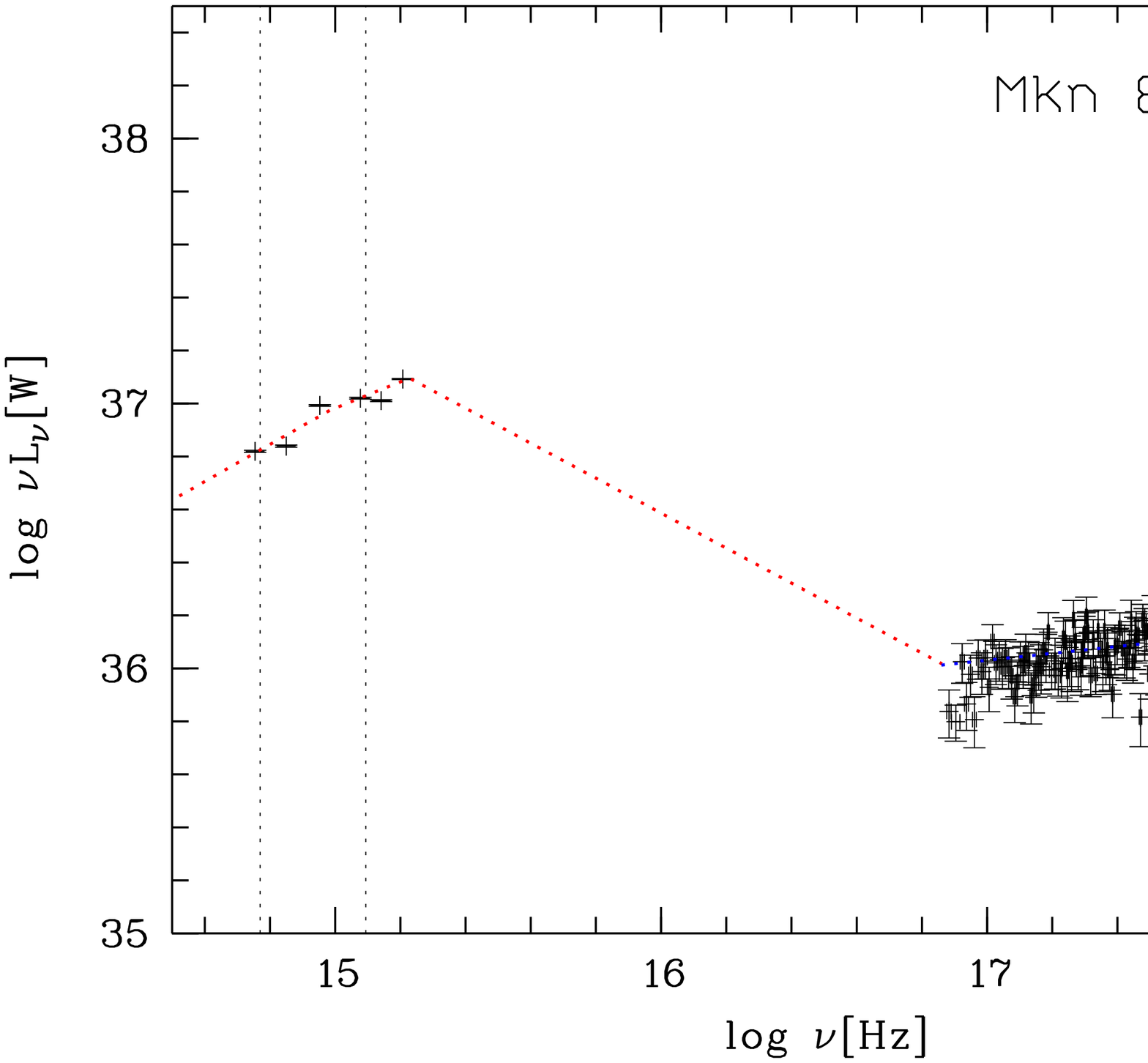}

\plotthree{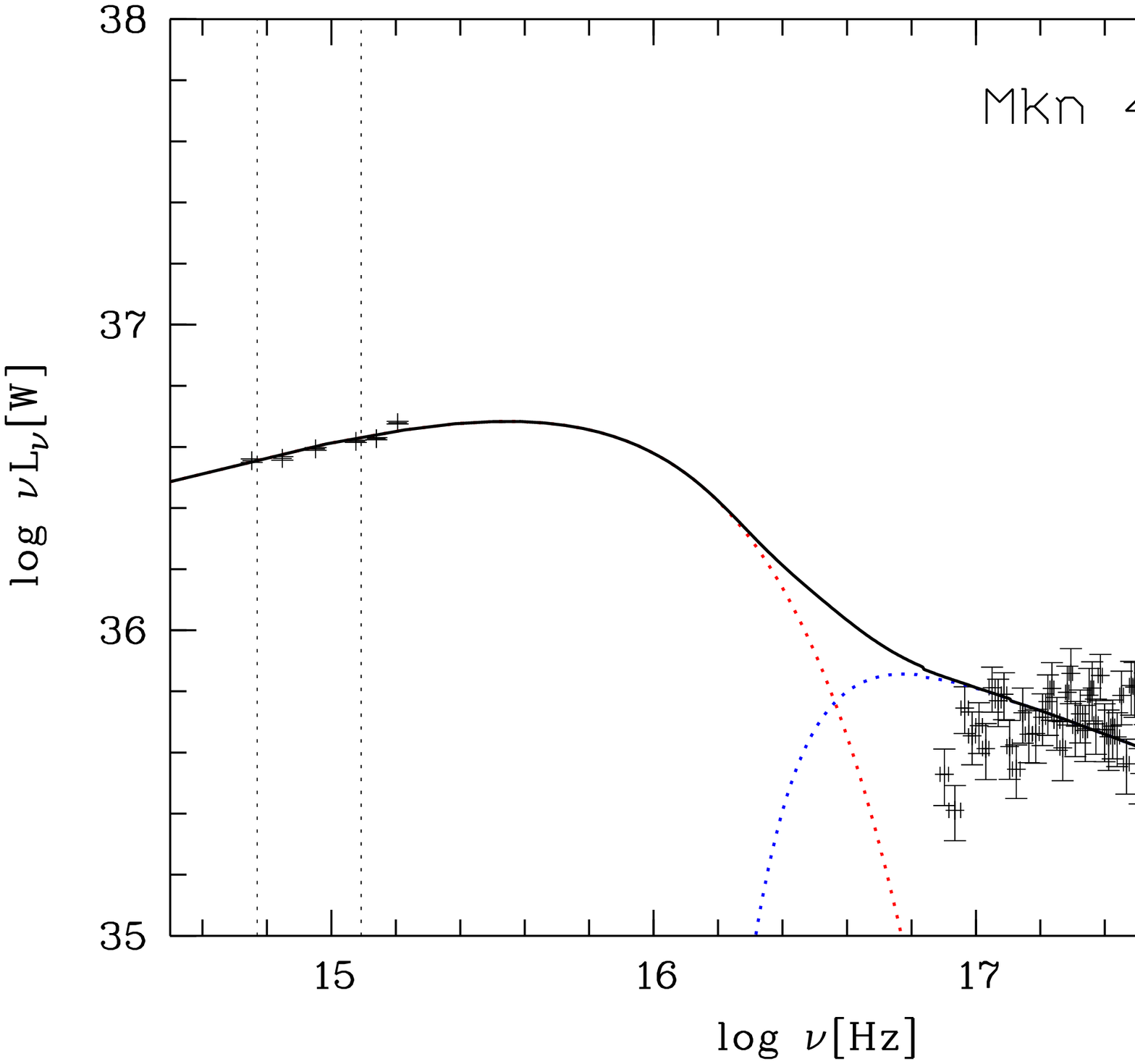}{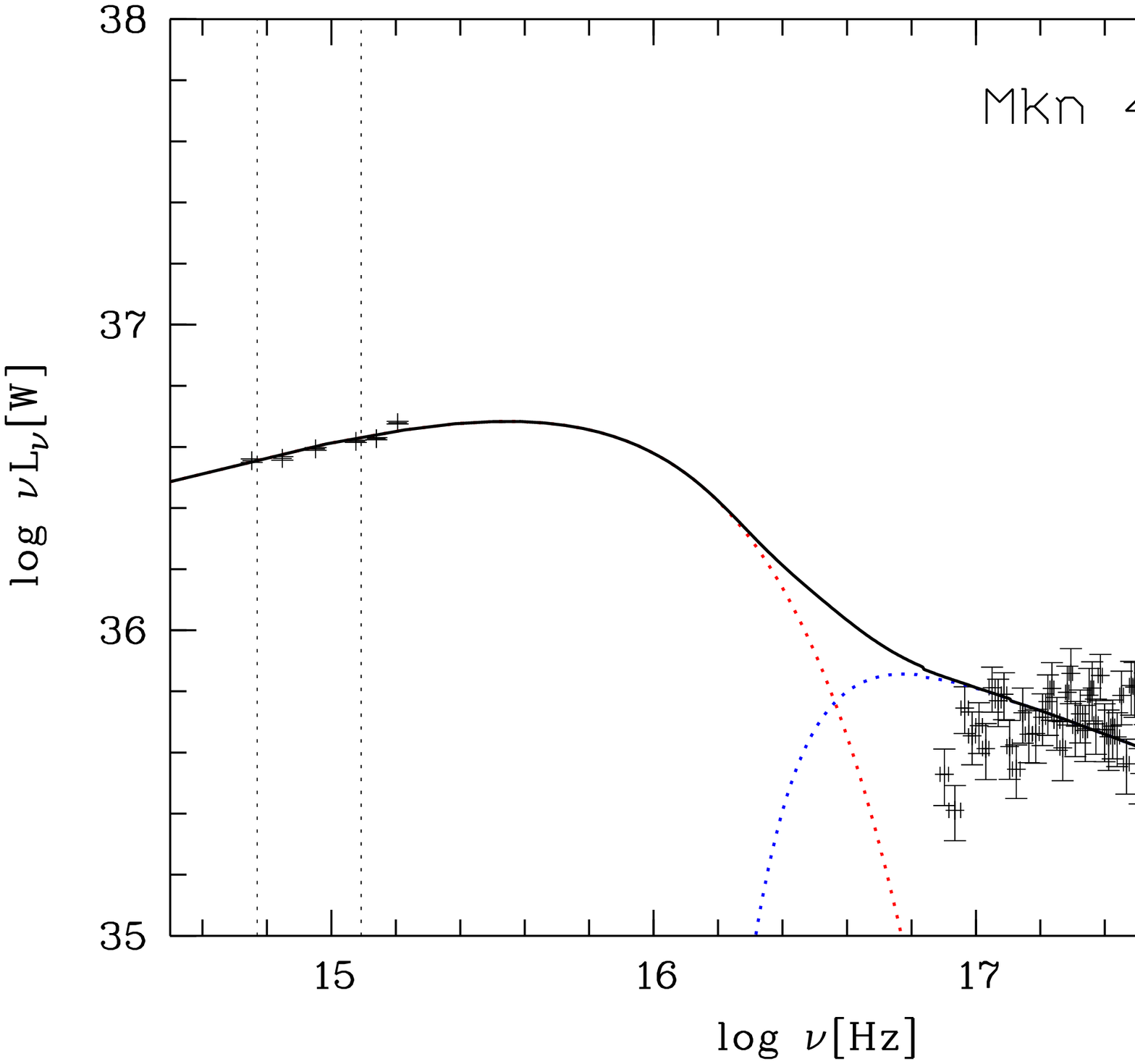}{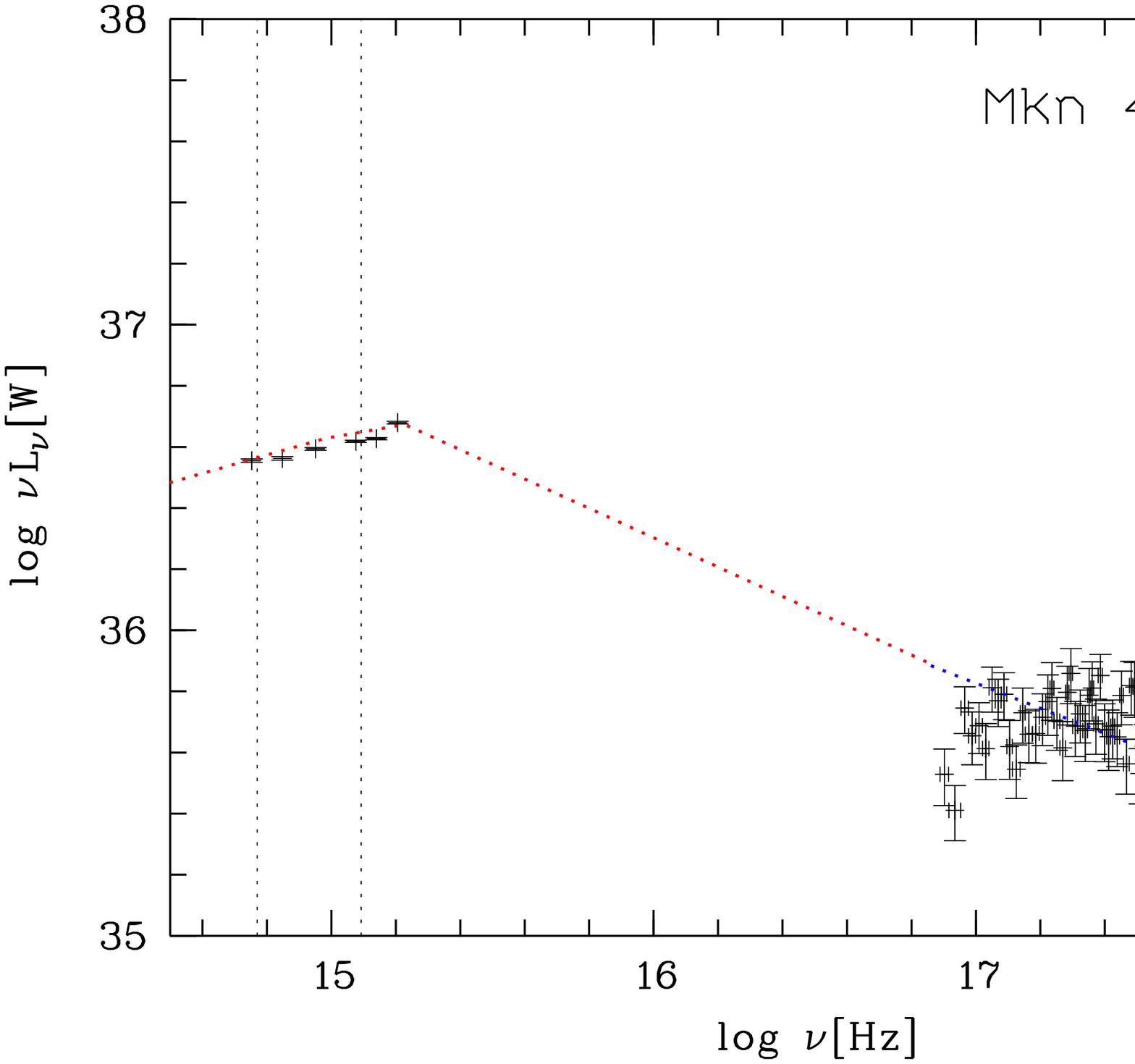}

\plotthree{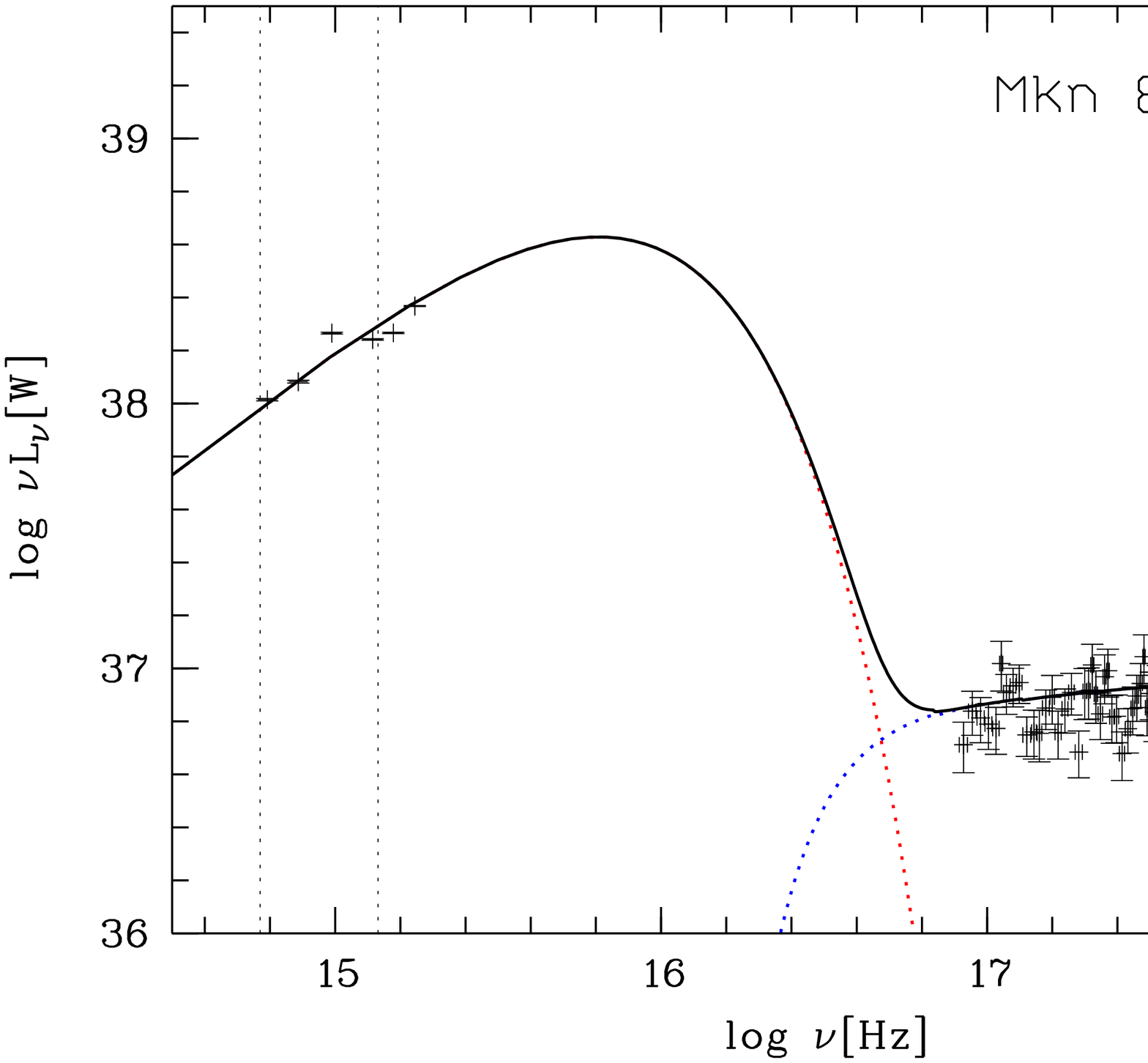}{f25_t.ps}{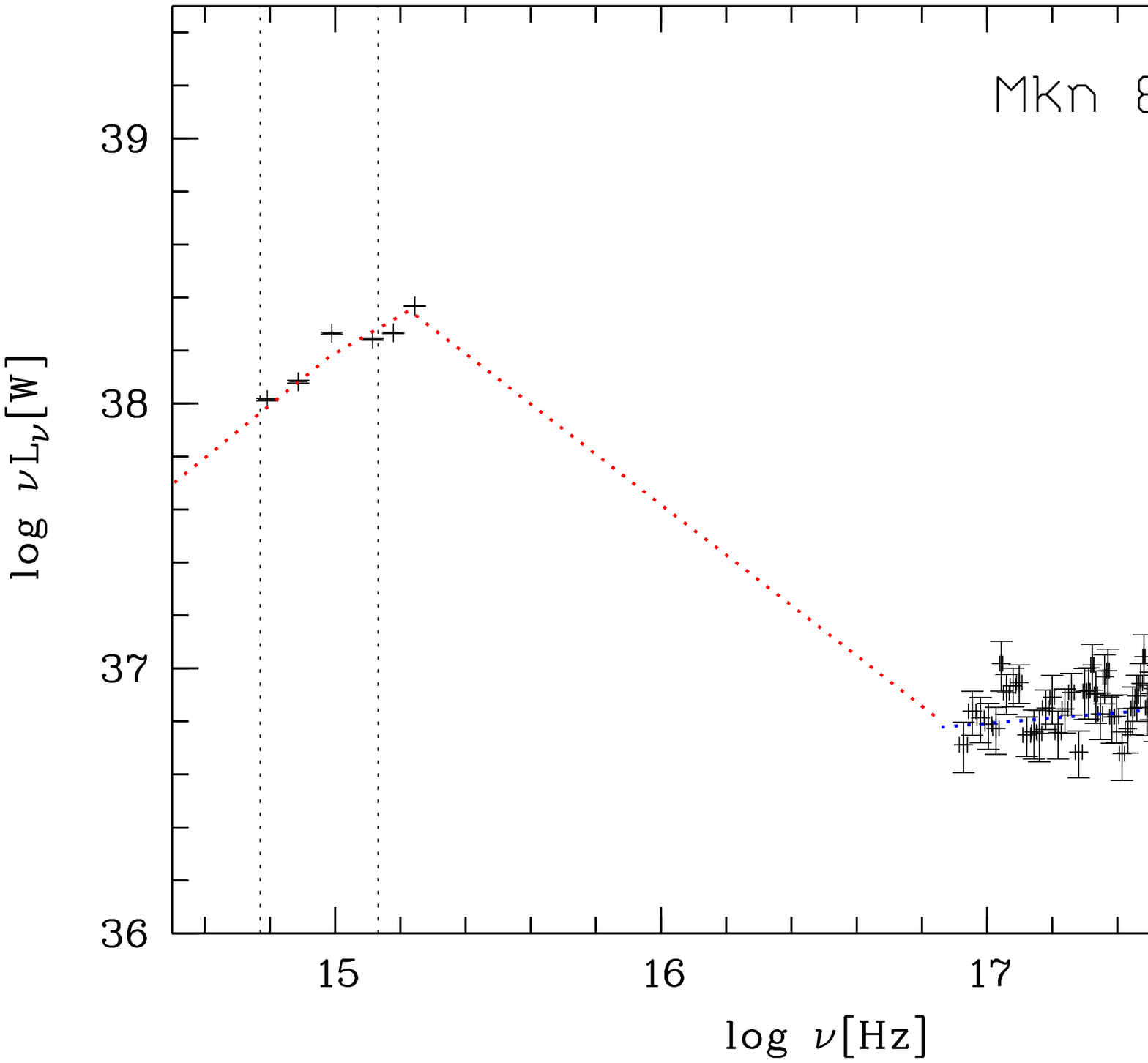}

\plotthree{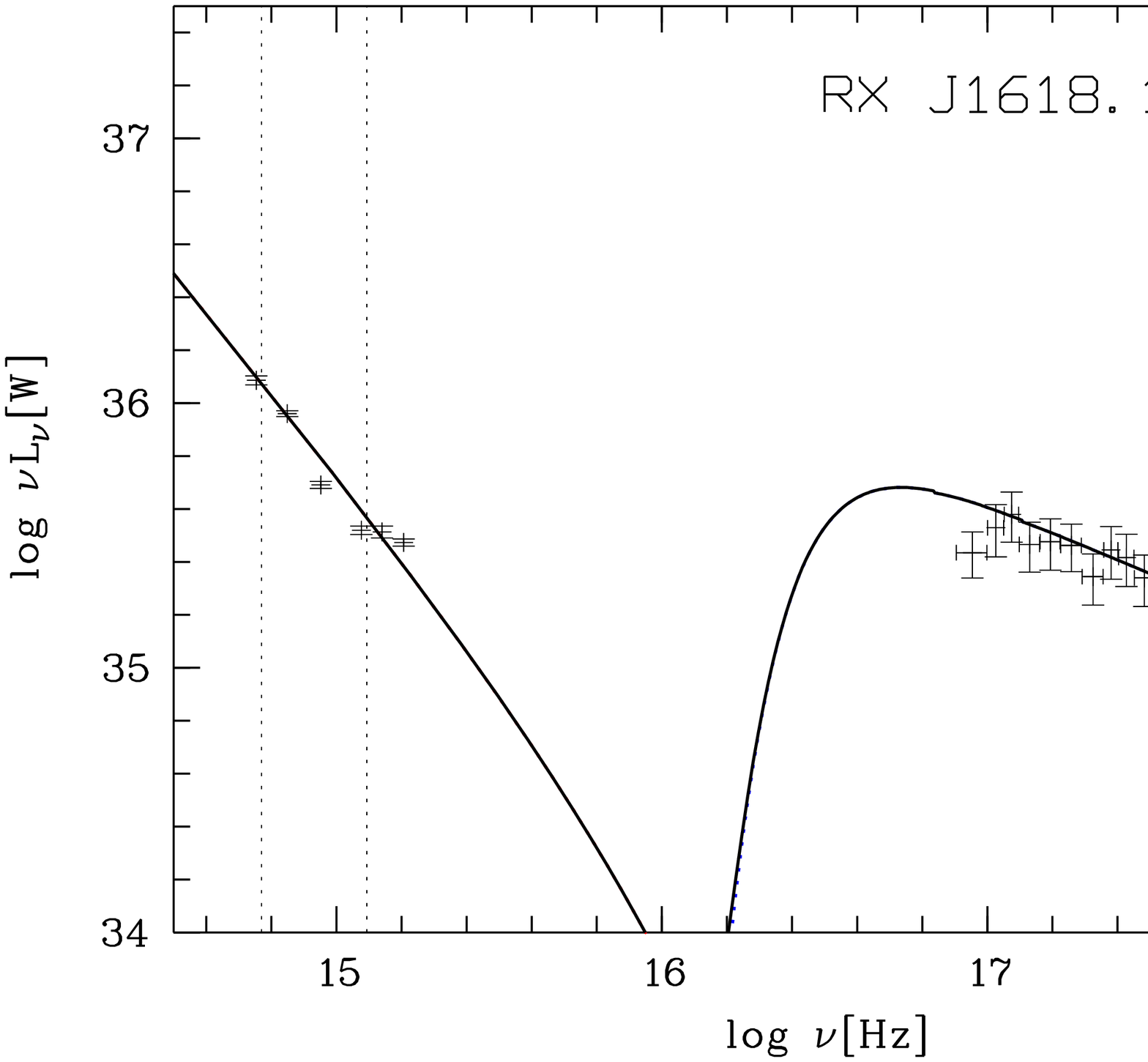}{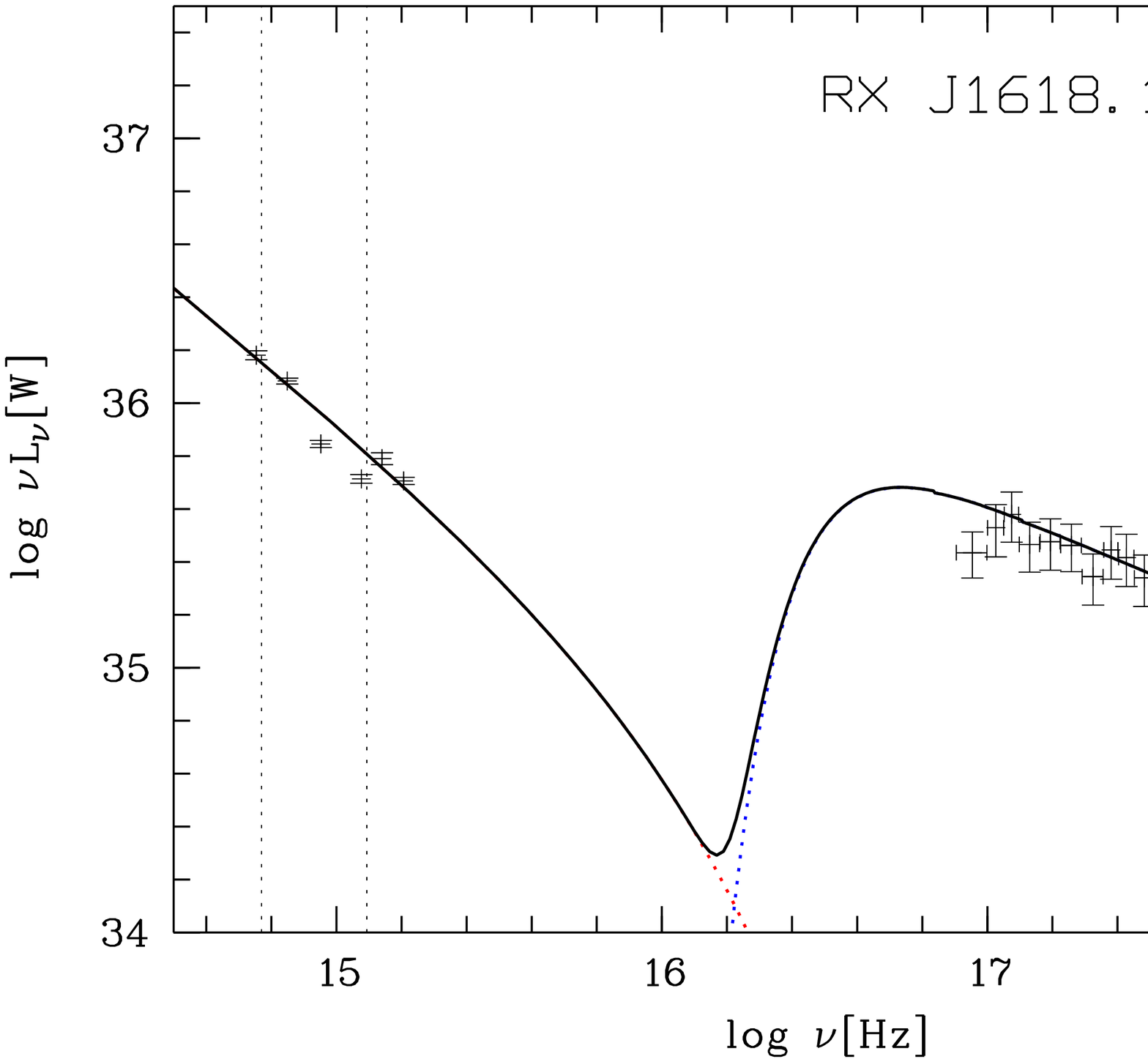}{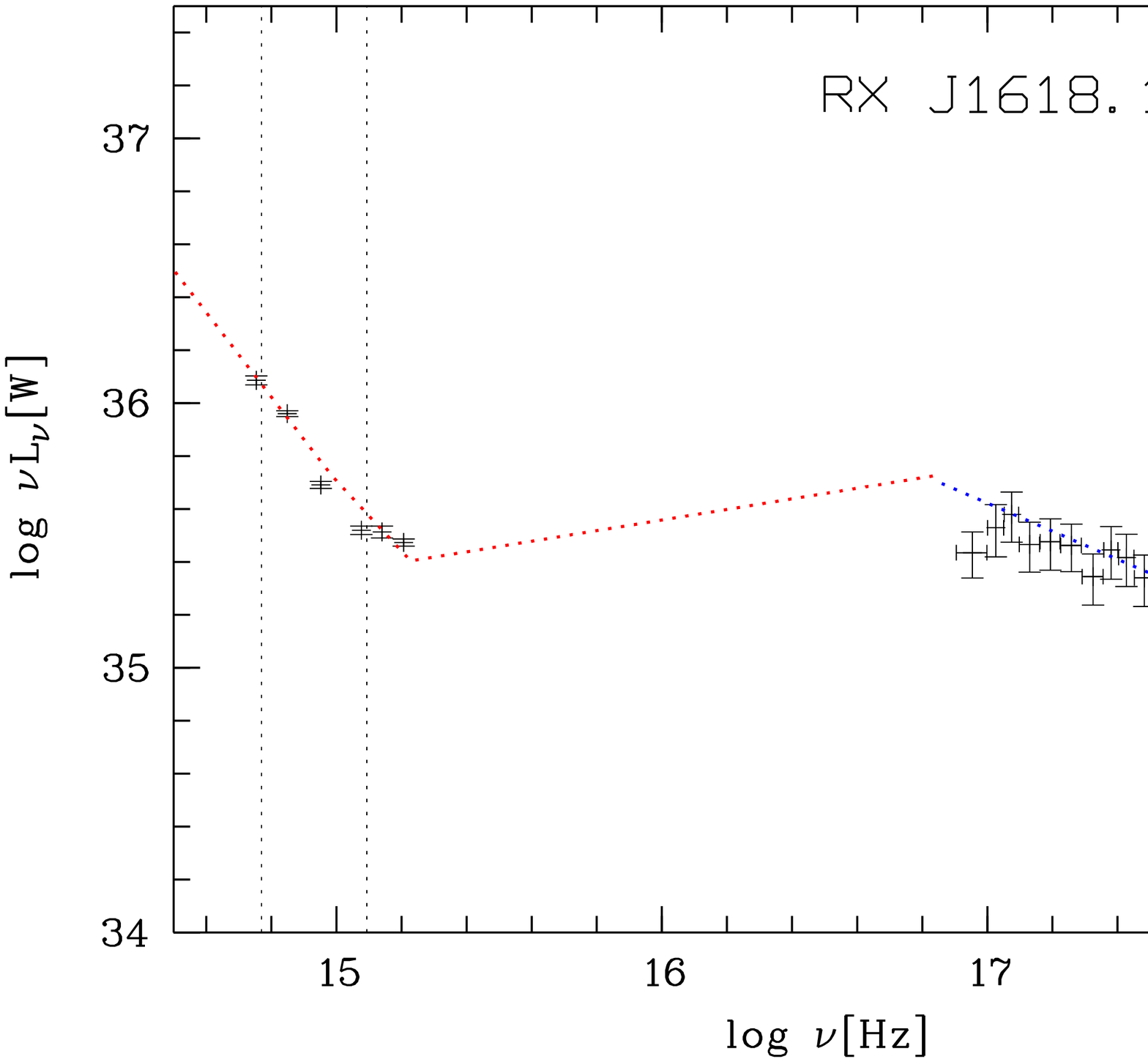}
\end{figure*}

\begin{figure*}
\epsscale{0.60}
\plotthree{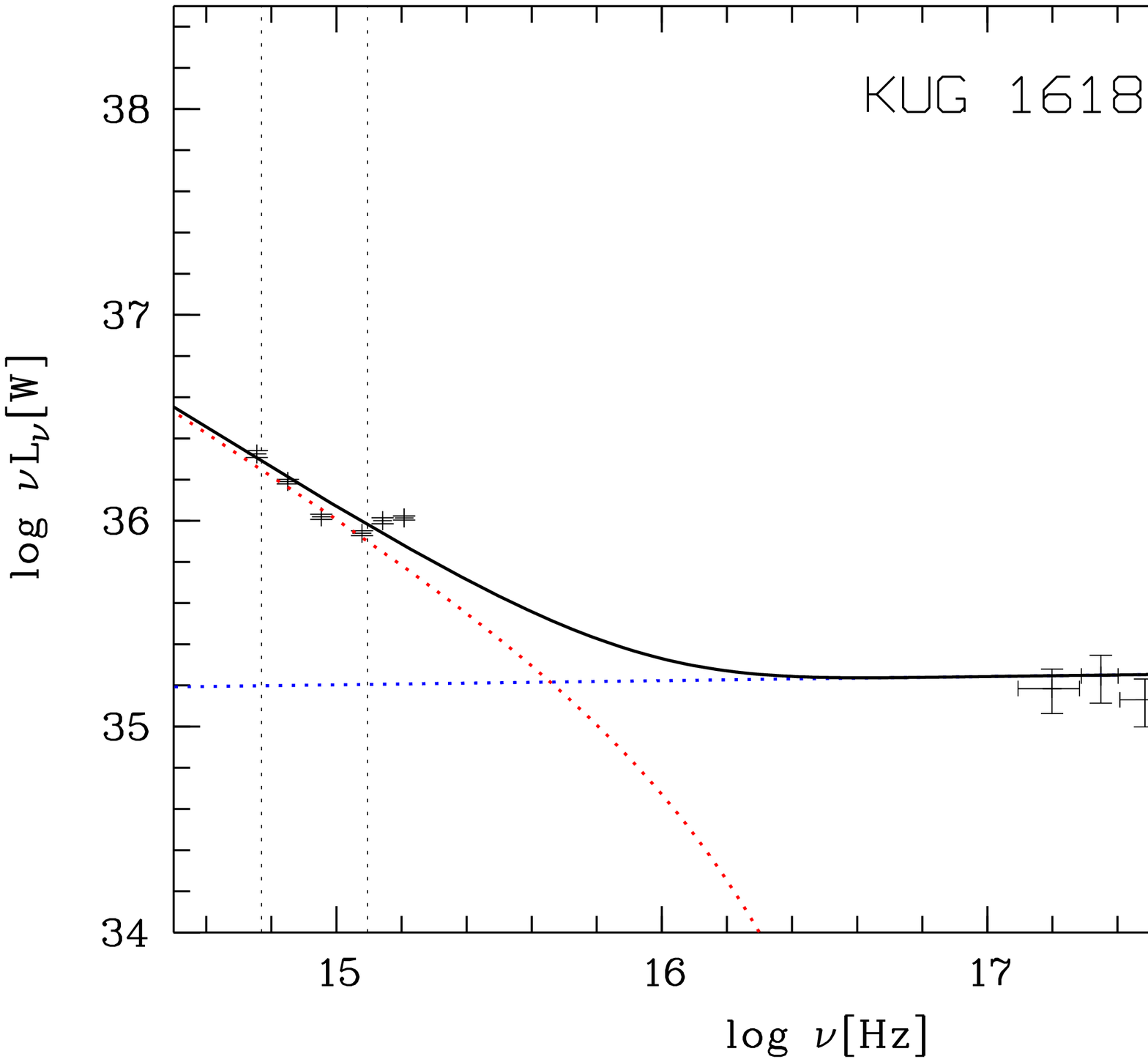}{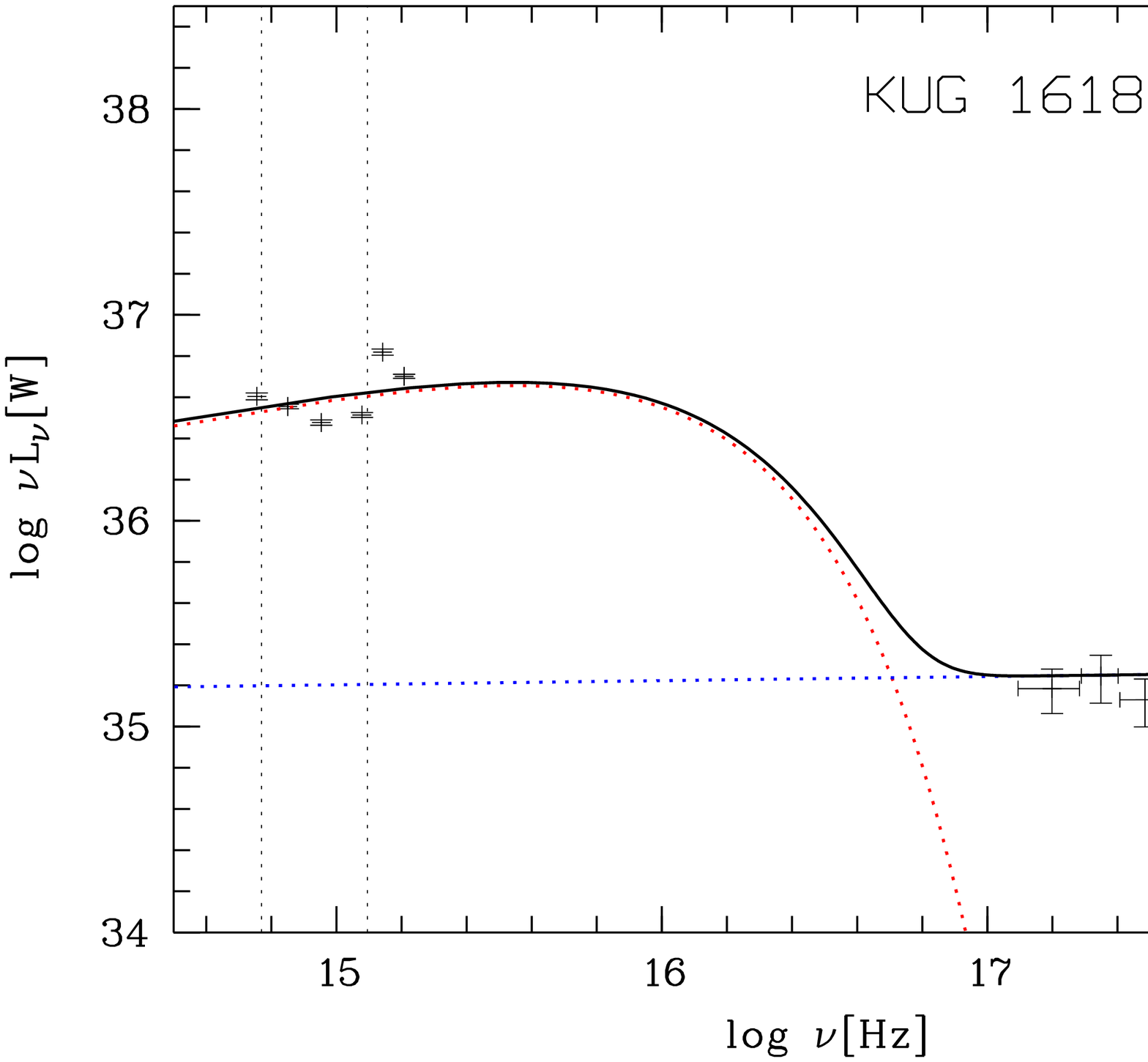}{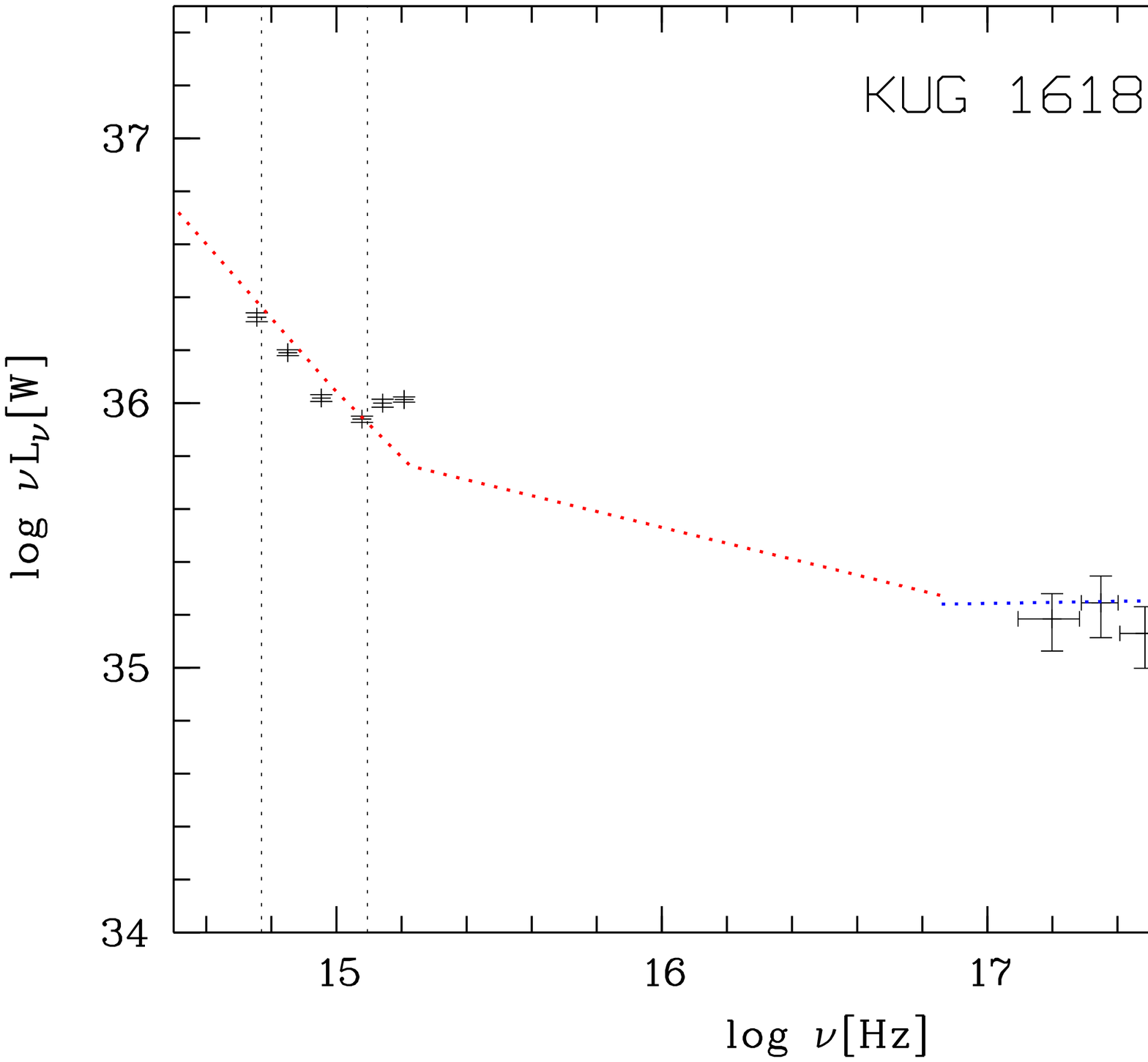}

\plotthree{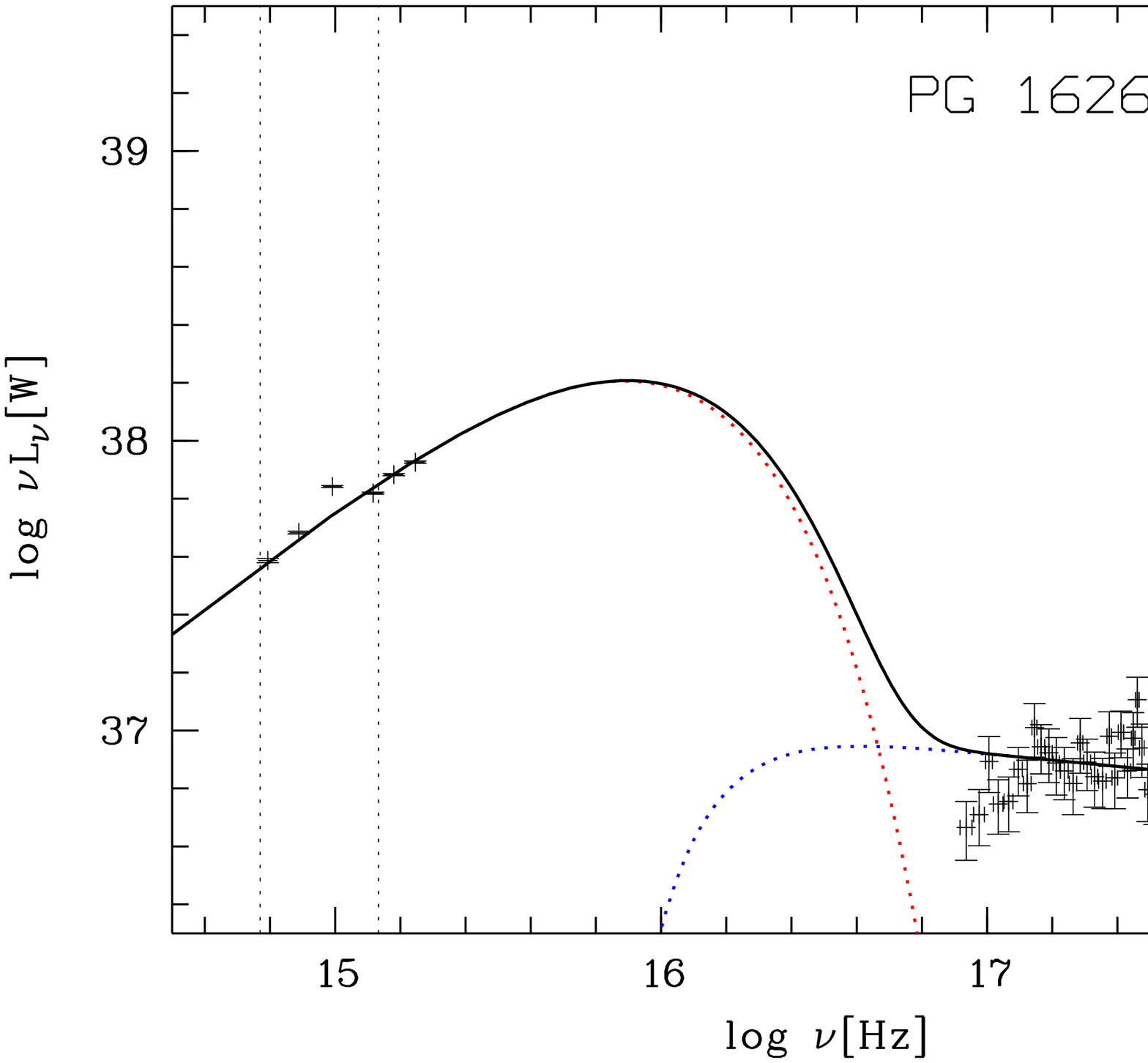}{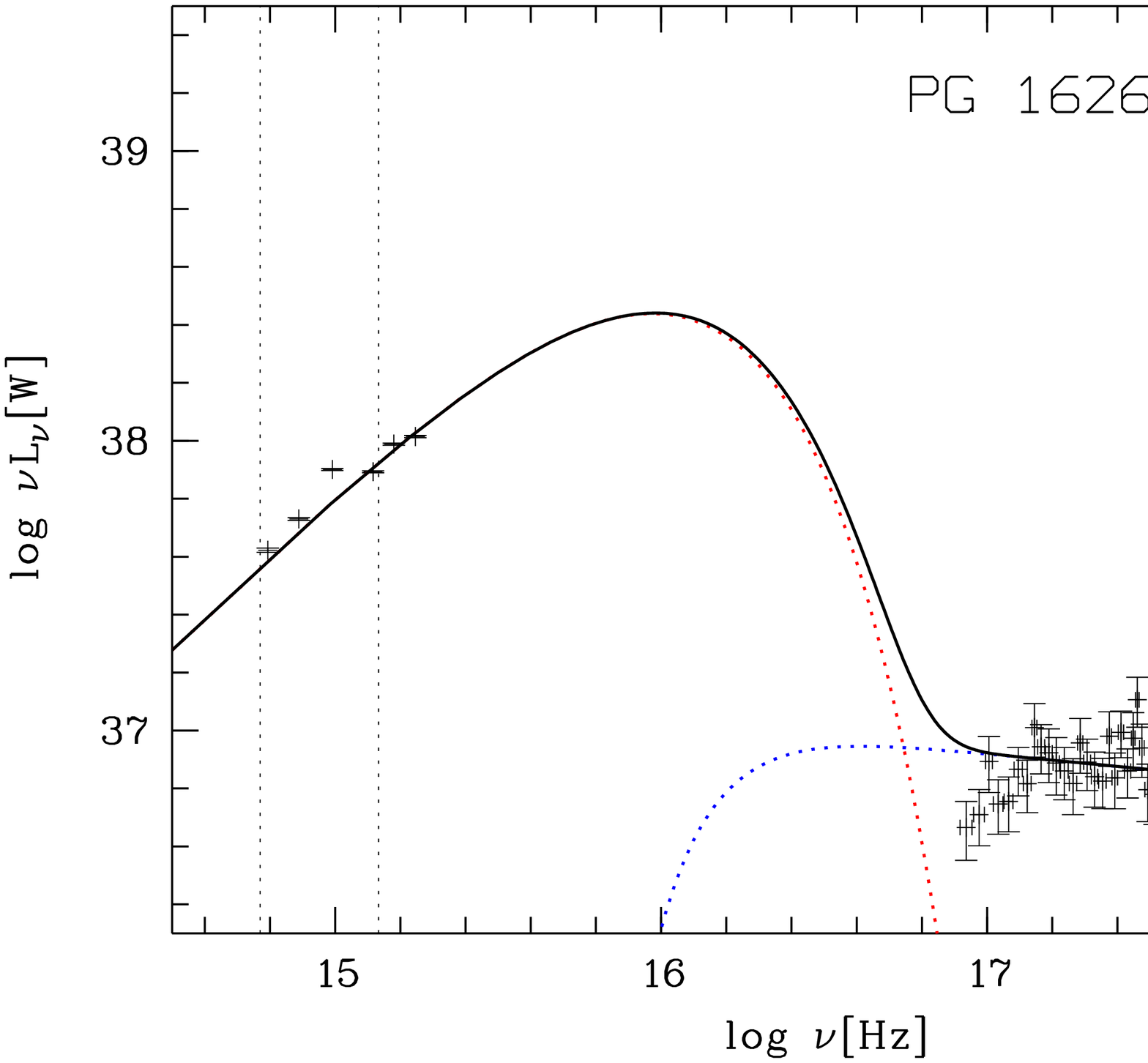}{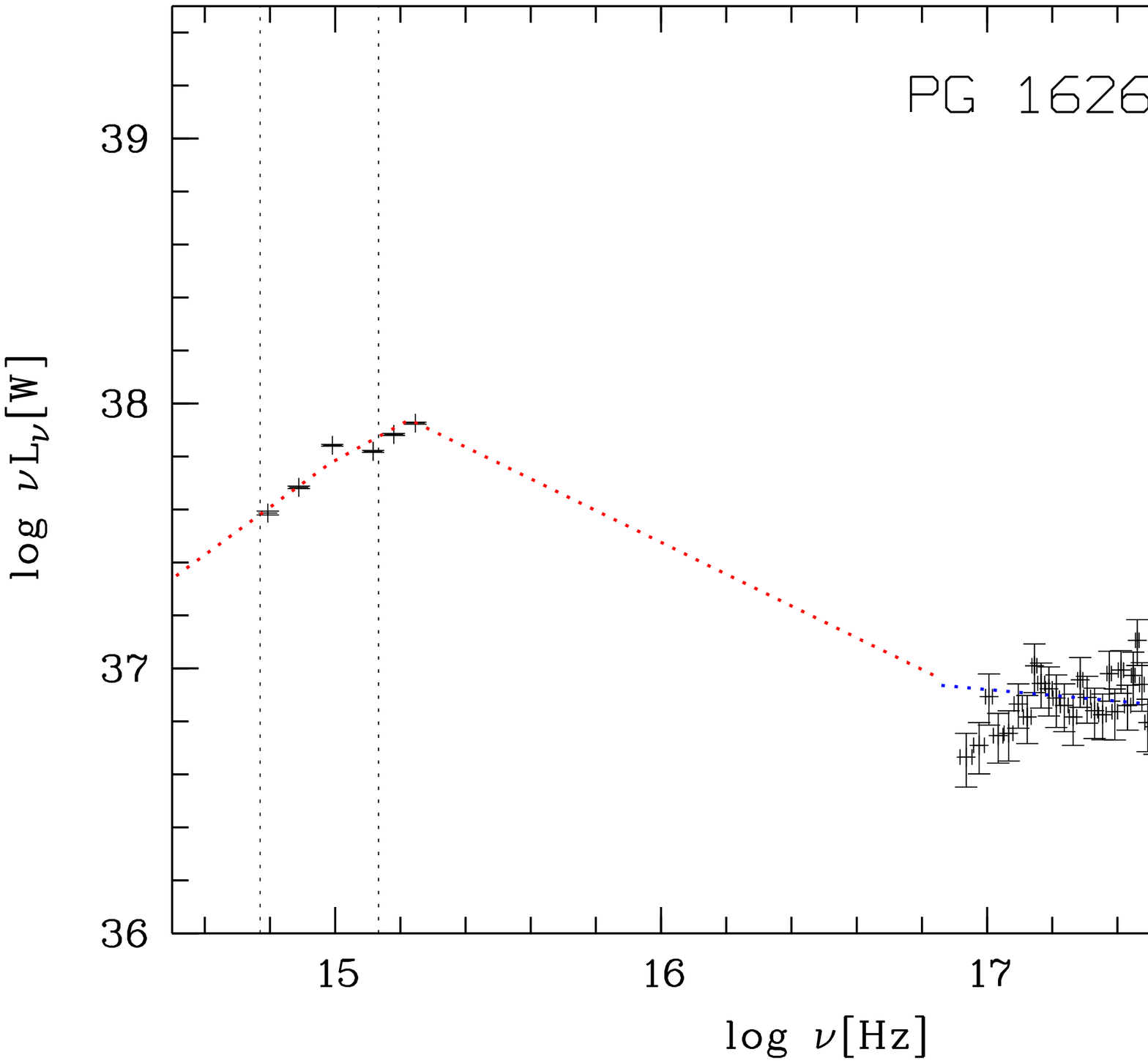}

\plotthree{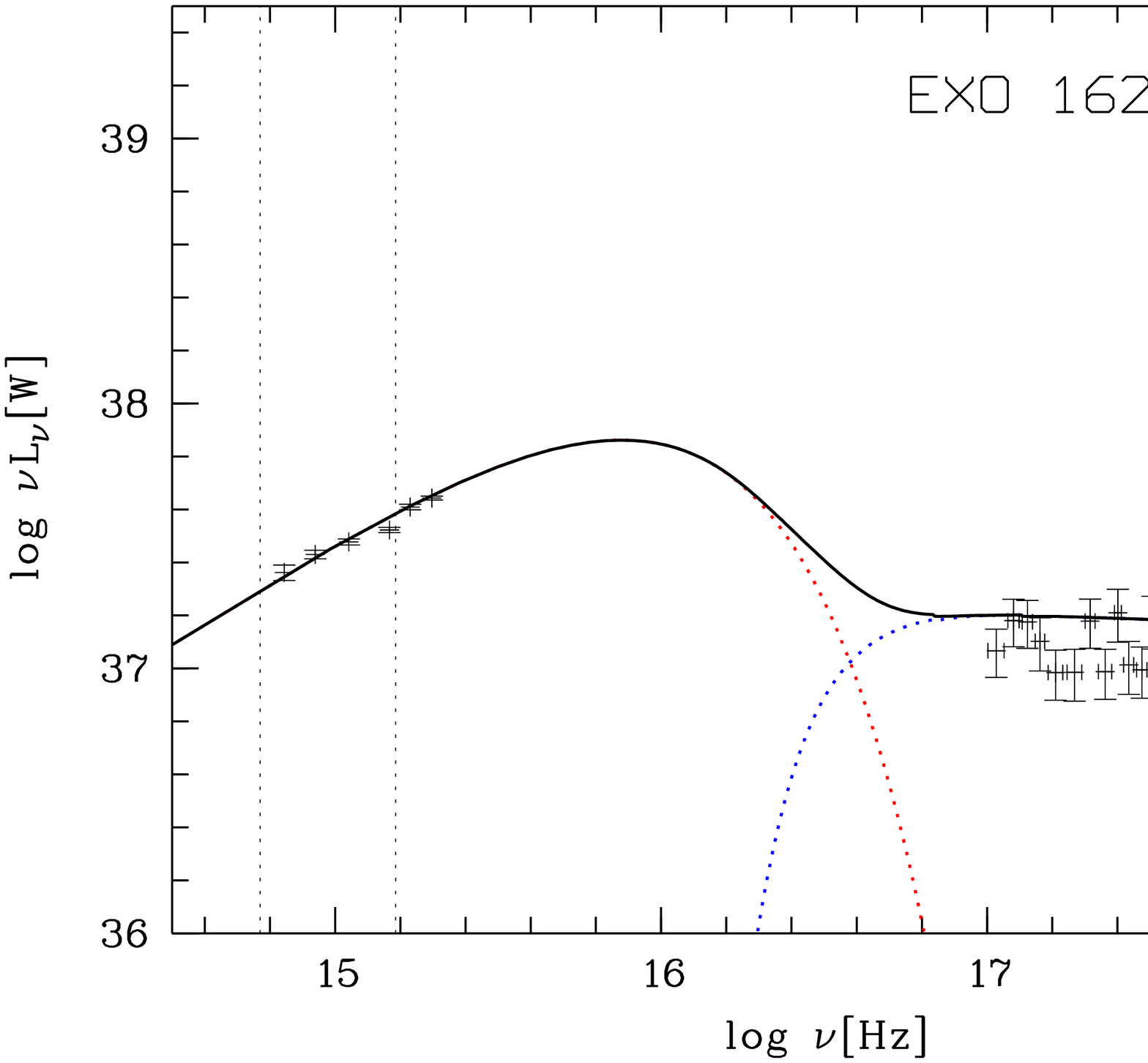}{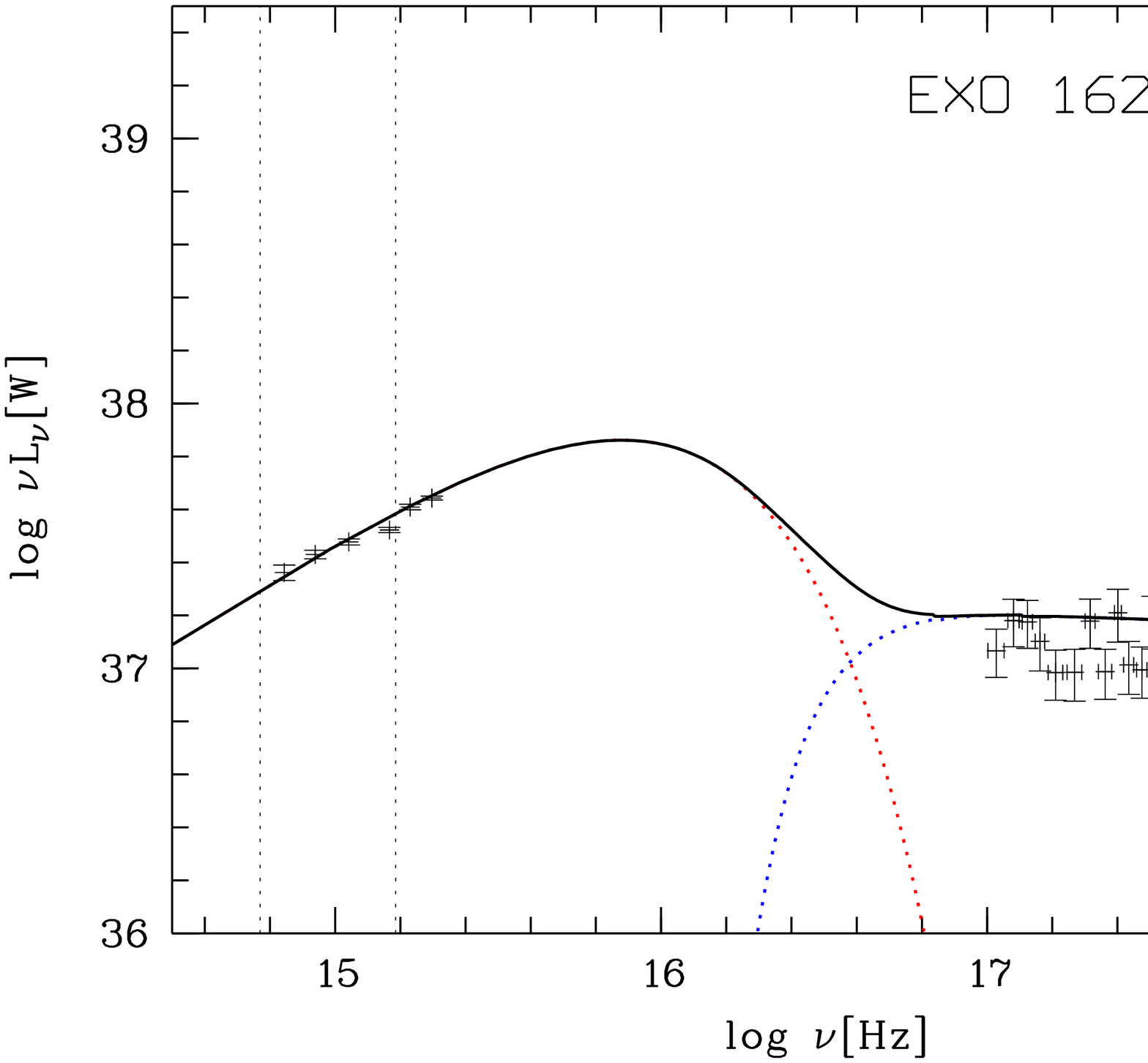}{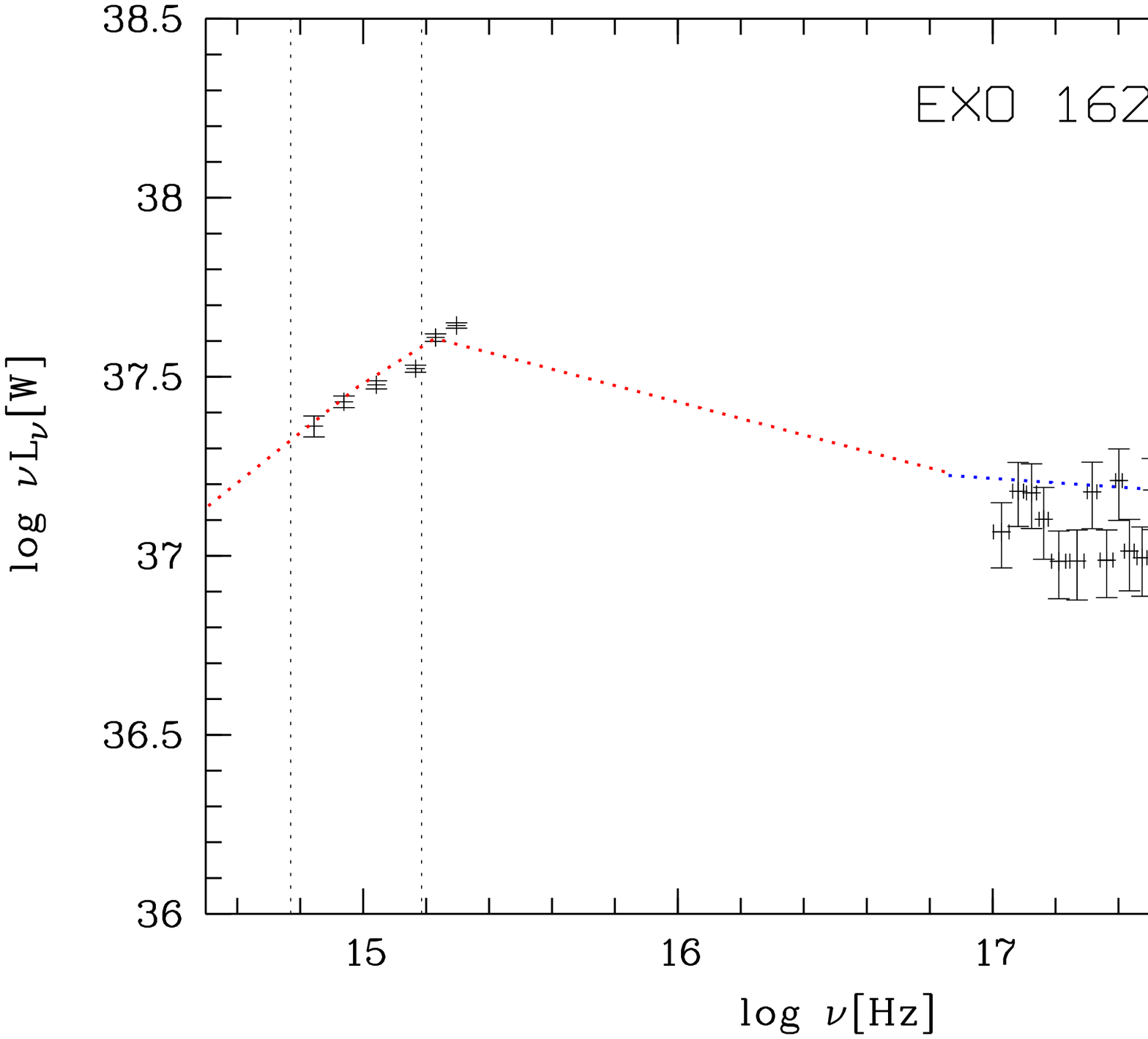}

\plotthree{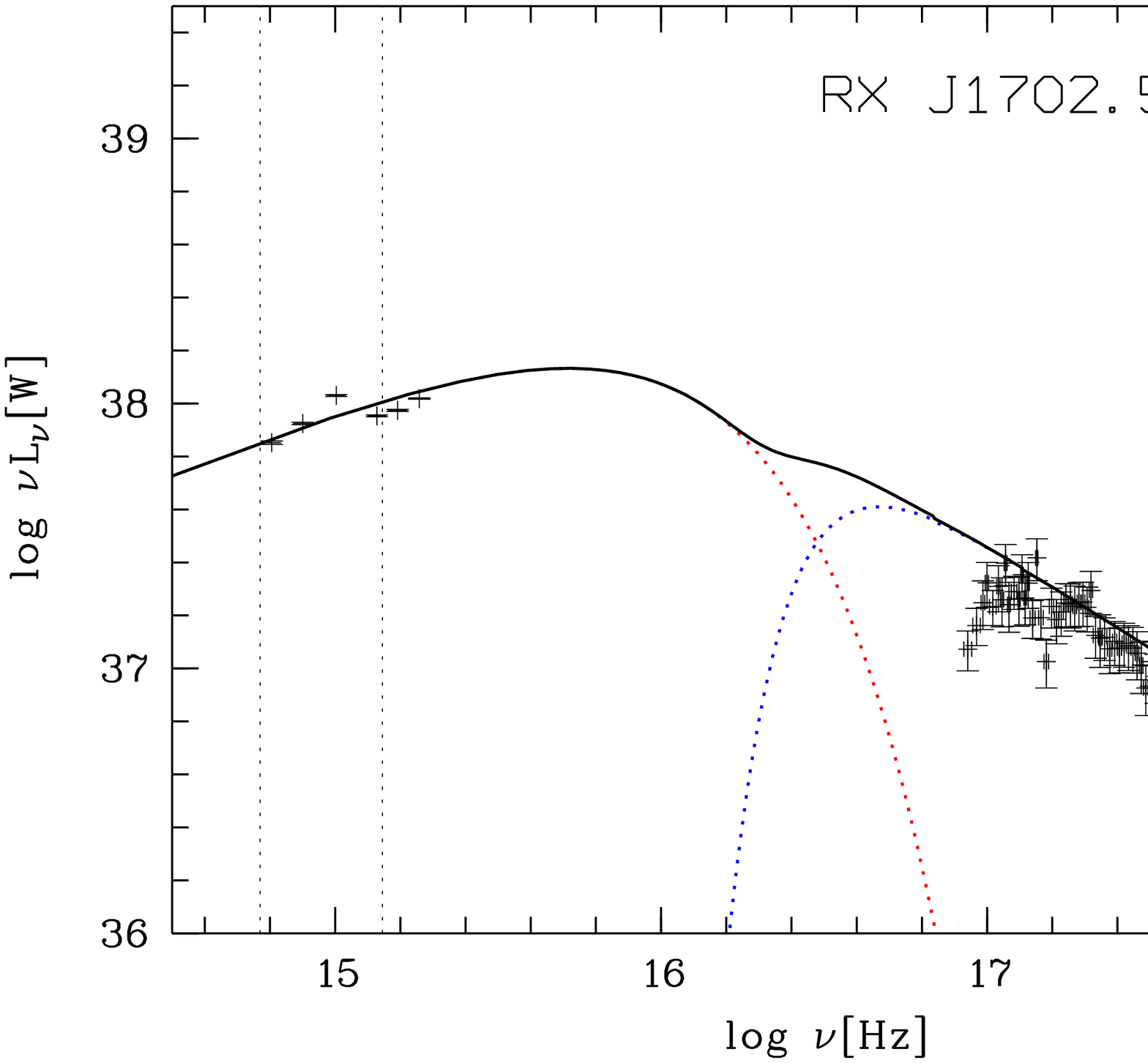}{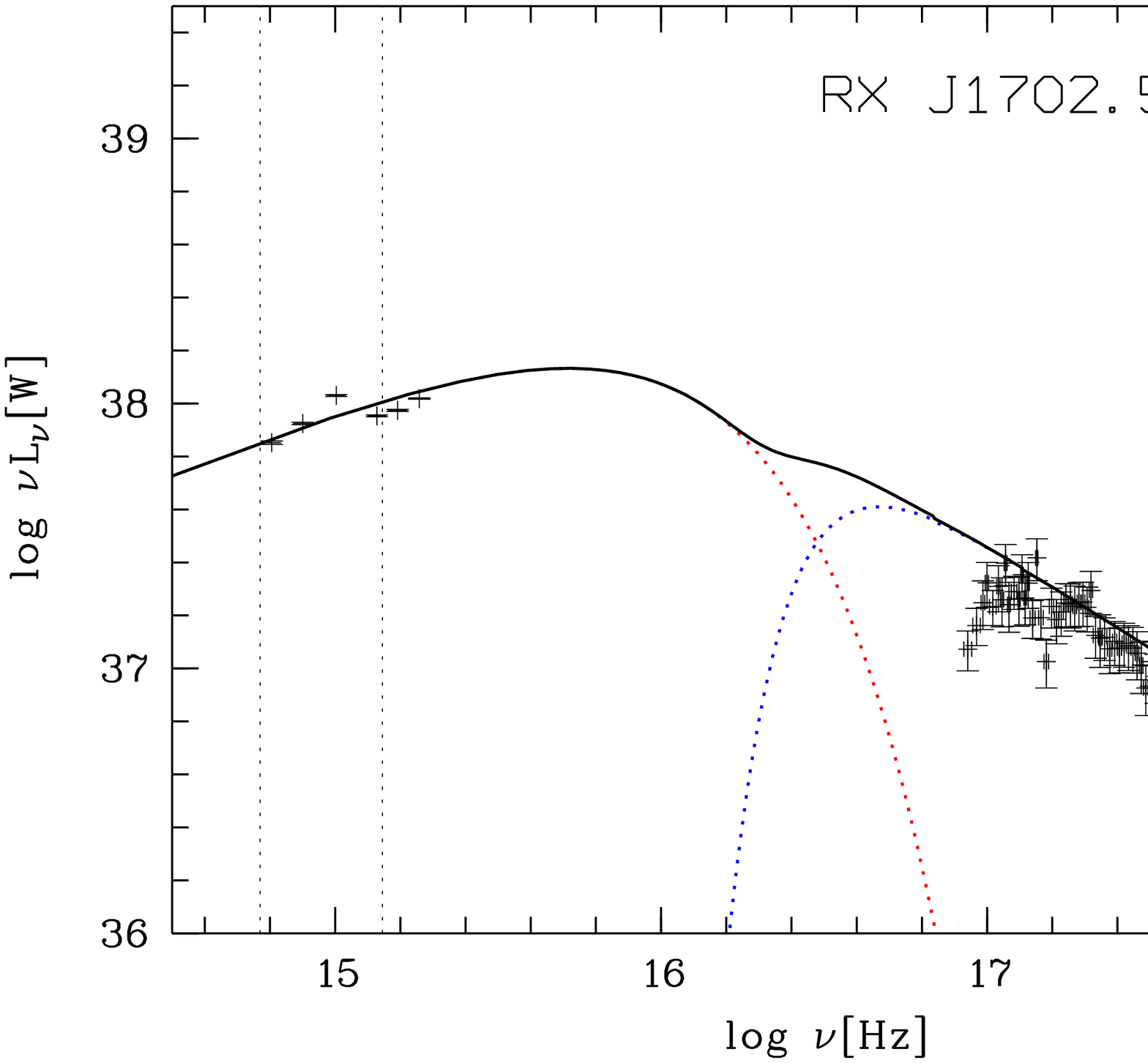}{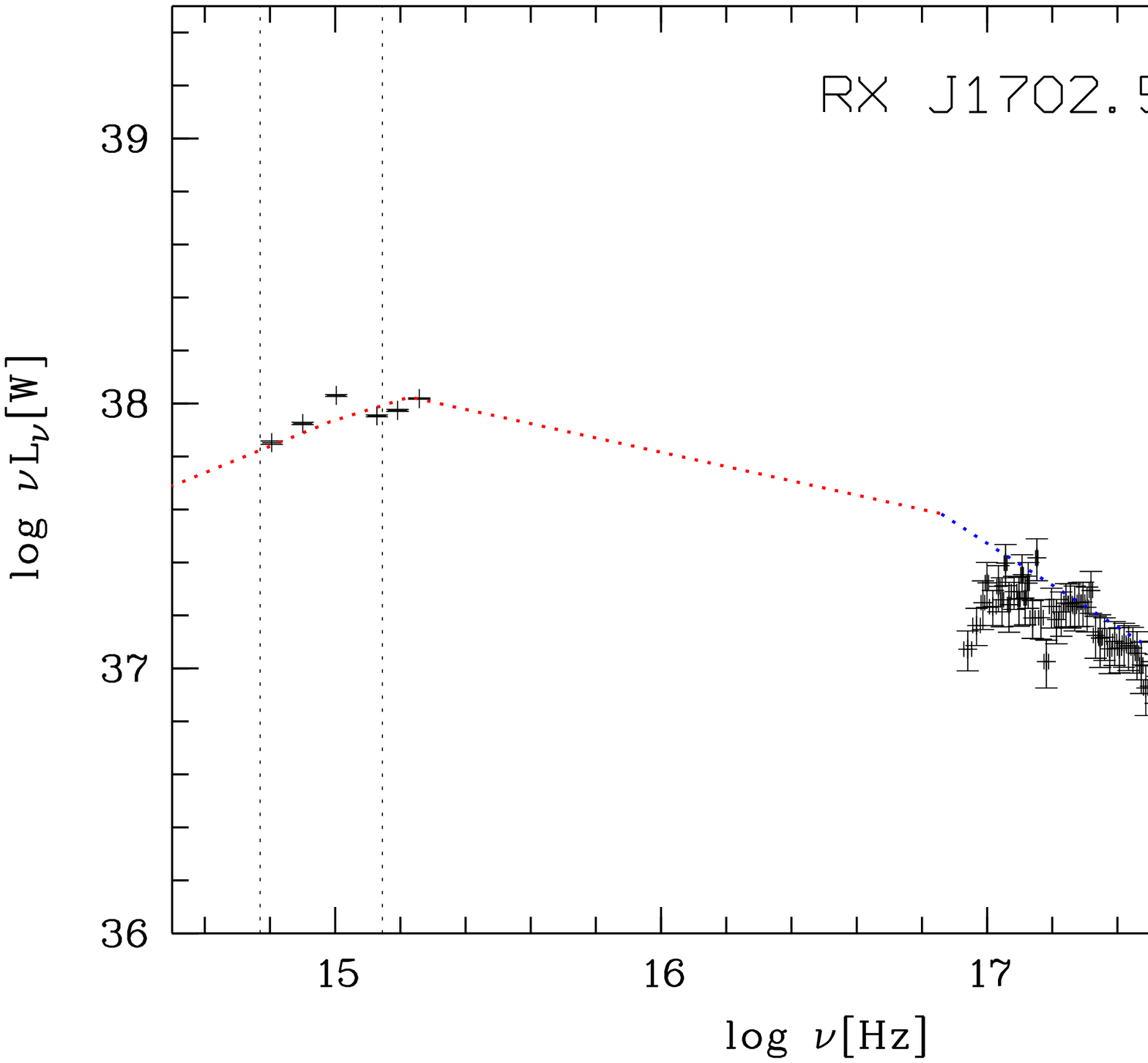}

\plotthree{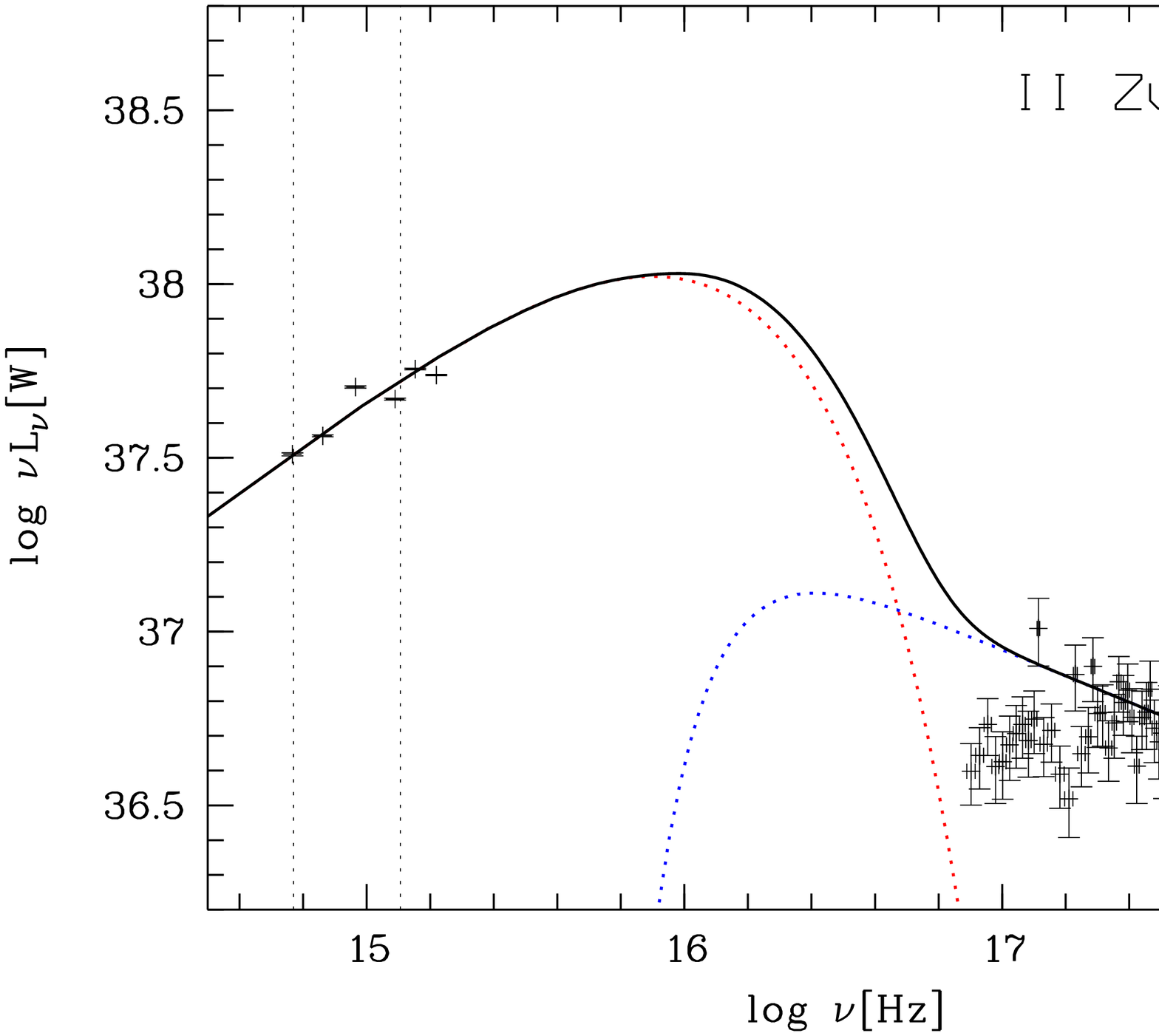}{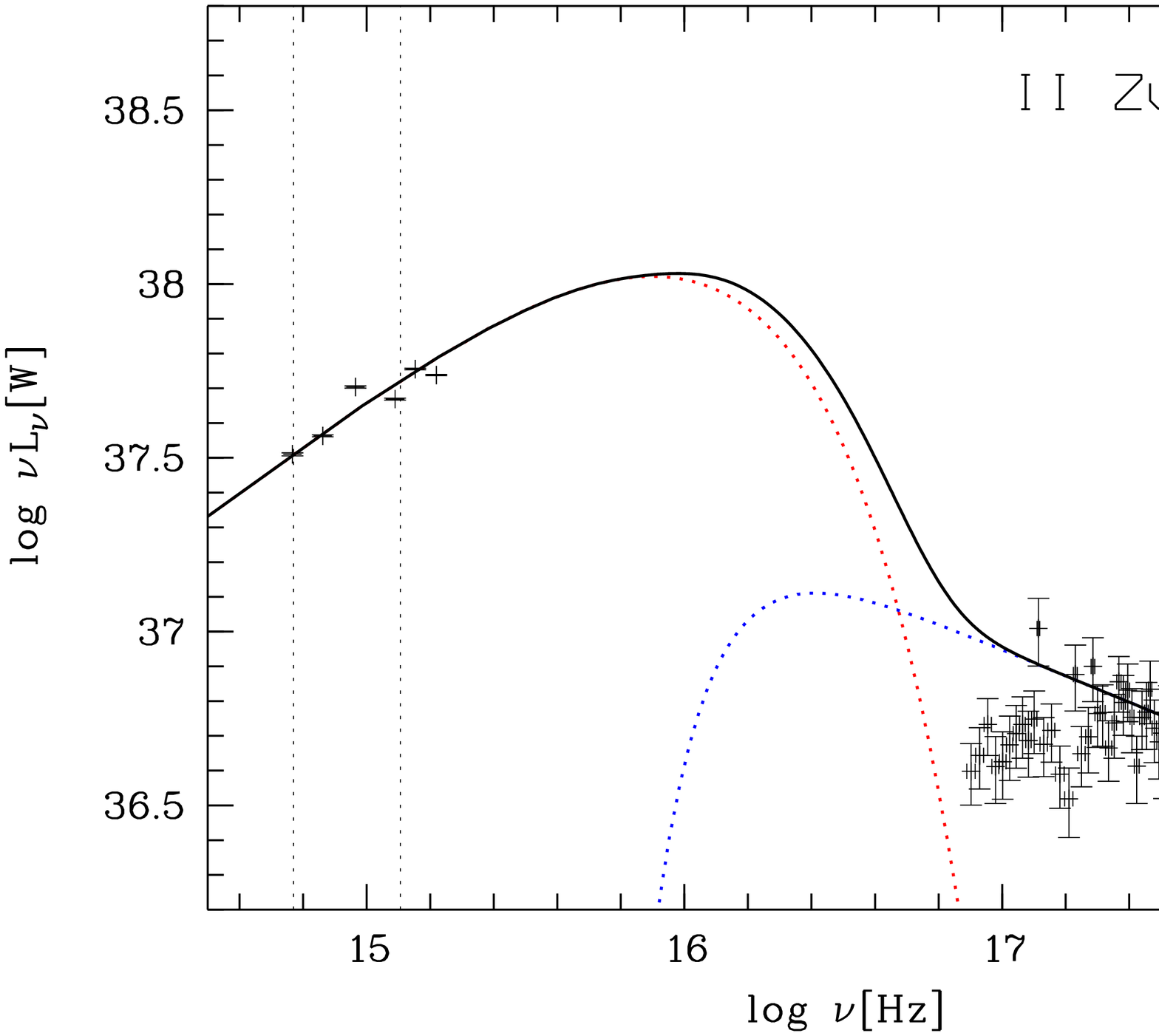}{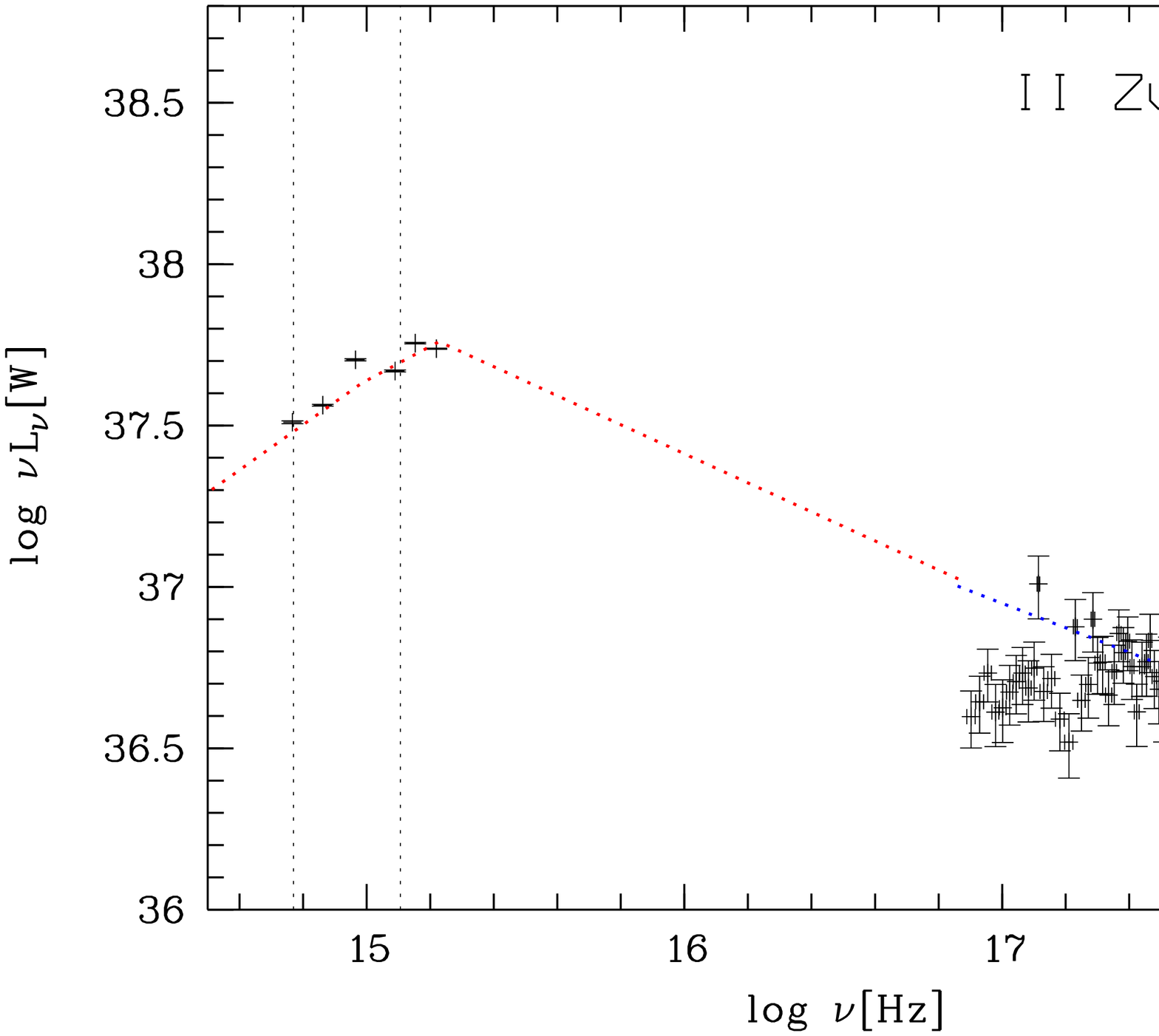}
\end{figure*}

\begin{figure*}
\epsscale{0.60}
\plotthree{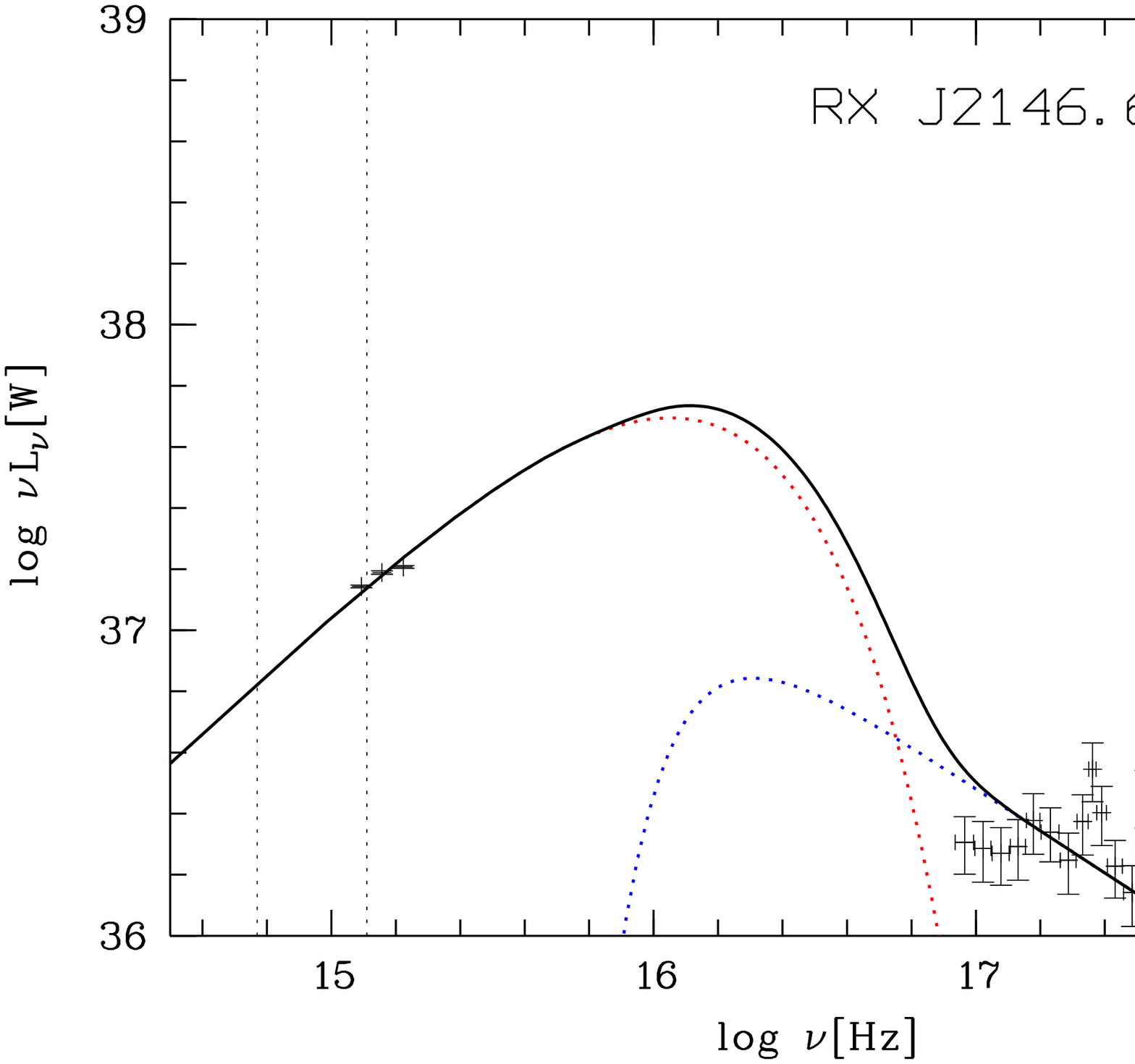}{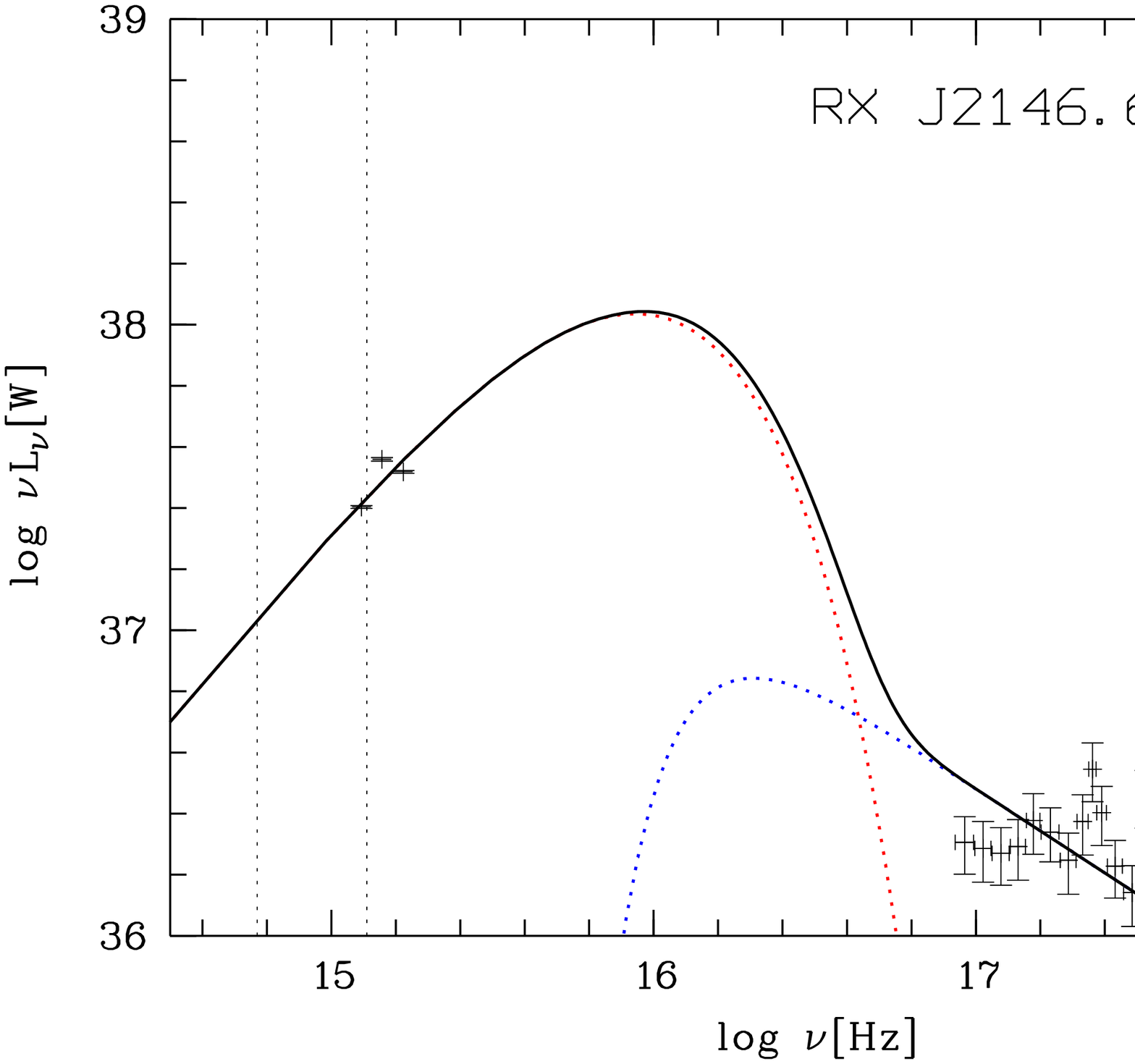}{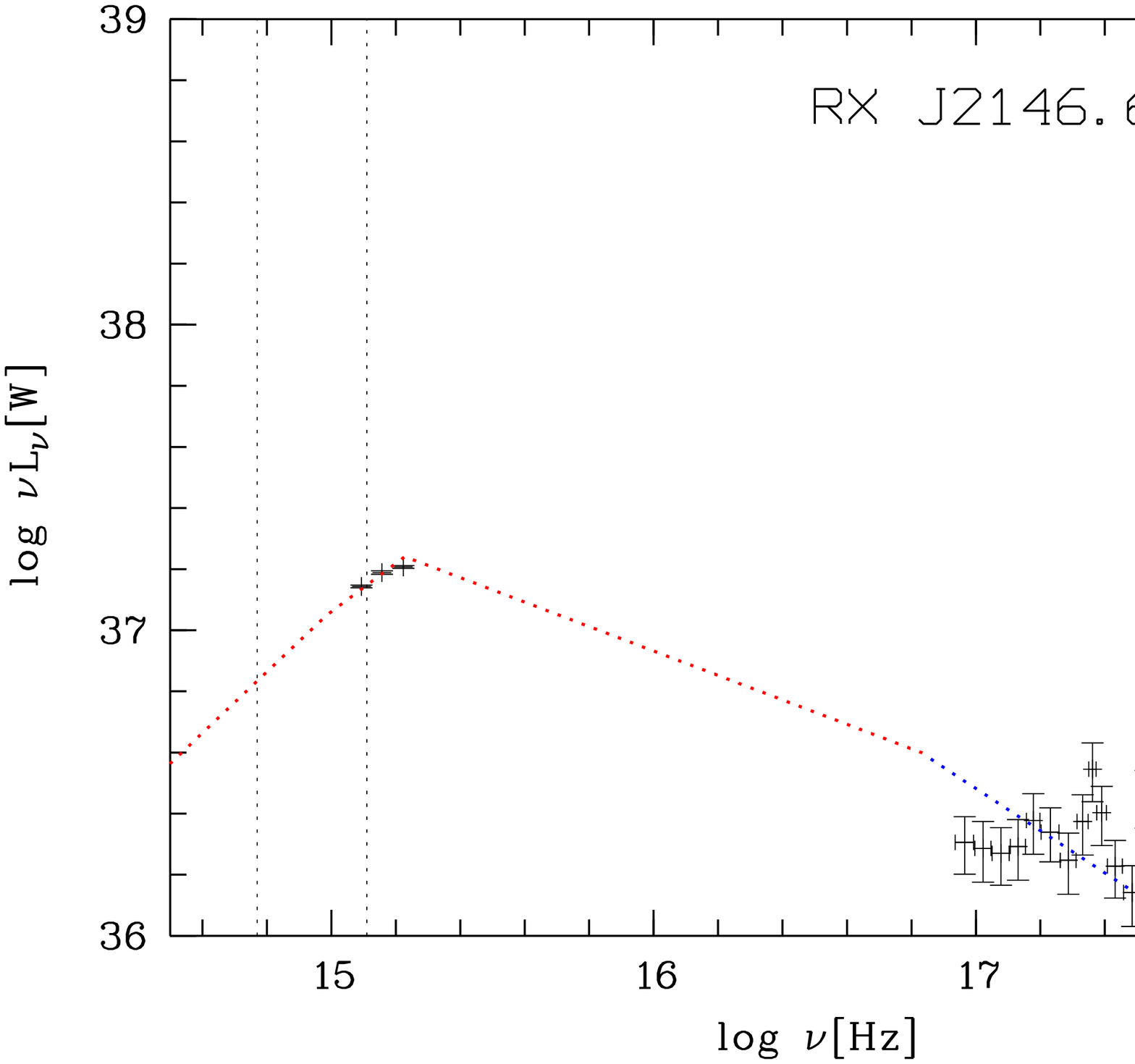}

\plotthree{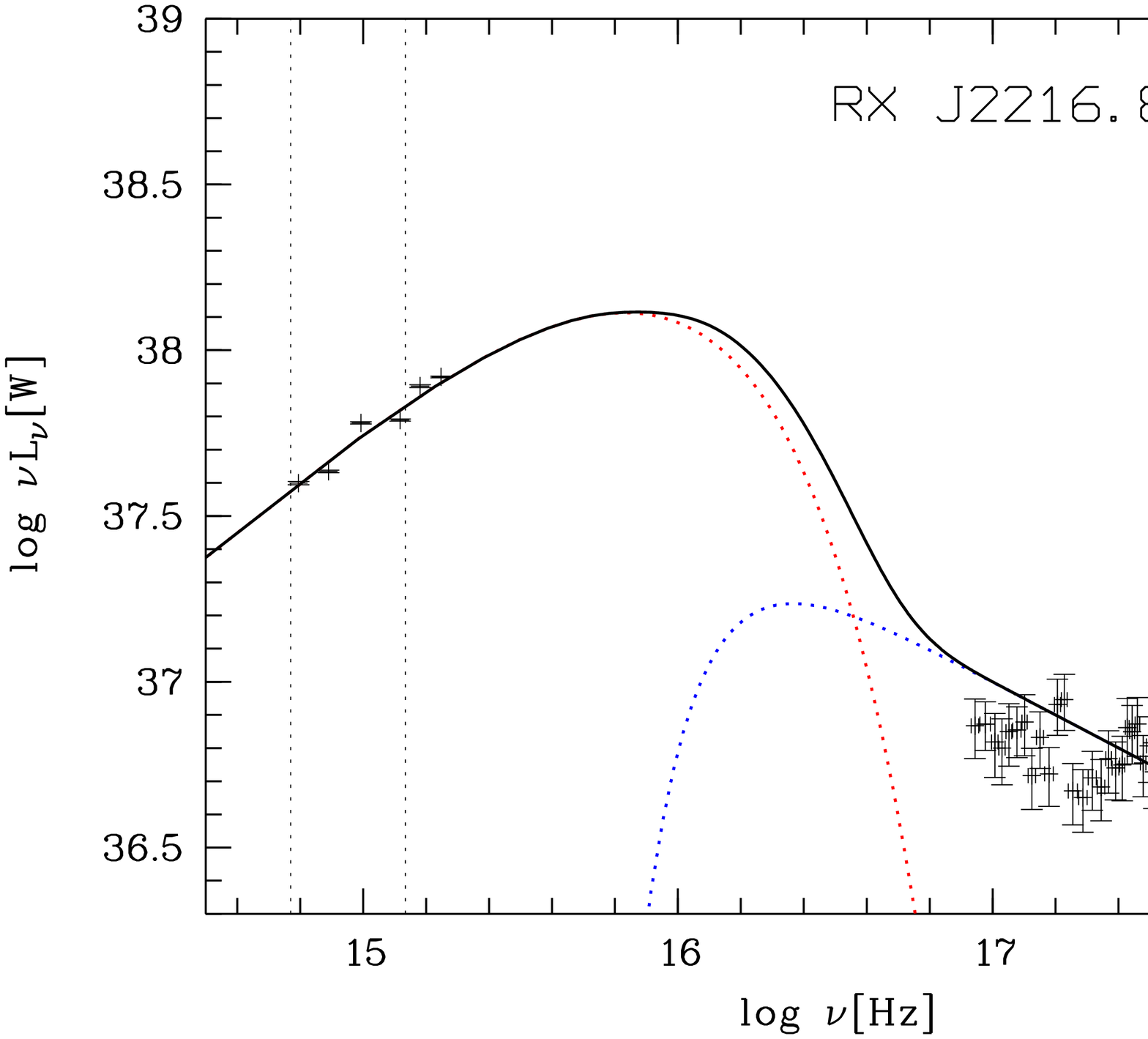}{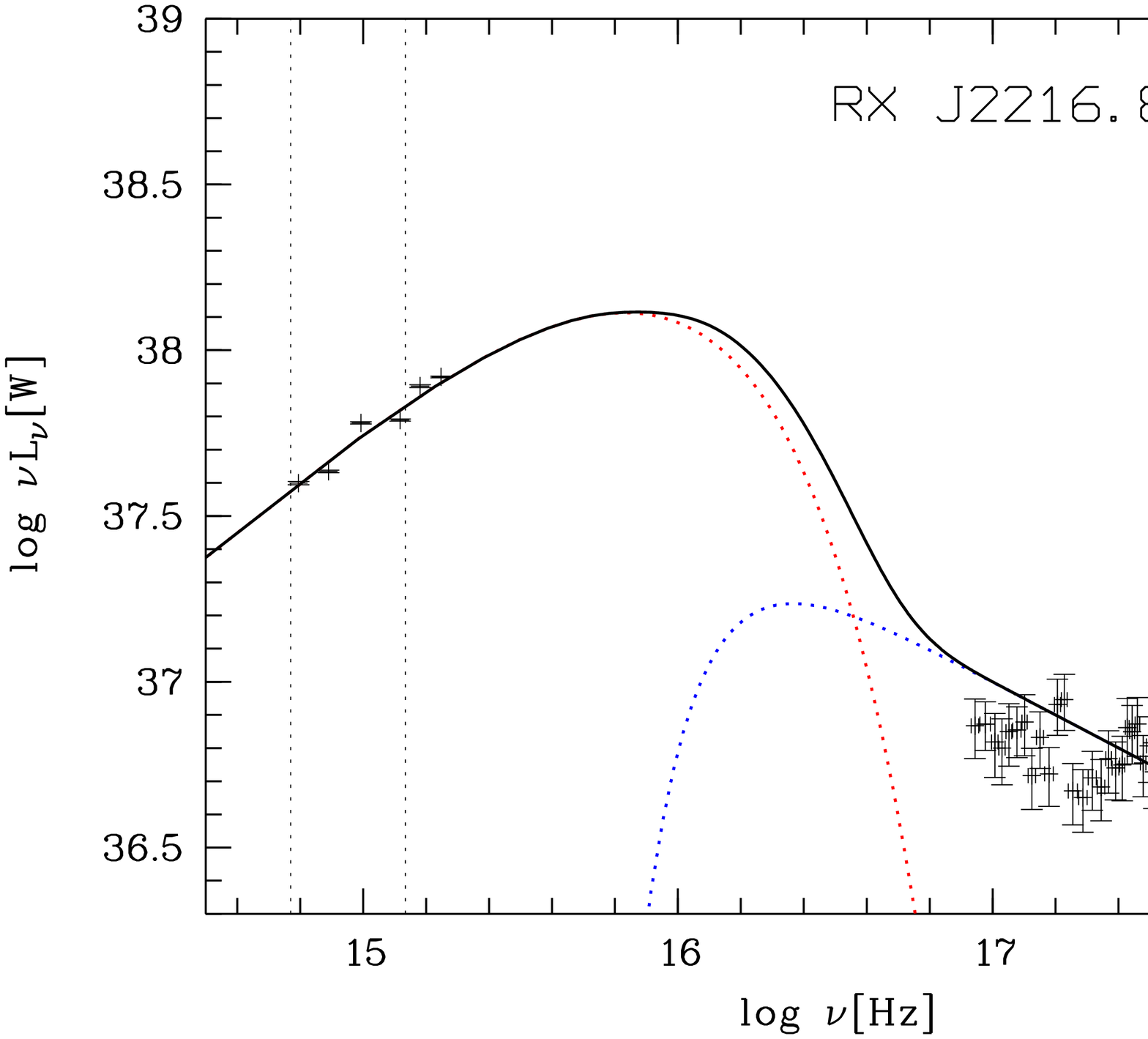}{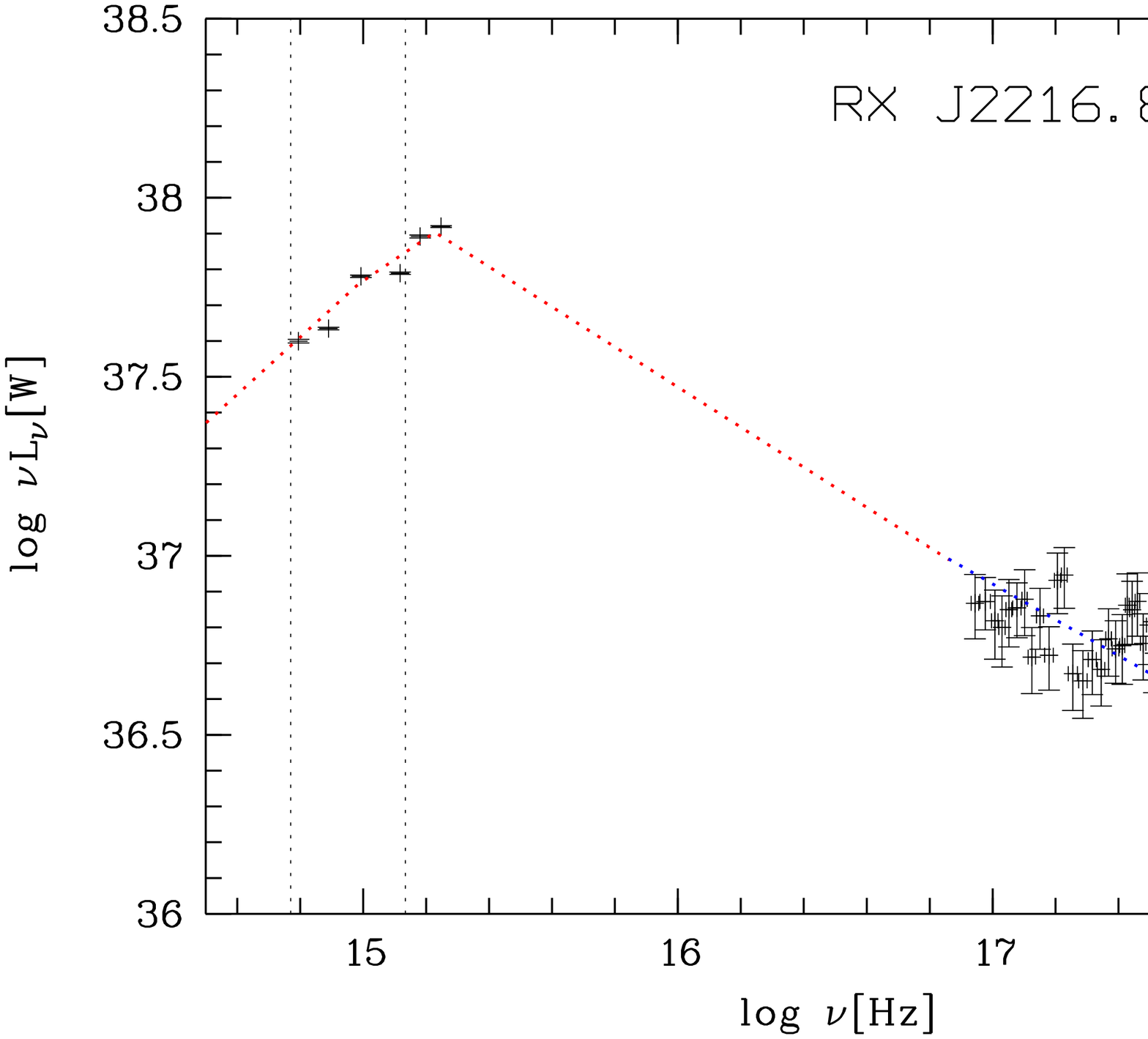}

\plotthree{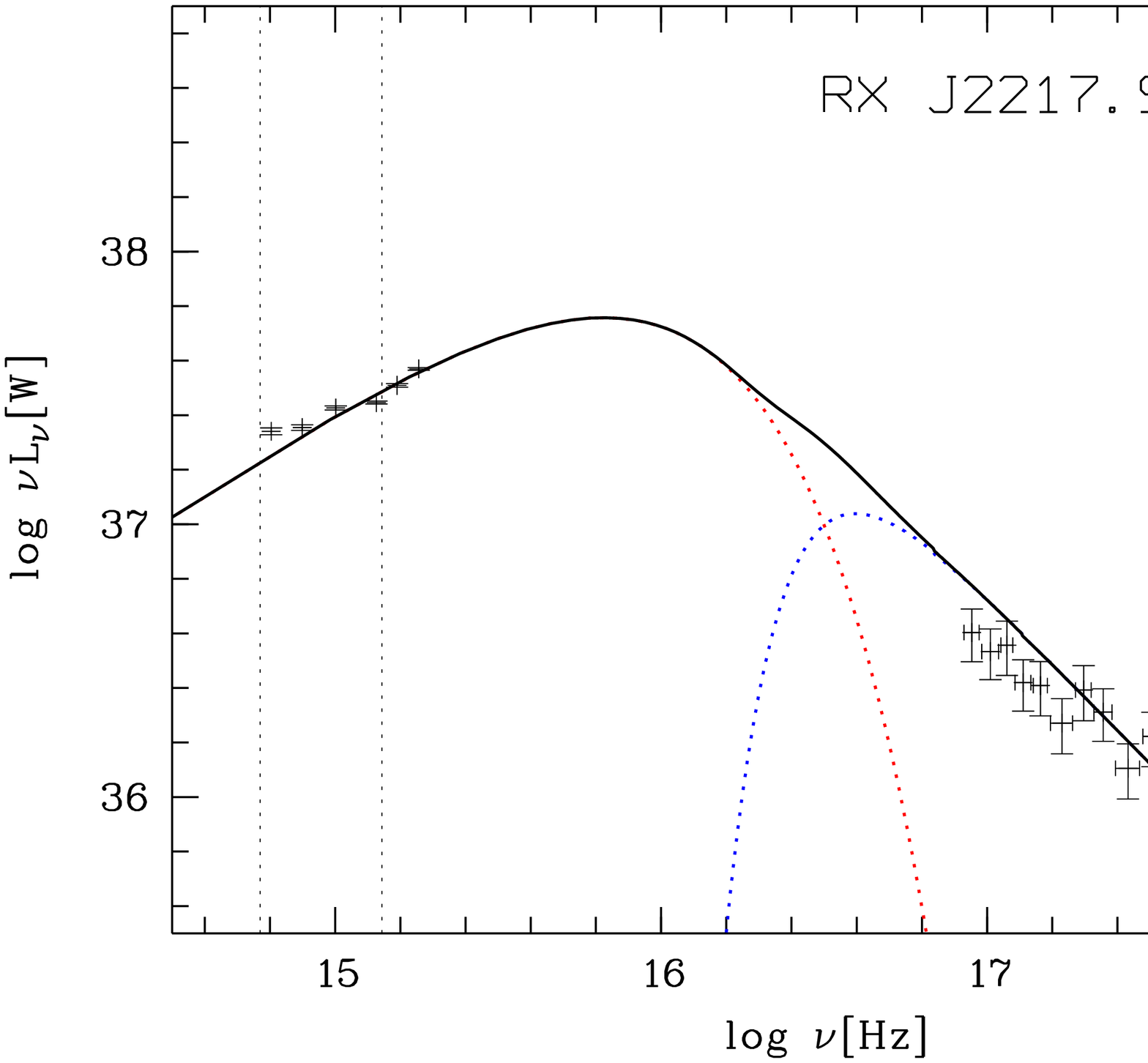}{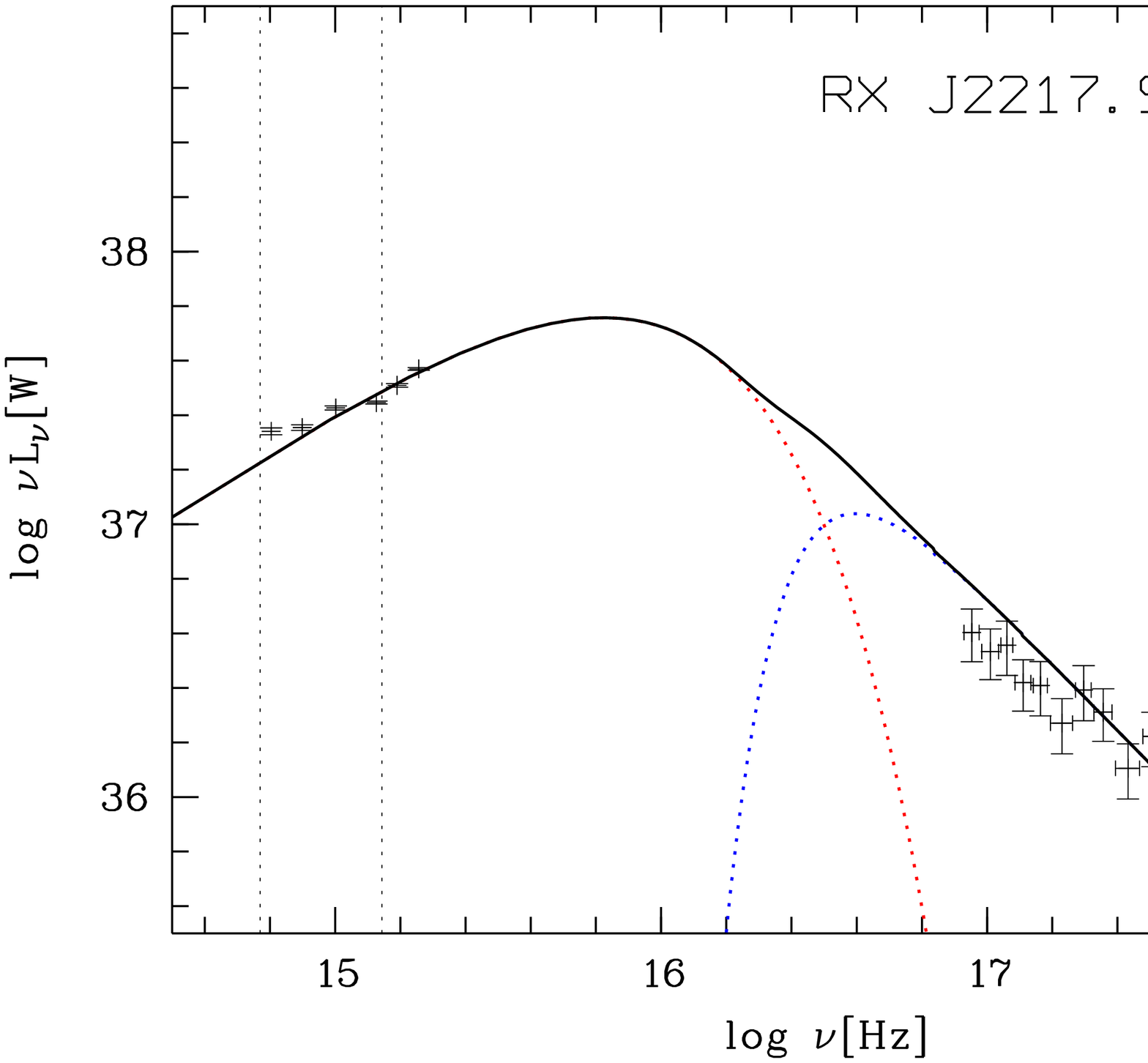}{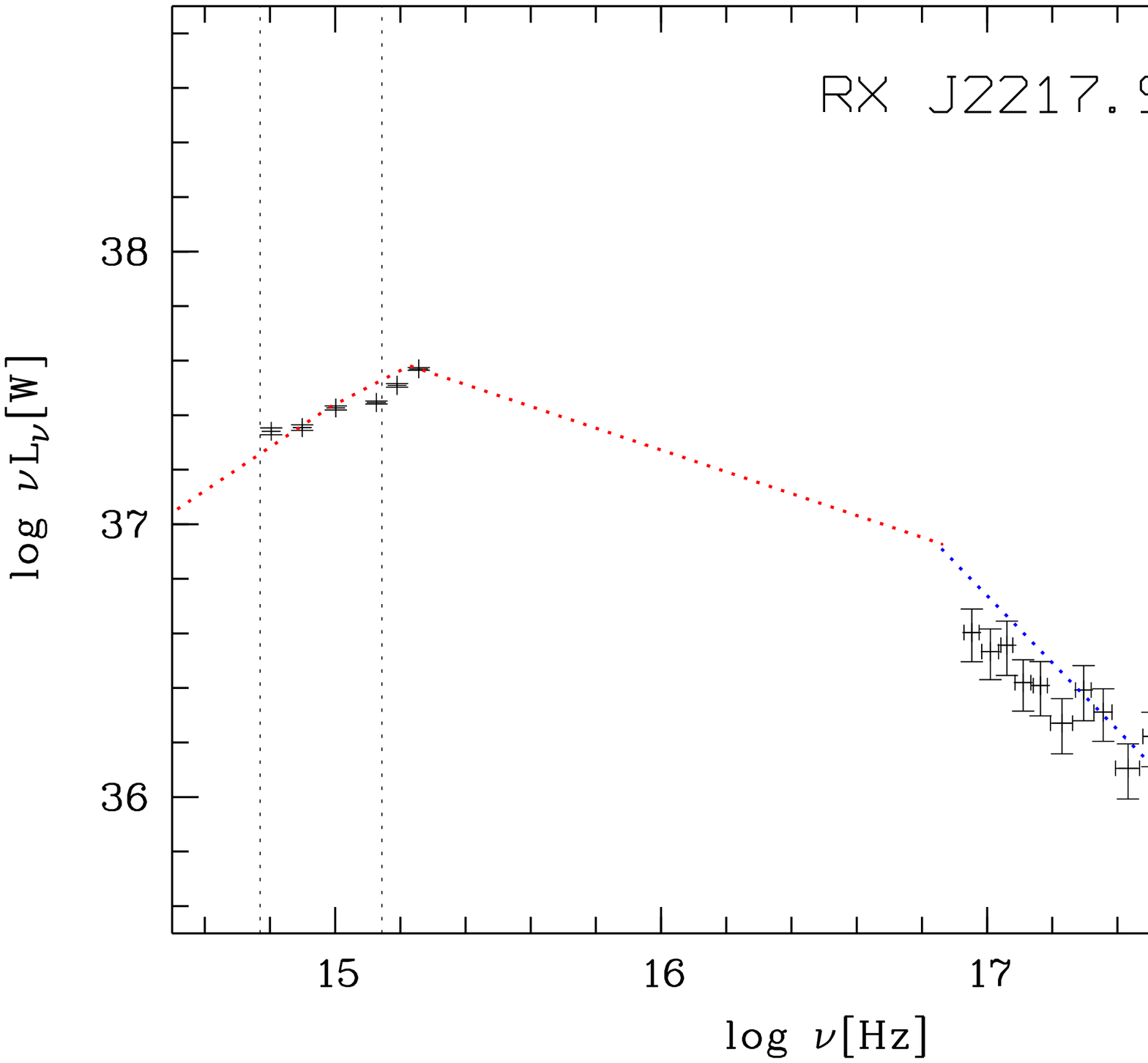}

\plotthree{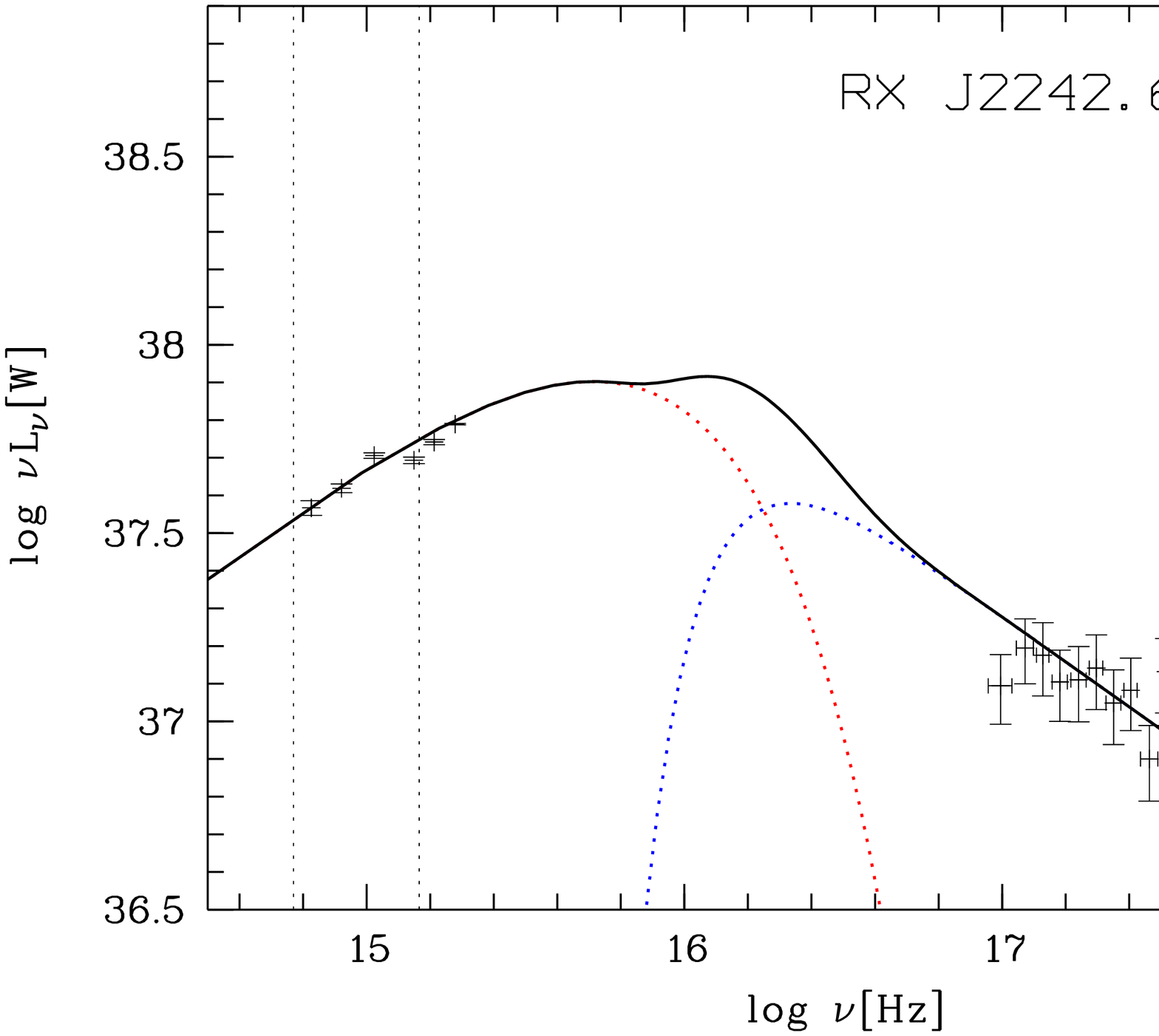}{f25_t.ps}{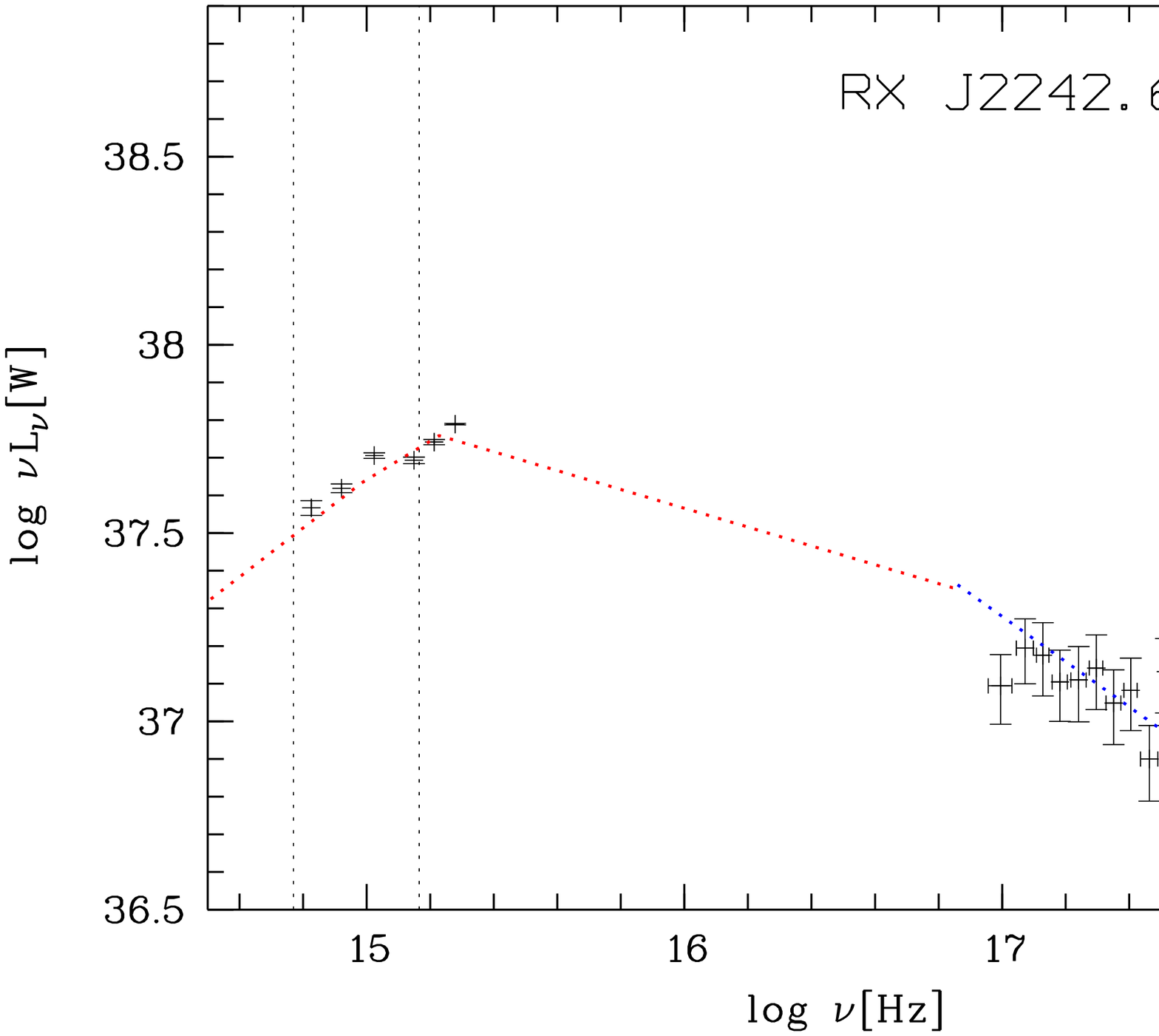}

\plotthree{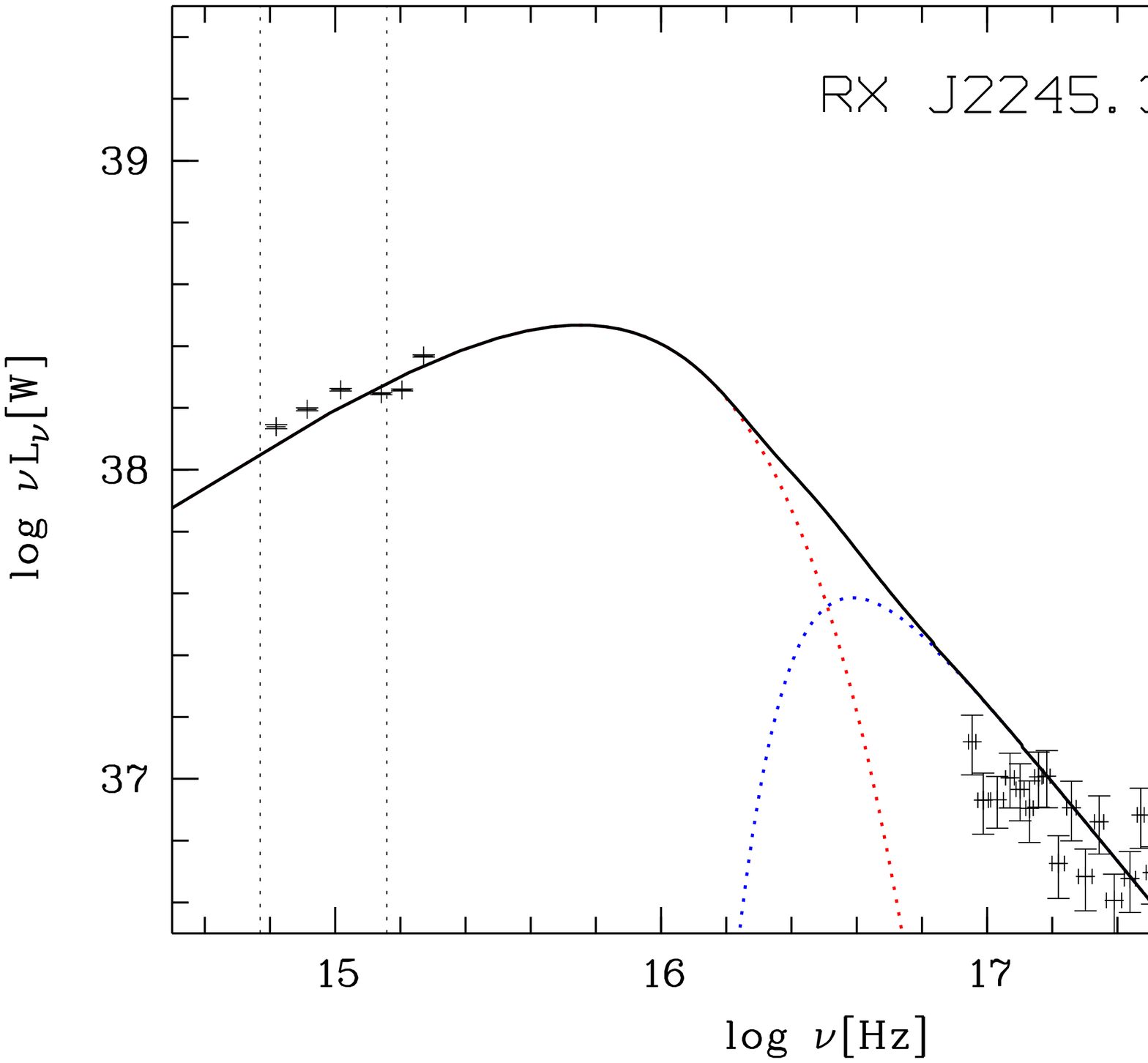}{f25_t.ps}{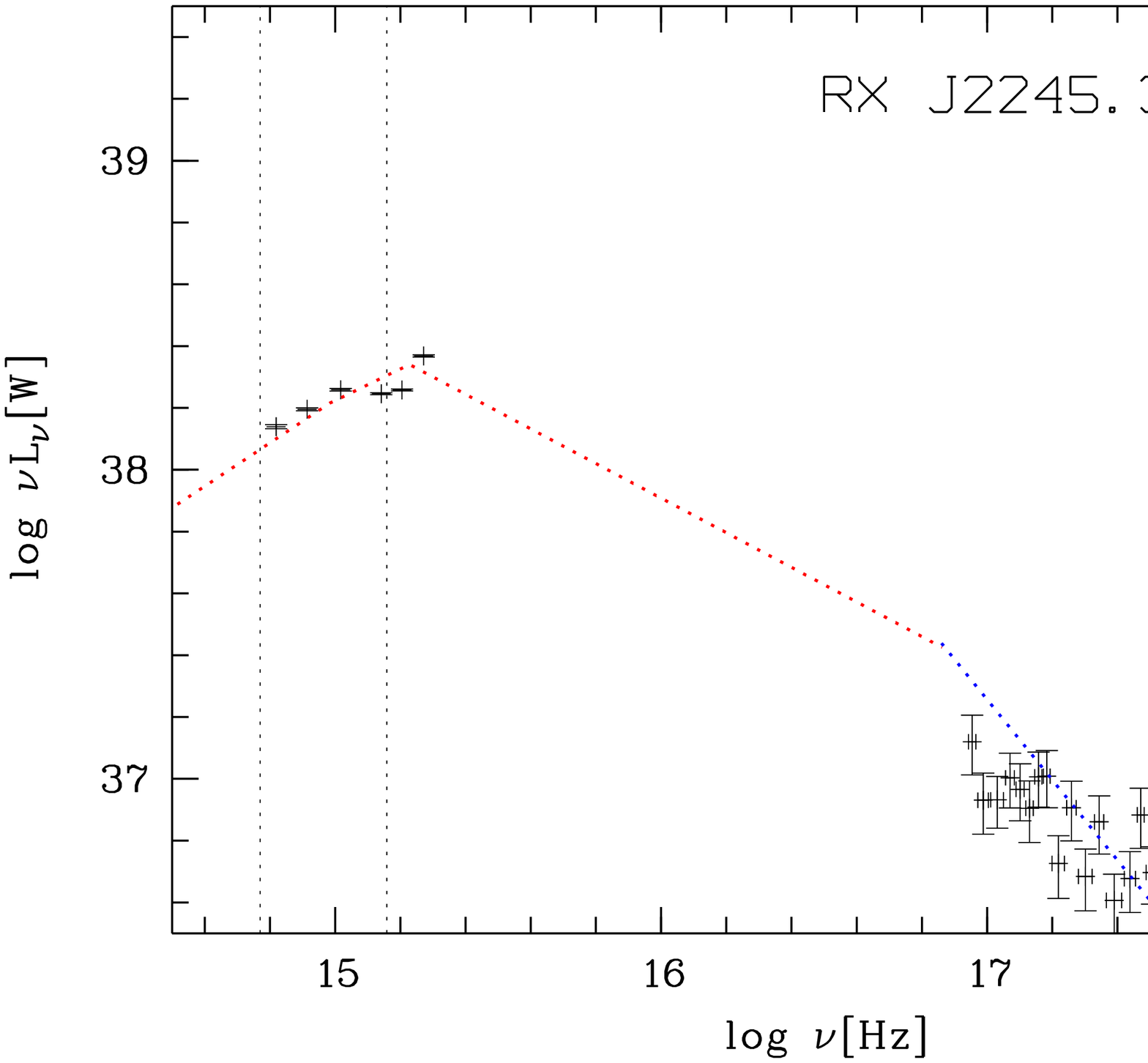}
\end{figure*}

\begin{figure*}
\epsscale{0.60}
\plotthree{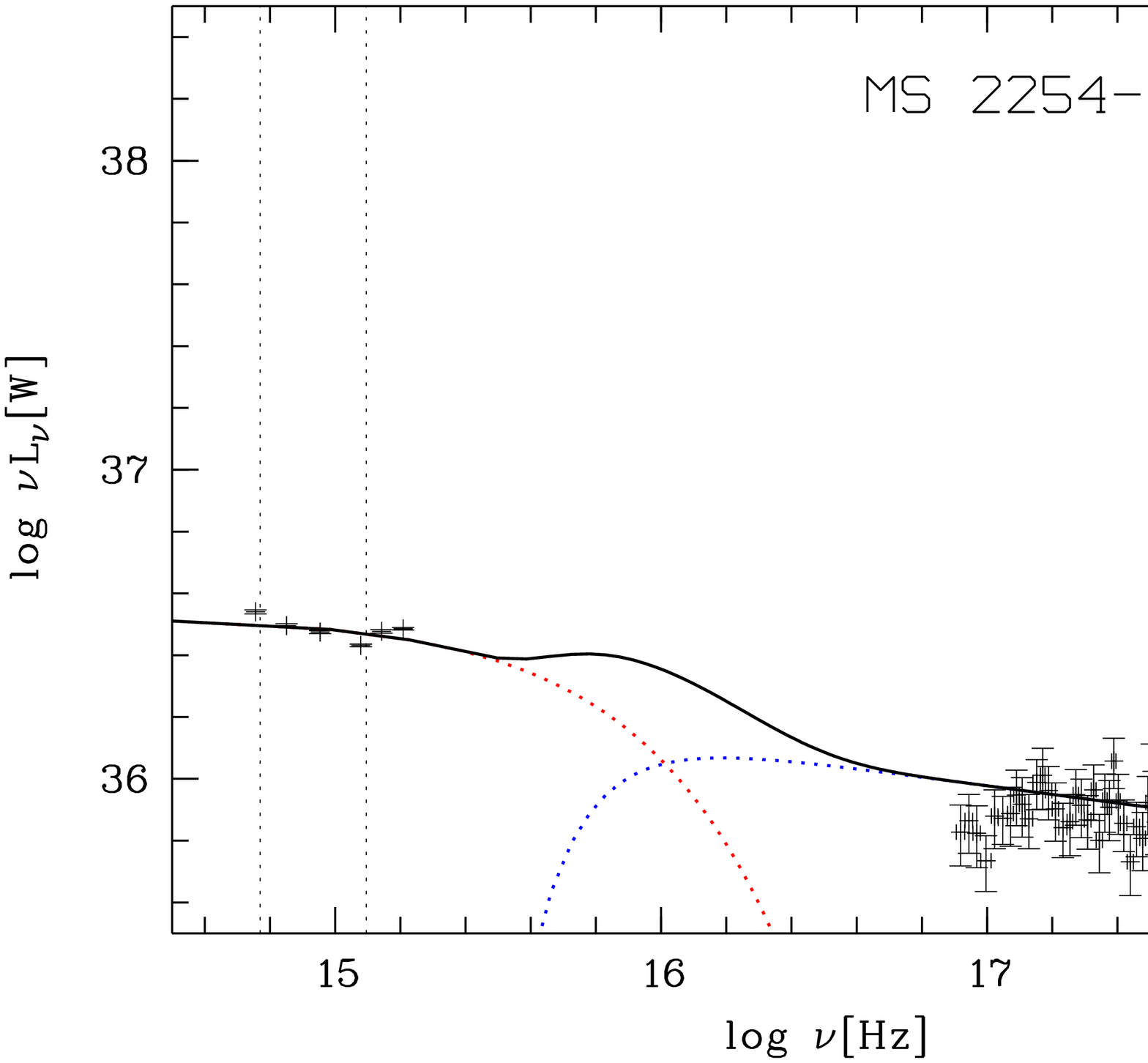}{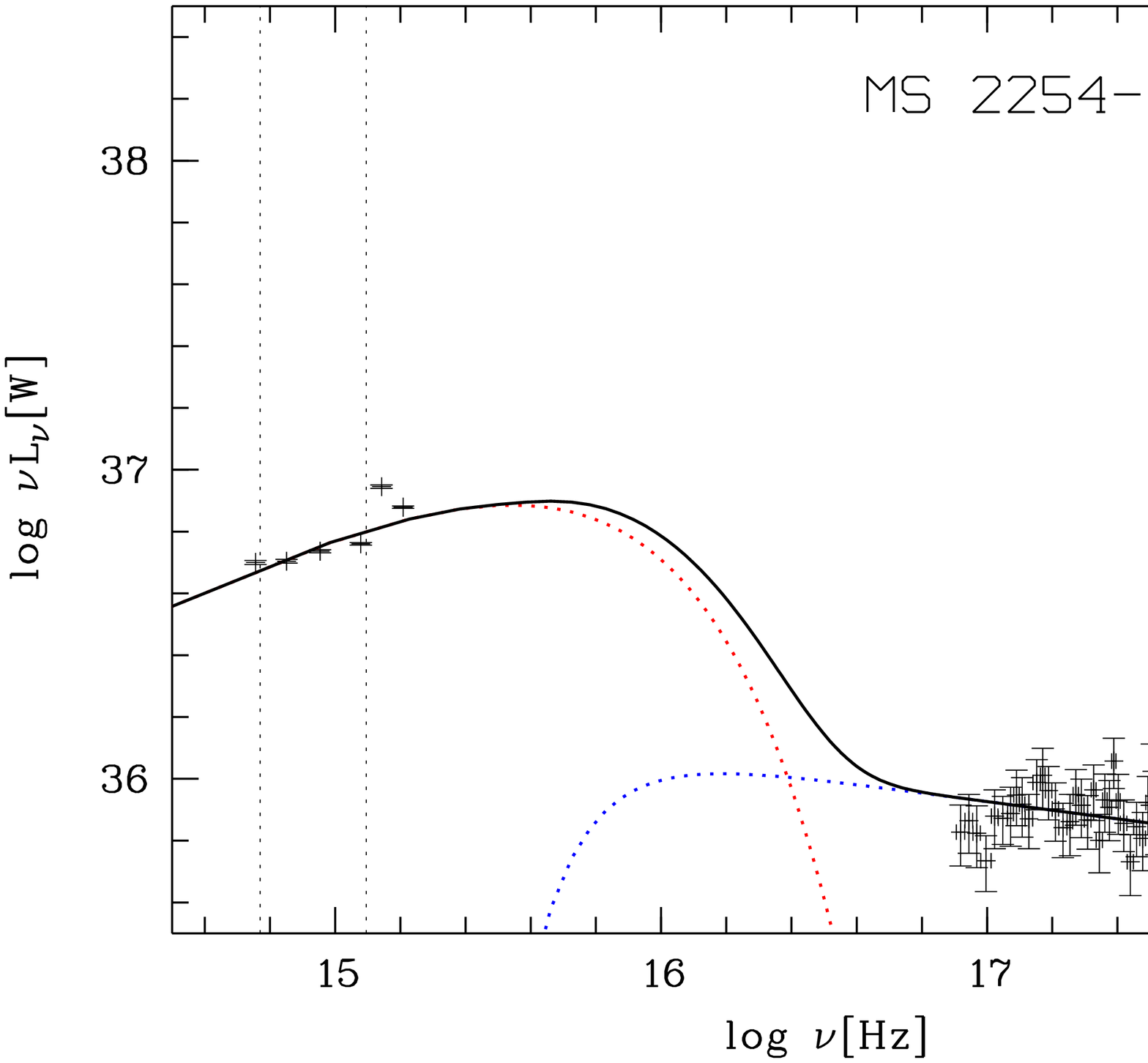}{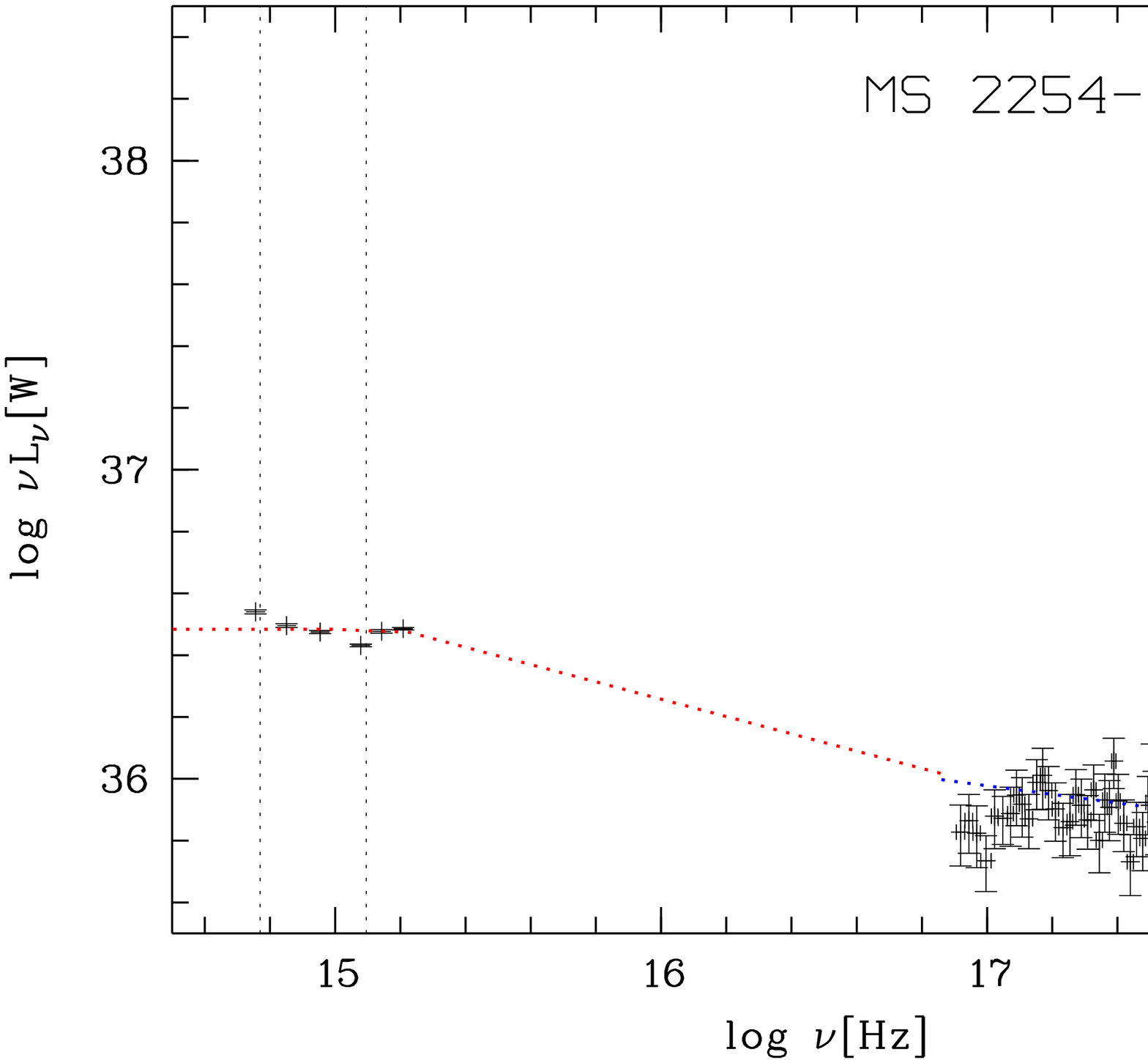}

\plotthree{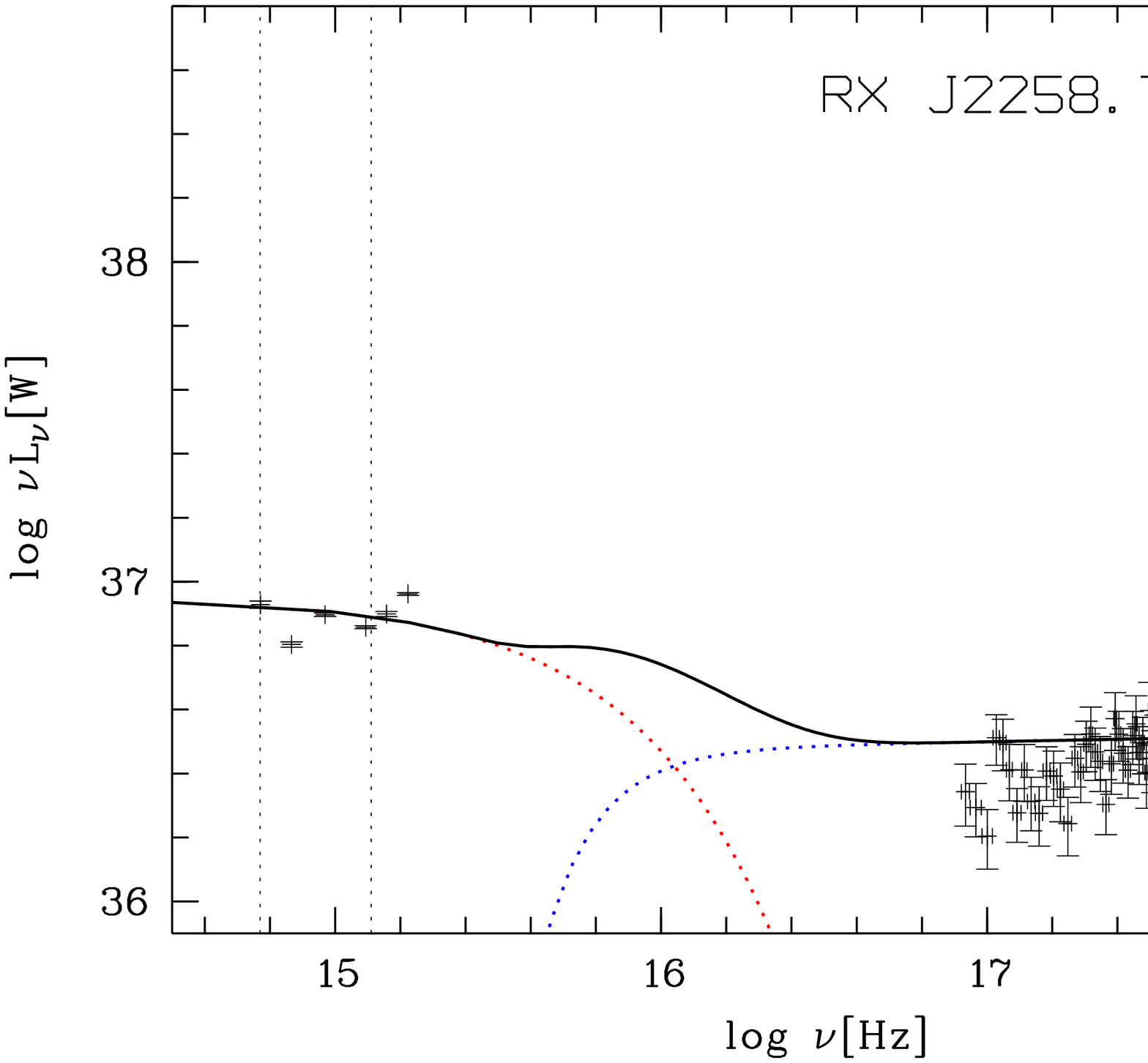}{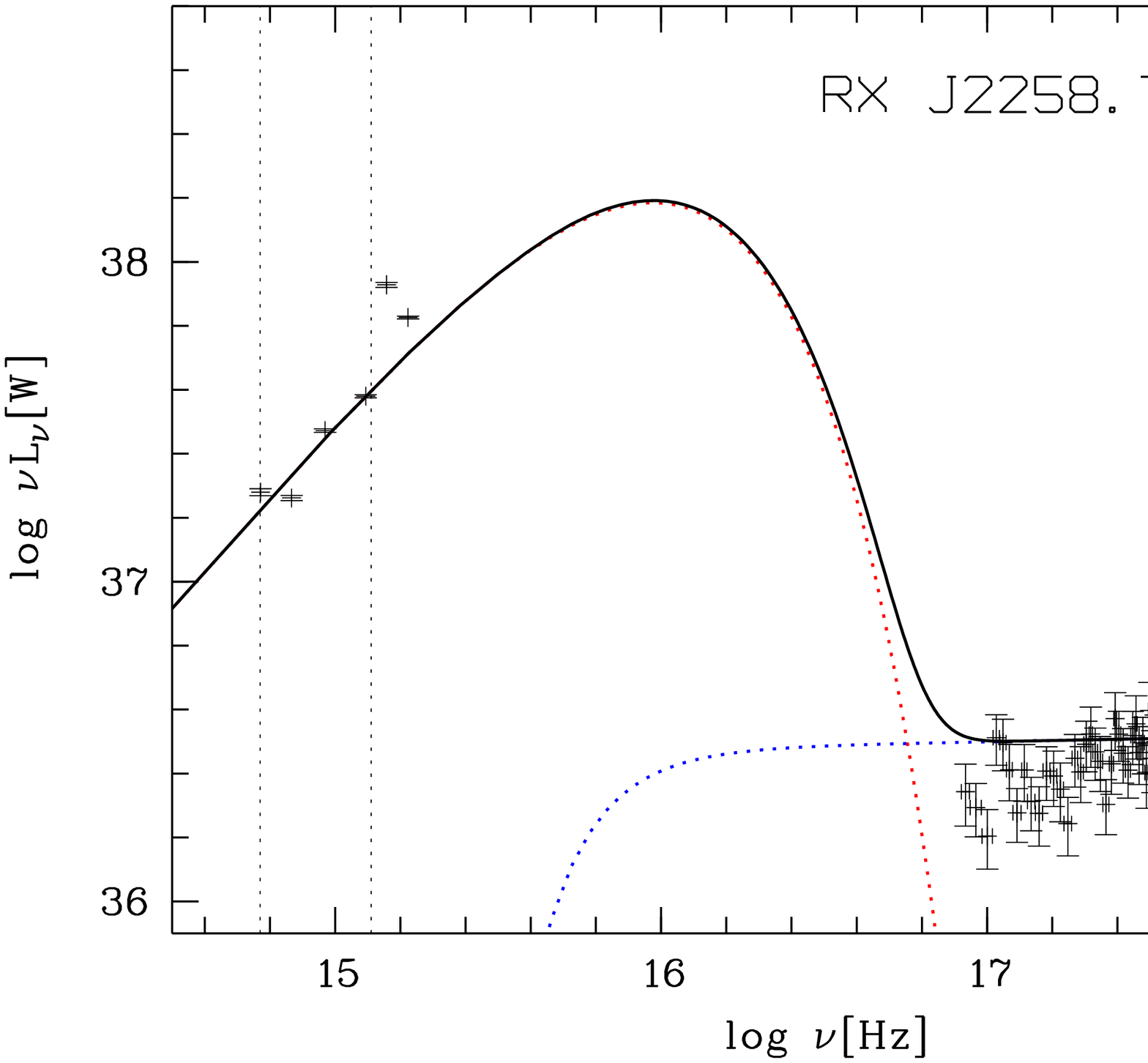}{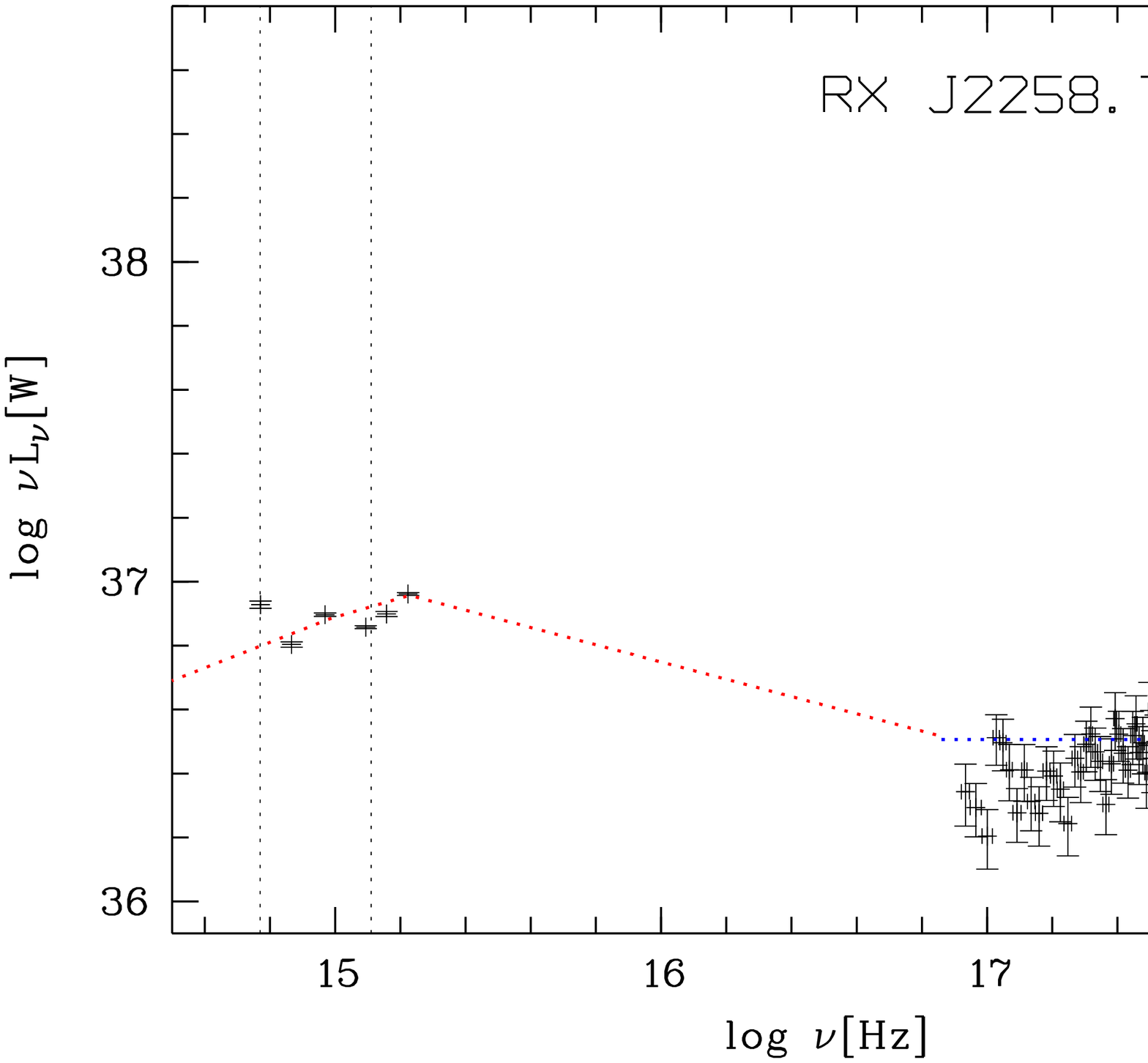}

\plotthree{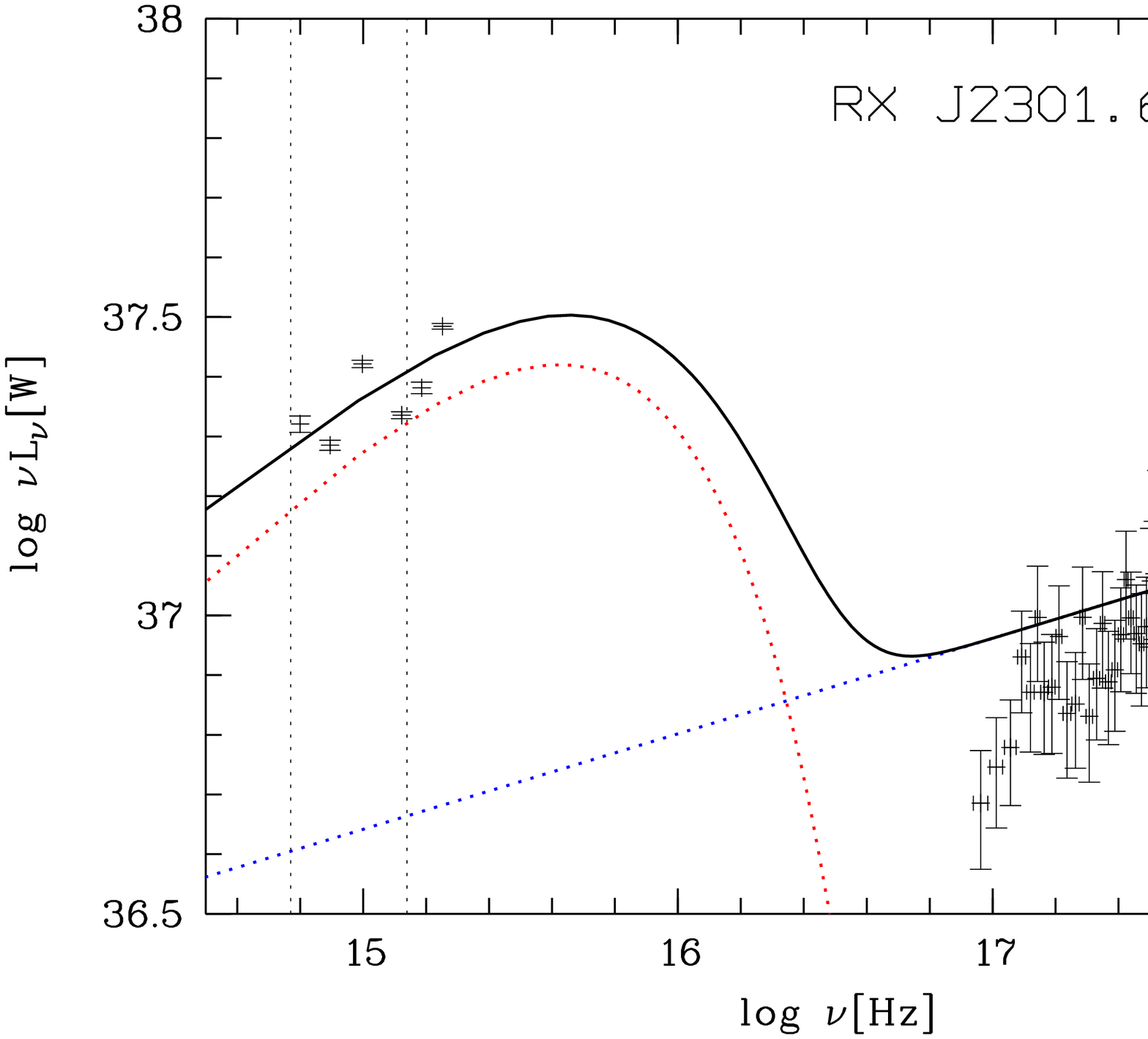}{f25_t.ps}{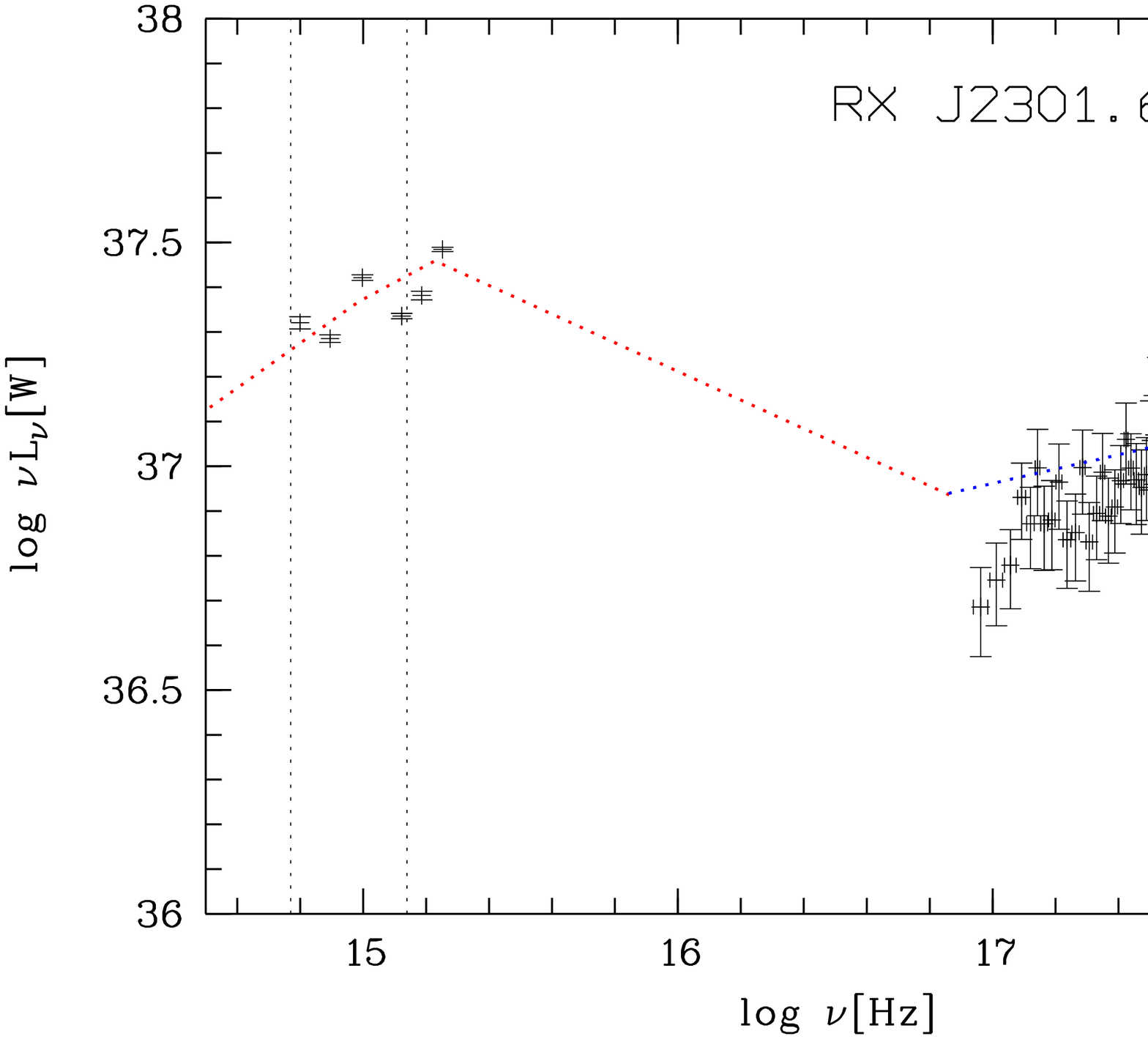}

\plotthree{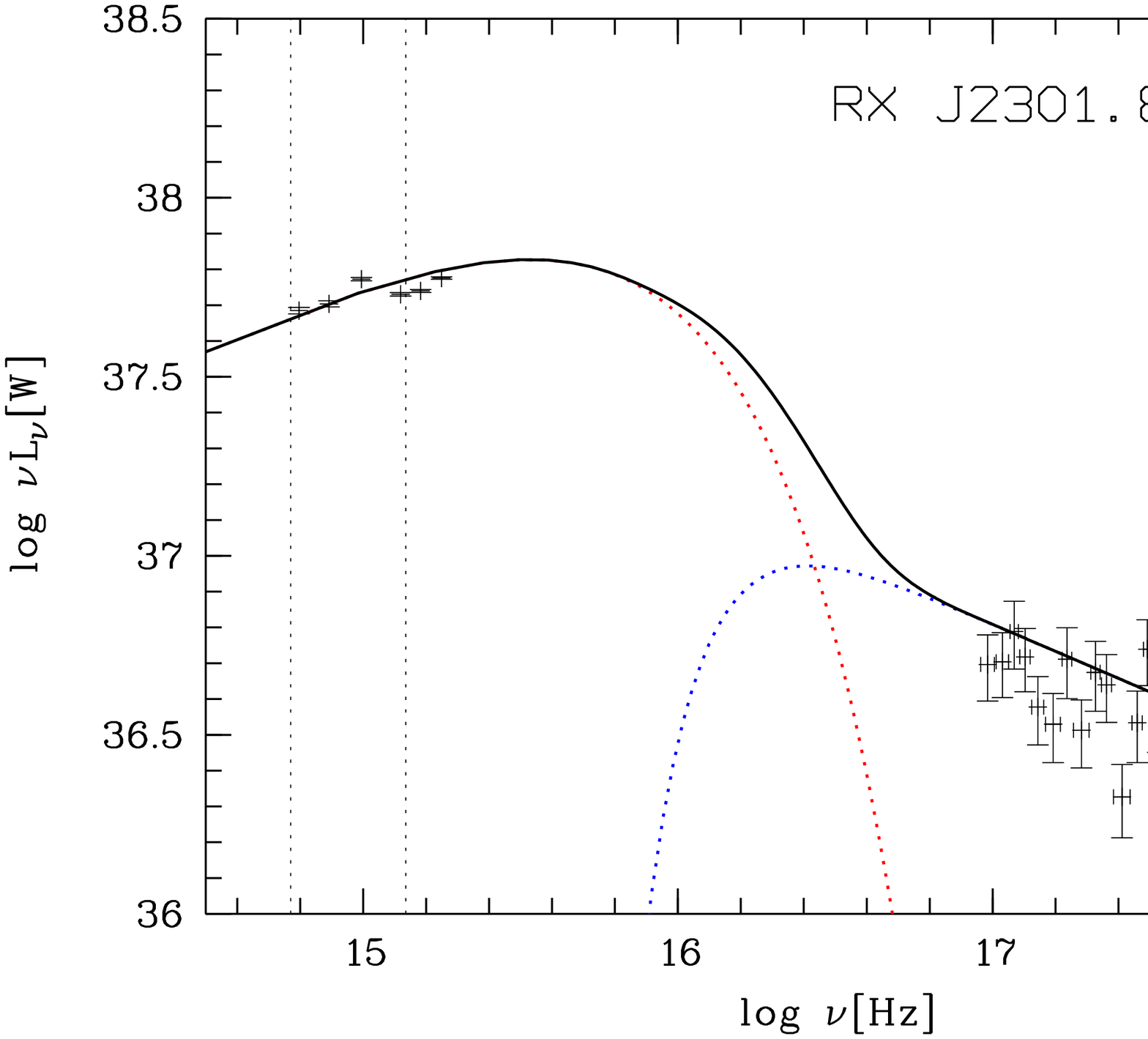}{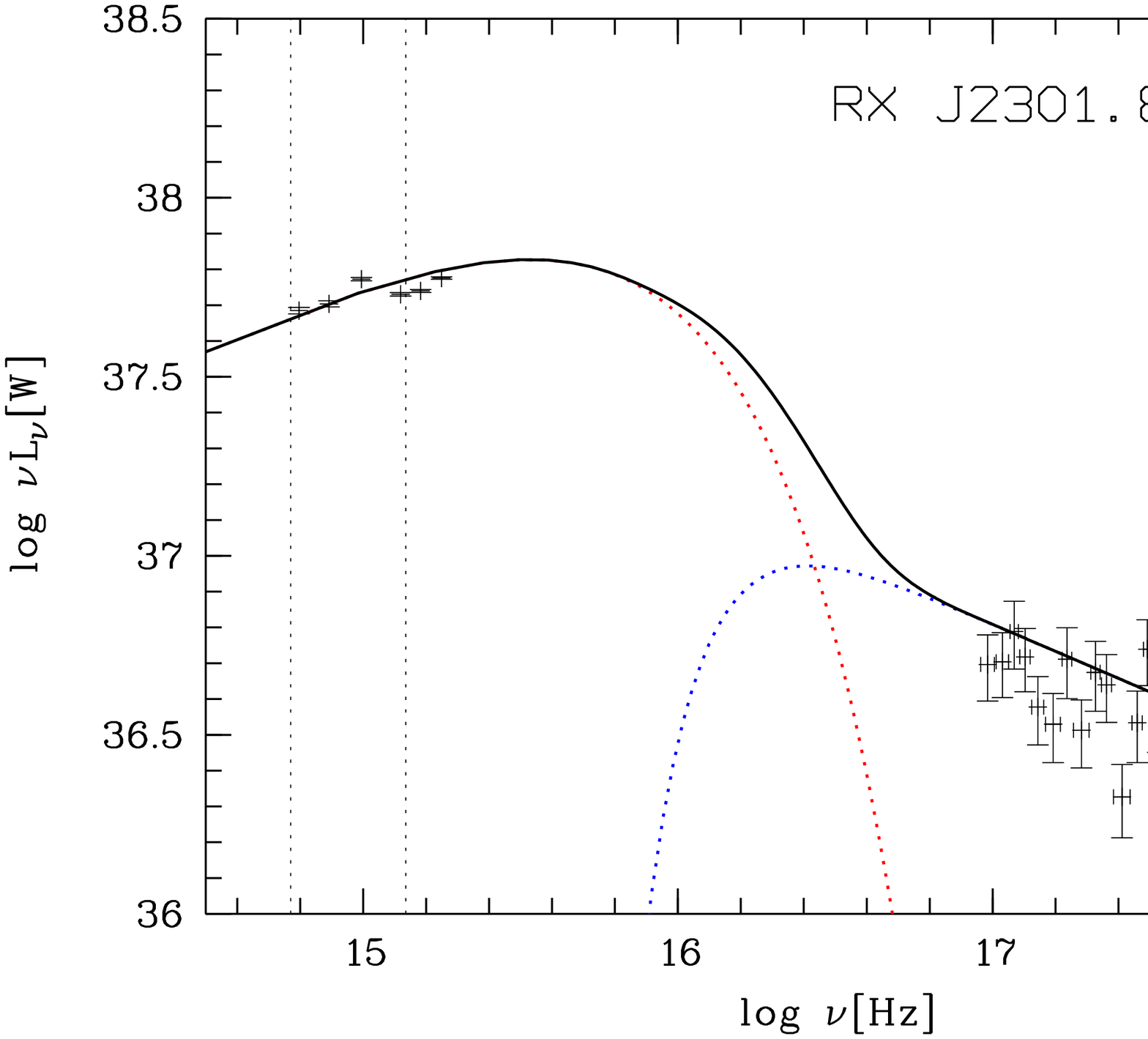}{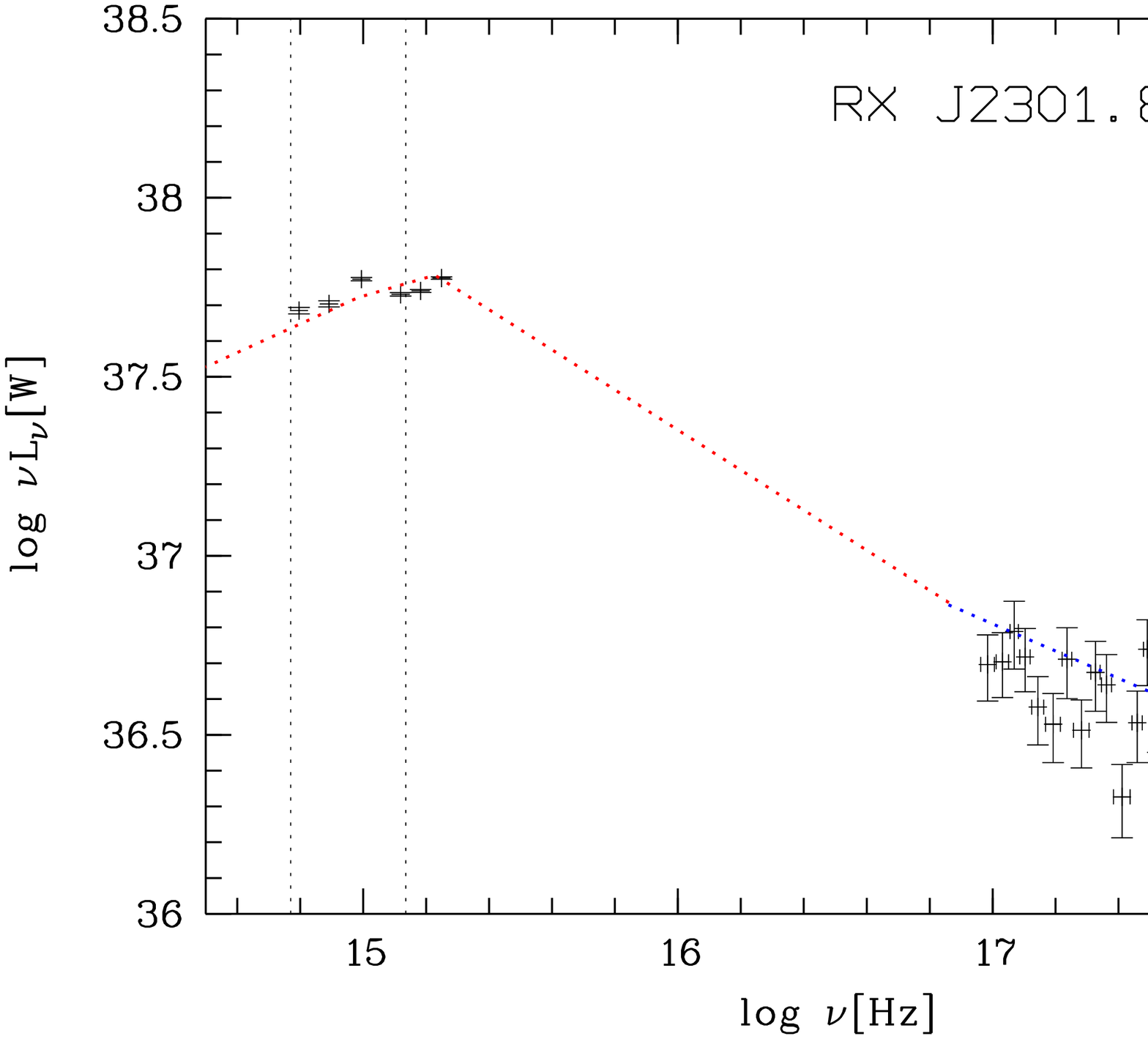}

\plotthree{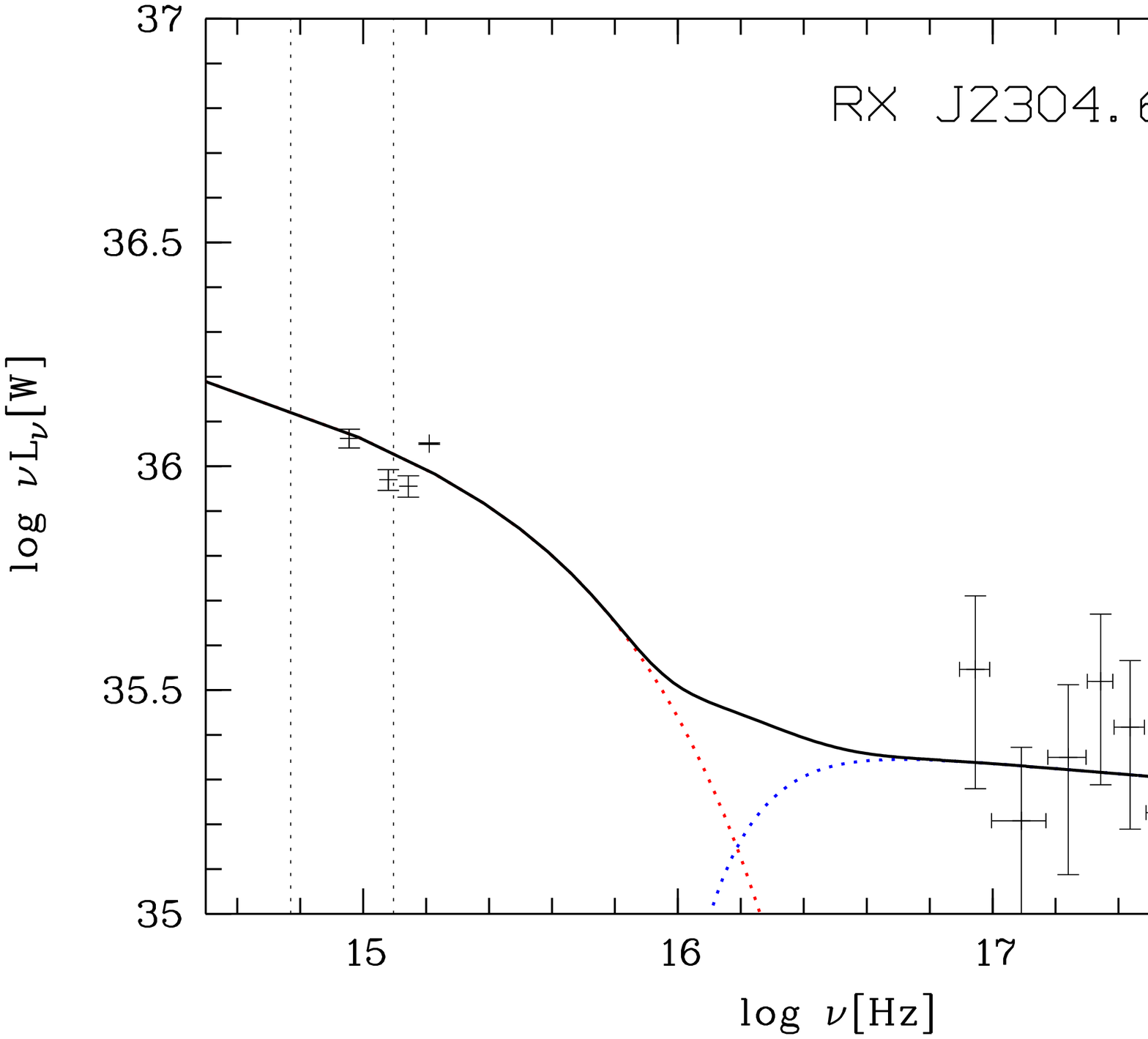}{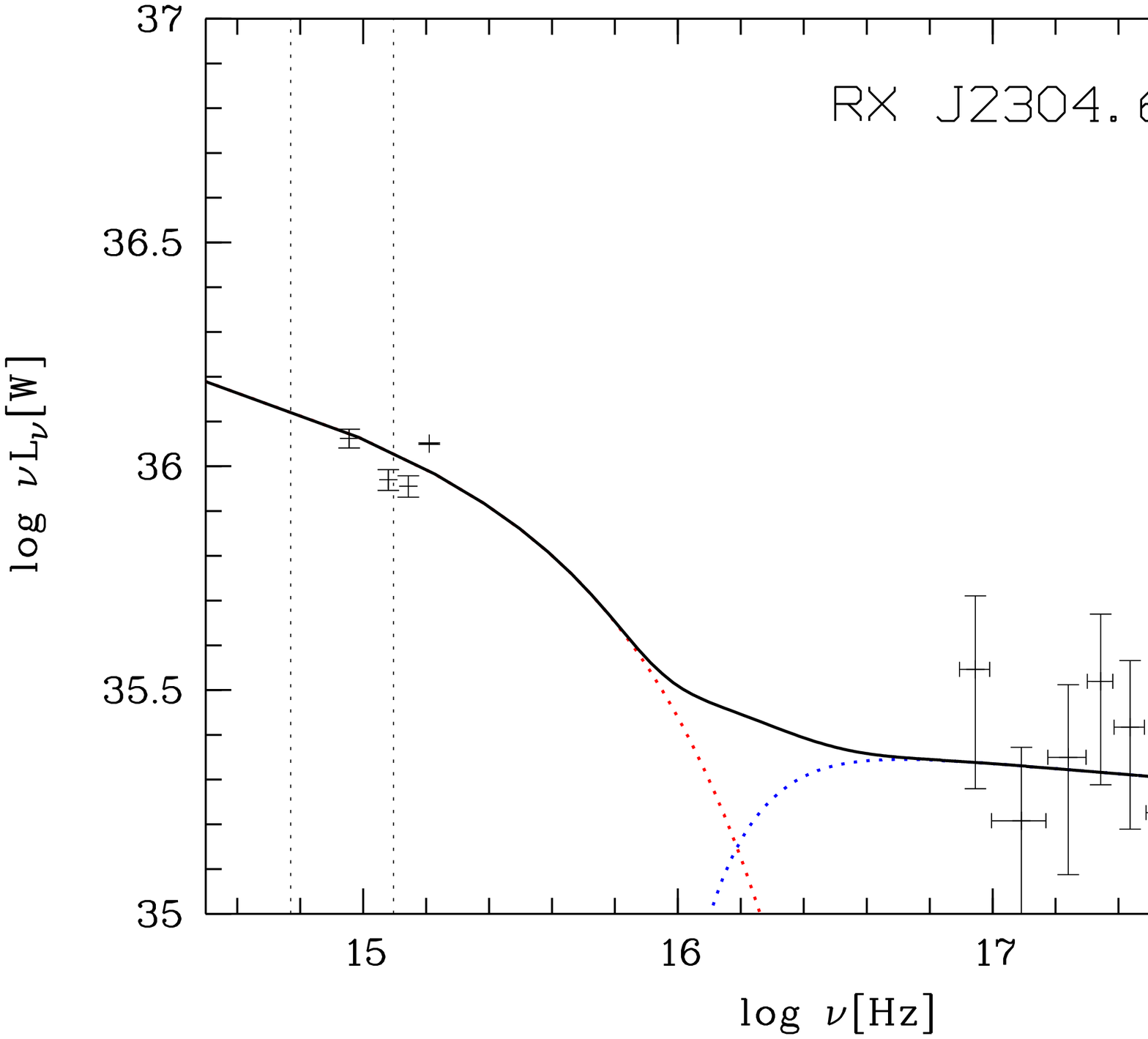}{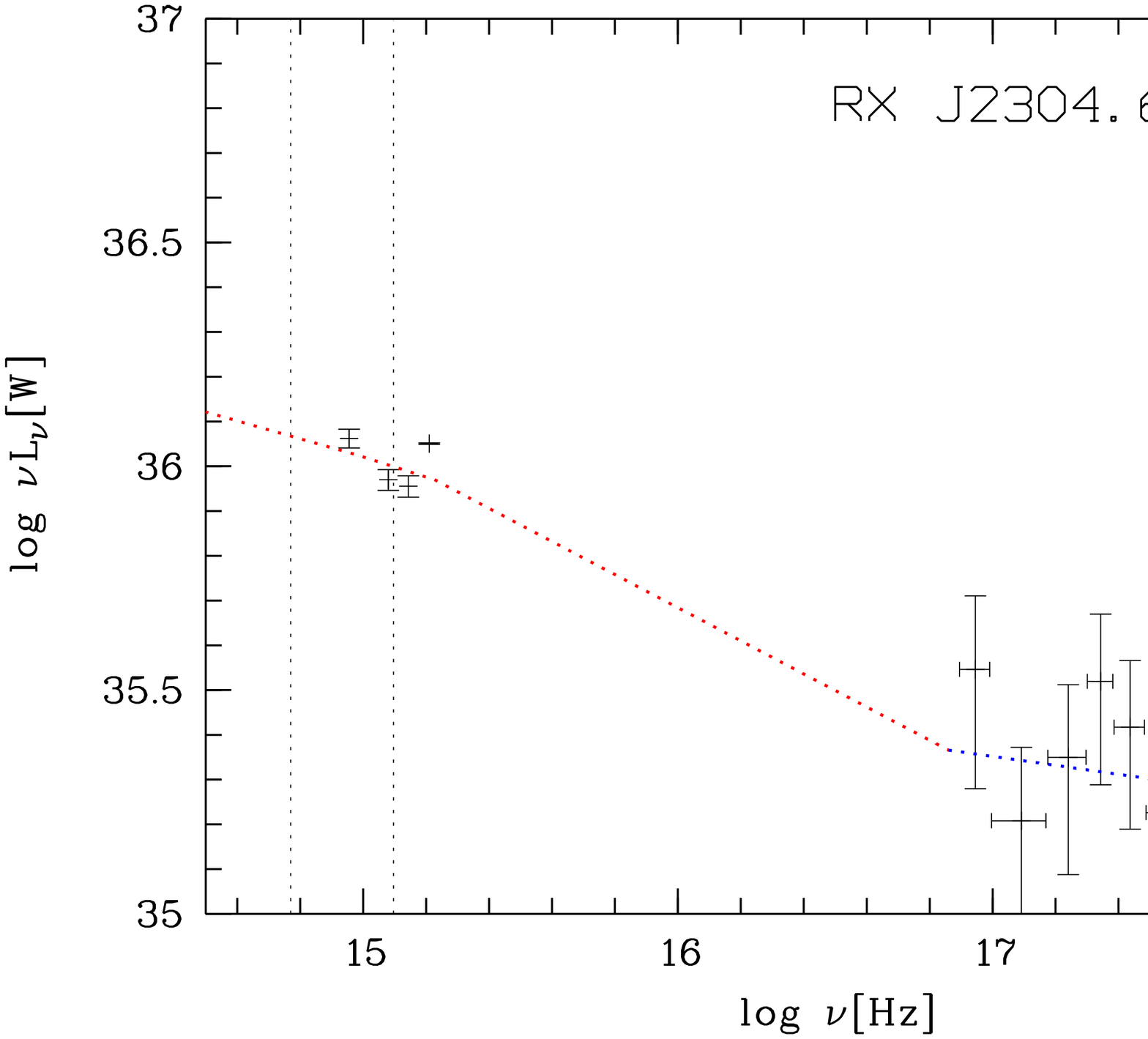}
\end{figure*}

\begin{figure*}
\epsscale{0.60}
\plotthree{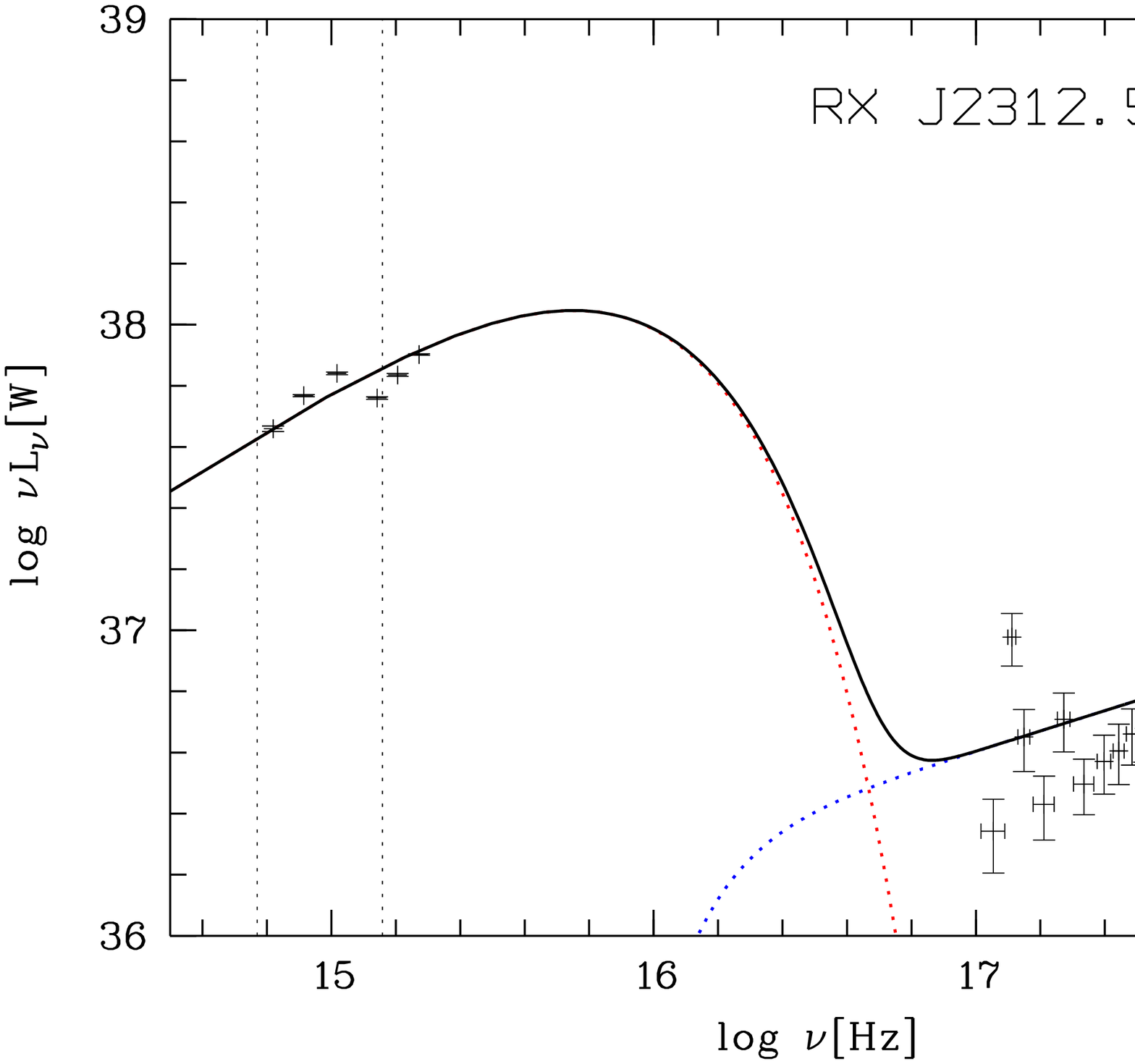}{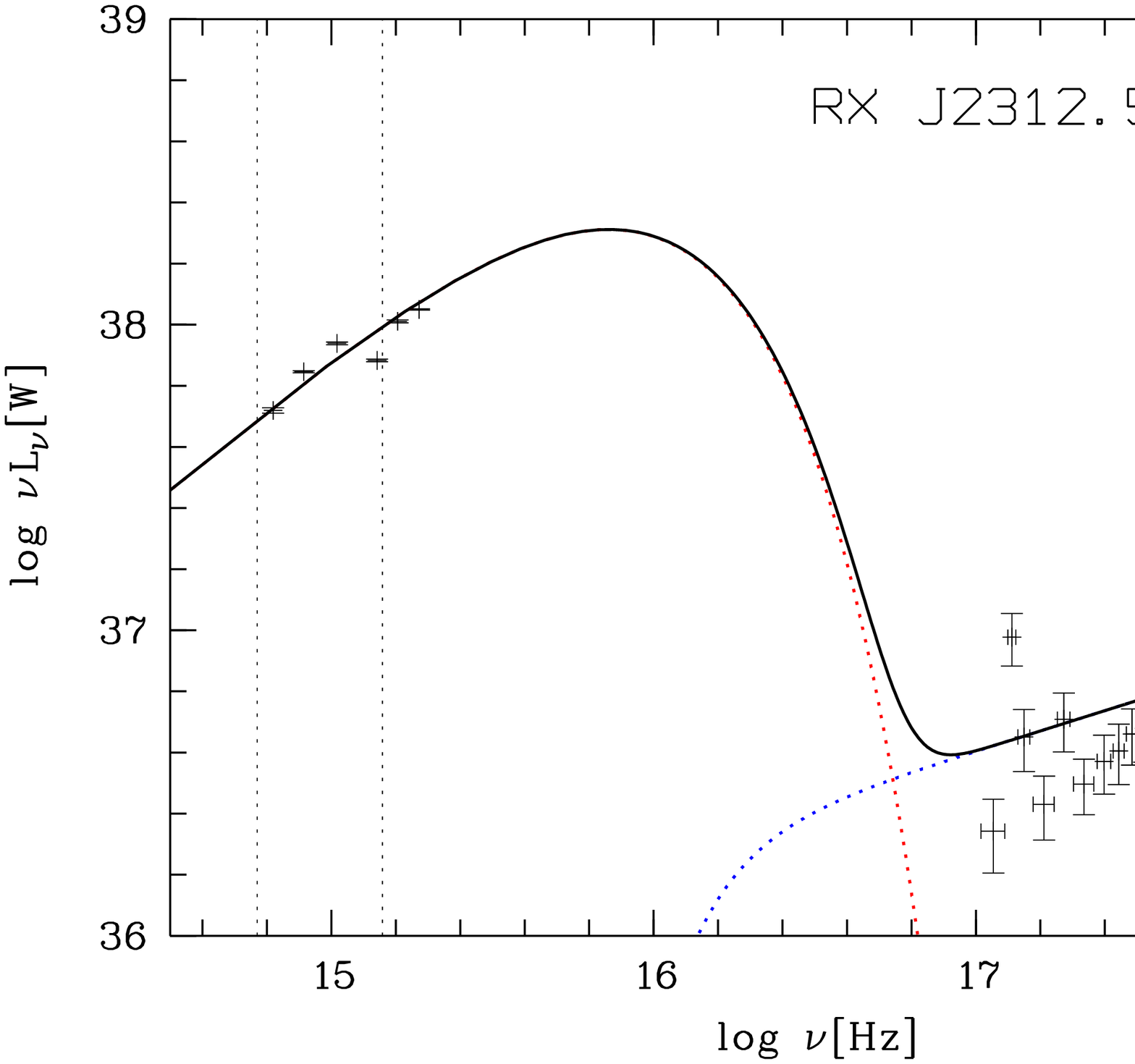}{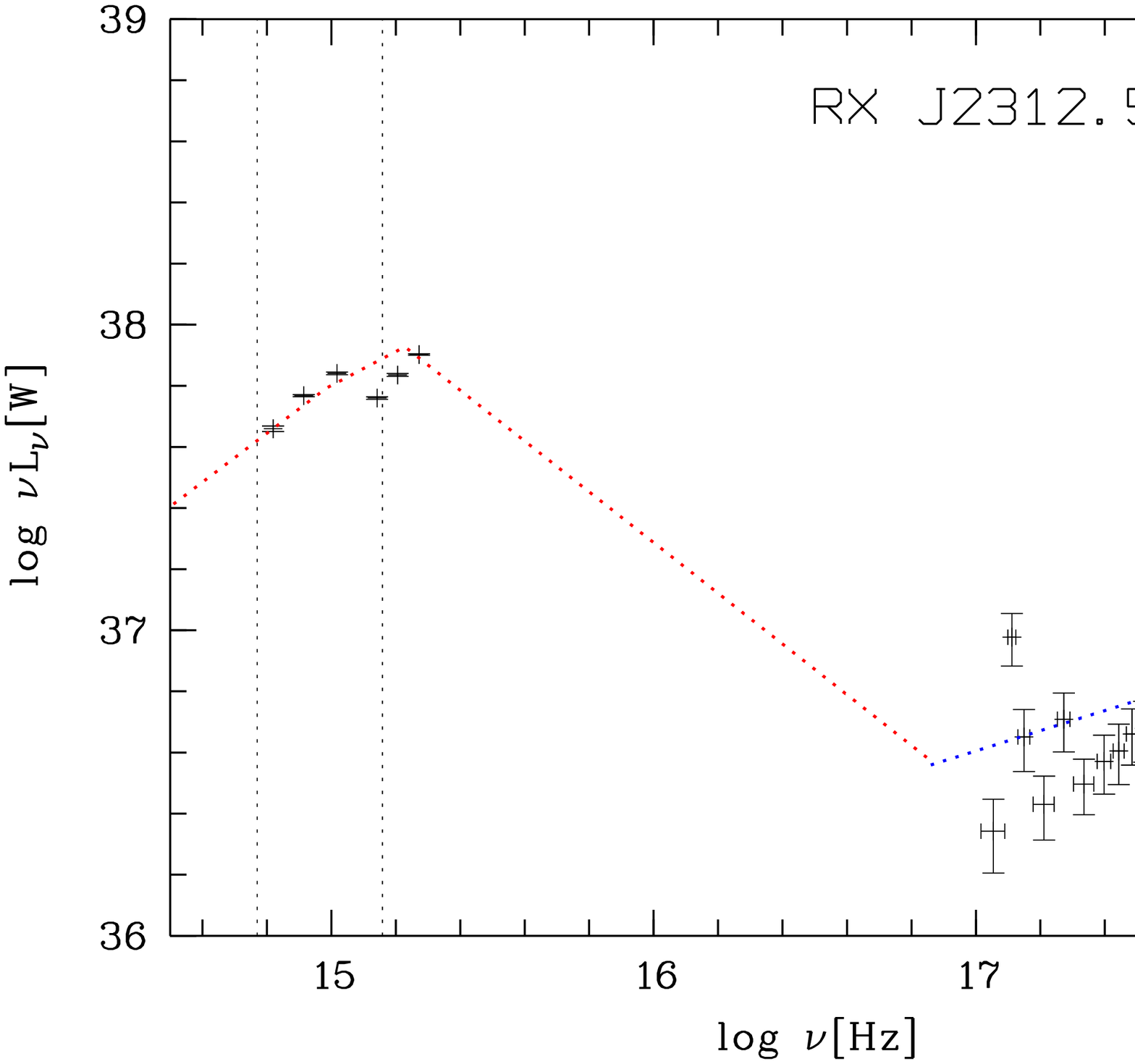}

\plotthree{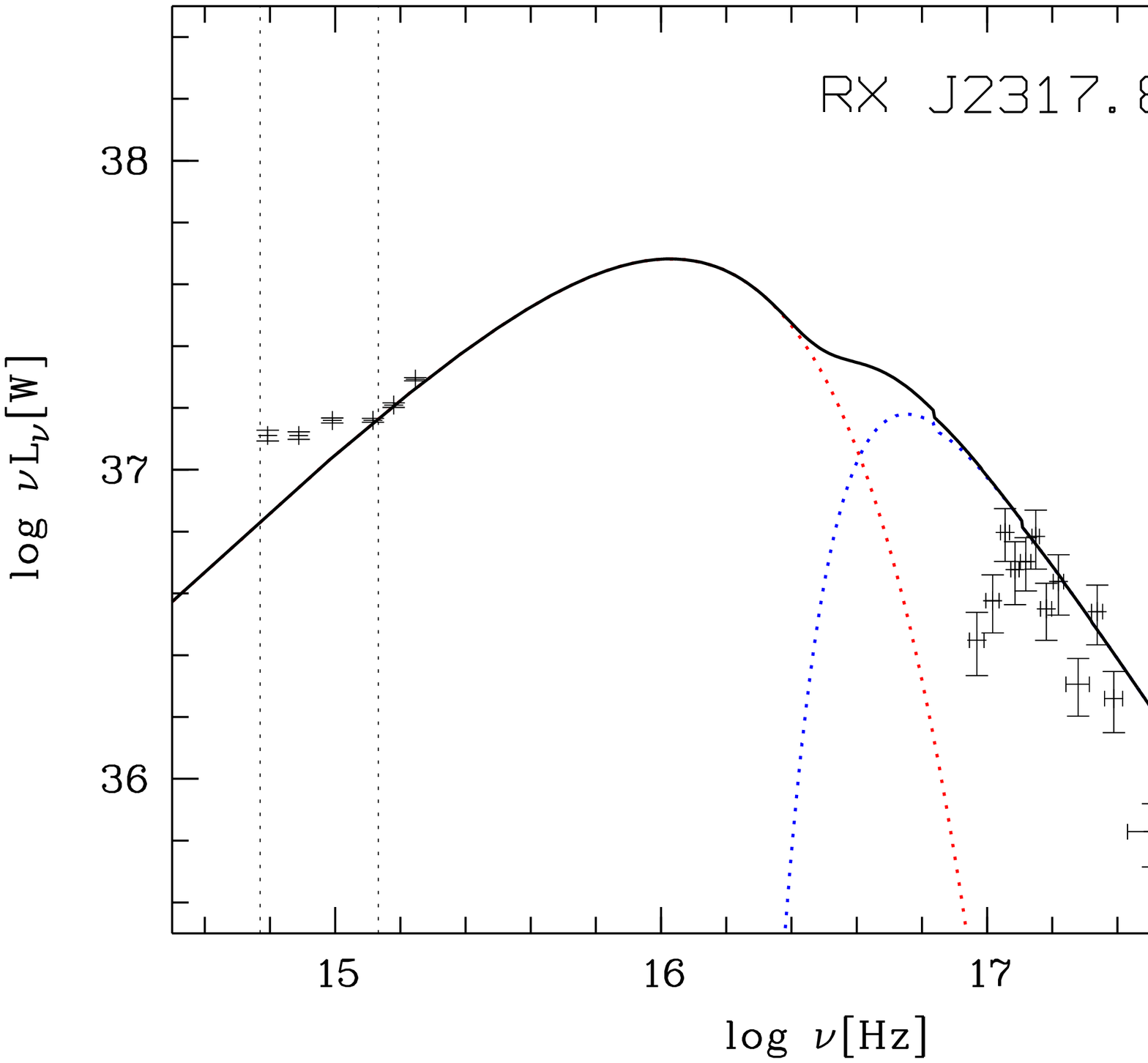}{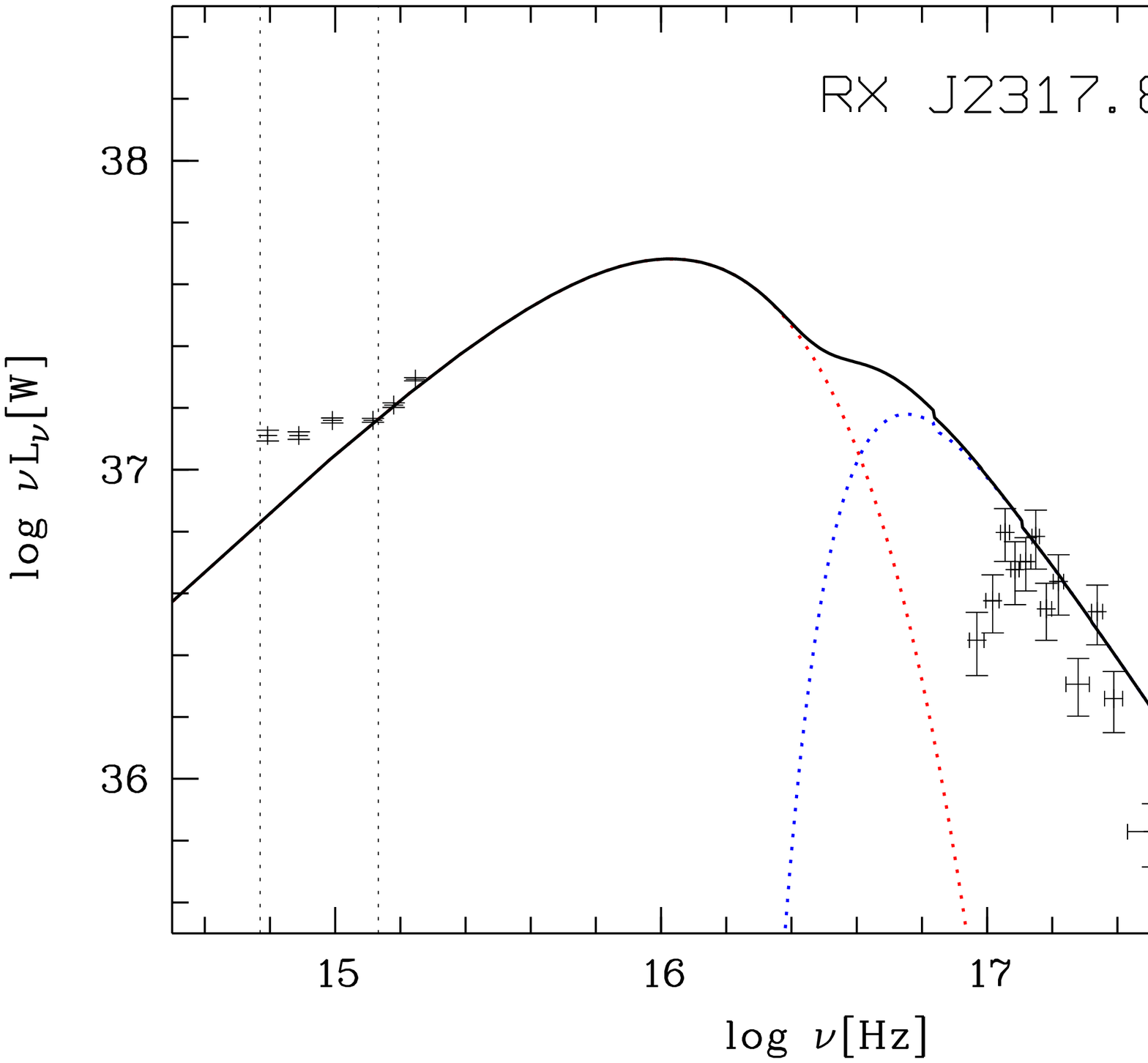}{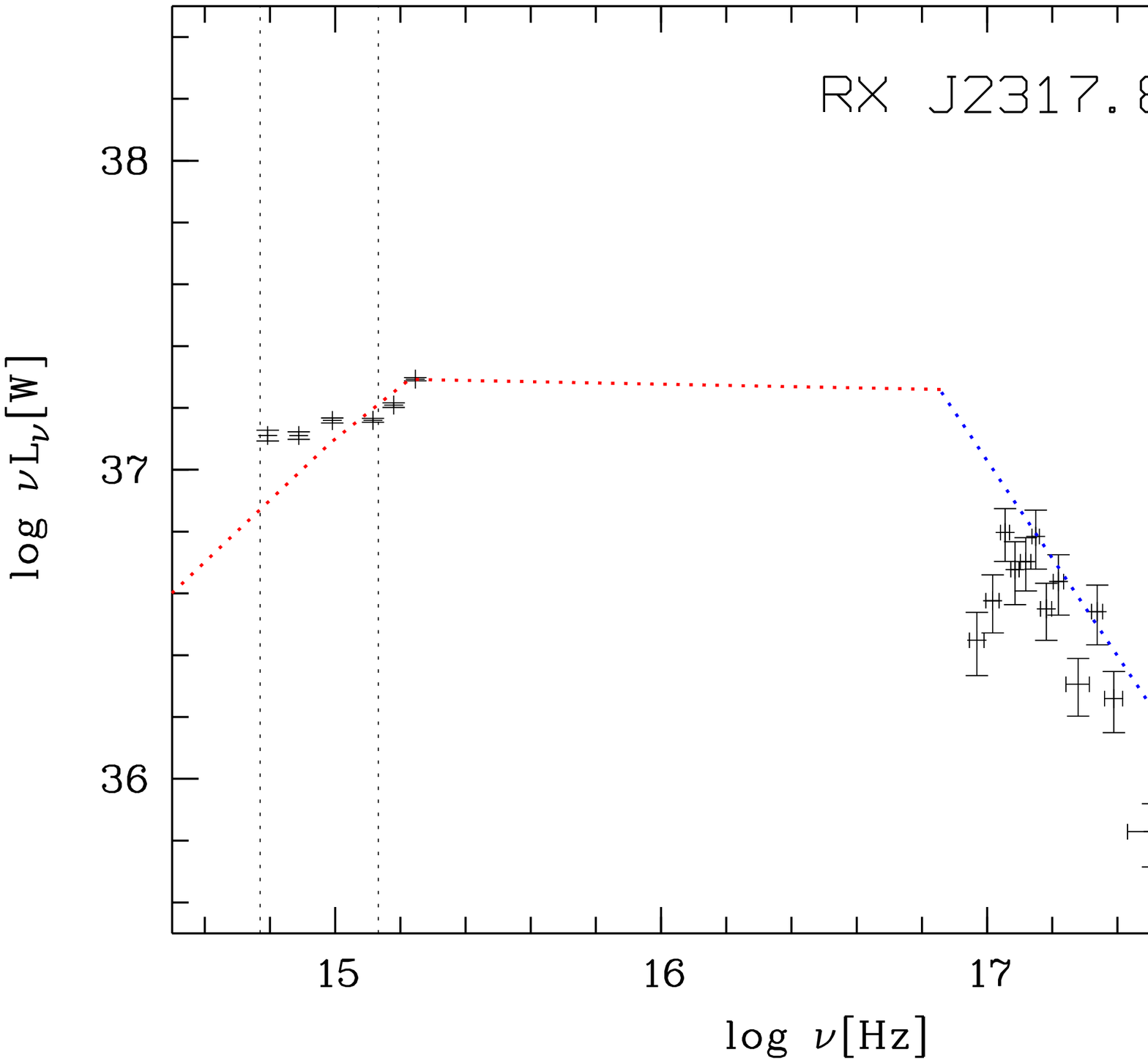}

\plotthree{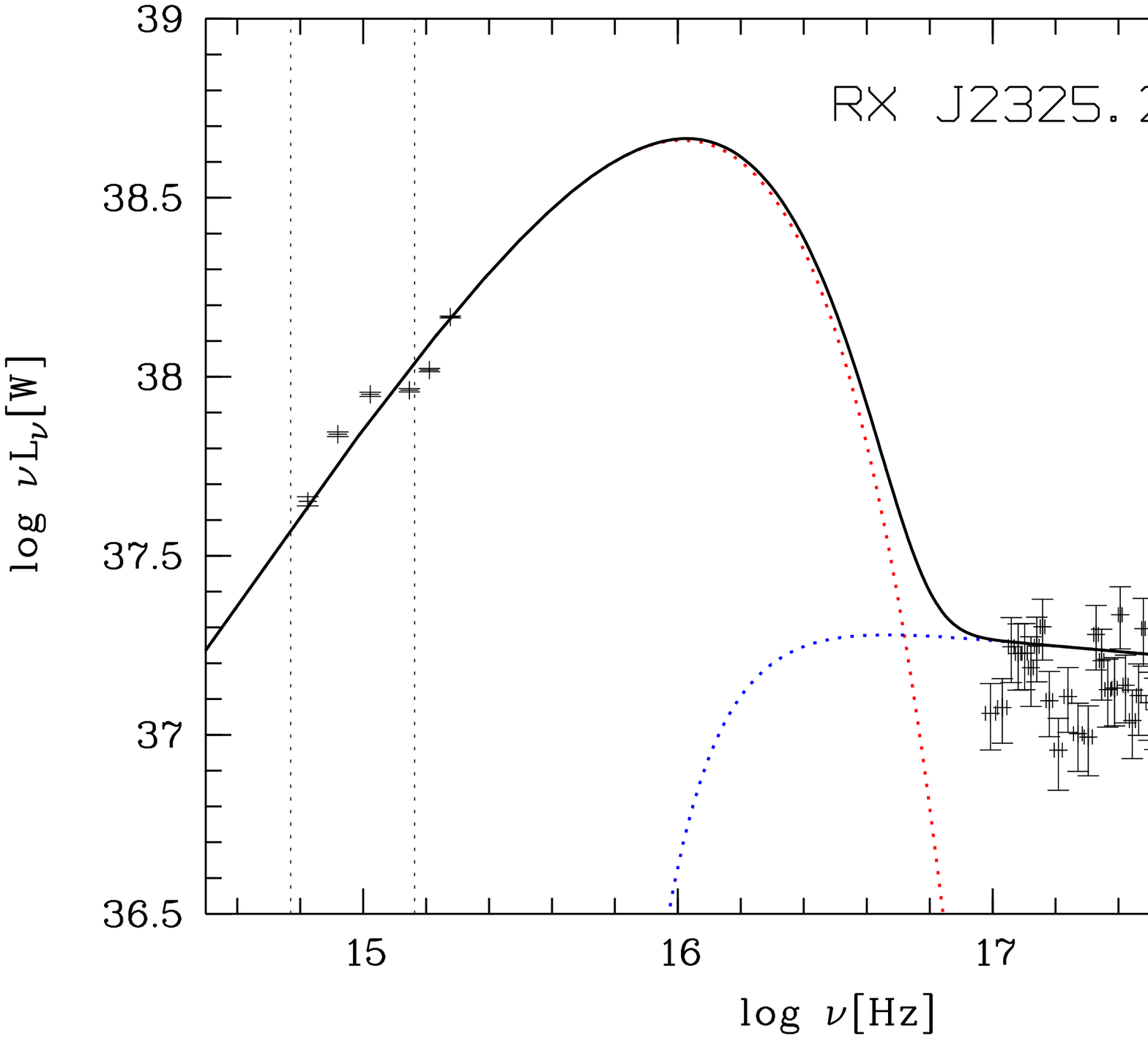}{f25_t.ps}{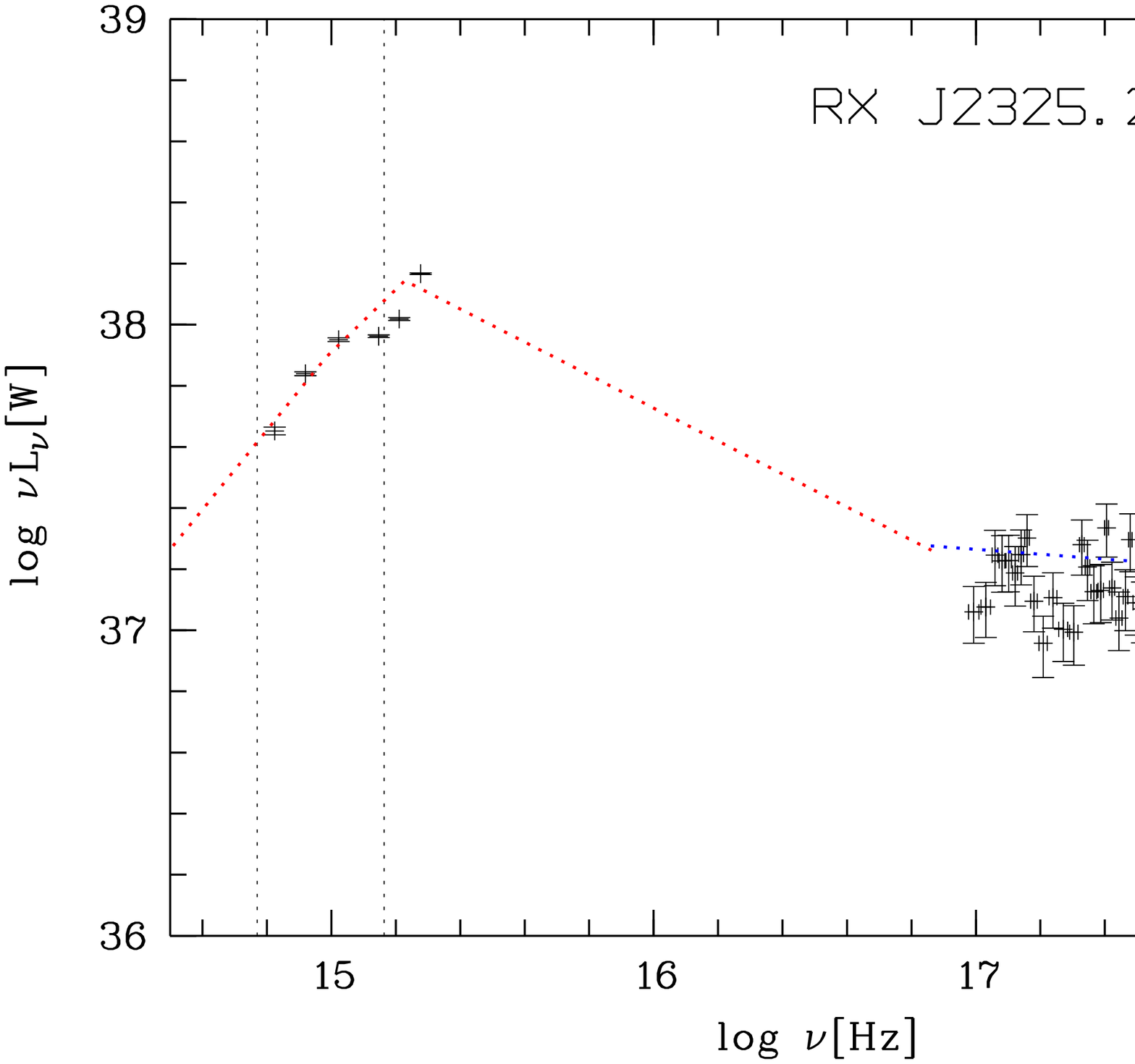}

\plotthree{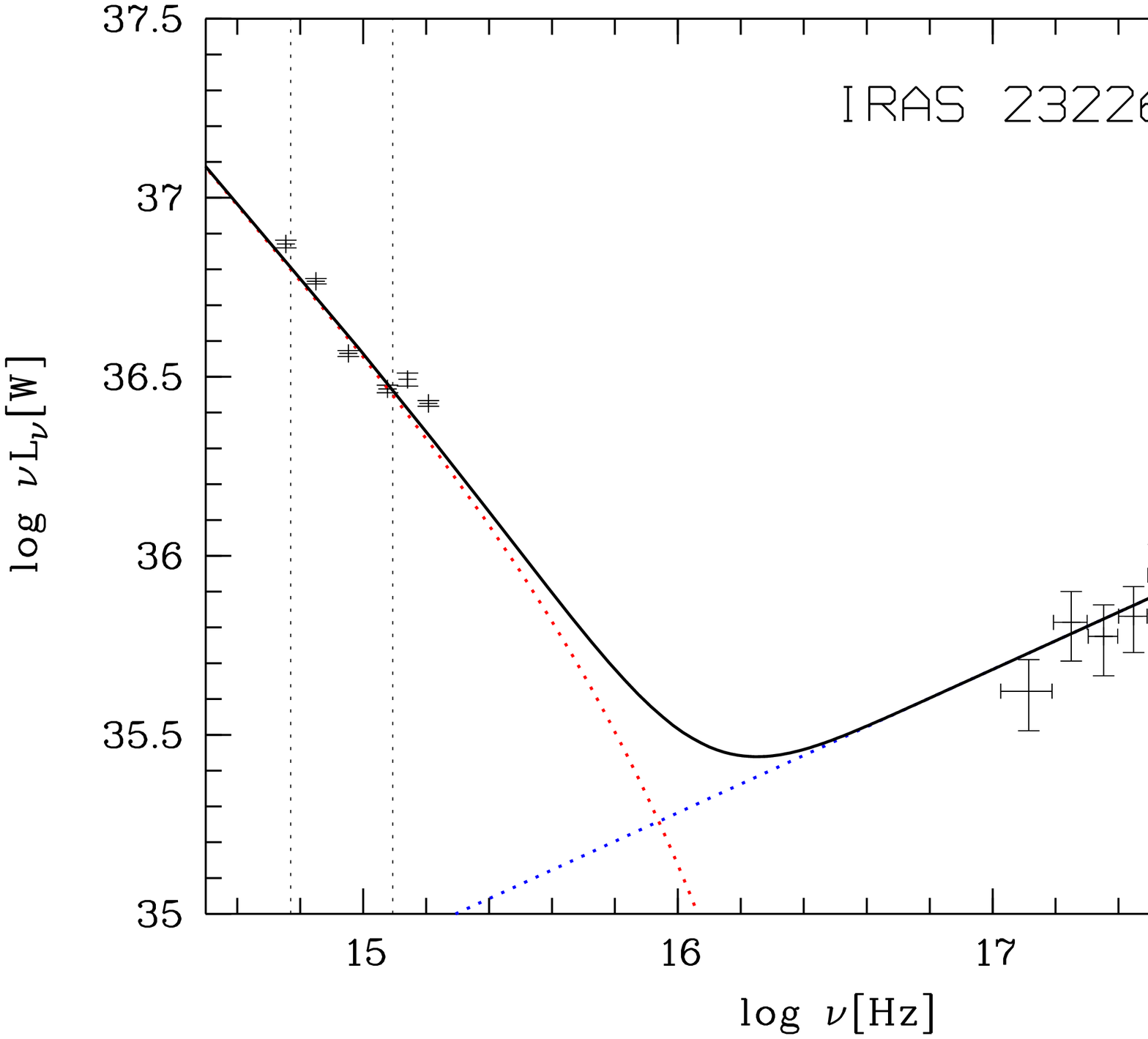}{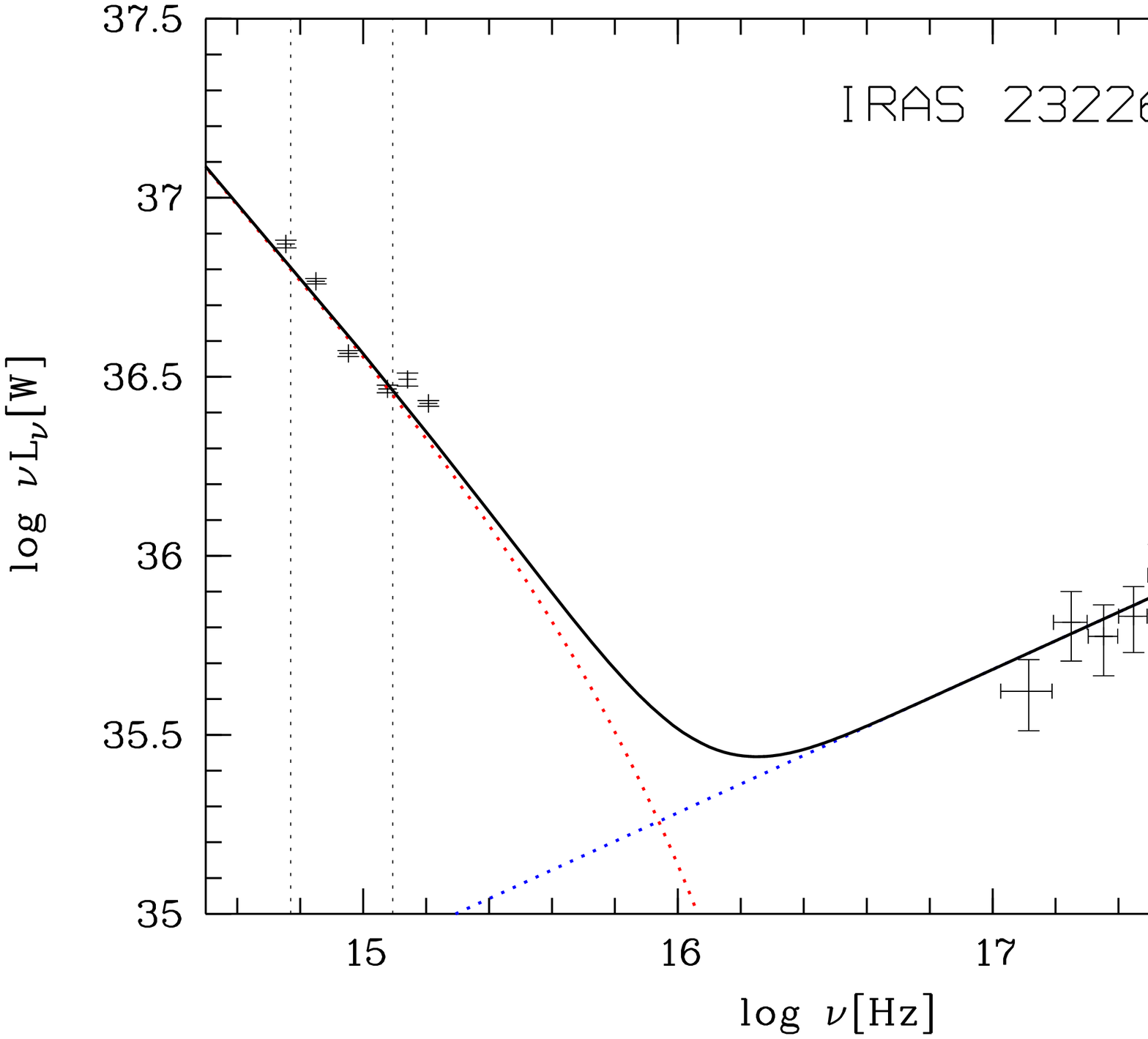}{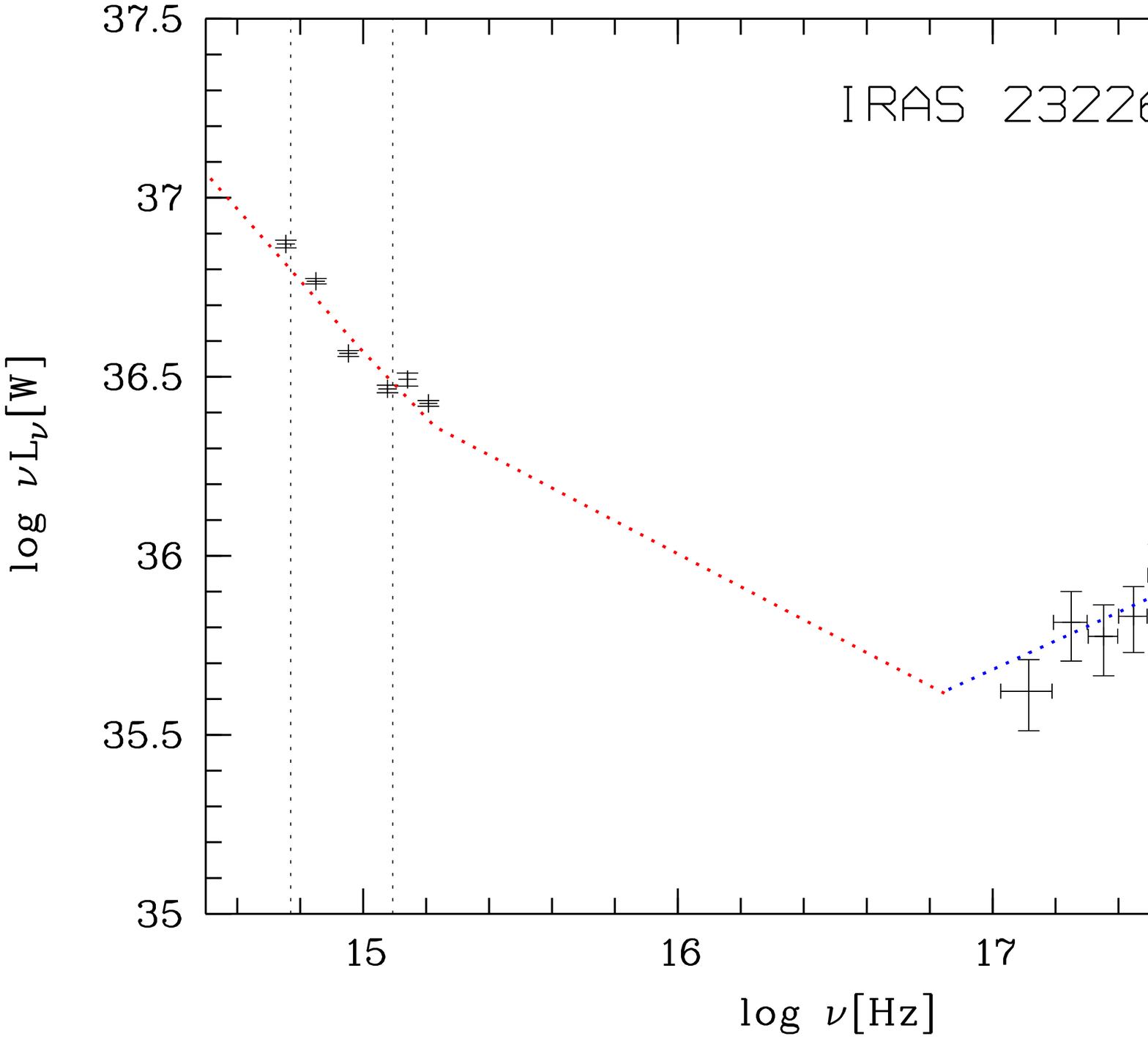}

\plotthree{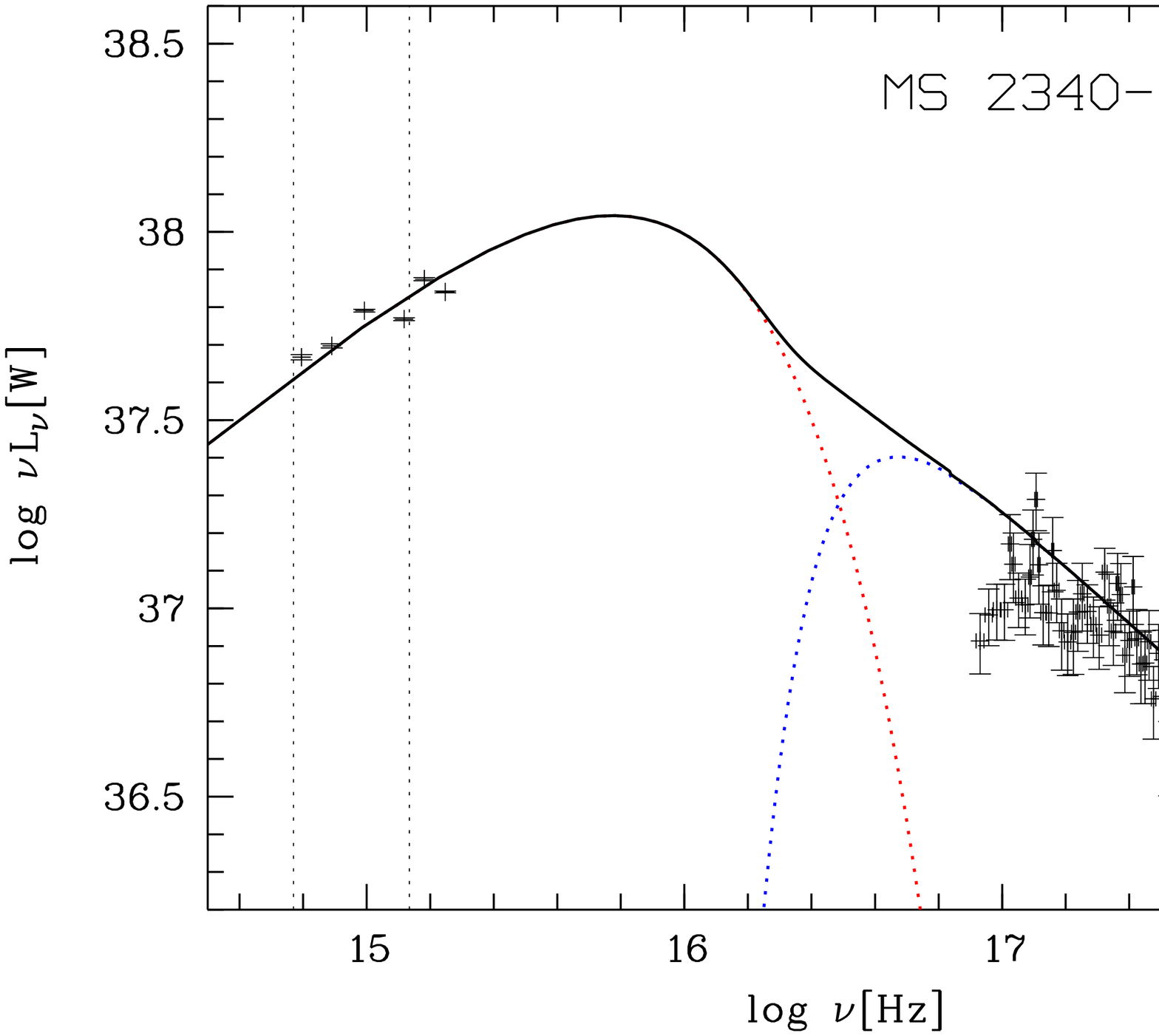}{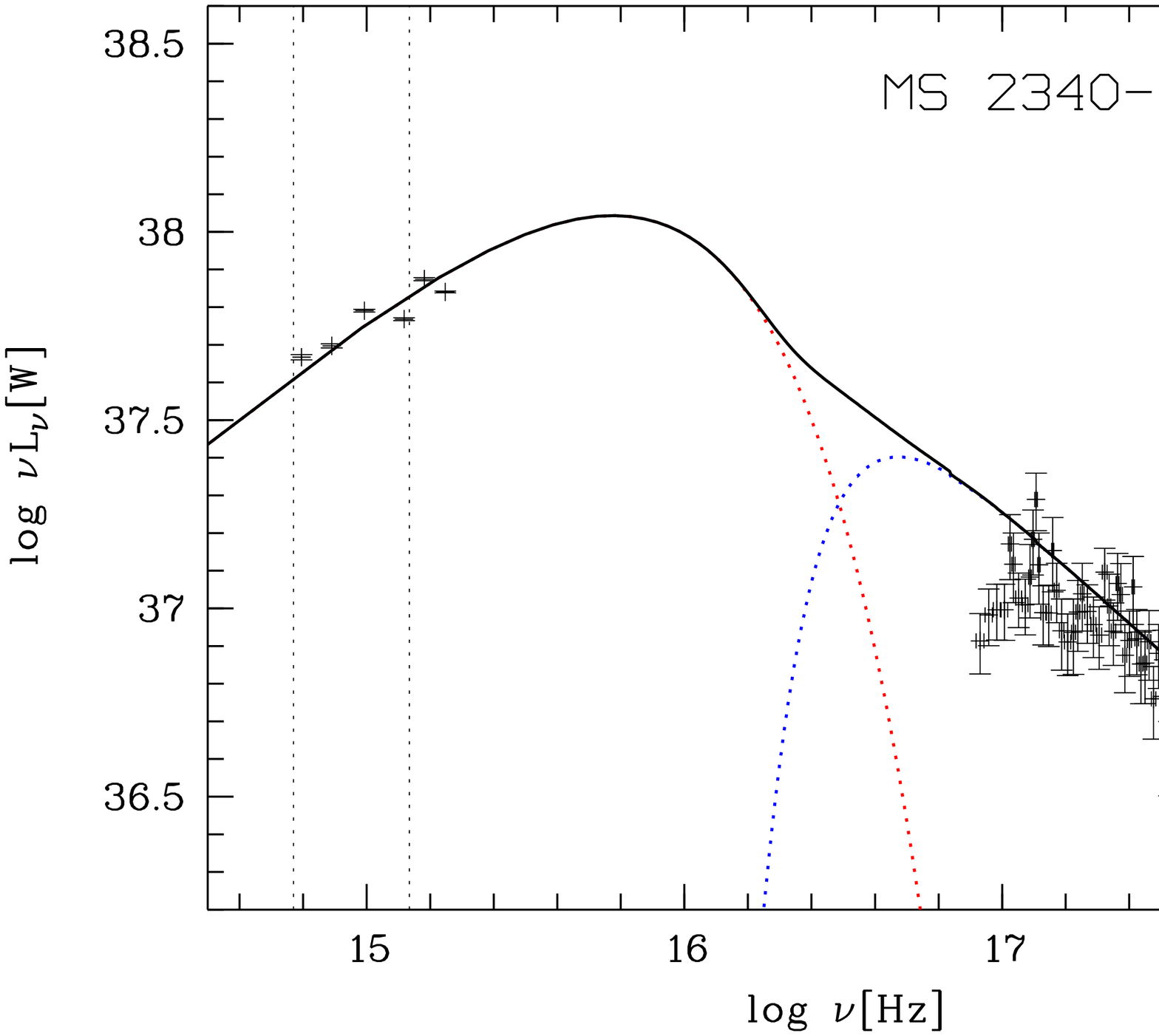}{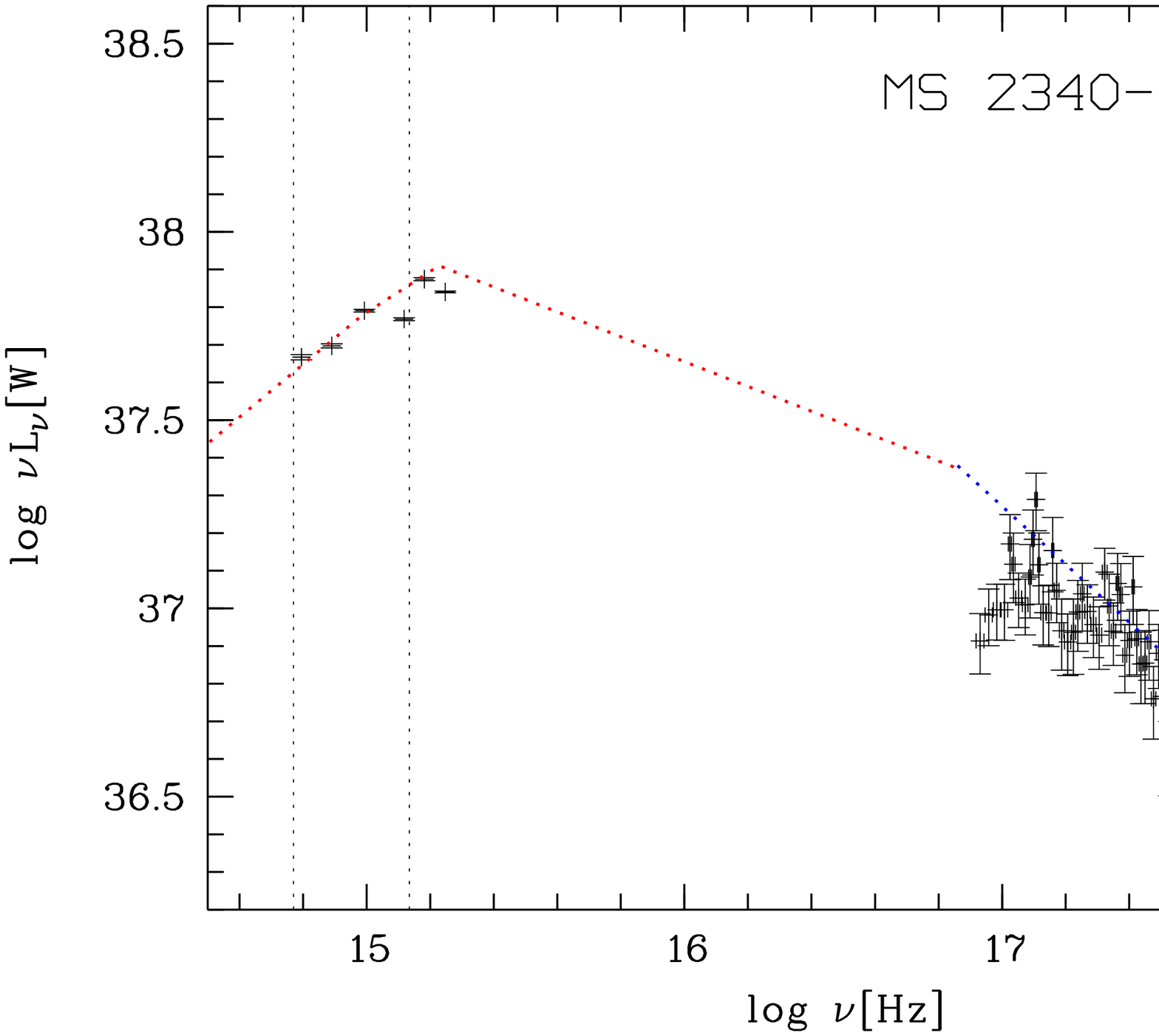}
\end{figure*}

\clearpage

\begin{figure*}
\epsscale{0.60}

\plotthree{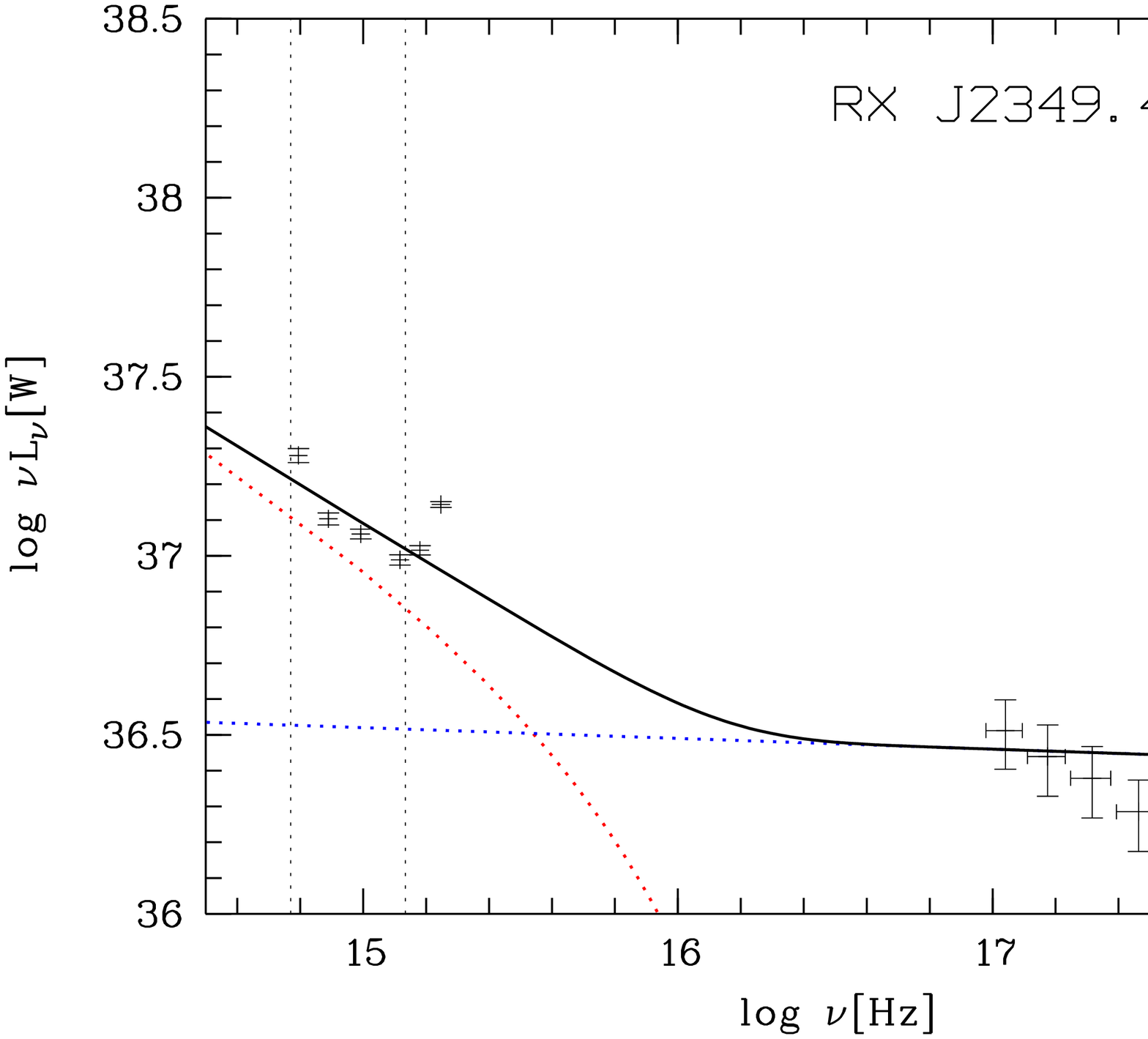}{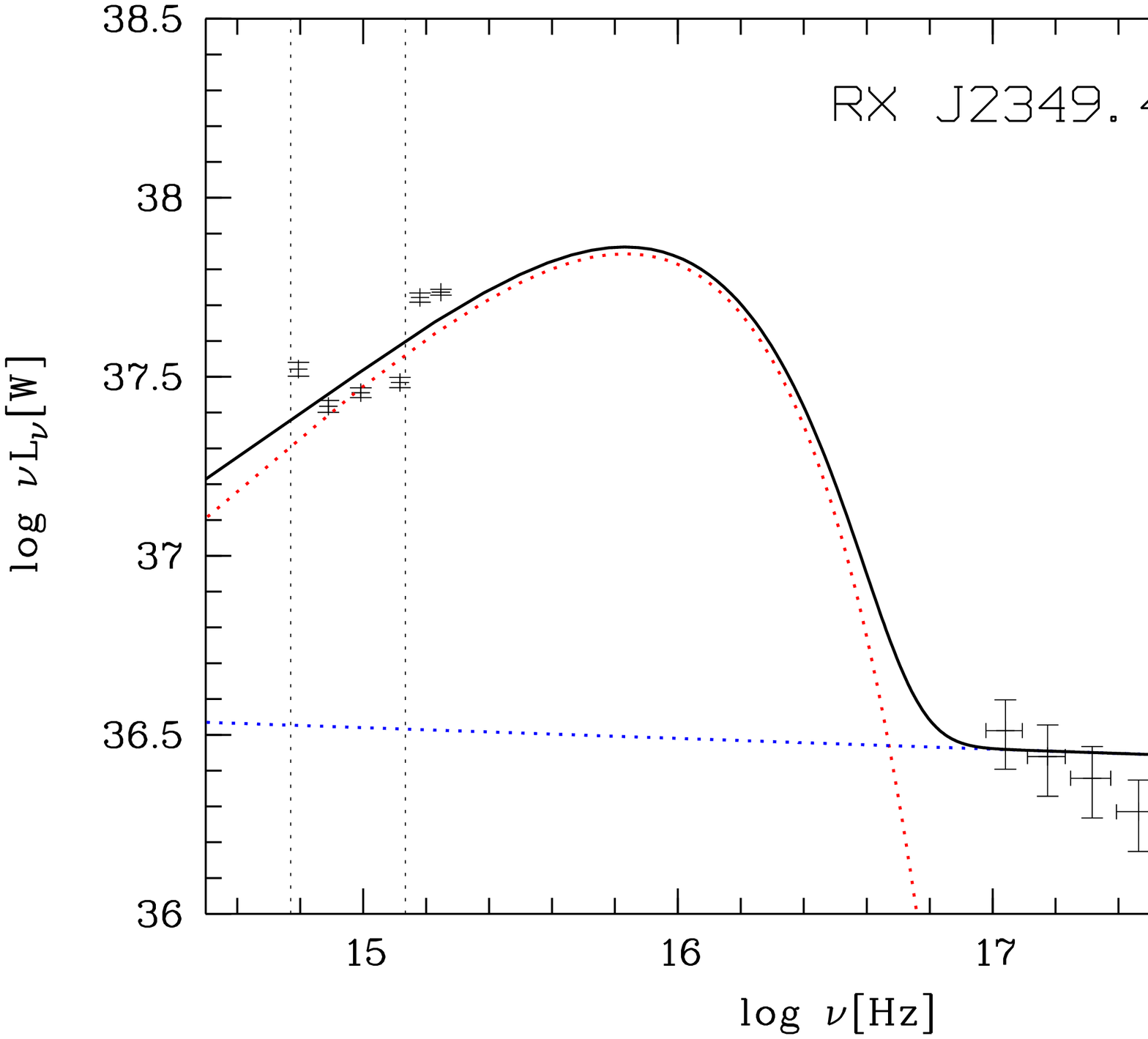}{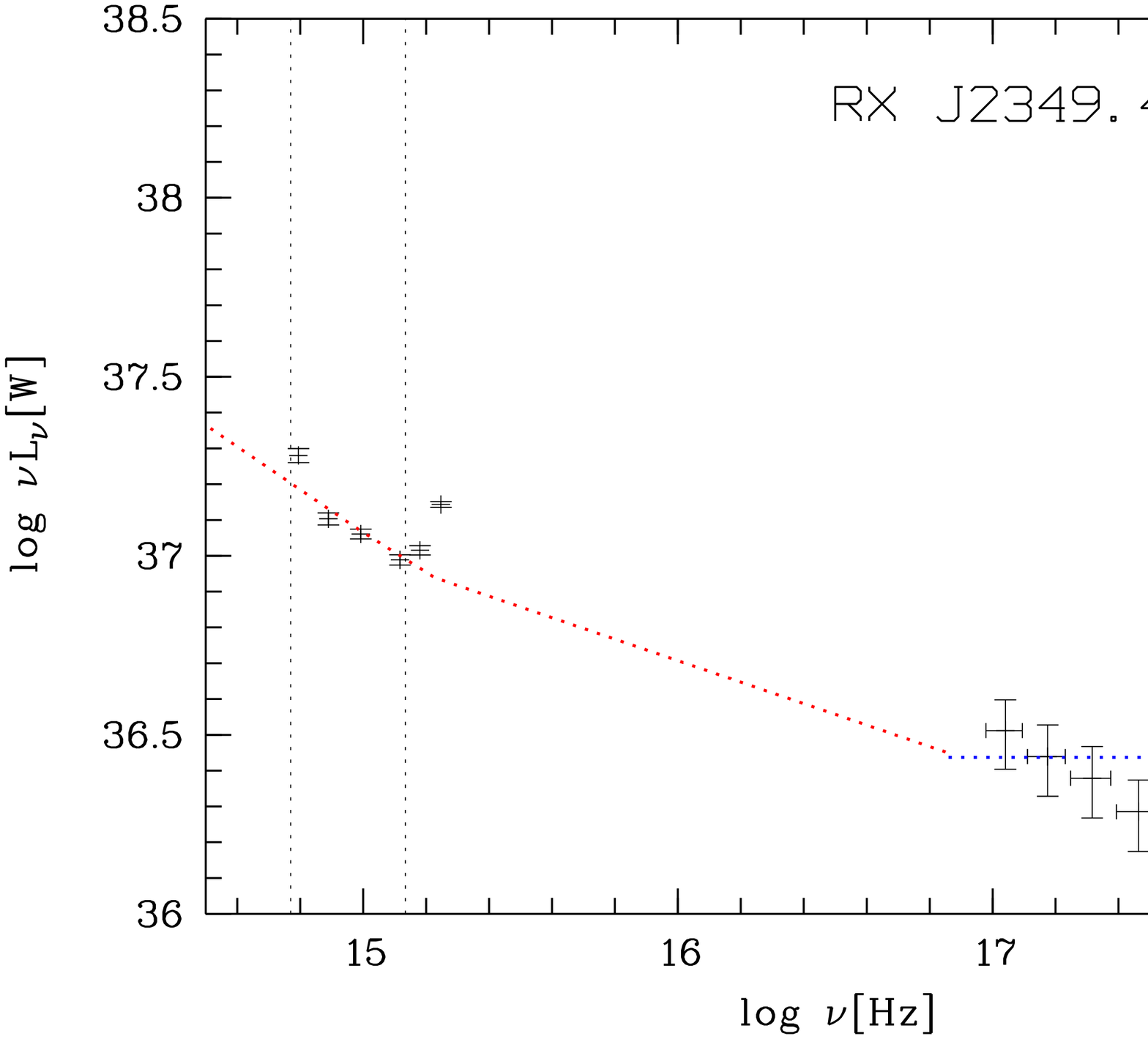}

\plotthree{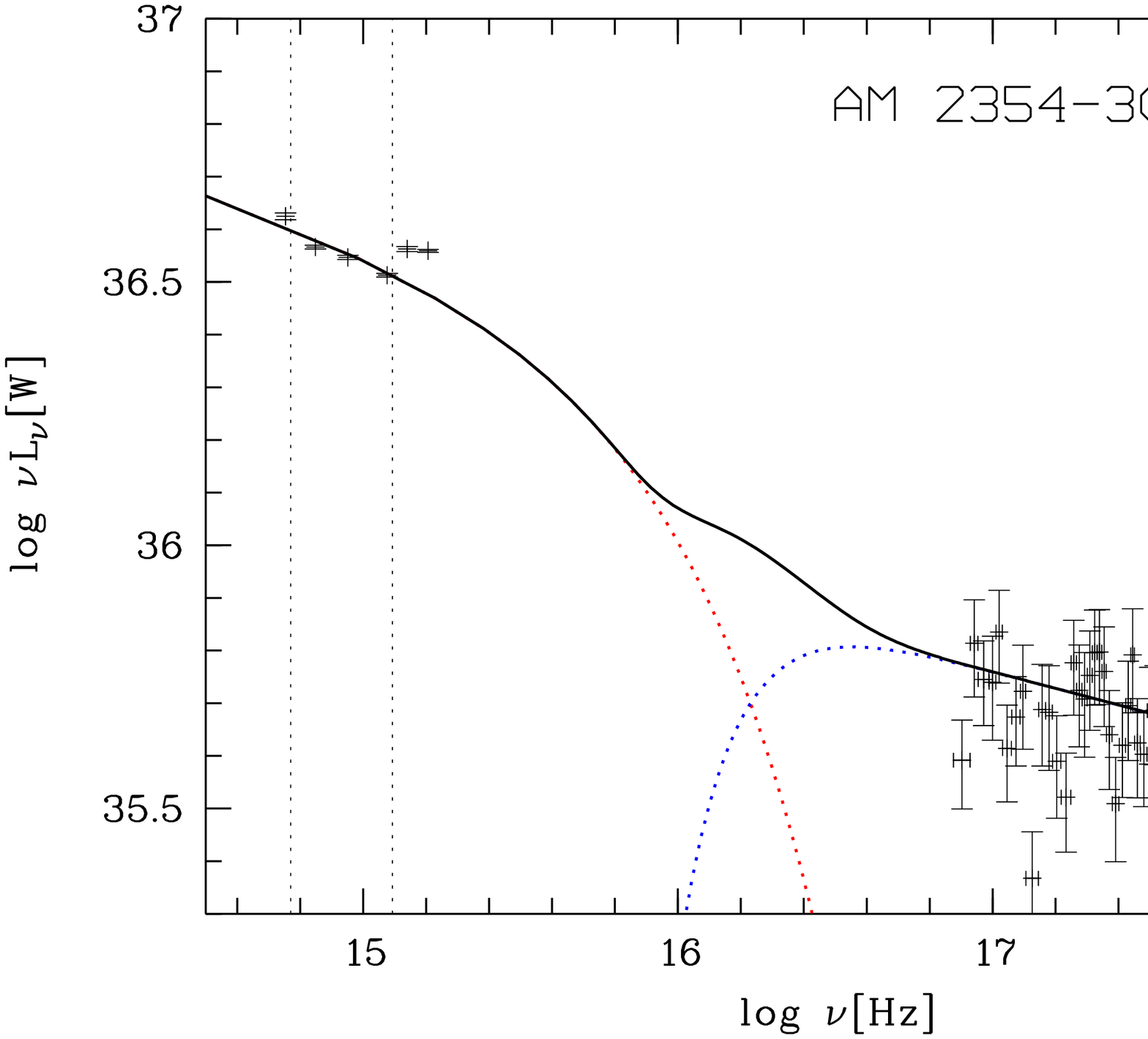}{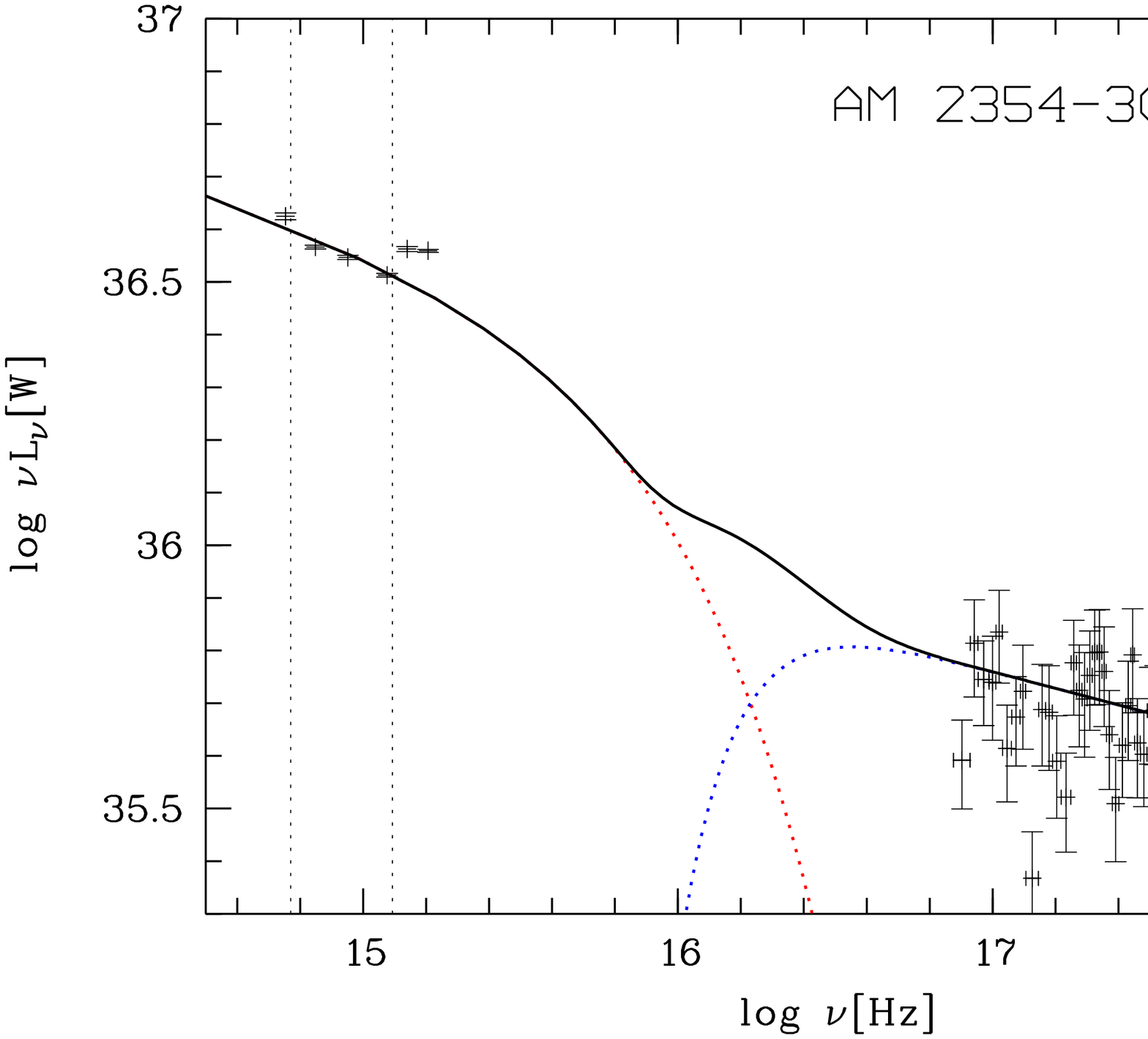}{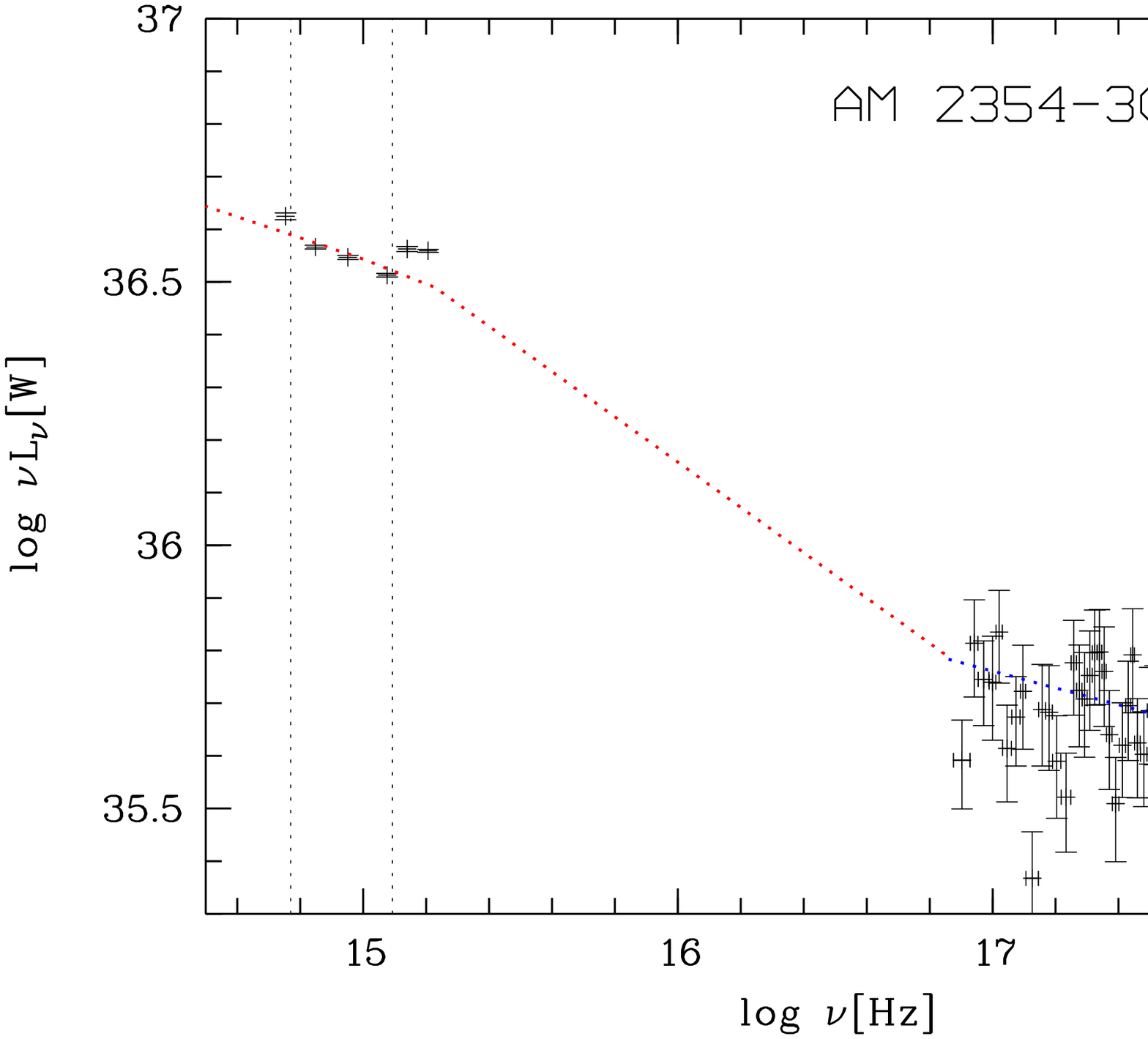}
\end{figure*}

\end{appendix}


\begin{thebibliography}{}
\bibitem[Arnaud et al.(1985)]{arnaud85} Arnaud, K.A., et al., 1985, \mnras, 217, 105
\bibitem[Arnaud (1996)]{arnaud96} Arnaud, K.~A., 1996, ASP
Conf.~Ser.~101: Astronomical Data Analysis Software and Systems V, 101, 17
\bibitem[Atlee \& Mathur(2009)]{atlee09} Atlee, D.W., \& Mathur, S., 2009, ApJ,
703, 1597
\bibitem[Bachev et al.(2009)]{bachev09} Bachev, R., Grupe, D., Boeva, S., Ovcharov, E., 
Valcheva, A., Semkov, E., Georgie, \& Gallo, L.C., 2009, \mnras, 399, 750
\bibitem[Barr \& Mushotzky (1986)]{barr86} Barr, P., \& Mushotzky, R.F., 1986, Nature, 330, 421
\bibitem[Barthelmy (2005)]{barthelmy05} Barthelmy, S.D., 2005, Space Science
Reviews, 120, 143
\bibitem[Bentz et al.(2006)]{bentz06} Bentz, M.C., Peterson, B.M., Pogge, R.W., Vestergaard, M., \& Onken, C.A., 
2006, \apj, 644, 133
\bibitem[Bentz et al.(2009)]{bentz09} Bentz, M.C., Peterson, B.M., Netzer, H., Pogge, R.W., \& Vestergaard, M., 2009, \apj, 697, 160
\bibitem[Beuermann (2008)]{beuermann08} Beuermann, K., 2008, \aap, 481, 919
\bibitem[Beuermann et al. (1999)]{beu99} Beuermann, K., Thomas, H.-C., Reinsch,
K., et al., 1999, \aap, 347, 47
\bibitem[Beuermann et al.(2006)]{beuermann06} Beuermann, K., Burwitz, V., \& Rauch, T., 2006, \aap, 458, 541
\bibitem[Boller et al. (1996)]{bol96} Boller, T., Brandt, W.N., \& Fink, H.H.,
1996, \aap, 305, 53
\bibitem[Boroson \& Green (1992)]{bor92} Boroson, T.A., \& Green, R.F., 1992, \apjs, 80, 109
\bibitem[Boroson (2002)]{boroson02} Boroson, T.A., 2002, \apj, 565, 78
\bibitem[Brandt et al.\ (1996)]{brandt96} Brandt, W.\ N., Fabian, A.\
  C., \& Pounds, K.\ A.\ 1996, \mnras, 278, 326
\bibitem[Brandt et al.(1997)]{brandt97} Brandt, W.N., Mathur, S., \& Elvis, M., 1997, \mnras, 285, L25
\bibitem[Brocksopp et al.(2006)]{brocksopp06} Brocksopp, C., Starling, R.L.C., Schady, P., Mason, K.O., Romero-Comenere, E., \& Puchnarewicz, E.M., 
2006, \mnras, 366, 953
\bibitem[Burrows et al. (2005)]{burrows04} Burrows, D., et al., 2005, Space
Science Reviews, 120, 165
\bibitem[Cardelli et al.(1989)]{cardelli89} Cardelli, J.A., Clayton, G.C., Mathis, J.S., 1989, \apj, 345, 245
\bibitem[Cohen (1983)]{cohen83} Cohen, R.D., 1983, \apj, 273, 489
\bibitem[Crummy et al.(2006)]{crummy06} Crummy, J., Fabian, A.C., Gallo, L.C., \& Ross, R.R., 2006, \mnras, 365, 1067
\bibitem[Czerny \& Elvis (1987)]{czerny87} Czerny, B., \& Elvis, M., 1987, \apj, 321, 305 
\bibitem[Dickey \& Lockman (1990)]{dic90} Dickey, J.M., \& Lockman, F.J., 1990,
\araa, 28, 215
\bibitem[Dietrich et al.(2009)]{dietrich09} Dietrich, M., Mathur, S., Grupe, D., \& Komossa, S., 2009, \apj, 696, 1998
\bibitem[Done et al.(2007)]{done07} Done, C., Sobolewska, M.A., Gierlinski, M., \& Schurch, \mnras 374, L15
\bibitem[Dong et al.(2008)]{dong08} Dong, X., Wang, T., Wang, J., Yuan, W., Zhou, H., Dai, H., \& Zhang, K., 2008,
\mnras, 383, 581
\bibitem[Dunn et al. (2006)]{dunn06} Dunn, J.P., Jackson, B., Deo, R.P.,
Farrington, C., Das, V., \& Crenshaw, D.M., 2006, \pasp, 118, 572
\bibitem[Elvis et al.(1994)]{elvis94} Elvis, M., et al., 1994, \apjs, 95, 1
\bibitem[Fabian et al.(2004)]{fabian04} Fabian, A.C., Miniutti, G., Gallo, L.C., Boller, T., Tanaka, Y., Vaughan, S., \& Ross, R.R., 2004, \mnras, 353, 1071
\bibitem[Fabian et al.(2009)]{fabian09} Fabian A.C., Vasudevan, R.V., Mushotzky, R.F., Reynolds, L.M., \& Winter, C.S., 2009,
\mnras, in press, arXiv:0901.0250
\bibitem[Gallo (2006)]{gallo06} Gallo, L.C., 2006, \mnras, 368, 479
\bibitem[Gallo et al. (2004)]{gallo04} Gallo, L.C., Tanaka, Y., Boller, T.,
Fabian, A.C., Vaughan, S., \& Brandt, W.N., 2004, \mnras, 353, 1064
\bibitem[Gehrels et al. (2004)]{gehrels04} Gehrels, N., et al., 2004, ApJ, 611,
1005
\bibitem[Ghosh et al.(2004)]{ghosh04} Ghosh, K.K., Swartz, D.A., Tennant, A.F., Wu, K., \& Ramsey, B.D., 2004, \apj, 607, L111
\bibitem[Gibson et al.(2008)]{gibson08} Gibson, R.R., Brandt, W.N., \& Schneider, D.P., 2008, 
\apj, 685, 773
\bibitem[Gierlinski \& Done (2004)]{gierlinski04} Gierlinski, M., \& Done, C., 2004, \mnras, 349, L7
\bibitem[Goad \& Page (2006)]{goad06} Goad, M.R., \& Page, K.L., 2006, 
Proceedings of the The X-ray Universe 2005 (ESA SP-604). 26-30 September 2005, 
El Escorial, Madrid, Spain. Editor: A. Wilson, p.621
\bibitem[Godet et al.,(2009)]{godet09} Godet, O., et al., 2009, \aap, 494, 775
\bibitem[Goodrich (1989)]{goodrich89} Goodrich, R.W., 1989, 342, 224
\bibitem[Green et al. (1993)]{green93} Green, A.R., McHardy, I.M., \& Lehto, H.J., \mnras, 265, 664
\bibitem[Grupe (2004)]{grupe04} Grupe, D., 2004, \aj, 127, 1799
\bibitem[Grupe et al. (1998a)]{gru98a} Grupe, D., Beuermann, K., Thomas, H.-C.,
Mannheim, K., \& Fink, H.H., 1998a, A\&A 330, 25
\bibitem[Grupe et al. (1998b)]{gru98b} Grupe, D., Wills, B.J., Wills, D.,
Beuermann, K., 1998b, \aap, 333, 827
\bibitem[Grupe et al.(1999)]{gru99} Grupe, D., Beuermann, K., Mannheim, K.,
\& Thomas, H.-C., 1999, \aap, 350, 805
\bibitem[Grupe et al. (2000)]{grupe00} Grupe, D., Leighly, K.M., Thomas, H.-C., \&
Laurent-Muehleisen, S.A., 2000, \aap, 356, 11
\bibitem[Grupe et al.(2001a)]{gru01} Grupe, D., Thomas, H.-C., \& Beuermann, K.,
2001a, \aap, 367, 470
\bibitem[Grupe et al.(2001b)]{grupe01b} Grupe, D., Thomas, H.-C., \& Leighly, K.M., 2001b
\aap, 369, 450
\bibitem[Grupe et al. (2004a)]{gru04a} Grupe, D., Wills, B.J., Leighly, K.M., \&
Meusinger, H., 2004a, \aj, 127, 156
\bibitem[Grupe et al. (2004b)]{grupe04b} Grupe, D., Leighly, K.M., Burwitz, V.,
Predehl, P., \& Mathur, S., 2004b, \aj, 128, 1524
\bibitem[Grupe  et al. (2004c)]{grupe04c} Grupe, D., Mathur, S., 
\& Komossa, S., 2004c, \aj, 127, 3161
\bibitem[Grupe et al. (2006a)]{grupe06} Grupe, D., Leighly, K.M., Komossa, S.,
Schady, P., O'Brien, P.T., Burrows, D.N., \& Nousek, J.A., 2006a, \aj, 132, 1189
\bibitem[Grupe et al.(2006b)]{grupe06b} Grupe, D., Mathur, S., Wilkes, B., \& Osmer, P., 2006b,
\aj, 131, 55
\bibitem[Grupe et al.(2007a)]{grupe07} Grupe, D., Schady, P., Leighly, K.M., Komossa, S.,
O'Brien, P.T., \& Nousek, J.A., 2007a, AJ, 133, 1988
\bibitem[Grupe et al (2007b)]{grupe07b} Grupe, D., Komossa, S., \& Gallo, L.C., 2007b, 
\apjl, 668, L111
\bibitem[Grupe et al.(2008a)]{grupe08} Grupe, D., Komossa, S., Gallo, L.C., Fabian, A.C., 
Larsson, J., Pradhan, A.K., Xu, D., \& Miniutti, G., 2008a, \apj,
681, 982
\bibitem[Grupe et al(2008b)]{grupe08b} Grupe, D., Leighly, K.M., \& Komossa, S., 2008, \aj, 136, 2343
\bibitem[Hill et al. (2004)]{hill04} Hill, J.E., et al., 2004, SPIE, 5165, 217
\bibitem[Hopkins et al.(2007)]{hopkins07} Hopkins, P.F., Gordon, R.T. \& Hernquist, L. 2007, \apj, 654, 731 
\bibitem[Iwasawa et al.\ (1999)]{iwasawa99} Iwasawa, K., Fabian, A.\ C., \&
  Nandra, K.\ 1999, \mnras, 307, 611
\bibitem[Jin et al.(2009)]{jin09} Jin, C., Done, C., Ward, M., Gierlinski, M., \& Mullaney, J., 2009, \mnras, 
in press, arXiv:0903.4698v1
\bibitem[Just et al.(2007)]{just07} Just, D.., Brandt, W.N., Shemmer, O., Steffen, A.T., Schneider, D.P., Chartas, G, \& Garmire, G.P., 2007,
\apj,665, 1004
\bibitem[Kaspi et al.(2000)]{kaspi00} Kaspi, S., Smith, P. S., Netzer, H.,
 Maoz, D., Jannuzi, B. T., \& Giveon, U., 2000, \apj, 533, 631
\bibitem[Kaspi et al.(2005)]{kaspi05} Kaspi, S., et al. 2005, \apj, 629, 61 
\bibitem[Kelly et al.(2008)]{kelly08} Kelly, B.C., Bechtold, J., Trump, J.R., Vestergaard, M., \& Siemiginowska, A., 
2008, \apjs, 176, 355
\bibitem[Kollatschny et al.(2001)]{kollat01} Kollatschny, W., Bischoff, K., Robinson, E.L., Welsh, W.F., \& Hill, G.J., 2001, \aap, 379, 125
\bibitem[Kollatschny et al.(2006)]{kollat06} Kollatschny, W., Zetzl, M., \& Dietrich, M., 2006, \aap, 454, 459
\bibitem[Komossa \& Fink (1997)]{komossa97} Komossa, S., \& Fink, H.H., 1997, Accretion Disks - New Aspects.
Lecture Notes in Physics, 487, 250
\bibitem[Komossa \& Bade (1998)]{koba98} Komossa, S., \& Bade,
N. 1998, \aap, 331, L49 
\bibitem[Komossa \& Meerschweinchen (2000)]{komossa00} Komossa, S., \& Meerschweinchen, 
J. 2000, \aap, 354, 411
\bibitem[Komossa et al.(2006)]{komossa06} Komossa, S., Voges, W., Xu, D., Mathur, S., Adorf, H.-M., Lemson, G., Duschl, W.J., \& Grupe, D., 2006,
\aj, 132, 531
\bibitem[Krongold et al.(2010)]{krongold09} Krongold, Y, Elvis, M., et al., 
2010, \apj, in press, arXiv:1001.1339
\bibitem[Lawrence \& Papadakis (1993)]{lawrence93} Lawrence, A., \& Papadakis, I., 1993, \apj, 414, L85
\bibitem[Leighly (1999)]{lei99a} Leighly, K.M., 1999, \apjs, 125, 297
\bibitem[Leighly (1999b)]{lei99b} Leighly, K.M., 1999b, \apjs, 125, 317
\bibitem[Leighly et al. (1997)]{leighly97} Leighly, K.M., Kay, L.\ E.,
  Wills, B.\ J., Wills, D.\ \& Grupe, D.\ 1997, \apj, 489L, 137
\bibitem[Leighly et al. (2007)]{leighly07} Leighly, K.M., Halpern, J.P., Jenkins, E.B.,
Grupe, D., Choi, J., \& Prescott, K.B.,  2007, \apj, 663, 103
\bibitem[Leighly et al.(2009)]{leighly09} Leighly, K.M., Hamann, F., Casebeer, D.A., \& Grupe, D., 2009, \apj, 701, 176
\bibitem[Lira et al.(1999)]{lira99} Lira, P., Lawrence, A., O'Brien, P., Johnson, R.A., 
Terlevich, R., \& Bannister, N., 1999, \mnras, 305, 109 
\bibitem[Liu et al.(2010)]{liu09} Liu, Y., et al., 2010, \apj, accepted,
arXiv:1001.0356
\bibitem[Longinotti et al.(2008)]{longinotti08} Longinotti, A.L., Nucita, A., Guainazzi,
M., et al., 2008, \aap, 484, 311
\bibitem[Lusso et al.(2010)]{lusso10} Lusso, E., et al., 2010, \aap, accepted,
arXiv:0912.4166v1 
\bibitem[Mannheim et al(1995)]{mannheim95} Mannheim, K., Schulte, M., Rachen, J., 1995, \aap, 303, L41
\bibitem[Marconi et al.(2004)]{marconi04} Marconi, A., Risaliti, G., Gilli, R., Hunt, L.K., Maiolino, R, \& Salvati, M., 
2004, \mnras, 351, 169
\bibitem[Marconi et al.(2008)]{marconi08} Marconi, A., et al. 2008, \apj, 678, 693 
\bibitem[Markwardt et al. (2005)]{markwardt05} Markwardt, C.B., Tueller, J.,
Skinner, G.K., Gehrels, N., Barthelmy, S.D., \& Mushotzky, R.F., 2005, \apj,
633, L77
\bibitem[Mason et al. (2001)]{mason01} Mason, K.O., et al., 2001, \aap, 365, L36
\bibitem[Mathur (2000)]{mat00} Mathur, S., 2000, \mnras, 314, L17
\bibitem[Nandra et al.(1997)]{nandra97} Nandra, K., George, I.M., Mushotzky, R.F., Turner, T.J., \& Yaqoob, T., 1997, \apj, 476, 70
\bibitem[O'Neill et al.(2005)]{oneill05} O'Neill, P.M., Nandra, K., Papadakis, I.E., \& Turner, T.J., 2005, \mnras, 358, 1405
\bibitem[Osterbrock \& Pogge (1985)]{oster85} Osterbrock, D.E. \& Pogge, R.W., 1985, ApJ, 297, 166
\bibitem[Pacciani et al.(2009)]{pacciani09} Pacciani, L., et al. 2009, \aap, 494, 49
\bibitem[Peterson et al.(2004)]{peterson04} Peterson, B., et al. 2004, \apj, 613, 682  
\bibitem[Pfeffermann et al.(1986)]{pfeffermann86} Pfeffermann, E., Briel, U.G., Hippmann, H., et al., 1986, SPIE, 733, 519
\bibitem[Poole et al.(2008)]{poole08} Poole, T.S., et al., 2008, \mnras, 383, 627
\bibitem[Pounds et al. (1995)]{pounds95} Pounds, K.A., Done, C., \& Osborne, J.P., 
1995, \mnras, 277, L5
\bibitem[Rodr\'iguez-Pascual et al. (1997)]{rodriguez97} Rodr\'iguez-Pascual, P. M.; Mas-Hesse, J. M.; Santos-Lleo, M., 1997, \aap, 327, 72
\bibitem[Roming et al. (2005)]{roming04} Roming, P.W.A., et al., 2005, Space
Science Reviews, 120, 95
\bibitem[Roming et al.(2009)]{roming09} Roming, P.W.A., et al., 2009, \apj, 690, 163
\bibitem[Ross et al.(1992)]{ross92} Ross, R.R., Fabian, A.C., \& Mineshige, S., 1992, \mnras, 58, 189
\bibitem[Schlegel et al.(1998)]{schlegel98} Schlegel, D.~J., Finkbeiner, D.~P.,
\& Davis, M.\ 1998, \apj, 500, 525
\bibitem[Schmidt (1963)]{schmidt63} Schmidt, M., 1963, Nature, 197, 1040
\bibitem[Shemmer et al.(2008)]{shemmer08} Shemmer, O., Brandt, W.N., Netzer, H., Maiolino, R., \& Kaspi, S., 2008, \apj, 682, 81
\bibitem[Shields (1978)]{shields78} Shields, G.A., 1978, Nature, 272, 706
\bibitem[Strateva et al. (2005)]{strateva05} Strateva, I.V., Brandt, W.N.,
Schneider, D.P., Vanden Berk, D.G., \& Vignali, C., 2005, \aj, 130, 387
\bibitem[Sulentic et al. (2000)]{sulentic00} Sulentic, J.W., Zwitter, T., Marziani,
P., \& Dultzin-Hacyan, D., 2000, \apj, 536, L5
\bibitem[Tananbaum et al. (1979)]{tananbaum79} Tananbaum, H., et al., 1979,
\apj, 234, L9
\bibitem[Thomas et al. (1998)]{thomas98} Thomas, H.-C., Beuermann, K., Reinsch, K.,
et al., 1998, \aap, 335, 467
\bibitem[Tueller et al.(2009)]{tueller09} Tueller, J., et al., 2009, \apjs, accepted, arXiv:0903.3037v1
\bibitem[Turner et al.(2001)]{tur01} Turner, M.J.L., Abbey, A., Arnaud, M., et
al., 2001, \aap, 365, L27
\bibitem[Turner \& Miller (2009)]{turner09} Turner, T.J., \& Miller, L., 2009, A\&ARv, 17, 47
\bibitem[Vasudevan \& Fabian(2009a)]{vasudevan08} Vasudevan, R.V., \& Fabian, A.C., 2009, \mnras, 392, 1124
\bibitem[Vasudevan et al.(2009b)]{vasudevan09} Vasudevan, R.V., Mushotzky, R.F.,
 Winter, L.M., \& Fabian, A.C., 2009, \mnras, 399, 1553
\bibitem[V\'eron-Cetty et al.(2007)]{veron07} V\'eron-Cetty, M.-P., V\'eron, P., Joly, M., \& Kollatschny, W., 2007, \aap, 475, 487
\bibitem[Voges et al. (1999)]{voges99} Voges, W., Aschenbach, B., Boller, T., et
al., 1999, \aap, 349, 389
\bibitem[Walter \& Fink (1993)]{walter93} Walter, R., \& Fink, H.H., 1993, \aap, 274, 105
\bibitem[Walter et al.(1994)]{walter94}  Walter, R., Orr, A.; Courvoisier, T. J.-L.; Fink, H. H.; Makino, F.; 
Otani, C.; Wamsteker, W., 1994, \aap, 285, 119
\bibitem[Wandel \& Petrosian (1988)]{wandel88} Wandel, A., \& Petrosian, V., 1988, \apj, 329, L11
\bibitem[Williams et al. (2004)]{williams04} Williams, R.J., Pogge, R.W., \& Mathur,
S., 2004, \apj, 610, 737
\bibitem[Wills et al. (1992)]{wills92} Wills, B.J., Wills, D., Evans, N.J., Natta,
A., Thompson, K.L., Breger, M., \& Sitko, M.L., 1992, \apj, 400, 96
\bibitem[Winter et al(2009)]{winter09} Winter, L.M., Mushotzky, R.F., Reynolds, C.S., \& Tueller, J., 2009, \apj, 690, 1322
\bibitem[Yuan et al. (1998)]{yuan98} Yuan, W., Siebert, J., Brinkmann, W.,
1998, \aap, 334, 498 
\bibitem[Young et al(2009)]{young09} Young, M., Elvis, M., \& Risaliti, G., 2009, \apjs, 183, 17
\bibitem[Zhou et al.(2006)]{zhou06} Zhou, H., Wang, T., Yuan, W., Lu, H., Dong, X., Wang, J., \& Lu, Y., 2006, \apjs, 166, 128
\end{thebibliography}
\end{document}